\documentclass[11pt]{article}
\newtheorem{theorem}{Theorem}[section]

\newcommand{\argmax}{\mathop\textrm{argmax}\limits}

\usepackage{tabularx}
\usepackage{times}
\usepackage{amsfonts}
\usepackage{graphicx}
\usepackage{amsmath}
\usepackage{amssymb}
\usepackage{algorithmic}
\usepackage{algorithm}
\usepackage{bm}
\usepackage{color}
\usepackage{float}
\usepackage{tabularx}
\usepackage{comment}
\usepackage{here}
\usepackage{multirow}
\usepackage{mwe}
\usepackage[export]{adjustbox}
\usepackage[labelfont=bf]{caption}

\newcommand{\RomanNumeralCaps}[1]{\MakeUppercase{\romannumeral #1}}
\usepackage{authblk}
\usepackage{url}
\title{Detecting Change Signs with Differential MDL Change Statistics for COVID-19 Pandemic Analysis}
\date{\vspace{-3ex}}

\author[1,*]{Kenji Yamanishi}
\author[2,*]{Linchuan Xu}
\author[1]{Ryo Yuki}
\author[1]{Shintaro Fukushima}
\author[1]{Chuan-hao Lin}
\affil[1]{Graduate School of Information Science and Technology, The University of Tokyo, Tokyo, 113-8656, Japan}
\affil[2]{Department of Computing, The Hong Kong Polytechnic University, Hung Hom, Kowloon, Hong Kong}

\affil[*]{yamanishi@mist.i.u-tokyo.ac.jp, linch.xu@polyu.edu.hk}

\begin{document}
\maketitle

\begin{abstract}
We are concerned with the issue of detecting changes and their signs from a data stream. 
For example, when given time series of 
COVID-19 cases in a region, we may raise early warning signals of outbreaks by detecting signs of changes in the cases.
We propose a novel methodology to address this issue.
The key idea is to employ a new information-theoretic notion, which we call the {\em  differential minimum description length change statistics}~(D-MDL), for measuring the scores of change sign.
We first give a fundamental theory for D-MDL.
We then demonstrate its effectiveness using synthetic datasets. 
We apply it to detecting early warning signals of the COVID-19 epidemic. 
We empirically demonstrate that D-MDL is able to raise early warning signals of events such as significant increase/decrease of cases.
Remarkably, for about $64\%$ of the events of significant increase of cases in 37 studied countries, our method can detect warning signals as early as nearly six days on average before the events, buying considerably long time for making responses. We further relate the warning signals to the basic reproduction number $R0$ and the timing of social distancing. 
The results showed that our method can effectively monitor the dynamics of $R0$, and confirmed the effectiveness of social distancing at containing the epidemic in a region.
We conclude that our method is a promising approach to the pandemic analysis from a data science viewpoint.
\end{abstract}

\section{Introduction}
\subsection{Motivation}
We address the issue of detecting changes and their signs 
in a data stream. 
For example, when given time series of the number of COVID-19 cases in a country, we may expect to warn the beginning of an epidemic by detecting changes and their signs.

Although change detection~\cite{page,hinkley,basseville} is a classical issue, it has remained open how signs of changes can be found.
In principle the degree of change at a given time point has been evaluated in terms of the discrepancy measure (e.g.. the Kullback-Leibler~(KL) divergence) between probability distributions of data before and after that time point~\cite{page,ty2006}. 
It is reasonable to think that the differentials of the KL divergence may be related to  signs of change.
This is because the first differential of the KL divergence is a velocity of change while its second differential is an acceleration of change.

The problem is here that in real cases, the KL-divergence and its differentials cannot be exactly calculated since the true distribution is unknown in advance.
A question lies in how we can estimate the discrepancy measure and their differentials from data when the parameter values are unknown.

This paper answers the above question from an information-theoretic viewpoint based on the {\em minimum description length} (MDL) principle~\cite{rissanen78}.
The MDL principle gives a strategy for evaluating the goodness of a probabilistic model in terms of  codelength required for encoding the data where a shorter codelength indicates a better model. 
We apply this principle to change detection where a shorter codelength indicates a more significant change.
Along this idea, we introduce the notion called the {\em differential MDL change statistics}~(D-MDL) for the measure of change signs.
We theoretically and empirically justify this notion, and then apply it to the COVID-19 pandemic analysis using open data sets. 

We are interested in how early our method can detect signs of outbreak of the COVID-19 in a region, and how the timing of social distancing events is evaluated in terms of the signs of outbreak.

\subsection{Significance of This Paper}
The significance of this paper is summarized as follows:

(1) {\em Proposal of D-MDL and its use for change sign detection.} 
We introduce a novel notion of D-MDL as an approximation of the differentials of KL-divergence associated with changes.
We propose algorithms for on-line change sign detection based on D-MDL.

(2) {\em Theoretical and empirical justification of D-MDL.}
We theoretically justify D-MDL in the hypothesis testing of change detection.
We consider the hypothesis tests which are equivalent with D-MDL scoring.
We derive upper bounds on the error probabilities for these tests to show that they converge exponentially to zero as sample size increases. 
The bounds on the error probabilities are used to determine a threshold for raising an alarm with D-MDL.
We also empirically justify D-MDL using synthetic datasets. 
We demonstrate that D-MDL outperforms existing change detection methods in terms of AUC for detecting the starting point of a gradual change.

(3) {\em Applications to COVID-19 pandemic analysis.}
On the basis of the theoretical and empirical advantages of D-MDL,
we apply D-MDL to COVID-19 pandemic analysis. 
We are mainly concerned with 
how early we are able to detect signs of outbreaks or the contraction of the epidemic for individual countries.
The results showed that for about $64\%$ of outbreaks in studied countries, our method can detect signs as early as about six days on average before the outbreaks. Considering the rapid spread, six days can earn us considerably long time for making responses, e.g., implementing control measures \cite{ kucharski2020}. 
Moreover, we analyze relations between the change detection results and social distancing events. One of findings is that
for individual countries, an average of about four changes/change signs detected before the implementation of social distancing correlates a significant decline from the peak of daily new cases by the end of April.

Change analysis is a pure data science methodology, which detects changes only using statistical models without using differential equations  about the time evolution. 
Meanwhile, SIR~(Susceptible Infected Recovered) model~\cite{siroriginal} is a typical simulation method which predicts the time evolution of infected population with physics model-based differential equations. Although the fitness of the SIR model or its variants to COVID-19 data was argued \cite{sir,seir}, the complicated situation of COVID-19 due to virus mutations~\cite{Wisem4857}, international interactions, highly variable responses from authorities, 
etc. does not necessarily make any simulation model perfect. 
Therefore, the basic reproduction number $R0$ \cite{r0} (a term in epidemiology, representing the average number of people who will contract a contagious disease from one person with that disease) estimated from the SIR model may not be precise. We empirically demonstrate that as a byproduct, the dynamics of $R0$ can be monitored by our methodology which only requires the information of daily new cases.
The data science approach then may
give new insights into epidemic analysis.

\vspace*{-0,2cm}
\subsection{Related Work}
There are plenty of work on change detection~\cite{page,hinkley,basseville,ty2006,guralnik,gavalda,fl2007,adams}. 
In many of them, the degree of change has been related to the discrepancy measure for two distributions before and after a time point, such as likelihood ratio, KL-divergence. However, there is no work on relating the differential information such as the velocity of the change to change sign detection.

Most of previous studies in change detection are concerned with detecting {\em abrupt changes}~\cite{basseville}.
In the scenario of concept drift~\cite{gama}, the issues of detecting various types of changes, including {\em incremental changes} and {\em gradual changes}, have been addressed. 
How to find signs of changes has been addressed in the scenarios of volatility shift detection~\cite{volatility}, gradual change detection~\cite{ym2016} and clustering change detection~\cite{hirai2018}. 
However, the notion of differential information  has never been related to change sign detection.

The MDL change statistics has been proposed as a test statistics in the  hypothesis testing for change detection~\cite{ym2016,yf2018}. 
It is defined as the difference between the total codelength required for encoding data for the non-change case and that for the change case at a specific time point $t$.
A number of data compression-based change statistics similar to it have also been proposed in  data mining \cite{keogh,vreeken,leeuwen}.
However, any differential variation of the compression-based change statistics has never been proposed. 

As for COVID-19 analysis, the effect of social distancing in Germany has been evaluated using the framework of change point analysis~\cite{germany}. 
There exist some work on prediction models with recurrent neural networks for COVID-19~(see e.g. \cite{rnn2}).
However, there is no work on machine learning approaches to detecting signs of outbreak for COVID-19.

The preliminary version of this paper appeared in the arxiv: \url{https://arxiv.org/abs/2007.15179}.

\section{Proposed Methods}
\subsection{Definitions of Changes and their Signs}
Let ${\mathcal X}$ be a domain, which is either discrete or continuous. Hereafter let ${\mathcal X}$ be discrete for the sake of the sake of notational simplicity. 
For  a random variable ${\bm x}\in {\mathcal X}$, 
let $p({\bm x};\theta)=p_{_{\theta}}({\bm x})$ be the probability mass function (or the probability density function in the continuous case) specified by a parameter $\theta$. 
Suppose that $\theta$ changes over time.
In the case when $\theta$ gradually changes over time, we are interested in detecting the starting point of that change.

Let us consider the discrete time $t$.
Let $\theta _{t}$ be the parameter value of $\theta$ at time $t$.
Let $D(p||q)$ denote the Kullback-Leibler~(KL) divergence between two probability mass functions $p$ and $q$:
\begin{eqnarray*}
D(p||q)=\sum _{{\bm x}}p({\bm x})\log \frac{p({\bm x})}{q({\bm x})}.
\end{eqnarray*}
We define the $0$th, $1$st, $2$nd change degrees at time $t$  as
\begin{align*}\label{changedegrees}
&\Phi _{t}^{(0)}\buildrel \rm def \over =D(p_{_{\theta _{t}}}||p_{_{\theta _{t-1}}}),\\
&\Phi _{t}^{(1)}\buildrel \rm def \over =\Phi _{t+1}^{(0)}-\Phi _{t}^{(0)}
=D(p_{_{\theta _{t+1}}}||p_{_{\theta _{t}}})-D(p_{_{\theta _{t}}}||p_{_{\theta _{t-1}}}),\\
&\Phi _{t}^{(2)}\buildrel \rm def \over =\Phi _{t}^{(1)}-\Phi _{t-1}^{(1)}=D(p_{_{\theta _{t+1}}}||p_{_{\theta _{t}}})-2D(p_{_{\theta _{t}}}||p_{_{\theta _{t-1}}})+D(p_{_{\theta _{t-1}}}||p_{_{\theta _{t-2}}}).
\end{align*}
When the parameter sequence 
 $\{\theta _{t}: t\in {\mathbb Z}\}$  is known, we can define the degree of changes at any given time point. 
We can think of $\Phi _{t}^{(0)}$ as the degree of change of the parameter value itself at time $t$.
We can think of $\Phi _{t}^{(1)}, \Phi _{t}^{(2)}$ as the {\em velocity of change} and the {\em acceleration of change} of the parameter at time $t$, respectively.
The velocity of change may take a higher value at the starting point of change (see Fig.\ref{symptom} in Section 6 in the supplementary material).
From this viewpoint we define the {\em $\alpha$th change sign degree} as $\Phi _{t}^{(\alpha)}\ (\alpha =1,2,\dots )$. 
However, the parameter values are not known in advance. The problem is how we can define the degree of changes when the true distributions are unknown.

\subsection{Differential MDL Change Statistics}
In the case where the true parameter value is unknown, the {\em  MDL change statistics} has been proposed to measure the change degree \cite{ym2016,yf2018} from a given data sequence.
Below we denote $x_{a},\dots , x_{b}=x_{a}^{b}$. In the case of $a=1$, we may drop off $a$ and write it as $x^{b}$. 

When the parameter $\theta$ is unknown, we may estimate it as $\hat{\theta}$ using the maximum likelihood estimation method  from a given sequence $x^{n}$.
I.e., $\hat{\theta}=\argmax _{\theta}p(x^{n};\theta).$
Note that the maximum likelihood function $p(x^{n};\hat{\theta})$ does not form a probability distribution of $x^{n}$ because $\sum _{x^{n}}p(x^{n};\hat{\theta})>1$. Thus we construct a {\em normalized maximum likelihood}~(NML) distribution ~\cite{rissanen1996} by 
\begin{eqnarray*}
p_{_{\rm NML}}(x^{n})\buildrel \rm def \over =\frac{\max _{\theta}p(x^{n};\theta )}{\sum _{y^{n}}\max _{\theta}p(y^{n};\theta )}
=\frac{\max _{\theta}p(x^{n};\theta )}{C_{n}}
\end{eqnarray*}
and consider the logarithmic loss for $x^{n}$ relative to it by
\begin{eqnarray}\label{nmllength}
L_{\rm NML}(x^{n})\buildrel \rm def \over =-\log p_{_{\rm NML}}(x^{n}),
\end{eqnarray}
which we call the {\em NML codelength}~\cite{rissanen1996},
where log means the natural logarithm and $C_{n}$ is called the {\em parametric complexity} defined as
\begin{eqnarray}\label{pc}
C_{n}\buildrel \rm def \over =\sum _{x^{n}_{1}}\max _{\theta }p(x^{n}_{1};\theta ).
\end{eqnarray}
It is known \cite{shtarkov} that (\ref{nmllength}) is the optimal codelength that achieves the Shtarkov's minimax regret in the case where the parameter value is unknown.  
It is known \cite{rissanen1996} that under some regularity condition for the model class, $C_{n}$ is asymptotically expanded as follows: 
\begin{eqnarray}\label{asymp}
C_{n}=\frac{d}{2}\log \frac{n}{2\pi}+\log \int \sqrt{|I(\theta )|}d\theta +o(1),
\end{eqnarray}
where $I(\theta )$ is the Fisher information matrix defined by $I(\theta )=\lim _{n\rightarrow \infty}\frac{1}{n}E_{\theta}[-\frac{\partial ^{2}\log p(X^{n}; \theta )}{\partial \theta \partial \theta ^{\top}}]$, $d$ is the dimensionality of $\theta$, and $\lim _{n\rightarrow \infty}o(1)=0$.

According to \cite{ym2016}, the {\em MDL change statistics} at time point $t$ is defined as follows:
\begin{eqnarray}\label{mdlcs}
\Psi _{t}^{(0)}&\buildrel \rm def \over =&\frac{1}{n}\{L_{_{\rm NML}}(x^{n}_{1})-
(L_{_{\rm NML}}(x^{t}_{1})+L_{_{\rm NML}}(x_{t+1}^{n}))\}.
\end{eqnarray}
The MDL change statistics is the difference between the NML codelength of a given data sequence for non-change and that for change at time $t$. It is a generalization of the likelihood ratio test~\cite{page}.

Therefore, by extending the change degrees $\Phi _{t}^{(0)}, \Phi _{t}^{(1)}, \Phi _{t}^{(2)},\dots $ to the cases where the true parameters are unknown, 
we newly introduce the following statistics:
\begin{eqnarray}
\Psi _{t}^{(1)}&\buildrel \rm def \over =&\Psi _{t+1}^{(0)}-\Psi _{t}^{(0)},\label{1stdiff}\\
\Psi _{t}^{(2)}&\buildrel \rm def \over =&\Psi _{t}^{(1)}-\Psi _{t-1}^{(1)}=\Psi _{t+1}^{(0)}-2\Psi _{t}^{(0)}+\Psi _{t-1}^{(0)},
\label{2nddiff}
\end{eqnarray}
$\Psi _{t}^{(\alpha)}$ corresponds to $\Phi _{t}^{(\alpha)}$.
We call $\Psi _{t}^{(\alpha)}$ the $\alpha$th {\em differential MDL change statistics}, abbreviated as the $\alpha$th D-MDL ($\alpha=0,1,2)$.
We think of  $\Psi _{t}^{(\alpha)}$ as the {\em $\alpha$th change sign degree} estimated from data.

For example, consider the uni-variate Gaussian distribution:
\begin{eqnarray}\label{gaussd}
p(x; \theta )=\frac{1}{\sqrt{2\pi}\sigma }\exp\left( -\frac{(x-\mu )^{2}}{2\sigma ^{2}}\right),
\end{eqnarray}
where $x\in {\mathbb R}$ and $\theta =(\mu , \sigma )$.
We assume $|\mu |< \mu _{\max}$ and $\sigma _{\min}<\sigma <\sigma _{\max}$
where  $\mu _{\max}<\infty$, $0<\sigma _{\min}, \sigma _{\max}<\infty $ are hyper parameters.
The $0$th D-MDL at time $t$ is calculated as
\begin{eqnarray}\label{gauss0}
  \Psi_t^{(0)}=\frac{1}{2n} \log \frac{\hat\sigma_0^{n}} {\hat\sigma_1^{t} \hat\sigma_2^{n-t}}
  + \frac{1}{n}\log \frac{C_{n}}{C_{t}C_{n-t}},
\end{eqnarray}
where $\hat\sigma_0, \hat\sigma_1$ and $\hat\sigma_2$ denote
the maximum likelihood~(ML) estimators calculated for $x_{1}^{n}, x_{1}^{t}$ and $x_{t+1}^{n}$, respectively.
$C_n$ is the normalizer of the NML, which is  calculated according to the study \cite{ym2016}, as
\[
  \log C_n=
  \frac12 \log \frac{16|\mu|_\mathrm{max}}{\pi\sigma^2_\mathrm{min}}
  +\frac{n}{2}\log \frac{n}{2\mathrm{e}}
  -\log \Gamma\left(\frac{n-1}2\right).
\]
The 1st and 2nd D-MDL are calculated according to  (\ref{1stdiff}) and  (\ref{2nddiff}) on the basis of (\ref{gauss0}).

\subsection{Hypothesis Testing for Change Detection}

\subsubsection{The $0$th D-MDL test}
We give rationale of D-MDL using the framework of hypothesis testing for change detection.
First suppose that a change point exists at $t$ or not. 
Let us consider the following hypothesis testing framework:
The null hypothesis $H_{0}$ is that there is no change point while the composite hypothesis $H_{1}$ is that $t$ is an only change point.
\begin{eqnarray*}
\begin{cases}
H_{0}: &x^{n}_{1}\sim p(X^{n};\theta _{0}),\\
H_{1}: &x_{1}^{t}\sim p(X^{t};\theta _{1}),\ \ x_{t+1}^{n}\sim p(X^{n-t};\theta _{2}),
\end{cases}
\end{eqnarray*}
where $\theta _{0},\theta _{1},\theta _{2}\ (\theta _{1}\neq \theta _{2})$ are all unknown.

With the MDL principle, the test statistics is given as follows: For an accuracy parameter $\epsilon >0$, 
\begin{eqnarray}
h_{0}(x^{n}; t, \epsilon )\buildrel \rm def \over = 
\Psi _{t}^{(0)}-\epsilon  ,
\end{eqnarray} 
where $\Psi _{t}^{(0)}$ is the $0$th D-MDL as in (\ref{mdlcs}).  
$H_{1}$ is accepted if $h_{0}(x^{n}; t, \epsilon )>0$, otherwise $H_{0}$ is accepted.
We call this test the {\em $0$th D-MDL test}.

We define {\em Type I error probability} (EP1) as the probability that the test accepts $H_1$  although $H_{0}$ is true (false alarm rate) while {\em Type II error probability} (EP2) as the one that the test accepts $H_{0}$ although $H_{1}$ is true (overlooking rate).
The following theorem justifies the use of the $0$th D-MDL in change detection.
\begin{theorem}\label{thm1}{\rm \cite{ym2016}}
Type I and II error probabilities for the $0$th D-MDL test are upper bounded as follows:
\begin{eqnarray}
EP1&< &\exp \left[ -n\left( \epsilon  -\frac{\log C_{n}}{n}\right) \right],\label{type10}\\
EP2&\leq &\exp\left[-n\left( d(p_{_{\rm NML}},p_{_{\theta _{1}*\theta _{2}}})-\frac{\log C_{t}C_{n-t}}{2n}-
\frac{\epsilon}{2} \right) \right],\label{type20}
\end{eqnarray}	
where $C_{n}$ is the parametric complexity as in (\ref{pc}) and 
\begin{eqnarray}
\hspace*{-0.5cm}
& &d(p,q)\buildrel \rm def \over =-\frac{1}{n}\log \left(\sum _{x^{n}}(p(x^{n})q(x^{n}))^{\frac{1}{2}}\right),\label{bhac}\\
\hspace*{-0.5cm}
& &p_{_{\rm NML}}(x^{n})=\frac{\max _{\theta }p(x^{n};\theta )}{\sum _{y^{n}}\max _{\theta}p(y^{n};\theta )},\ \ 
p_{_{\theta _{1}*\theta _{2}}}(x^{n})=p(x^{t}_{1};\theta _{1})p(x_{t+1}^{n};\theta _{2}).\nonumber
\end{eqnarray}
\end{theorem}
This theorem shows that Type I and II error probabilities in  (\ref{type10}) and (\ref{type20}) converge to zero exponentially in $n$ as $n$ increases for some appropriate $\epsilon$. We see that the error exponents depend on the parametric complexities of the model class as well as the Bhattacharyya distance in  (\ref{bhac}) between the null and composite hypotheses.
In this sense the $0$th MDL test is effective in change point detection. 

\subsubsection{The $1$st D-MDL test}
Next we give a hypothesis testing setting equivalent with the $1$st D-MDL scoring.
We consider the situation 
where a change point exists at time either $t$ or $t+1$.
Let us consider the following hypotheses: The null hypothesis $H_{0}$ is that the change point is $t$ while the composite one $H_{1}$ is that it is $t+1$.
\begin{eqnarray*}
\begin{cases}
H_{0}: &x^{t}_{1}\sim p(X^{t};\theta _{0}),\ \ x^{n}_{t+1}\sim p(X^{n-t};\theta _{1}), \\
H_{1}: &x_{1}^{t+1}\sim p(X^{t+1};\theta _{2}),\ \ x_{t+2}^{n}\sim p(X^{n-t-1};\theta _{3}),
\end{cases}
\end{eqnarray*}
where $\theta _{0},\theta _{1},\theta _{2},\theta _{3}\ (\theta _{0}\neq \theta _{1},\ \theta _{2}\neq \theta _{3})$ are all unknown.

We consider the following test statistics: For an accuracy parameter $\epsilon >0$, {\small 
\begin{equation*}\label{mdl1}
h_{1}(x^{n}; t, \epsilon )\buildrel \rm def \over 
=\frac{1}{n}\left\{\left(L_{_{\rm NML}}(x^{t}_{1})+L_{_{\rm NML}}(x_{t+1}^{n})\right)
- (L_{_{\rm NML}}(x_{1}^{t+1})+L_{_{\rm NML}}(x_{t+2}^{n}))\right\}-\epsilon, 
\end{equation*}}
which compares the NML codelength for $H_{0}$ with that for $H_{1}.$
We accept $H_{1}$ if $h_{1}(x^{n}; t, \epsilon )>0$, otherwise we accept $H_{0}$.
We call this test the {\em $1$st D-MDL test}.
We easily see 
\begin{eqnarray}
h_{1}(x^{n}; t, \epsilon )=\Psi _{t}^{(1)}-\epsilon 
=\Psi _{t+1}^{(0)}-\Psi_{t}^{(0)}-\epsilon,\label{hyp1}
\end{eqnarray}
where $\Psi _{t}^{(1)}$ is the $1$st D-MDL.
This implies that the 1st D-MDL test is equivalent with testing whether the 1st D-MDL is larger than $\epsilon$ or not. Thus the basic performance of discrimination with the 1st D-MDL can be reduced to that of the 1st D-MDL test.

The following theorem shows the basic property of the $1$st D-MDL test.
\begin{theorem}\label{thm2}
Type I and II error probabilities for the $1$st D-MDL test are upper bounded as follows:
\begin{align}
&EP1< \exp \left[ -n\left( \epsilon  -\frac{\log C_{t}C_{n-t}}{n}\right) \right],\label{type11}\\
&EP2\leq \exp\left[-n\left( d(p_{_{\rm NML}(t)},p_{_{\theta _{2}*\theta _{3}}})-\frac{\log C_{t+1}C_{n-t-1}}{2n}-
\frac{\epsilon}{2} \right) \right],\label{type21}
\end{align}
where $C_{n}$ is the parametric complexity as in (\ref{pc}), $d$ is the Bhattacharyya distance as in (\ref{bhac}) and 
\begin{eqnarray*}
p_{_{\rm NML}(t)}(x^{n})&=&\frac{\max _{\theta }p(x^{t}_{1};\theta )} {\sum _{y^{t}_{1}}\max _{\theta}p(y^{t}_{1};\theta )}\cdot \frac{\max _{\theta }p(x^{n-t}_{t+1};\theta )}{\sum _{y^{n-t}_{t+1}}\max _{\theta}p(y^{n-t}_{t+1};\theta )},\\
p_{_{\theta _{2}*\theta _{3}}}(x^{n})&=&p(x^{t+1}_{1};\theta _{2})p(x_{t+2}^{n};\theta _{3}).
\end{eqnarray*}
\end{theorem}
(The proof is in Section 4 in the supplementary material.)\\ 
This theorem shows that Type I and II error probabilities in (\ref{type11}) and  (\ref{type21}) converge to zero exponentially in $n$ as $n$ increases where the error exponents are related to the parametric complexities for the hypotheses as well as the Bhattacharyya distance  between the null and composite hypotheses.
Type I error probability in (\ref{type11}) will be used for determining a threshold of the alarm in Sec.2.5.

\subsubsection{The $2$nd D-MDL test}

Next we consider a hypothesis testing setting equivalent with the 2nd D-MDL scoring. 
Suppose that change points exists either at time $t$ or at $t-1$ and $t+1$. 
\begin{eqnarray*}
\begin{cases}
H_{0}: &x^{t}_{1}\sim p(X^{t};\theta _{0}),\  \ x_{t+1}^{n}\sim p(X^{n-t};\theta _{1}),\\
H_{1}: &x^{t-1}_{1}\sim p(X^{t-1};\theta _{2}),\ \ x_{t}x_{t+1}\sim p(X^{2};\theta _{3}), \ x_{t+2}^{n}\sim p(X^{n-t-1};\theta _{4}),
\end{cases}
\end{eqnarray*}
where $\theta _{0},\theta _{1},\theta _{2},\theta _{3},\theta _{4},\ (\theta _{0}\neq \theta _{1}, \theta _{2}\neq \theta_{3}\neq \theta _{4})$ are all unknown.
$H_{0}$ is the hypothesis that a change happens at time $t$ while $H_{1}$ is the hypothesis that two changes happen at time $t-1$ and $t+1.$
In $H_{0}$, $t$ is a single change point while in $H_{1},$ $t$ is an inflection point between two close change points. Thus it tests whether time $t$ is a change point or a transition point of close changes.

The test statistics is: For an accuracy parameter $\epsilon >0$, 
\begin{align}\label{test2nd}
&h_{2}(x^{n}; t, \epsilon )\buildrel \rm def \over = \frac{1}{n}\left\{
\left( L_{_{\rm NML}}(x_{1}^{t})+L_{_{\rm NML}}(x_{t+1}^{n}) \right)\right.\nonumber \\
& \ \ \ \ \ \ \ \ \ -\left. \left( L_{_{\rm NML}}(x_{1}^{t-1})+L_{_{\rm NML}}(x_{t}x_{t+1})+L_{_{\rm NML}}(x_{t+2}^{n}) \right)\right\} -\epsilon .
\end{align}
We accept $H_{1}$  if $h_{2}(x^{n}; t, \epsilon )>0$, otherwise accept $H_{0}$.
We call this test the {\em $2$nd MDL test}.

Under the assumption 
$(1/n)L_{_{\rm NML}}(x^{t+1}_{1})\approx (1/n)( L_{_{\rm NML}}(x_{1}^{t-1})+ L_{_{\rm NML}}(x_{t}x_{t+1}))$ and $ (1/n)L_{_{\rm NML}}(x^{n}_{t})\approx \frac{1}{n}(L_{_{\rm NML}}(x_{t}x_{t+1})+L_{_{\rm NML}}(x_{t+2}^{n})),$
we have
\begin{eqnarray}
\Psi ^{(2)}_{t}\approx 2h_{2}(x^{n};t, \epsilon) +2\epsilon . \label{hyp2}
\end{eqnarray}
This implies that the $2$nd D-MDL test is equivalent with testing whether the 2nd D-MDL is larger than $2\epsilon$ or not. Thus the basic performance of discrimination with the 2nd D-MDL can be reduced to that of the 2nd D-MDL test.

The following theorem shows the basic property of the $2$nd D-MDL test.
\begin{theorem}
Type I and II error probabilities for the 2nd D-MDL test are upper bounded as follows:
\begin{align}
&PE1< \exp \left[ -n\left( \epsilon  -\frac{\log C_{t}C_{n-t}}{n}\right) \right],\label{type12}\\
&PE2\leq \exp\left[-n\left( d(p_{_{\rm NML(t)}},p_{_{\theta _{2}*\theta _{3} *\theta _{4}}})-\frac{\log C_{t-1}C_{2}C_{n-t+1}}{2n}-
\frac{\epsilon}{2} \right) \right],\label{type22}
\end{align}
where $C_{n}$ is the parametric complexity as in (\ref{pc}), $d$ is the Bhattacharyya distance as in (\ref{bhac}) and 
\begin{eqnarray*}
p_{_{\rm NML}(t)}(x^{n})&=&\frac{\max _{\theta }p(x^{t};\theta )} {\sum _{y^{t}}\max _{\theta}p(y^{t};\theta )}\cdot \frac{\max _{\theta }p(x^{n-t};\theta )}{\sum _{y^{n-t}}\max _{\theta}p(y^{n-t};\theta )},\\
p_{_{\theta _{2}*\theta _{3}*\theta _{4}}}(x^{n})&=&p(x^{t-1}_{1};\theta _{2})p(x_{t}x_{t+1};\theta _{3})p(x_{t+2}^{n};\theta _{4}).
\end{eqnarray*}
\end{theorem}
This theorem can be proven similarly with Theorem 2.2.
Type I probability in (\ref{type12}) will be used for determining the threshold of the change sign alarm in Sec.2.5. 

\subsection{Sequential Change Sign Detection with D-MDL}
In the previous sections, we considered how to measure the change sign scores at a specific time point $t$. In order to detect change signs sequentially for the case where there exist multiple change points, 
we can conduct sequential change sign detection using D-MDL in a similar manner with \cite{ym2016}. We give two variants of the sequential algorithms. One is the sequential D-MDL algorithm with {\em fixed windowing} while the other is that with {\em adaptive windowing}.
In the former, we prepare a local window of fixed size to calculate D-MDL at the center of the window. We then slide the window to obtain a sequence of D-MDL change scores as with \cite{ym2016}.
We raise an alarm when the score exceeds the predetermined threshold $\beta$. The algorithm is summarized in Algorithm 1:

\begin{algorithm}
\label{parameterchangedetection}
\caption{
{ Sequential D-MDL algorithm with fixed windowing}
}
\begin{algorithmic}
\STATE
{\bf Given:} $2h$: window size, $T$: data length, 
$\beta$: threshold parameter
\FORALL{$t=h+1, \dots,T-h+1$}
\STATE Input $x_{t-h},\dots ,x_{t+h}.$
\STATE Calculate a change score $\Psi ^{(\alpha )}_{t}=h_{\alpha }(x_{t-h}^{t+h}, h, 0)$ 
$(\alpha =0, 1,2)$ at  $t$ by sliding the window. 
\STATE Make an alarm if and only if $\Psi _{t}^{(\alpha )}>\beta $.
\ENDFOR
\end{algorithmic}
\end{algorithm}


In the study \cite{kmy2017}, the sequential algorithm with adaptive windowing (SCAW) was proposed by combining the $0$th D-MDL with ADWIN algorithm~\cite{gavalda}
where the window grows until the maximum of the MDL change statistics in the window exceeds a threshold. Once it exceeds the threshold, we drop the data earlier than the time point where the maximum is achieved and the window shrinks. Then the process restarts. 
It outputs the size of window whenever a change point is detected.

According to the study \cite{kmy2017}, the threshold $\epsilon _{w}$ for $\Psi^{(0)}$ is set so that the total number of  false alarms is finite. This is set as follows: for some parameter $\delta >0$, when the parameter  is $d$-dimensional, 
\begin{eqnarray}\label{windowsize}
\epsilon _{w}=(2+d/2+\delta )\log w +\log (1/\delta ).
\end{eqnarray}

\subsection{Hierarchical Sequential D-MDL Algorithm}
Practically, we combine the algorithm with adaptive windowing for the 0th D-MDL and the algorithms with fixed windowing for the 1st and 2nd D-MDL.
We call this algorithm the {\em hierarchical sequential D-MDL algorithm}.
It is designed as follows. 
We first output not only a $0$th D-MDL score but also a window size with the $0$th D-MDL with adaptive windowing and raise an alarm when the window shrinks, i.e.,(\ref{windowsize}) is satisfied for some time in the window. We then output the 1st and 2nd D-MDL scores using the window produced by the 0th D-MDL and raise alarms when for some time in the window, the $1$st or $2$nd D-MDL exceeds the threshold so as to expect the 1st and 2nd D-MDL to detect change signs before the window shrinkage. 

In this algorithm, the threshold $\epsilon _{w}^{(1)}$ for the 1st D-MDL $\Psi ^{(1)}_{t}$ is determined so that Type I error probability in (\ref{type11}) is less than the confidence parameter $\delta _{1}$.
That is, from (\ref{type11}) and (\ref{asymp}), letting $\epsilon _{w}^{(1)}=\epsilon w ,$
\begin{eqnarray*}
{\rm Type\ I\ prob,}&<&
\exp(-\epsilon _{w}^{(1)}+(d/2)\log (w/2)\times 2)\leq \delta_{1}.
\end{eqnarray*}
This yields
\begin{equation}\label{threshold1}
    \epsilon_{w}^{(1)}\geq d\log (w/2)+\log (1/\delta _{1}).
\end{equation}
We employ the righthand side of (\ref{threshold1}) as the threshold of an alert of the 1st D-MDL.

The threshold $\epsilon ^{(2)}_{w}=\epsilon w$ for the 2nd D-MDL $\Psi ^{(2)}_{t}$ can also be derived similarly with the 1st one. Note that by  (\ref{hyp2}), the threshold is 2 times the accuracy parameter for the hypothesis testing.
Letting $\delta _{2}$ be the confidence parameter, by (\ref{type12}), Type I error probability is less than $\delta _{2}$ if the following inequality holds:
\begin{equation}\label{threshold2}
    \epsilon_{w}^{(2)}\geq 2(d\log (w/2)+\log (1/\delta _{2})).
\end{equation}
We employ the righthand side of (\ref{threshold2}) as the threshold of an alert of the 2nd D-MDL.
In practice, $\delta_{1}$ and $\delta _{2}$ are estimated from data (see Sec. 4.2).
The hierarchical sequential D-MDL algorithm is summarized in Algorithm 2:

\begin{algorithm}
\label{parameterchangedetection2}
\caption{
Hierarchical Sequential D-MDL Algorithm
}
\begin{algorithmic}
\STATE
{\bf Given:} $T$: data length, $\{\epsilon _{w}^{(\alpha)}\}$: threshold parameters $\delta_0, \delta_1, \delta_2$ 
\STATE $W\leftarrow \emptyset$
\FORALL{$t=1, \dots, T$} 
\STATE $W\leftarrow W\cup x_{t}$ 
\IF 
 {$\max _{t\in W}\Psi ^{(0 )}_{t}=\max _{t\in W}h_{0}(x_{\rm start}^{|W|}, t,0) >\epsilon _{|W|}^{(0)}/|W|$} 
\STATE Drop the data earlier than $\argmax_{t\in W} \Psi_{t}^{(0 )}(W)$ 
\STATE Raise a 0th D-MDL alert{(a change point alert})\\
\ENDIF
\IF
{ for some $t\in W$, $\Psi ^{(1 )}_{t}=h_{1}(x_{\rm start}^{|W|}, t,0) >\epsilon _{|W|}^{(1)}/|W|$}
\STATE  Raise a 1st D-MDL alert ({a velocity change alert})
\ENDIF 
\IF {for some $t\in W$, $\Psi ^{(2 )}_{t}=h_{2}(x_{\rm start}^{|W|}, t,0) >\epsilon _{|W|}^{(2)}/|W|$}
\STATE  Raise a 2nd D-MDL alert {(a acceleration change alert})
\ENDIF
\STATE Output $|W|$.
\ENDFOR
\end{algorithmic}
\end{algorithm}

\section{Result \RomanNumeralCaps{1}: Experiments with Synthetic Data}

\subsection{Datasets}
To evaluate how well D-MDL performs 
for abrupt/gradual change detection, 
we consider two cases; multiple mean change detection and 
multiple variance one. 

In the case of multiple mean change detection, we constructed datasets as follows:
each datum was independently drawn from the Gaussian distribution 
$\mathcal{N}(\mu_{t}, 1)$ where the mean $\mu_{t}$ abruptly/gradually changed over time according to the following rule: In the case of abrupt changes, 
\begin{align*}
\mu_{t} &= 0.3 \sum_{i=1}^{9} (10-i) H(n-1000i),
\label{eq:data_generation_multiple_jumping_mean} 
\end{align*}
where $H(x)$ is the Heaviside step function that takes $1$ if $x> 0$ otherwise $0$. 
In the case of gradual changes, $H$ is replaced with the following continuous function:
\begin{eqnarray*}
S(x) = \begin{cases}
       0 & (x < 0), \\
       x/300 & ( 0 \leq x < 300), \\
       1 & (x \geq 300).
       \end{cases} \nonumber
\label{eq:graduall_changing_mean}
\end{eqnarray*}

In the case of multiple variance change detection, 
each datum was independently drawn from the Gaussian distribution $\mathcal{N}(0, \sigma_{t}^{2})$ where the variance $\sigma _{t}^{2}$ abruptly/gradually changed over time according to the following rule:
In the case of abrupt changes, 
\begin{eqnarray*}
\log{ \sigma_{t} } = 0.1 \sum_{i=1}^{9} (10-i) H(n-1000i).
\label{eq:data_generation_multiple_jumping_variance} 
\end{eqnarray*}
In the case of gradual changes, $H$ is replaced with $S$ as with the multiple mean changes.

We define a {\em sign of a gradual change} as the starting point of that change.
In all the datasets, change points for abrupt changes and change signs for gradual changes were set at nine points: $t=1000$, $2000$, $\dots$, $9000$. 

\begin{table}[!th]
\small
\caption{\textbf{Average AUC scores $\pm$ standard deviation on the synthetic datasets. MMC and MVC stand for multiple-mean-changing and multiple-variance-changing, respectively.}}
\label{table:combination}
\begin{center}
\begin{tabular}{lcccc}
\hline
& \multicolumn{2}{c}{MMC datasets} & \multicolumn{2}{c}{MVC datasets}\\
\hline
      & Abrupt & Gradual  & Abrupt & Gradual \\
\hline
BOCPD     & $0.55 \pm 0.06$  
          & $0.42 \pm 0.04$ 
          & $0.57 \pm 0.02$  
          & $0.35 \pm 0.03$\\
CF        & $0.59 \pm 0.03$  
          & $0.51 \pm 0.03$ 
          & $0.61 \pm 0.02$ 
          & $0.51 \pm 0.02$\\
ADWIN2    & $0.50 \pm 0.00$
          & $0.54 \pm 0.02$ 
          & $0.50 \pm 0.00$  
          & $0.46 \pm 0.02$\\
D-MDL (0th) & ${\bf 0.92 \pm 0.02}$  
          & $0.61 \pm 0.04$ 
          & ${\bf 0.83 \pm 0.03}$
          & $0.52 \pm 0.05$\\
D-MDL (1st) & $0.48 \pm 0.01$  
          & ${\bf 0.62 \pm 0.02}$ 
          & $0.27 \pm 0.02$  
          & ${\bf 0.53 \pm 0.02}$\\
D-MDL (2nd) & $0.49 \pm 0.01$  
          & $0.62 \pm 0.00$ 
          & $0.49 \pm 0.00$ 
          & $0.53 \pm 0.00$\\
\hline
\label{table:synthetic}
\end{tabular}
\end{center}
\end{table}

\subsection{Evaluation Metric}
For any change detection algorithm that outputs change scores for all time points, letting $\beta$ be a threshold parameter, 
we convert change-point scores $\{ s_{t} \}$ 
into binary alarms $\{ a_{t} \}$ as follows:
\begin{eqnarray*}
a_{t} = \begin{cases}
        1 & (s_{t} > \beta), \\
        0 & (\mathrm{otherwise}).
        \end{cases}
\end{eqnarray*}

By varying $\beta$, 
we evaluate the change detection algorithms 
in terms of benefit and false alarm rate defined as follows: 
Let $T$ be a maximum tolerant delay of change detection. 
When the change truly starts from $t^{\ast}$, 
we define {\em benefit} of an alarm at time $t$ as 
\begin{eqnarray*}
b(t; t^{\ast}) =
  \begin{cases}
  1 - \frac{ | t-t^{\ast} | }{T} & ( 0 \leq |t - t^{\ast} | < T), \\
  0 & (\mathrm{otherwise}),
  \end{cases}
\end{eqnarray*}
where $t^{\ast}$ is a change point for abrupt change, while it is a sign for gradual change.

The total benefit of alarm sequence $a_{0}^{n-1}$ is calculated as 
\begin{eqnarray*}
B(a_{0}^{n-1}) = \sum_{k=0}^{n-1} a_{k} b(k; t^{\ast}).
\end{eqnarray*}
The number of {\em false alarms} is calculated as 
\begin{eqnarray*}
N(a_{0}^{n-1}) = \sum_{k=0}^{n-1} a_{k} \Theta (b(k; t^{\ast}) = 0).
\end{eqnarray*}
where $\Theta (t)$ takes 1 if and only if $t$ is true, otherwise $0$. 
We evaluate the performance of any algorithm in terms of AUC~(Area under curve) of the graph 
of the total benefit $B / \sup_{\beta} B$, 
against the false alarm rate (FAR) $N / \sup_{\beta} N$, 
with $\beta$ varying. 

\subsection{Methods for Comparison}
In order to conduct the sequential D-MDL algorithm, we employed the univariate Gaussian distribution whose probability density function is given by (\ref{gaussd}). 

We employed three change detection methods for comparison: \\
 (1) {\em Bayesian online change point detection}
 (BOCPD)~\cite{adams}: 
    A retrospective Bayesian online change detection method. 
    It originally calculates the posterior of run length. 
    We modified it to compute a change score 
    by taking the expectation of the reciprocal of run length 
    with respect to the posterior. \\ 
    (2) {\em ChangeFinder} (CF)~\cite{ty2006}: 
 A state-of-the-art method of abrupt change detection. \\
 (3) {\em ADWIN2}~\cite{gavalda}: 
    A detection method with adaptive windowing. \\
\ We conducted the sequential D-MDL algorithms with fixed window size in order to investigate their most basic performance in terms of the AUC metric.
The sequential D-MDL algorithm with adaptive windowing outputs the window size rather than the D-MDL values themselves, hence in order to evaluate the effectiveness of the magnitude of D-MDL, the sequential D-MDL with fixed windowing is a better target for the comparison.
All of CF, BOCPD, and ADWIN2 had some parameters, 
which we determined from 5 training sequences 
so that the AUC scores were made the largest. 

\subsection{Results}

The results are summarized in Table \ref{table:combination}.
We see that both for the datasets, in the case of the abrupt changes, the $0$th D-MDL performed best, while for the gradual changes, the $1$st D-MDL performed best and the $2$nd D-MDL performed worse than the $1$st but better than the $0$th.
That matches our intuition.
Because the $0$th D-MDL was designed so that it could detect abrupt changes while the $1$st one was designed so that it could detect starting points of gradual changes.

\section{Result \RomanNumeralCaps{2}: Applications to COVID-19 Outbreak Analysis}
We define {\em outbreak} as a significant increase in the number of cases in a country. We note that to contain the spread of COVID-19, many countries have enacted social distancing policies, e.g., stay-at-home order, closing non-essential services, and limiting travel. We thus relate the results of our change detection to social distancing events.

We are mainly concerned with the following two questions:\\
1.  How early are the outbreak signs detected prior to outbreaks? \\
2.  How are the outbreaks/outbreak signs related to the social distancing events?

As a byproduct, the dynamics of the basic reproduction number $R0$ \cite{r0} can be monitored, which can serve as supplementary information to the value of $R0$ estimated from the SIR model \cite{r0-sir}.

 
\subsection{Data Source}
We studied the data provided by European Centre for Disease Prevention and Control (ECDC) which can be accessed through the link https://www.ecdc.europa.eu/en/publications-data/download-todays-data-geographic-distribution-covid-19-cases-worldwide. In this paper, we focused on the first wave because the situations become very complicated in later waves, e.g., virus mutations \cite{Wisem4857}, people being tired of social distancing and the mixture of two waves in the transition period. In particular, we studied 37 countries with no less than 10,000 cumulative cases by Apr. 30 since some countries started to ease the social distancing around the date. More details can be found in Section 1 of the supplementary material.

\subsection{Data Modeling}
We studied two data models by considering the value of $R0$, which by definition is the product of transmissibility, the average contact rate between susceptible and infected individuals, and the duration of infectiousness \cite{r0-sir}. At the initial phase of an epidemic, $R0$ is larger than one \cite{r0}. And the cumulative cases may grow exponentially \cite{anderson1992, chowell2016}. We thus employed the Malthusian growth model \cite{malthus1992} because it is widely used for characterizing the early phase of an epidemic \cite{chowell2016}. In particular, the cumulative cases at time $t$, $C(t)$, grows according to the following equation:
\begin{align}
    C(t) = C(0)\exp (rt),
\end{align}
where $C(0)$ is the number of cases at the start of an epidemic, and $r$ is the growth rate of daily new cases. In the experiments, we took the logarithm of $C(t)$ to obtain the linear regression of the logarithm growth with respect to time as follows:
\begin{align}
    \log C(t) = rt + \log C(0).
\end{align}
We modeled the residual error of the linear regression using the univariate Gaussian. See Section 5 in the supplementary file for the detail of calculation of the MDL change statistics for this model.
When a change is detected in the modeling of the residual error, we examine the increase/decrease in the coefficient of the linear regression, i.e., $r$. We expect to detect changes in the parameter of the {\em exponential modeling} to monitor the increase/decrease of $R0$ because $R0-1$ is proportional to $r$ \cite{anderson1992}.

In later phases, the exponential growth pattern may not hold. For instance, when $R0 < 1$, daily new cases would continue to decline and cease to exist \cite{r0}. Considering the complicated real scenarios, epidemic models with certain assumptions on the growth rate or $R0$ may not fit an epidemic at a given time. Therefore, we employed the univariate Gaussian model as in  (\ref{gaussd}) to directly fit the number of daily cases, without assuming any patterns of the growth. The change in the parameter of the {\em Gaussian modeling} may reveal the relation between one and $R0$, i.e., $R0 > 1$ when daily cases increase significantly or $R0 < 1$ when daily cases decrease significantly. 

We conducted the hierarchical sequential D-MDL algorithm as in Sec. 2.6. 
The confidence parameter $\delta_0$ 
was set to be 
$0.05$. 
$\delta_{1}$ and $\delta_{2}$ 
were determined as follows:
we calculated the D-MDL scores around the time when the initial warning was announced by an authority;
we determined $\delta _{1},\delta _{2}$ so that the score was the threshold.
For example, the initial warning raised by the government of Japan which called for voluntary event cancellation was on Feb.~$27$ \cite{jpr0}.


\begin{figure}[H]   
\centering
\begin{tabular}{cc}
            \vspace{-0.3cm}
            \textbf{a} & \includegraphics[keepaspectratio, height=3.3cm, valign=T]
			{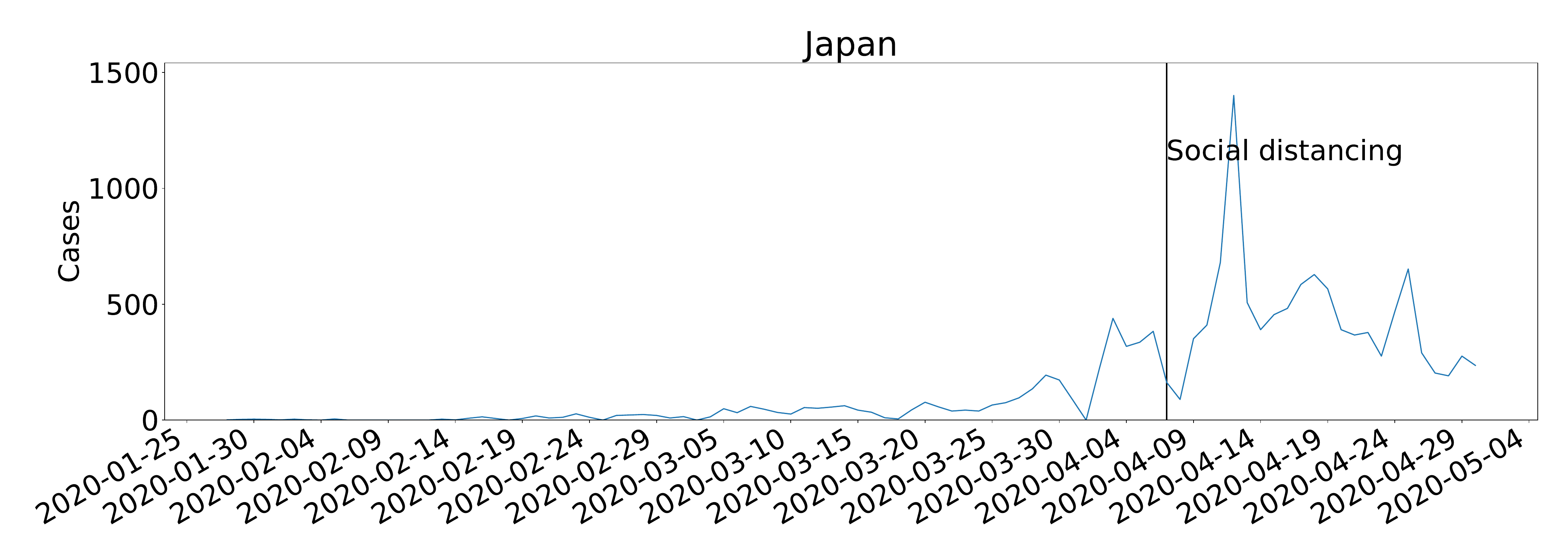} \\
	        \vspace{-0.3cm}
            \textbf{b} & \includegraphics[keepaspectratio, height=3.3cm, valign=T]
			{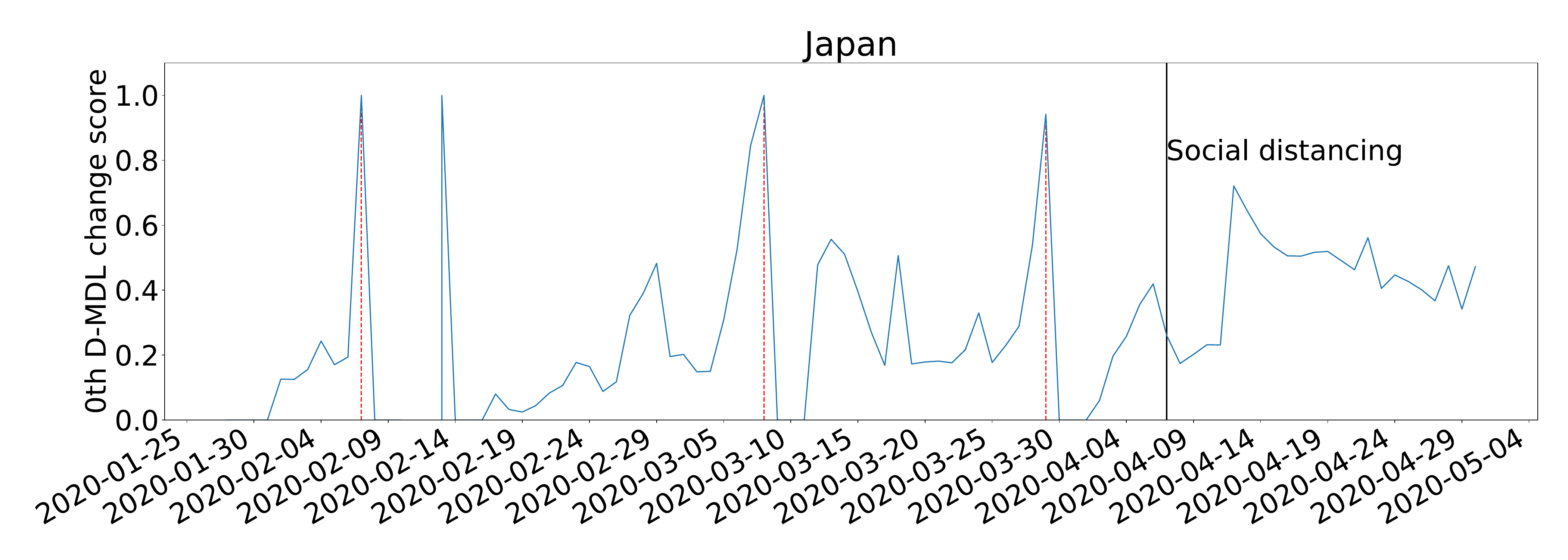}  \\
			\vspace{-0.3cm}
 	        \textbf{c} & \includegraphics[keepaspectratio, height=3.3cm, valign=T]
			{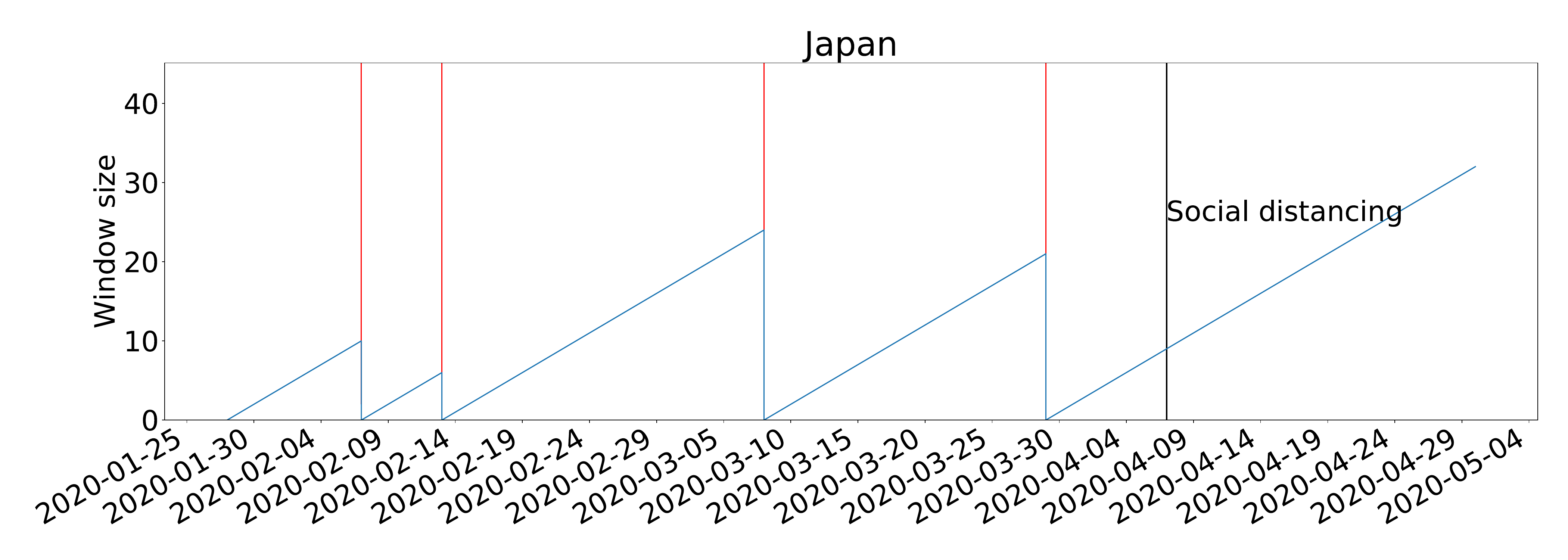} \\
			\vspace{-0.3cm}
		    \textbf{d} & \includegraphics[keepaspectratio, height=3.3cm, valign=T]
			{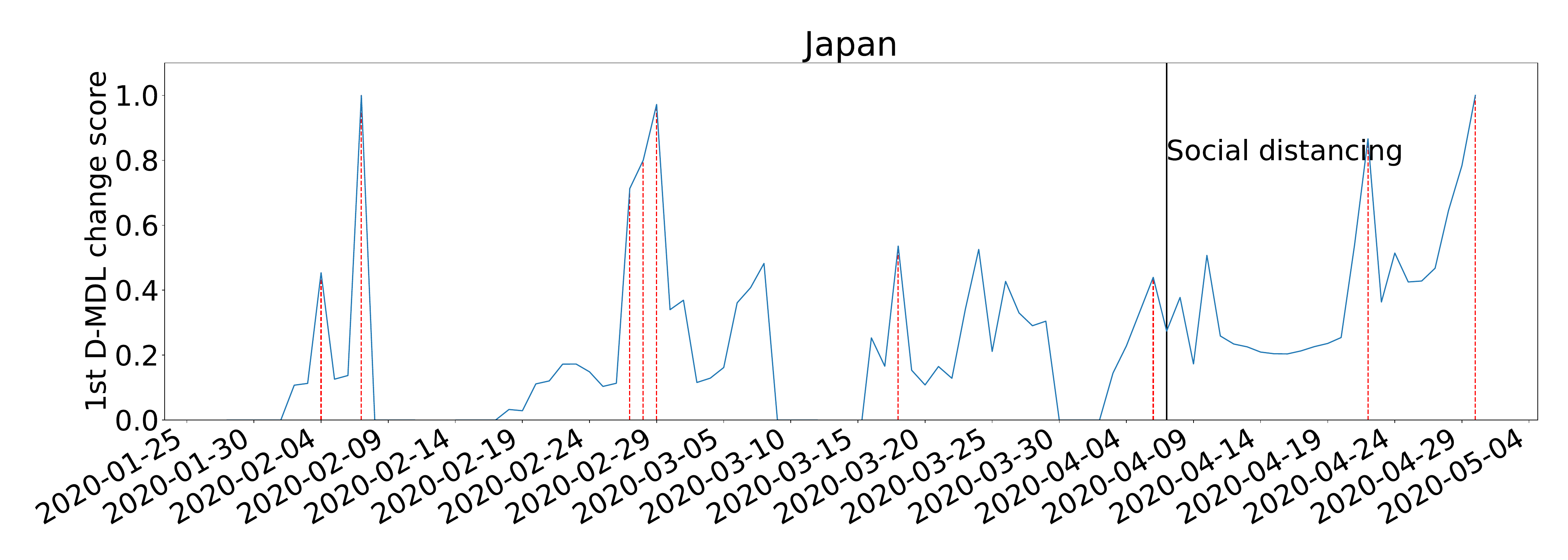}\\
		    \vspace{-0.35cm}
		    \textbf{e} & \includegraphics[keepaspectratio, height=3.3cm, valign=T]
			{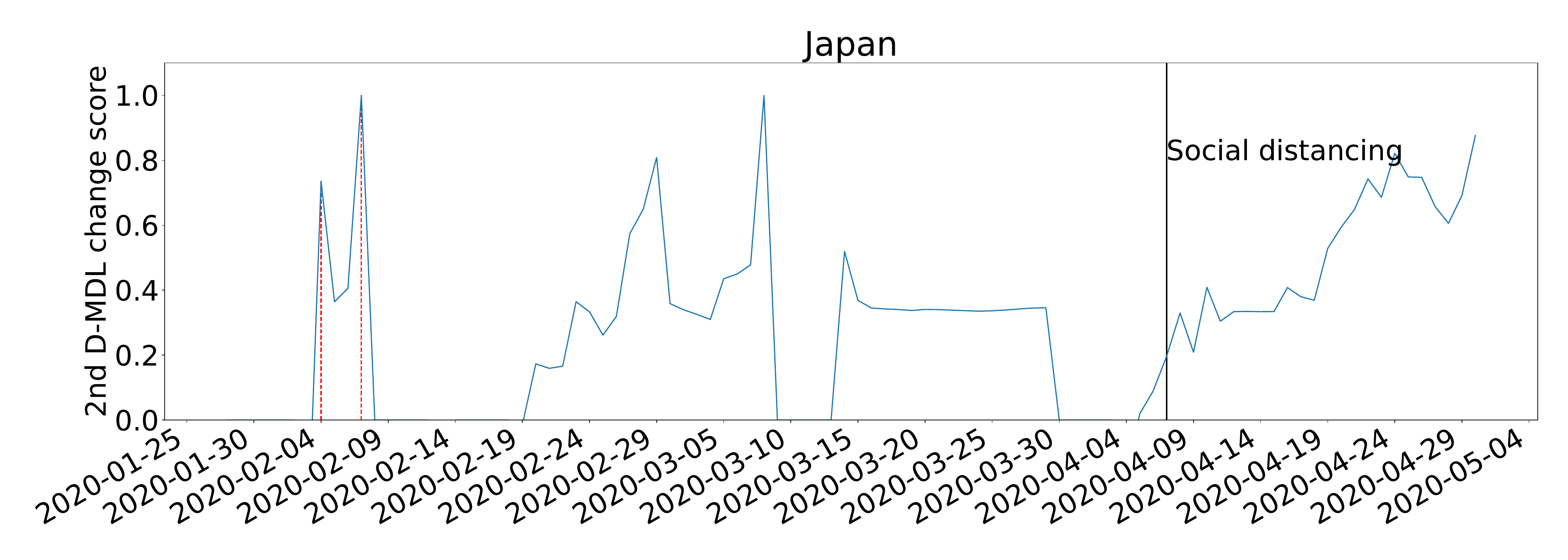}
		\end{tabular}
		\caption{\textbf{The results for Japan with the Gaussian modeling.} The date on which the social distancing was implemented is marked by a solid line in black. \textbf{a,} the number of daily new cases. \textbf{b,} the change scores produced by the 0th D-MDL where the line in blue denotes values of scores and dashed lines in red mark alarms. \textbf{c,} the window sized for the sequential D-MDL algorithm with adaptive windowing where lines in red mark the shrinkage of windows. \textbf{d,} the change scores produced by the 1st D-MDL. \textbf{e,} the change scores by the 2nd D-MDL. In all figures the negative scores are omitted.
		}
		\label{fig:Japan_Gaussian}
\end{figure} 

\begin{figure}[H]   
\centering
\begin{tabular}{cc}
\vspace{-0.3cm}
		    \textbf{a} & \includegraphics[keepaspectratio, height=3.3cm, valign=T]
			{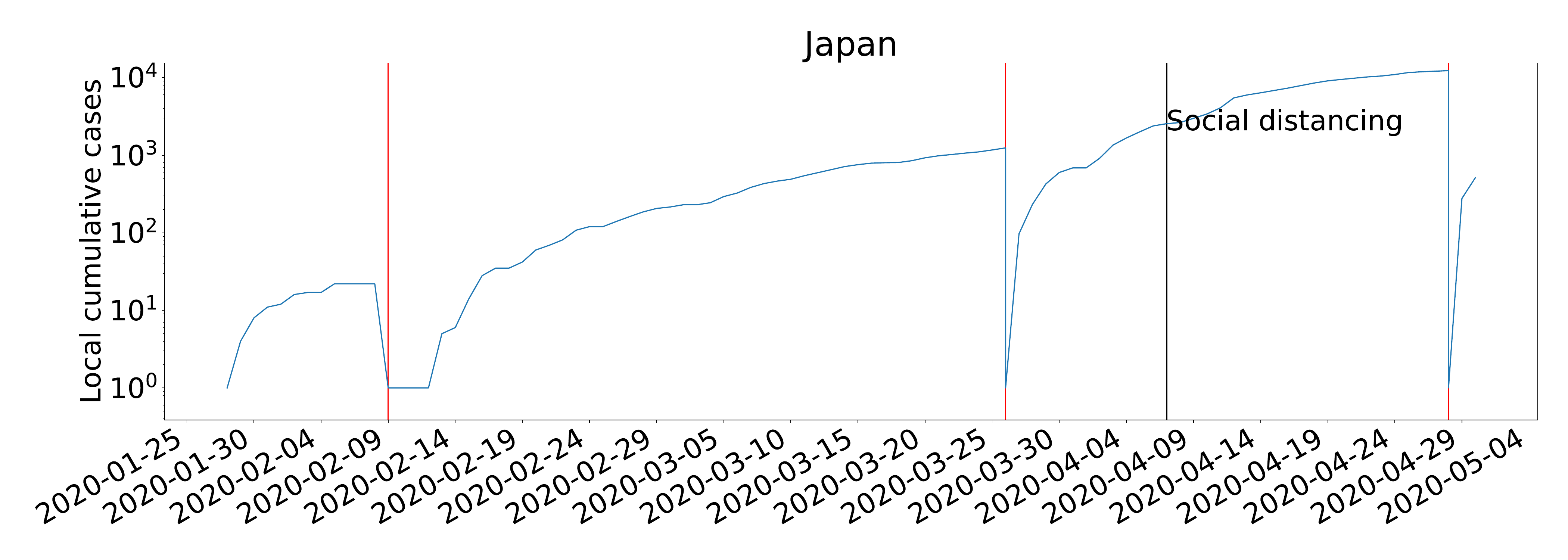} 	\\
	        \vspace{-0.3cm}
            \textbf{b} & \includegraphics[keepaspectratio, height=3.3cm, valign=T]
			{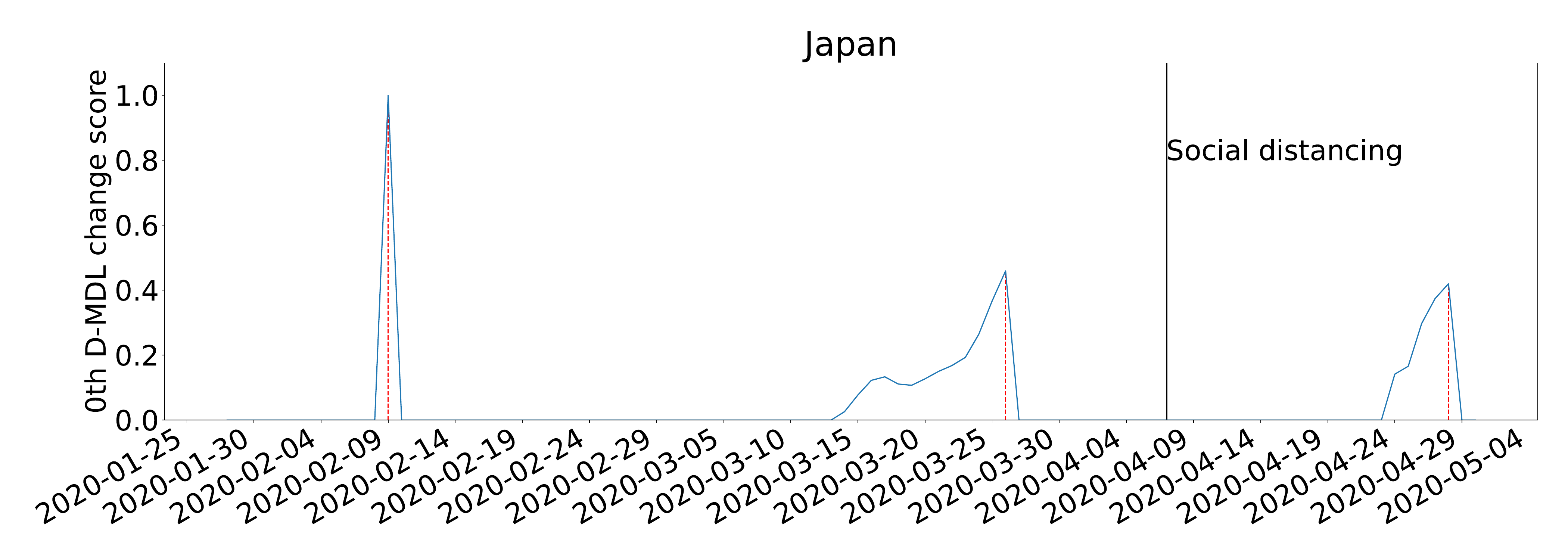}  	\\
            \vspace{-0.3cm}
            \textbf{c} & \includegraphics[keepaspectratio, height=3.3cm, valign=T]
			{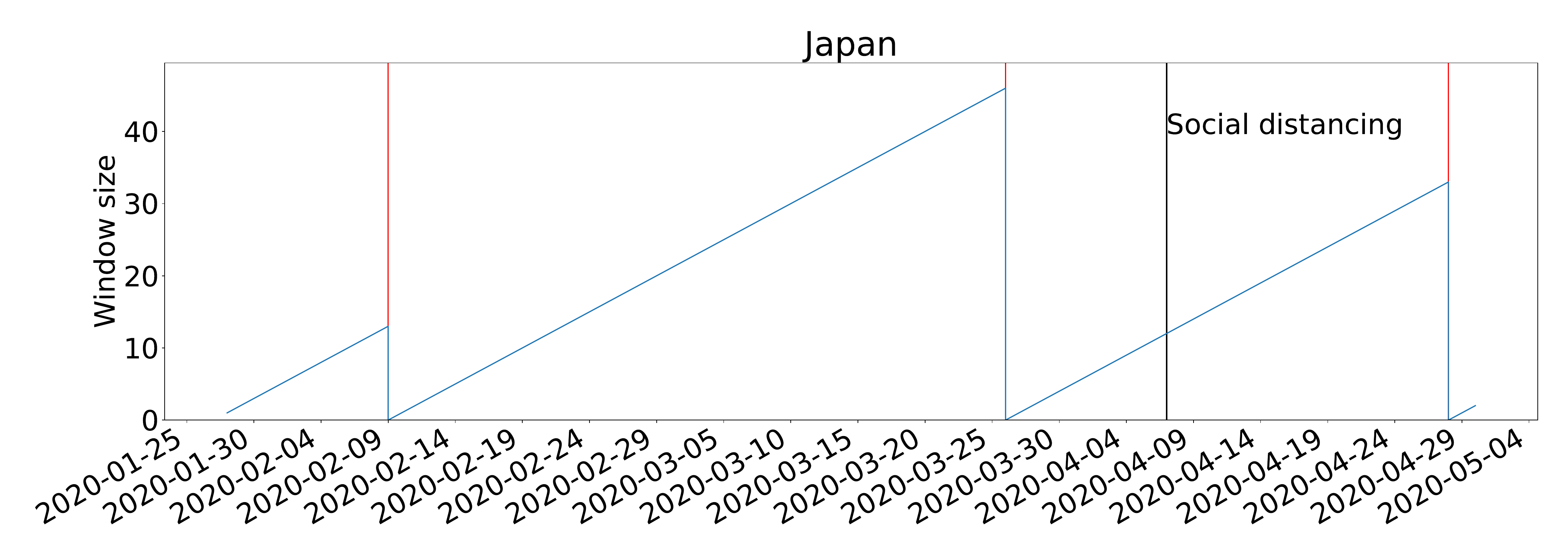} \\
	    	\vspace{-0.3cm}
		    \textbf{d} & \includegraphics[keepaspectratio, height=3.3cm, valign=T]
			{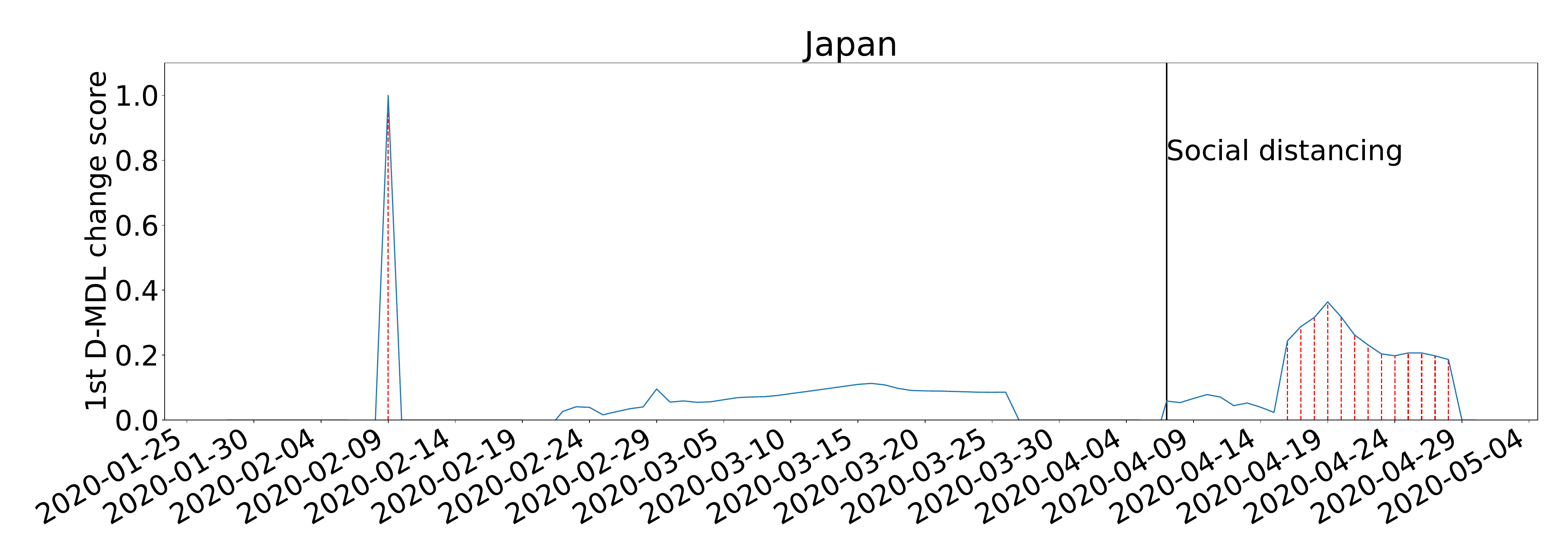} \\
		    \vspace{-0.3cm}
		    \textbf{e} & \includegraphics[keepaspectratio, height=3.3cm, valign=T]
			{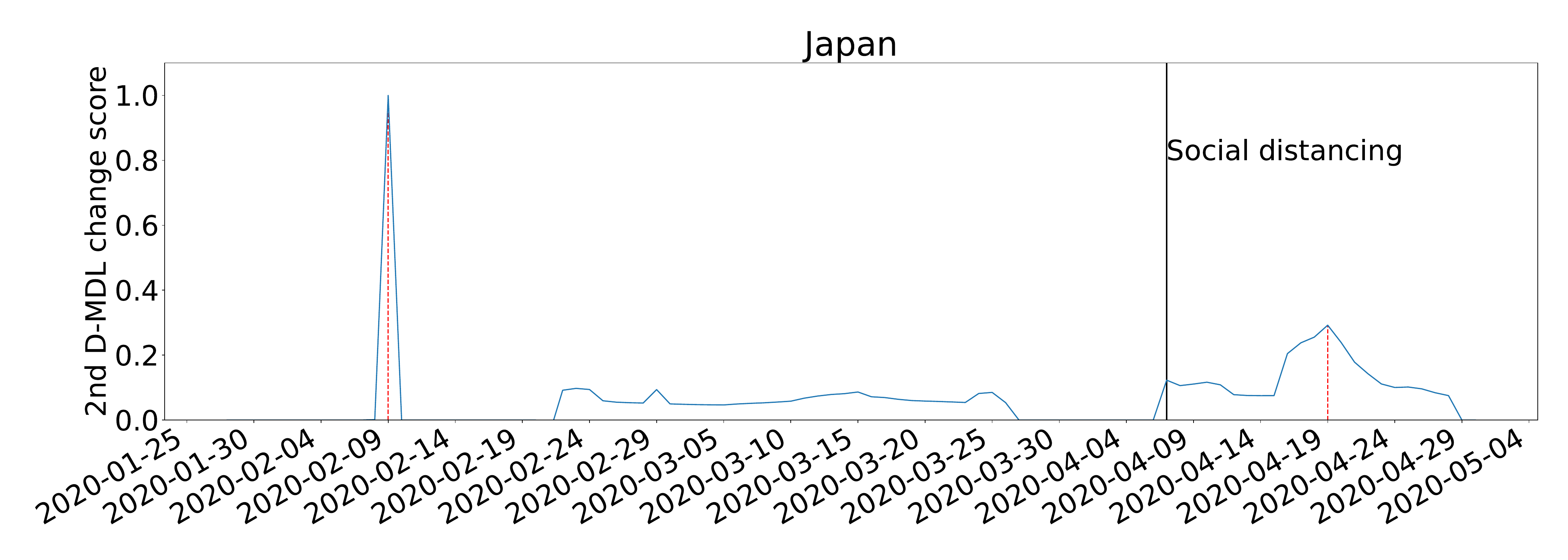}
		 \end{tabular}
		
		    \caption{
		    \textbf{The results for Japan with the exponential modeling.} The label "Local cumulative cases" in subfigure (a) means that the cumulative cases only accumulate daily cases from each starting date of change detection.
		    }
		\label{fig:Japan_Exp}
\end{figure} 

\subsection{Case Study}
We present one representative case study of Japan due to space consideration. 
State of emergency as the social distancing event was issued on Apr. 7. The results are presented in Fig. \ref{fig:Japan_Gaussian} and Fig. \ref{fig:Japan_Exp} for the Gaussian modeling and the exponential modeling, respectively. Change scores were normalized into [0, 1]. The data of Japan did not include the confirmed cases from ?Diamond Princess?.

With the Gaussian modeling, there were several alarms raised before the social distancing event. For each alarm raised by the 0th D-MDL, the interpretation can be a statistically significant increase in cases, with reference to Fig. \ref{fig:Japan_Gaussian}(a). Hereafter, a change that is detected by the 0th D-MDL and that corresponds to the increase of cases is regarded as an outbreak, which instantiates our definition of outbreak. The outbreak detection is the classic change detection. We further relate it to $R0$. Around the dates of the alarms, $R0 > 1$ was considered since we can confirm that the new infections resulted from community transmission. Correspondingly, $R0$ was estimated around 2.5 in early March by an epidemiological study \cite{jpr0}. When the 0th D-MDL raised an alarm, the window size shrank to zero. Before that, both the 1st and the 2nd D-MDL raised alarms, which are interpreted as the changes in the velocity and the acceleration of the increase of cases, respectively. We can conclude that the 1st and the 2nd D-MDL were able to detect the signs of the outbreak by examining the velocity and the acceleration of the spread. The sign detection is the new concept with which we propose to supplement the classic change detection. The 0th D-MDL raised no alarms about outbreaks after the event. We think the social distancing played a critical role in containing the spread because it can significantly suppress $R0$ through reducing the contact rate. The 1st D-MDL still raised alarms, which were about signs of decreases in the cases.

As for the exponential modeling, there were alarms raised by the 0th D-MDL both before and after the social distancing event. By looking at the growth pattern of local cumulative cases in Fig. \ref{fig:Japan_Exp}(a), we can see that all the alarms were about the cessations of the exponential growth. Moreover, we checked that the alarms were associated with decreases in the coefficient of the linear regression. Therefore, we concluded that all the alarms indicated the significantly decreases in $R0$. Although the last two alarms were raised on Mar. 26 and Apr. 28, the dates as the change points were within the windows as of Mar. 26 and Apr. 28, and were identified as Mar. 12 and Apr. 18, respectively. There was an epidemiological study \cite{jpr0} which showed the effectiveness of the initial warning announced on Feb. 27 at reducing $R0$. As a result, it demonstrates that our method can effectively identify the decrease in $R0$ around Mar. 12. According to the result, our method identified another decrease in $R0$ around Apr. 18, which we think was mainly due to the social distancing event on Apr. 7. Therefore, our method based on the exponential modeling also confirmed that social distancing was very effective at containing the spread. The alarms raised by the 1st and 2nd D-MDL demonstrate the capability of the sign detection.

As a comparison, the Gaussian modeling was effective at estimating the relation between one and $R0$ while the exponential modeling was able to monitor the change in the value of $R0$. The two models form a complementary relation on monitoring the dynamics of $R0$. For instance, for Japan, the Gaussian modeling showed that the value of $R0$ reminded at a value larger than one, and the exponential modeling showed that its value decreased during the studied period. Due to the difference in the modeling, the changes detected by the 0th D-MDL were at different dates between the Gaussian modeling and the exponential modeling. In terms of sign detection, both the Gaussian modeling and the exponential modeling are effective.


\subsection{Summarization on Individual Countries}
This section summarizes several statistics about the change detection results in Table \ref{sta_gaussian}, and presents two interesting observations. The first is about how early the signs can be detected prior to changes. For the countries studied, there were 106 and 54 changes in total detected by the Gaussian modeling and the exponential modeling, respectively. There were more changes detected by the Gaussian modeling because daily cases would significantly change with either $R0>1$ or $R0<1$ while it may take relatively longer time for significant changes in $R0$. The number of changes whose signs were detected by either the 1st or the 2nd D-MDL is 68 and 26 for the Gaussian modeling and the exponential modeling, respectively, representing high detection rates. For each change whose signs were detected, we measured the time difference between the earliest sign alarm and the change alarm. For the Gaussian modeling which can detect outbreaks, the time difference in terms of the number of days is 6.25 (mean) $\pm$ 6.04 (standard deviation). Considering the fast spread, six days can buy us considerably long time to prepare for an outbreak, and even to avoid a potential outbreak. 

For the Gaussian modeling, the 1st D-MDL detected signs for 65 changes and the 2nd D-MDL detected signs for 27 changes. The smaller number for the 2nd D-MDL might be because the 1st D-MDL is better at detecting starting points of gradual changes, and is consistent with results on the synthetic datasets as in Table \ref{table:synthetic}.
The number of days before which the 1st D-MDL detected signs was 6.35 $\pm$ 5.91, and the number for the 2nd D-MDL was 5.56 $\pm$ 6.50. Note that not all the changes allowed for sign detection since the 1st D-MDL and the 2nd D-MDL sign detection require one more and two more data points in the window than the 0th D-MDL, respectively. The number of changes allowing for a 1st D-DML sign was 88 while the number for a 2nd D-DML sign was 81. Hence, it turned out that some changes occurred too quickly before signs can be detected. The analysis of the results obtained by the exponential modeling is similar and omitted for space consideration. 

\begin{table*}[t]
\small
\centering
\caption{\textbf{Summarization of statistics where changes represent the alarms raised by the 0th D-MDL and signs are alarms raised by either the 1st or the 2nd D-MDL.}}\label{tab:statistics}
  \begin{tabular}{p{11cm}cc} \hline
    Measurement & Gaussian & Exponential \\ \hline
    Total number of changes  & $106$ & 54\\
    Number/percentage of changes whose signs were detected by either the 1st or the 2nd D-MDL & $68$/$64\%$ & 26/$48\%$\\
    Number of days before which the first sign was detected by either the 1st or the 2nd D-MDL for a change & $6.25\pm 6.04$ & $11.27\pm 7.72$\\
    Total number of changes that allowed for the 1st/2nd D-MDL sign detection & $88$/$81$ & 53/53\\
    Number of changes whose signs were detected by the 1st/2nd D-MDL & $65$/$27$ & 26/6\\
    Number of days before which the first 1st D-MDL sign was detected for a change & $6.35\pm 5.91$ & $11.27\pm 7.72$\\
    Number of days before which the first 2nd D-MDL sign was detected for a change & $5.56\pm 6.50$ & $5.17\pm 5.67$\\    
    Number of changes and signs before the event for the downward countries & $4.30\pm 2.79$ & --\\
    Number of changes and signs before the event for the non-downward countries & $5.96\pm 4.22$ & --\\
    Number of days from event's date to the first downward change's date for downward countries & $30.00\pm 8.28$ & -- \\
    Number of days from event's date to Apr. 30 for non-downward countries & $36.54\pm 7.28$ & --\\
    Number of decreasing changes and signs for the downward countries & -- & $10.60\pm 6.67$\\
    Number of decreasing changes and signs the non-downward countries & -- & $9.96\pm 9.65$\\
    \hline
  \end{tabular}
  \label{sta_gaussian}
\end{table*}

Second, we observed that on average, countries responding faster in terms of a smaller number of alarms raised by the Gaussian modeling before the social distancing event saw a quicker contraction of daily cases. As of Apr. 30, the curve of daily cases in many countries had been flatten, and even started to be downward. Therefore, alarms for declines in the number of daily cases from the global peak number were raised for ten countries including Austria, China, Germany, Iran, Italy, Netherlands, South Korea, Spain, Switzerland, and Turkey. These countries are referred to as {\em downward countries}. 
In total, the number of all kinds of alarms raised before the event for downward countries was 4.30 $\pm$ 2.79 while it was 5.96 $\pm$ 4.22 for other countries. 
Therefore, if the social distancing is a viable option, it is suggested that the action should better be taken before it is late, e.g., later than four alarms. We further measured that it took an average of 30 days to suppress the spread if prompt social distancing policies were enacted. 
By contrast, the average number of days from the social distancing event to Apr. 30 was nearly 37 for non-downward countries, which is considerably more than the time used for suppressing the spread in downward countries.

The results of the exponential modeling confirmed the above observation. In particular, changes and their signs which corresponded to decreases in $R0$ for the downward countries were more than that for the non-downward countries as shown Table \ref{sta_gaussian}. 


\section{Conclusion}
This paper has proposed a novel methodology for detecting signs of changes from a data stream. 
The key idea is to use the differential MDL change statistics (D-MDL) as a sign score. 
We have theoretically justified D-MDL using the hypothesis testing framework and have empirically justified the sequential D-MDL algorithm using the synthetic data.
We have applied D-MDL to the COVID-19 pandemic analysis. 
We have observed that the $0$th D-MDL can find change points related to outbreaks and that the $1$st and $2$nd D-MDL were able to detect their signs several days earlier than them. 
We have further related the change points to the dynamics of the basic reproduction number $R0$.
This analysis is a new promising approach to the pandemic analysis from the view of data science.

Future work includes studying the second wave and third wave which are more complicated situations than the first wave.

\end{document}


\setlength{\abovedisplayskip}{2pt}
\setlength{\belowdisplayskip}{2pt}

\maketitle

The supplementary materials contain the following information:
\begin{itemize}
    \item Section 1 presents the information of studied data.
    \item Section 2 presents some details about the implementation of our Hierarchical Sequential D-MDL Algorithm presented in Algorithm 2.
    \item Section 3 presents the proof of Theorem 2.2. 
    \item Section 4 gives the details of calculation of the MDL change statistics for the linear regression model.
    \item Section 5 introduces additional experiment results.   
\end{itemize}

\section{Data Information}
We employed the data provided by European Centre for Disease Prevention and Control (ECDC) via https://www.ecdc.europa.eu/en/publications-data/download-todays-data-geographic-distribution-covid-19-cases-worldwide.
For information, there are 37 countries that had no less than 10,000 cases in total by Apr. 30, including Austria, Belarus, Belgium, Brazil, Canada, Chile, China, Ecuador, France, Germany, India, Indonesia, Iran, Ireland, Israel, Italy, Japan, Mexico, Netherlands, Pakistan, Peru, Poland, Portugal, Qatar, Romania, Russia, Saudi Arabia, Singapore, South Korea, Spain, Sweden, Switzerland, Turkey, Ukraine, United Arab Emirates, United Kingdom, United States of America in alphabetic order.

We collected the date on which the social distancing was implemented from the information listed in the IHME COVID-19 predictions via \url{https://covid19.healthdata.org/united-kingdom}. If a certain country is not listed in the website, we referred to the Wikipedia page for the COVID-19 pandemic of the country, e.g., the COVID-19 pandemic in South Korea \url{https://en.wikipedia.org/wiki/COVID-19_pandemic_in_South_Korea}. For information, the dates are Austria: 2020-3-16, Belarus: 2020-4-9, Belgium: 2020-3-18, Brazil: 2020-3-24, Canada: 2020-3-17, Chile: 2020-3-26, China: 2020-1-23, France: 2020-3-17, Germany: 2020-3-16, India: 2020-3-25, Indonesia: 2020-4-6, Iran: 2020-3-24, Ireland: 2020-3-12, Israel: 2020-3-15, Italy: 2020-3-9, Japan: 2020-4-7, Mexico: 2020-3-23, Netherlands: 2020-3-15, Pakistan: 2020-3-24, Peru: 2020-3-16, Poland: 2020-3-24, Portugal: 2020-3-19, Qatar: 2020-3-23, Romania: 2020-3-23, Russia: 2020-3-30, Saudi Arabia: 2020-3-24, Singapore: 2020-4-7, South Korea: 2020-2-25, Spain: 2020-3-13, Sweden: 2020-3-24, Switzerland: 2020-3-16, Turkey: 2020-3-21, Ukraine: 2020-3-25, United Arab Emirates: 2020-3-31, United Kingdom: 2020-3-24, United States of America: 2020-3-19. Ecuador was excluded from the list above because the social distancing is introduced to be related to changes incurred by declines in the number of cases and there was a very large number of cases in the initial phase of the epidemic in Ecuador. The large number might be an outlier due to the data collection procedure, and would make any changes after that date downward changes. But we still studied the change/change sign detection for Ecuador.

\section{Implementation Details}
There are three hyper-parameters in our Hierarchical Sequential D-MDL Algorithm, which are $\delta_0, \delta_1$ and $\delta_2$ for specifying the threshold of changes in the 0th D-MDL, the 1st D-MDL, and the 2nd D-MDL, respectively.  
$\delta_0$ was set to be $0.05$. $\delta_{1}$ and $\delta_{2}$ were determined as follows:
we calculated the D-MDL scores around the time when the initial warning was announced by an authority;
we determined $\delta _{1},\delta _{2}$ so that the score was the threshold. If the resulting $\delta_{1},\delta_{2}$ were larger than 1, they were set to be 0.99 because of the concept of confidence parameter. In particular, we calculated $\delta_{1}$ and $\delta_{2}$ by using the initial warning in Japan, and applied them to all the other countries. The initial warning for Japan was set on Feb.~$27$, when the government called for voluntary event cancellation. 

We determined the starting date of change detection for each country as the date whose past ten days saw an average of at least one case. This is because few cases in a country may be imported from other countries, and may have nothing to do with local community transmission. We implemented our algorithm in Python 3. Detailed instructions about how to conduct the experiments presented in this paper with our implementation are available at \url{https://github.com/IbarakikenYukishi/differential-mdl-change-statistics}. Moreover, we developed an online detection system based on our methodology which can perform change/sign detection everyday and raise corresponding alarms. The system can be accessed through \url{ https://ibarakikenyukishi.github.io/d-mdl-html/index.html}.

\section{Proof of Theorem 2.2}
Let the maximum likelihood estimator of $\theta _{i}$ be $\hat{\theta}_{i}\ (i=0,1,2,3)$.
Let us define the event  as
\begin{eqnarray*}
h_{1}(x^{n}_{1}; t, \epsilon )&=&-\log p(x^{t}_{1};\hat{\theta}_{0}(x^{t}_{1}))-\log p(x^{n}_{t+1};\hat{\theta}_{1}(x^{n}_{t+1}))+\log C_{t}+\log C_{n-t} \nonumber \\
& &+\log p(x^{t+1}_{1};\hat{\theta}_{2}(x^{t+1}_{1}))+\log p(x^{n}_{t+2};\hat{\theta}_{3}(x^{n}_{t+2}))-\log C_{t+1}-\log C_{n-t-1}-n\epsilon >0.
\end{eqnarray*}
Equivalently, 
\begin{eqnarray}\label{event1}
 & &p(x^{t}_{1};\hat{\theta}_{0}(x^{t}_{1}))p(x^{n}_{t+1};\hat{\theta}_{1}(x^{n}_{t+1}))\nonumber \\
& &<\frac{ p(x^{t+1}_{1};\hat{\theta}_{2}(x^{t+1}_{1}))}{ C_{t+1}}\cdot 
\frac{ p(x^{n}_{t+2};\hat{\theta}_{3}(x^{n}_{t+2}))}{ C_{n-t-1}}\cdot 
\exp \left(-n \left(\epsilon -\frac{\log C_{t}C_{n-t}}{n} \right) \right).
\end{eqnarray}
Type I error probability is evaluated as follows:
Let the true parameter be $\theta _{0}^{*}$ and $\theta _{1}^{*}$.
\begin{eqnarray*}
& &\sum_{x^{n}_{1}\cdots (\ref{event1})}p(x^{t}_{1};\theta _{0}^{*})p(x^{n}_{t+1};\theta _{1}^{*})\\
& &<\sum_{x^{n}\cdots (\ref{event1})}\frac{ p(x^{t+1}_{1};\hat{\theta}_{2}(x^{t+1}_{1}))}{ C_{t+1}}\cdot 
\frac{ p(x^{n}_{t+2};\hat{\theta}_{3}(x^{n}_{t+2}))}{ C_{n-t-1}}\cdot 
\exp \left(-n \left(\epsilon -\frac{\log C_{t}C_{n-t}}{n} \right) \right)
\\
& &\leq \left(\sum_{x^{t+1}_{1}}\frac{ p(x^{t+1}_{1};\hat{\theta}_{2}(x^{t+1}_{1}))}{ C_{t+1}}\right) \left( \sum_{x^{n}_{t+2}}\frac{ p(x^{n}_{t+2};\hat{\theta}_{3}(x^{n}_{t+2}))}{ C_{n-t-1}}\right)\cdot 
\exp \left(-n \left(\epsilon -\frac{\log C_{t}C_{n-t}}{n} \right) \right)\\
& &= \exp \left(-n \left(\epsilon -\frac{\log C_{t}C_{n-t}}{n} \right) \right),
\end{eqnarray*}
where we have used the following relations:
\begin{eqnarray*}
\sum_{x^{t+1}_{1}}\frac{ p(x^{t+1}_{1};\hat{\theta}_{2}(x^{t+1}_{1}))}{ C_{t+1}}= 1,\ \ \ \ \sum_{x^{n}_{t+2}}\frac{ p(x^{n}_{t+2};\hat{\theta}_{3}(x^{n}_{t+2}))}{ C_{n-t-1}}= 1.
\end{eqnarray*}

Next we evaluate Type II error probability.
Let us define the event as 
\begin{eqnarray}\label{event2}
h_{1}(x^{n}_{1},t,\epsilon )\leq 0
\end{eqnarray}
and let $p_{_{\rm NML}(t)}(x^{n}_{1})\buildrel \rm def \over =(p(x^{t}_{1};\hat{\theta}_{0}(x^{t}_{1})/C_{t})(p(x^{n}_{t+2};\hat{\theta}_{1}(x^{n}_{t+1}))/C_{n-t})$. 
Let the true parameter be $\theta _{2}^{*}$ and $\theta _{3}^{*}$.
Then under the event (\ref{event2}), 
\begin{eqnarray*}
-\log p_{_{\rm NML}(t)}(x^{n}_{1})
& \leq &-\log p(x^{t+1}_{1}; \hat{\theta} _{2}(x^{t+1}_{1}))-\log p(x^{n}_{t+2}; \hat{\theta} _{3}(x^{n}_{t+2}))+\log C_{t+1}+\log C_{n-t-1}+n\epsilon \\
& \leq &-\log p(x^{t+1}_{1}; \theta _{2}^{*})-\log p(x^{n}_{t+2}; \theta _{3}^{*})+\log C_{t+1} C_{n-t-1}+n\epsilon 
\end{eqnarray*}
This implies 
\begin{eqnarray*}
1\leq \left(\frac{p_{_{\rm NML}}(x^{n}_{1})}{p(x^{t+1}_{1};\theta _{2}^{*})p(x^{n}_{t+2};{\theta}_{3}^{*})}\right)^{\frac{1}{2}}
\exp \left( \frac{1}{2}(\log C_{t+1}C_{n-t-1} +n\epsilon )\right)
\end{eqnarray*}
The Type II error probability is upper-bounded as follows:
\begin{eqnarray*}
& &\sum _{x^{n}_{1}\cdots (\ref{event2})}p(x^{t+1}_{1};\theta ^{*}_{2})
p(x^{n}_{t+2};\theta ^{*}_{3})\\
& &\leq \sum _{x^{n}_{1}\cdots (\ref{event2})}(p(x^{t+1}_{1};\theta ^{*}_{2})
p(x^{n}_{t+2};\theta ^{*}_{3}))\left(\frac{p_{_{\rm NML}}(x^{n}_{1})}{p(x^{t+1}_{1};\theta _{2}^{*})p(x^{n}_{t+2};{\theta}_{3}^{*})}\right)^{\frac{1}{2}}
\exp \left( \frac{1}{2}(\log C_{t+1}C_{n-t-1} +n\epsilon )\right)\\
& &\leq \sum _{x^{n}_{1}}\left(p_{_{\rm NML}(t)}(x^{n}_{1})(p(x^{t+1}_{1};\theta _{2}^{*})p(x^{n}_{t+2};\theta _{3}^{*})\right)^{\frac{1}{2}}\exp \left( \frac{1}{2}(\log C_{t+1}C_{n-t-1} +n\epsilon )\right)\\
& &=\exp \left(-n\left(d(p_{_{\rm NML}(t)},p_{\theta_{2}^{*}}*p_{\theta _{3}^{*}})-\frac{\log C_{t+1}C_{n-t-1}}{2n}-\frac{\epsilon}{2}\right)\right),
\end{eqnarray*}
where $d$ and $p_{\theta_{2}^{*}}*p_{\theta _{3}^{*}}$ are defined as in Theorem 2.1 and 2.2. 
This completes the proof.
\hspace*{\fill}$\Box$\\

\section{MDL change statistics for linear regression}
We consider the linear regression model defined as follows:
\begin{align*}
&X^{n}=(x_{1},\dots  x_{n})^{\top}=W_{n}^{\top}\beta+\epsilon,\ \ 
\epsilon\sim \mathcal{N}(0,\sigma ^{2}I_n),\\
&W_{n}=\left(
\begin{array}{ccccc}1&1&1&...&1\\
 1&2&3&...&n
\end{array}
 \right)^\top \in \mathbb{R}^{n\times 2},\  \beta \in \mathbb{R}^2,
\end{align*}
Let us define the class of linear regression by 
\begin{align*}
&{\mathcal P}=\left\{p(X^{n};\theta )=\frac{1}{(\sqrt{2\pi} \sigma)^{d}}\exp\left( -\frac{||X-W_{n}^\top \beta ||^{2}}{2\sigma ^{2}}\right):\right. \\
& \ \left. \theta =(\beta ,\sigma ^{2}) \in \mathbb{R}^{3},\ n=1,2,\dots \right\}.
\end{align*}
Then the NML codelength for $x^{n}$ relative to this class is
\begin{equation*}\label{1}
L_{_{\rm NML}}(x^{n})=\frac{n}{2}\log \hat{\sigma}^{2}+\log \frac{R}{\sigma _{min}^{2}}-\log \Gamma \left(\frac{n}{2}-1\right)+\frac{n}{2}\log (n\pi),
\end{equation*}
where $\sigma_{\rm min}$ and $R$ are hyper-parameters  determined so that 
$\hat{\sigma}_i^2\ge\sigma_{\rm min}^2$ and $
\|\hat{\beta}\| \le nR$.
$\hat{\sigma }^{2}_{t}$ is a maximum likelihood estimator of $\sigma ^{2}$ from $x^{t}_{1}$ and so on.
Thus the MDL change statistics at time $t$ is
\begin{eqnarray*}\label{2}
\Phi _{t}&=&\frac{h}{2}\log \frac{\sigma ^{2}_{h}}{\sigma _{t}\sigma _{h-t}}-\log \frac{R}{\sigma ^{2}_{min}}\\
& &\ \ \ \ \ -\log \frac{\Gamma(h/2-1)}{\Gamma(t/2-1)\Gamma ((h-t)/2-1)}\\
& &\ \ \ \ \ +\frac{1}{2}(h\log h-t\log t-(h-t)\log (h-t))?\nonumber 
\end{eqnarray*}
We used this formula for the exponential growth model in Section 4.2 letting $x_{t}=\log C(t)$ for the number of cumulative cases $C(t)$.

\section{Additional Experiment Results}
\subsection{Case Study of South Korea}
\begin{figure}[H] 
\centering
\begin{tabular}{cc}
            \vspace{-0.3cm}
		 	\textbf{a} & \includegraphics[keepaspectratio, height=2.8cm, valign=T]
		{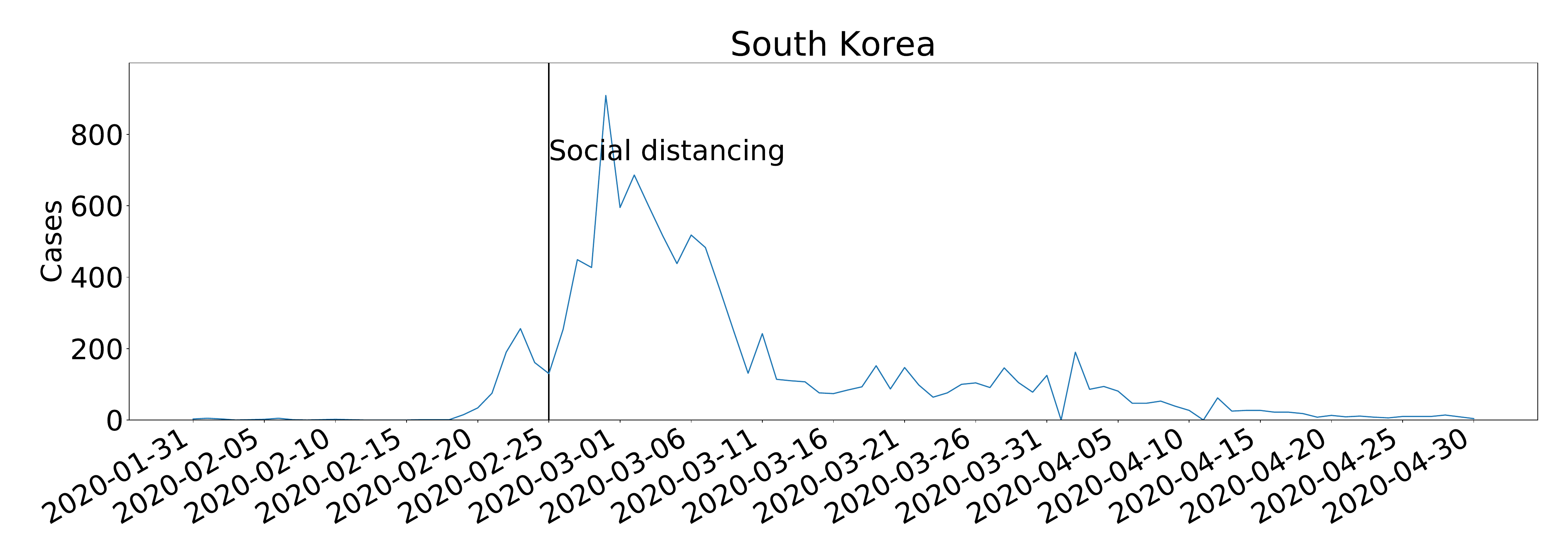} \\
			\vspace{-0.3cm}
	 	    \textbf{b} & \includegraphics[keepaspectratio, height=2.8cm, valign=T]
		{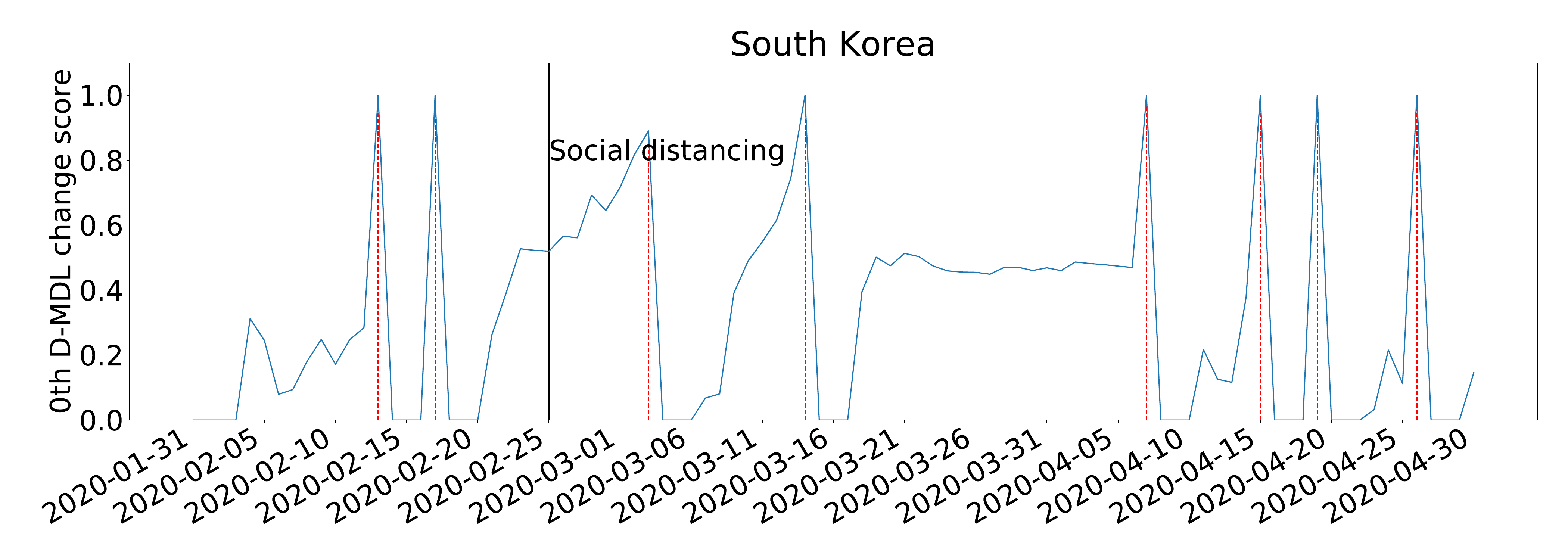}   \\
	        \vspace{-0.3cm}
			\textbf{c} & \includegraphics[keepaspectratio, height=2.8cm, valign=T]
		{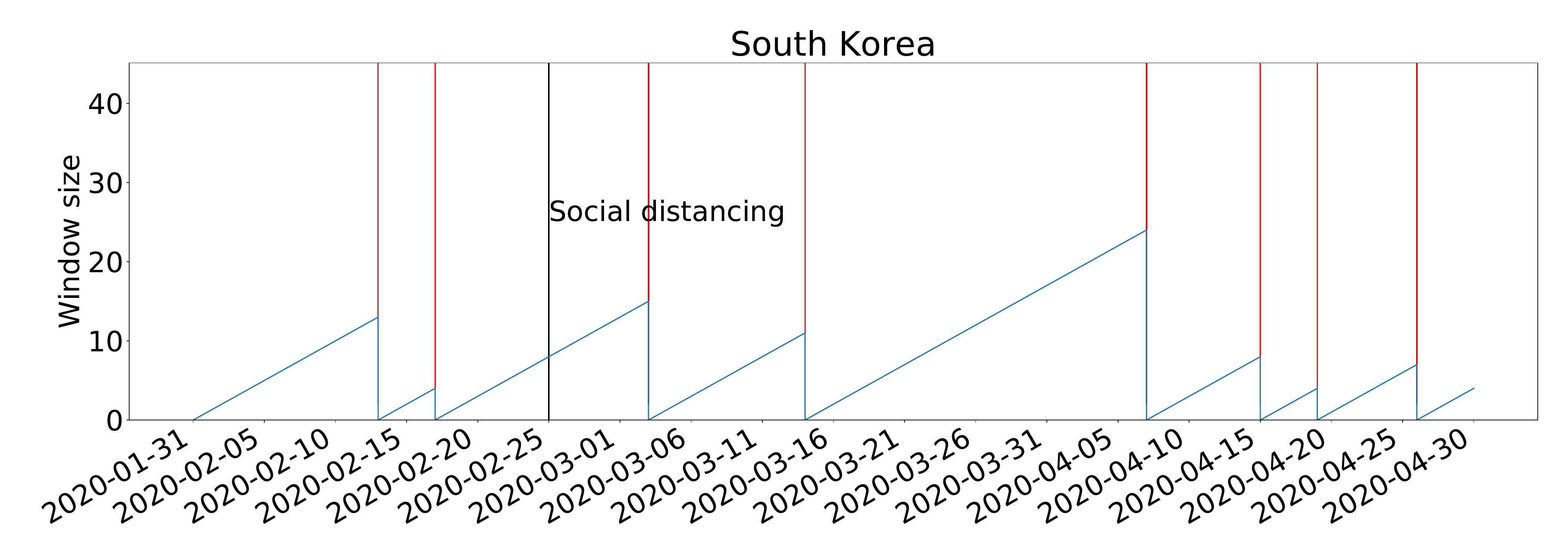} \\
		    \vspace{-0.3cm}
			\textbf{d} & \includegraphics[keepaspectratio, height=2.8cm, valign=T]
		{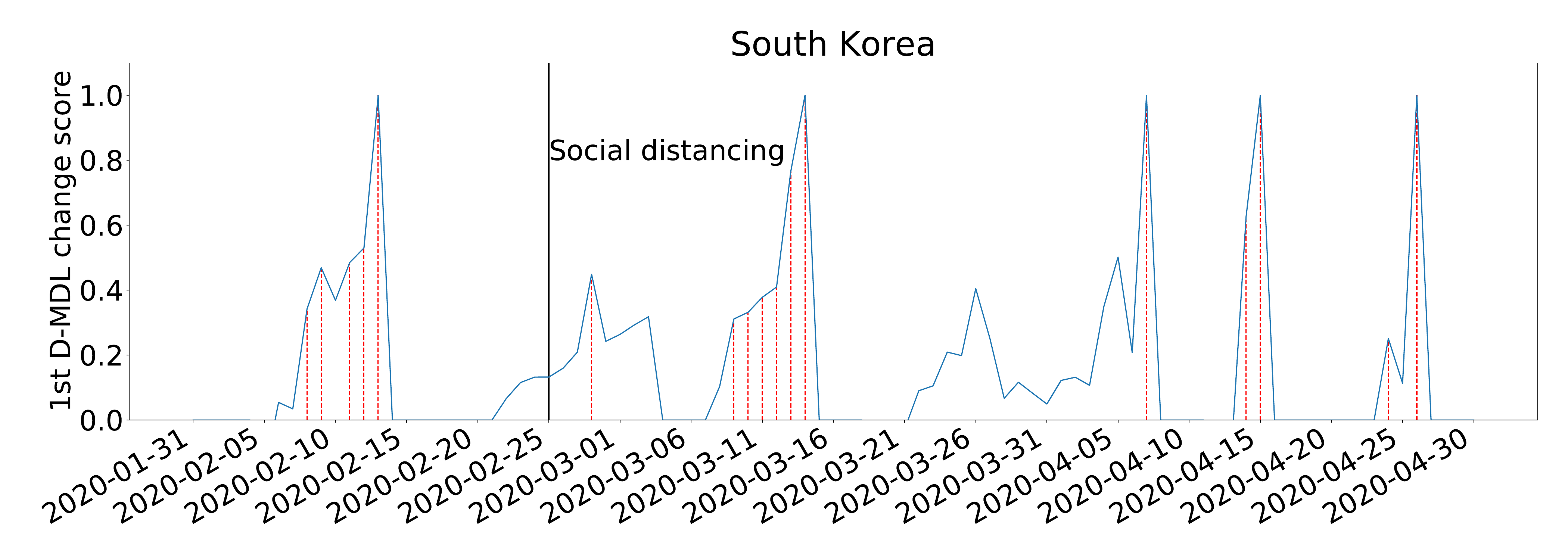} \\
		    \vspace{-0.3cm}
			\textbf{e} & \includegraphics[keepaspectratio, height=2.8cm, valign=T]
		{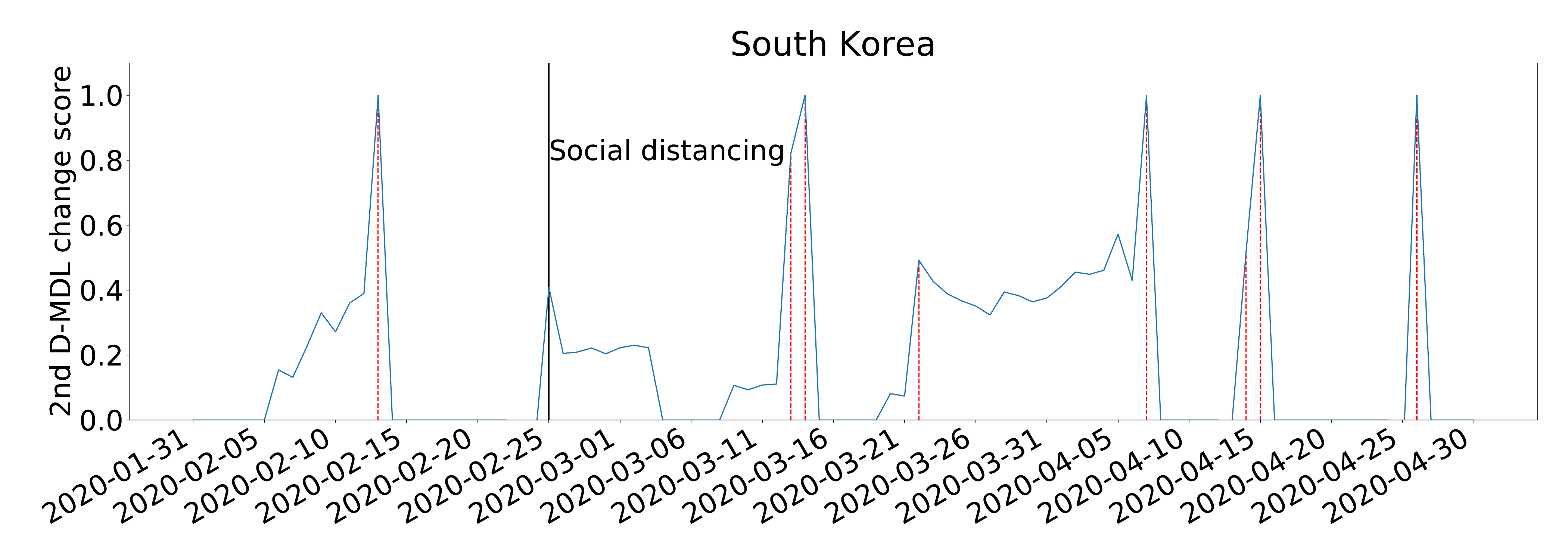} \\
		\end{tabular}
			\caption{\textbf{The results for South Korea with Gaussian modeling.} The date on which the social distancing was implemented is marked by a solid line in black. \textbf{a,} the number of daily new cases. \textbf{b,} the change scores produced by the 0th D-MDL where the line in blue denotes values of scores and dashed lines in red mark alarms. \textbf{c,} the window sized for the sequential D-MDL algorithm with adaptive windowing where lines in red mark the shrinkage of windows. \textbf{d,} the change scores produced by the 1st D-MDL. \textbf{e,} the change scores produced by the 2nd D-MDL. 
			}
	\label{fig:South_Korea_gaussian}
\end{figure}

\begin{figure}[H]  
\centering
\begin{tabular}{cc}
            \vspace{-0.3cm}
			\textbf{a} & \includegraphics[keepaspectratio, height=2.8cm, valign=T]
			{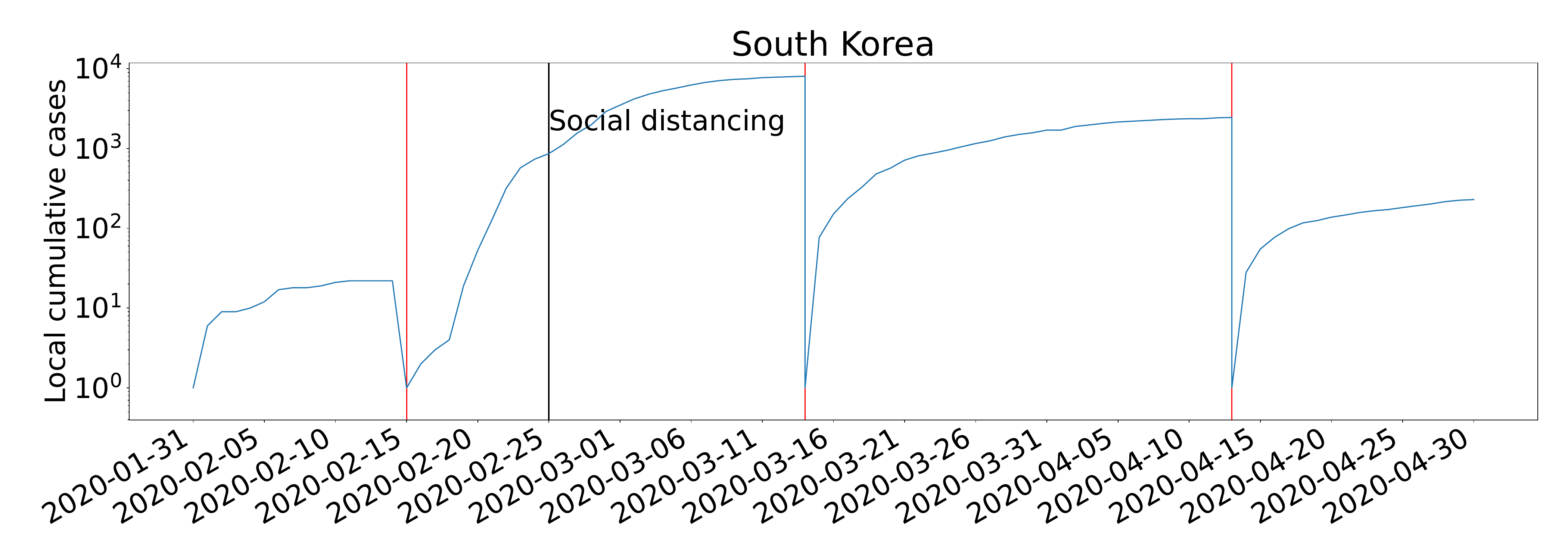} \\
	        \vspace{-0.3cm}
            \textbf{b} & \includegraphics[keepaspectratio, height=2.8cm, valign=T]
			{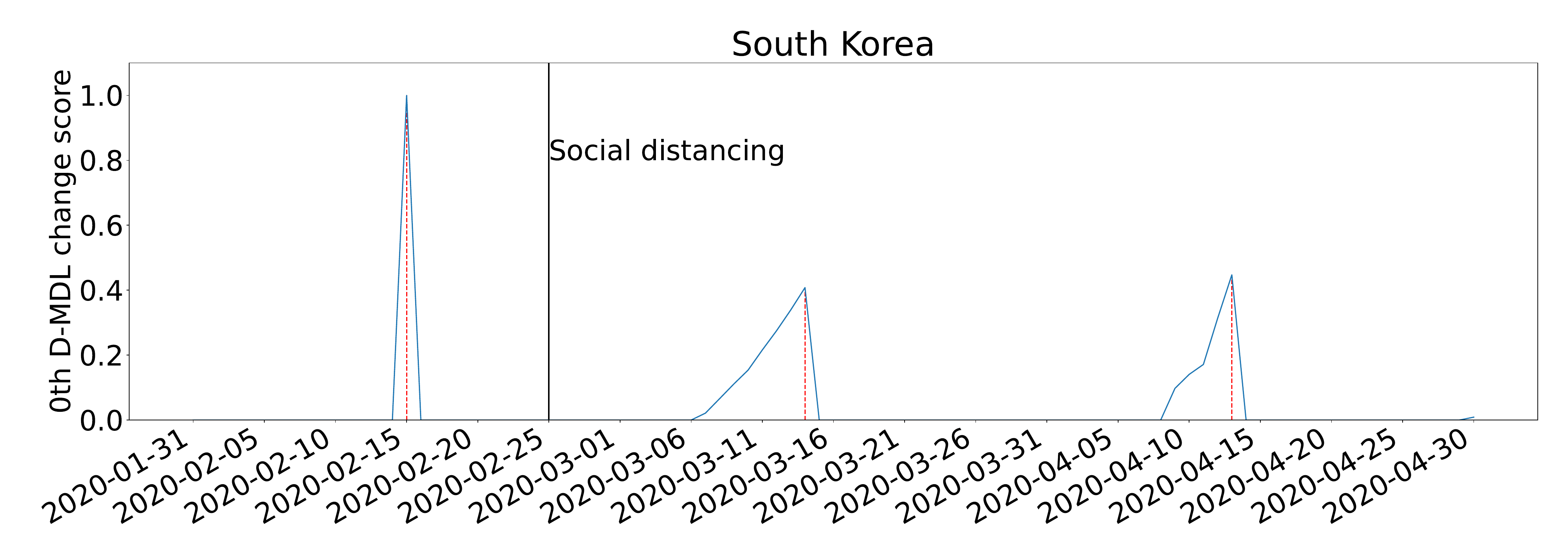}   \\
            \vspace{-0.3cm}
            \textbf{c} & \includegraphics[keepaspectratio, height=2.8cm, valign=T]
			{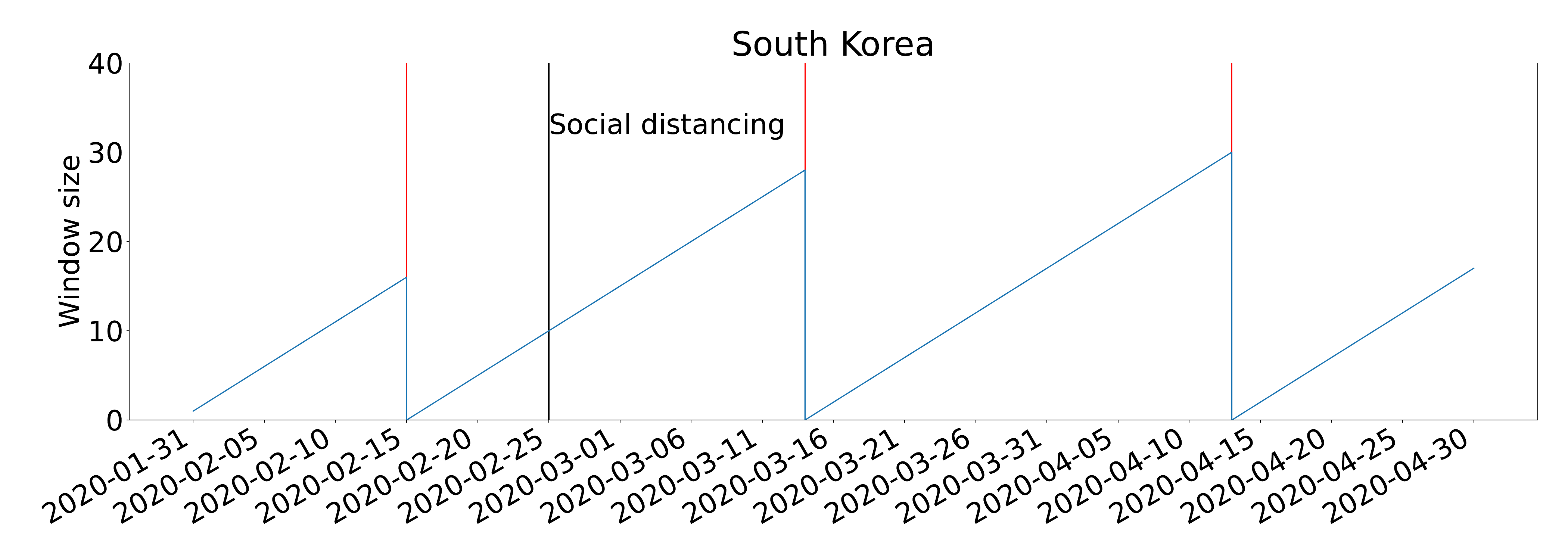} \\
			\vspace{-0.3cm}
			\textbf{d} & \includegraphics[keepaspectratio, height=2.8cm, valign=T]
			{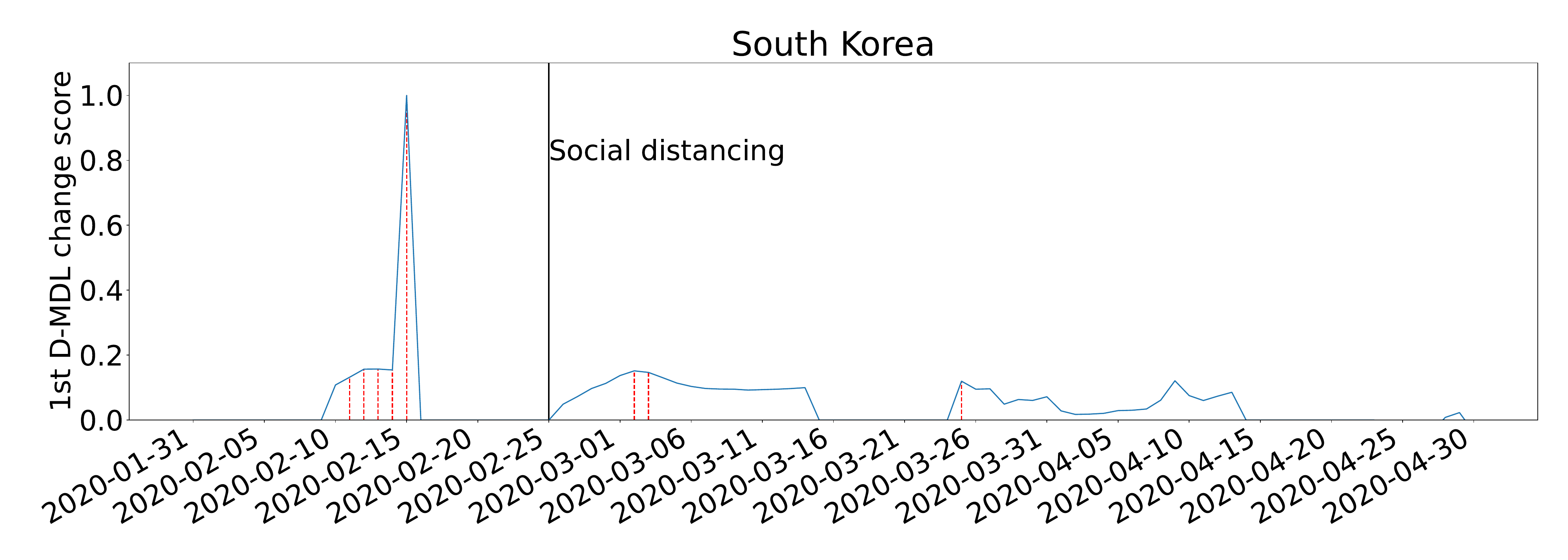} \\
			\vspace{-0.3cm}
			\textbf{e} & \includegraphics[keepaspectratio, height=2.8cm, valign=T]
			{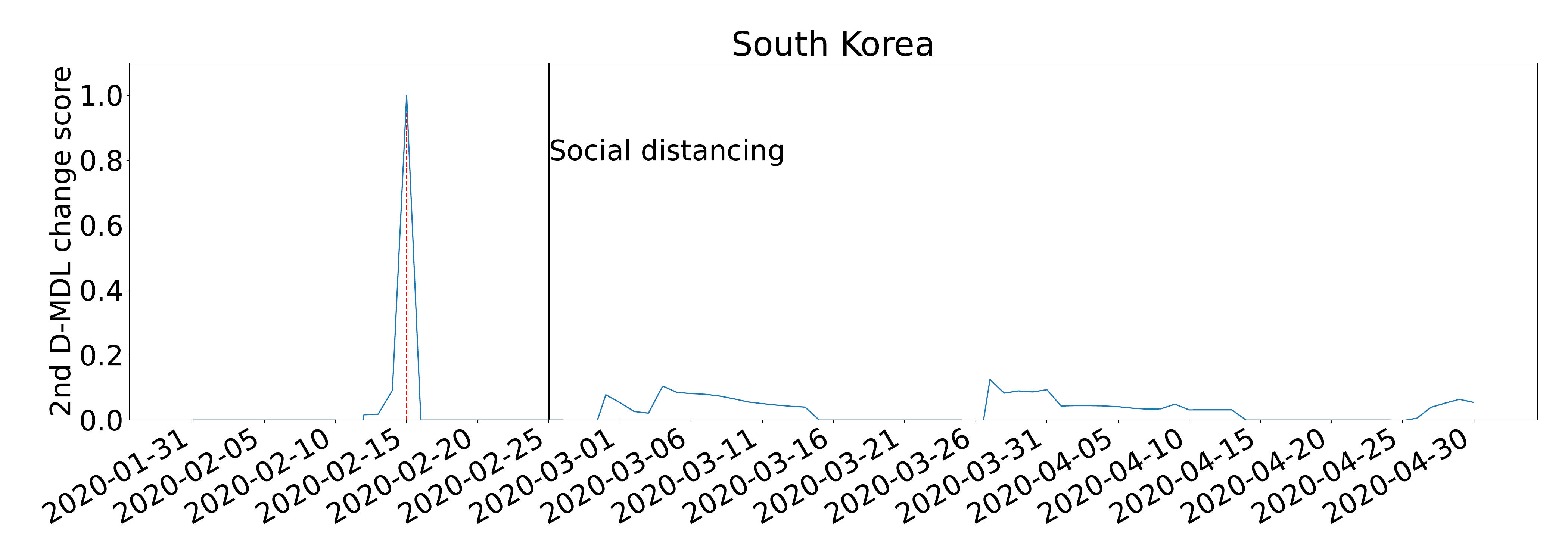} \\
		\end{tabular}
			\caption{\textbf{The results for South Korea with exponential modeling.} 
			The date on which the social distancing was implemented is marked by a solid line in black. \textbf{a,} the number of cumulative cases. \textbf{b,} the change scores produced by the 0th D-MDL where the line in blue denotes values of scores and dashed lines in red mark alarms. \textbf{c,} the window sized for the sequential D-MDL algorithm with adaptive windowing where lines in red mark the shrinkage of windows. \textbf{d,} the change scores produced by the 1st D-MDL. \textbf{e,} the change scores produced by the 2nd D-MDL.
			}
			\label{fig:South_Korea_exp}
\end{figure}

The date of the implementation of social distancing was considered as Feb. 25 from which many non-essential services were closed. We present results in Fig. \ref{fig:South_Korea_gaussian} and Fig. \ref{fig:South_Korea_exp} for the Gaussian modeling and the exponential modeling, respectively. 

With the Gaussian modeling, there were several alarms raised before the social distancing event. Around the dates of the alarms, $R0 > 1$ was considered since we can confirm that the new infections resulted from community transmission. Correspondingly, $R0$ was estimated at 1.5 by an epidemiological study \cite{skr0}. When the 0th D-MDL raised an alarm, the window size shrank to zero. Before that, both the 1st and the 2nd D-MDL raised alarms, which are interpreted as the changes in the velocity and the acceleration of the increase of cases, respectively. We can conclude that the 1st and the 2nd D-MDL were able to detect the signs of the outbreak by examining the velocity and the acceleration of the spread. 

The 0th D-MDL raised several alarms after the event, and the latest ones corresponded to decreases of cases. It is not difficult to tell that the corresponding $R0$ was less than one. We think that the social distancing played a critical role in containing the spread because it can suppress $R0$ through reducing the contact rate, which was supported by studies \cite{skr0, park2020, greenstone2020, wilder2020}. Both the 1st and the 2nd D-MDL again demonstrated the capability of early sign detection.

As for the exponential modeling, there were alarms raised by the 0th D-MDL both before and after the social distancing event. By looking at the growth pattern of local cumulative cases in Fig. \ref{fig:South_Korea_exp}(a), we can see that all the alarms were about the cessations of the exponential growth. Moreover, we checked that the alarms were associated with decreases in the coefficient of the linear regression. Therefore, we concluded that all the alarms indicated the significantly decreases in $R0$. The alarms raised by the 1st and 2nd D-MDL demonstrate the capability of the sign detection.

\subsection{Results for All the Studied Countries}
This section presents the results for all the studied countries with both the Gaussian modeling and the exponential modeling. Since the interpretation for each country is similar and there are many countries, we omit the analyses. Please refer to the cases studies of Japan and South Korea for more information. It is worth mentioning that there were only two alarms raised by the 0th D-MDL about increases of $R0$, which were in Germany and Singapore, respectively. Note that $R0$ is an inherent property of COVID-19 \cite{r0-sir}, and its value would not increase unless there are virus mutations which lead to increases in the transmissibility of COVID-19.  Accordingly, $R0$ would decrease at most time as the susceptible individuals decrease and social distancing events increase. After checking the figure of daily cases, we conclude that the increase of $R0$ in Germany may be because of a significant increase in testing, and the increase in Singapore may be because daily cases before the alarm were mostly imported from other countries.

\newpage

\begin{figure}[H] 
\centering
\begin{tabular}{cc}
		 	\textbf{a} & \includegraphics[keepaspectratio, height=3.3cm, valign=T]
			{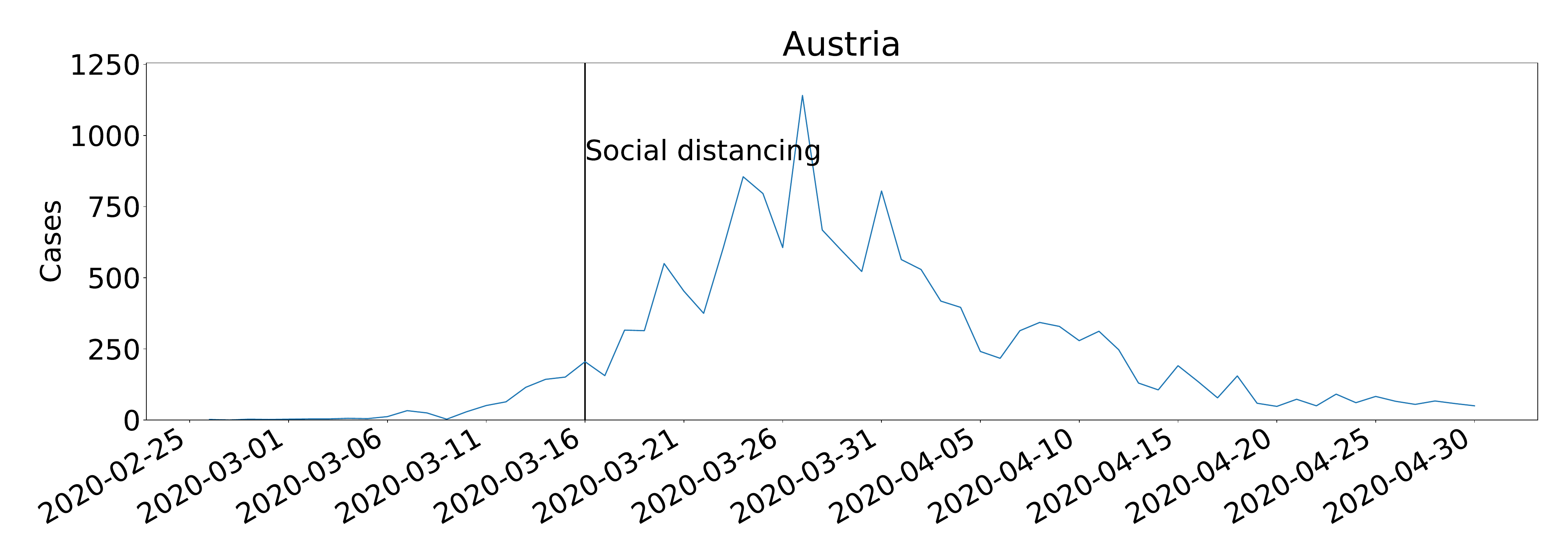} \\
			\vspace{-0.35cm}
	 	    \textbf{b} & \includegraphics[keepaspectratio, height=3.3cm, valign=T]
			{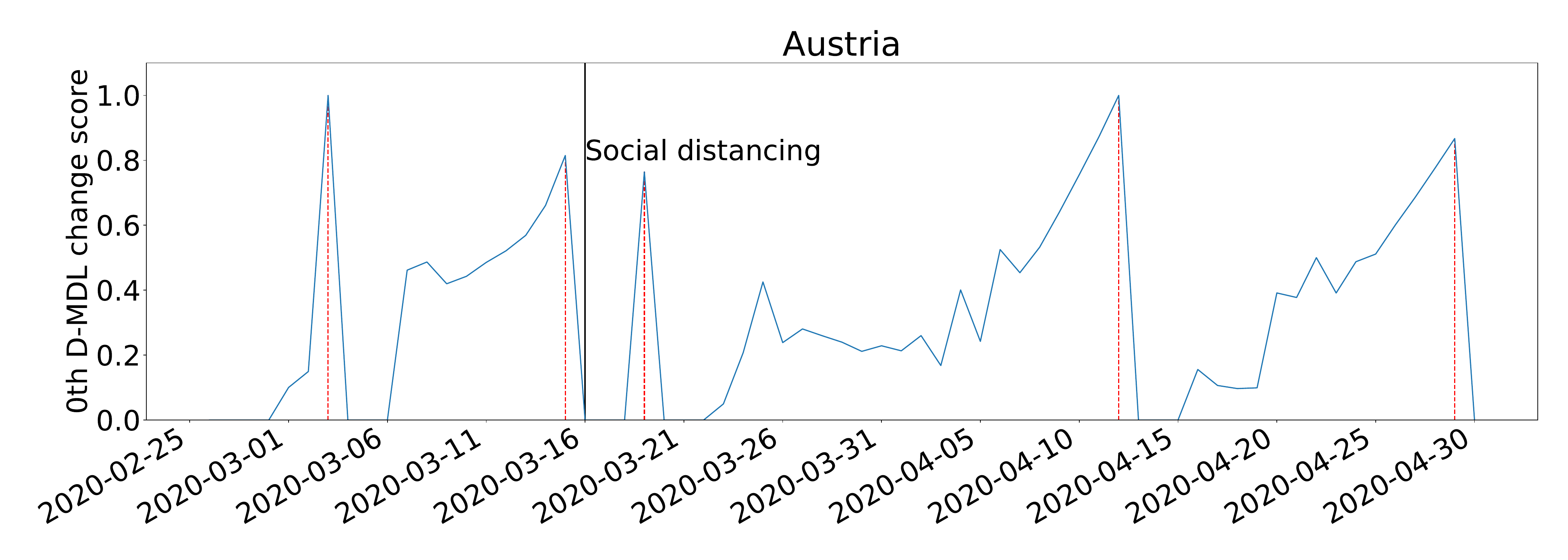}   \\
	        \vspace{-0.35cm}
			\textbf{c} & \includegraphics[keepaspectratio, height=3.3cm, valign=T]
			{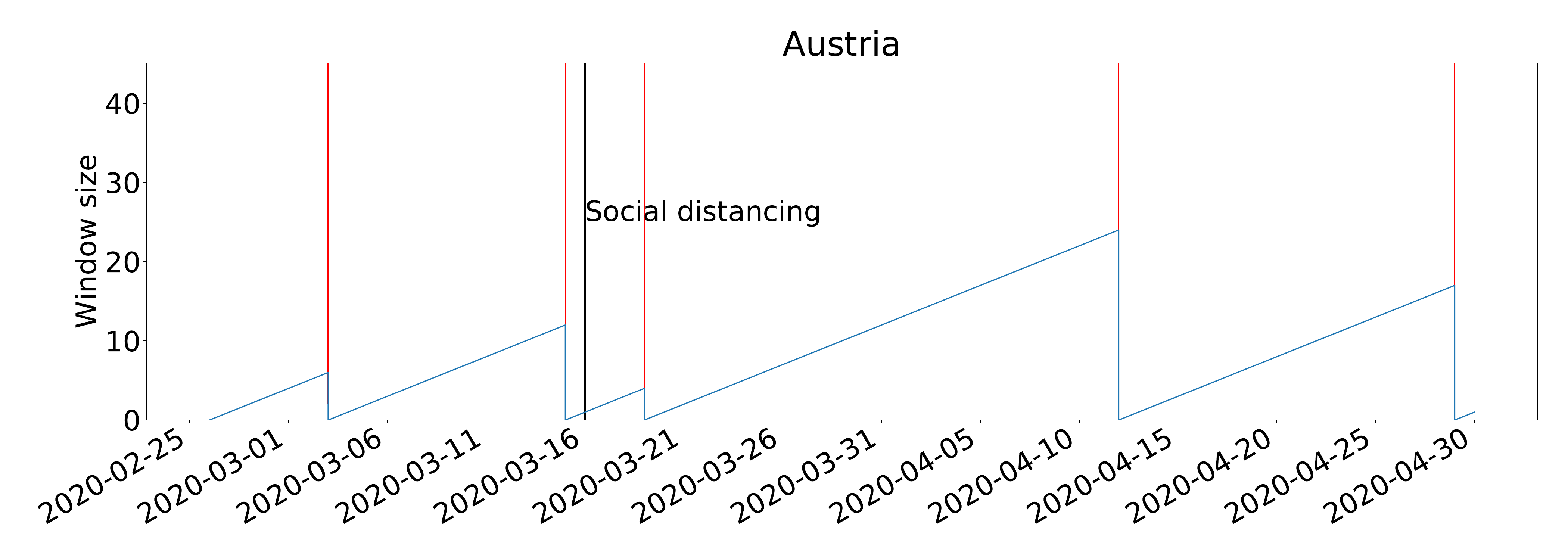} \\
		    \vspace{-0.35cm}
			\textbf{d} & \includegraphics[keepaspectratio, height=3.3cm, valign=T]
			{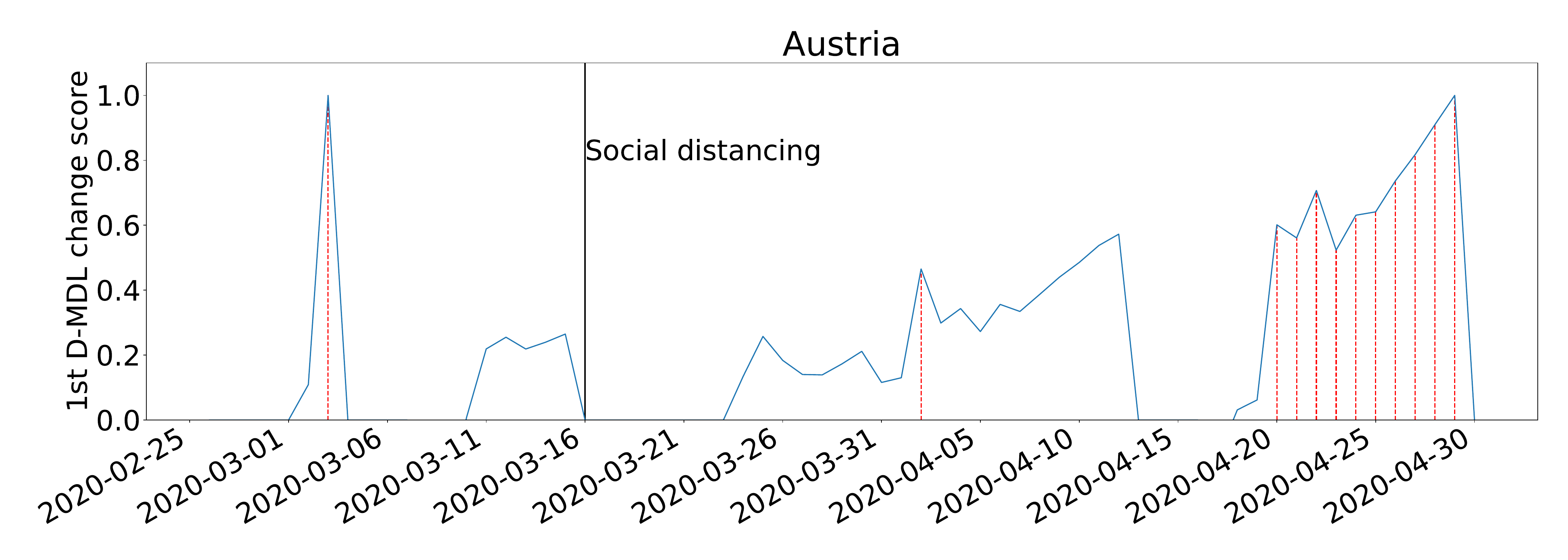} \\
		    \vspace{-0.35cm}
			\textbf{e} & \includegraphics[keepaspectratio, height=3.3cm, valign=T]
			{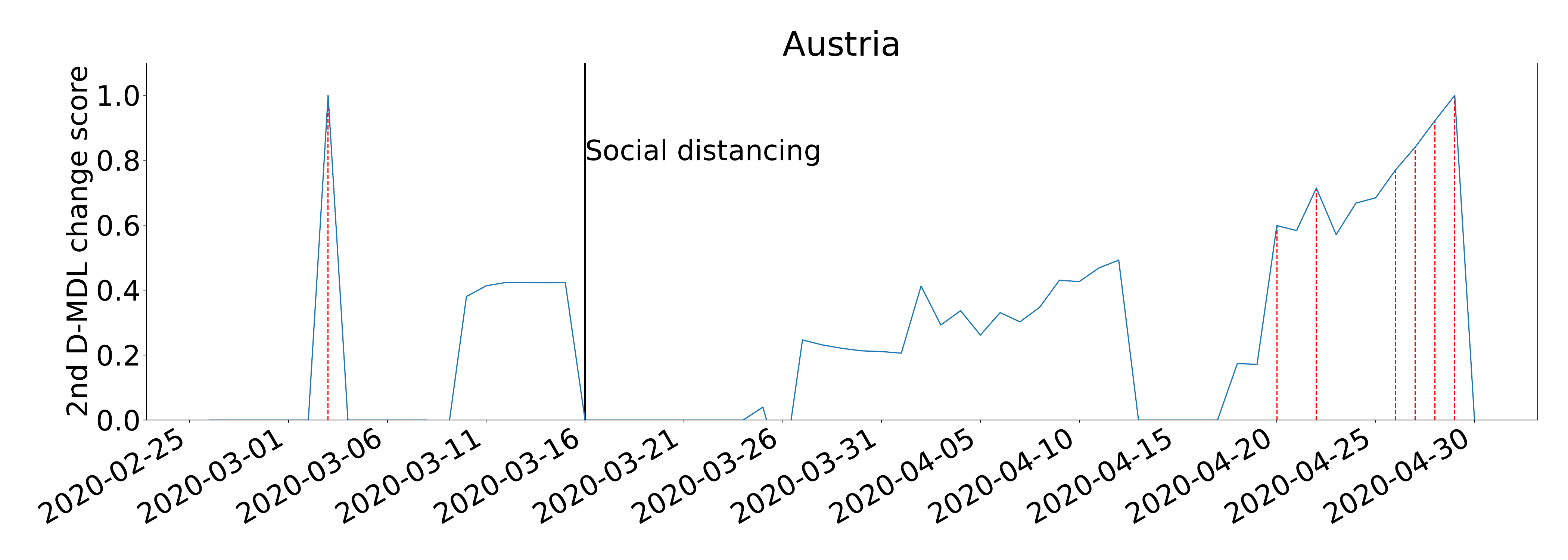} \\
		\end{tabular}
			\caption{\textbf{The results for Austria with Gaussian modeling.} The date on which the social distancing was implemented is marked by a solid line in black. \textbf{a,} the number of daily new cases. \textbf{b,} the change scores produced by the 0th M-DML where the line in blue denotes values of scores and dashed lines in red mark alarms. \textbf{c,} the window sized for the sequential D-DML algorithm with adaptive window where lines in red mark the shrinkage of windows. \textbf{d,} the change scores produced by the 1st D-MDL. \textbf{e,} the change scores produced by the 2nd D-MDL.}
\end{figure}

\begin{figure}[H]  
\centering
\begin{tabular}{cc}
			\textbf{a} & \includegraphics[keepaspectratio, height=3.3cm, valign=T]
			{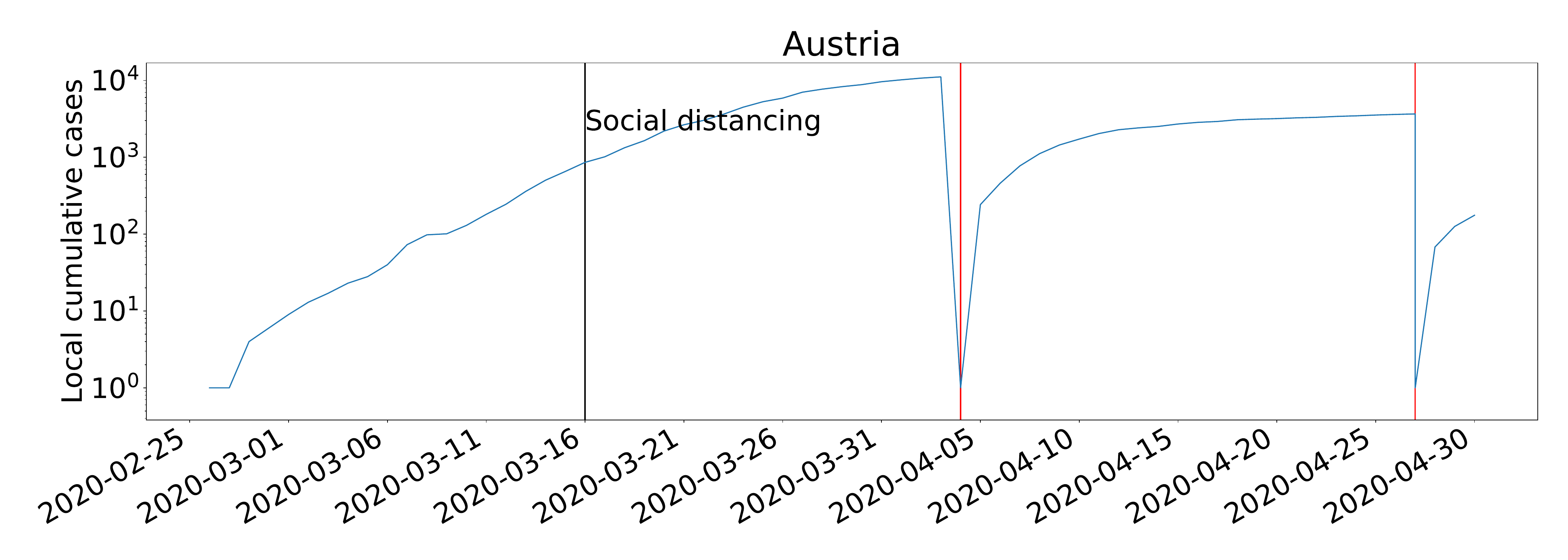} \\
	        \vspace{-0.35cm}
            \textbf{b} & \includegraphics[keepaspectratio, height=3.3cm, valign=T]
			{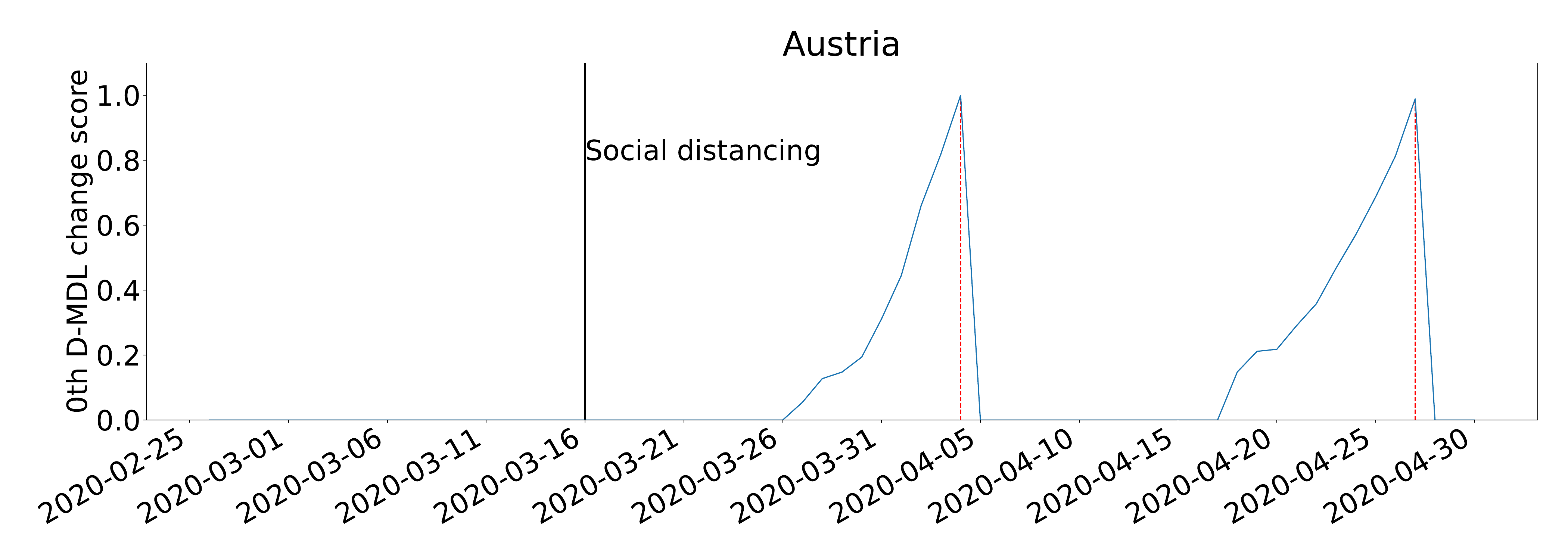}   \\
            \vspace{-0.35cm}
            \textbf{c} & \includegraphics[keepaspectratio, height=3.3cm, valign=T]
			{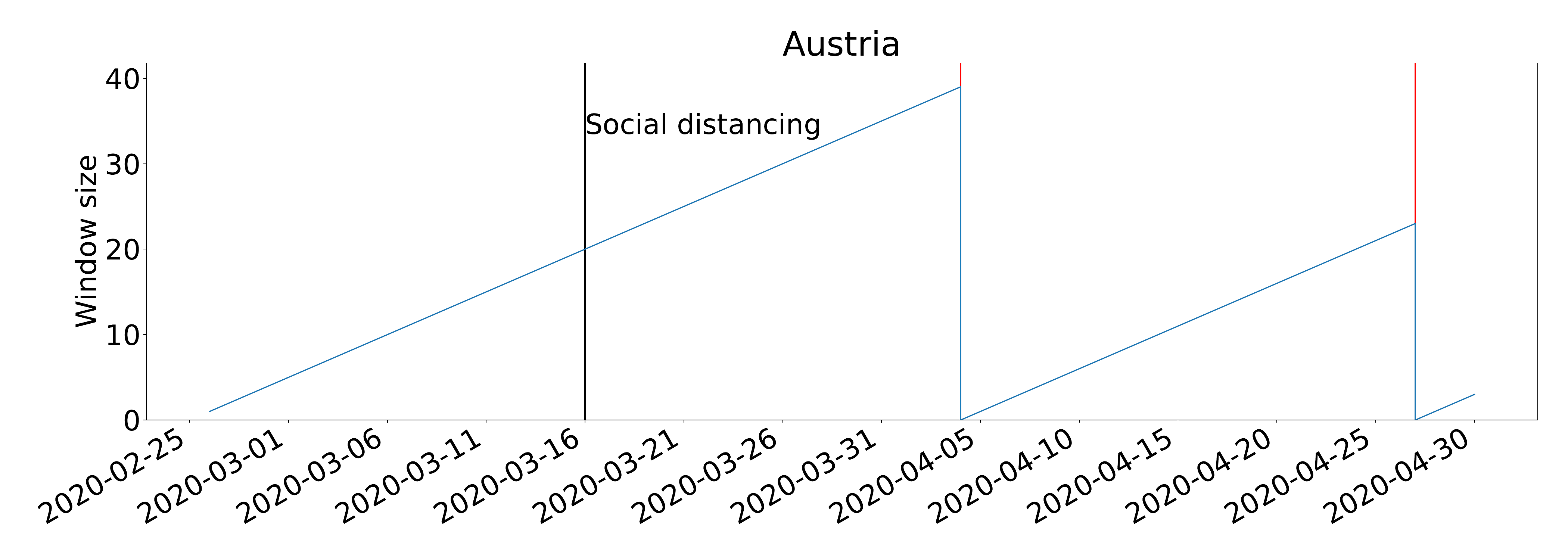} \\
			\vspace{-0.35cm}
			\textbf{d} & \includegraphics[keepaspectratio, height=3.3cm, valign=T]
			{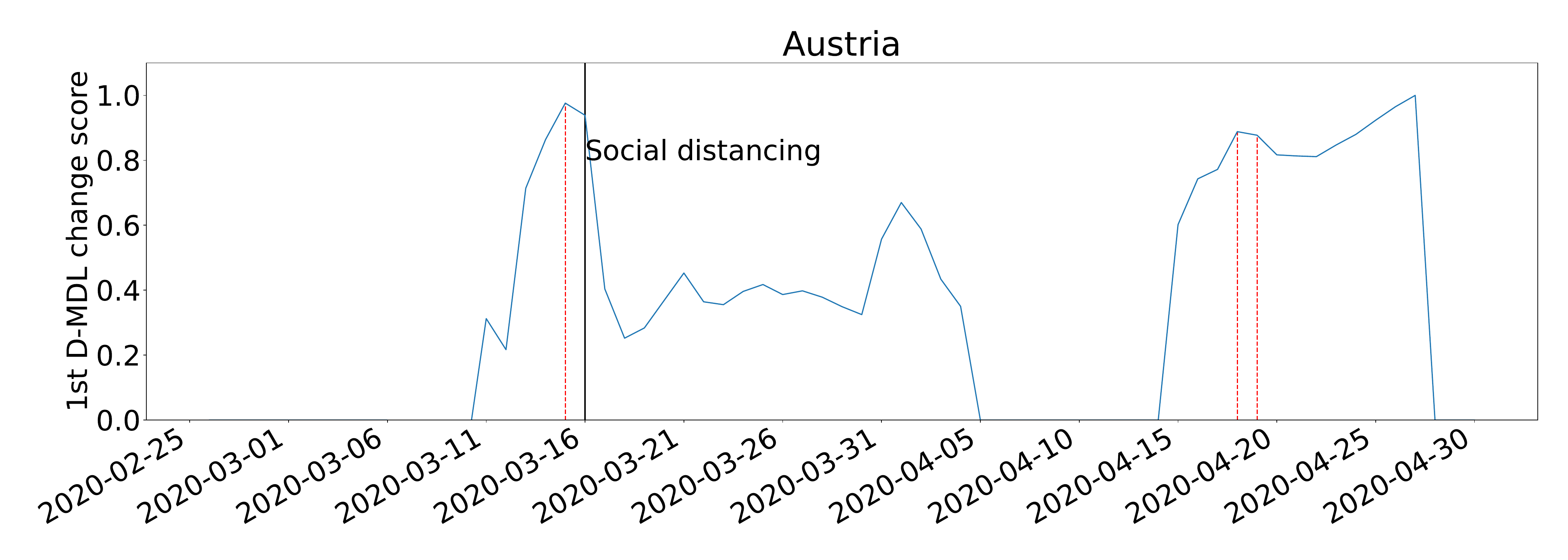} \\
			\vspace{-0.35cm}
			\textbf{e} & \includegraphics[keepaspectratio, height=3.3cm, valign=T]
			{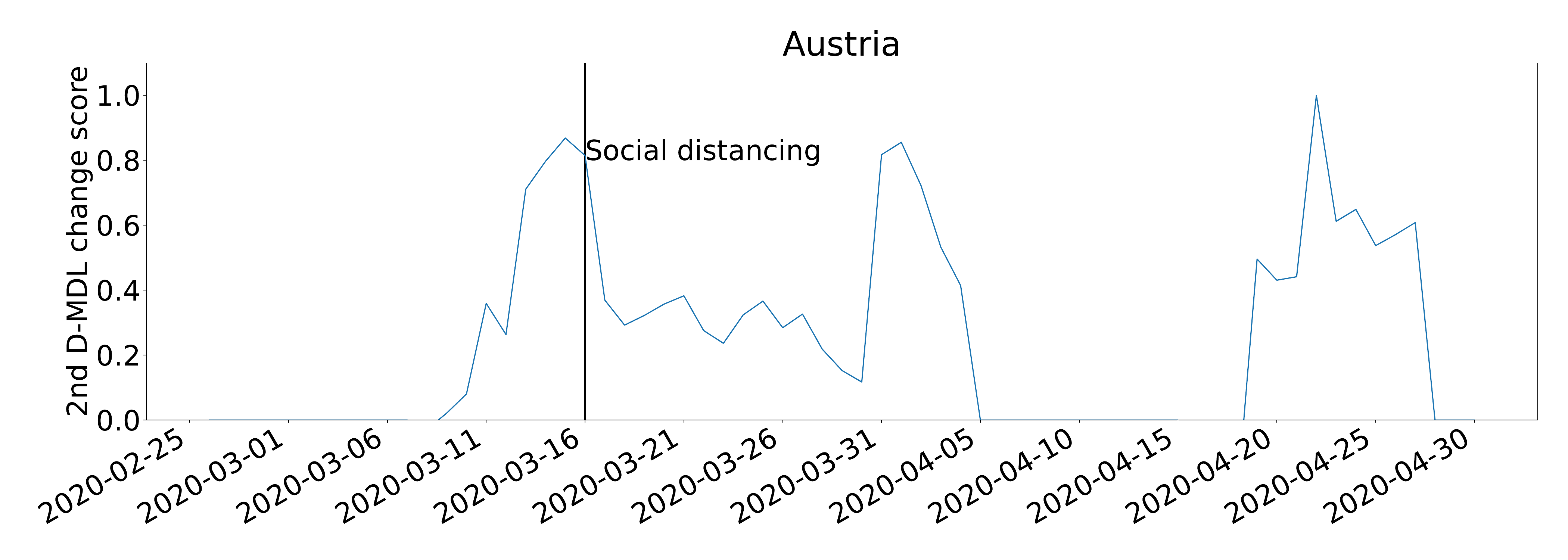} \\
		\end{tabular}
			\caption{\textbf{The results for Austria with exponential modeling.} The date on which the social distancing was implemented is marked by a solid line in black. \textbf{a,} the number of cumulative cases. \textbf{b,} the change scores produced by the 0th M-DML where the line in blue denotes values of scores and dashed lines in red mark alarms. \textbf{c,} the window sized for the sequential D-DML algorithm with adaptive window where lines in red mark the shrinkage of windows. \textbf{d,} the change scores produced by the 1st D-MDL. \textbf{e,} the change scores produced by the 2nd D-MDL.}
\end{figure}

\begin{figure}[H] 
\centering
\begin{tabular}{cc}
		 	\textbf{a} & \includegraphics[keepaspectratio, height=3.3cm, valign=T]
			{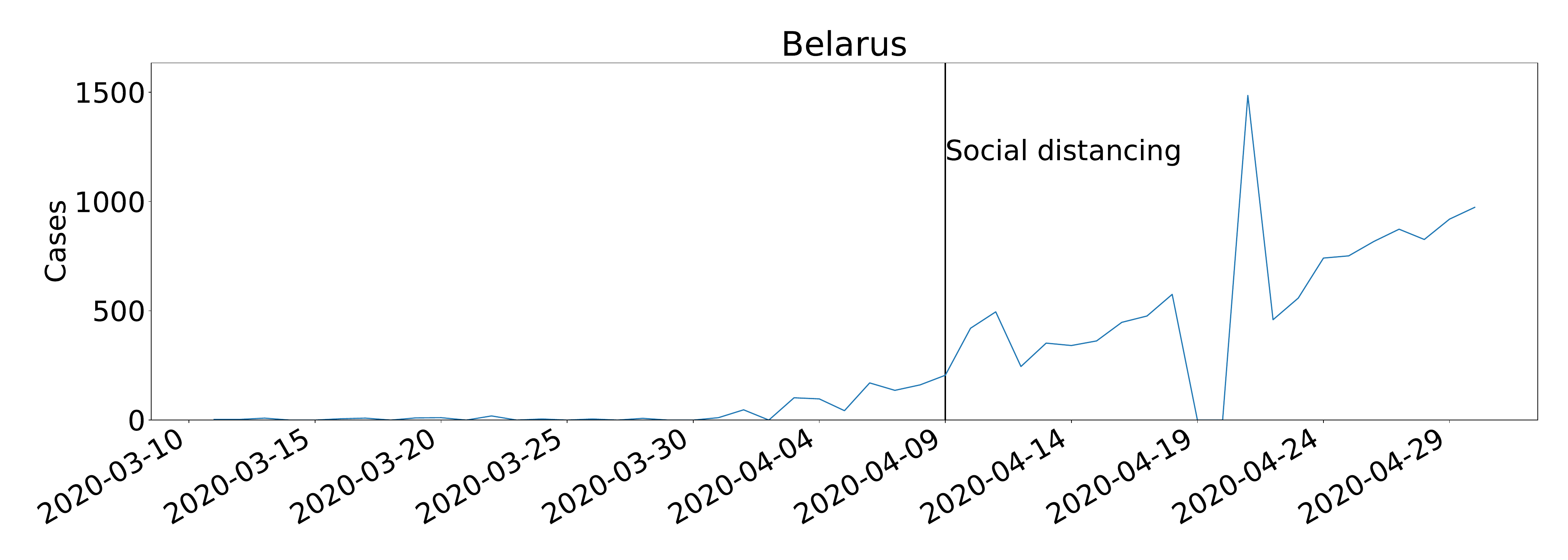} \\
			\vspace{-0.35cm}
	 	    \textbf{b} & \includegraphics[keepaspectratio, height=3.3cm, valign=T]
			{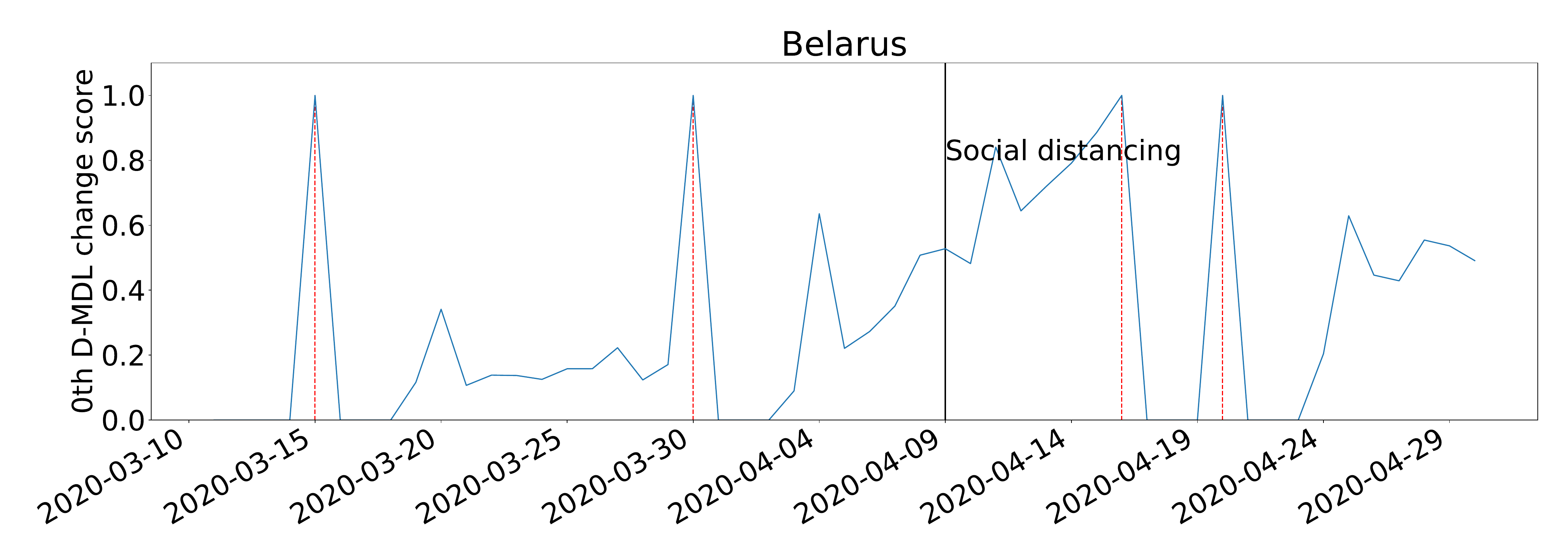}   \\
	        \vspace{-0.35cm}
			\textbf{c} & \includegraphics[keepaspectratio, height=3.3cm, valign=T]
			{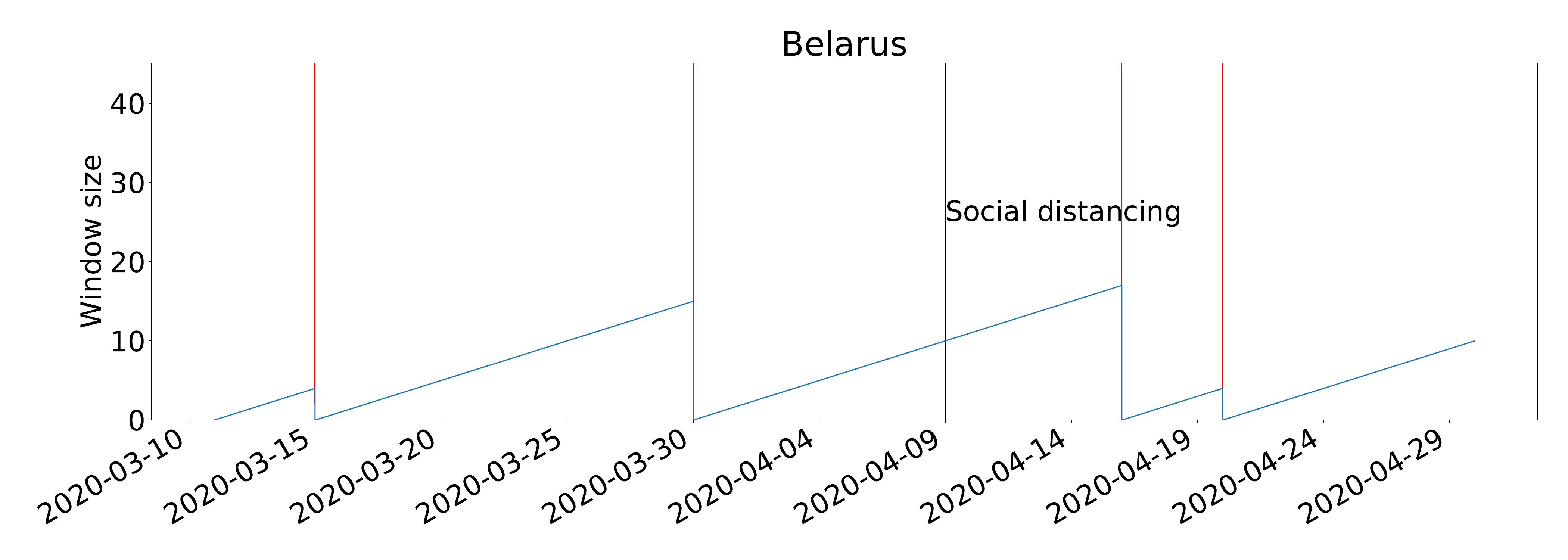} \\
		    \vspace{-0.35cm}
			\textbf{d} & \includegraphics[keepaspectratio, height=3.3cm, valign=T]
			{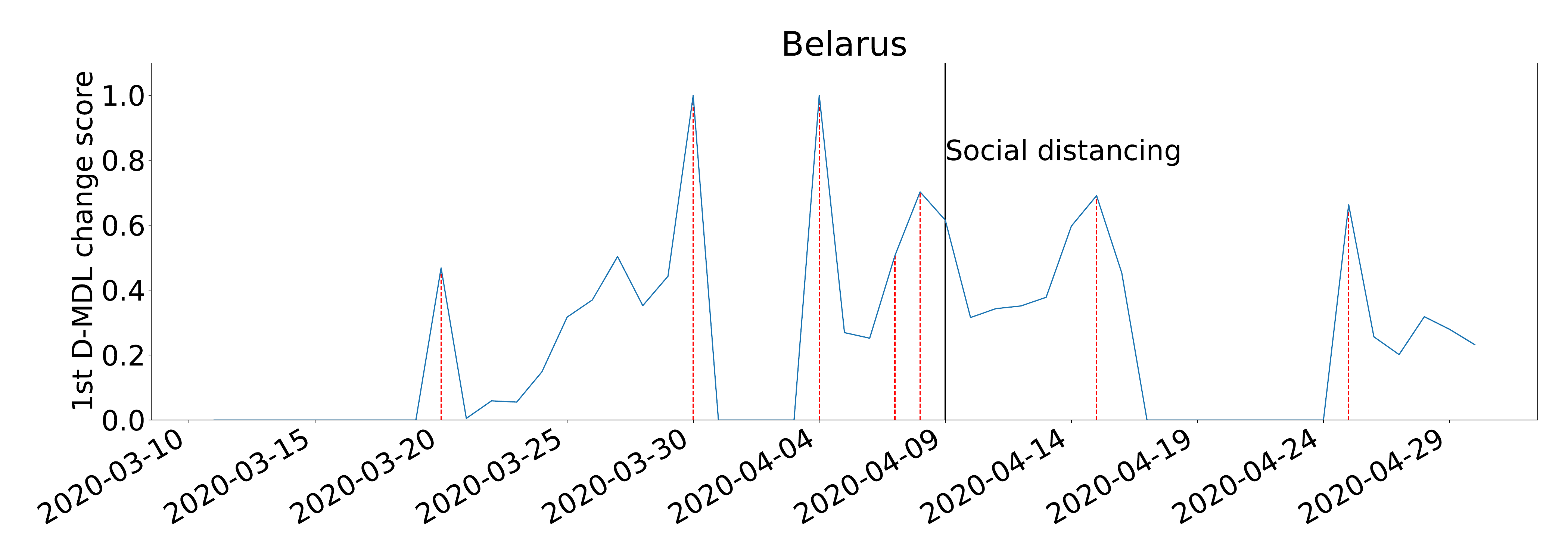} \\
		    \vspace{-0.35cm}
			\textbf{e} & \includegraphics[keepaspectratio, height=3.3cm, valign=T]
			{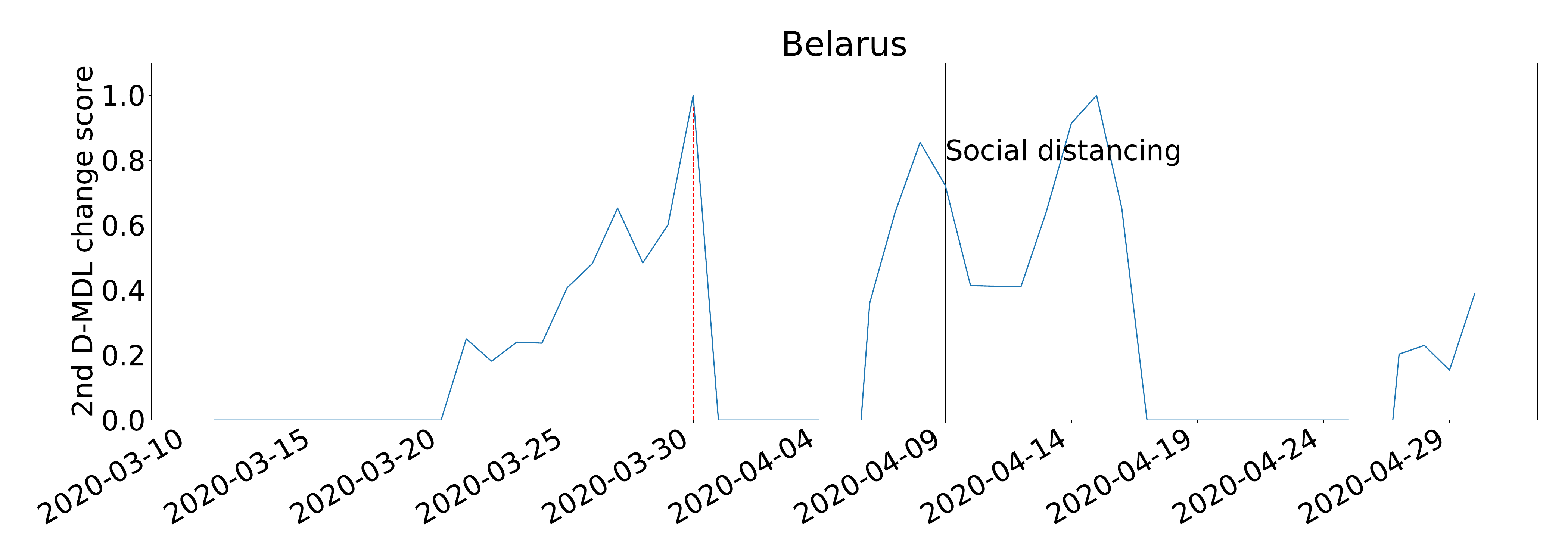} \\
		\end{tabular}
			\caption{\textbf{The results for Belarus with Gaussian modeling.} The date on which the social distancing was implemented is marked by a solid line in black. \textbf{a,} the number of daily new cases. \textbf{b,} the change scores produced by the 0th M-DML where the line in blue denotes values of scores and dashed lines in red mark alarms. \textbf{c,} the window sized for the sequential D-DML algorithm with adaptive window where lines in red mark the shrinkage of windows. \textbf{d,} the change scores produced by the 1st D-MDL. \textbf{e,} the change scores produced by the 2nd D-MDL.}
\end{figure}

\begin{figure}[H]  
\centering
\begin{tabular}{cc}
			\textbf{a} & \includegraphics[keepaspectratio, height=3.3cm, valign=T]
			{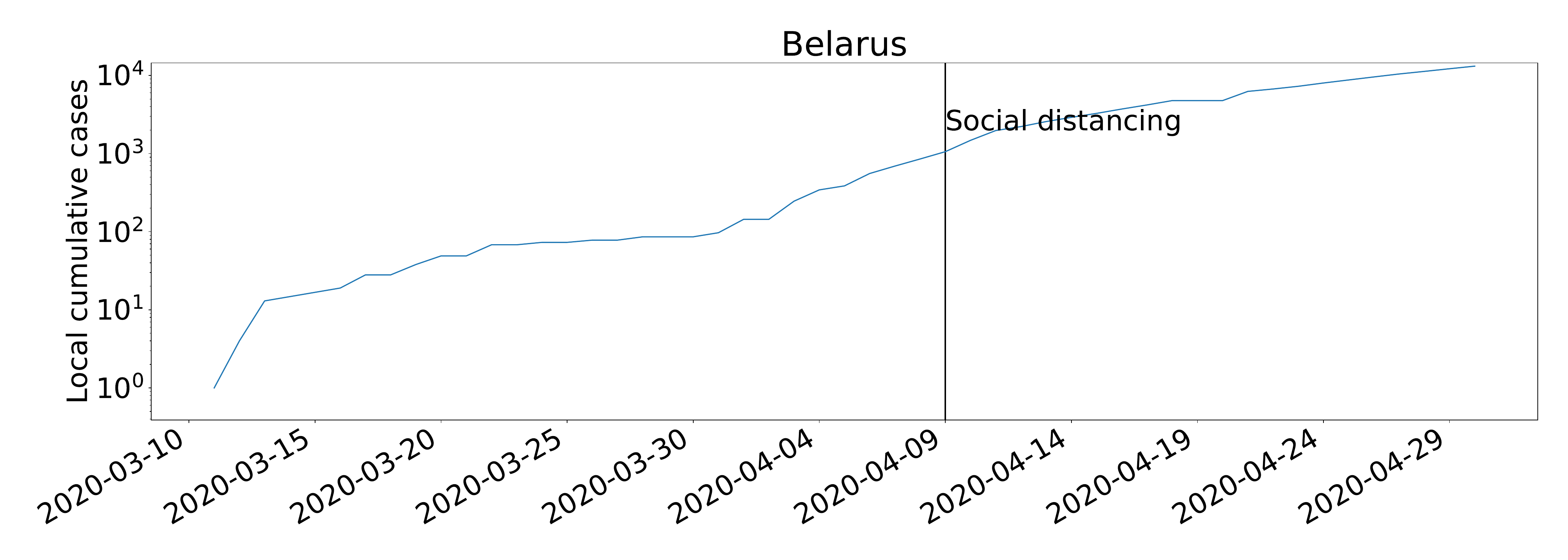} \\
	        \vspace{-0.35cm}
            \textbf{b} & \includegraphics[keepaspectratio, height=3.3cm, valign=T]
			{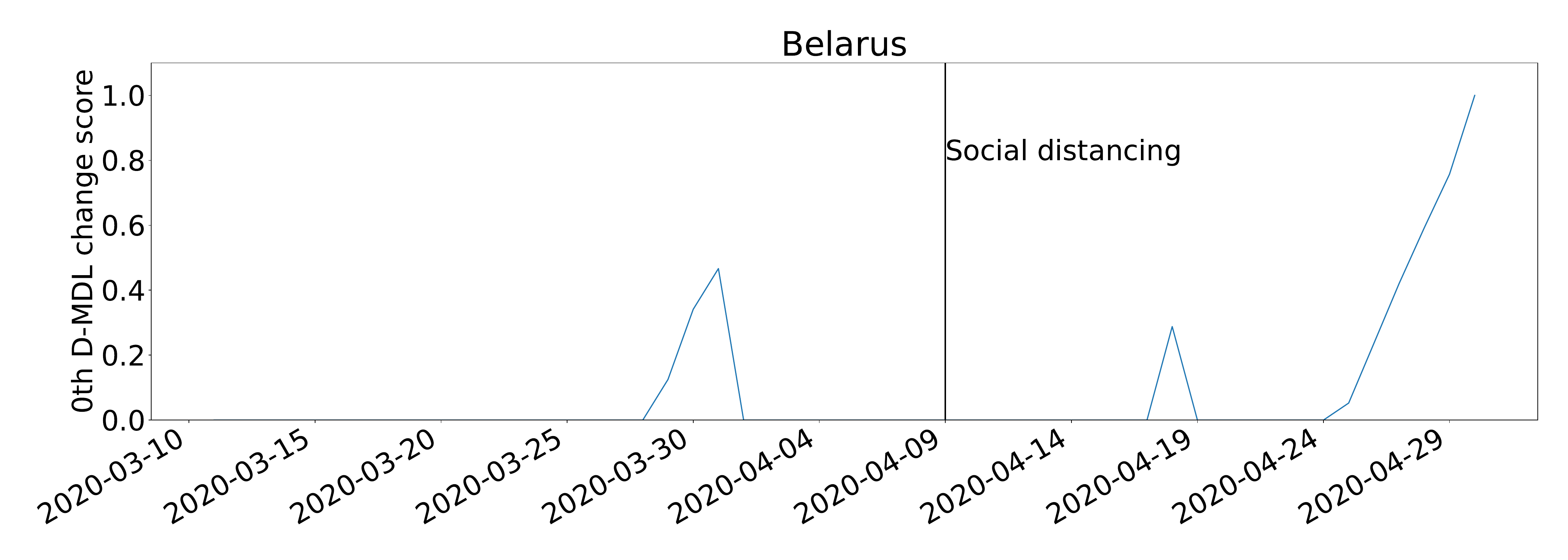}   \\
            \vspace{-0.35cm}
            \textbf{c} & \includegraphics[keepaspectratio, height=3.3cm, valign=T]
			{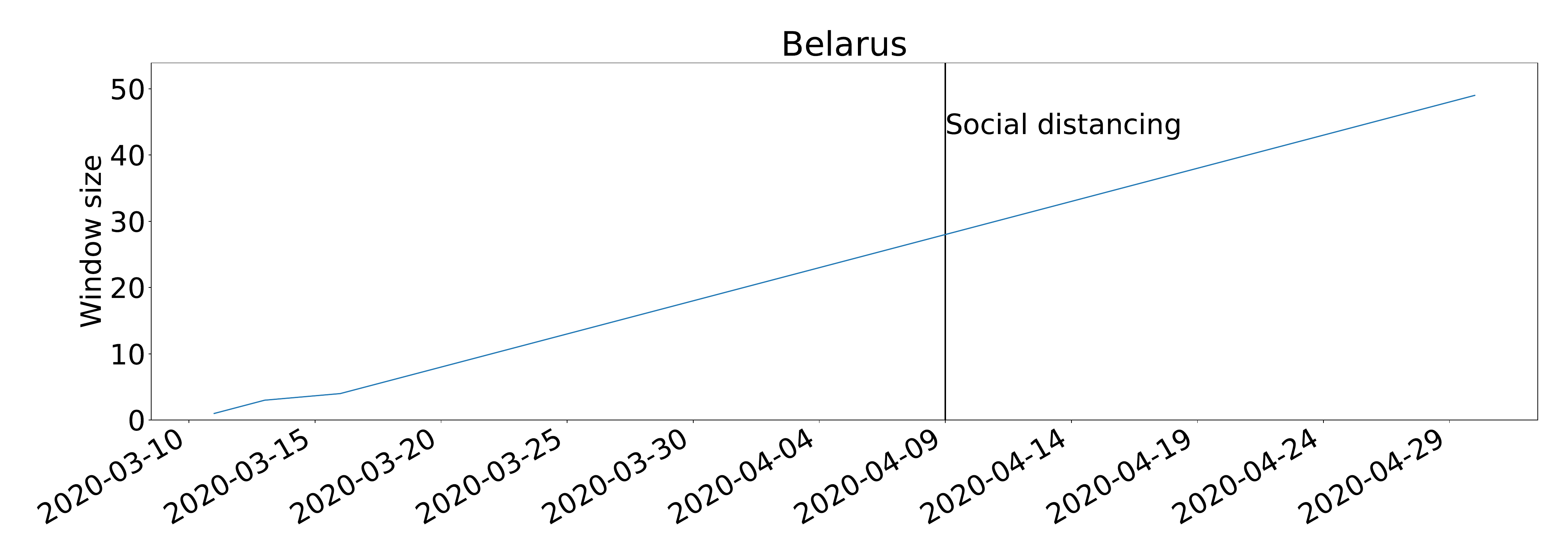} \\
			\vspace{-0.35cm}
			\textbf{d} & \includegraphics[keepaspectratio, height=3.3cm, valign=T]
			{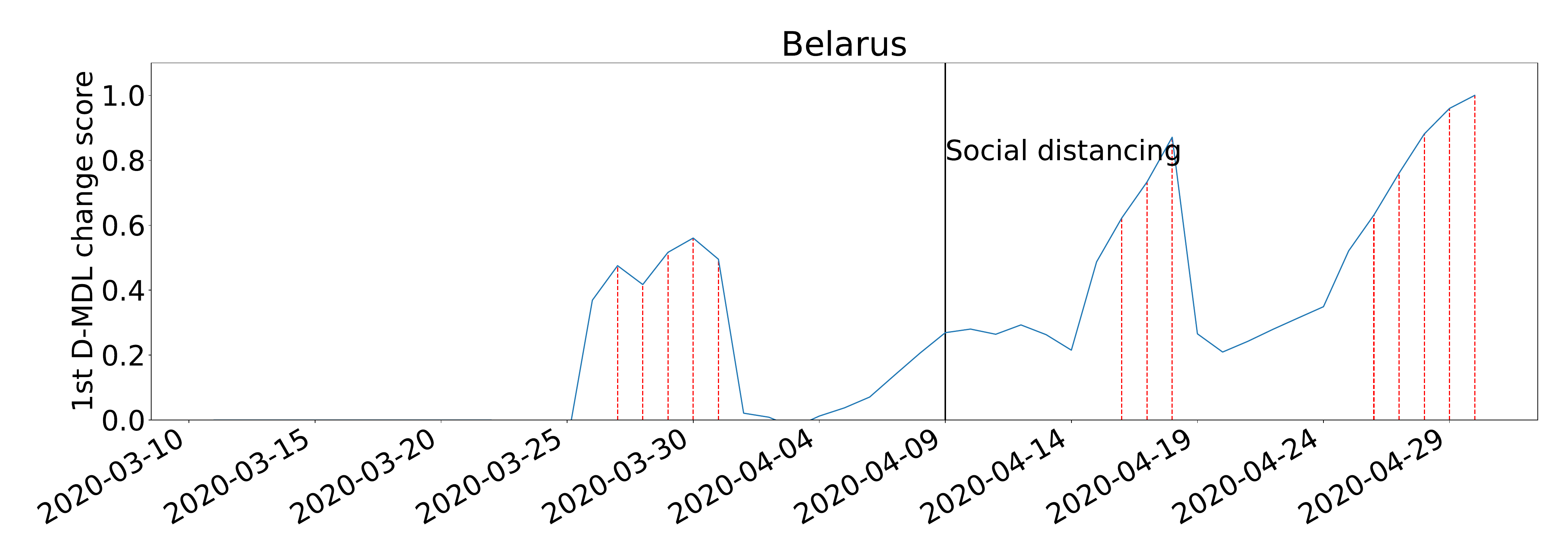} \\
			\vspace{-0.35cm}
			\textbf{e} & \includegraphics[keepaspectratio, height=3.3cm, valign=T]
			{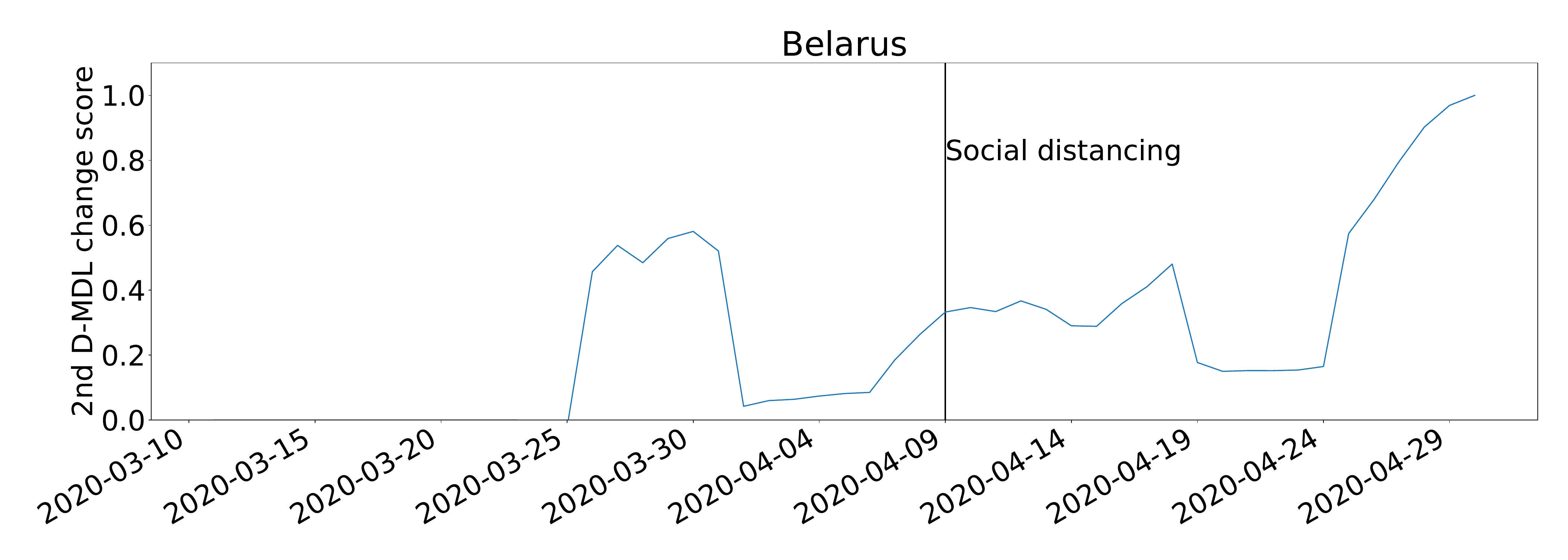} \\
		\end{tabular}
			\caption{\textbf{The results for Belarus with exponential modeling.} The date on which the social distancing was implemented is marked by a solid line in black. \textbf{a,} the number of cumulative cases. \textbf{b,} the change scores produced by the 0th M-DML where the line in blue denotes values of scores and dashed lines in red mark alarms. \textbf{c,} the window sized for the sequential D-DML algorithm with adaptive window where lines in red mark the shrinkage of windows. \textbf{d,} the change scores produced by the 1st D-MDL. \textbf{e,} the change scores produced by the 2nd D-MDL.}
\end{figure}

\begin{figure}[H] 
\centering
\begin{tabular}{cc}
		 	\textbf{a} & \includegraphics[keepaspectratio, height=3.3cm, valign=T]
			{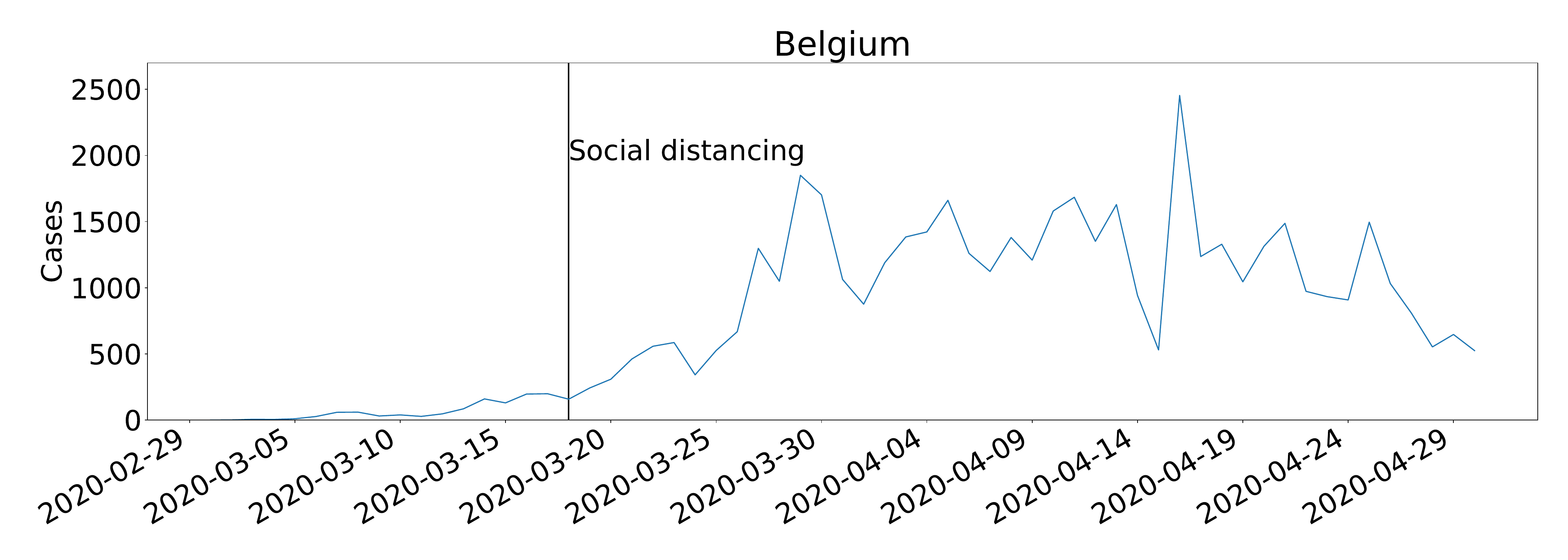} \\
			\vspace{-0.35cm}
	 	    \textbf{b} & \includegraphics[keepaspectratio, height=3.3cm, valign=T]
			{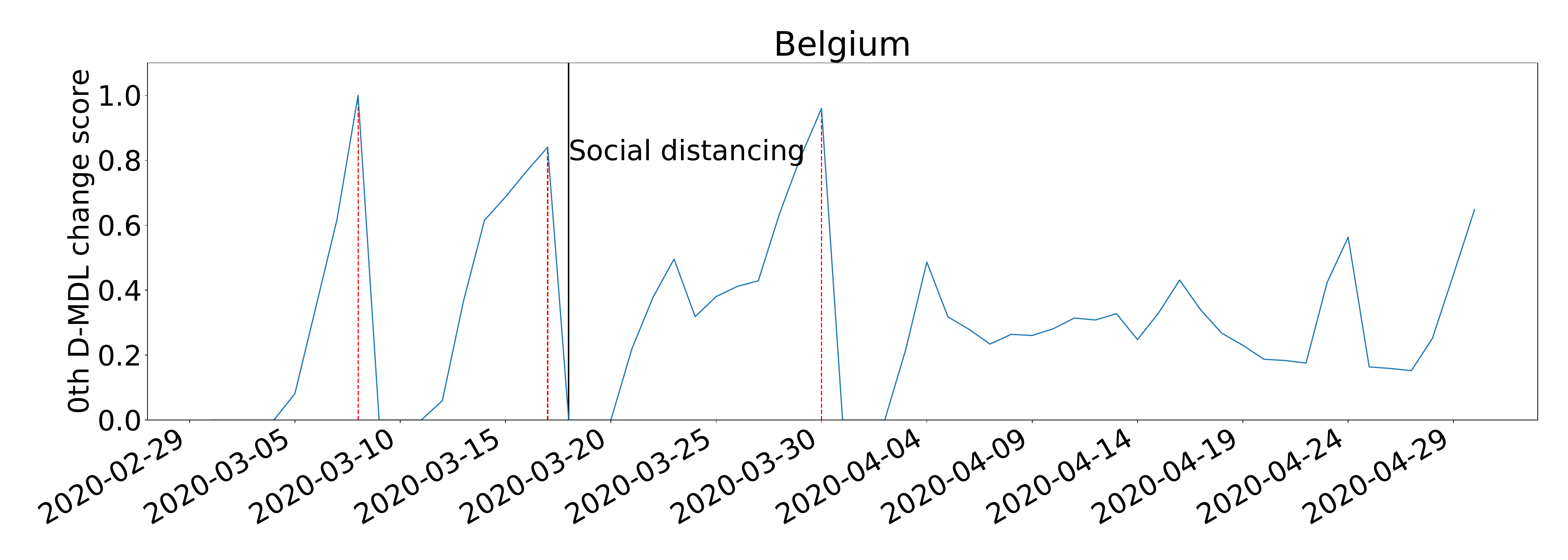}   \\
	        \vspace{-0.35cm}
			\textbf{c} & \includegraphics[keepaspectratio, height=3.3cm, valign=T]
			{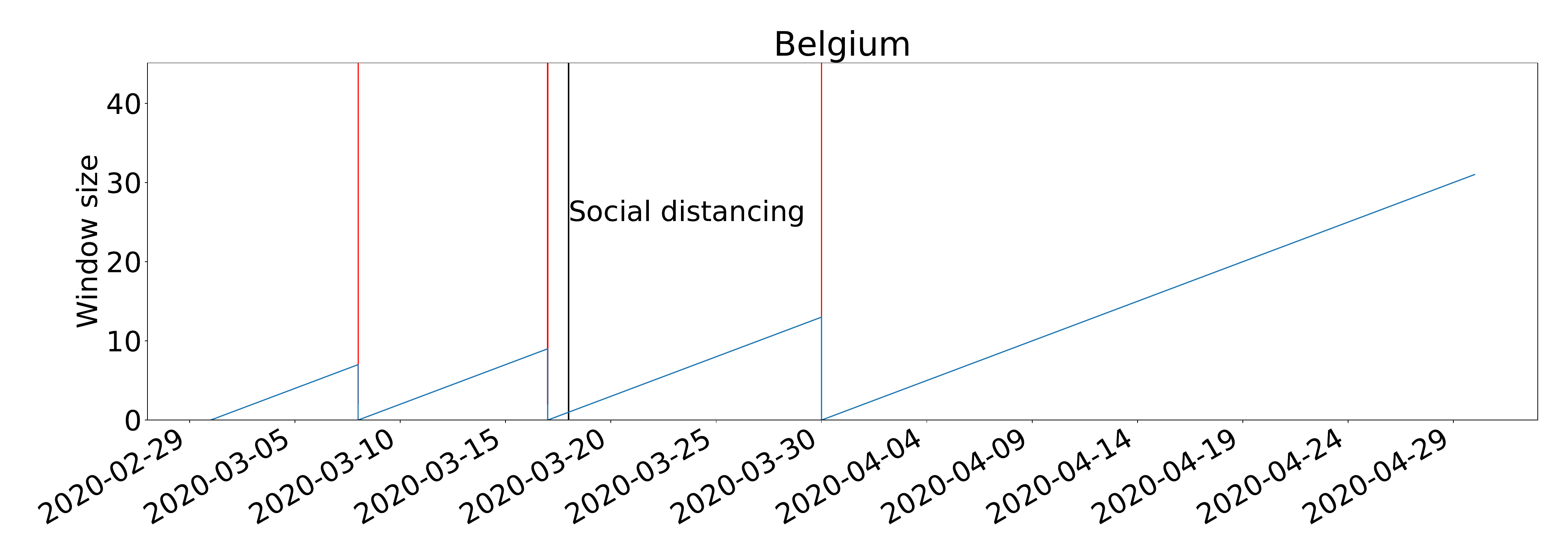} \\
		    \vspace{-0.35cm}
			\textbf{d} & \includegraphics[keepaspectratio, height=3.3cm, valign=T]
			{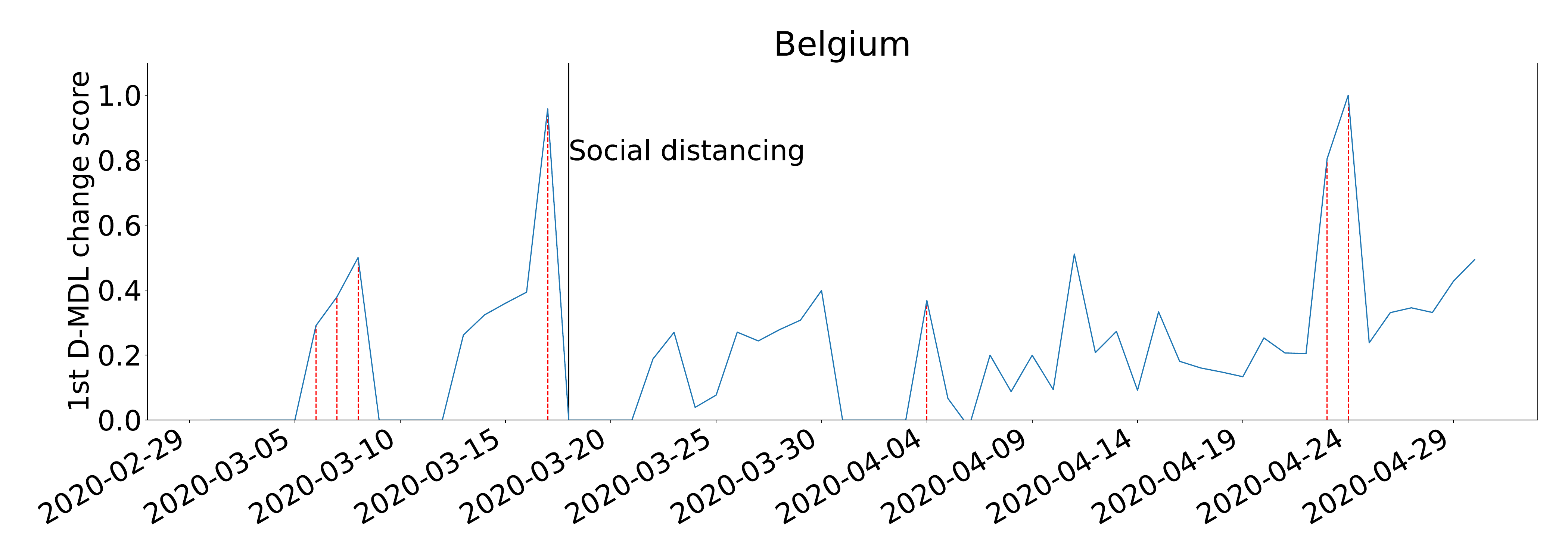} \\
		    \vspace{-0.35cm}
			\textbf{e} & \includegraphics[keepaspectratio, height=3.3cm, valign=T]
			{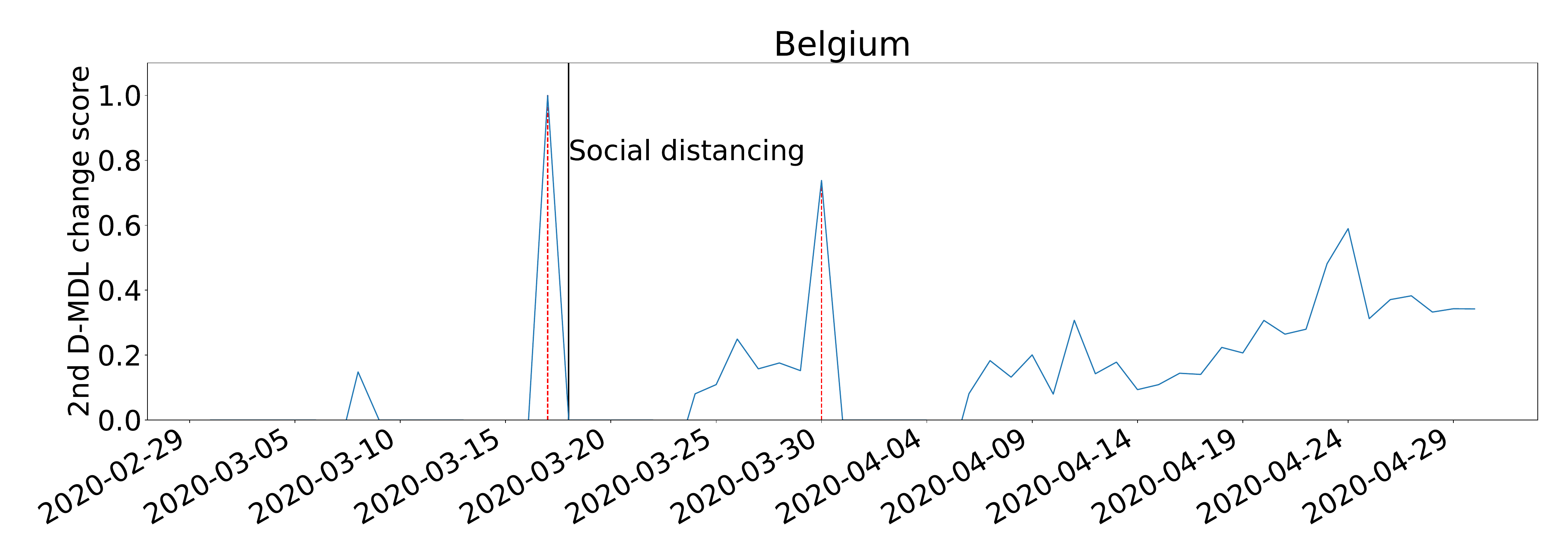} \\
		\end{tabular}
			\caption{\textbf{The results for Belgium with Gaussian modeling.} The date on which the social distancing was implemented is marked by a solid line in black. \textbf{a,} the number of daily new cases. \textbf{b,} the change scores produced by the 0th M-DML where the line in blue denotes values of scores and dashed lines in red mark alarms. \textbf{c,} the window sized for the sequential D-DML algorithm with adaptive window where lines in red mark the shrinkage of windows. \textbf{d,} the change scores produced by the 1st D-MDL. \textbf{e,} the change scores produced by the 2nd D-MDL.}
\end{figure}

\begin{figure}[H]  
\centering
\begin{tabular}{cc}
			\textbf{a} & \includegraphics[keepaspectratio, height=3.3cm, valign=T]
			{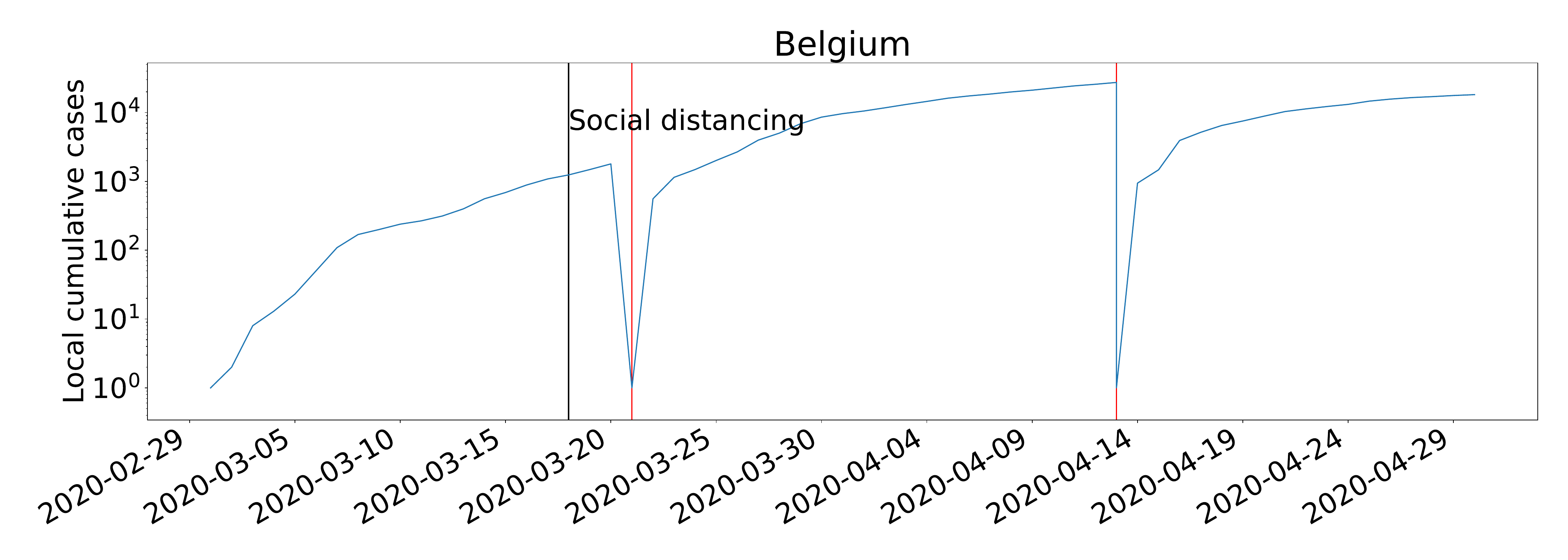} \\
	        \vspace{-0.35cm}
            \textbf{b} & \includegraphics[keepaspectratio, height=3.3cm, valign=T]
			{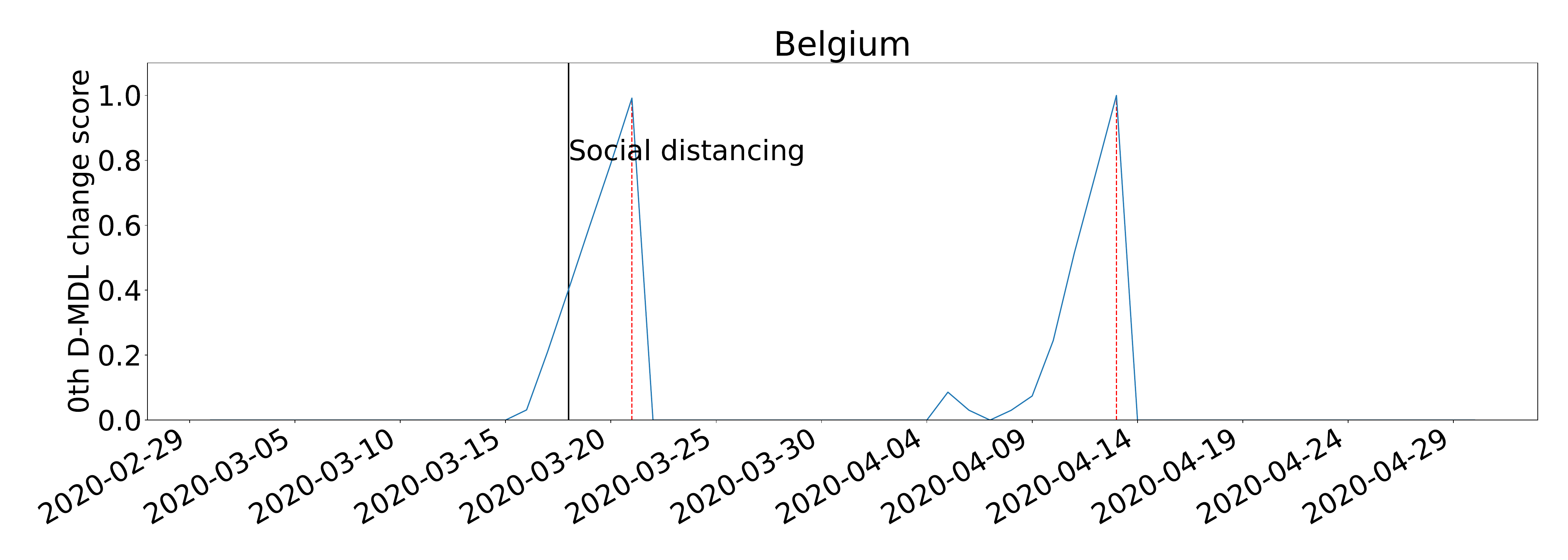}   \\
            \vspace{-0.35cm}
            \textbf{c} & \includegraphics[keepaspectratio, height=3.3cm, valign=T]
			{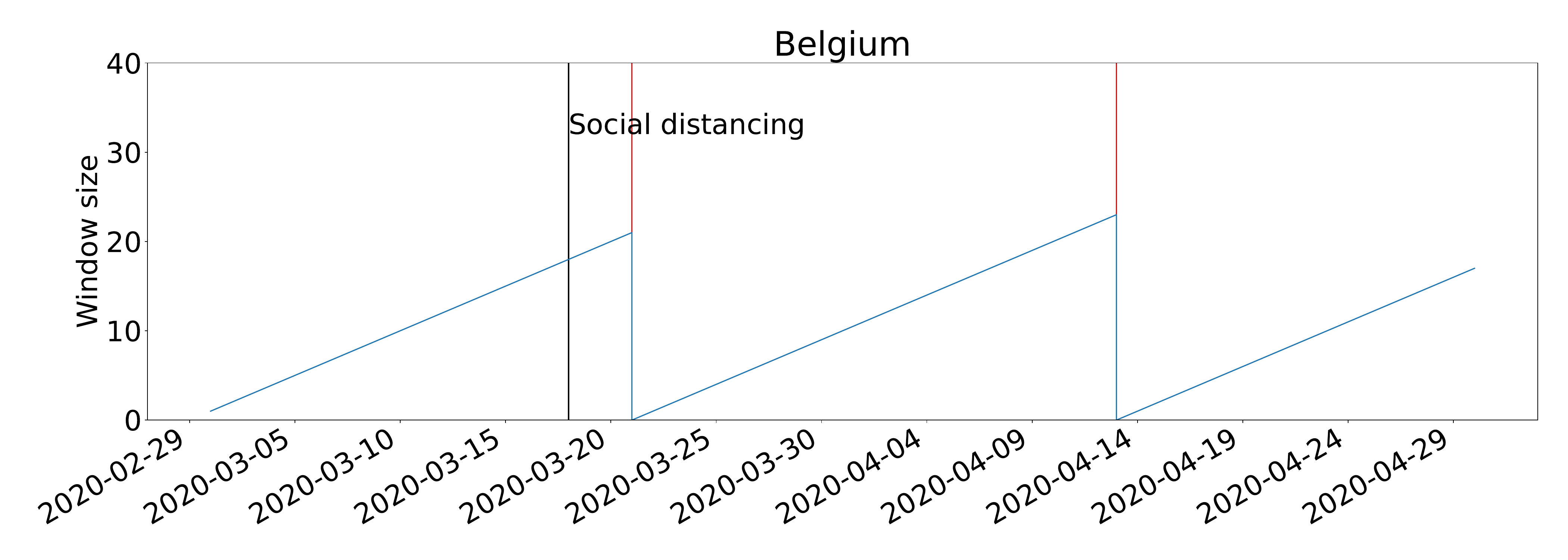} \\
			\vspace{-0.35cm}
			\textbf{d} & \includegraphics[keepaspectratio, height=3.3cm, valign=T]
			{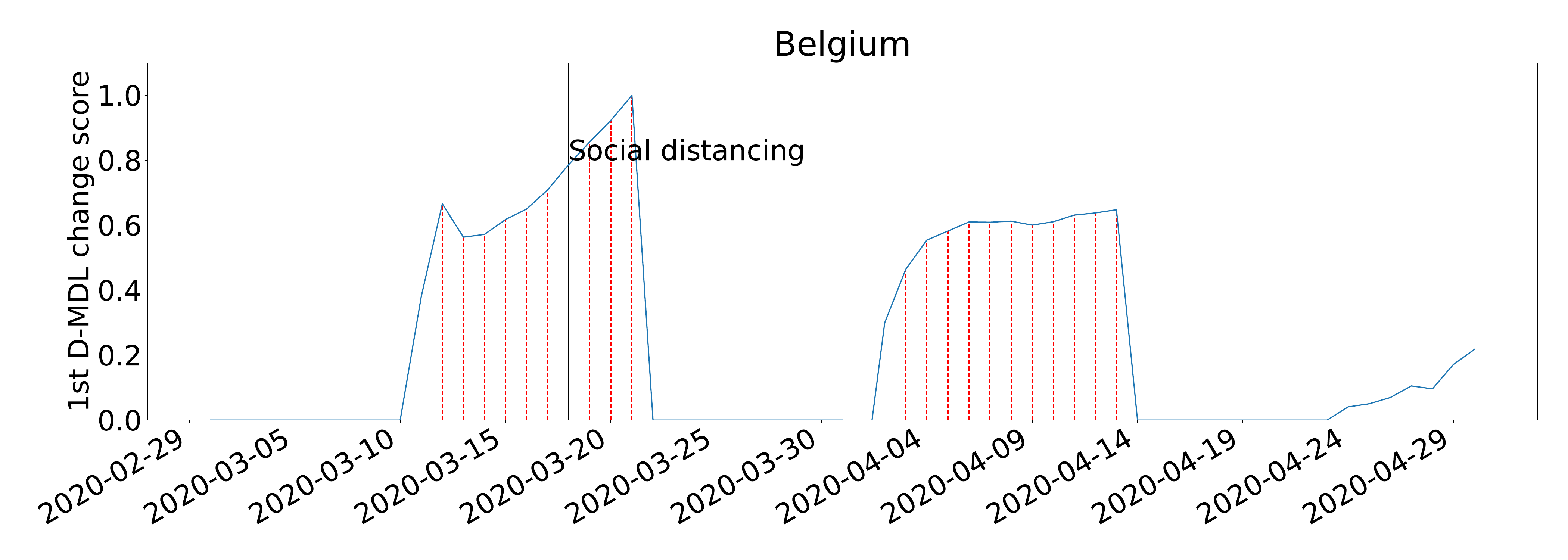} \\
			\vspace{-0.35cm}
			\textbf{e} & \includegraphics[keepaspectratio, height=3.3cm, valign=T]
			{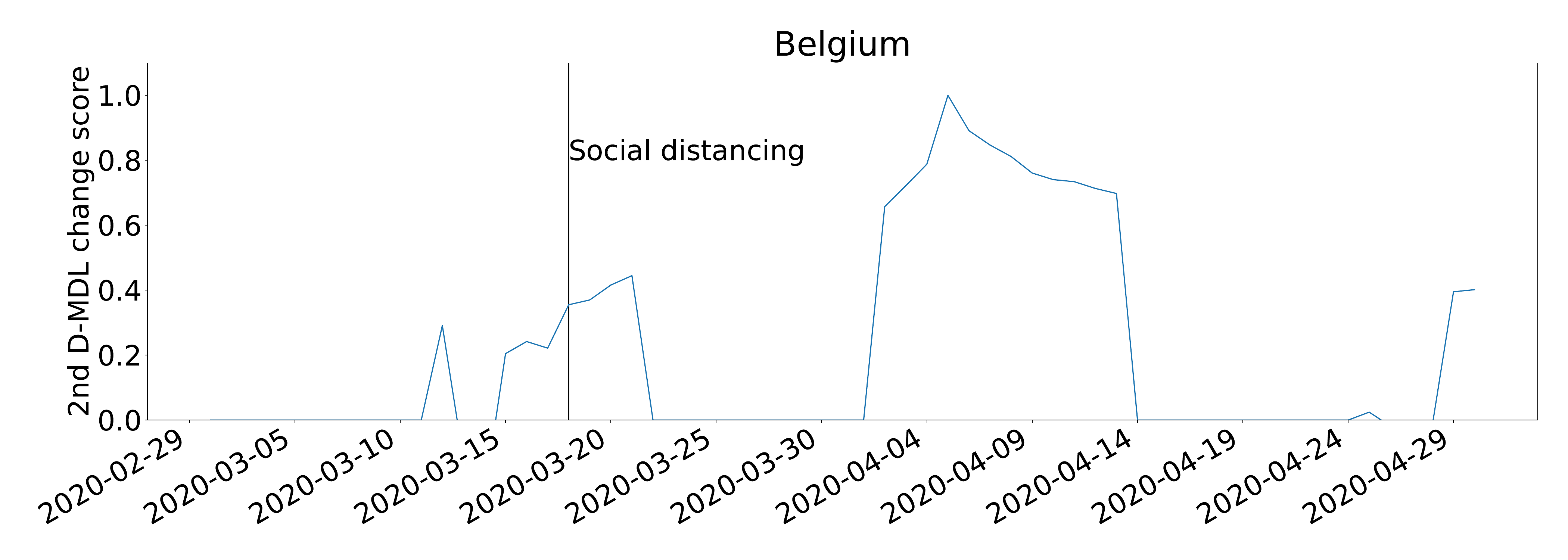} \\
		\end{tabular}
			\caption{\textbf{The results for Belgium with exponential modeling.} The date on which the social distancing was implemented is marked by a solid line in black. \textbf{a,} the number of cumulative cases. \textbf{b,} the change scores produced by the 0th M-DML where the line in blue denotes values of scores and dashed lines in red mark alarms. \textbf{c,} the window sized for the sequential D-DML algorithm with adaptive window where lines in red mark the shrinkage of windows. \textbf{d,} the change scores produced by the 1st D-MDL. \textbf{e,} the change scores produced by the 2nd D-MDL.}
\end{figure}

\begin{figure}[H] 
\centering
\begin{tabular}{cc}
		 	\textbf{a} & \includegraphics[keepaspectratio, height=3.3cm, valign=T]
			{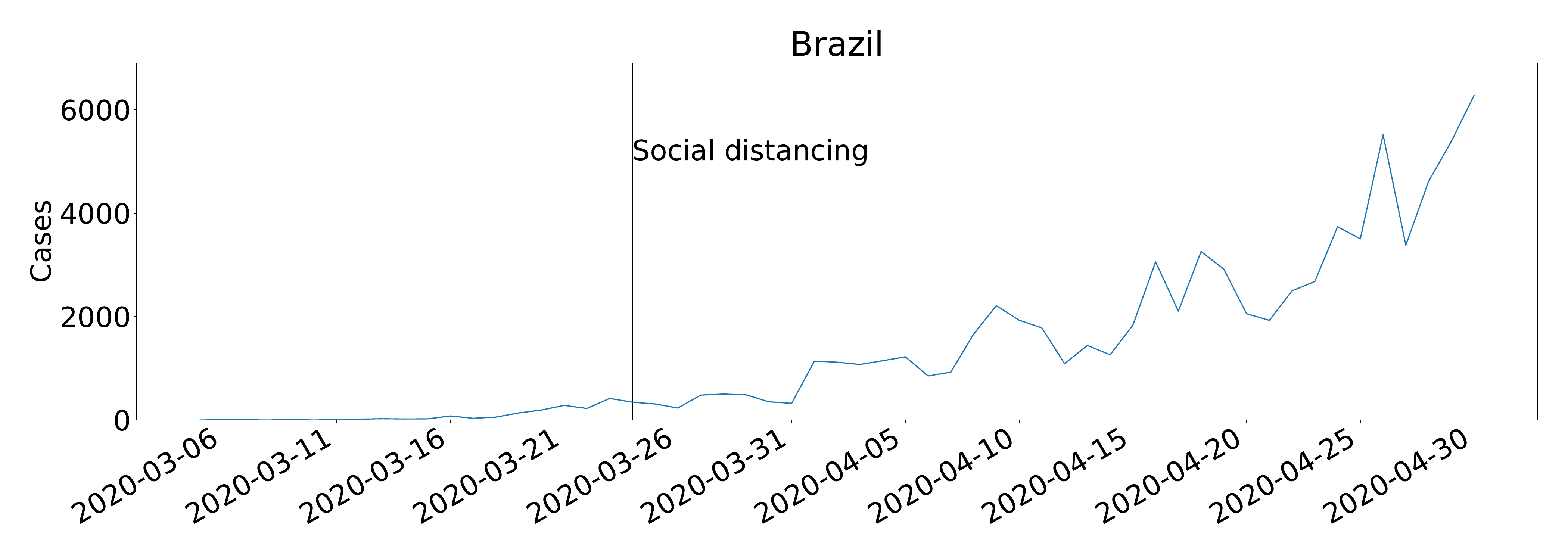} \\
			\vspace{-0.35cm}
	 	    \textbf{b} & \includegraphics[keepaspectratio, height=3.3cm, valign=T]
			{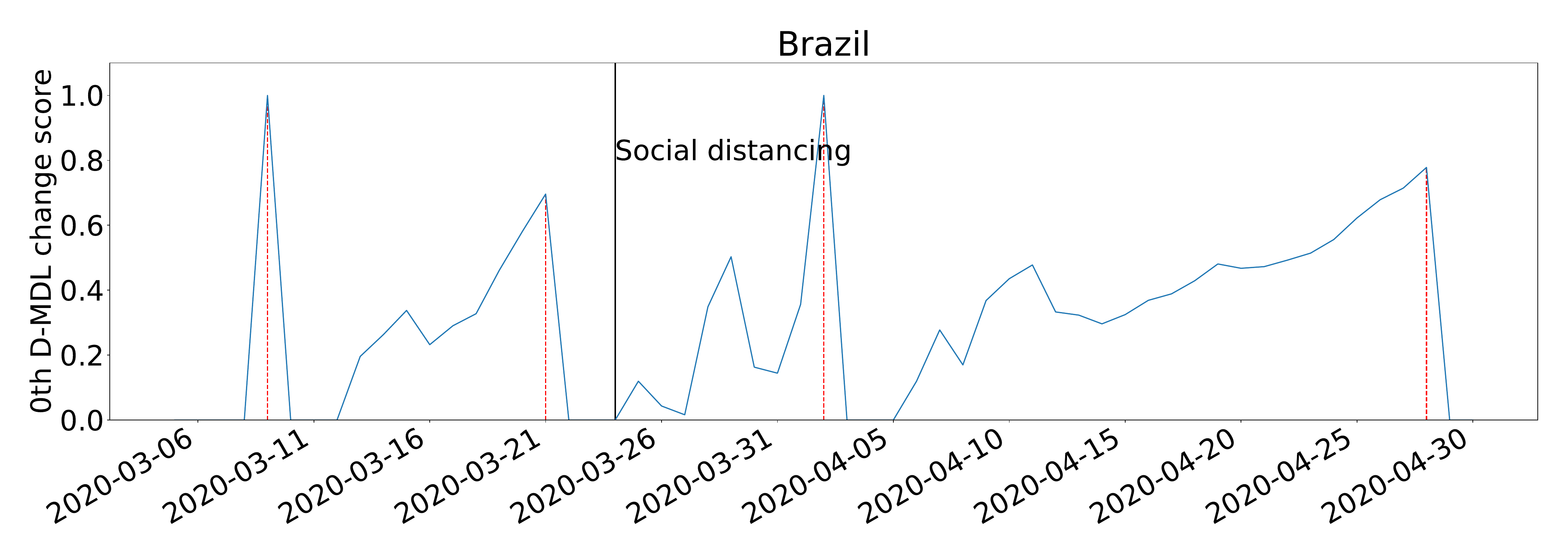}   \\
	        \vspace{-0.35cm}
			\textbf{c} & \includegraphics[keepaspectratio, height=3.3cm, valign=T]
			{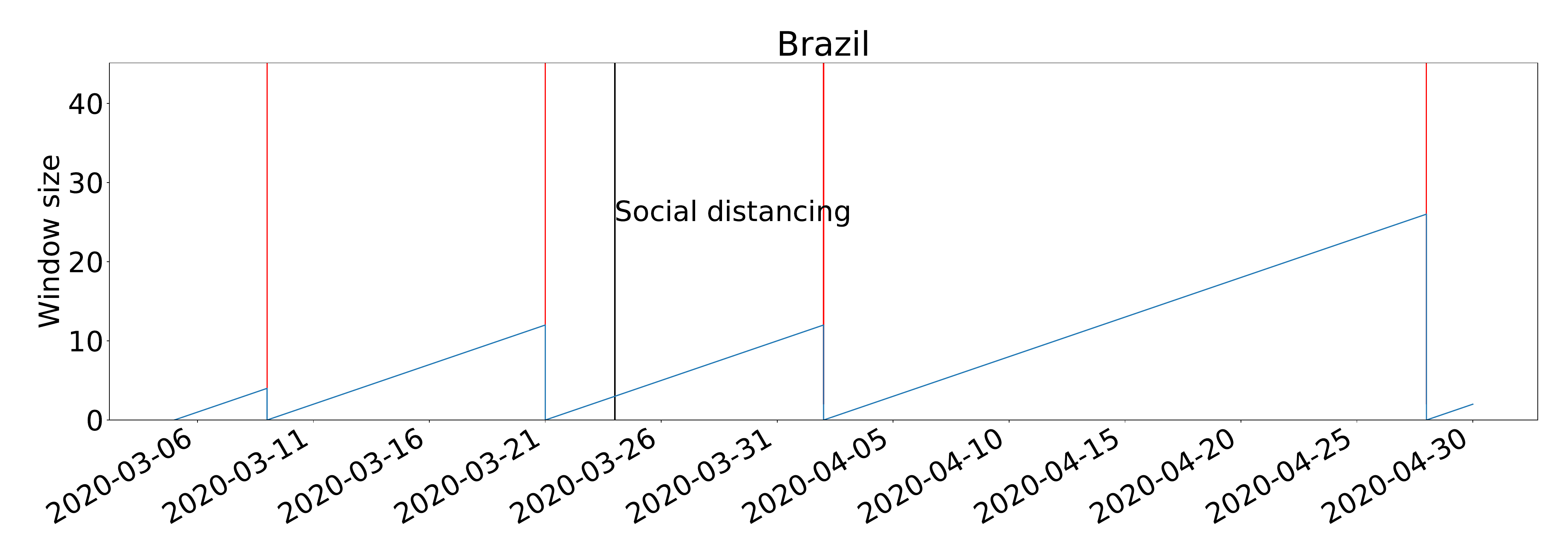} \\
		    \vspace{-0.35cm}
			\textbf{d} & \includegraphics[keepaspectratio, height=3.3cm, valign=T]
			{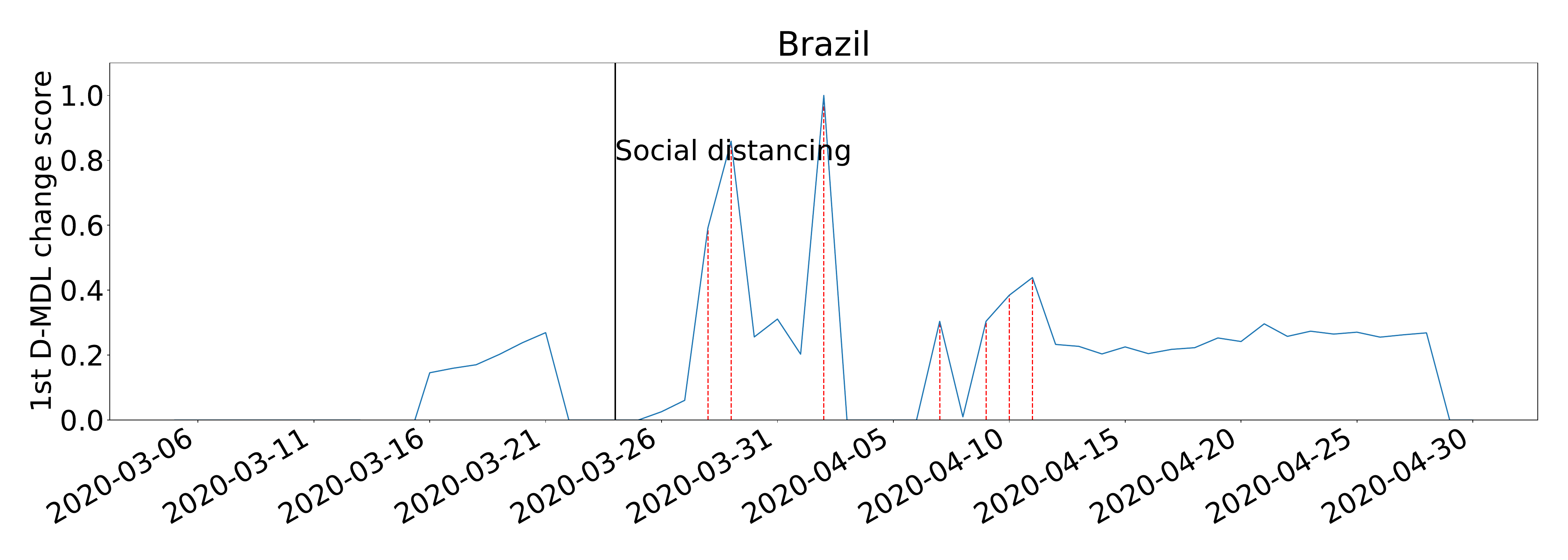} \\
		    \vspace{-0.35cm}
			\textbf{e} & \includegraphics[keepaspectratio, height=3.3cm, valign=T]
			{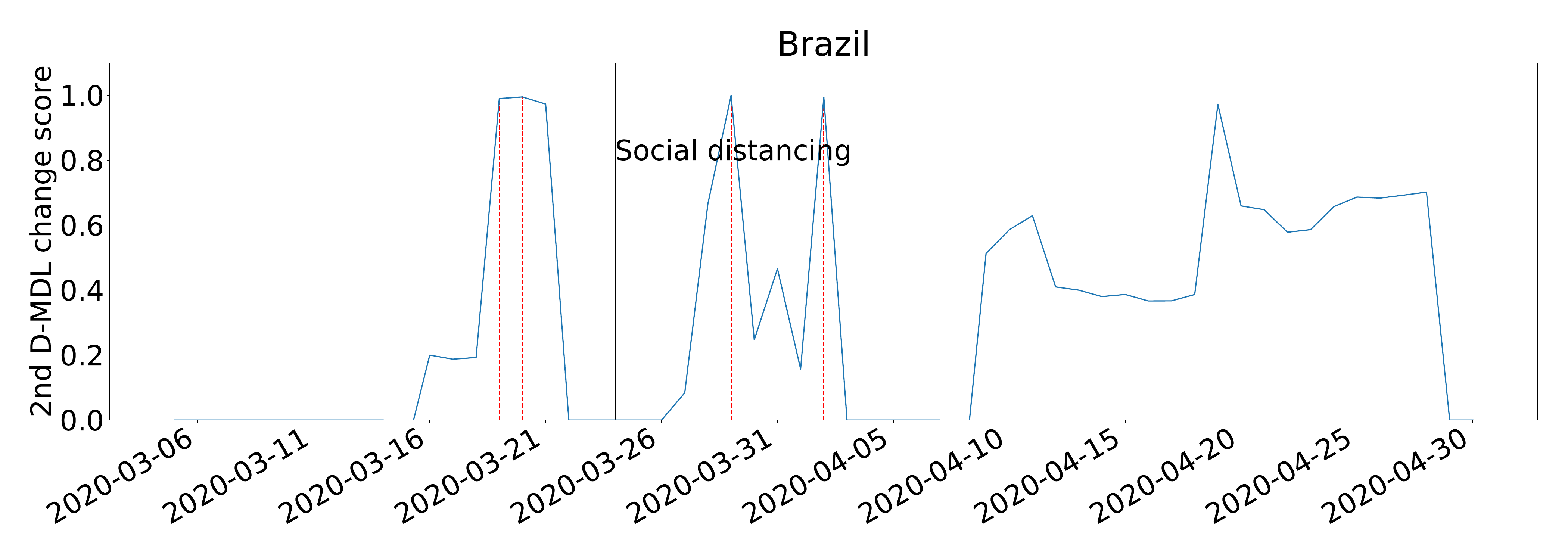} \\
		\end{tabular}
			\caption{\textbf{The results for Brazil with Gaussian modeling.} The date on which the social distancing was implemented is marked by a solid line in black. \textbf{a,} the number of daily new cases. \textbf{b,} the change scores produced by the 0th M-DML where the line in blue denotes values of scores and dashed lines in red mark alarms. \textbf{c,} the window sized for the sequential D-DML algorithm with adaptive window where lines in red mark the shrinkage of windows. \textbf{d,} the change scores produced by the 1st D-MDL. \textbf{e,} the change scores produced by the 2nd D-MDL.}
\end{figure}

\begin{figure}[H]  
\centering
\begin{tabular}{cc}
			\textbf{a} & \includegraphics[keepaspectratio, height=3.3cm, valign=T]
			{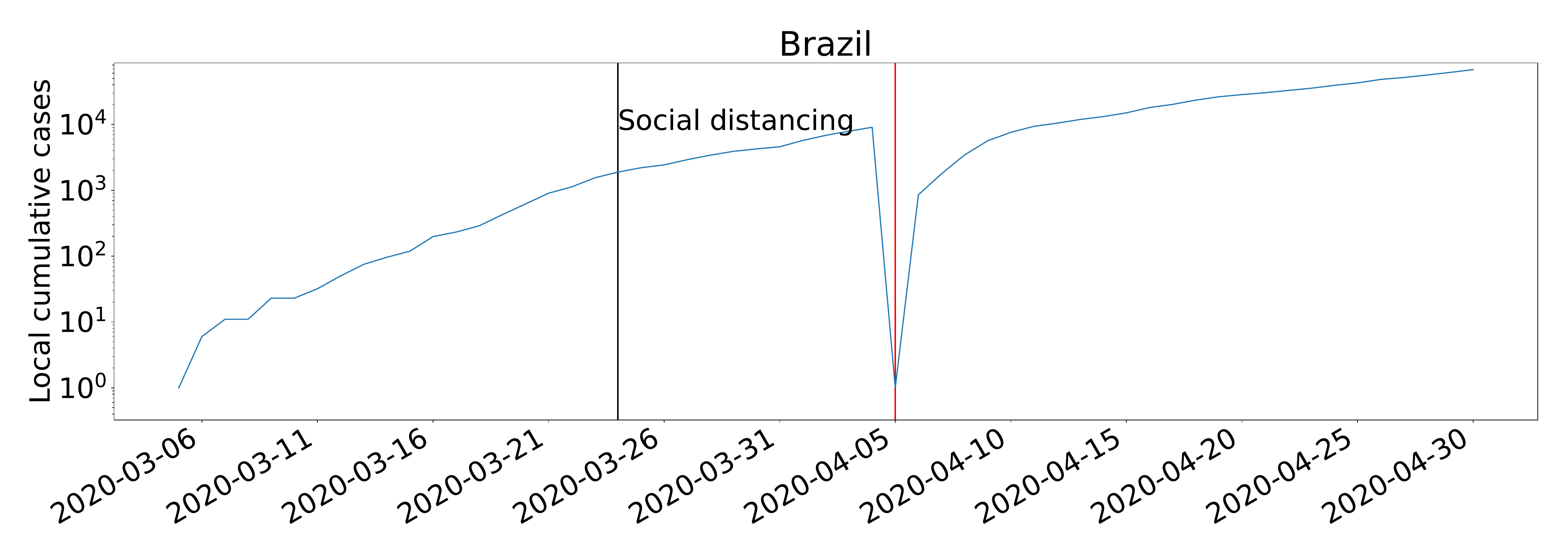} \\
	        \vspace{-0.35cm}
            \textbf{b} & \includegraphics[keepaspectratio, height=3.3cm, valign=T]
			{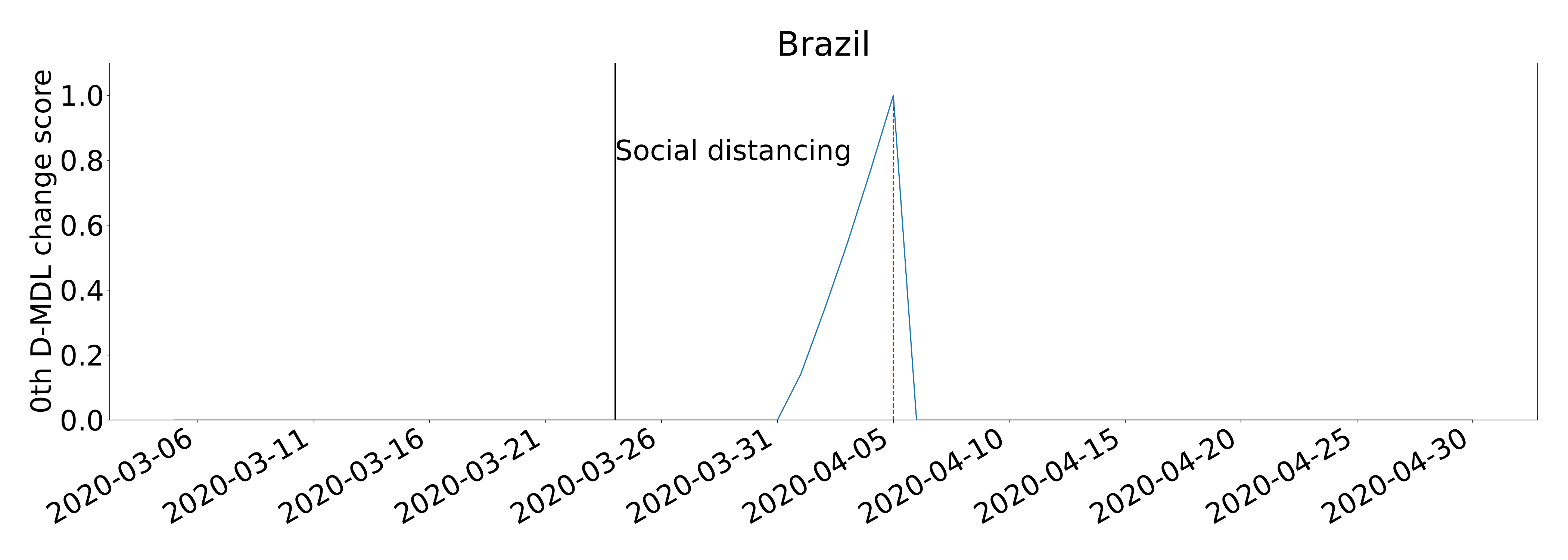}   \\
            \vspace{-0.35cm}
            \textbf{c} & \includegraphics[keepaspectratio, height=3.3cm, valign=T]
			{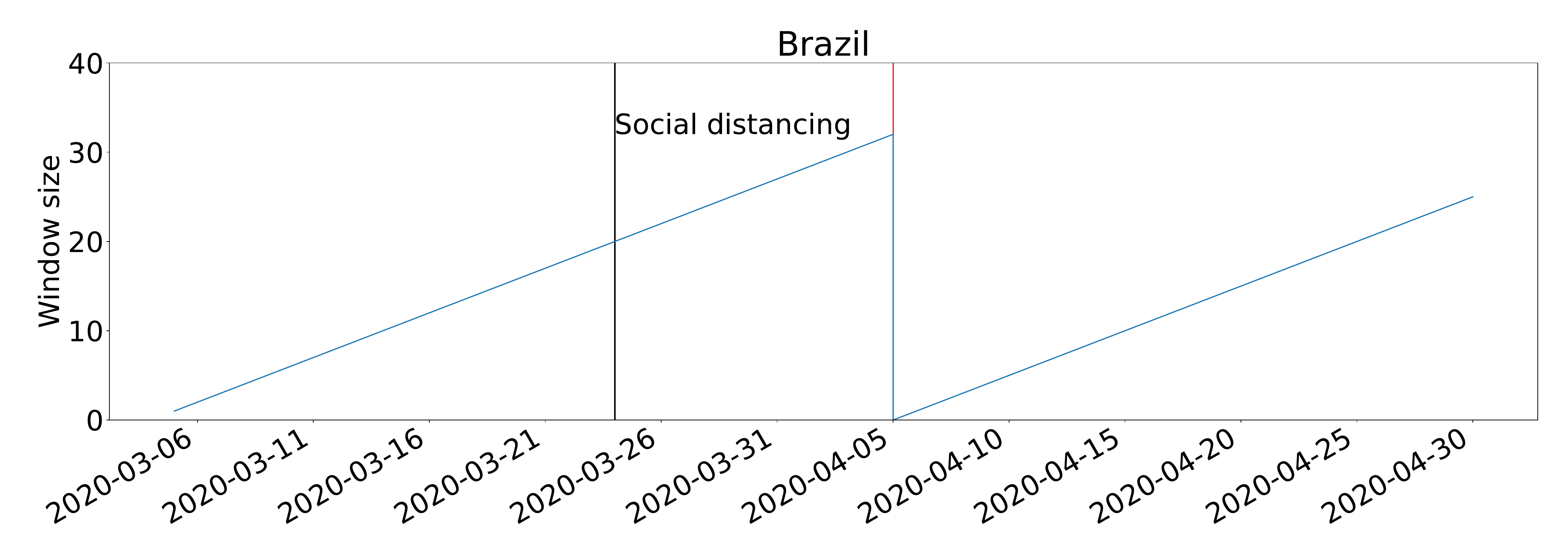} \\
			\vspace{-0.35cm}
			\textbf{d} & \includegraphics[keepaspectratio, height=3.3cm, valign=T]
			{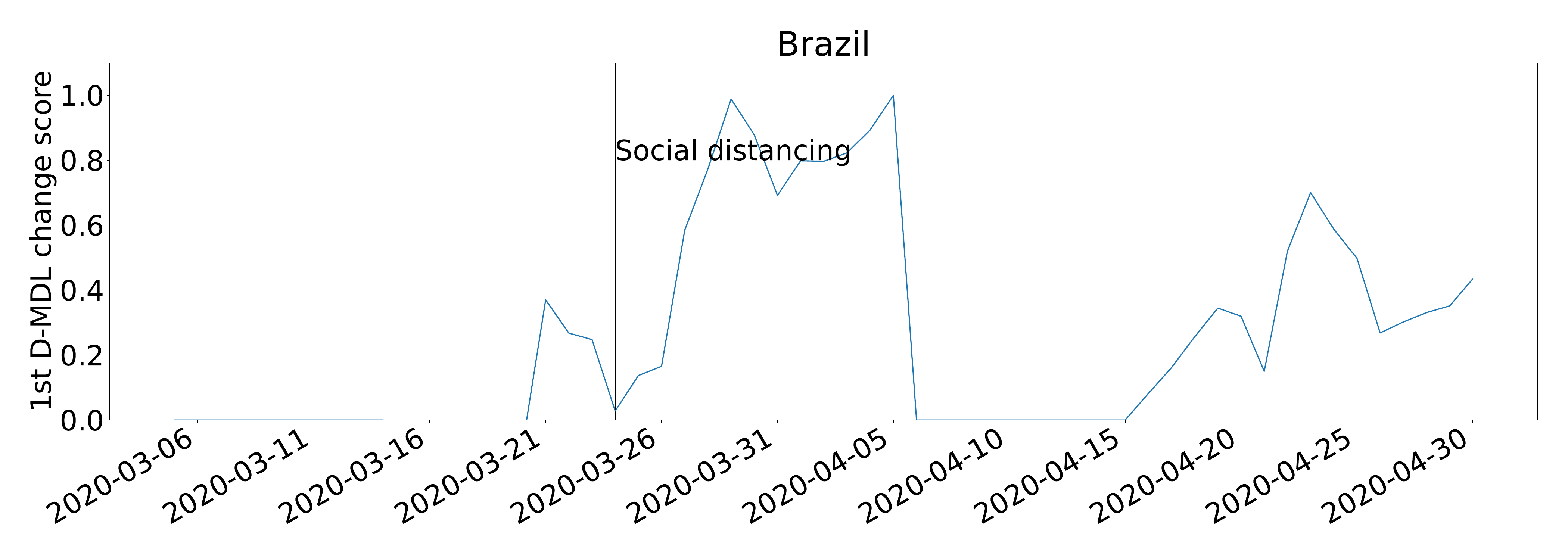} \\
			\vspace{-0.35cm}
			\textbf{e} & \includegraphics[keepaspectratio, height=3.3cm, valign=T]
			{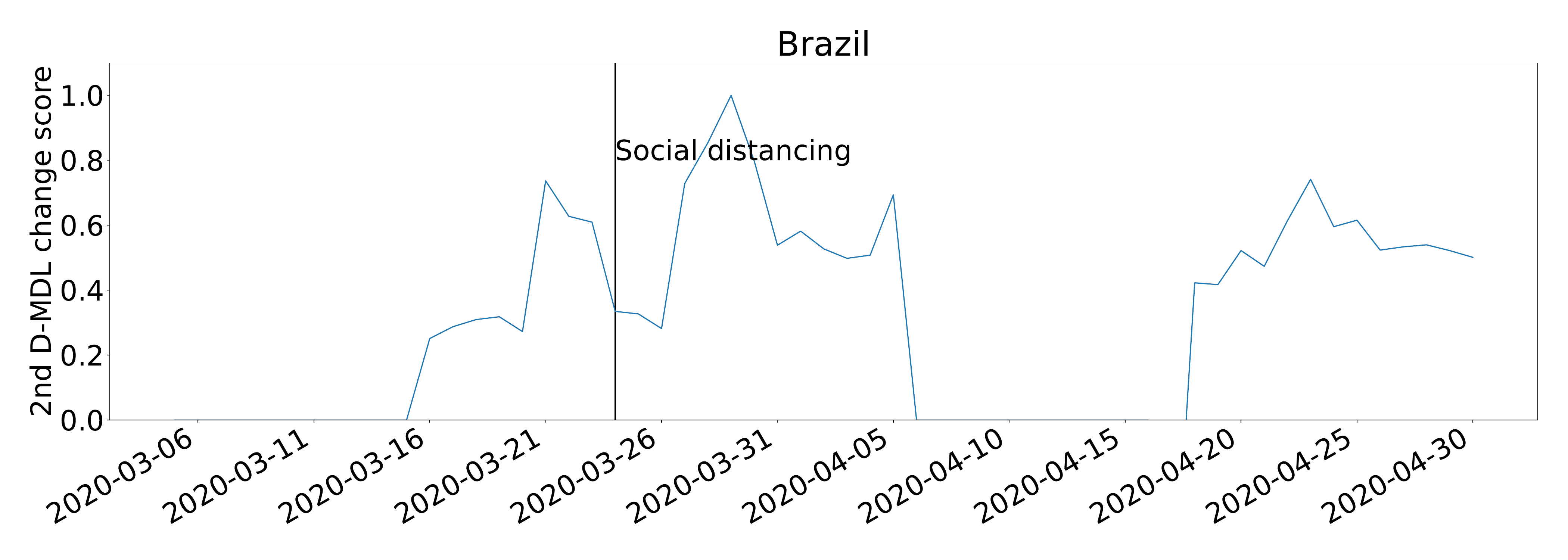} \\
		\end{tabular}
			\caption{\textbf{The results for Brazil with exponential modeling.} The date on which the social distancing was implemented is marked by a solid line in black. \textbf{a,} the number of cumulative cases. \textbf{b,} the change scores produced by the 0th M-DML where the line in blue denotes values of scores and dashed lines in red mark alarms. \textbf{c,} the window sized for the sequential D-DML algorithm with adaptive window where lines in red mark the shrinkage of windows. \textbf{d,} the change scores produced by the 1st D-MDL. \textbf{e,} the change scores produced by the 2nd D-MDL.}
\end{figure}

\begin{figure}[H] 
\centering
\begin{tabular}{cc}
		 	\textbf{a} & \includegraphics[keepaspectratio, height=3.3cm, valign=T]
			{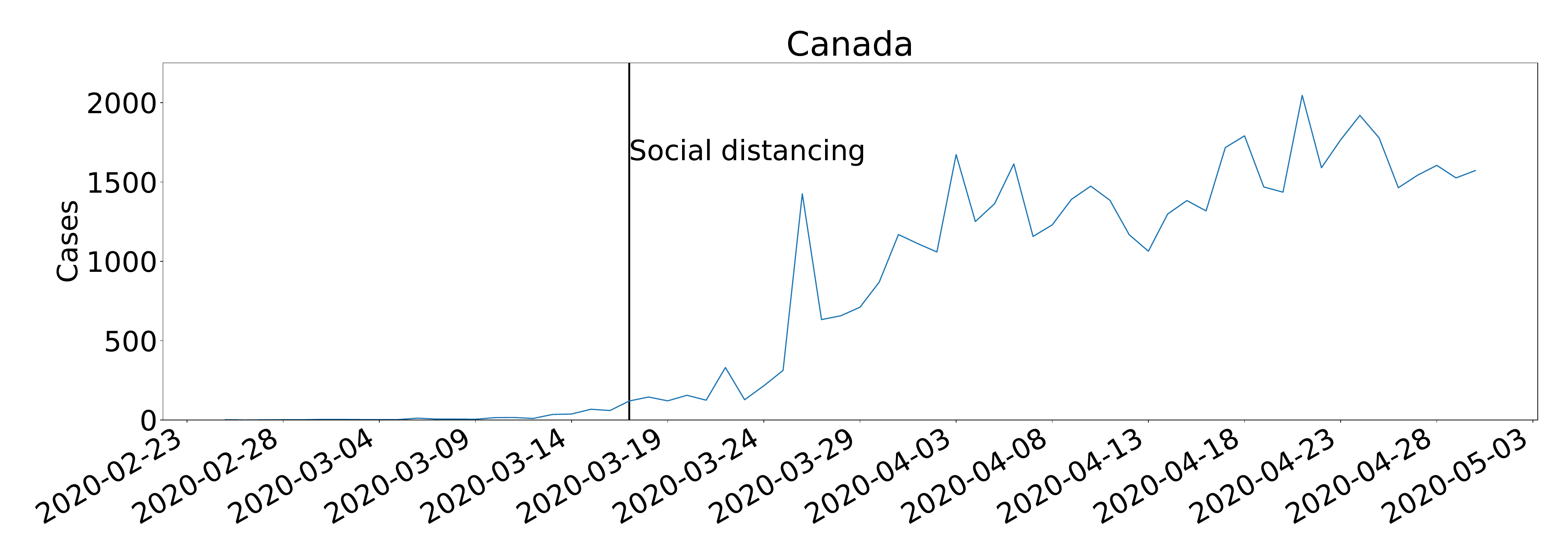} \\
			\vspace{-0.35cm}
	 	    \textbf{b} & \includegraphics[keepaspectratio, height=3.3cm, valign=T]
			{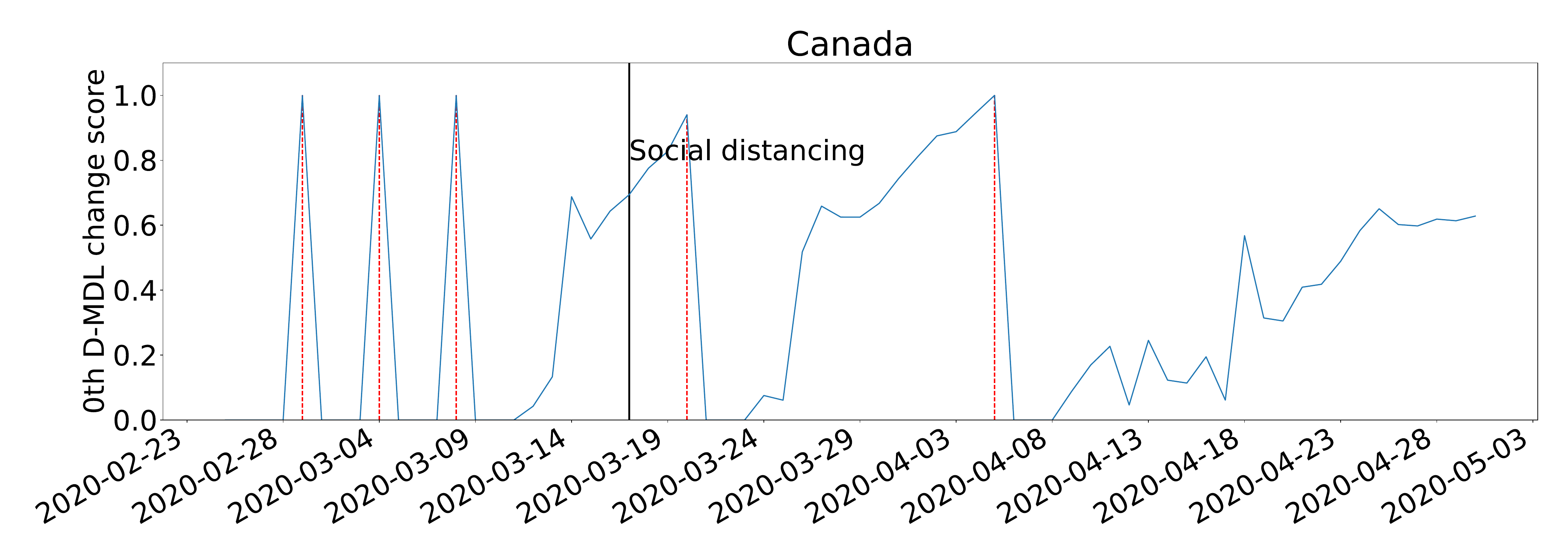}   \\
	        \vspace{-0.35cm}
			\textbf{c} & \includegraphics[keepaspectratio, height=3.3cm, valign=T]
			{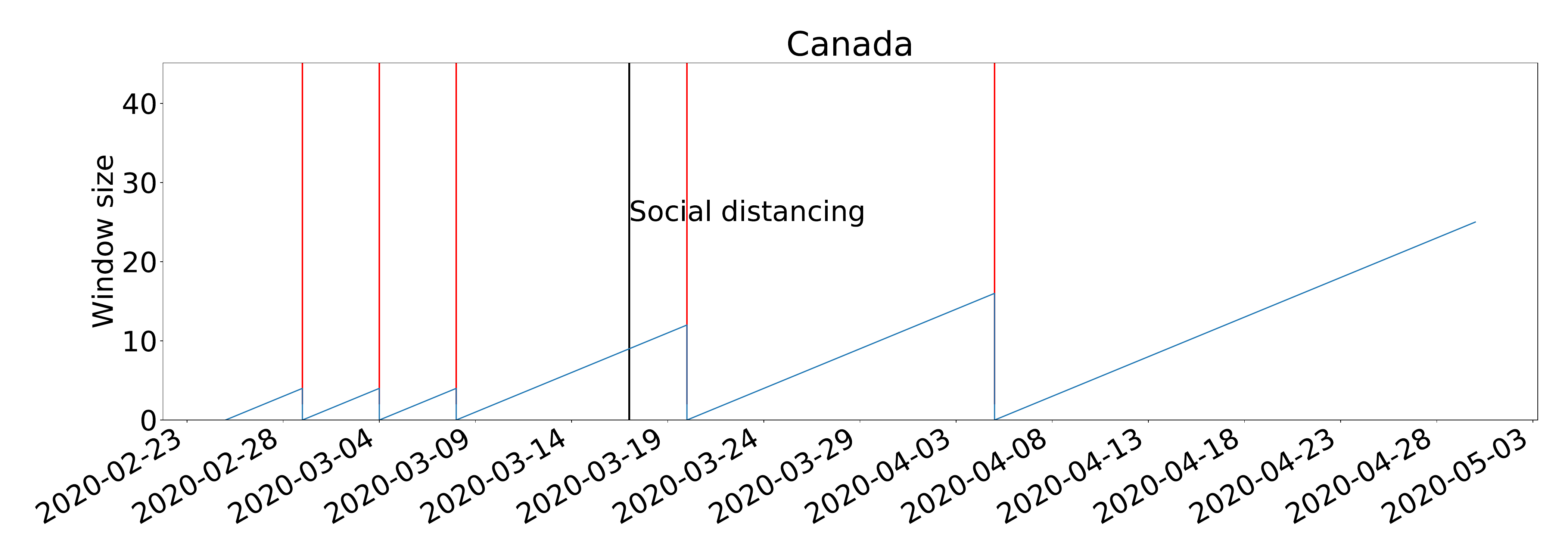} \\
		    \vspace{-0.35cm}
			\textbf{d} & \includegraphics[keepaspectratio, height=3.3cm, valign=T]
			{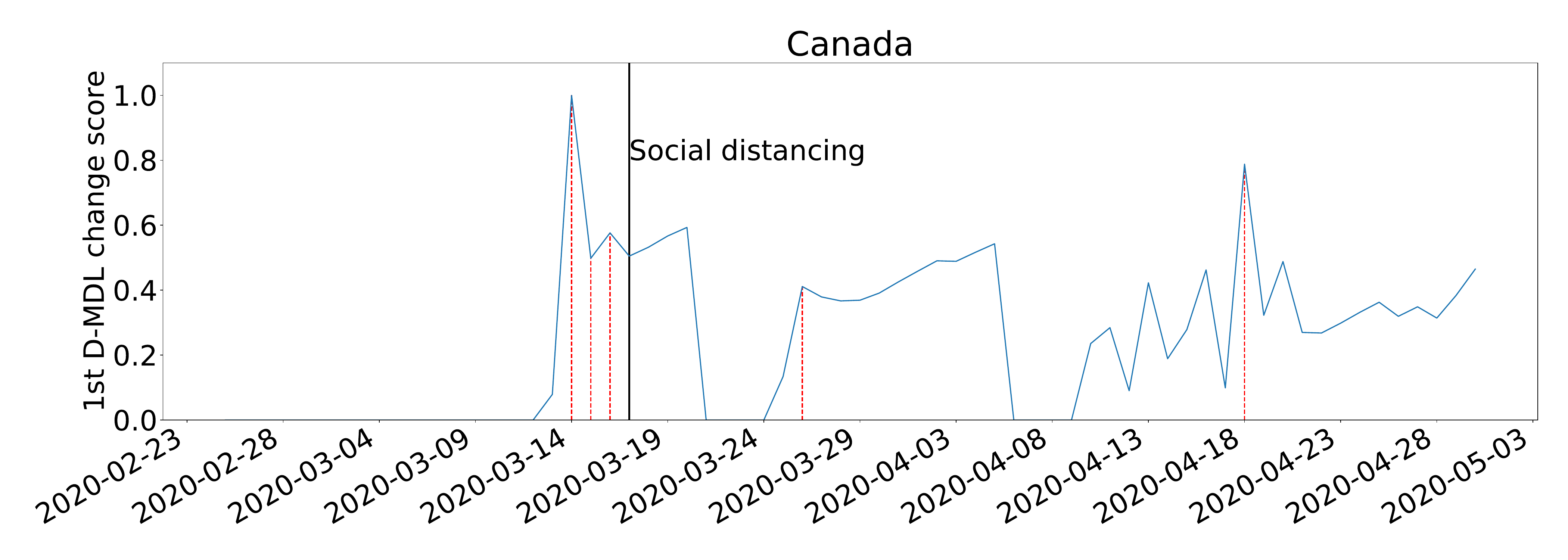} \\
		    \vspace{-0.35cm}
			\textbf{e} & \includegraphics[keepaspectratio, height=3.3cm, valign=T]
			{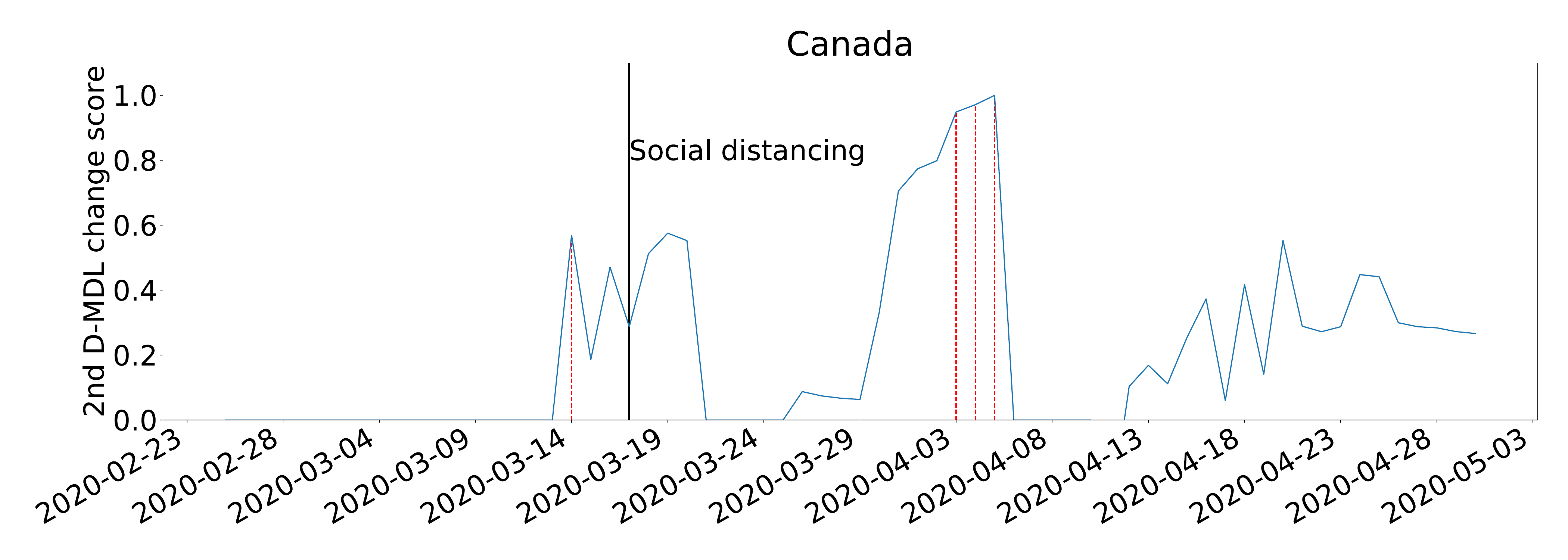} \\
		\end{tabular}
			\caption{\textbf{The results for Canada with Gaussian modeling.} The date on which the social distancing was implemented is marked by a solid line in black. \textbf{a,} the number of daily new cases. \textbf{b,} the change scores produced by the 0th M-DML where the line in blue denotes values of scores and dashed lines in red mark alarms. \textbf{c,} the window sized for the sequential D-DML algorithm with adaptive window where lines in red mark the shrinkage of windows. \textbf{d,} the change scores produced by the 1st D-MDL. \textbf{e,} the change scores produced by the 2nd D-MDL.}
\end{figure}

\begin{figure}[H]  
\centering
\begin{tabular}{cc}
			\textbf{a} & \includegraphics[keepaspectratio, height=3.3cm, valign=T]
			{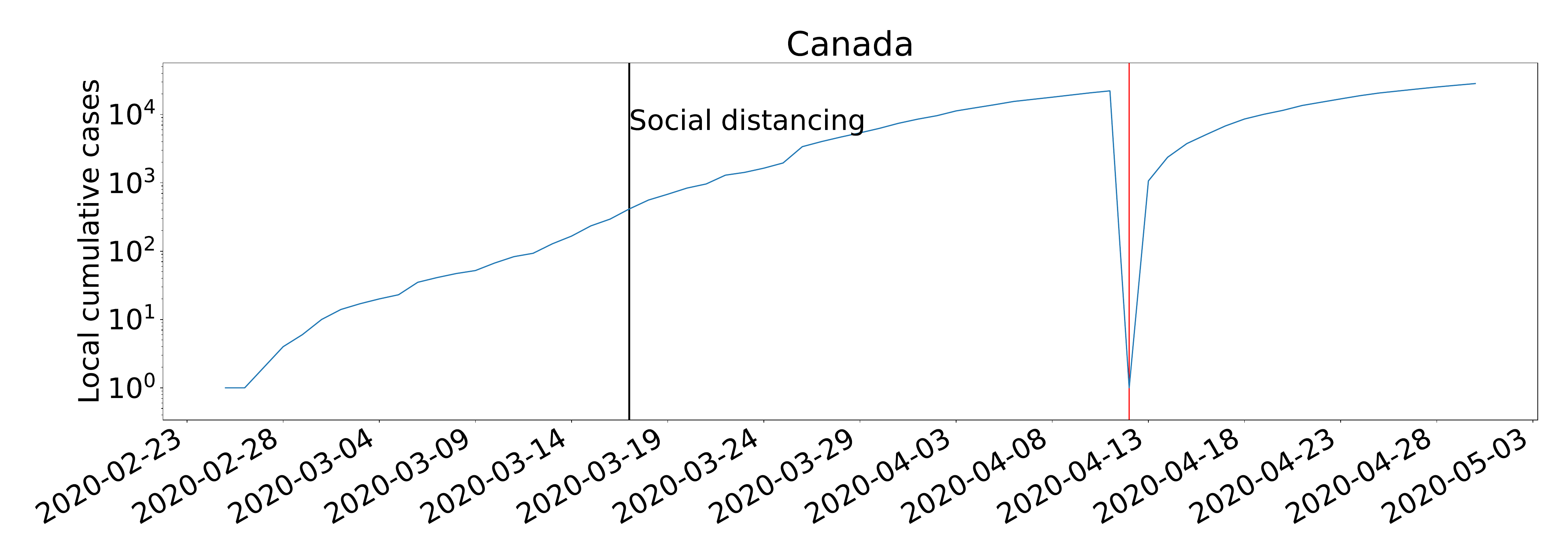} \\
	        \vspace{-0.35cm}
            \textbf{b} & \includegraphics[keepaspectratio, height=3.3cm, valign=T]
			{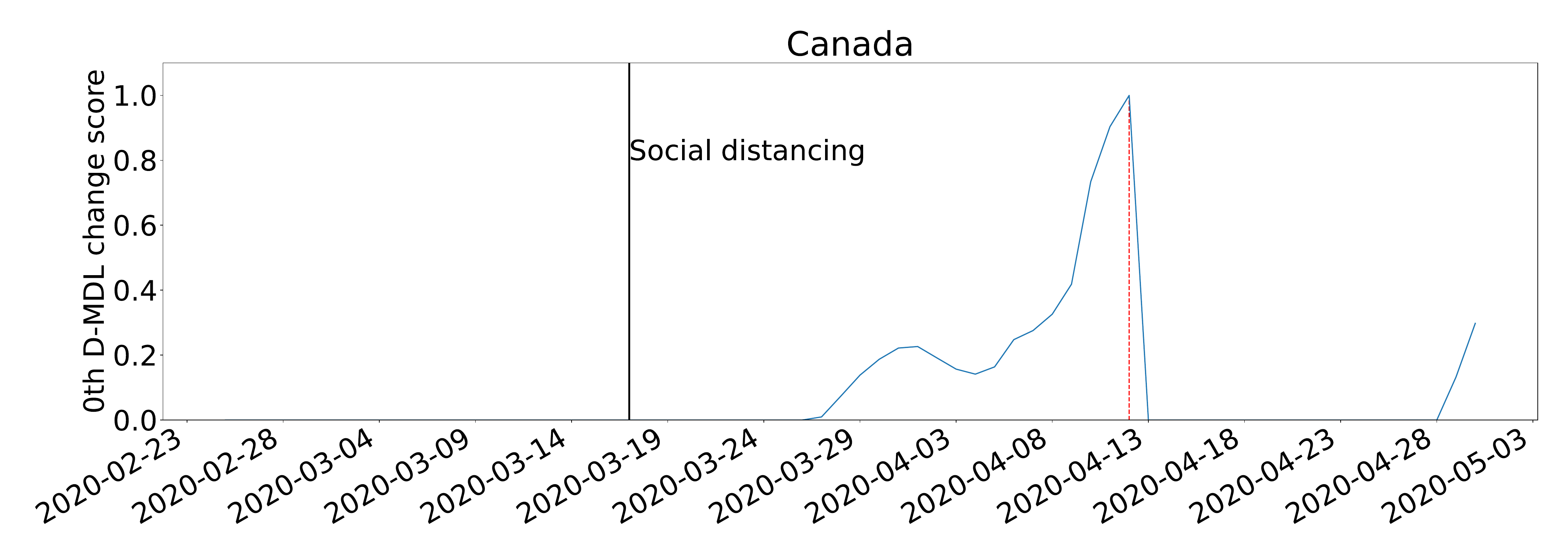}   \\
            \vspace{-0.35cm}
            \textbf{c} & \includegraphics[keepaspectratio, height=3.3cm, valign=T]
			{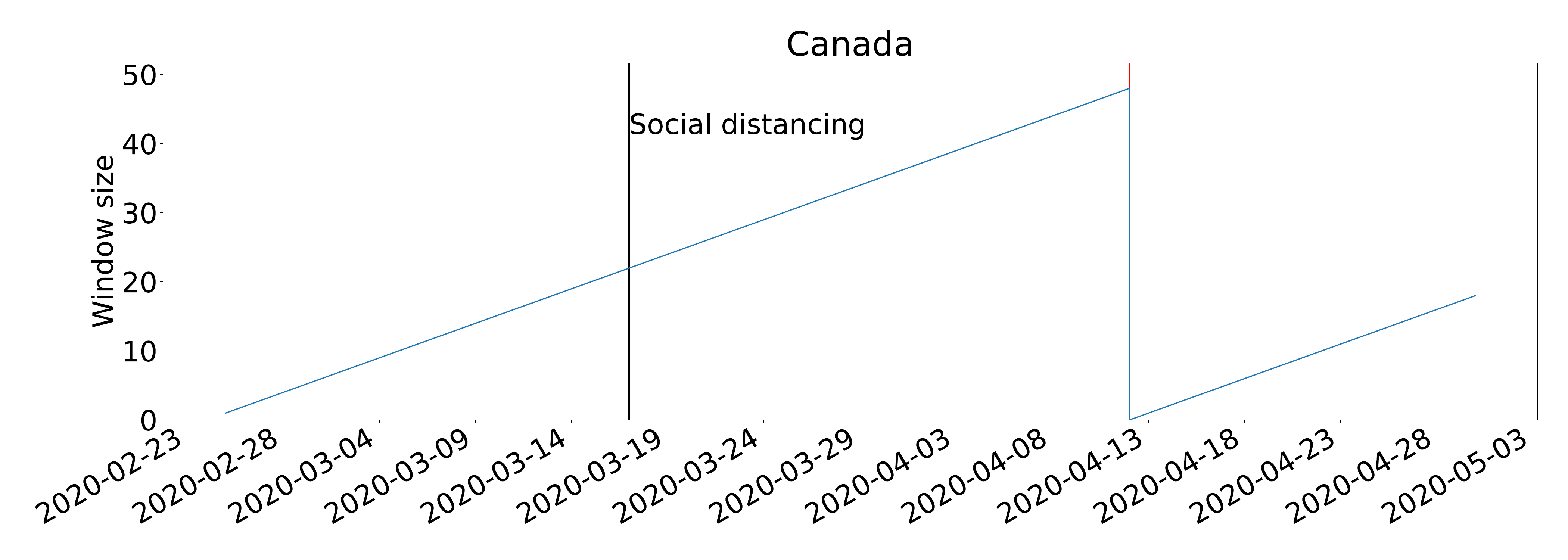} \\
			\vspace{-0.35cm}
			\textbf{d} & \includegraphics[keepaspectratio, height=3.3cm, valign=T]
			{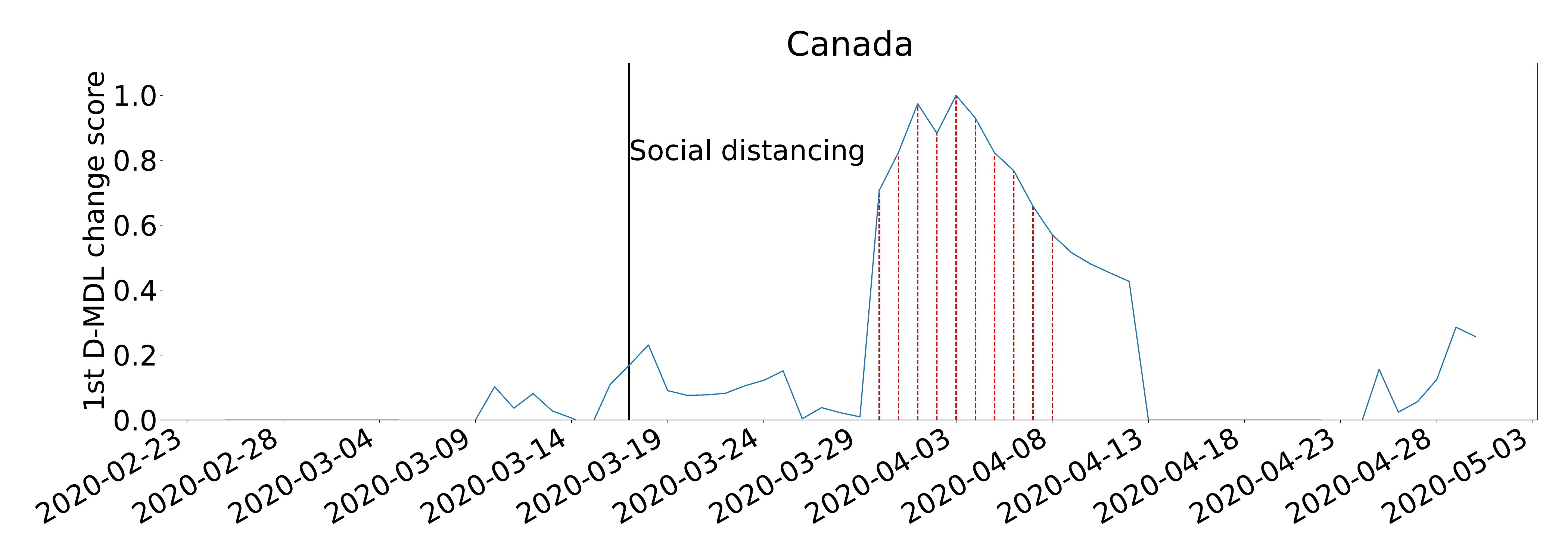} \\
			\vspace{-0.35cm}
			\textbf{e} & \includegraphics[keepaspectratio, height=3.3cm, valign=T]
			{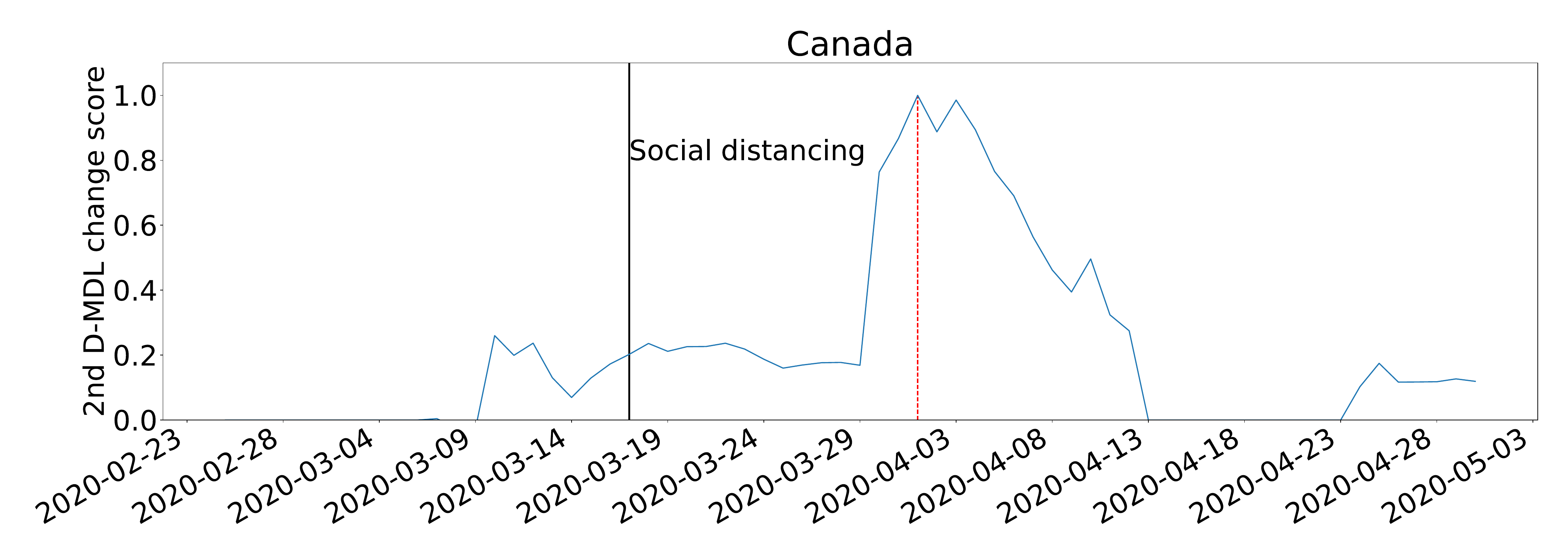} \\
		\end{tabular}
			\caption{\textbf{The results for Canada with exponential modeling.} The date on which the social distancing was implemented is marked by a solid line in black. \textbf{a,} the number of cumulative cases. \textbf{b,} the change scores produced by the 0th M-DML where the line in blue denotes values of scores and dashed lines in red mark alarms. \textbf{c,} the window sized for the sequential D-DML algorithm with adaptive window where lines in red mark the shrinkage of windows. \textbf{d,} the change scores produced by the 1st D-MDL. \textbf{e,} the change scores produced by the 2nd D-MDL.}
\end{figure}

\begin{figure}[H] 
\centering
\begin{tabular}{cc}
		 	\textbf{a} & \includegraphics[keepaspectratio, height=3.3cm, valign=T]
			{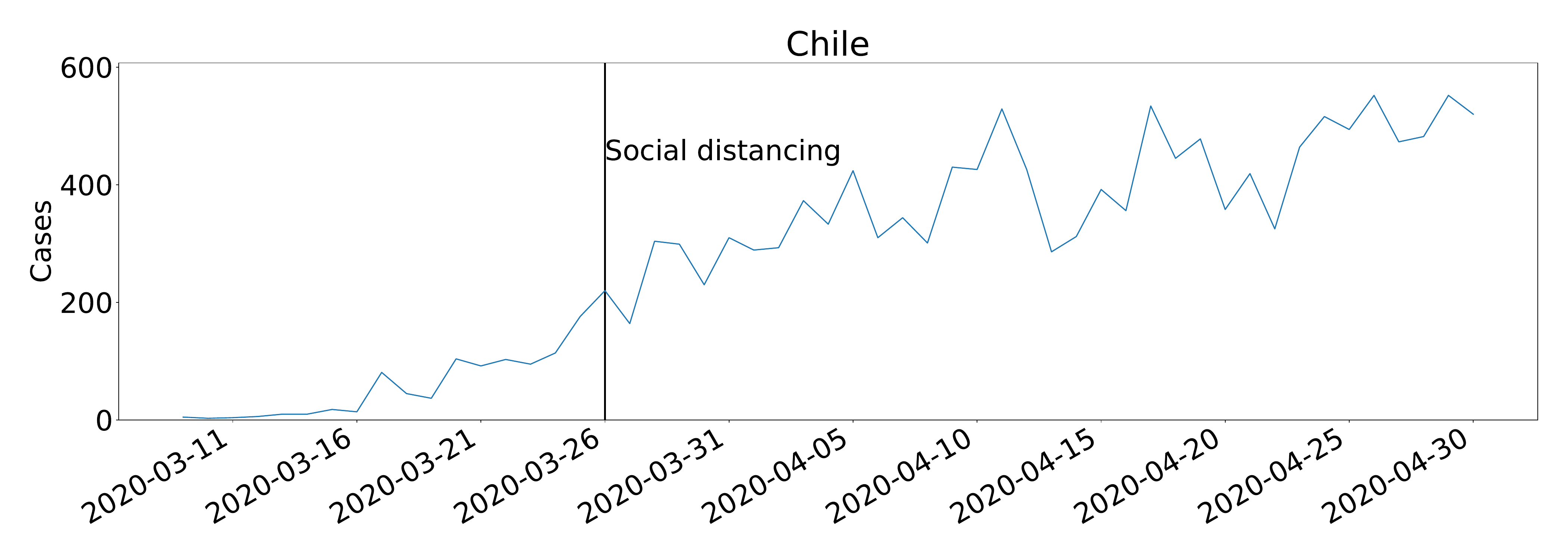} \\
			\vspace{-0.35cm}
	 	    \textbf{b} & \includegraphics[keepaspectratio, height=3.3cm, valign=T]
			{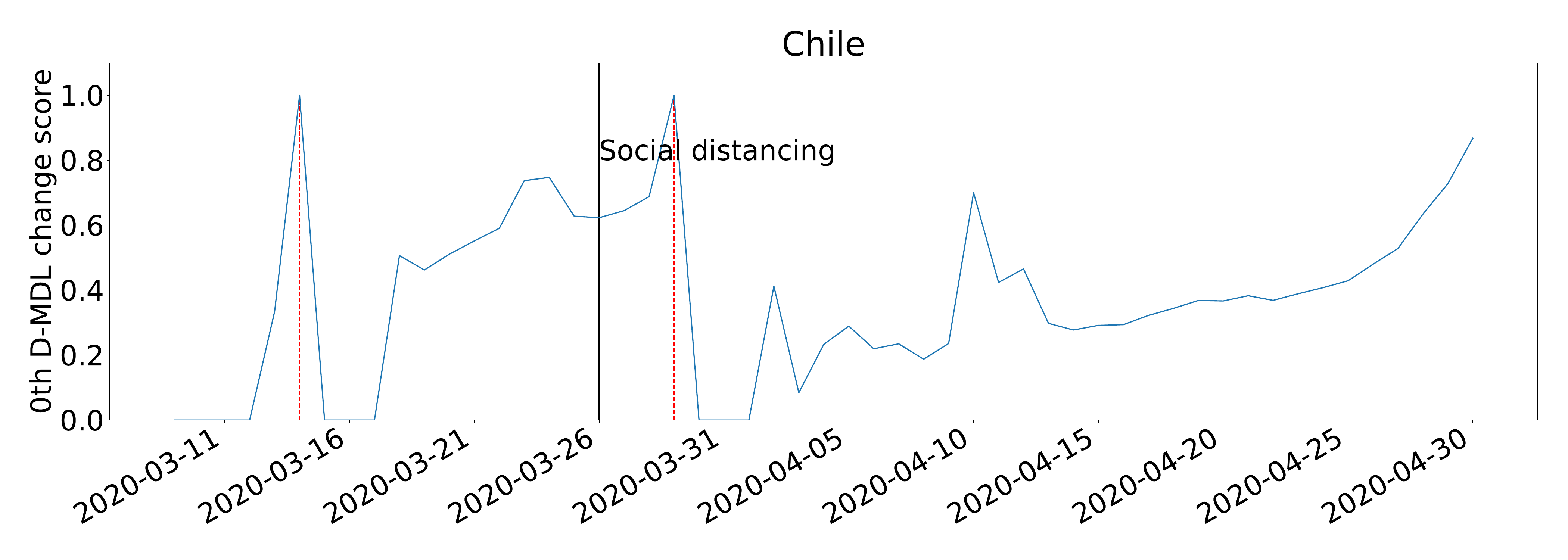}   \\
	        \vspace{-0.35cm}
			\textbf{c} & \includegraphics[keepaspectratio, height=3.3cm, valign=T]
			{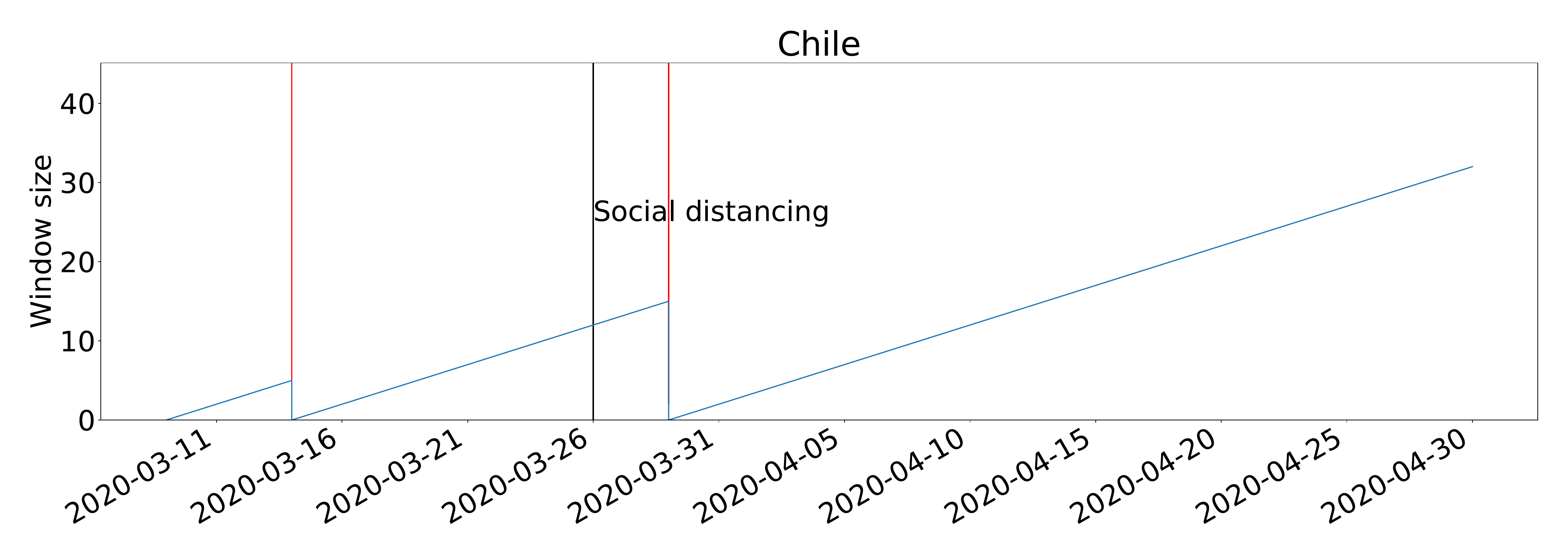} \\
		    \vspace{-0.35cm}
			\textbf{d} & \includegraphics[keepaspectratio, height=3.3cm, valign=T]
			{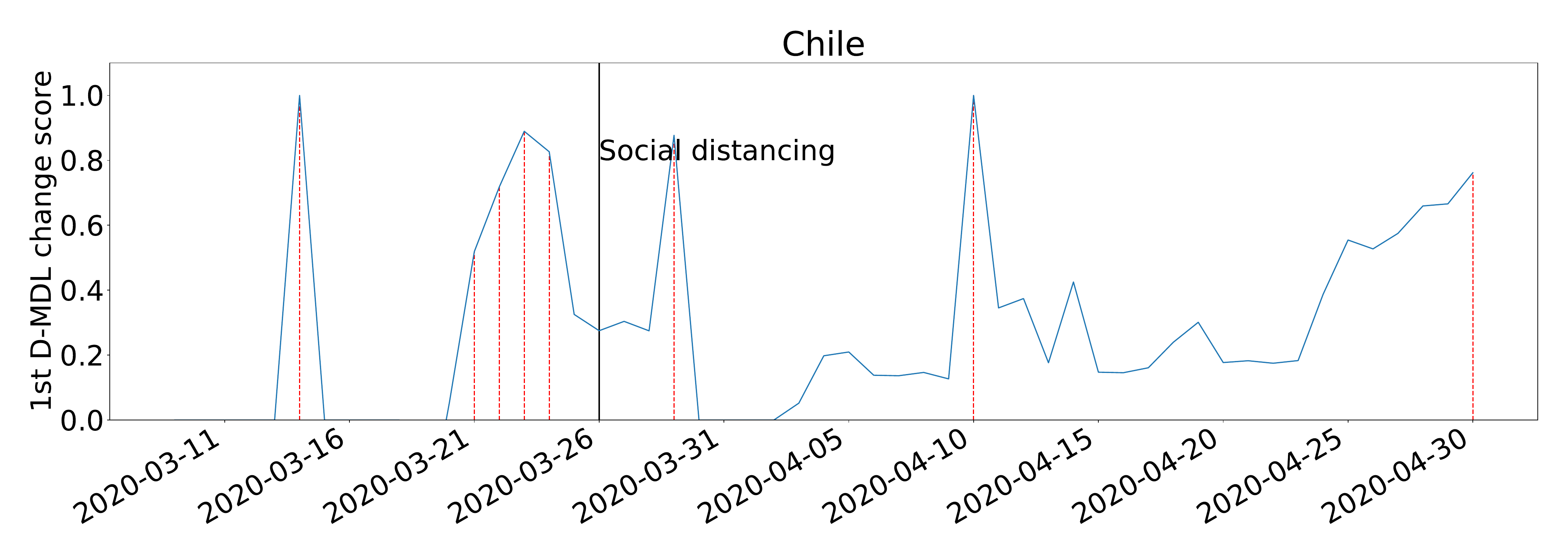} \\
		    \vspace{-0.35cm}
			\textbf{e} & \includegraphics[keepaspectratio, height=3.3cm, valign=T]
			{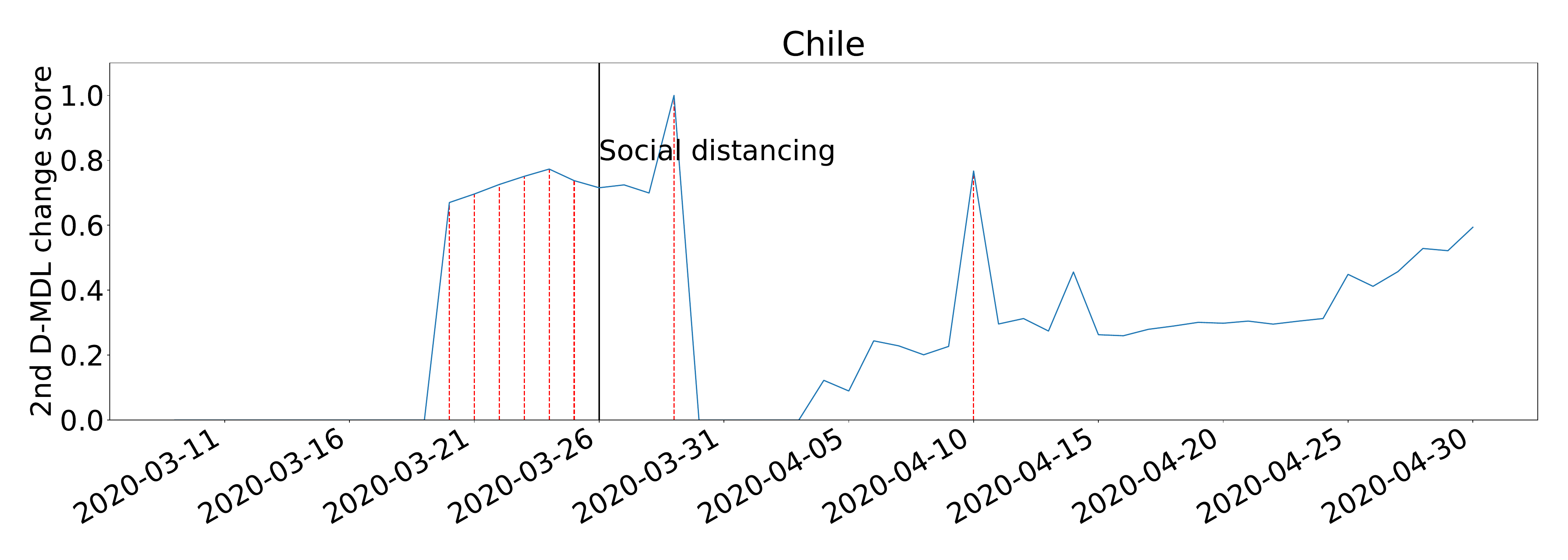} \\
		\end{tabular}
			\caption{\textbf{The results for Chile with Gaussian modeling.} The date on which the social distancing was implemented is marked by a solid line in black. \textbf{a,} the number of daily new cases. \textbf{b,} the change scores produced by the 0th M-DML where the line in blue denotes values of scores and dashed lines in red mark alarms. \textbf{c,} the window sized for the sequential D-DML algorithm with adaptive window where lines in red mark the shrinkage of windows. \textbf{d,} the change scores produced by the 1st D-MDL. \textbf{e,} the change scores produced by the 2nd D-MDL.}
\end{figure}

\begin{figure}[H]  
\centering
\begin{tabular}{cc}
			\textbf{a} & \includegraphics[keepaspectratio, height=3.3cm, valign=T]
			{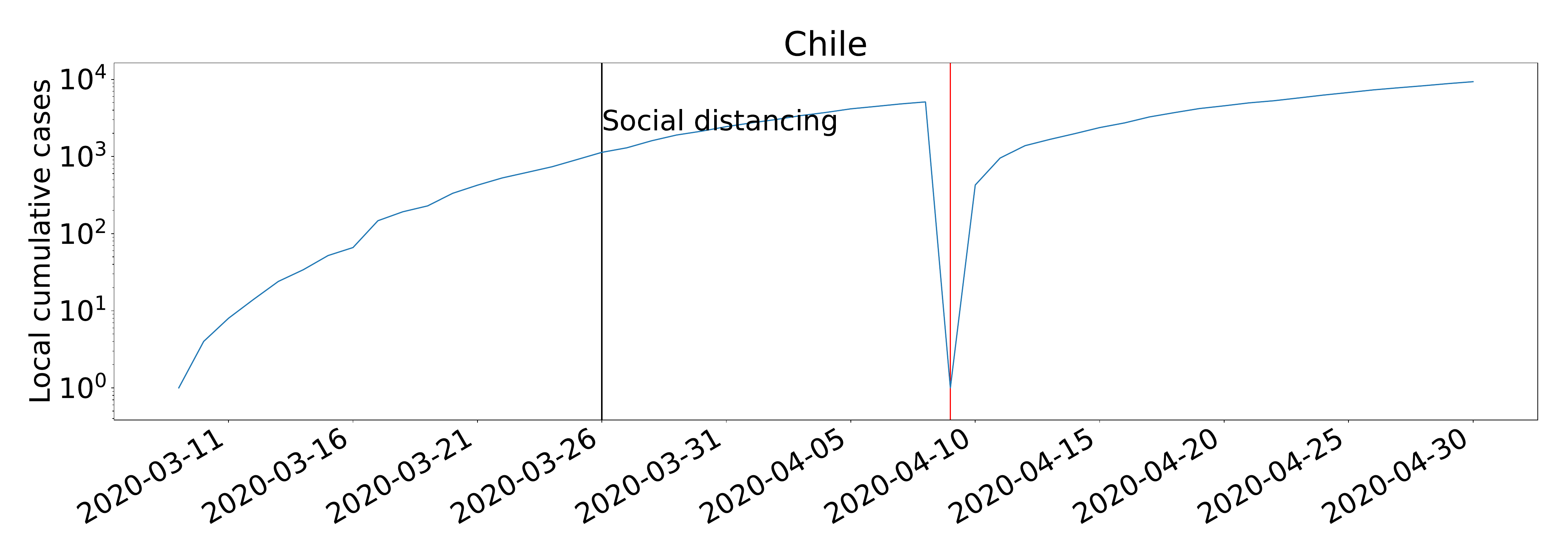} \\
	        \vspace{-0.35cm}
            \textbf{b} & \includegraphics[keepaspectratio, height=3.3cm, valign=T]
			{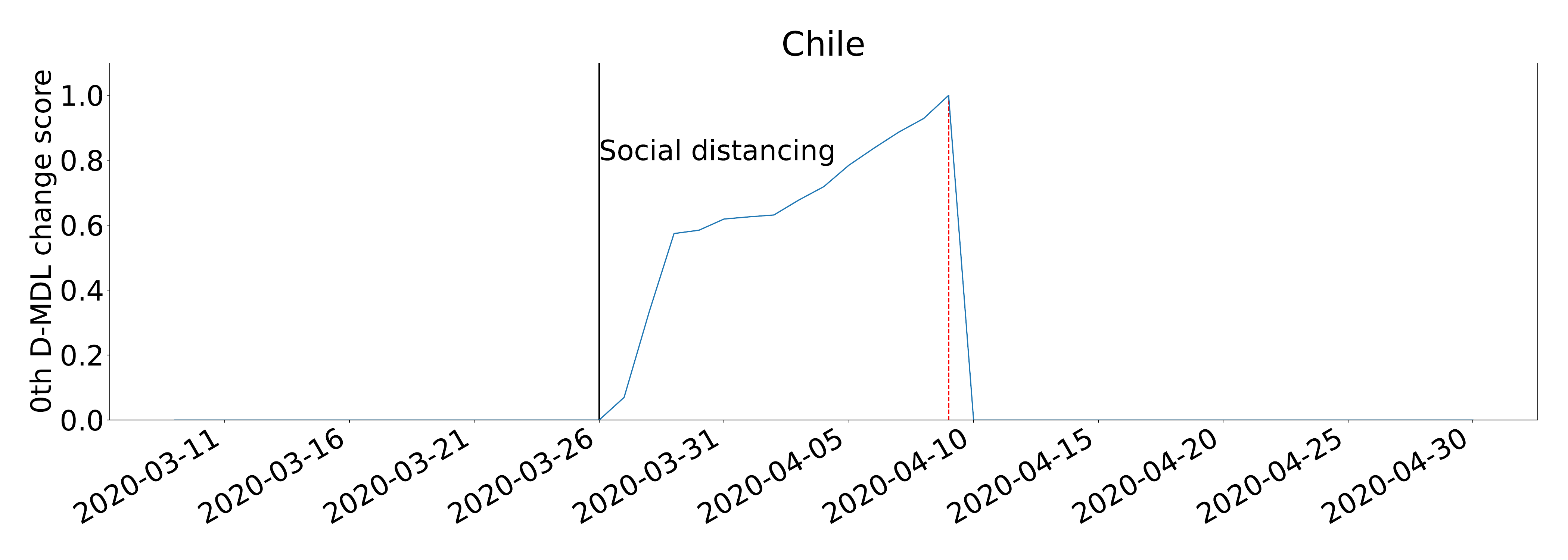}   \\
            \vspace{-0.35cm}
            \textbf{c} & \includegraphics[keepaspectratio, height=3.3cm, valign=T]
			{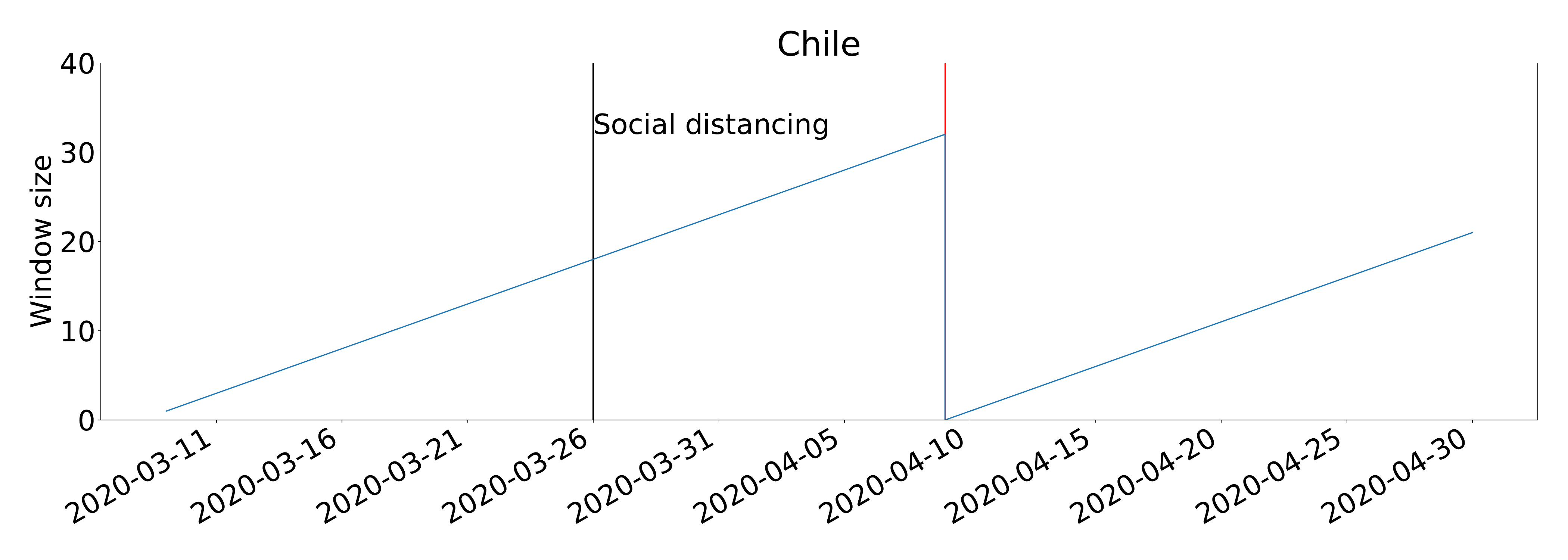} \\
			\vspace{-0.35cm}
			\textbf{d} & \includegraphics[keepaspectratio, height=3.3cm, valign=T]
			{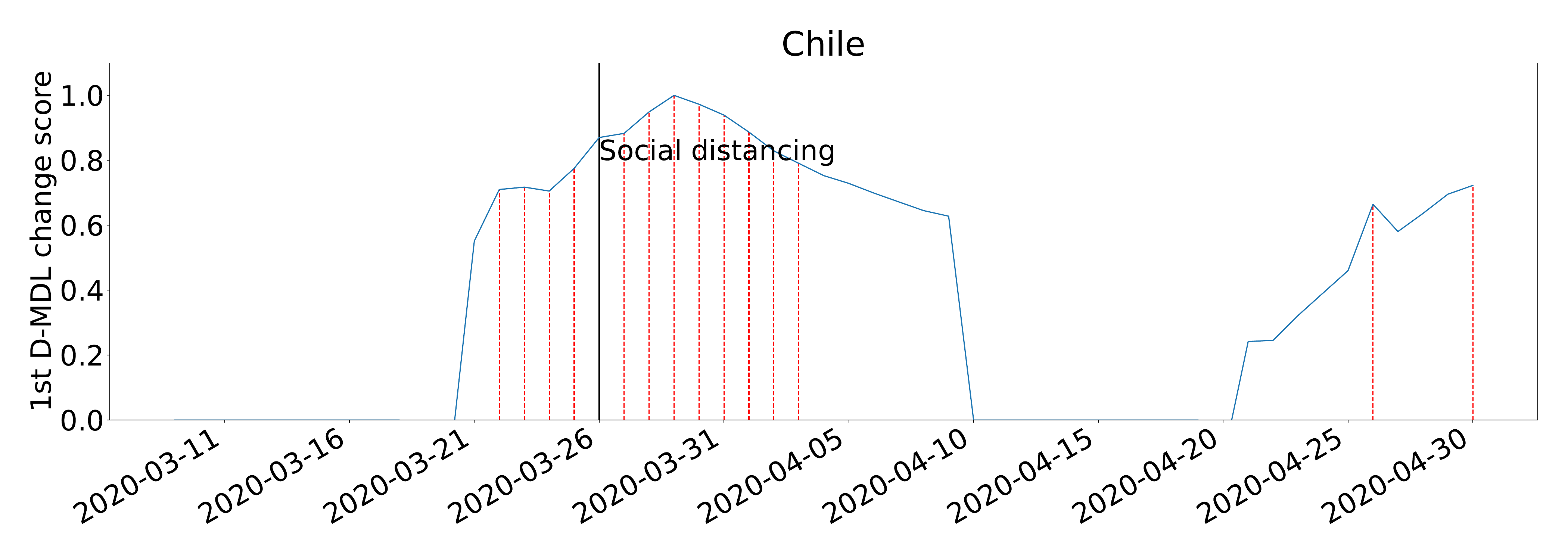} \\
			\vspace{-0.35cm}
			\textbf{e} & \includegraphics[keepaspectratio, height=3.3cm, valign=T]
			{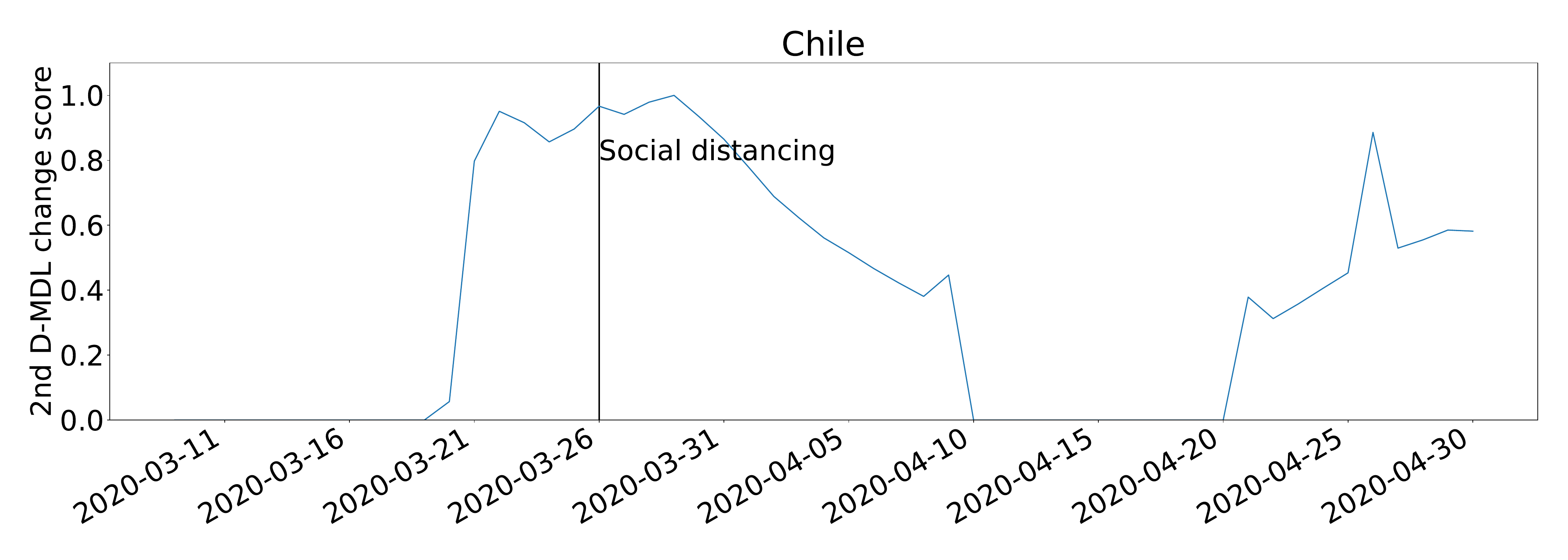} \\
		\end{tabular}
			\caption{\textbf{The results for Chile with exponential modeling.} The date on which the social distancing was implemented is marked by a solid line in black. \textbf{a,} the number of cumulative cases. \textbf{b,} the change scores produced by the 0th M-DML where the line in blue denotes values of scores and dashed lines in red mark alarms. \textbf{c,} the window sized for the sequential D-DML algorithm with adaptive window where lines in red mark the shrinkage of windows. \textbf{d,} the change scores produced by the 1st D-MDL. \textbf{e,} the change scores produced by the 2nd D-MDL.}
\end{figure}

\begin{figure}[H] 
\centering
\begin{tabular}{cc}
		 	\textbf{a} & \includegraphics[keepaspectratio, height=3.3cm, valign=T]
			{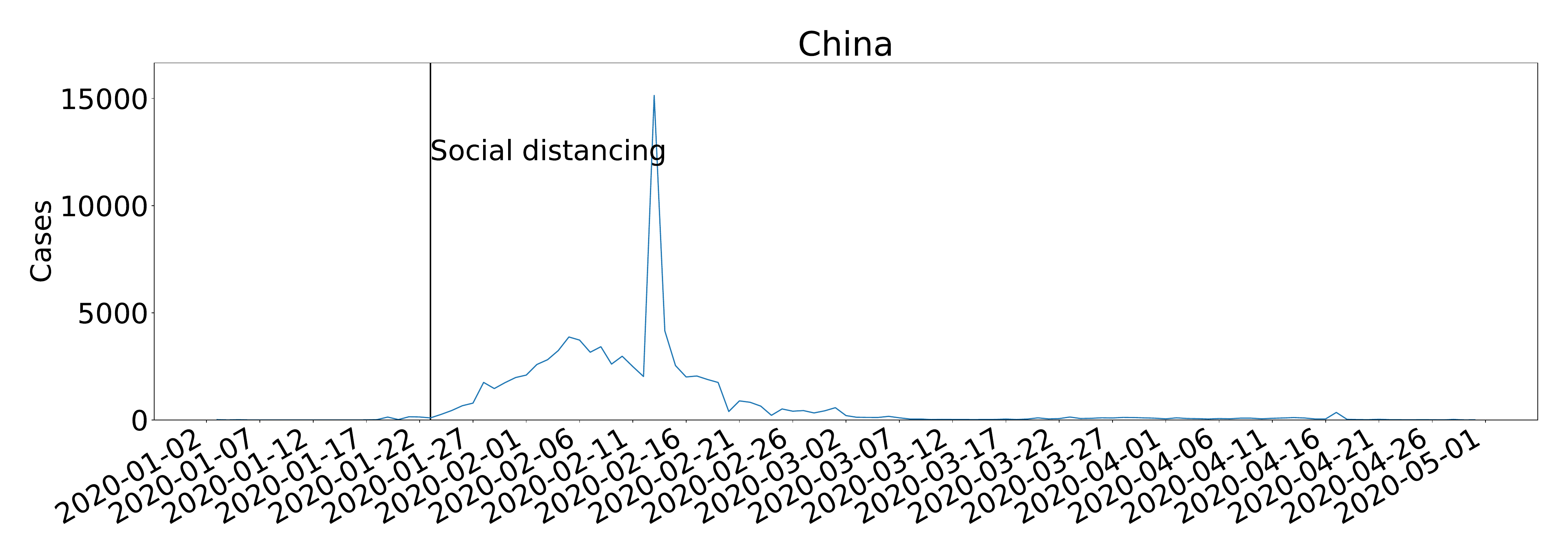} \\
			\vspace{-0.35cm}
	 	    \textbf{b} & \includegraphics[keepaspectratio, height=3.3cm, valign=T]
			{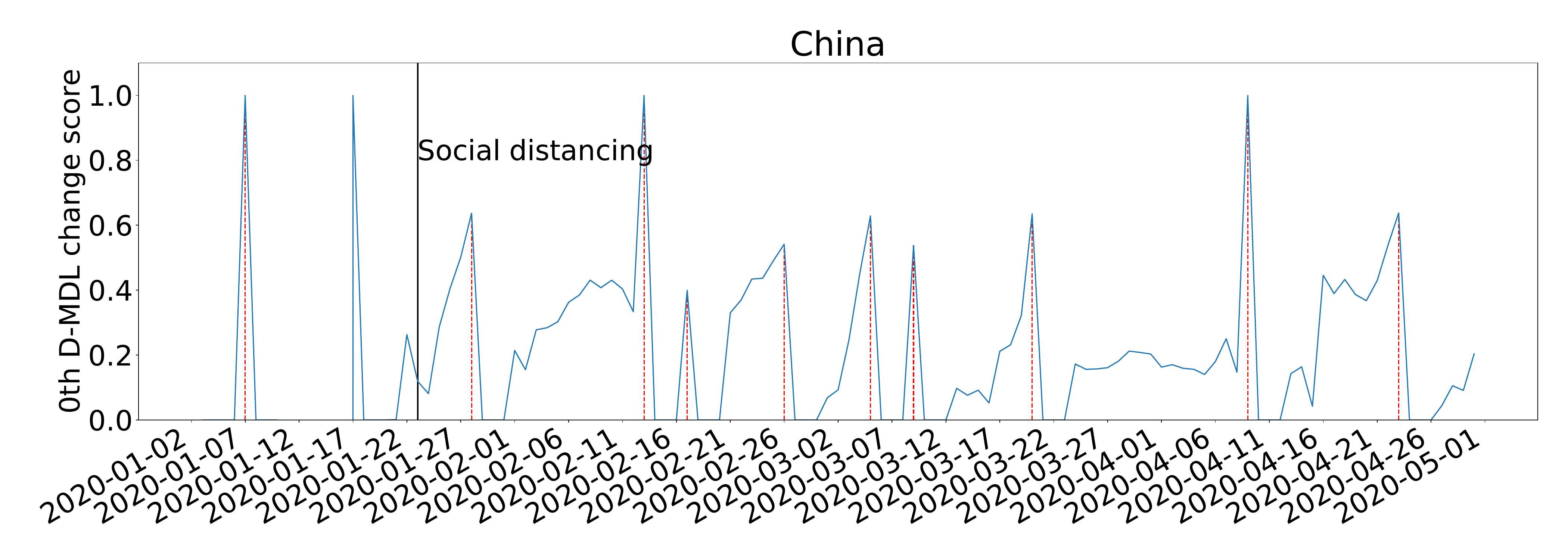}   \\
	        \vspace{-0.35cm}
			\textbf{c} & \includegraphics[keepaspectratio, height=3.3cm, valign=T]
			{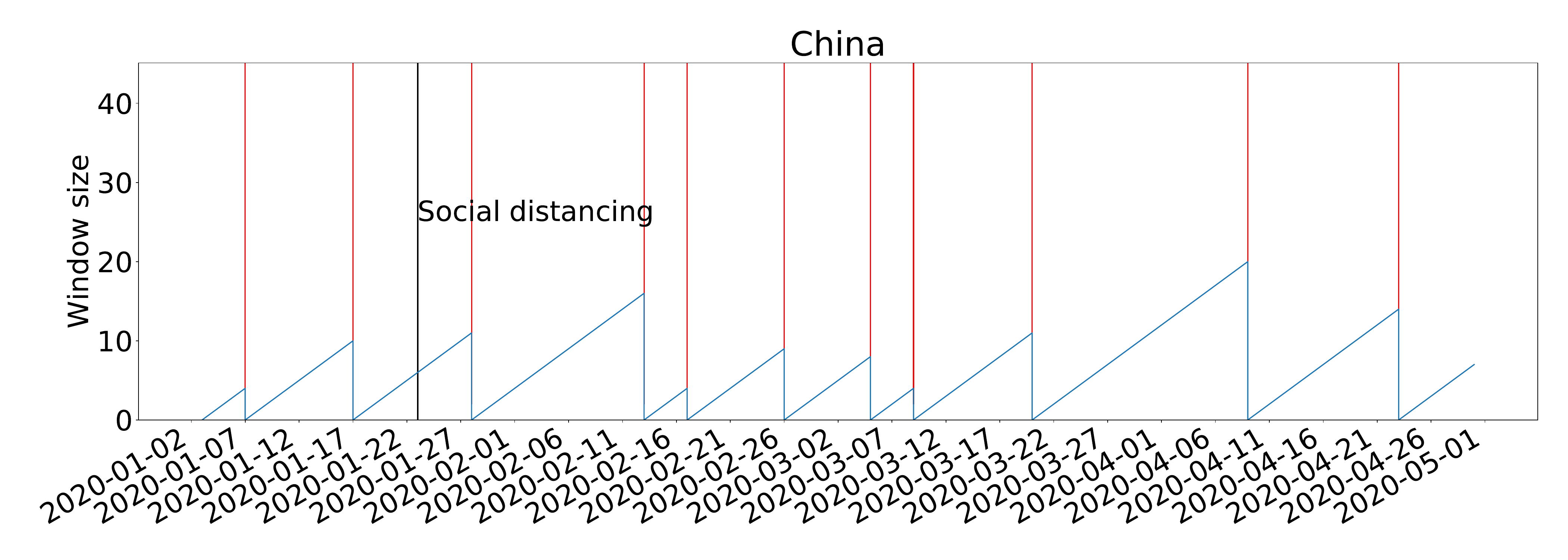} \\
		    \vspace{-0.35cm}
			\textbf{d} & \includegraphics[keepaspectratio, height=3.3cm, valign=T]
			{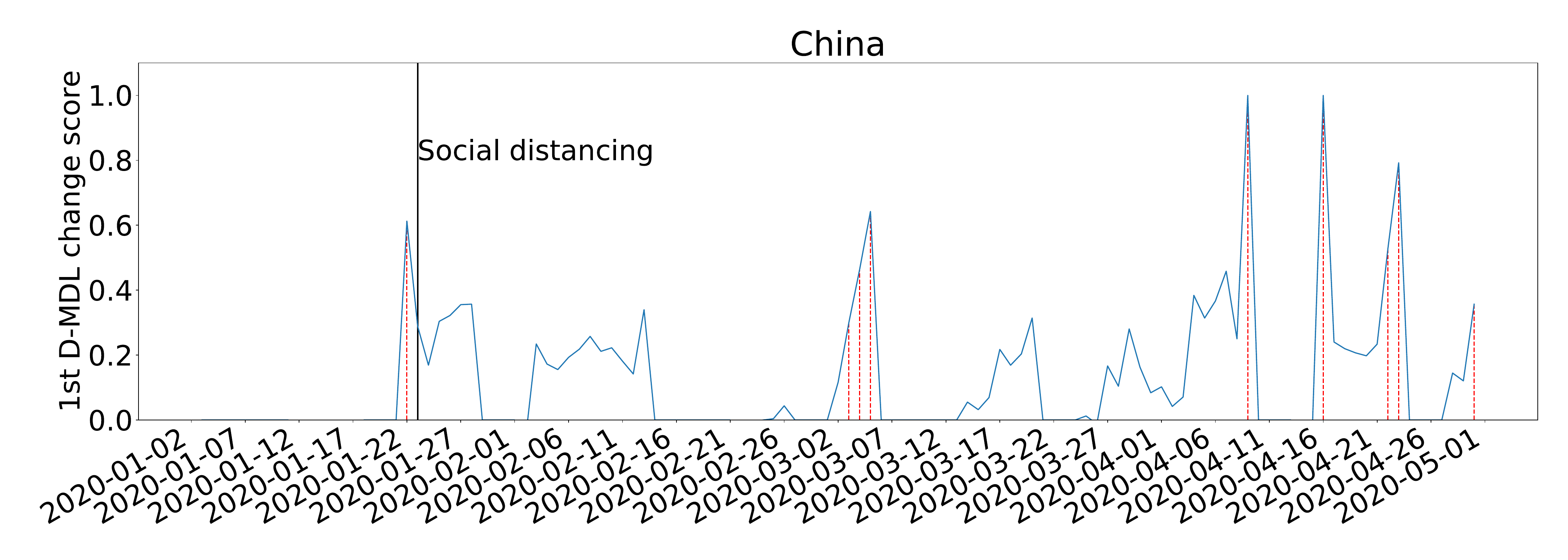} \\
		    \vspace{-0.35cm}
			\textbf{e} & \includegraphics[keepaspectratio, height=3.3cm, valign=T]
			{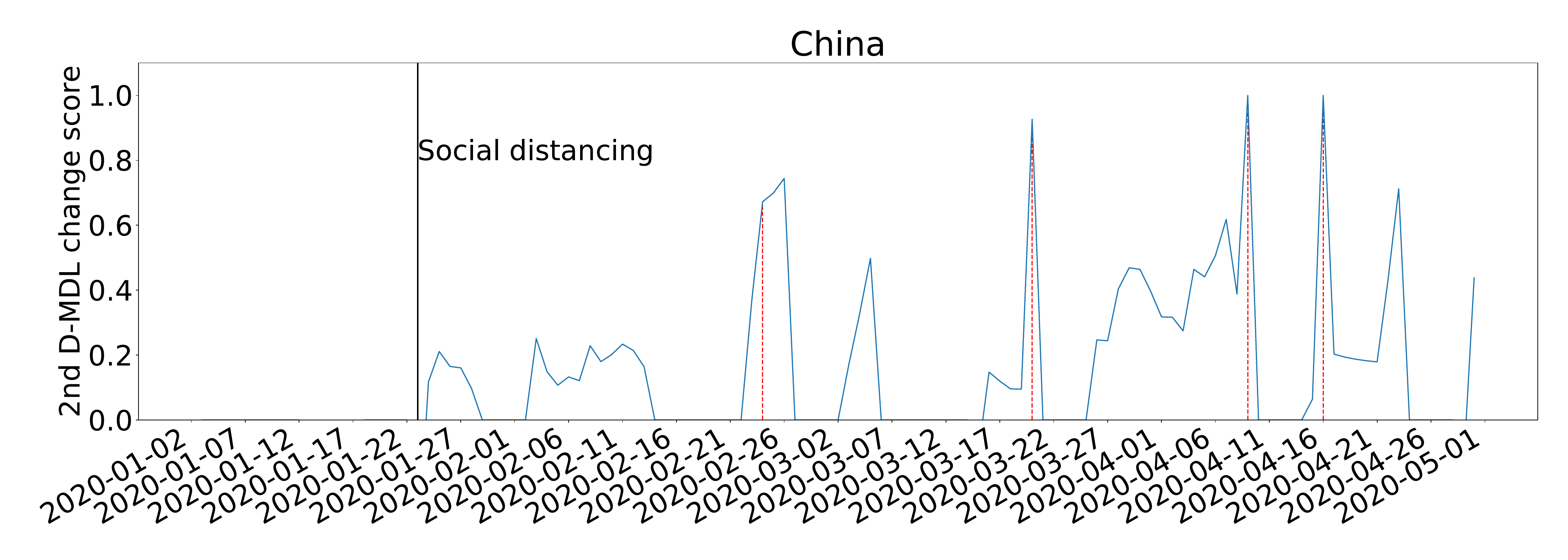} \\
		\end{tabular}
			\caption{\textbf{The results for China with Gaussian modeling.} The date on which the social distancing was implemented is marked by a solid line in black. \textbf{a,} the number of daily new cases. \textbf{b,} the change scores produced by the 0th M-DML where the line in blue denotes values of scores and dashed lines in red mark alarms. \textbf{c,} the window sized for the sequential D-DML algorithm with adaptive window where lines in red mark the shrinkage of windows. \textbf{d,} the change scores produced by the 1st D-MDL. \textbf{e,} the change scores produced by the 2nd D-MDL.}
\end{figure}

\begin{figure}[H]  
\centering
\begin{tabular}{cc}
			\textbf{a} & \includegraphics[keepaspectratio, height=3.3cm, valign=T]
			{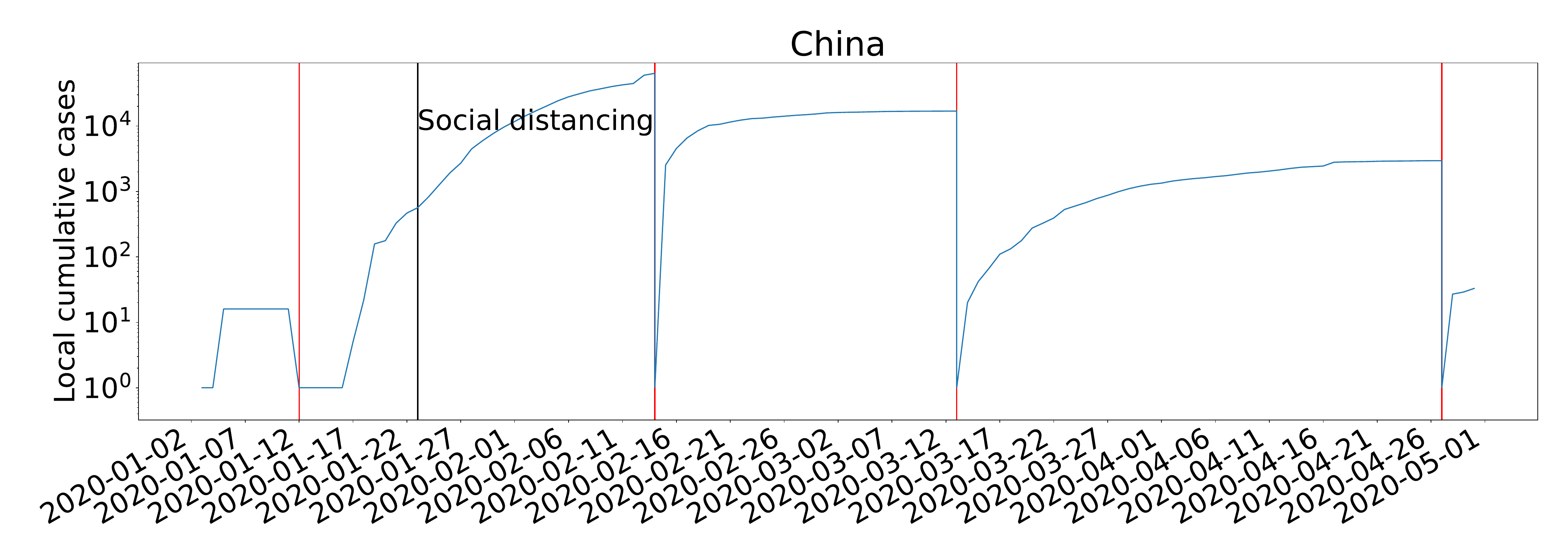} \\
	        \vspace{-0.35cm}
            \textbf{b} & \includegraphics[keepaspectratio, height=3.3cm, valign=T]
			{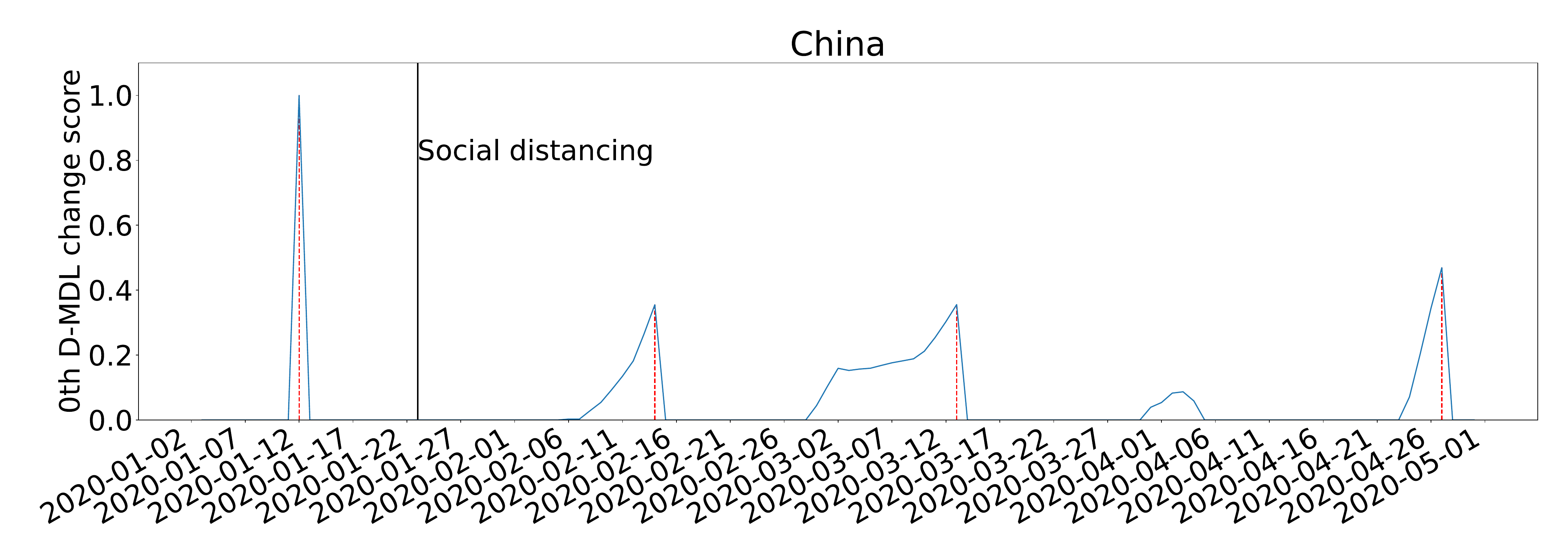}   \\
            \vspace{-0.35cm}
            \textbf{c} & \includegraphics[keepaspectratio, height=3.3cm, valign=T]
			{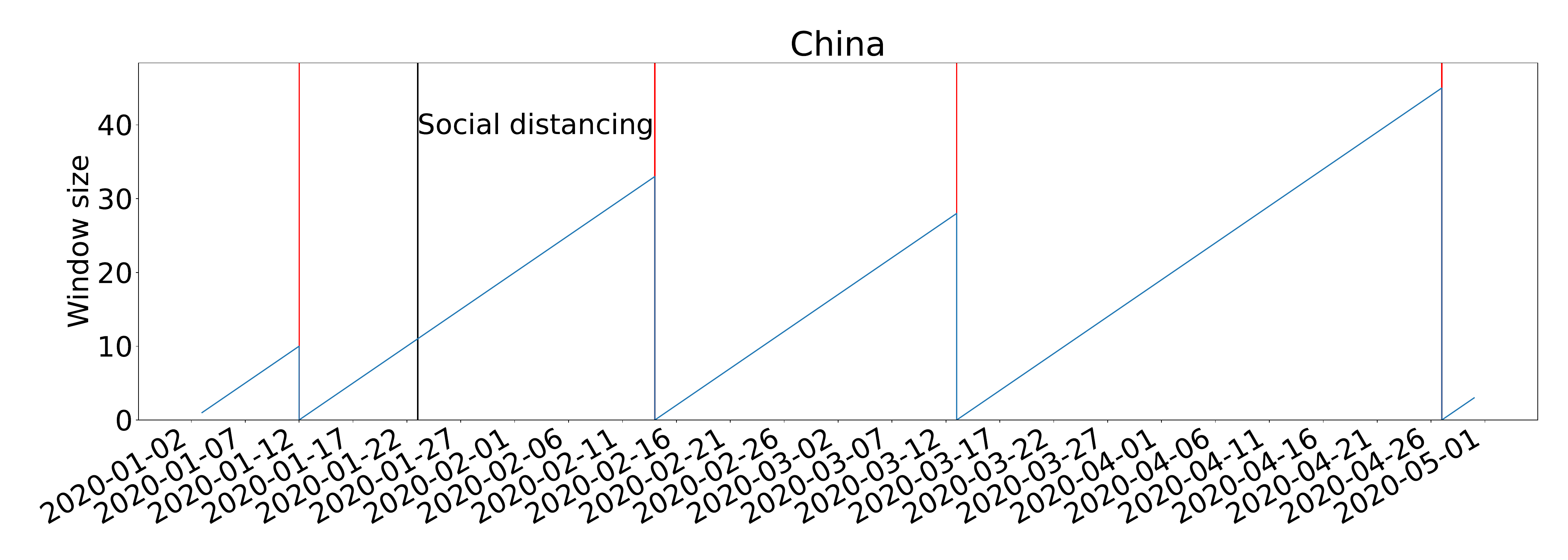} \\
			\vspace{-0.35cm}
			\textbf{d} & \includegraphics[keepaspectratio, height=3.3cm, valign=T]
			{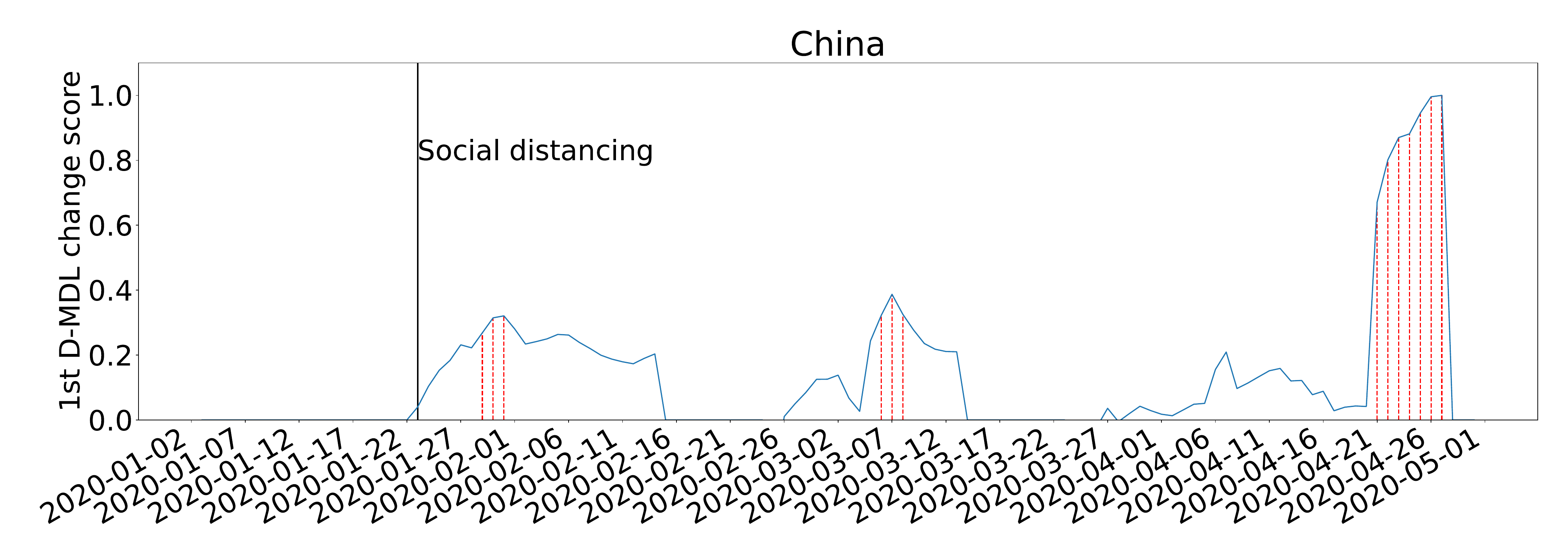} \\
			\vspace{-0.35cm}
			\textbf{e} & \includegraphics[keepaspectratio, height=3.3cm, valign=T]
			{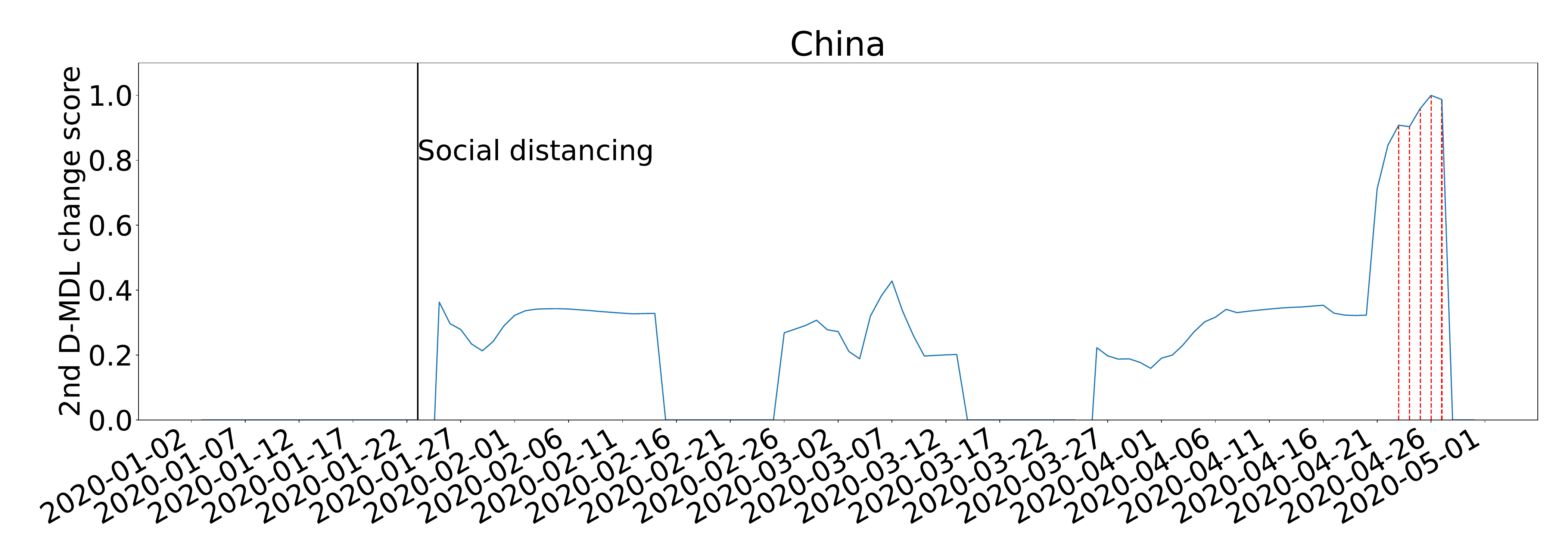} \\
		\end{tabular}
			\caption{\textbf{The results for China with exponential modeling.} The date on which the social distancing was implemented is marked by a solid line in black. \textbf{a,} the number of cumulative cases. \textbf{b,} the change scores produced by the 0th M-DML where the line in blue denotes values of scores and dashed lines in red mark alarms. \textbf{c,} the window sized for the sequential D-DML algorithm with adaptive window where lines in red mark the shrinkage of windows. \textbf{d,} the change scores produced by the 1st D-MDL. \textbf{e,} the change scores produced by the 2nd D-MDL.}
\end{figure}

\begin{figure}[H] 
\centering
\begin{tabular}{cc}
		 	\textbf{a} & \includegraphics[keepaspectratio, height=3.3cm, valign=T]
			{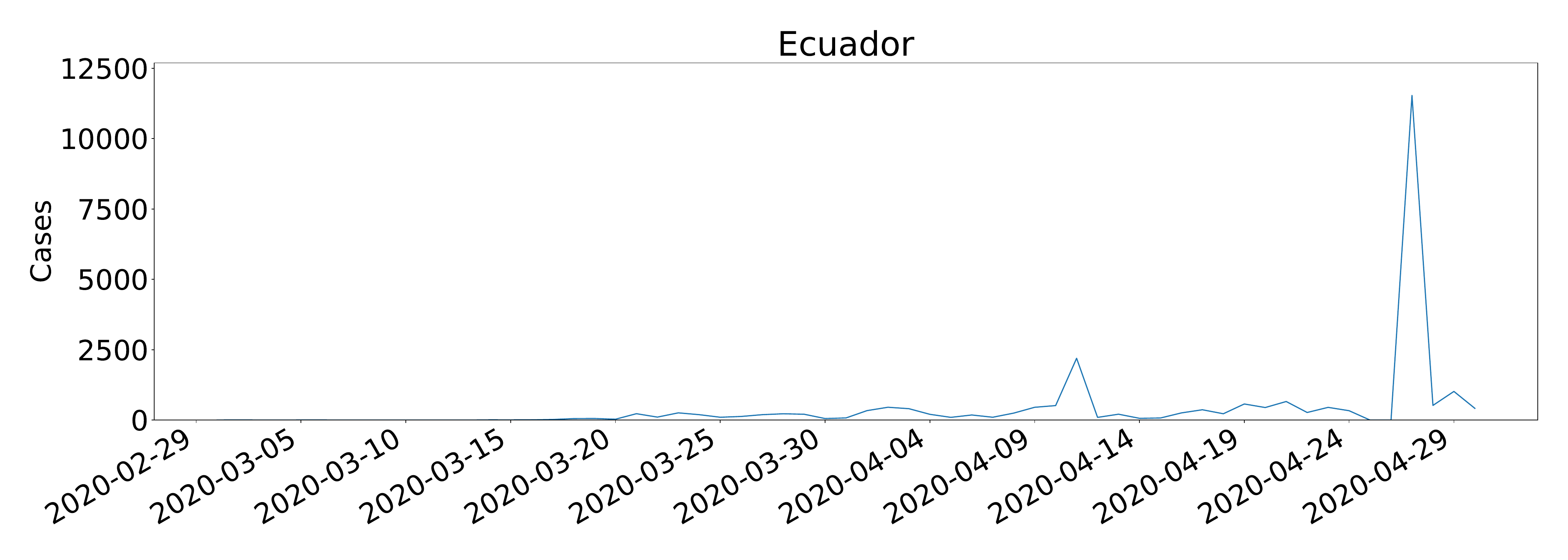} \\
			\vspace{-0.35cm}
	 	    \textbf{b} & \includegraphics[keepaspectratio, height=3.3cm, valign=T]
			{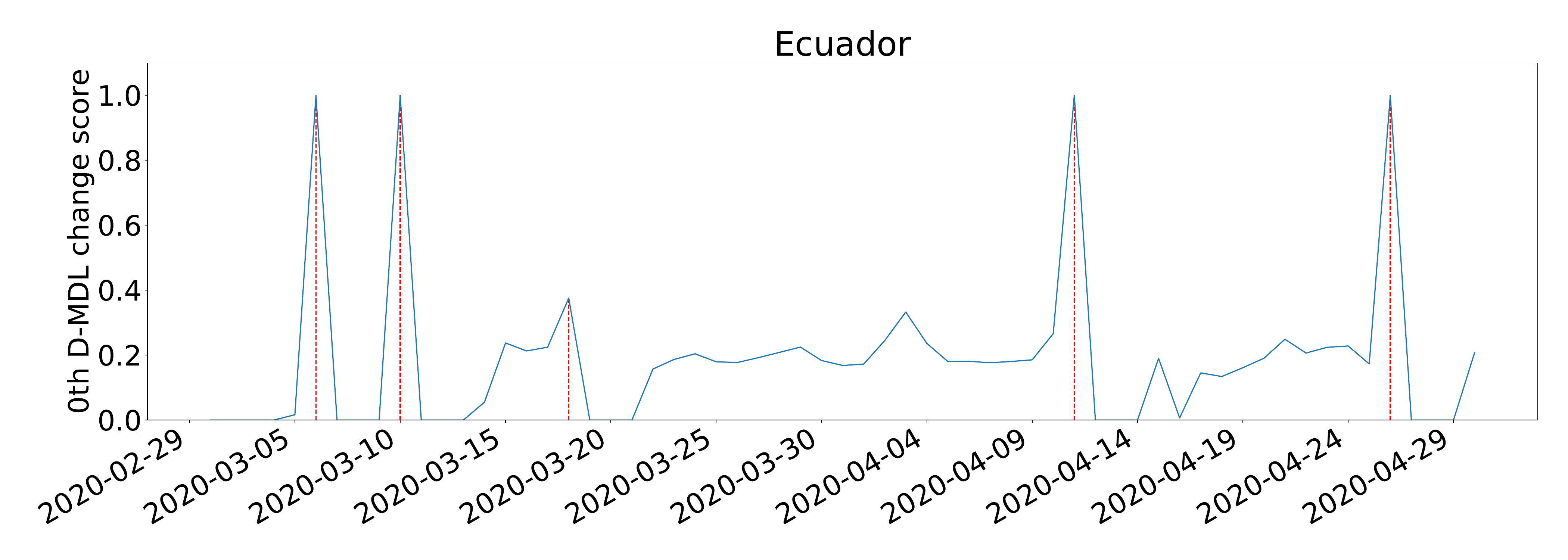}   \\
	        \vspace{-0.35cm}
			\textbf{c} & \includegraphics[keepaspectratio, height=3.3cm, valign=T]
			{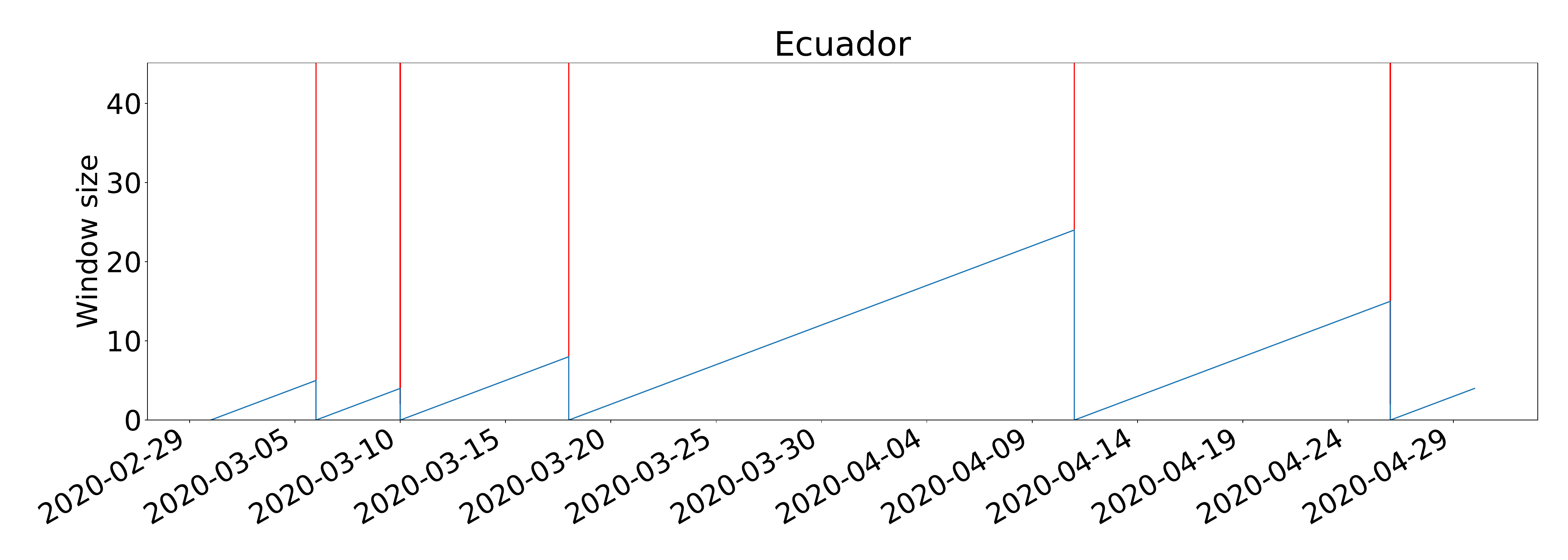} \\
		    \vspace{-0.35cm}
			\textbf{d} & \includegraphics[keepaspectratio, height=3.3cm, valign=T]
			{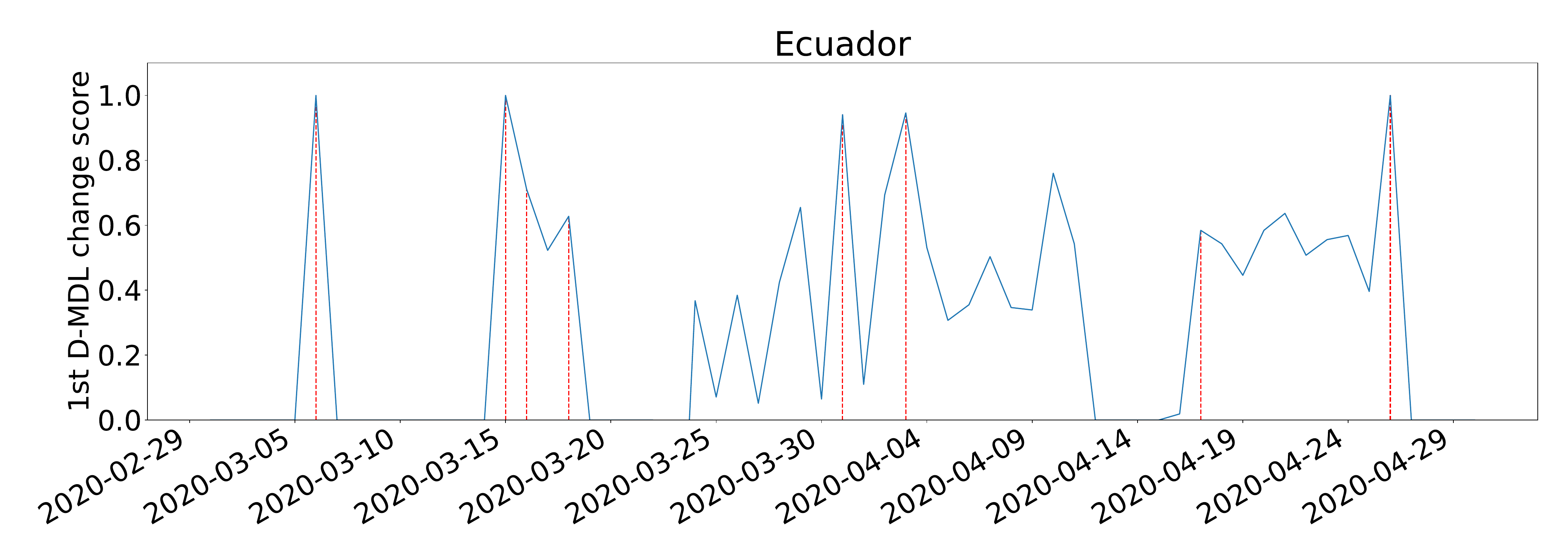} \\
		    \vspace{-0.35cm}
			\textbf{e} & \includegraphics[keepaspectratio, height=3.3cm, valign=T]
			{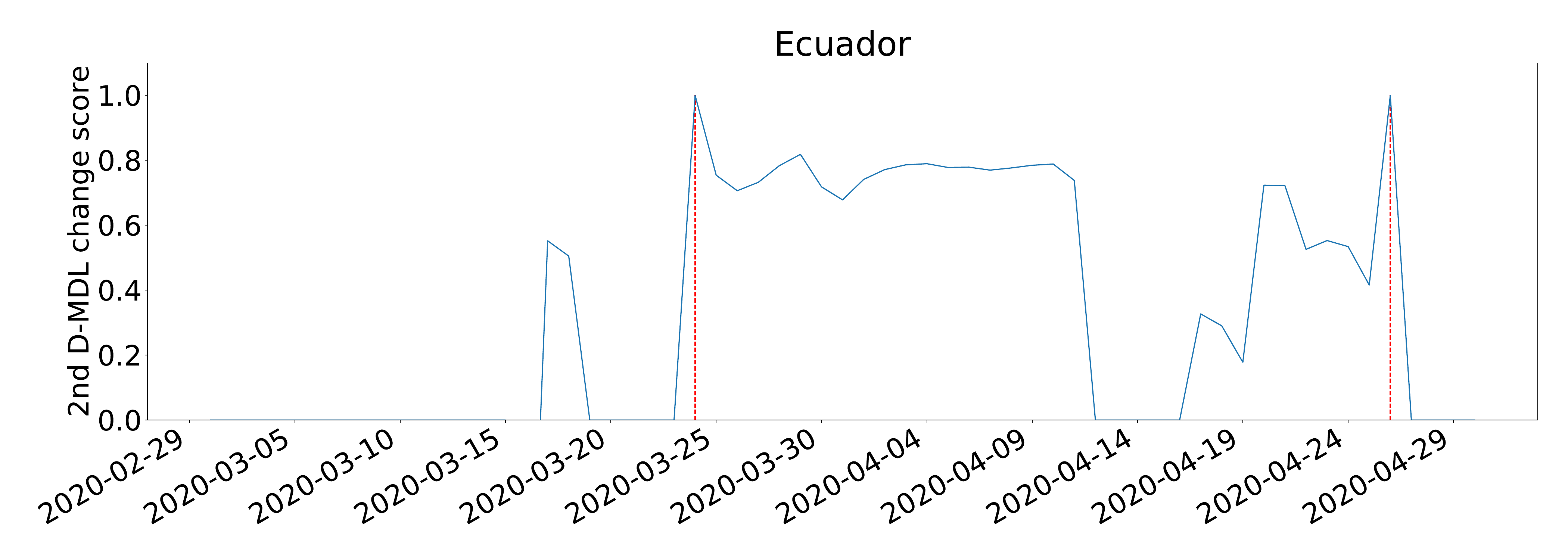} \\
		\end{tabular}
			\caption{\textbf{The results for Ecuador with Gaussian modeling.} The date on which the social distancing was implemented is marked by a solid line in black. \textbf{a,} the number of daily new cases. \textbf{b,} the change scores produced by the 0th M-DML where the line in blue denotes values of scores and dashed lines in red mark alarms. \textbf{c,} the window sized for the sequential D-DML algorithm with adaptive window where lines in red mark the shrinkage of windows. \textbf{d,} the change scores produced by the 1st D-MDL. \textbf{e,} the change scores produced by the 2nd D-MDL.}
\end{figure}

\begin{figure}[H]  
\centering
\begin{tabular}{cc}
			\textbf{a} & \includegraphics[keepaspectratio, height=3.3cm, valign=T]
			{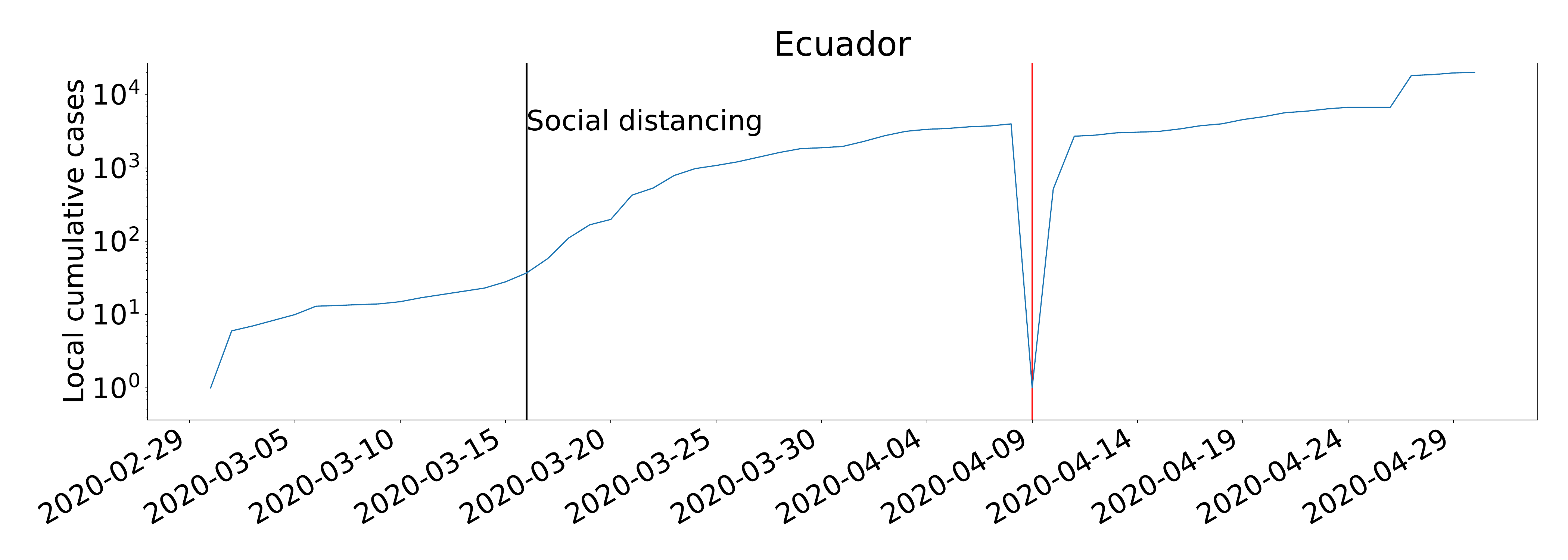} \\
	        \vspace{-0.35cm}
            \textbf{b} & \includegraphics[keepaspectratio, height=3.3cm, valign=T]
			{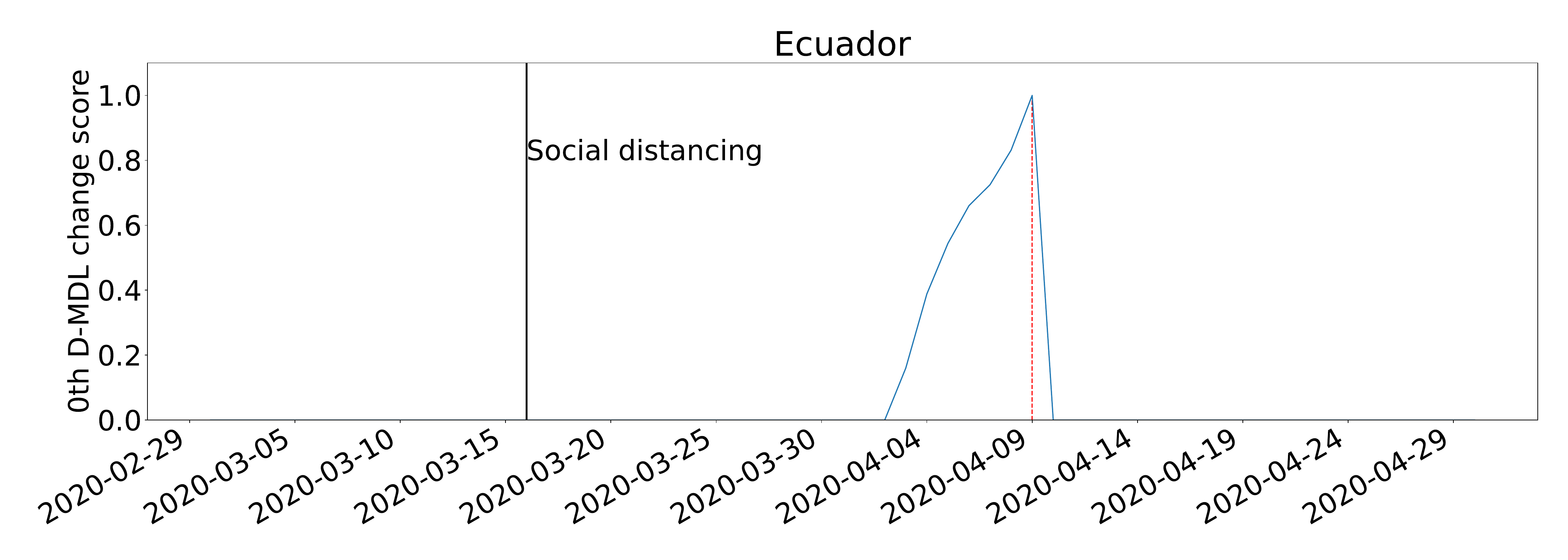}   \\
            \vspace{-0.35cm}
            \textbf{c} & \includegraphics[keepaspectratio, height=3.3cm, valign=T]
			{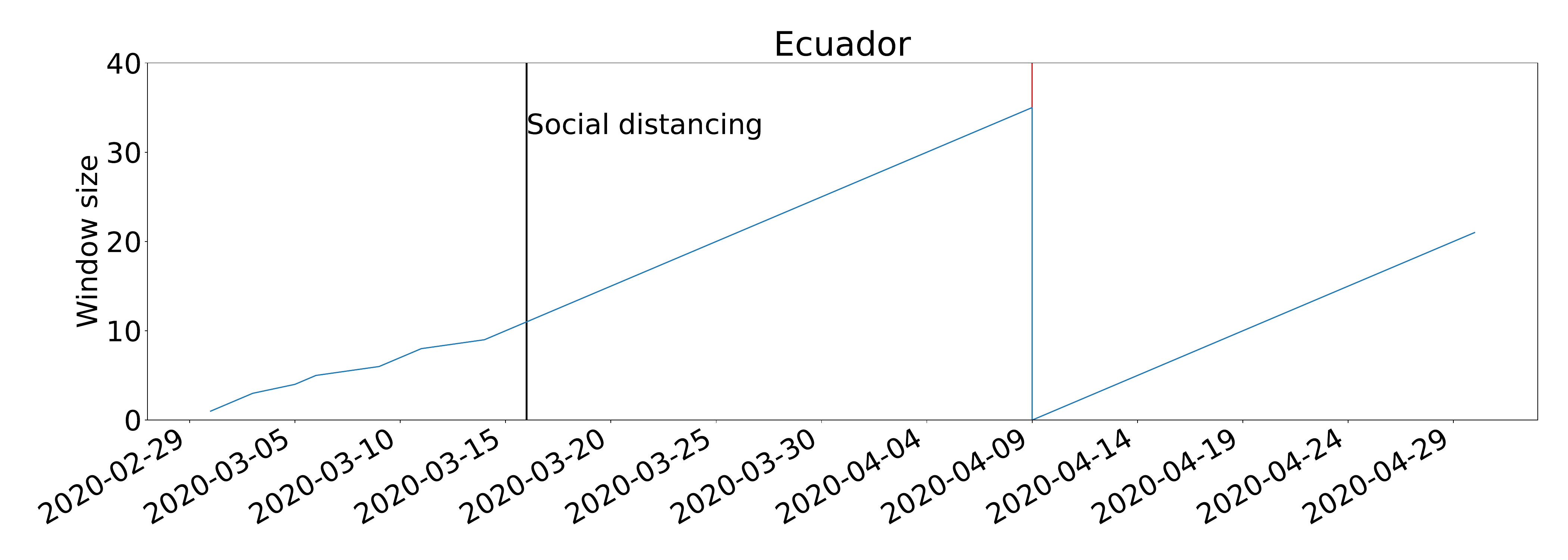} \\
			\vspace{-0.35cm}
			\textbf{d} & \includegraphics[keepaspectratio, height=3.3cm, valign=T]
			{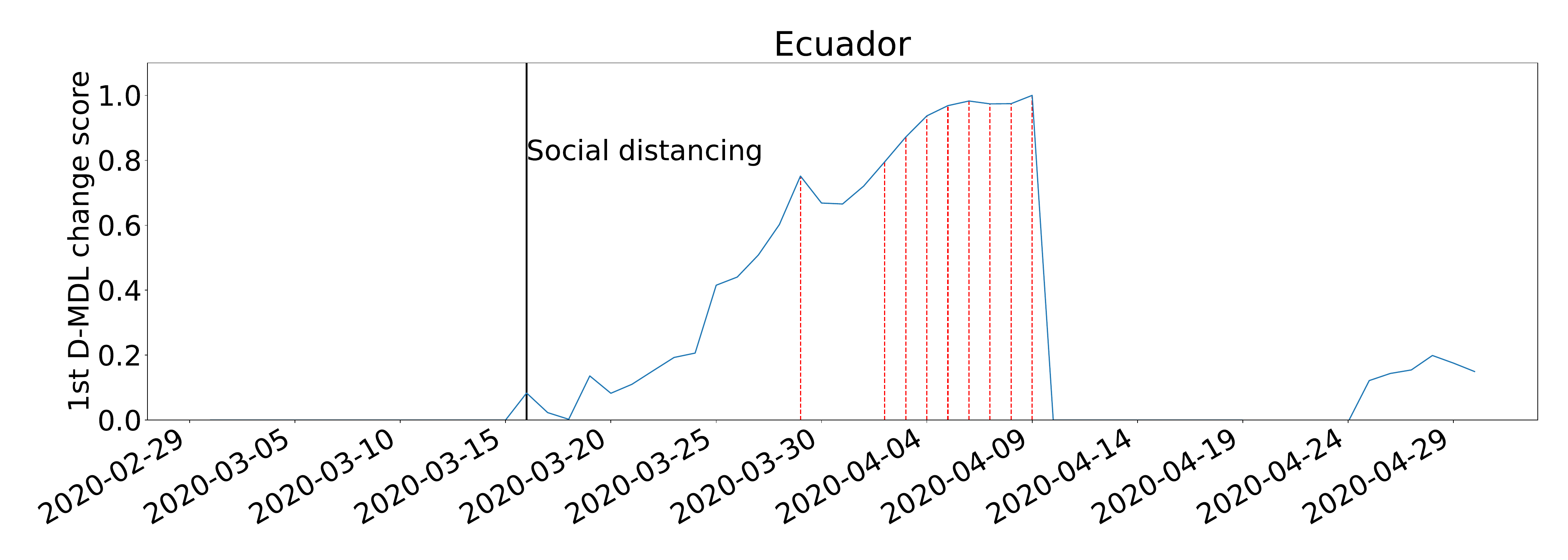} \\
			\vspace{-0.35cm}
			\textbf{e} & \includegraphics[keepaspectratio, height=3.3cm, valign=T]
			{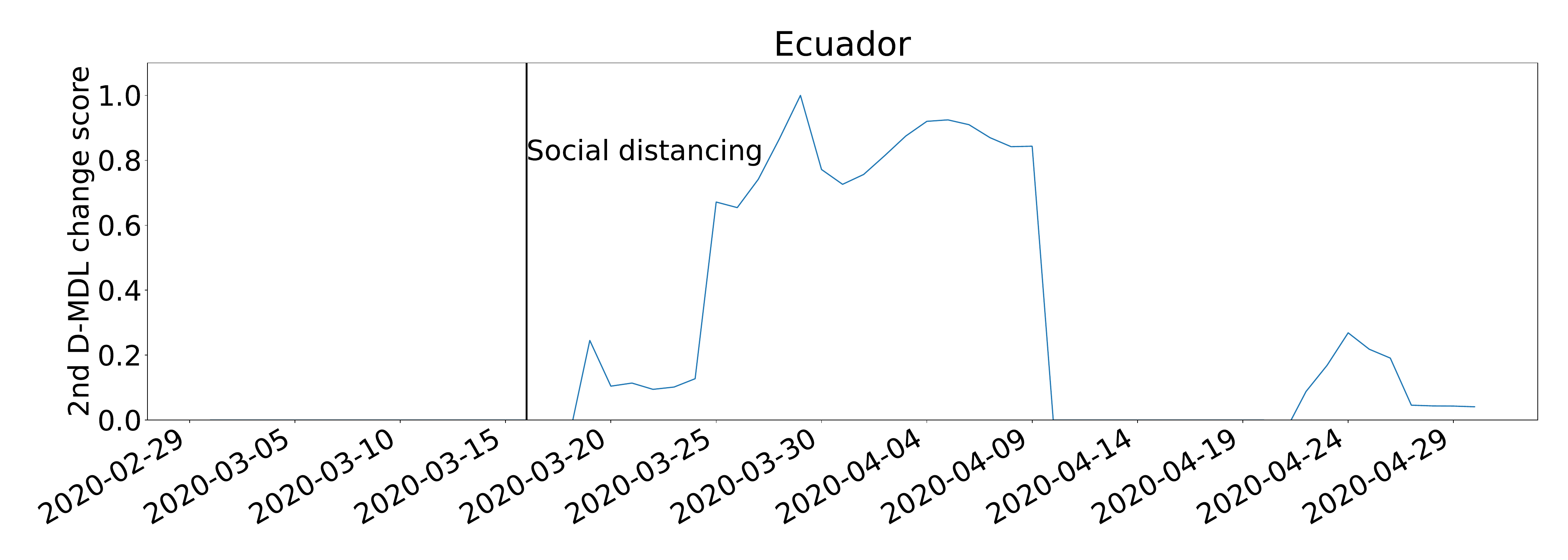} \\
		\end{tabular}
			\caption{\textbf{The results for Ecuador with exponential modeling.} The date on which the social distancing was implemented is marked by a solid line in black. \textbf{a,} the number of cumulative cases. \textbf{b,} the change scores produced by the 0th M-DML where the line in blue denotes values of scores and dashed lines in red mark alarms. \textbf{c,} the window sized for the sequential D-DML algorithm with adaptive window where lines in red mark the shrinkage of windows. \textbf{d,} the change scores produced by the 1st D-MDL. \textbf{e,} the change scores produced by the 2nd D-MDL.}
\end{figure}

\begin{figure}[H] 
\centering
\begin{tabular}{cc}
		 	\textbf{a} & \includegraphics[keepaspectratio, height=3.3cm, valign=T]
			{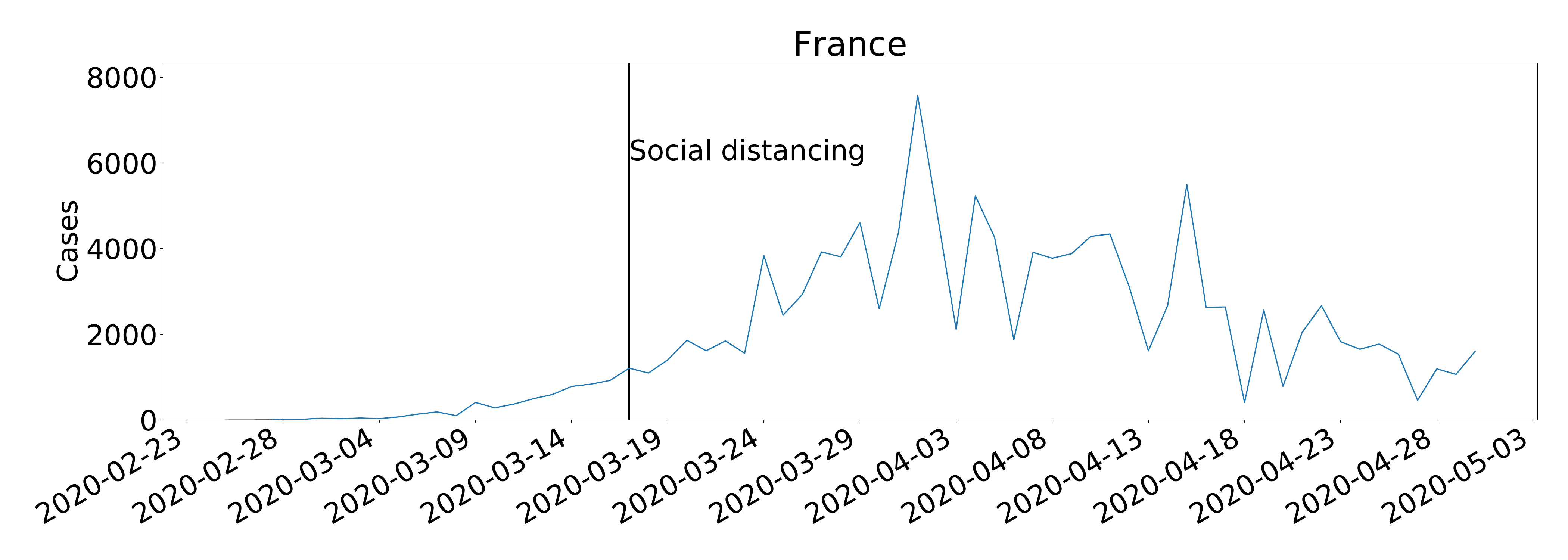} \\
			\vspace{-0.35cm}
	 	    \textbf{b} & \includegraphics[keepaspectratio, height=3.3cm, valign=T]
			{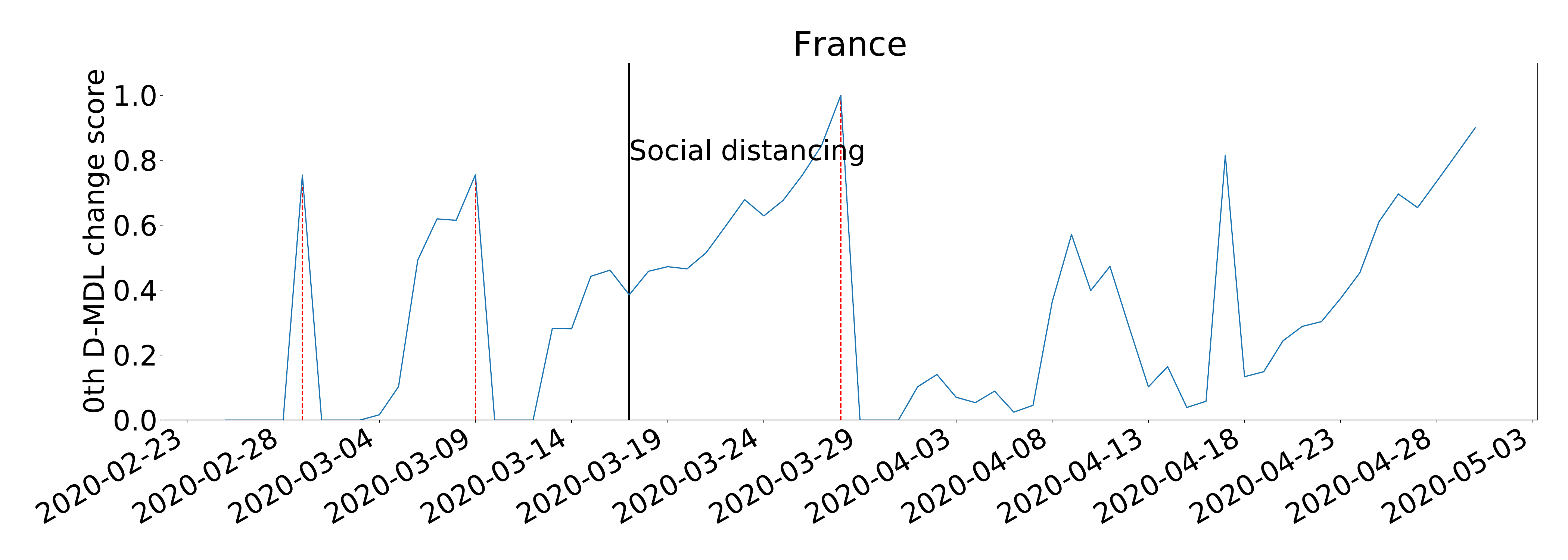}   \\
	        \vspace{-0.35cm}
			\textbf{c} & \includegraphics[keepaspectratio, height=3.3cm, valign=T]
			{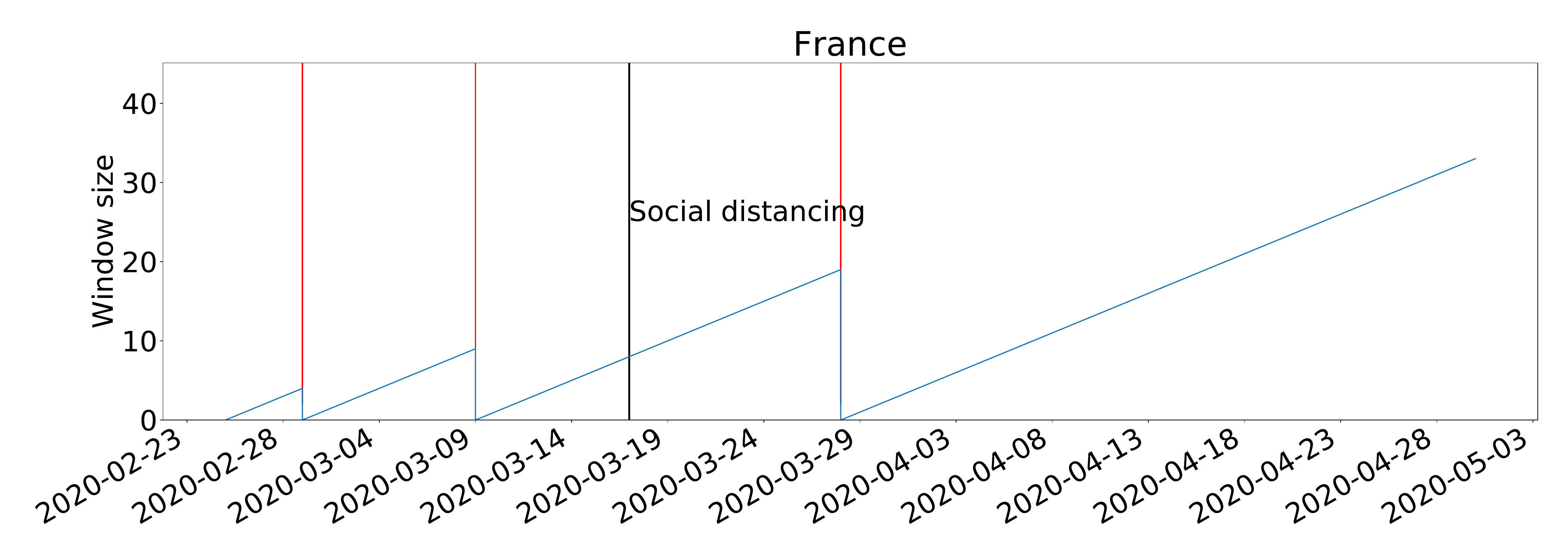} \\
		    \vspace{-0.35cm}
			\textbf{d} & \includegraphics[keepaspectratio, height=3.3cm, valign=T]
			{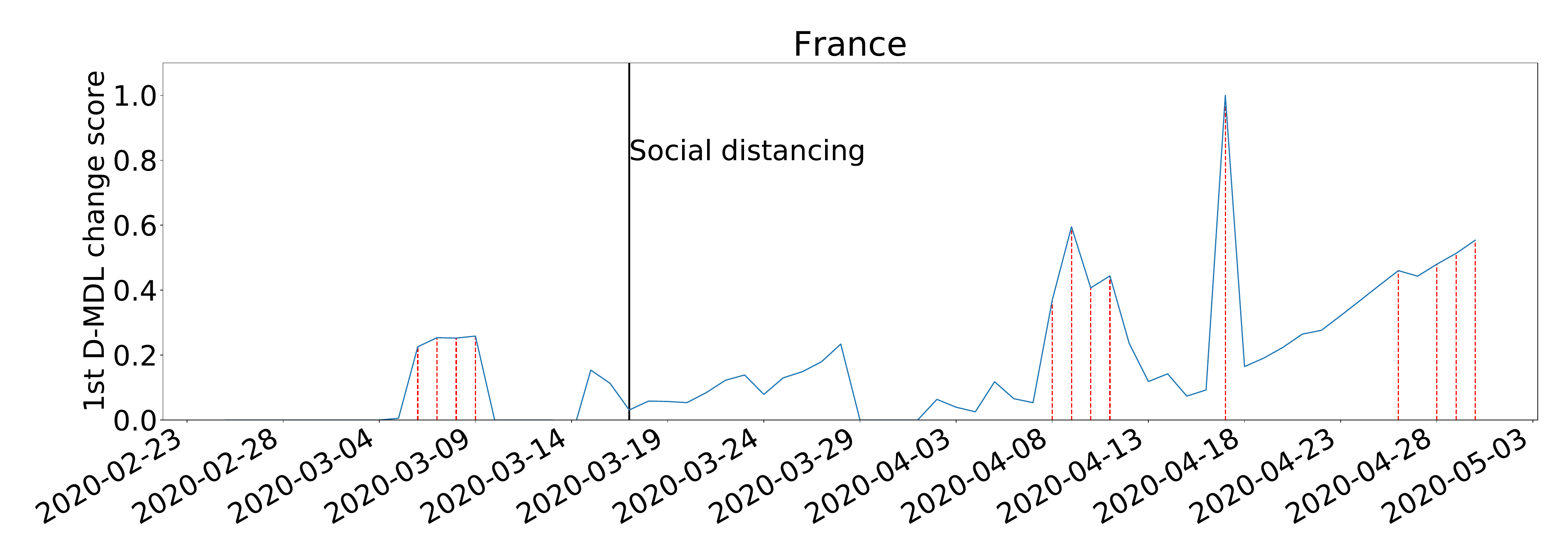} \\
		    \vspace{-0.35cm}
			\textbf{e} & \includegraphics[keepaspectratio, height=3.3cm, valign=T]
			{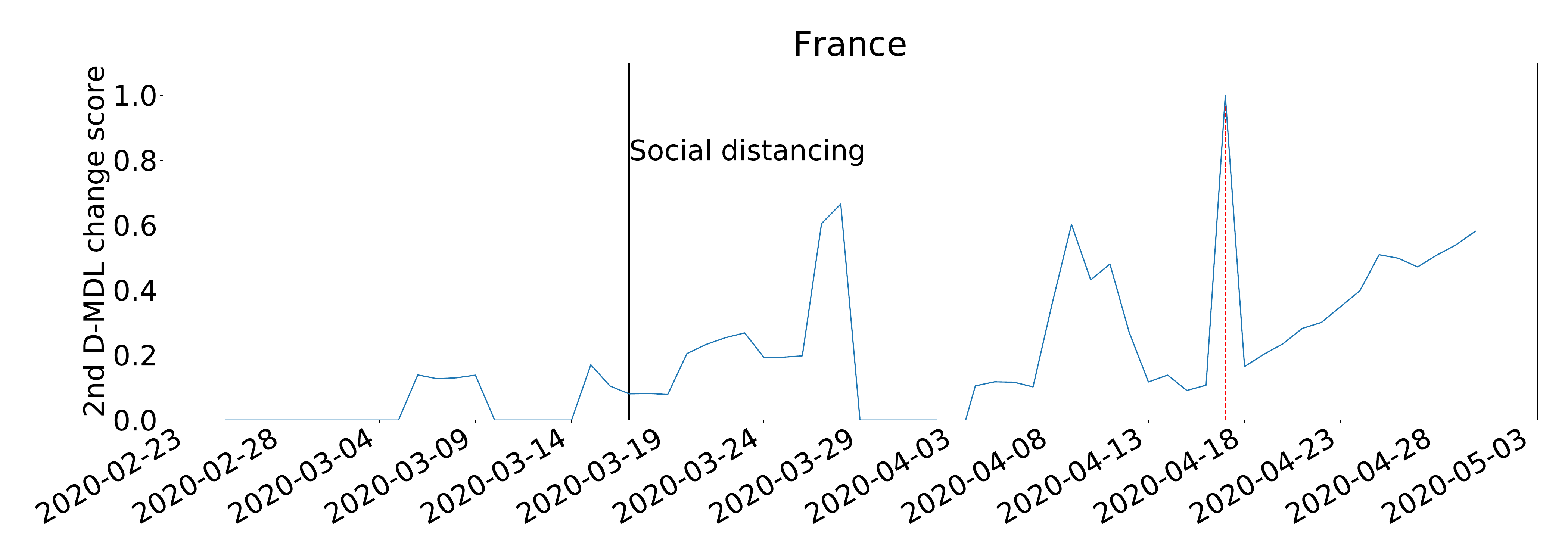} \\
		\end{tabular}
			\caption{\textbf{The results for France with Gaussian modeling.} The date on which the social distancing was implemented is marked by a solid line in black. \textbf{a,} the number of daily new cases. \textbf{b,} the change scores produced by the 0th M-DML where the line in blue denotes values of scores and dashed lines in red mark alarms. \textbf{c,} the window sized for the sequential D-DML algorithm with adaptive window where lines in red mark the shrinkage of windows. \textbf{d,} the change scores produced by the 1st D-MDL. \textbf{e,} the change scores produced by the 2nd D-MDL.}
\end{figure}

\begin{figure}[H]  
\centering
\begin{tabular}{cc}
			\textbf{a} & \includegraphics[keepaspectratio, height=3.3cm, valign=T]
			{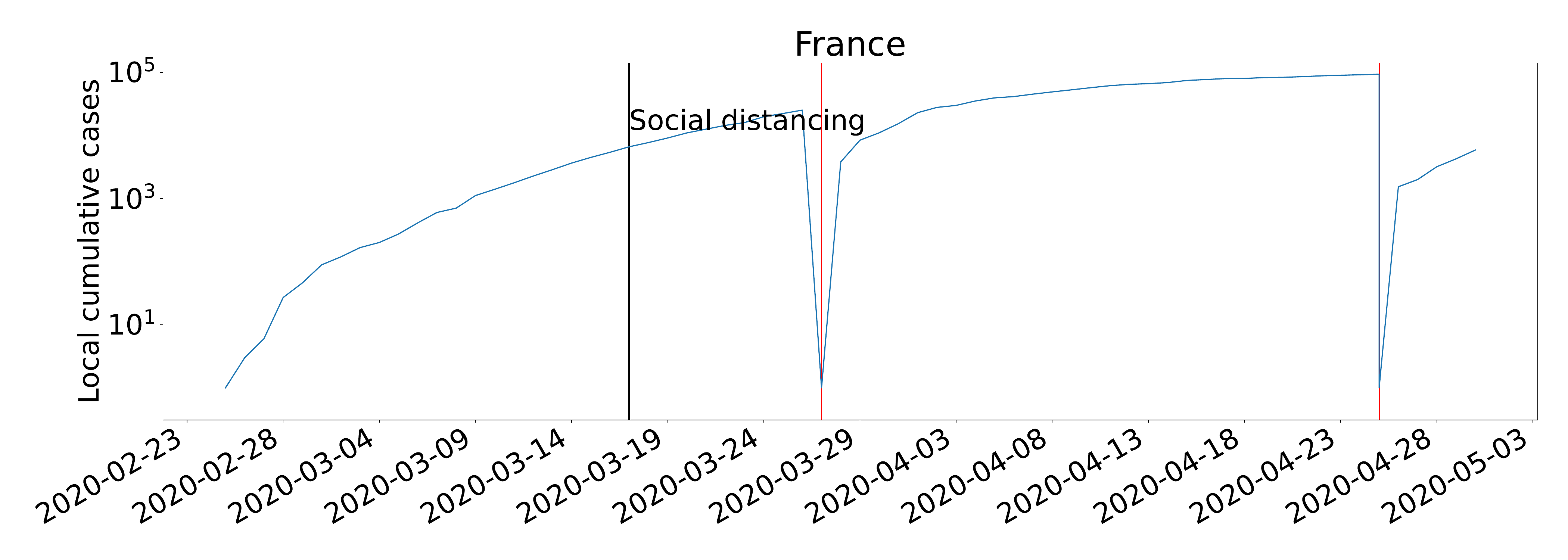} \\
	        \vspace{-0.35cm}
            \textbf{b} & \includegraphics[keepaspectratio, height=3.3cm, valign=T]
			{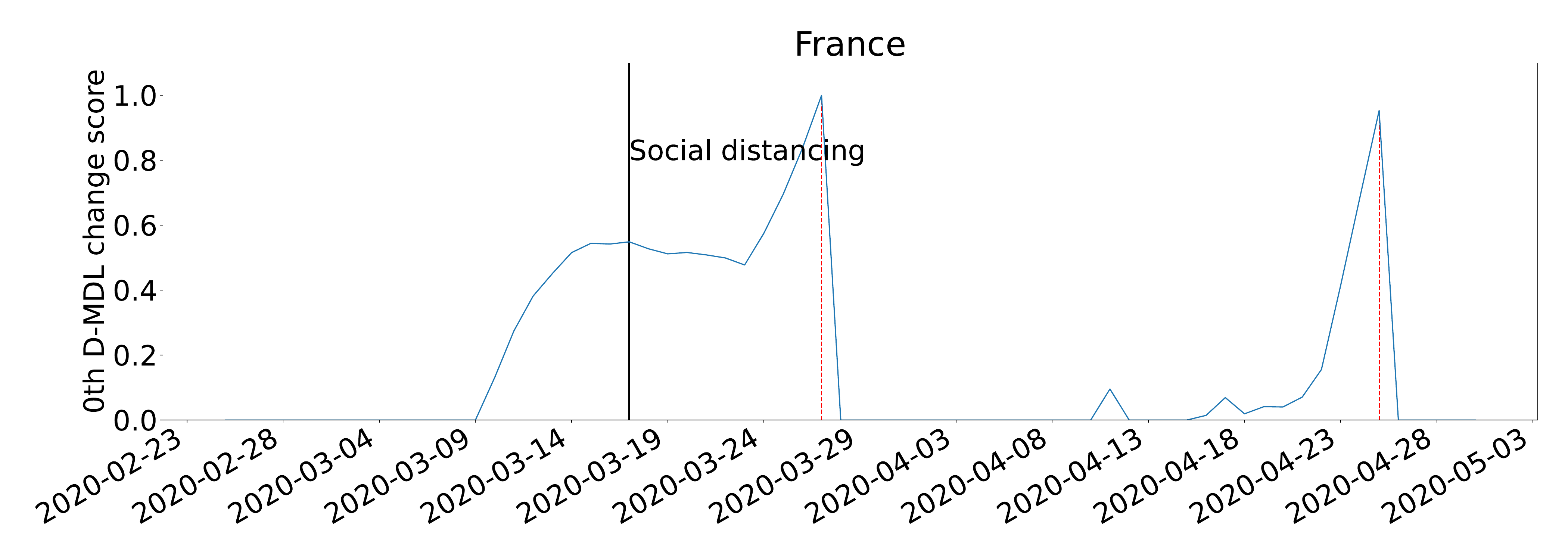}   \\
            \vspace{-0.35cm}
            \textbf{c} & \includegraphics[keepaspectratio, height=3.3cm, valign=T]
			{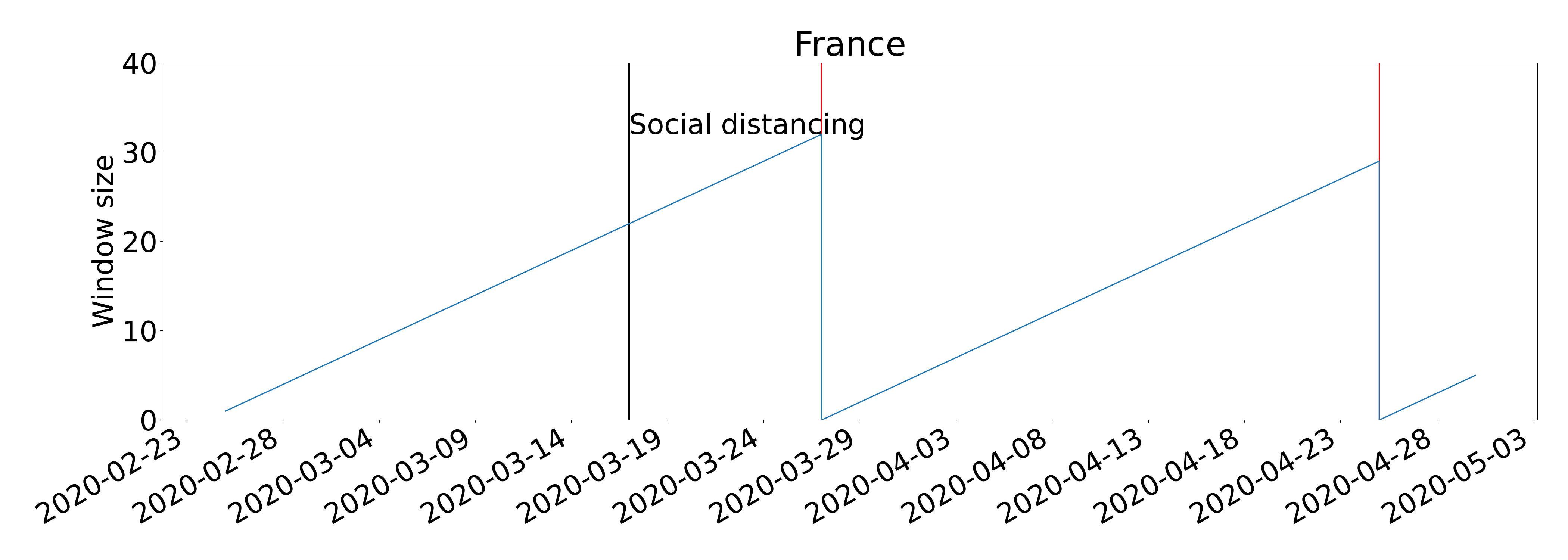} \\
			\vspace{-0.35cm}
			\textbf{d} & \includegraphics[keepaspectratio, height=3.3cm, valign=T]
			{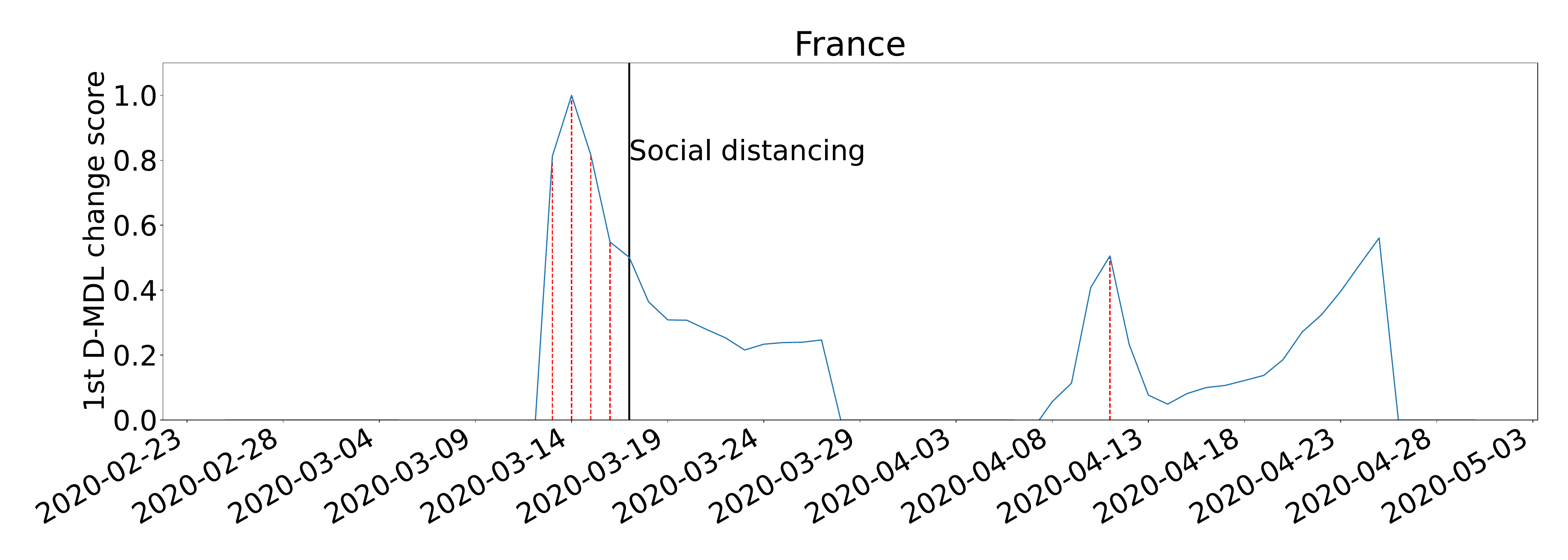} \\
			\vspace{-0.35cm}
			\textbf{e} & \includegraphics[keepaspectratio, height=3.3cm, valign=T]
			{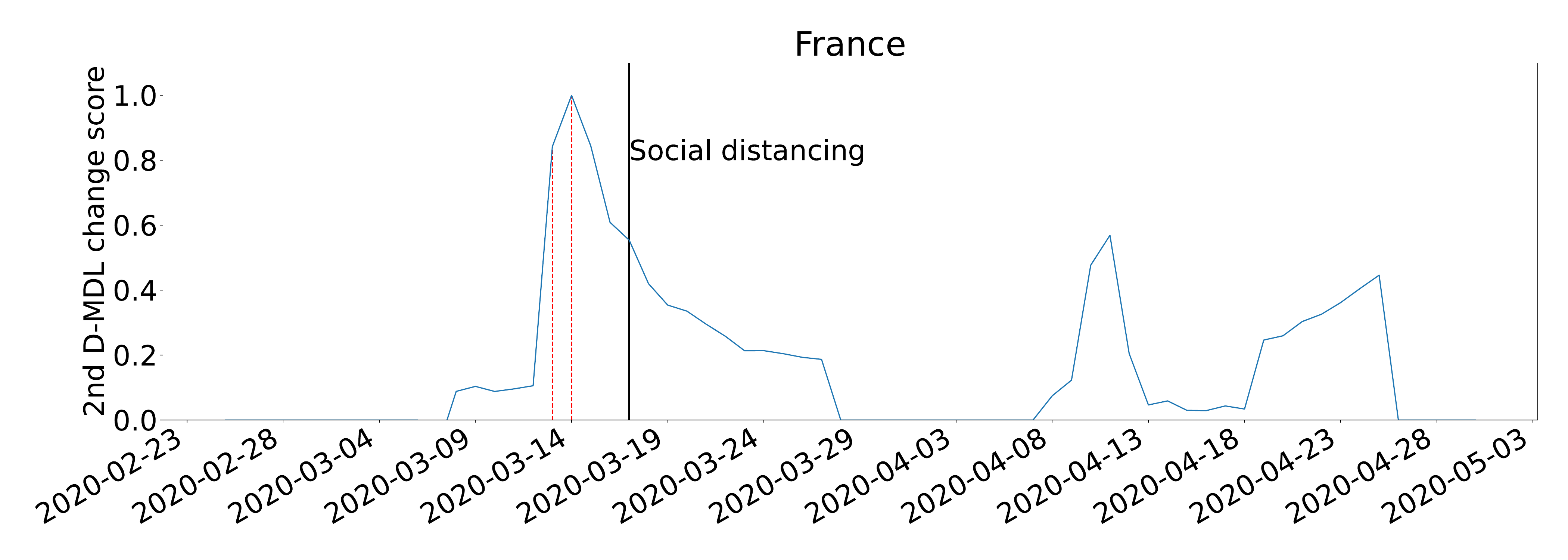} \\
		\end{tabular}
			\caption{\textbf{The results for France with exponential modeling.} The date on which the social distancing was implemented is marked by a solid line in black. \textbf{a,} the number of cumulative cases. \textbf{b,} the change scores produced by the 0th M-DML where the line in blue denotes values of scores and dashed lines in red mark alarms. \textbf{c,} the window sized for the sequential D-DML algorithm with adaptive window where lines in red mark the shrinkage of windows. \textbf{d,} the change scores produced by the 1st D-MDL. \textbf{e,} the change scores produced by the 2nd D-MDL.}
\end{figure}

\begin{figure}[H] 
\centering
\begin{tabular}{cc}
		 	\textbf{a} & \includegraphics[keepaspectratio, height=3.3cm, valign=T]
			{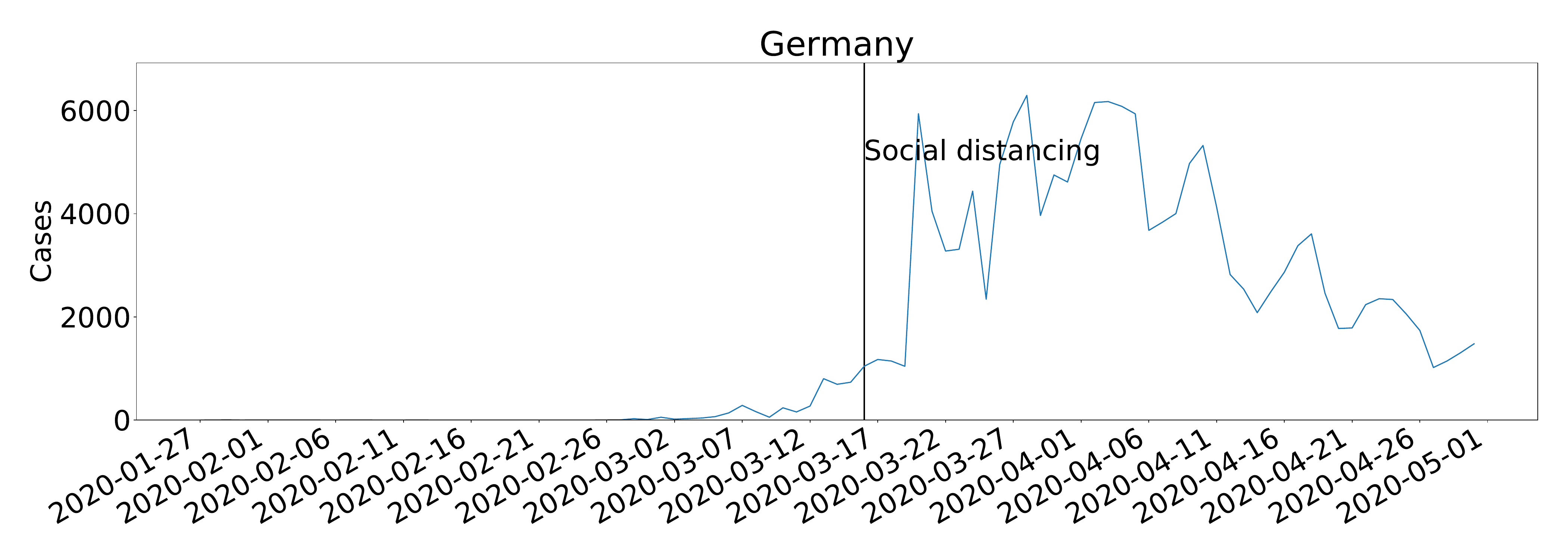} \\
			\vspace{-0.35cm}
	 	    \textbf{b} & \includegraphics[keepaspectratio, height=3.3cm, valign=T]
			{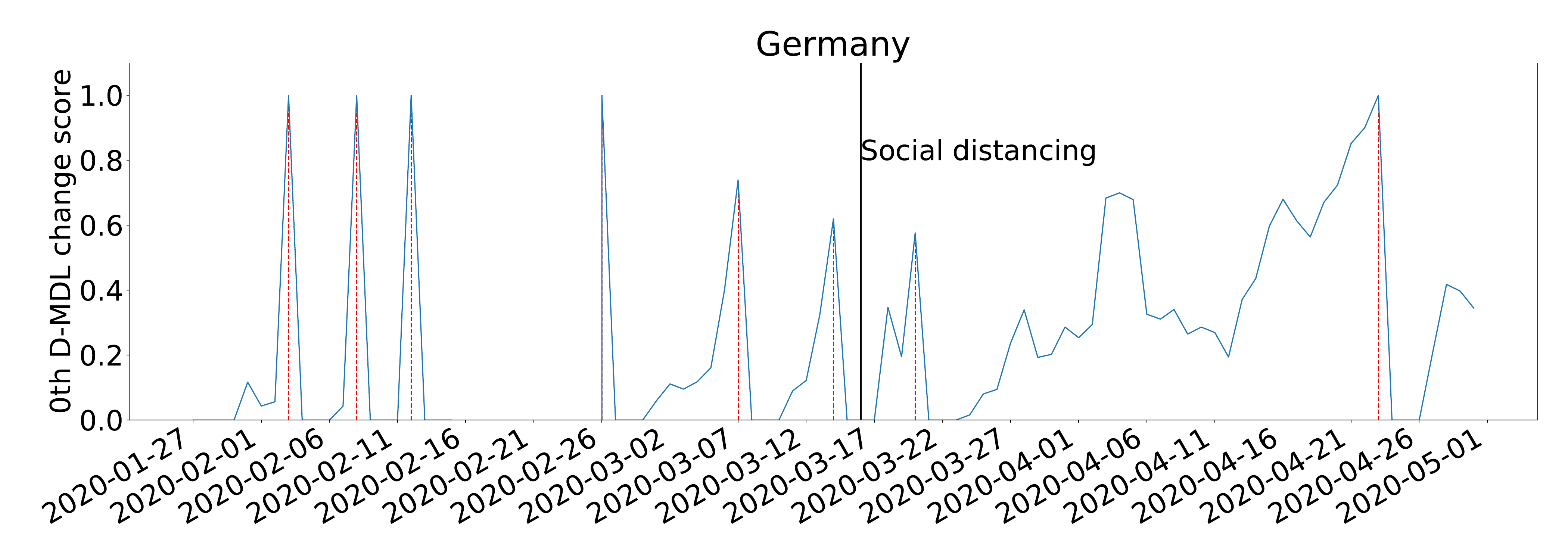}   \\
	        \vspace{-0.35cm}
			\textbf{c} & \includegraphics[keepaspectratio, height=3.3cm, valign=T]
			{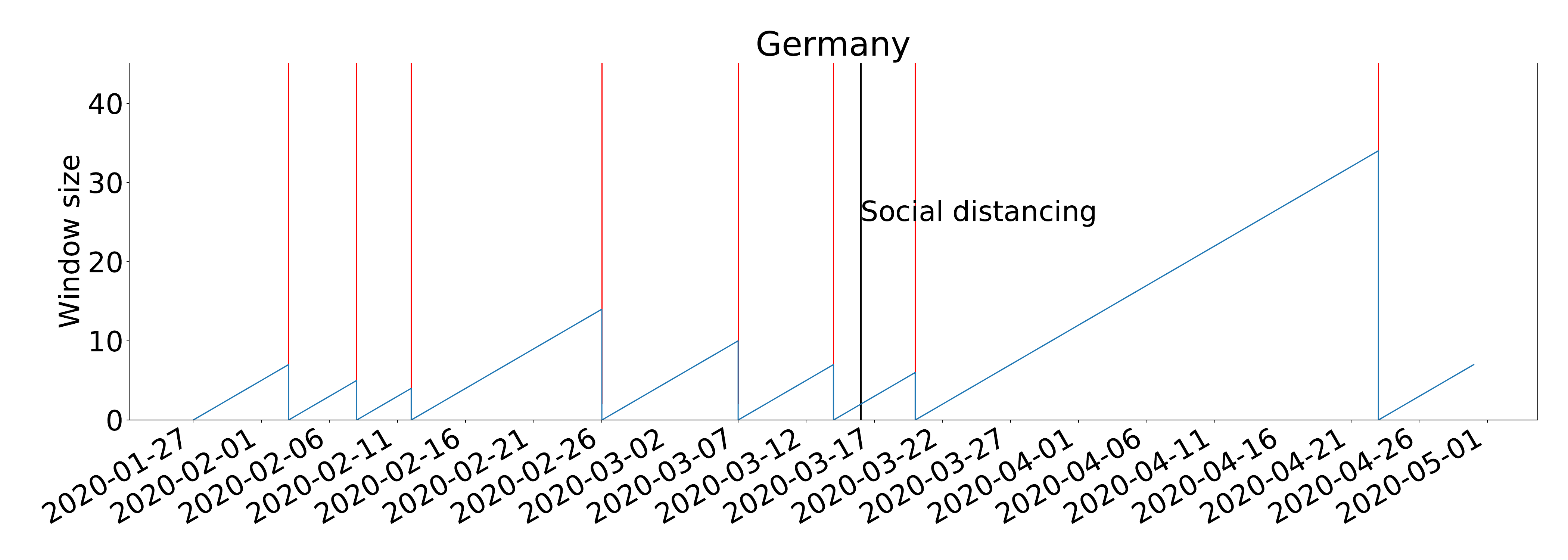} \\
		    \vspace{-0.35cm}
			\textbf{d} & \includegraphics[keepaspectratio, height=3.3cm, valign=T]
			{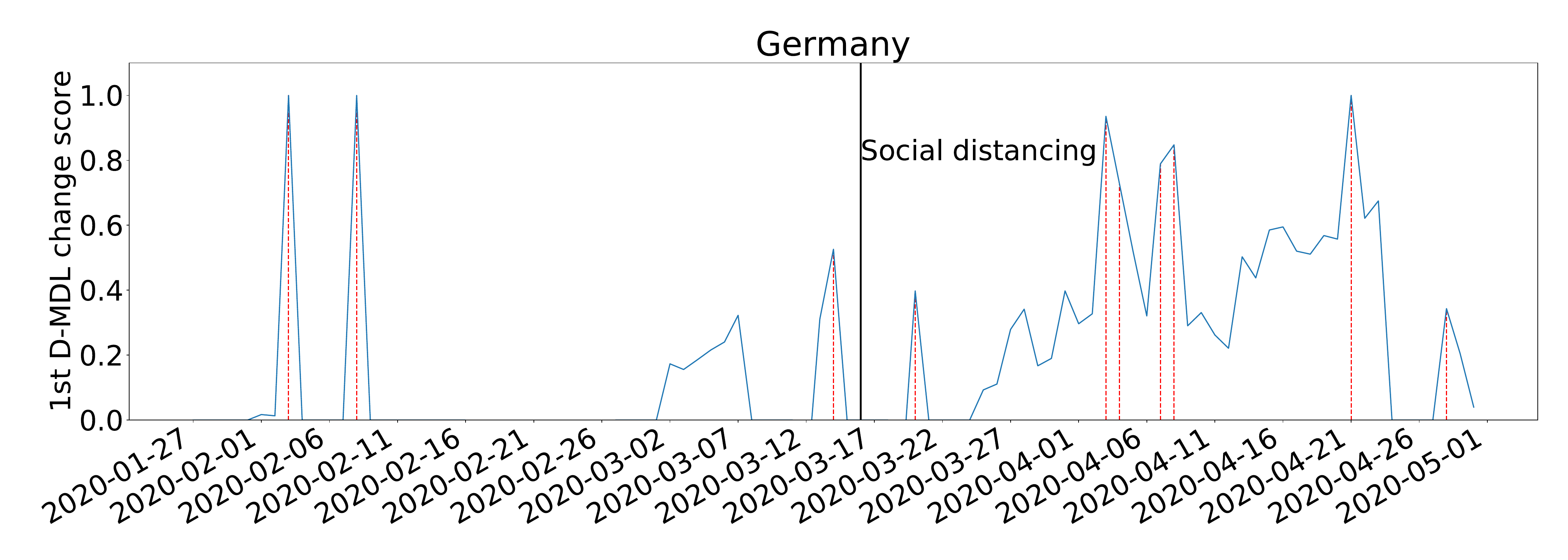} \\
		    \vspace{-0.35cm}
			\textbf{e} & \includegraphics[keepaspectratio, height=3.3cm, valign=T]
			{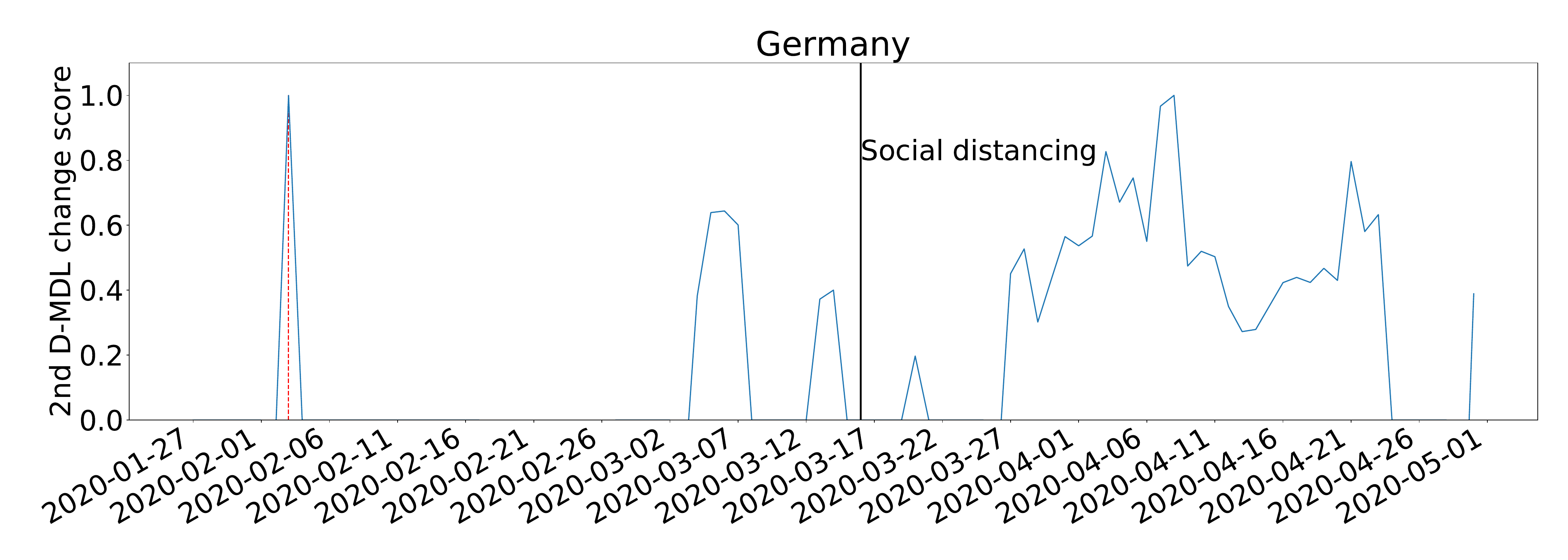} \\
		\end{tabular}
			\caption{\textbf{The results for Germany with Gaussian modeling.} The date on which the social distancing was implemented is marked by a solid line in black. \textbf{a,} the number of daily new cases. \textbf{b,} the change scores produced by the 0th M-DML where the line in blue denotes values of scores and dashed lines in red mark alarms. \textbf{c,} the window sized for the sequential D-DML algorithm with adaptive window where lines in red mark the shrinkage of windows. \textbf{d,} the change scores produced by the 1st D-MDL. \textbf{e,} the change scores produced by the 2nd D-MDL.}
\end{figure}

\begin{figure}[H]  
\centering
\begin{tabular}{cc}
			\textbf{a} & \includegraphics[keepaspectratio, height=3.3cm, valign=T]
			{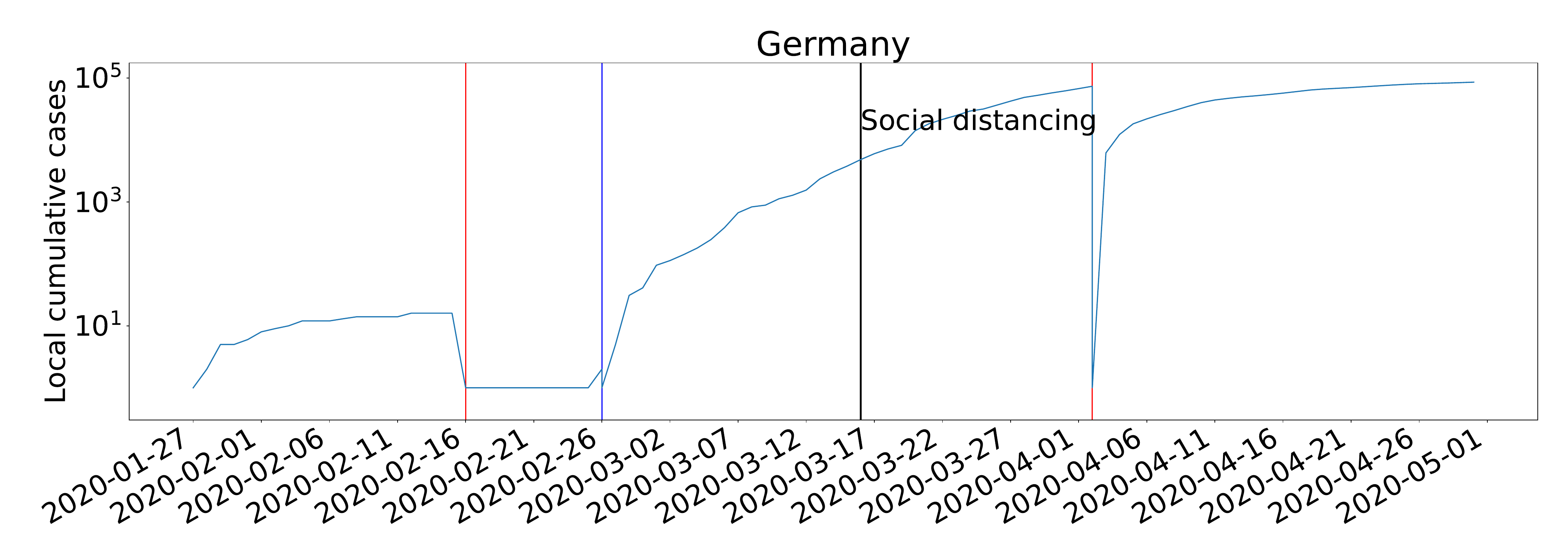} \\
	        \vspace{-0.35cm}
            \textbf{b} & \includegraphics[keepaspectratio, height=3.3cm, valign=T]
			{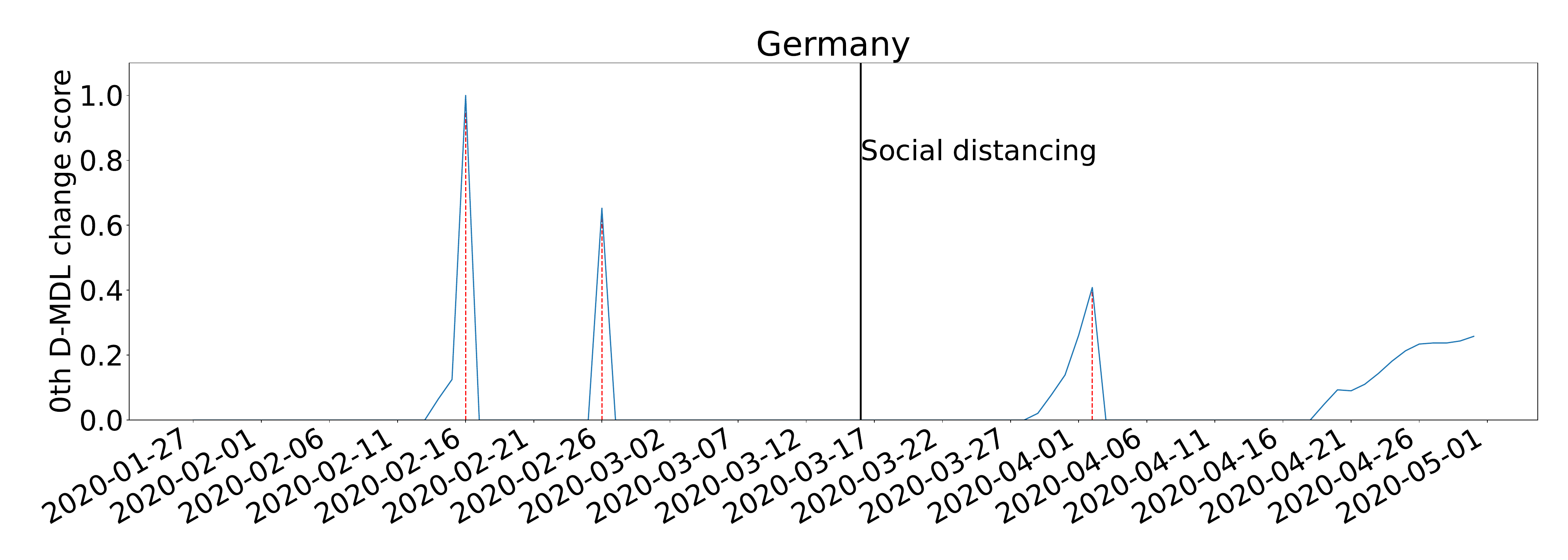}   \\
            \vspace{-0.35cm}
            \textbf{c} & \includegraphics[keepaspectratio, height=3.3cm, valign=T]
			{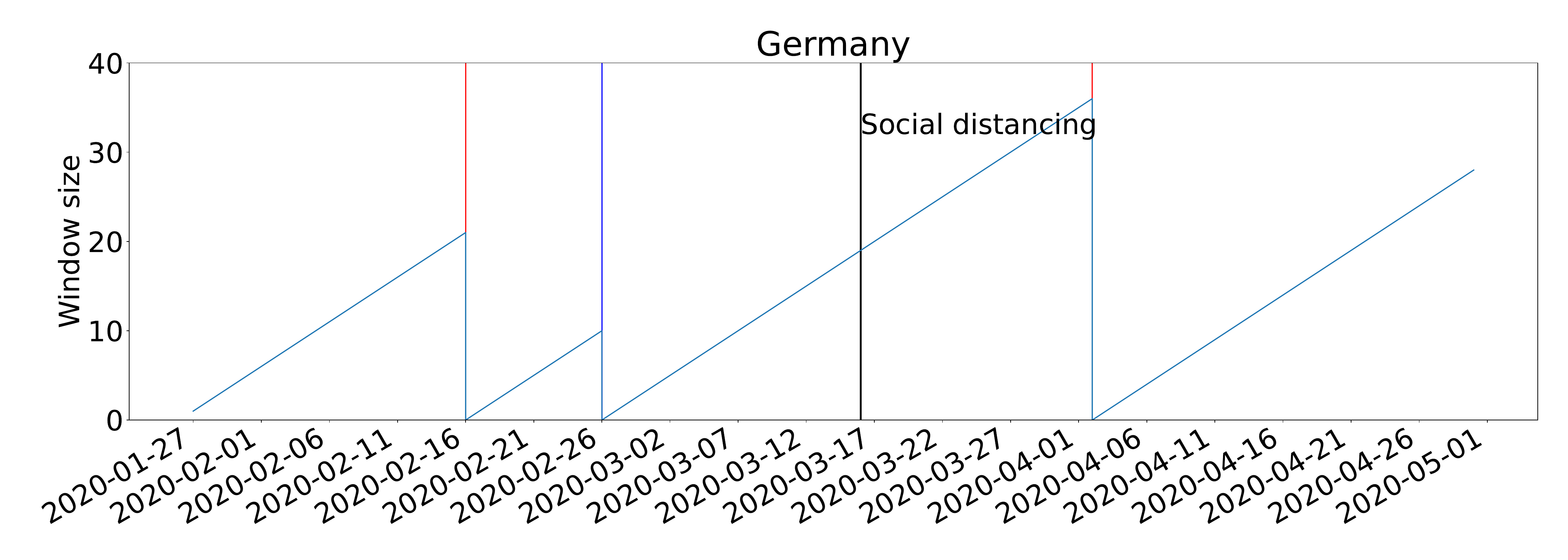} \\
			\vspace{-0.35cm}
			\textbf{d} & \includegraphics[keepaspectratio, height=3.3cm, valign=T]
			{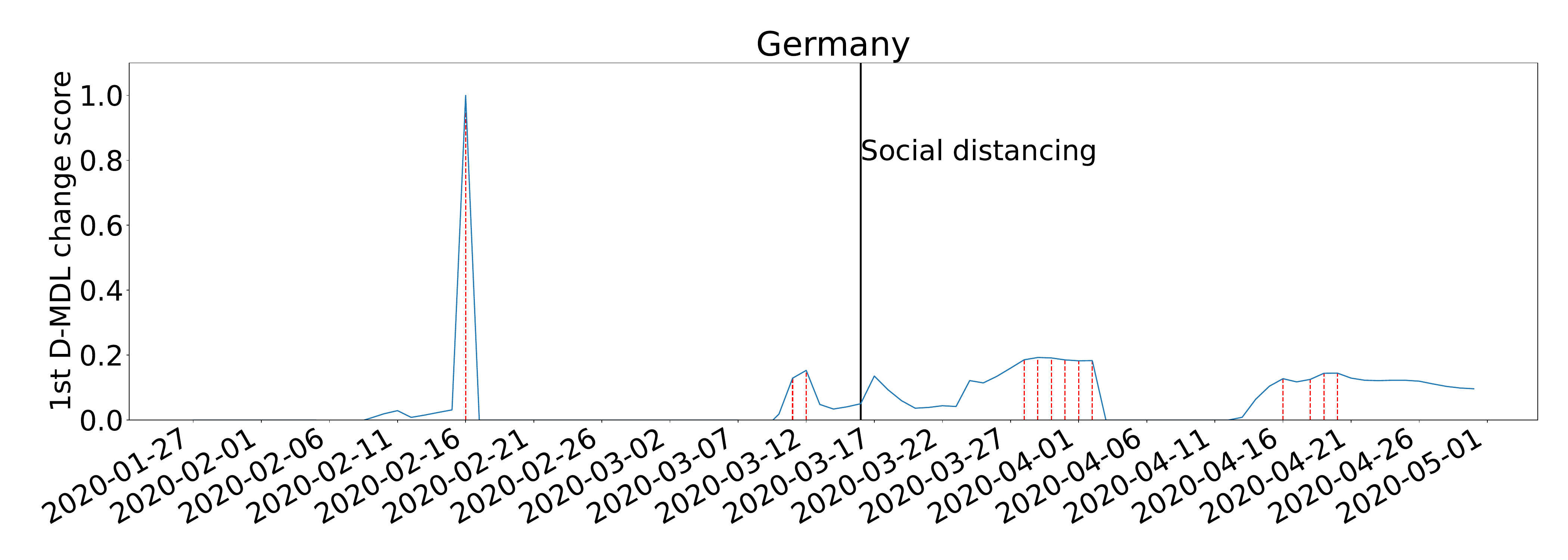} \\
			\vspace{-0.35cm}
			\textbf{e} & \includegraphics[keepaspectratio, height=3.3cm, valign=T]
			{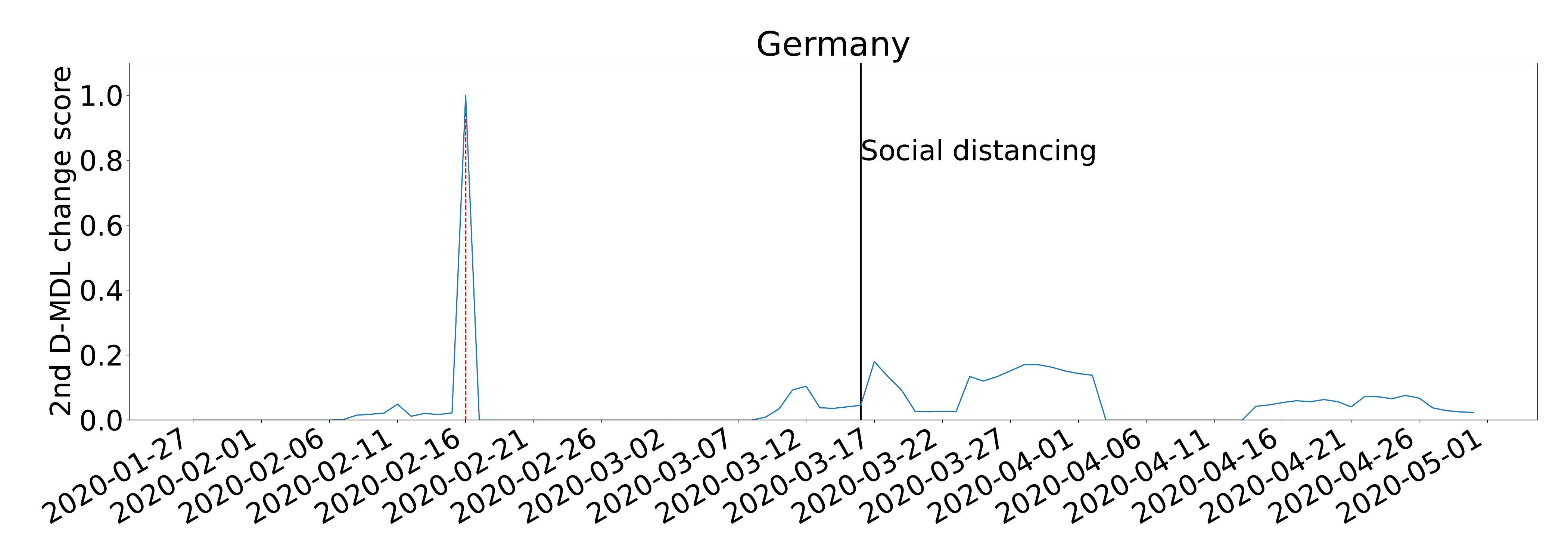} \\
		\end{tabular}
			\caption{\textbf{The results for Germany with exponential modeling.} The date on which the social distancing was implemented is marked by a solid line in black. \textbf{a,} the number of cumulative cases. \textbf{b,} the change scores produced by the 0th M-DML where the line in blue denotes values of scores and dashed lines in red mark alarms. \textbf{c,} the window sized for the sequential D-DML algorithm with adaptive window where lines in red mark the shrinkage of windows. \textbf{d,} the change scores produced by the 1st D-MDL. \textbf{e,} the change scores produced by the 2nd D-MDL.}
\end{figure}

\begin{figure}[H] 
\centering
\begin{tabular}{cc}
		 	\textbf{a} & \includegraphics[keepaspectratio, height=3.3cm, valign=T]
			{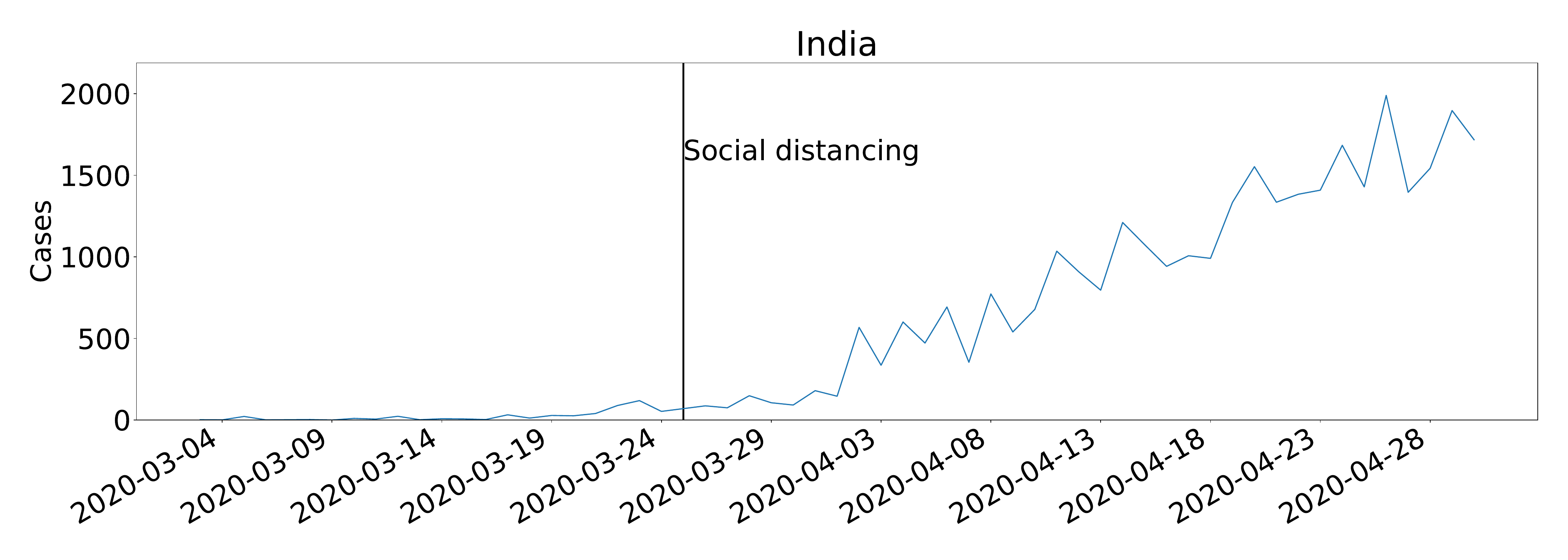} \\
			\vspace{-0.35cm}
	 	    \textbf{b} & \includegraphics[keepaspectratio, height=3.3cm, valign=T]
			{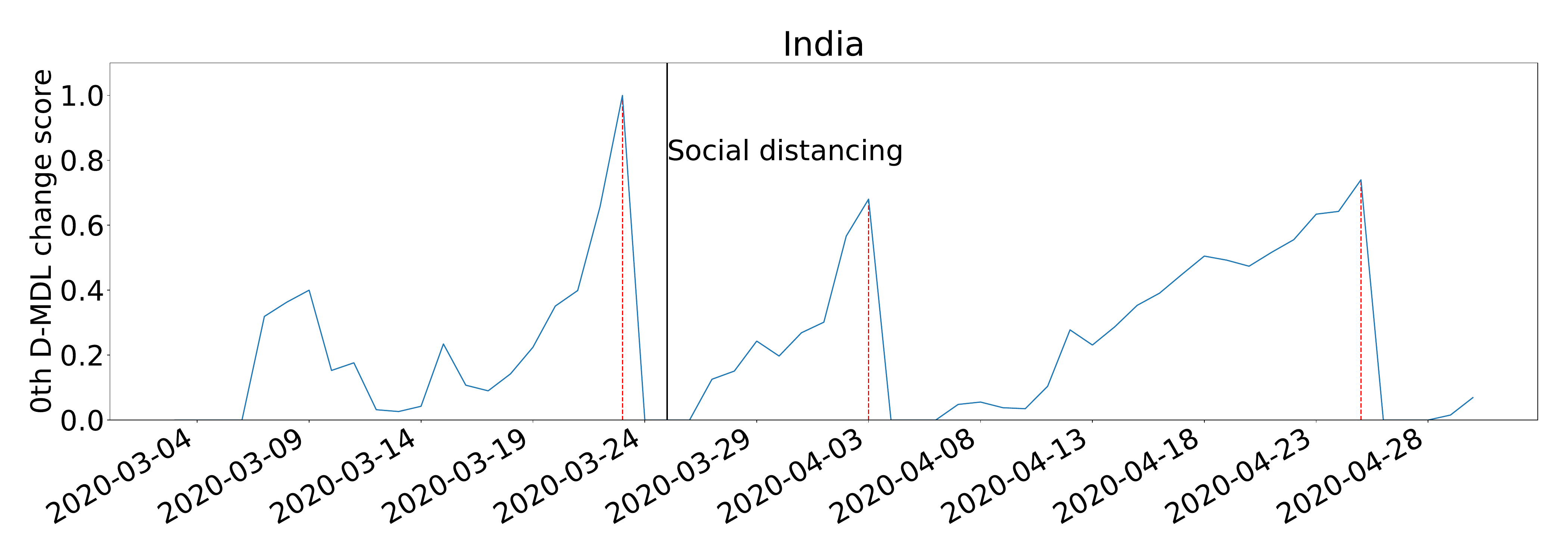}   \\
	        \vspace{-0.35cm}
			\textbf{c} & \includegraphics[keepaspectratio, height=3.3cm, valign=T]
			{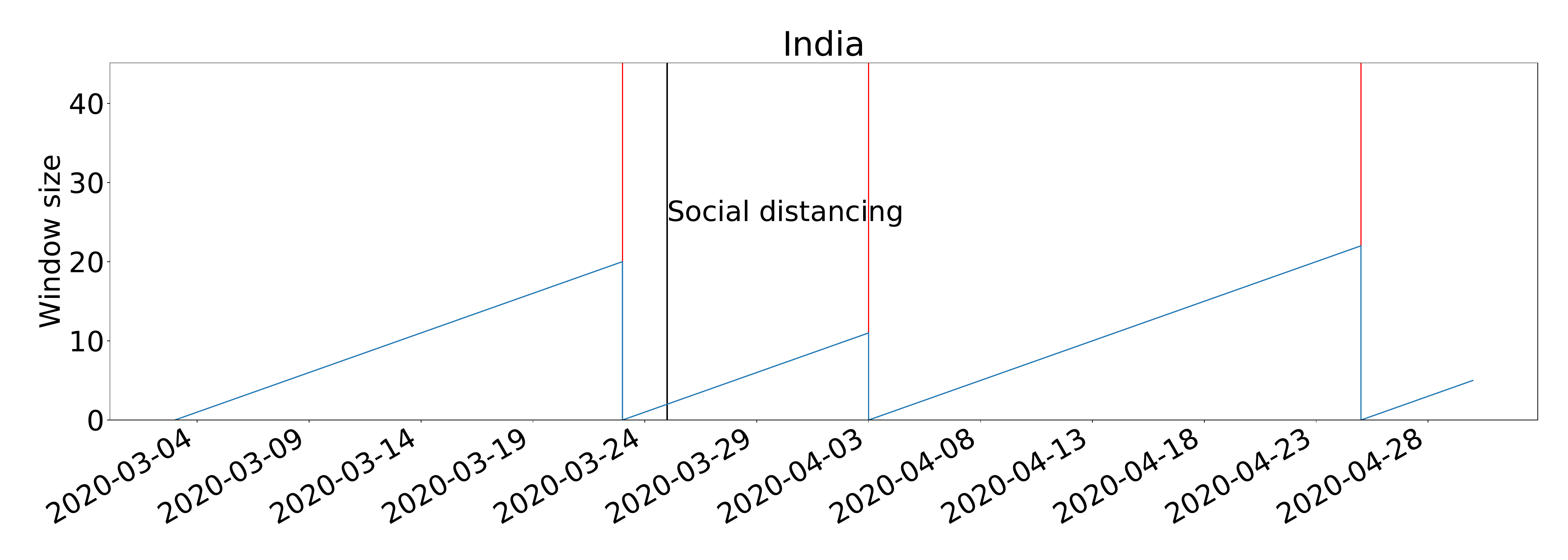} \\
		    \vspace{-0.35cm}
			\textbf{d} & \includegraphics[keepaspectratio, height=3.3cm, valign=T]
			{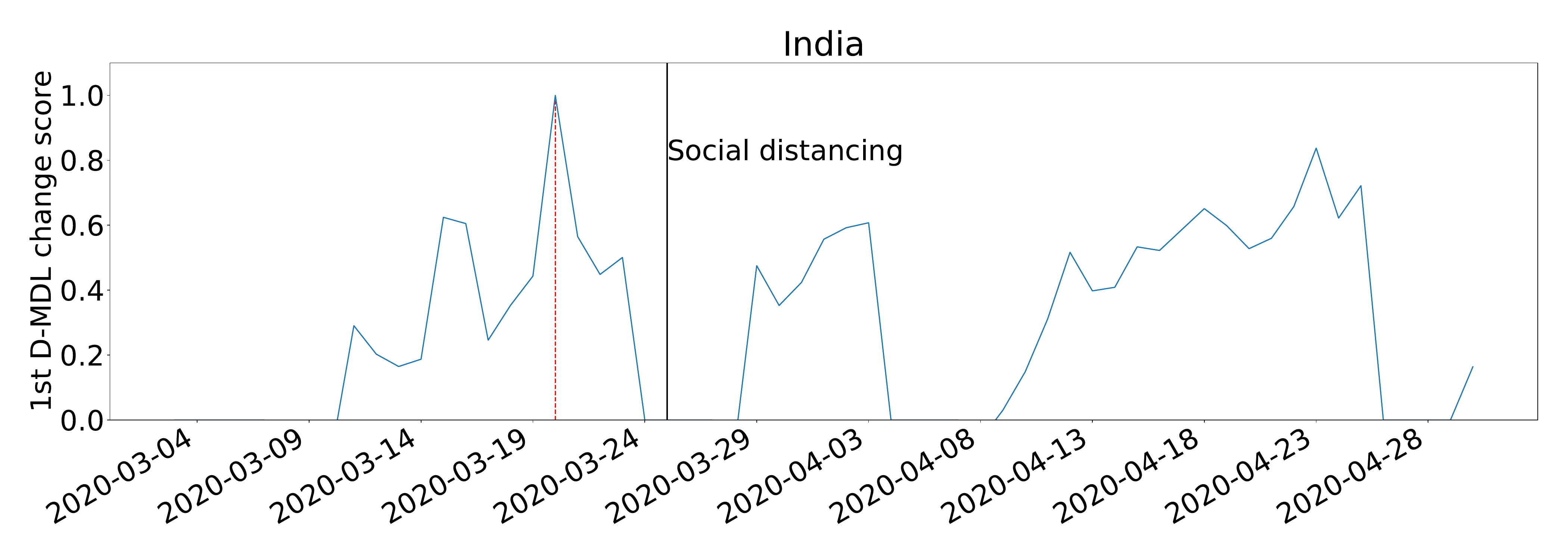} \\
		    \vspace{-0.35cm}
			\textbf{e} & \includegraphics[keepaspectratio, height=3.3cm, valign=T]
			{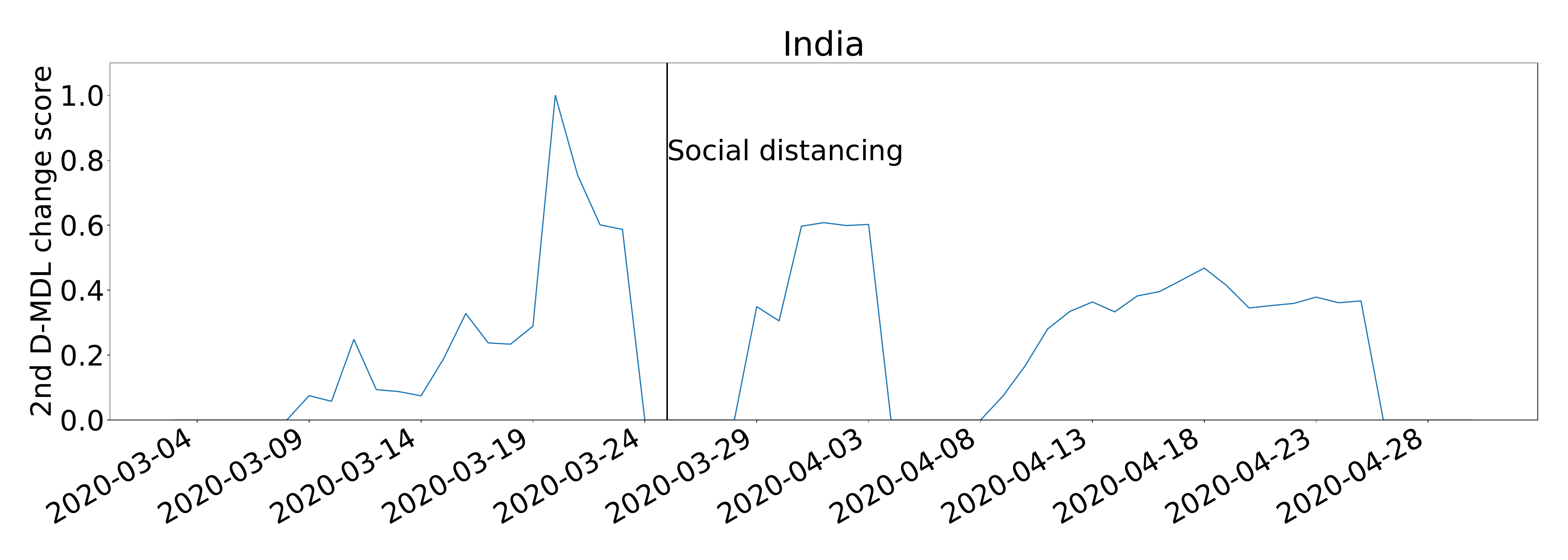} \\
		\end{tabular}
			\caption{\textbf{The results for India with Gaussian modeling.} The date on which the social distancing was implemented is marked by a solid line in black. \textbf{a,} the number of daily new cases. \textbf{b,} the change scores produced by the 0th M-DML where the line in blue denotes values of scores and dashed lines in red mark alarms. \textbf{c,} the window sized for the sequential D-DML algorithm with adaptive window where lines in red mark the shrinkage of windows. \textbf{d,} the change scores produced by the 1st D-MDL. \textbf{e,} the change scores produced by the 2nd D-MDL.}
\end{figure}

\begin{figure}[H]  
\centering
\begin{tabular}{cc}
			\textbf{a} & \includegraphics[keepaspectratio, height=3.3cm, valign=T]
			{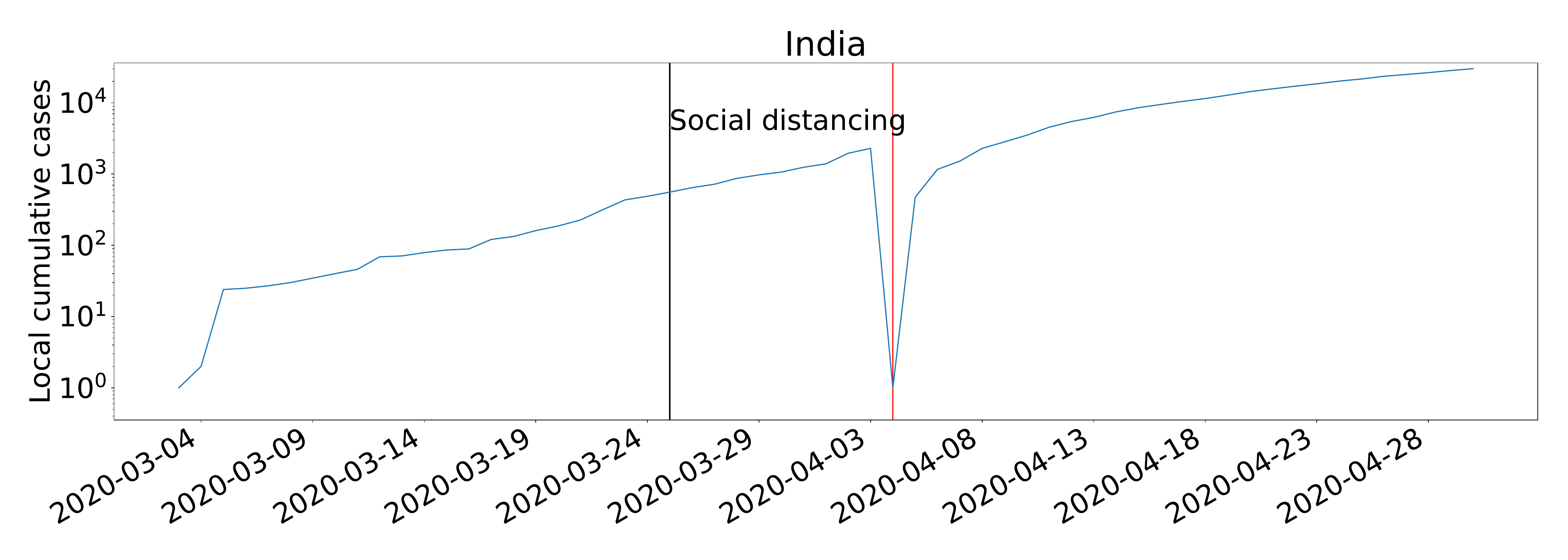} \\
	        \vspace{-0.35cm}
            \textbf{b} & \includegraphics[keepaspectratio, height=3.3cm, valign=T]
			{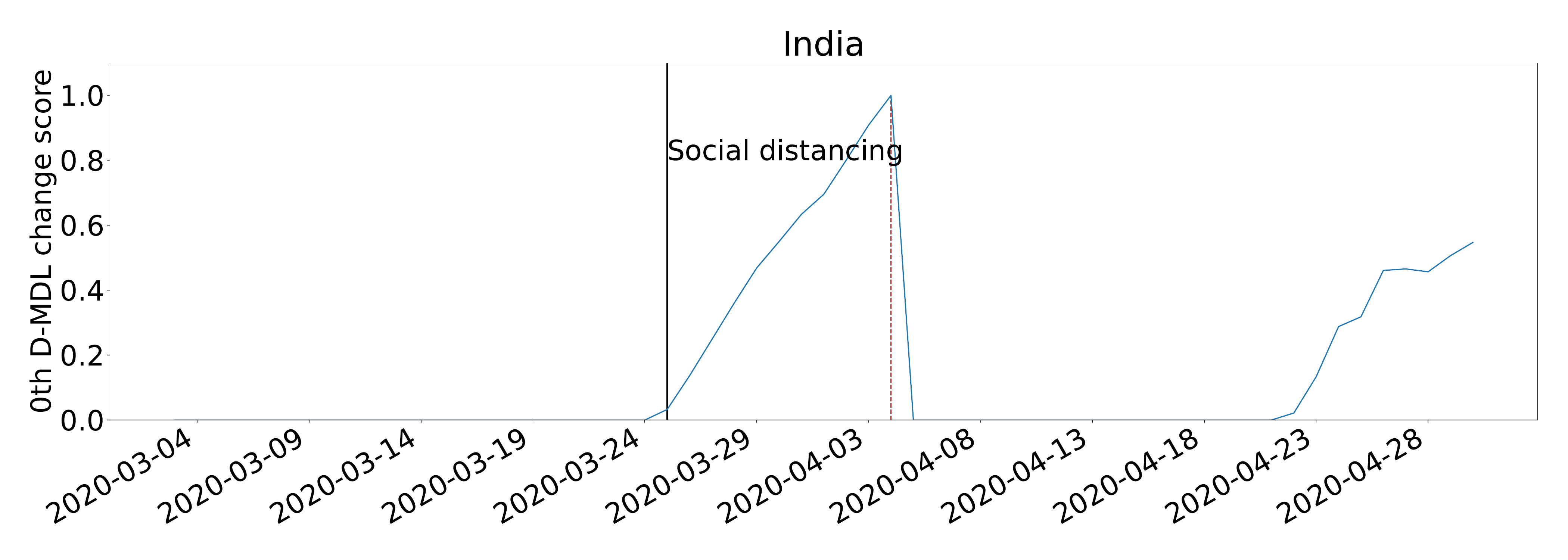}   \\
            \vspace{-0.35cm}
            \textbf{c} & \includegraphics[keepaspectratio, height=3.3cm, valign=T]
			{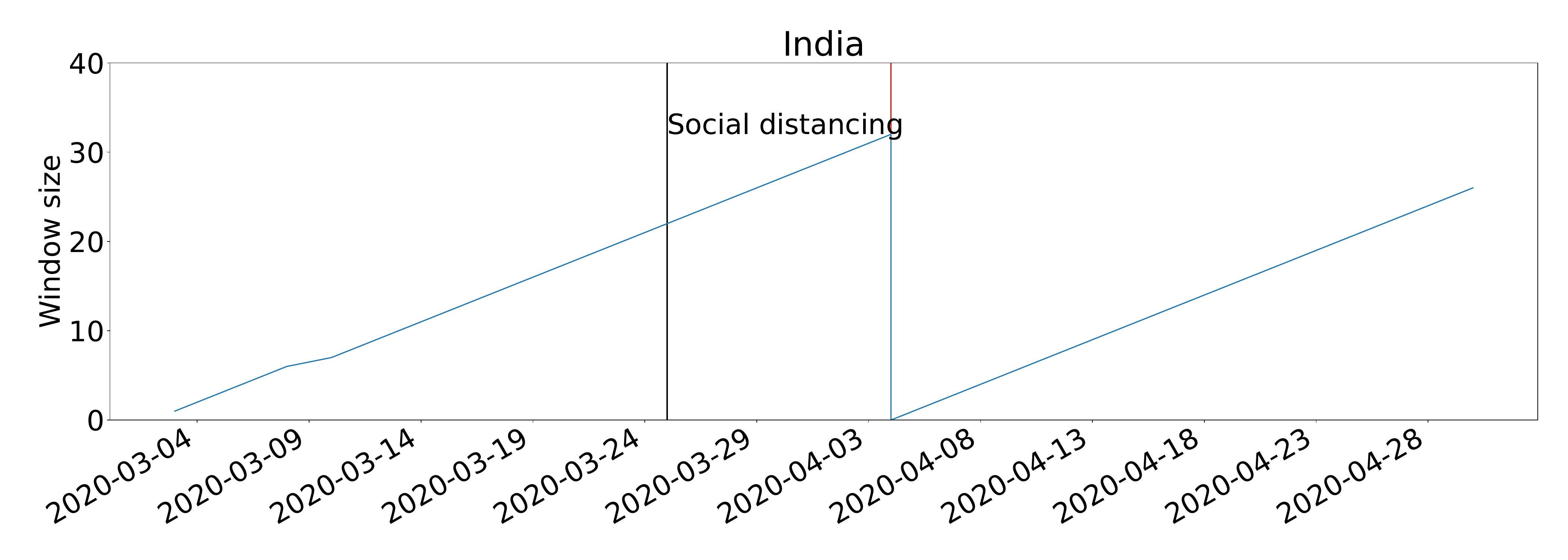} \\
			\vspace{-0.35cm}
			\textbf{d} & \includegraphics[keepaspectratio, height=3.3cm, valign=T]
			{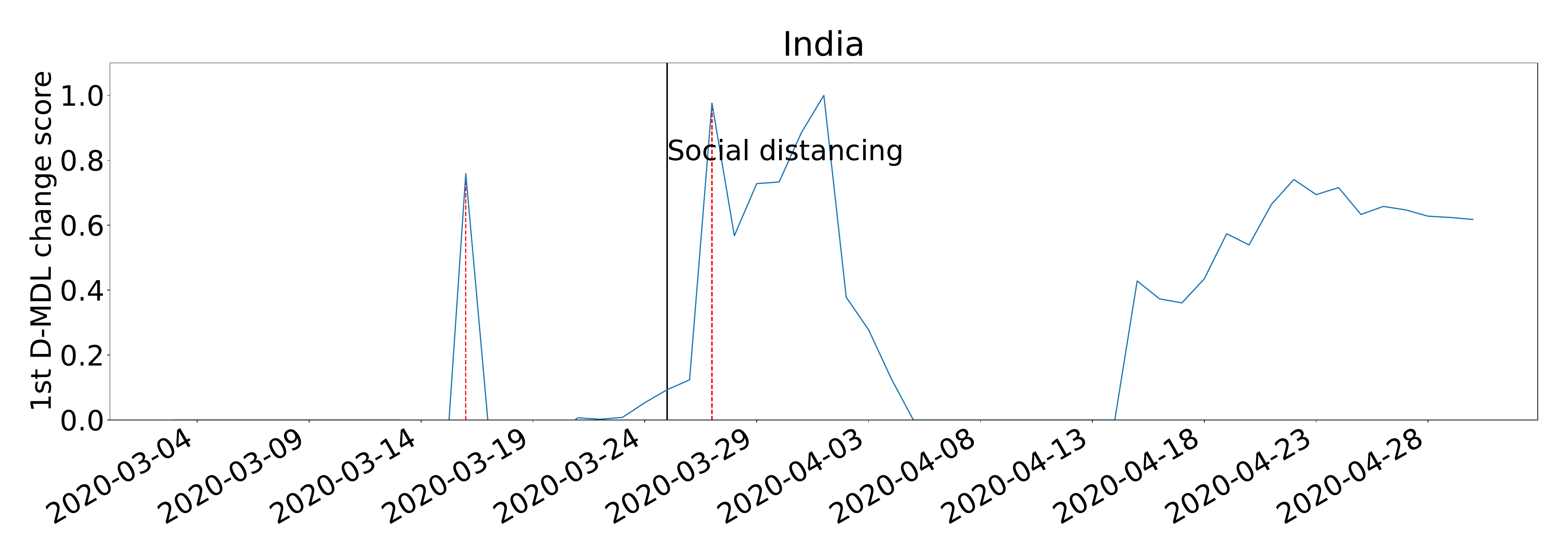} \\
			\vspace{-0.35cm}
			\textbf{e} & \includegraphics[keepaspectratio, height=3.3cm, valign=T]
			{./images_exp/South_Korea_2_score.pdf} \\
		\end{tabular}
			\caption{\textbf{The results for India with exponential modeling.} The date on which the social distancing was implemented is marked by a solid line in black. \textbf{a,} the number of cumulative cases. \textbf{b,} the change scores produced by the 0th M-DML where the line in blue denotes values of scores and dashed lines in red mark alarms. \textbf{c,} the window sized for the sequential D-DML algorithm with adaptive window where lines in red mark the shrinkage of windows. \textbf{d,} the change scores produced by the 1st D-MDL. \textbf{e,} the change scores produced by the 2nd D-MDL.}
\end{figure}

\begin{figure}[H] 
\centering
\begin{tabular}{cc}
		 	\textbf{a} & \includegraphics[keepaspectratio, height=3.3cm, valign=T]
			{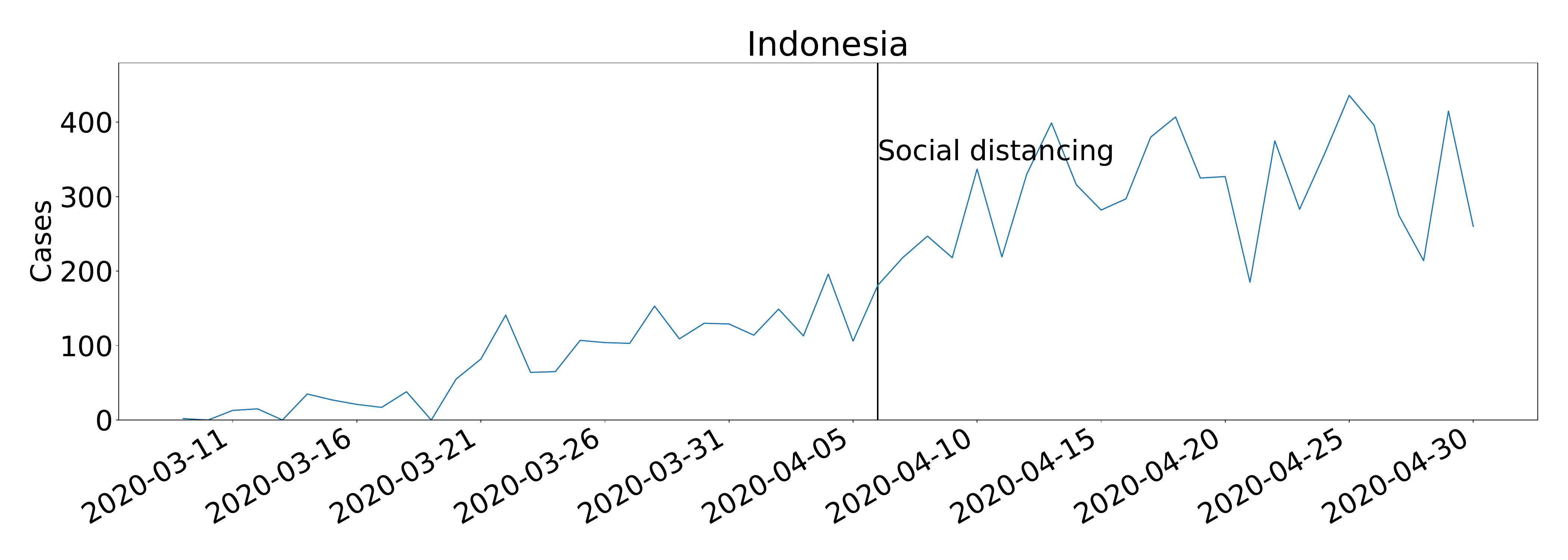} \\
			\vspace{-0.35cm}
	 	    \textbf{b} & \includegraphics[keepaspectratio, height=3.3cm, valign=T]
			{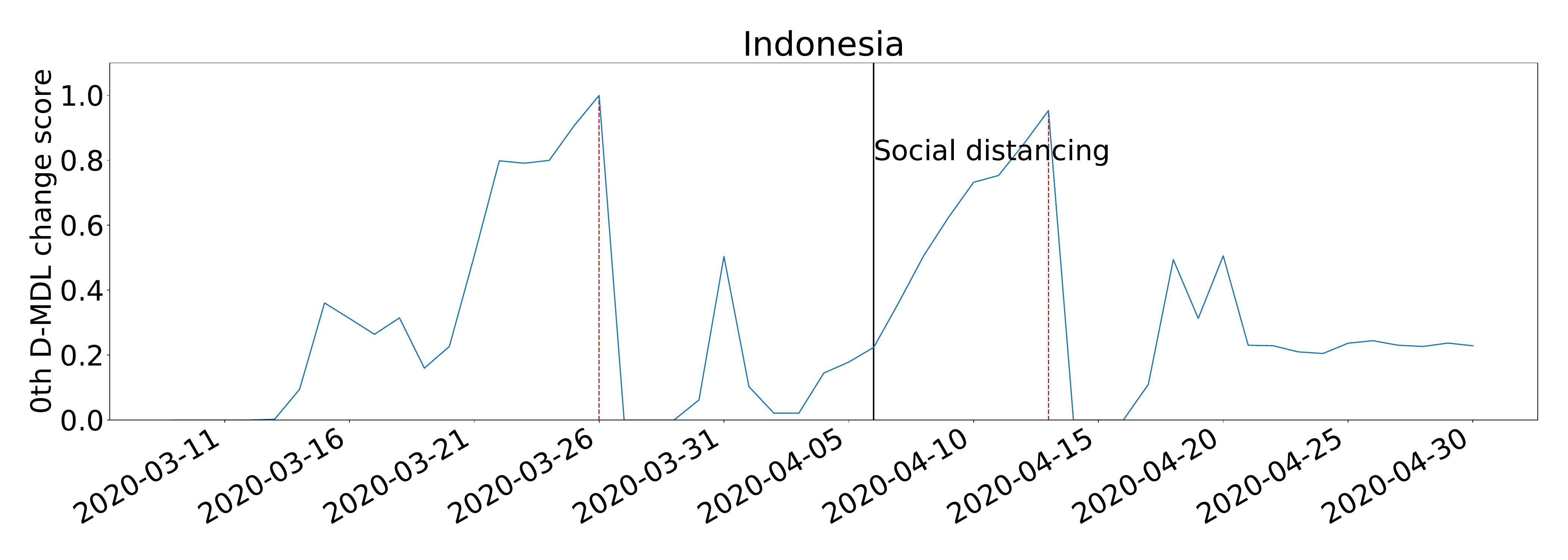}   \\
	        \vspace{-0.35cm}
			\textbf{c} & \includegraphics[keepaspectratio, height=3.3cm, valign=T]
			{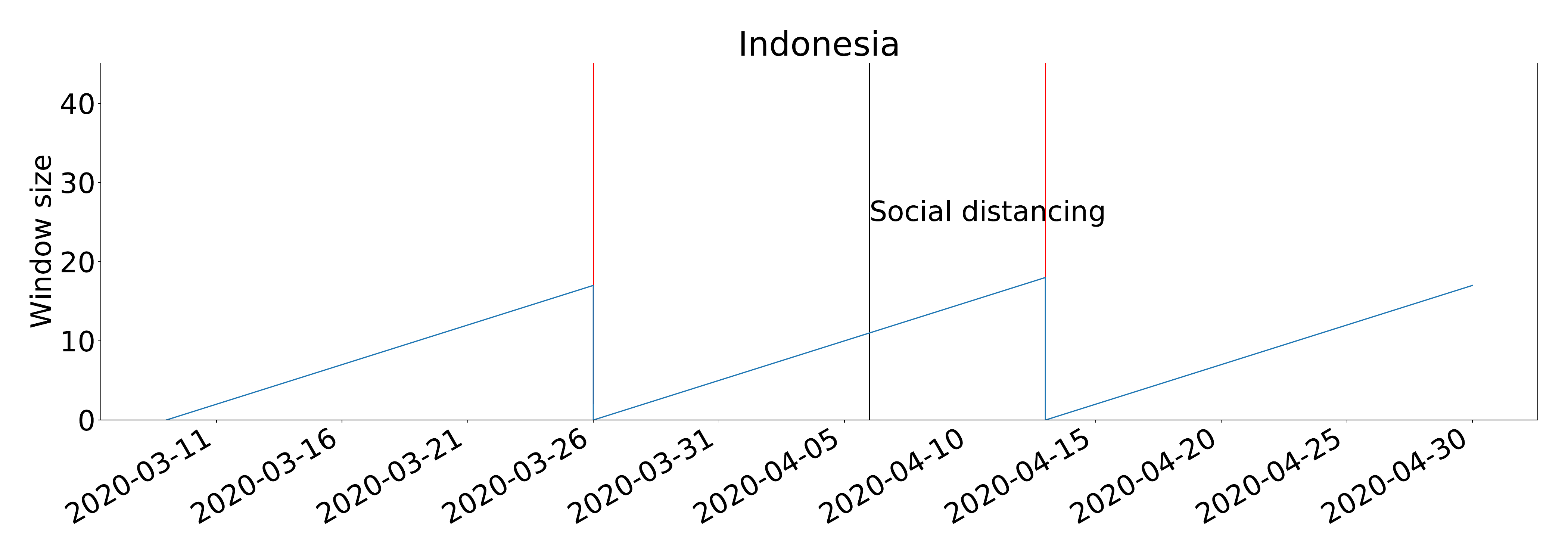} \\
		    \vspace{-0.35cm}
			\textbf{d} & \includegraphics[keepaspectratio, height=3.3cm, valign=T]
			{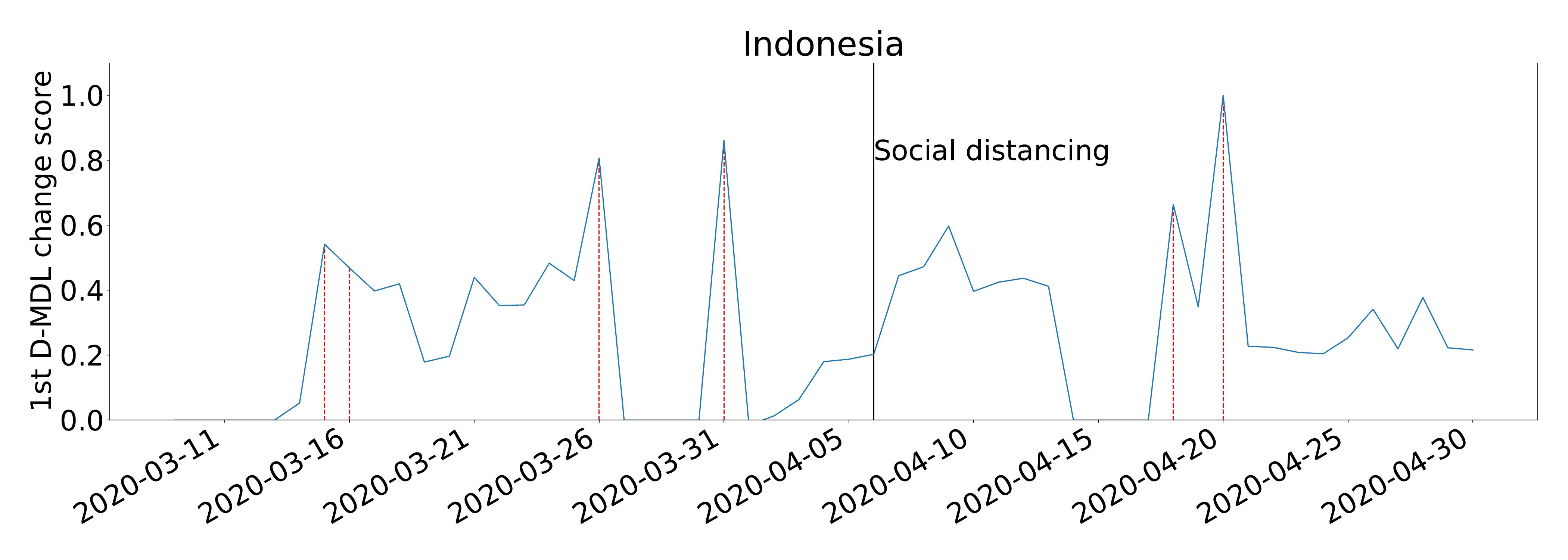} \\
		    \vspace{-0.35cm}
			\textbf{e} & \includegraphics[keepaspectratio, height=3.3cm, valign=T]
			{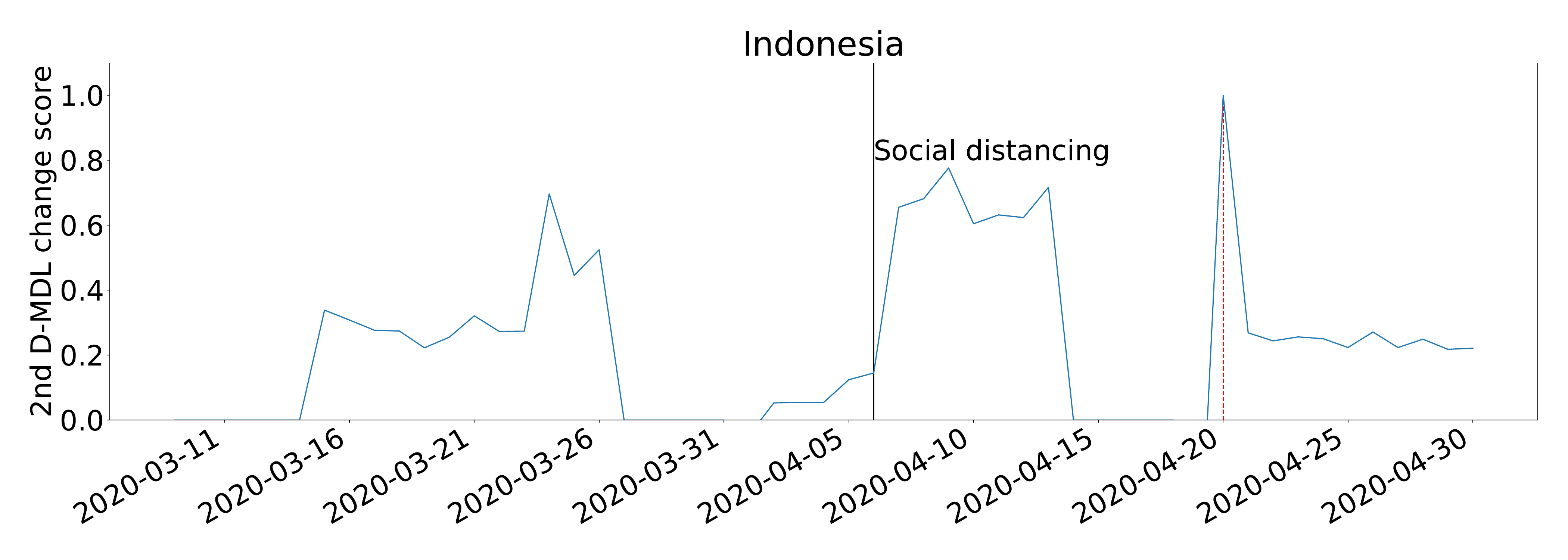} \\
		\end{tabular}
			\caption{\textbf{The results for Indonesia with Gaussian modeling.} The date on which the social distancing was implemented is marked by a solid line in black. \textbf{a,} the number of daily new cases. \textbf{b,} the change scores produced by the 0th M-DML where the line in blue denotes values of scores and dashed lines in red mark alarms. \textbf{c,} the window sized for the sequential D-DML algorithm with adaptive window where lines in red mark the shrinkage of windows. \textbf{d,} the change scores produced by the 1st D-MDL. \textbf{e,} the change scores produced by the 2nd D-MDL.}
\end{figure}

\begin{figure}[H]  
\centering
\begin{tabular}{cc}
			\textbf{a} & \includegraphics[keepaspectratio, height=3.3cm, valign=T]
			{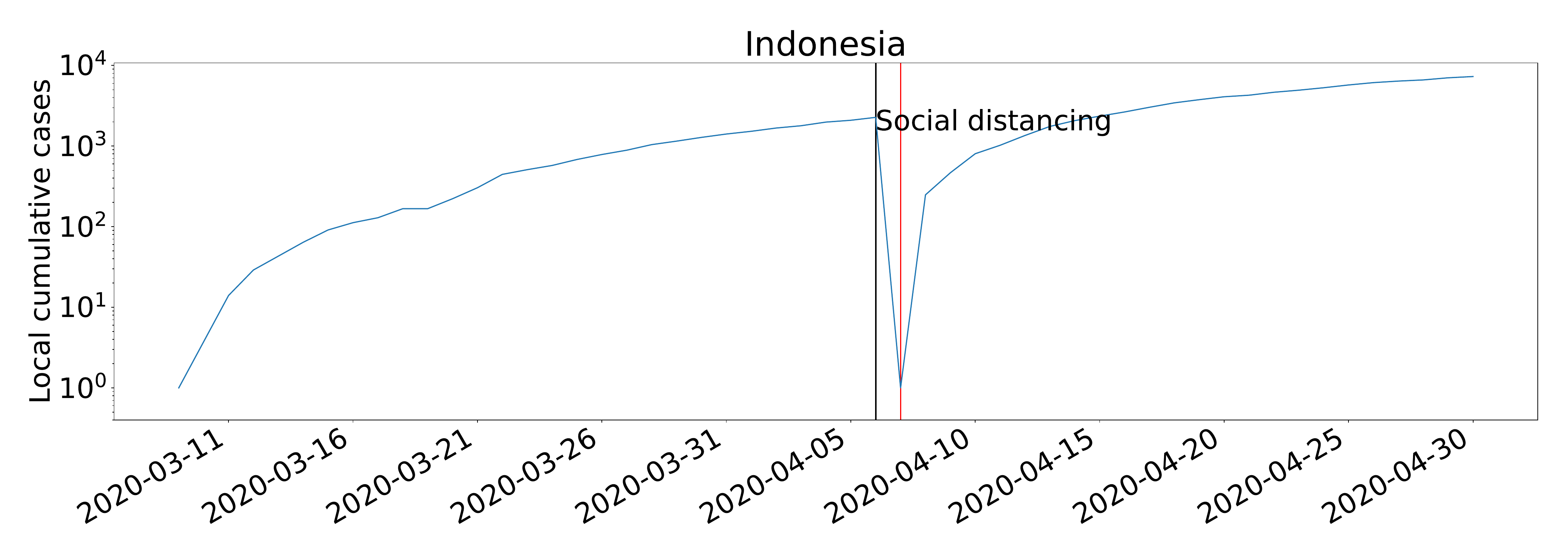} \\
	        \vspace{-0.35cm}
            \textbf{b} & \includegraphics[keepaspectratio, height=3.3cm, valign=T]
			{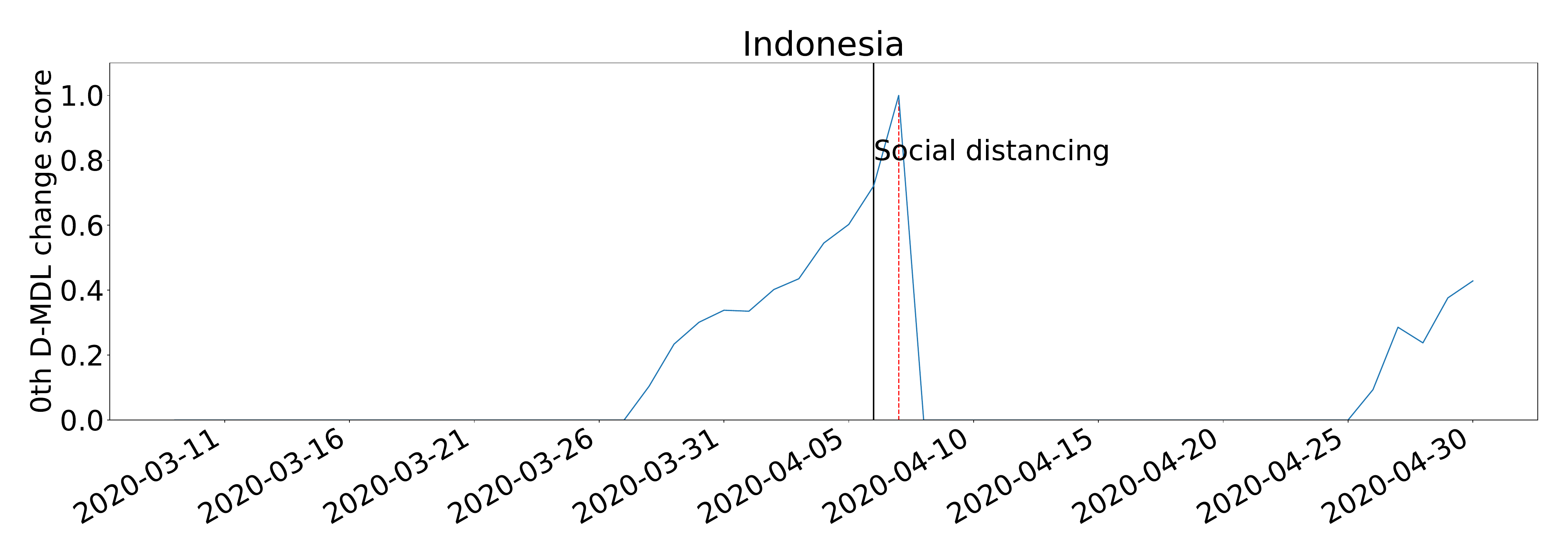}   \\
            \vspace{-0.35cm}
            \textbf{c} & \includegraphics[keepaspectratio, height=3.3cm, valign=T]
			{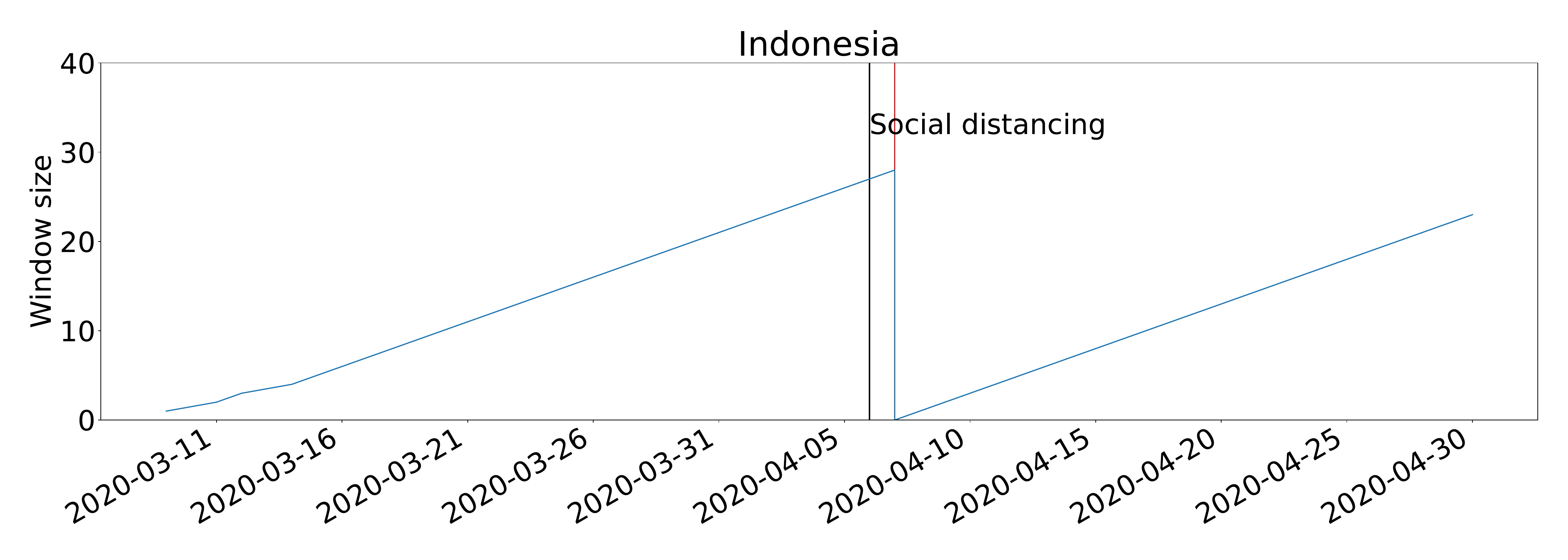} \\
			\vspace{-0.35cm}
			\textbf{d} & \includegraphics[keepaspectratio, height=3.3cm, valign=T]
			{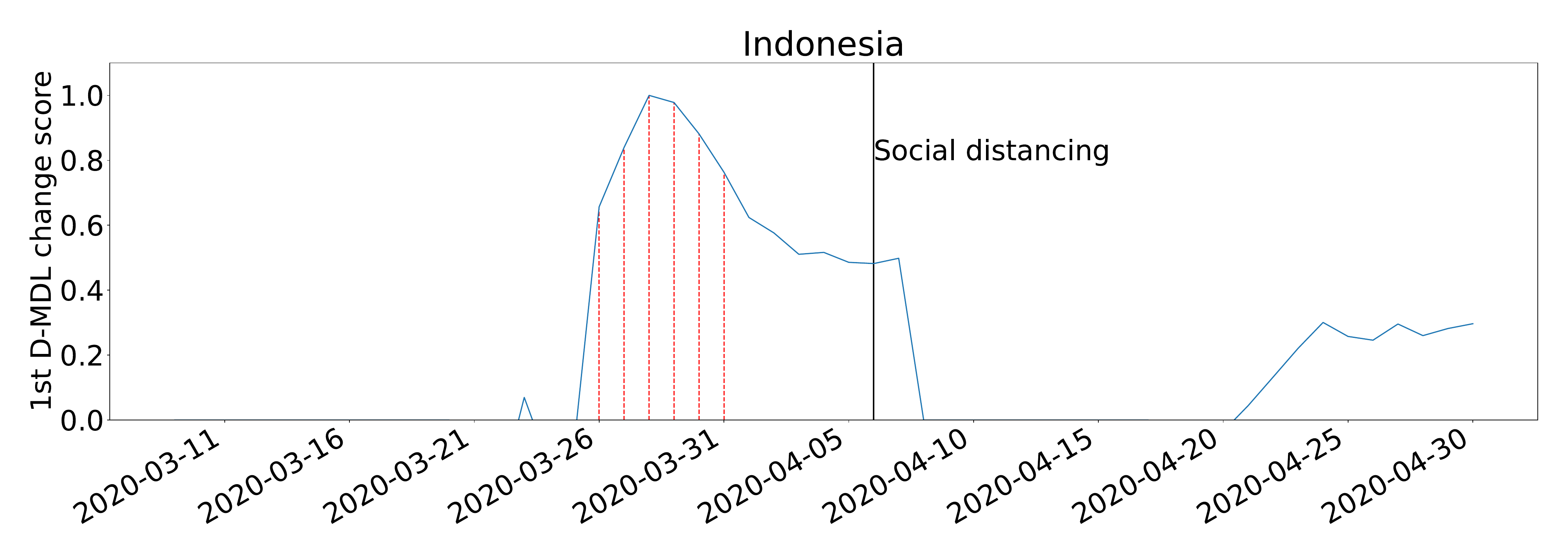} \\
			\vspace{-0.35cm}
			\textbf{e} & \includegraphics[keepaspectratio, height=3.3cm, valign=T]
			{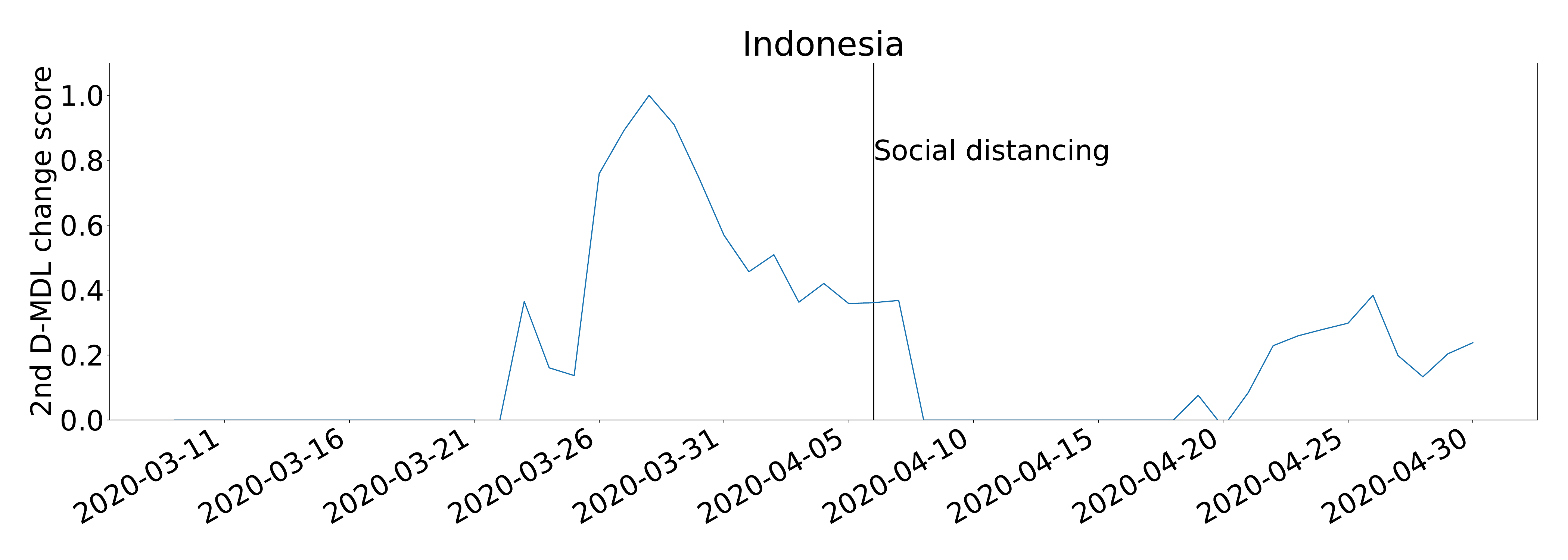} \\
		\end{tabular}
			\caption{\textbf{The results for Indonesia with exponential modeling.} The date on which the social distancing was implemented is marked by a solid line in black. \textbf{a,} the number of cumulative cases. \textbf{b,} the change scores produced by the 0th M-DML where the line in blue denotes values of scores and dashed lines in red mark alarms. \textbf{c,} the window sized for the sequential D-DML algorithm with adaptive window where lines in red mark the shrinkage of windows. \textbf{d,} the change scores produced by the 1st D-MDL. \textbf{e,} the change scores produced by the 2nd D-MDL.}
\end{figure}

\begin{figure}[H] 
\centering
\begin{tabular}{cc}
		 	\textbf{a} & \includegraphics[keepaspectratio, height=3.3cm, valign=T]
			{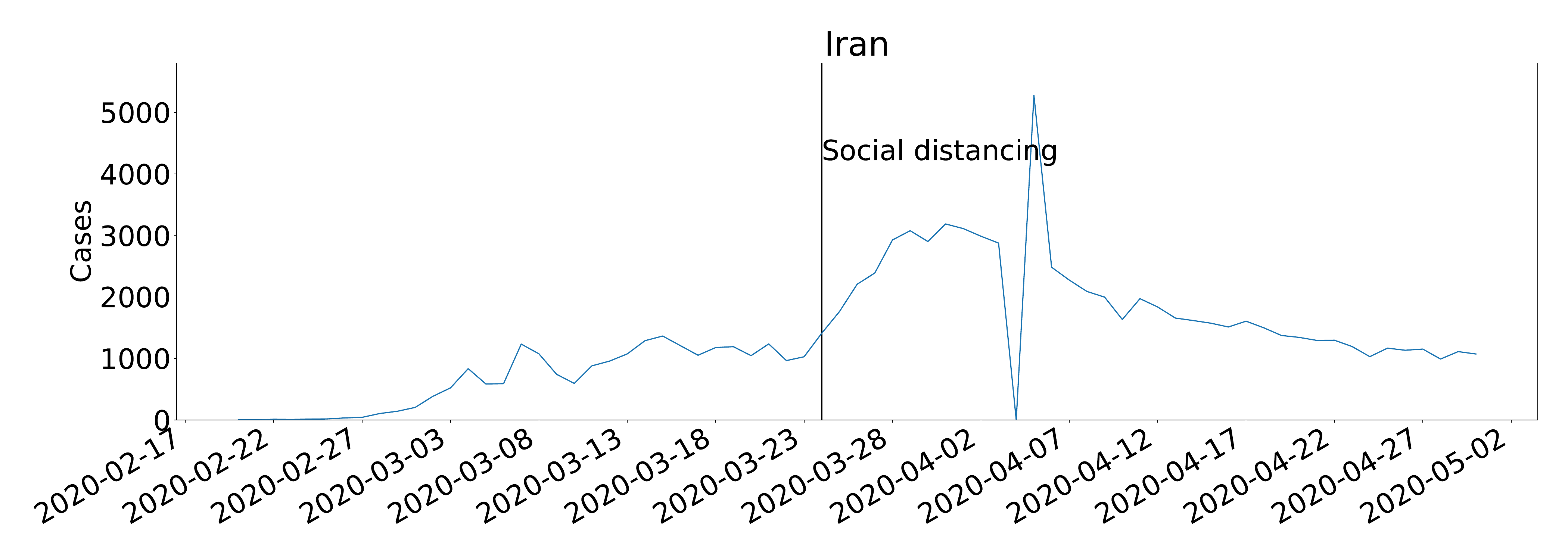} \\
			\vspace{-0.35cm}
	 	    \textbf{b} & \includegraphics[keepaspectratio, height=3.3cm, valign=T]
			{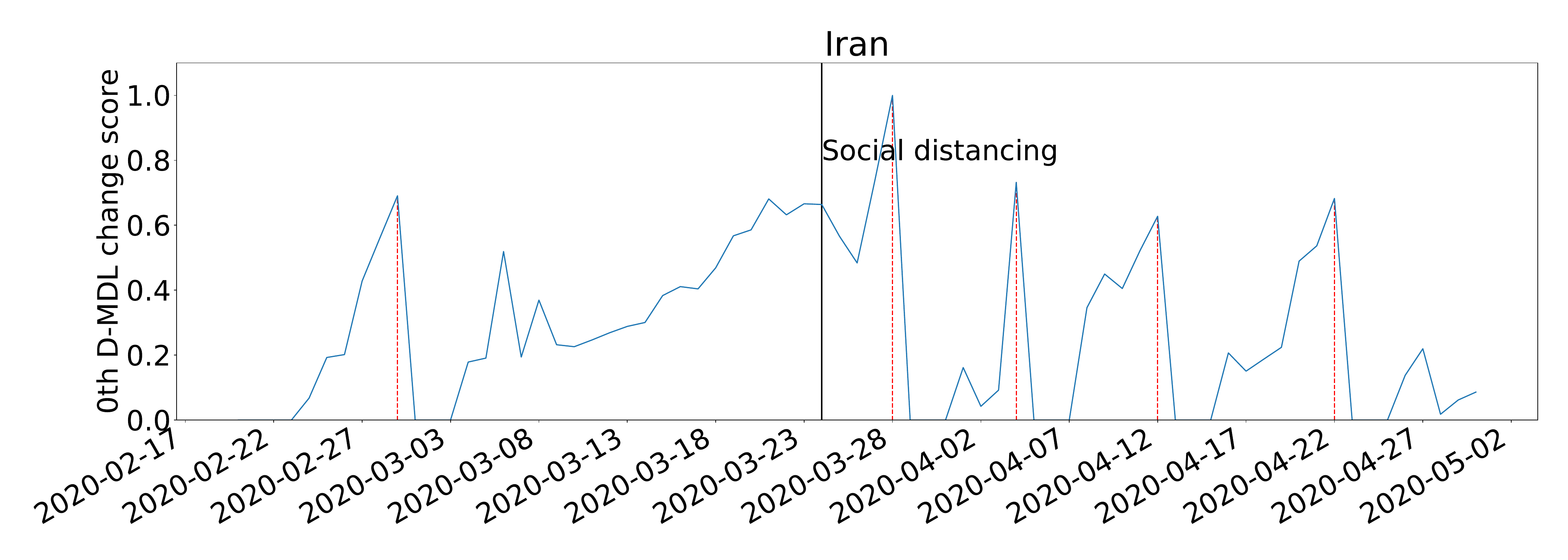}   \\
	        \vspace{-0.35cm}
			\textbf{c} & \includegraphics[keepaspectratio, height=3.3cm, valign=T]
			{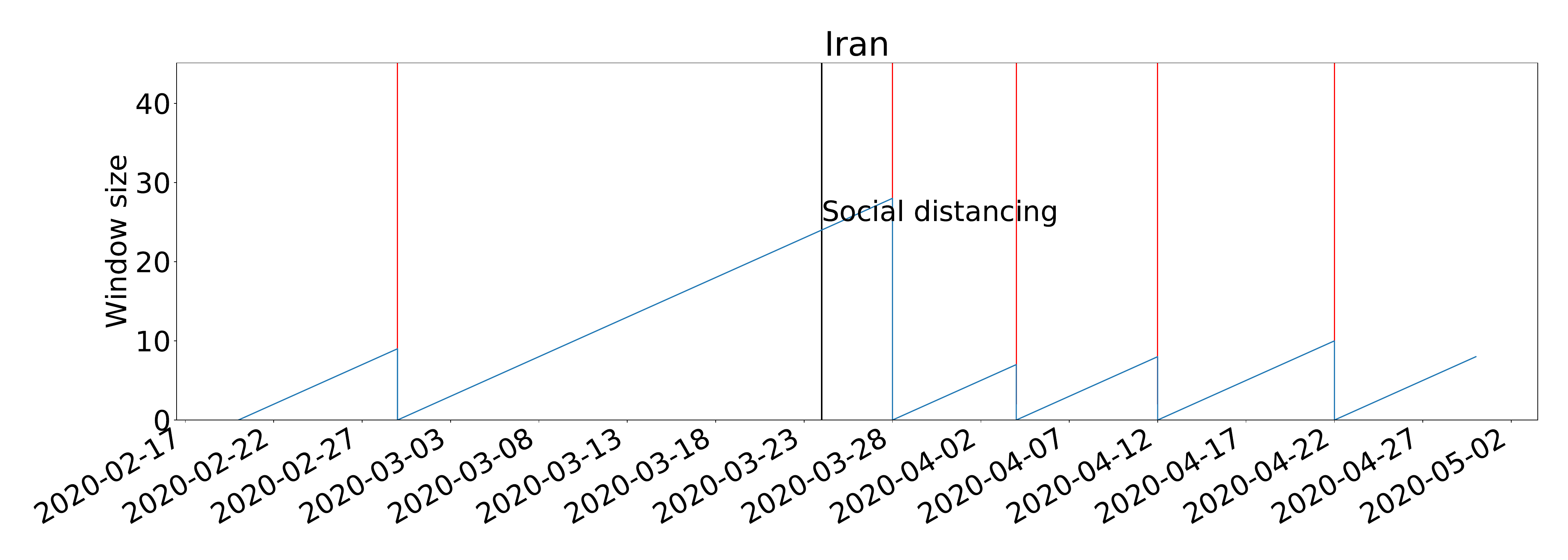} \\
		    \vspace{-0.35cm}
			\textbf{d} & \includegraphics[keepaspectratio, height=3.3cm, valign=T]
			{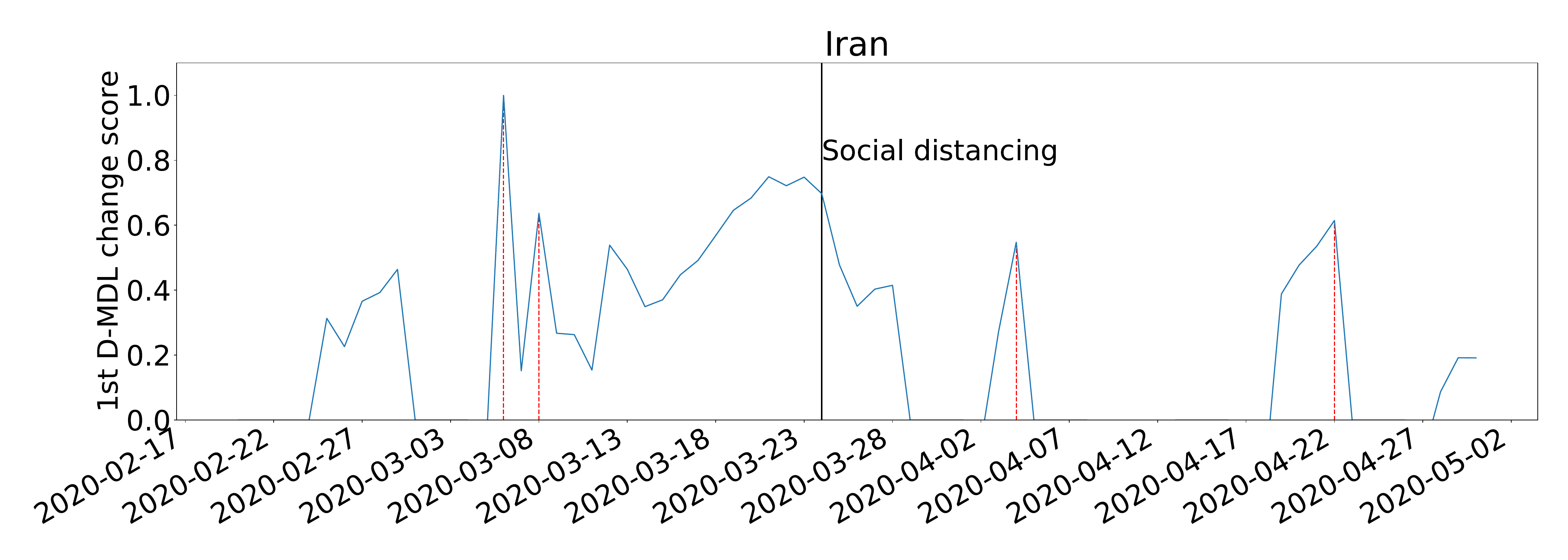} \\
		    \vspace{-0.35cm}
			\textbf{e} & \includegraphics[keepaspectratio, height=3.3cm, valign=T]
			{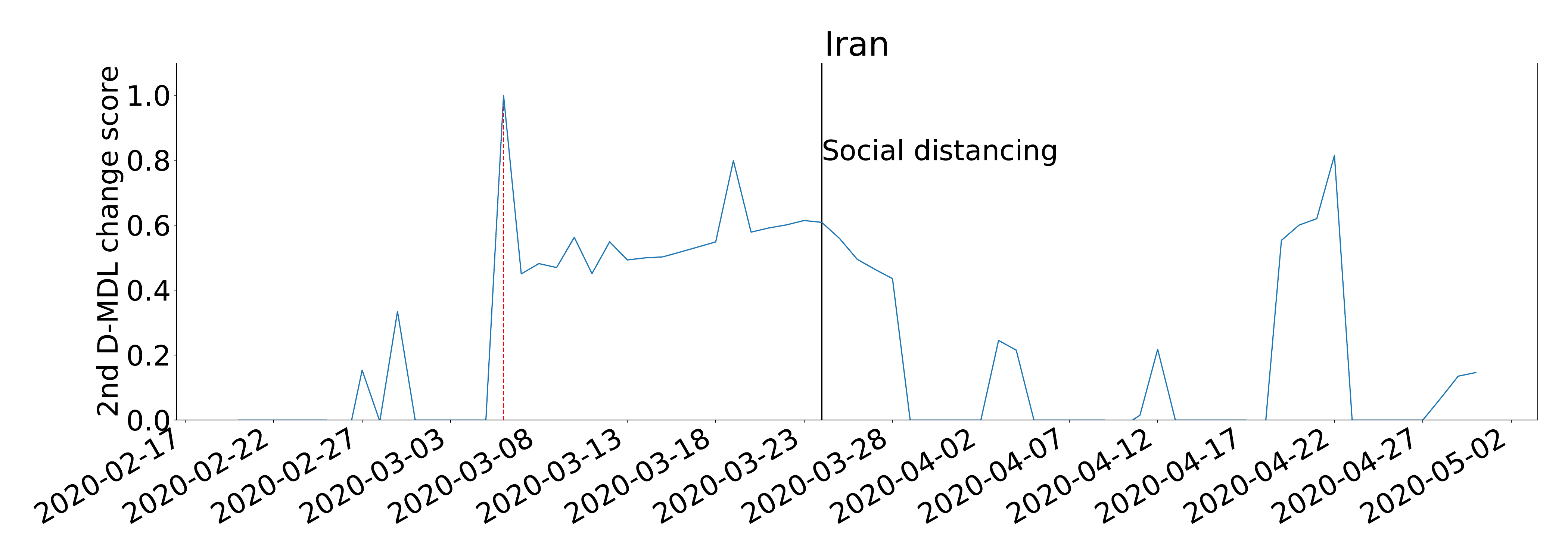} \\
		\end{tabular}
			\caption{\textbf{The results for Iran with Gaussian modeling.} The date on which the social distancing was implemented is marked by a solid line in black. \textbf{a,} the number of daily new cases. \textbf{b,} the change scores produced by the 0th M-DML where the line in blue denotes values of scores and dashed lines in red mark alarms. \textbf{c,} the window sized for the sequential D-DML algorithm with adaptive window where lines in red mark the shrinkage of windows. \textbf{d,} the change scores produced by the 1st D-MDL. \textbf{e,} the change scores produced by the 2nd D-MDL.}
\end{figure}

\begin{figure}[H]  
\centering
\begin{tabular}{cc}
			\textbf{a} & \includegraphics[keepaspectratio, height=3.3cm, valign=T]
			{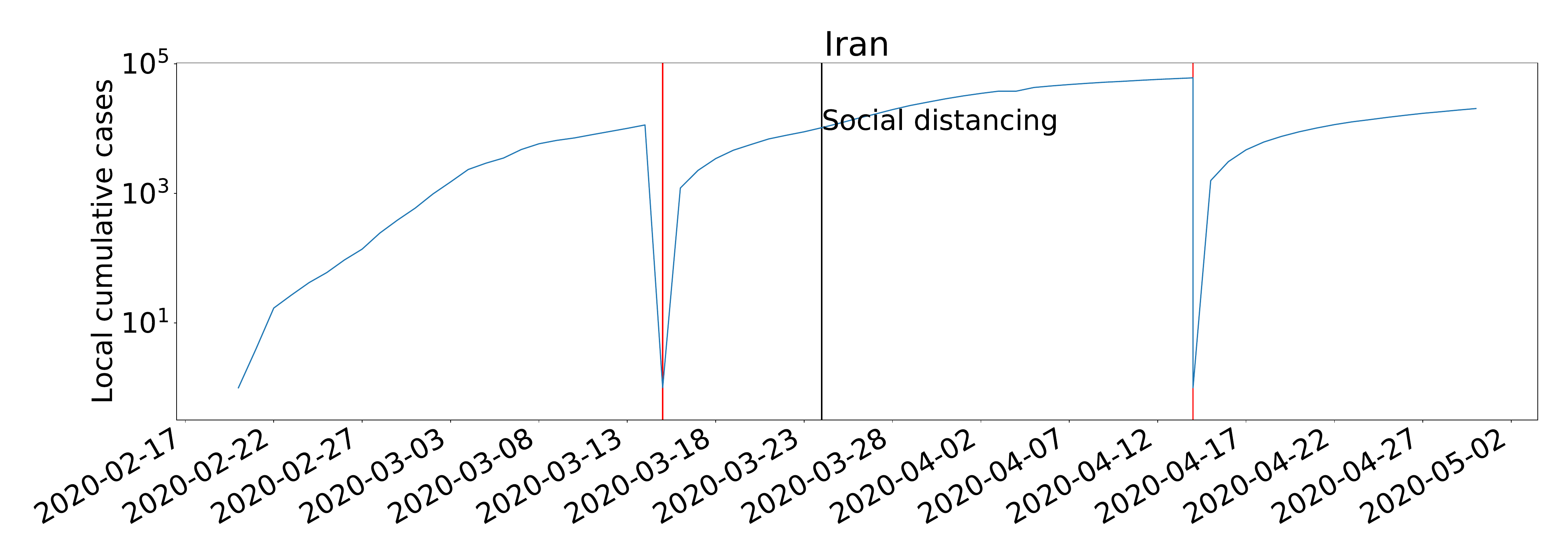} \\
	        \vspace{-0.35cm}
            \textbf{b} & \includegraphics[keepaspectratio, height=3.3cm, valign=T]
			{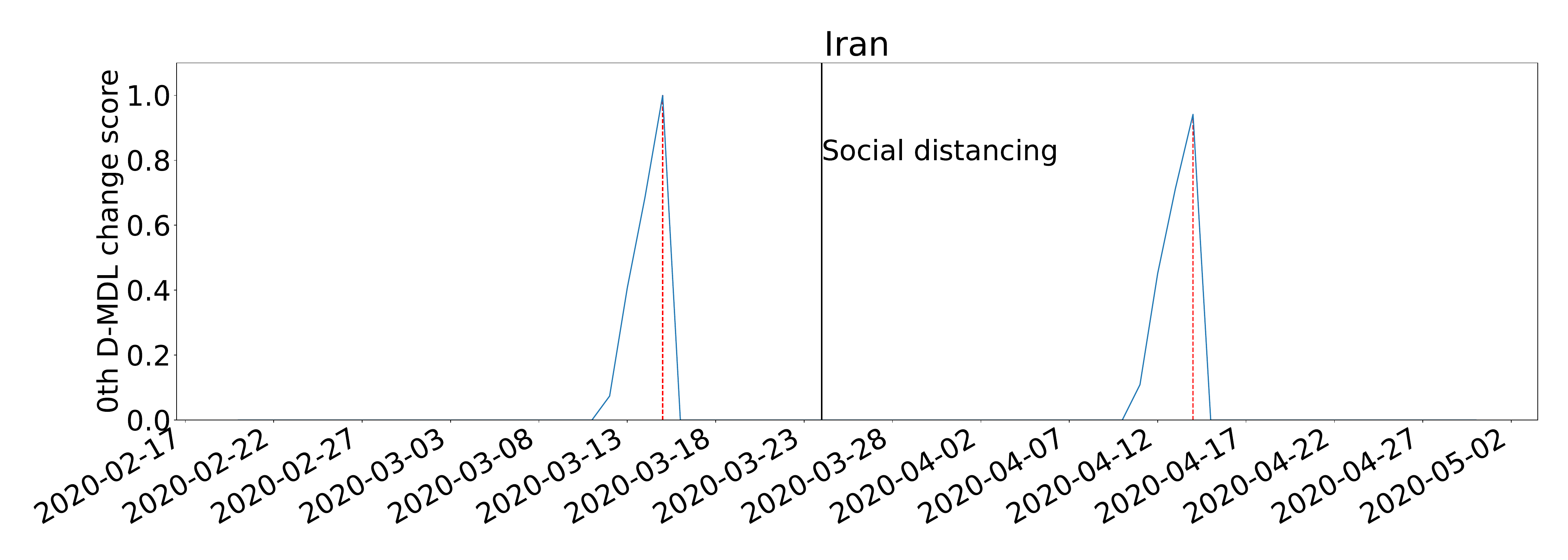}   \\
            \vspace{-0.35cm}
            \textbf{c} & \includegraphics[keepaspectratio, height=3.3cm, valign=T]
			{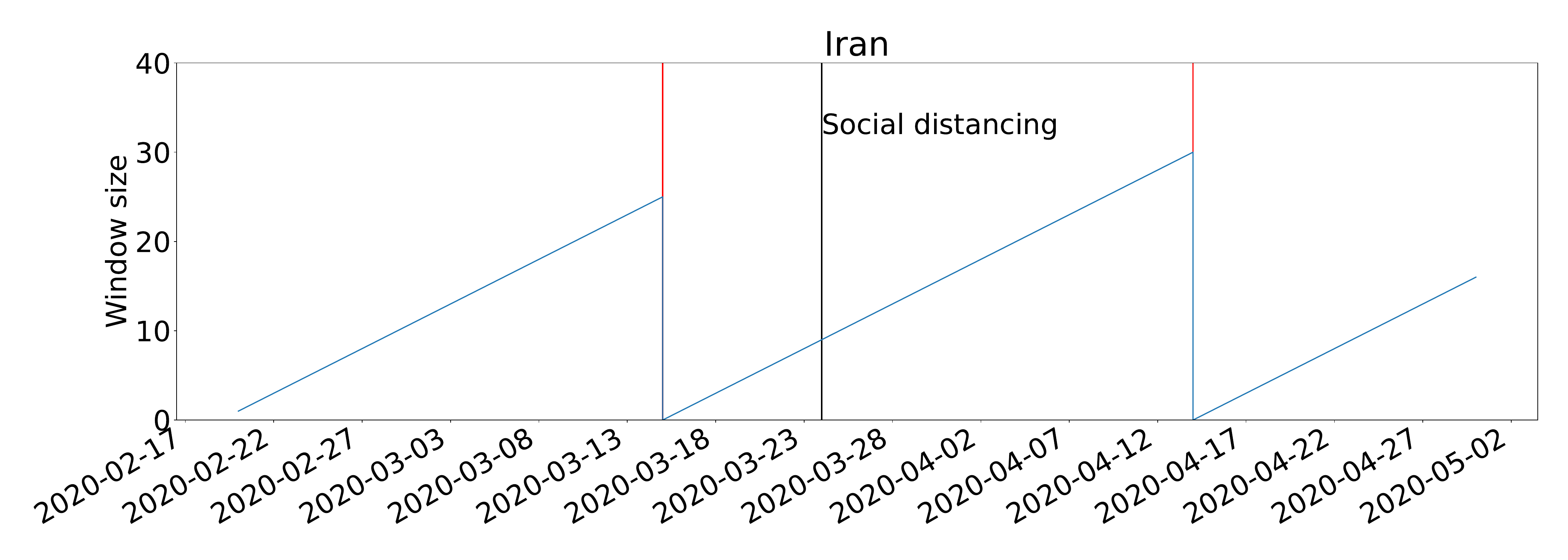} \\
			\vspace{-0.35cm}
			\textbf{d} & \includegraphics[keepaspectratio, height=3.3cm, valign=T]
			{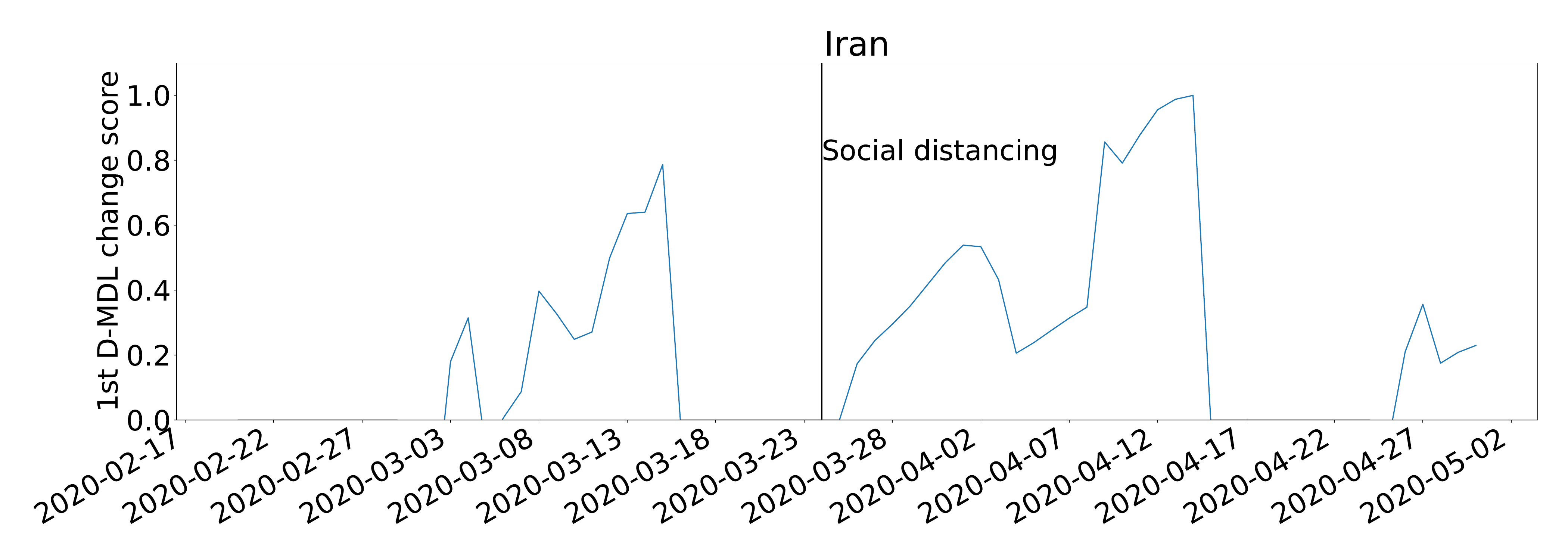} \\
			\vspace{-0.35cm}
			\textbf{e} & \includegraphics[keepaspectratio, height=3.3cm, valign=T]
			{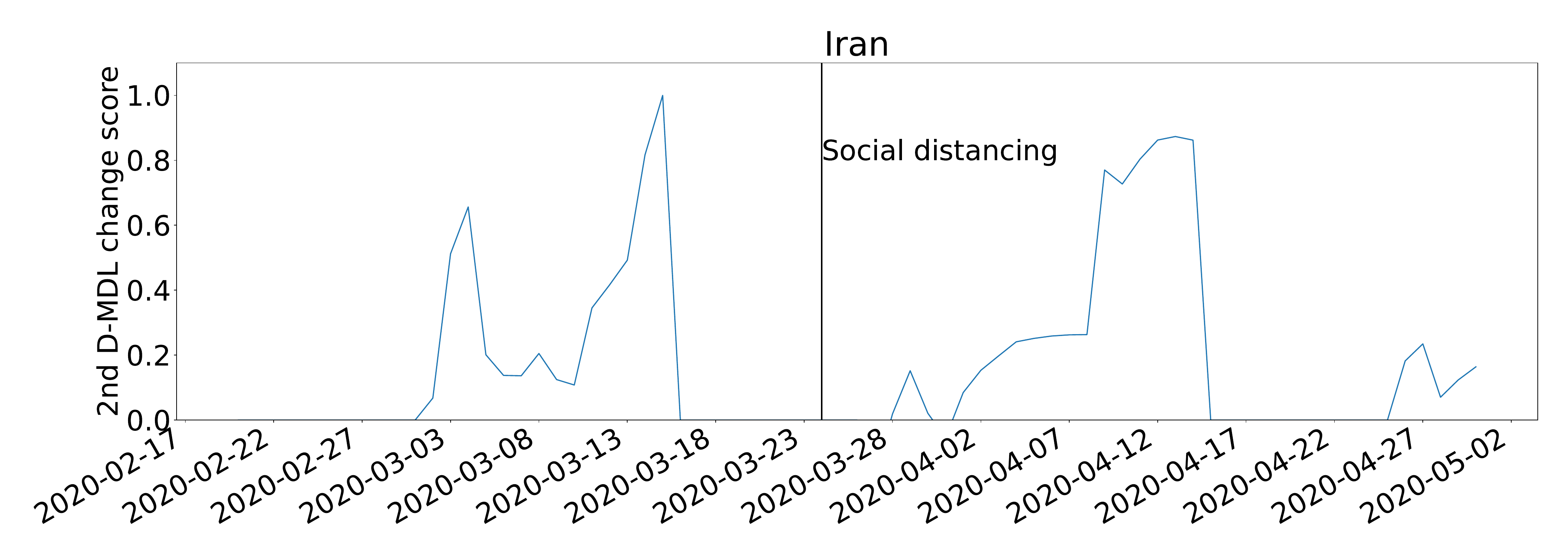} \\
		\end{tabular}
			\caption{\textbf{The results for Iran with exponential modeling.} The date on which the social distancing was implemented is marked by a solid line in black. \textbf{a,} the number of cumulative cases. \textbf{b,} the change scores produced by the 0th M-DML where the line in blue denotes values of scores and dashed lines in red mark alarms. \textbf{c,} the window sized for the sequential D-DML algorithm with adaptive window where lines in red mark the shrinkage of windows. \textbf{d,} the change scores produced by the 1st D-MDL. \textbf{e,} the change scores produced by the 2nd D-MDL.}
\end{figure}

\begin{figure}[H] 
\centering
\begin{tabular}{cc}
		 	\textbf{a} & \includegraphics[keepaspectratio, height=3.3cm, valign=T]
			{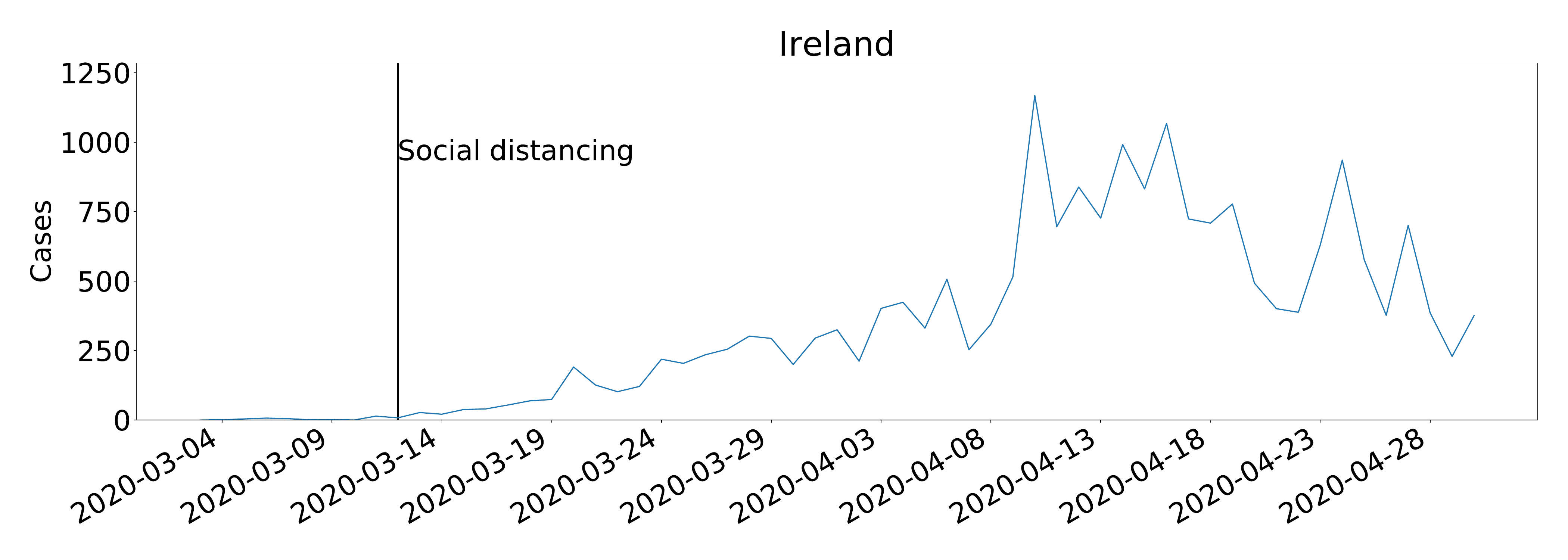} \\
			\vspace{-0.35cm}
	 	    \textbf{b} & \includegraphics[keepaspectratio, height=3.3cm, valign=T]
			{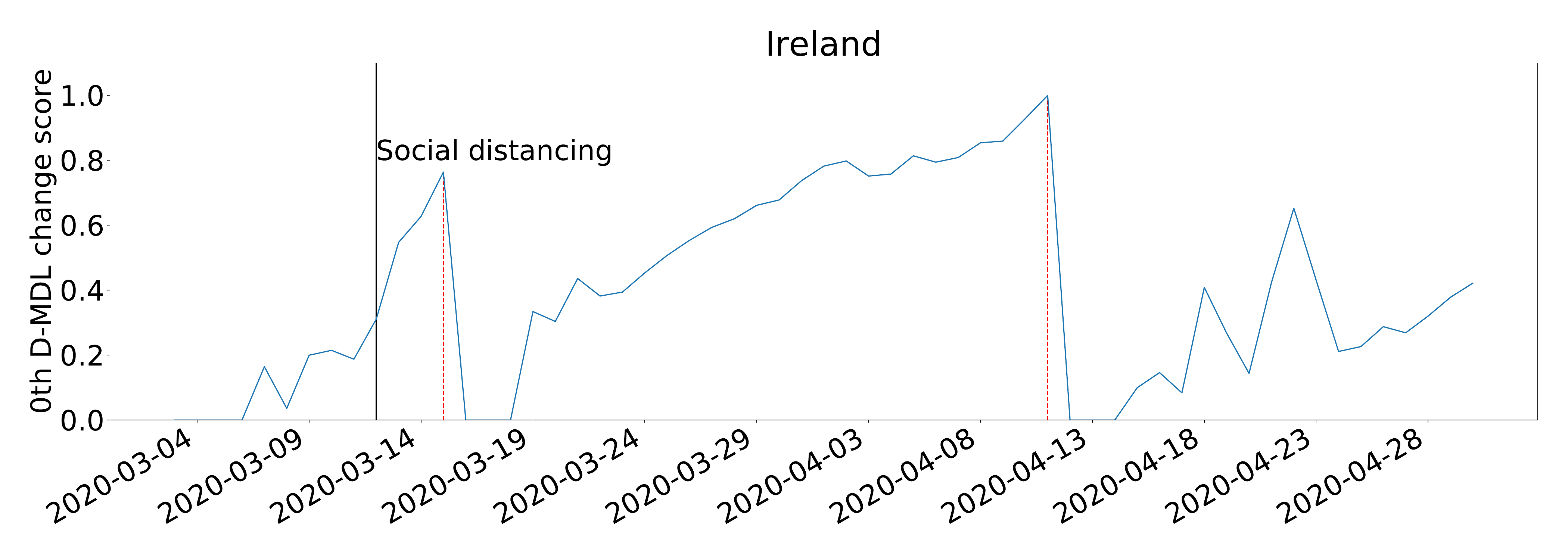}   \\
	        \vspace{-0.35cm}
			\textbf{c} & \includegraphics[keepaspectratio, height=3.3cm, valign=T]
			{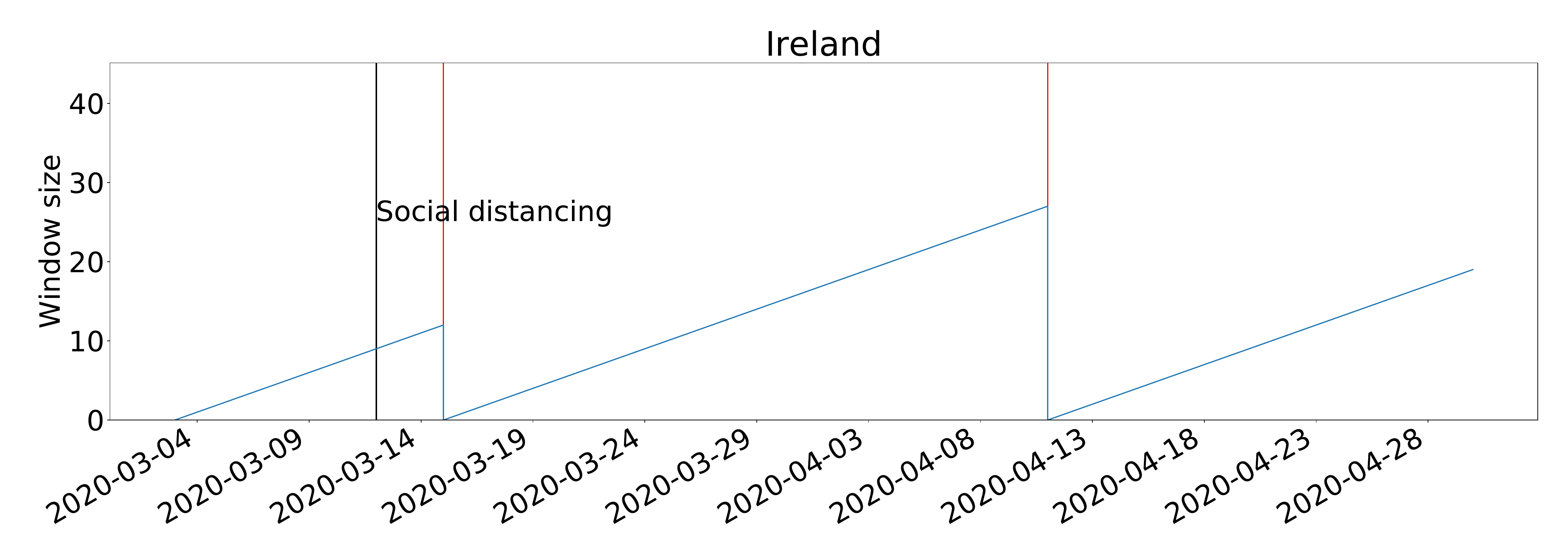} \\
		    \vspace{-0.35cm}
			\textbf{d} & \includegraphics[keepaspectratio, height=3.3cm, valign=T]
			{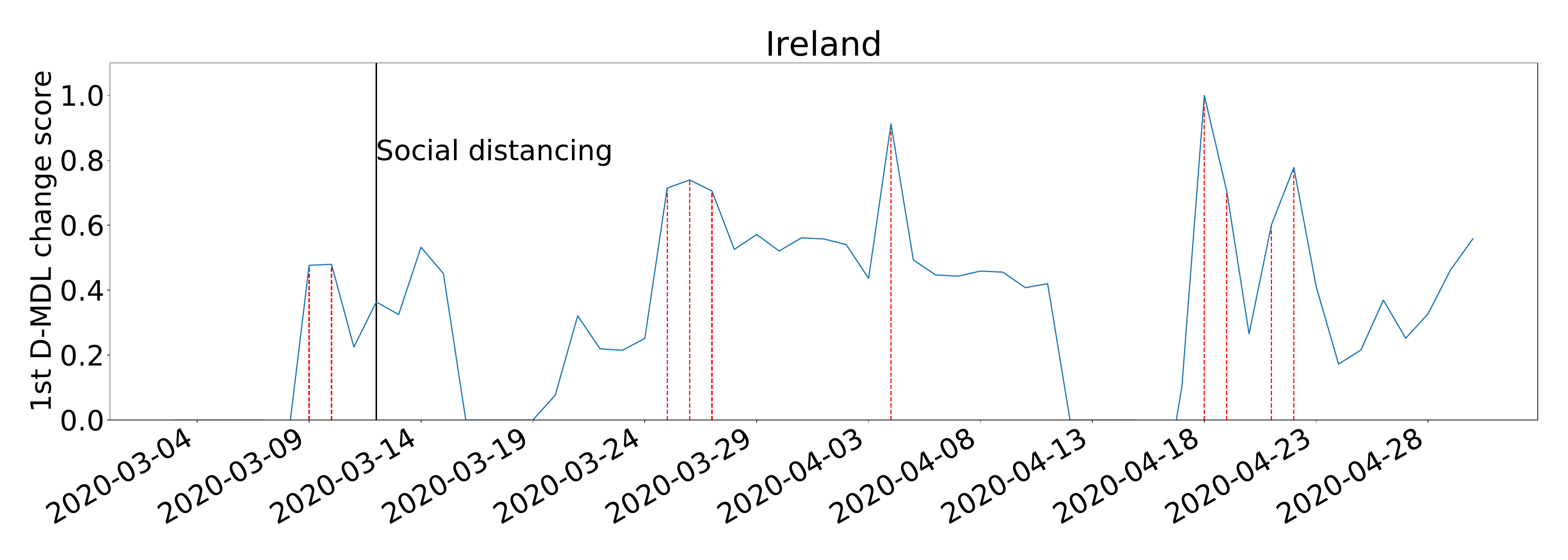} \\
		    \vspace{-0.35cm}
			\textbf{e} & \includegraphics[keepaspectratio, height=3.3cm, valign=T]
			{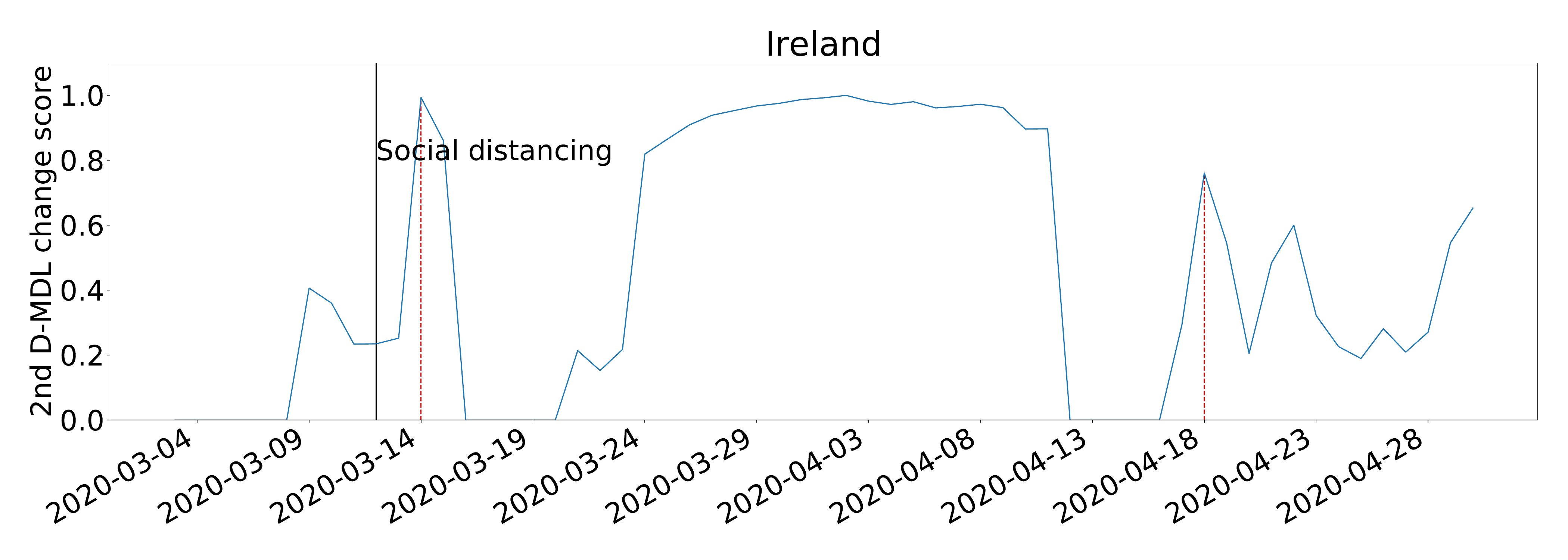} \\
		\end{tabular}
			\caption{\textbf{The results for Ireland with Gaussian modeling.} The date on which the social distancing was implemented is marked by a solid line in black. \textbf{a,} the number of daily new cases. \textbf{b,} the change scores produced by the 0th M-DML where the line in blue denotes values of scores and dashed lines in red mark alarms. \textbf{c,} the window sized for the sequential D-DML algorithm with adaptive window where lines in red mark the shrinkage of windows. \textbf{d,} the change scores produced by the 1st D-MDL. \textbf{e,} the change scores produced by the 2nd D-MDL.}
\end{figure}

\begin{figure}[H]  
\centering
\begin{tabular}{cc}
			\textbf{a} & \includegraphics[keepaspectratio, height=3.3cm, valign=T]
			{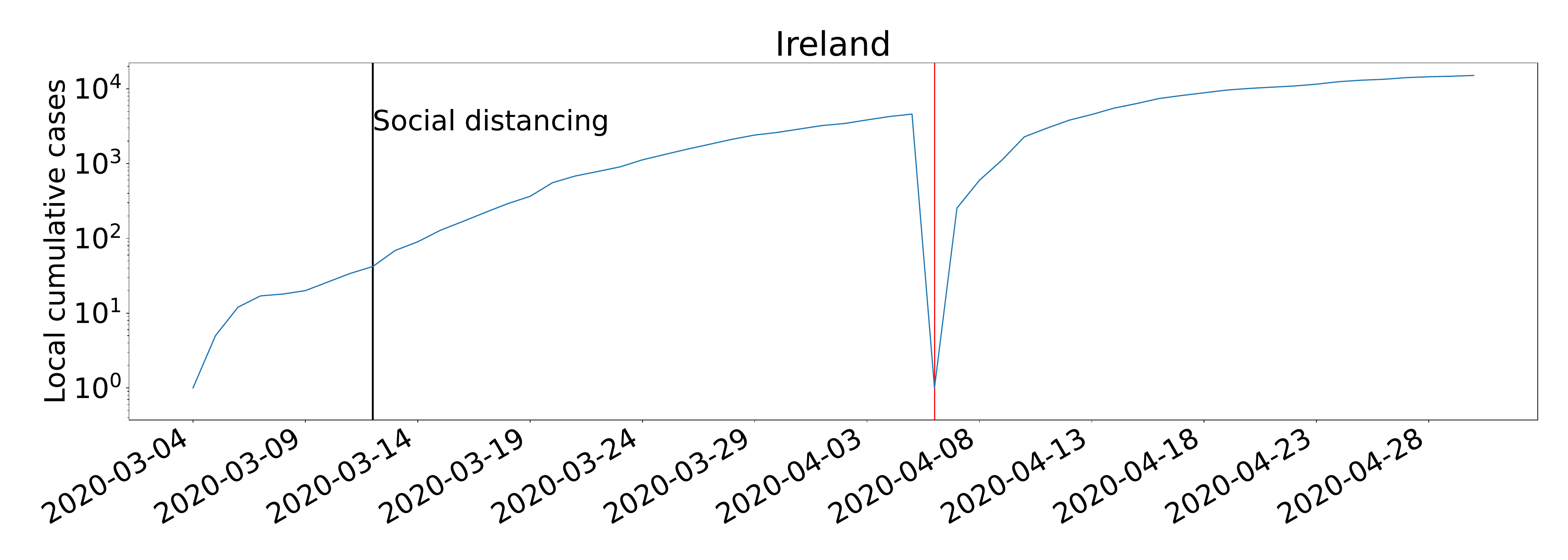} \\
	        \vspace{-0.35cm}
            \textbf{b} & \includegraphics[keepaspectratio, height=3.3cm, valign=T]
			{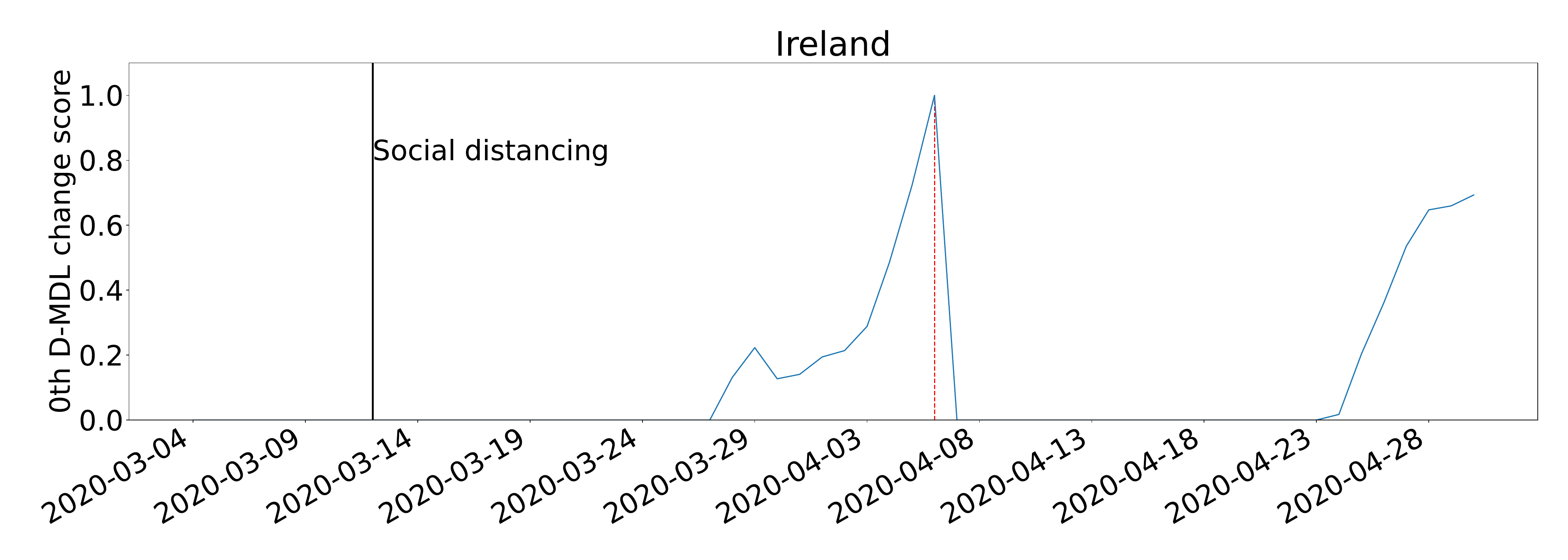}   \\
            \vspace{-0.35cm}
            \textbf{c} & \includegraphics[keepaspectratio, height=3.3cm, valign=T]
			{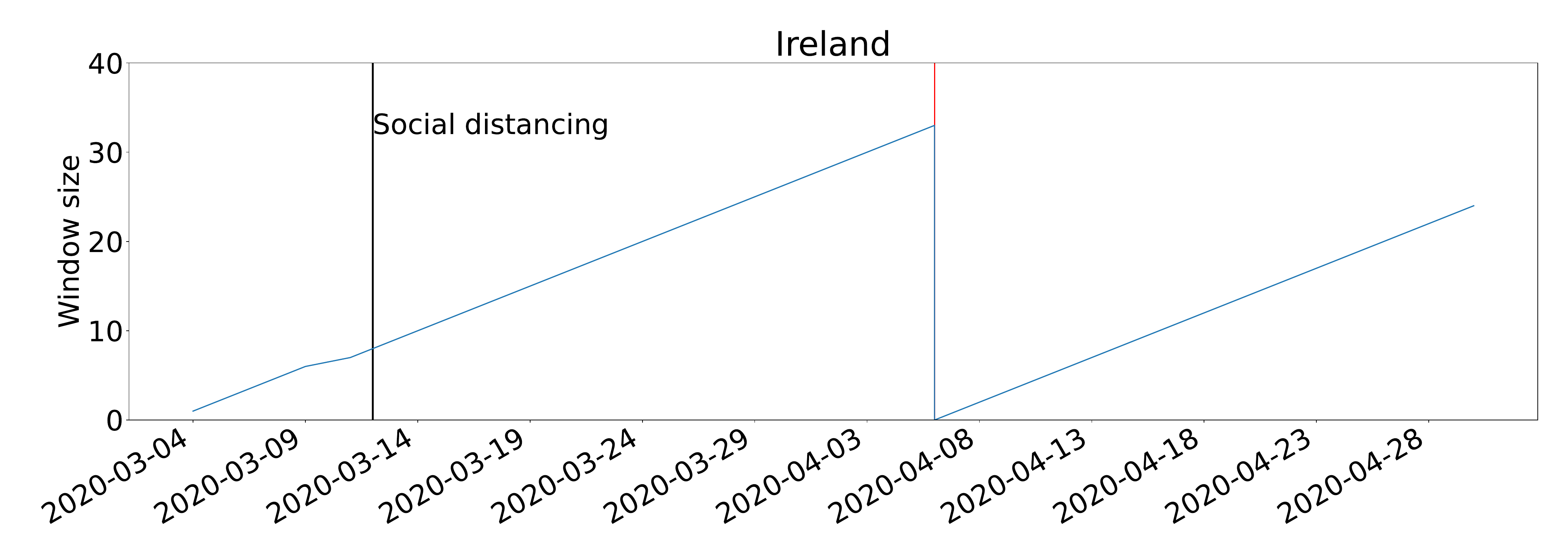} \\
			\vspace{-0.35cm}
			\textbf{d} & \includegraphics[keepaspectratio, height=3.3cm, valign=T]
			{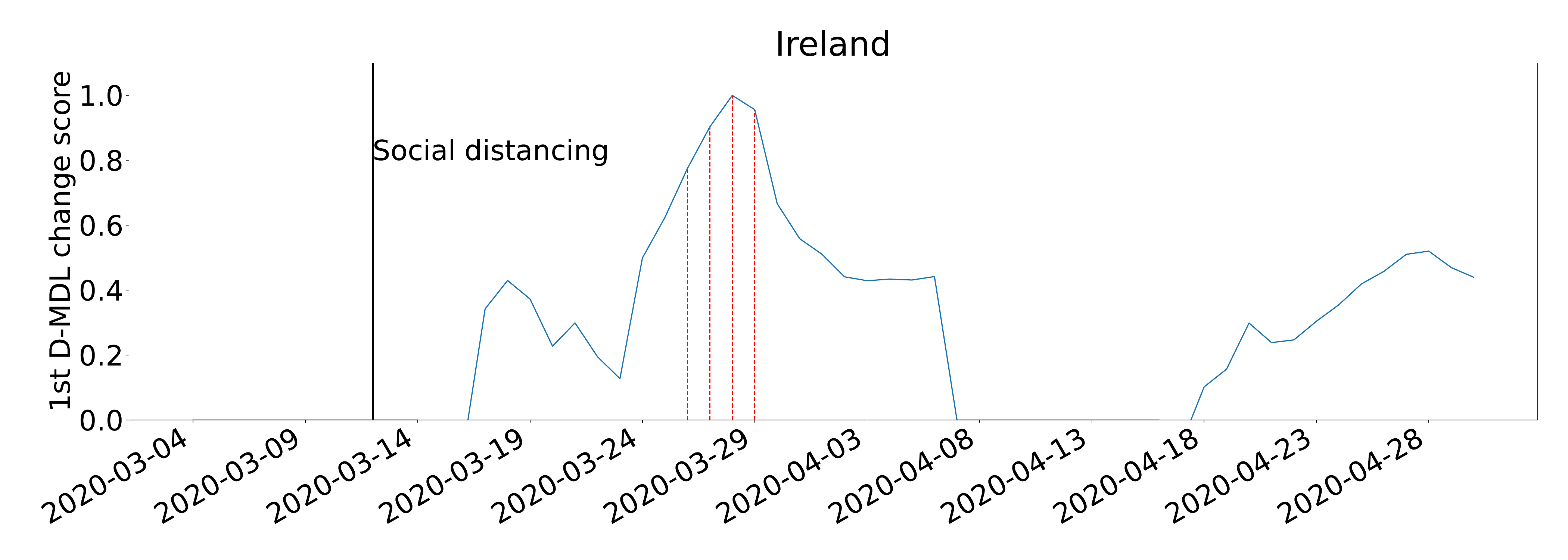} \\
			\vspace{-0.35cm}
			\textbf{e} & \includegraphics[keepaspectratio, height=3.3cm, valign=T]
			{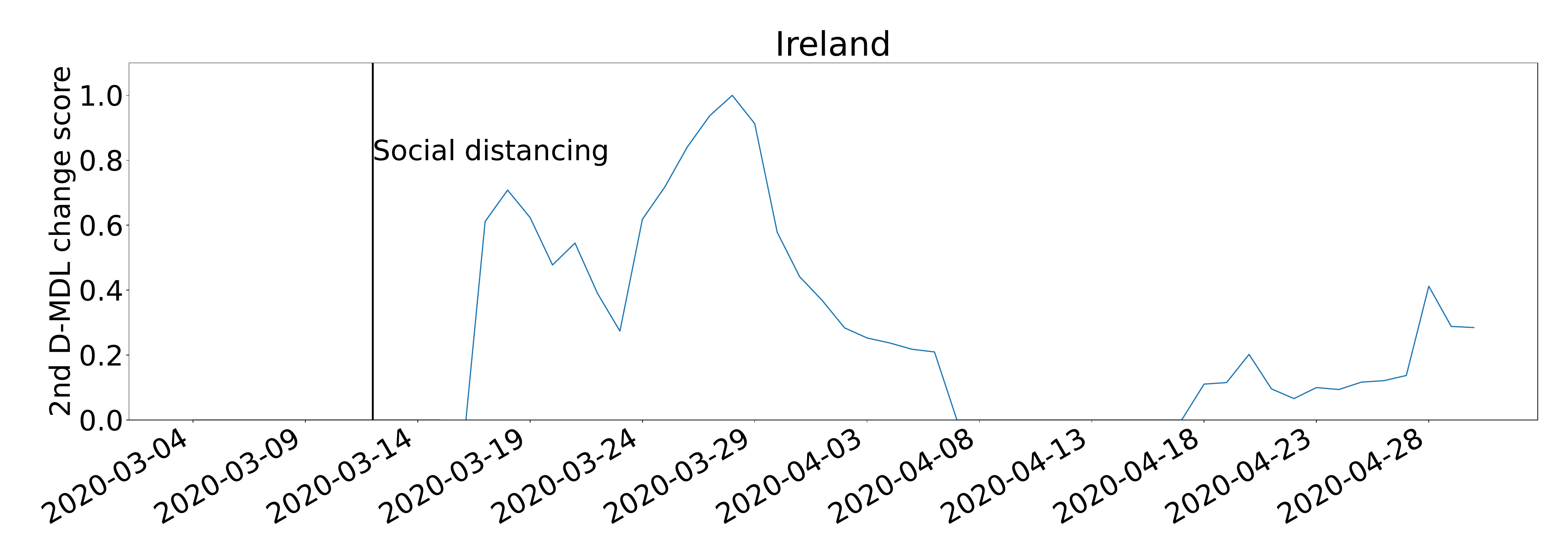} \\
		\end{tabular}
			\caption{\textbf{The results for Ireland with exponential modeling.} The date on which the social distancing was implemented is marked by a solid line in black. \textbf{a,} the number of cumulative cases. \textbf{b,} the change scores produced by the 0th M-DML where the line in blue denotes values of scores and dashed lines in red mark alarms. \textbf{c,} the window sized for the sequential D-DML algorithm with adaptive window where lines in red mark the shrinkage of windows. \textbf{d,} the change scores produced by the 1st D-MDL. \textbf{e,} the change scores produced by the 2nd D-MDL.}
\end{figure}

\begin{figure}[H] 
\centering
\begin{tabular}{cc}
		 	\textbf{a} & \includegraphics[keepaspectratio, height=3.3cm, valign=T]
			{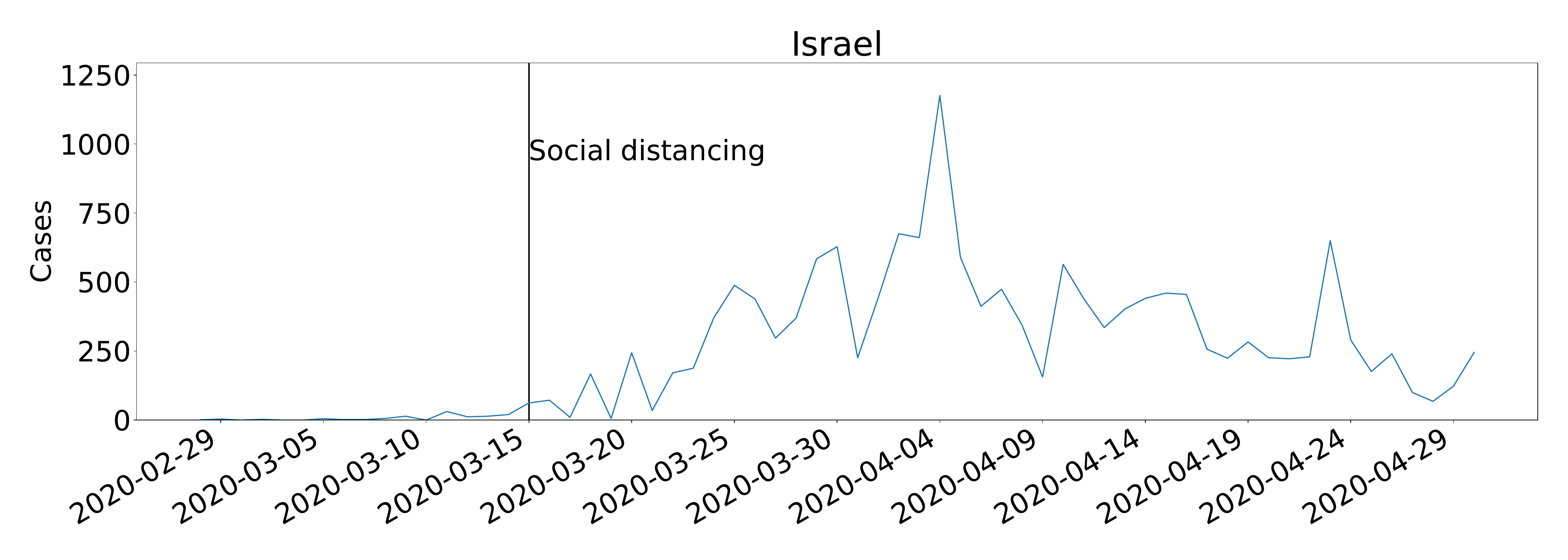} \\
			\vspace{-0.35cm}
	 	    \textbf{b} & \includegraphics[keepaspectratio, height=3.3cm, valign=T]
			{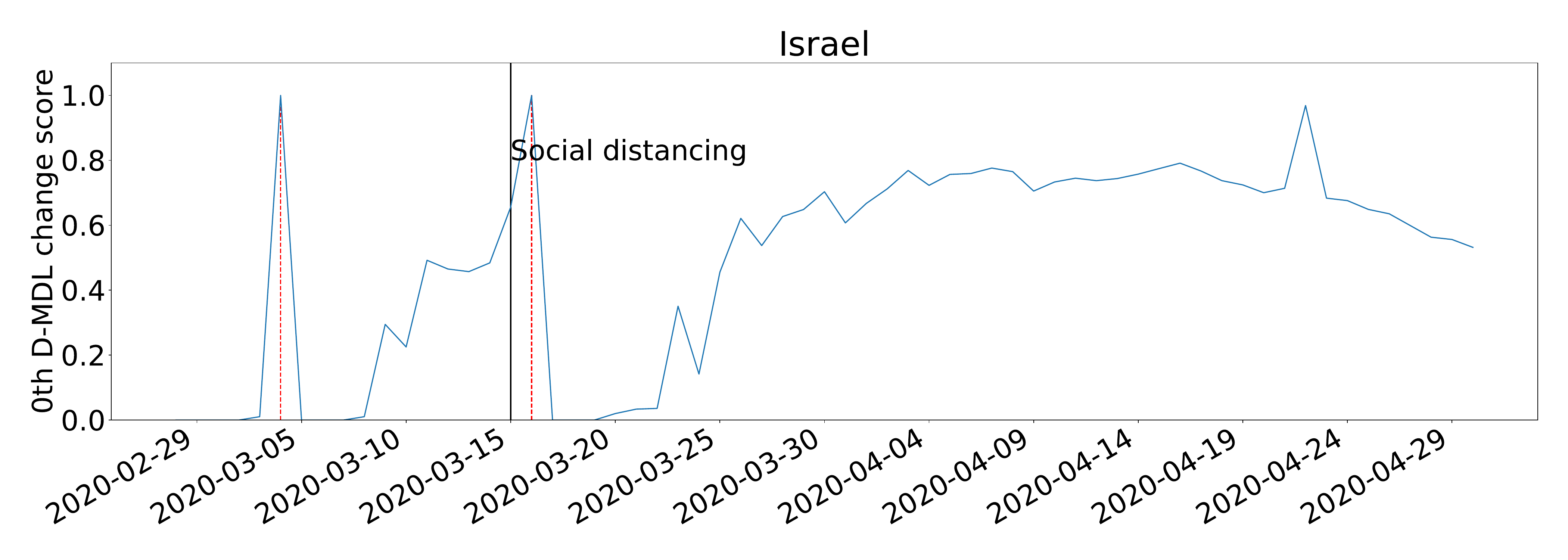}   \\
	        \vspace{-0.35cm}
			\textbf{c} & \includegraphics[keepaspectratio, height=3.3cm, valign=T]
			{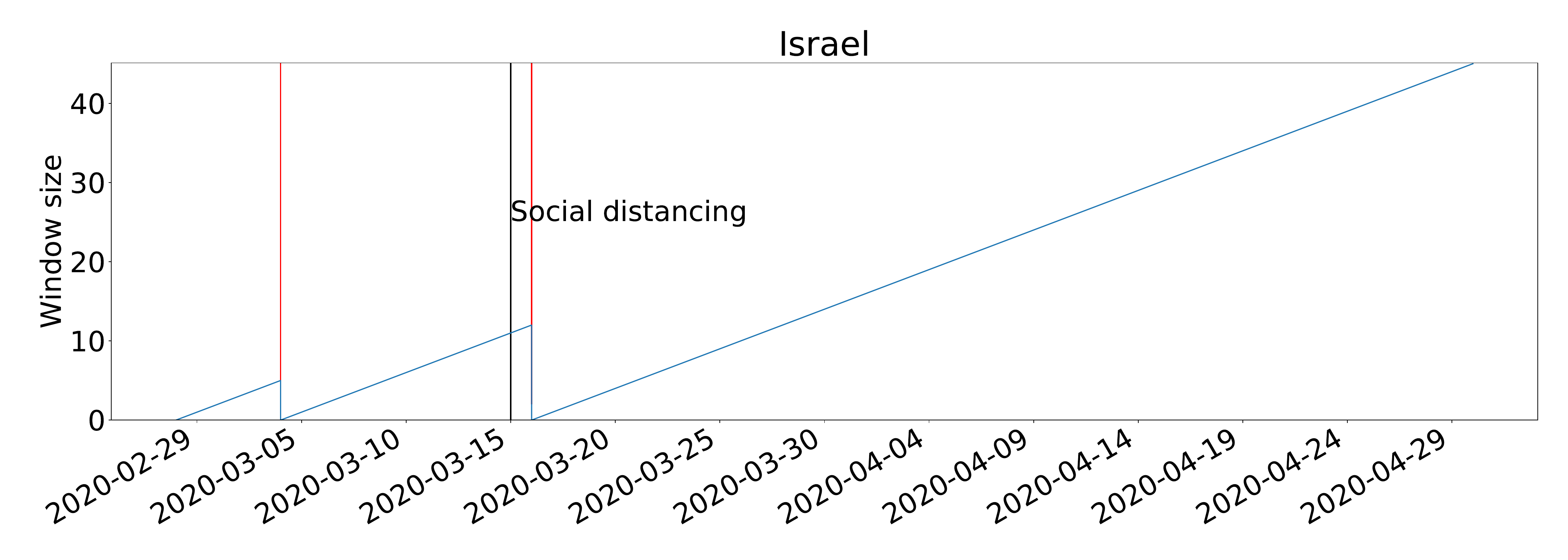} \\
		    \vspace{-0.35cm}
			\textbf{d} & \includegraphics[keepaspectratio, height=3.3cm, valign=T]
			{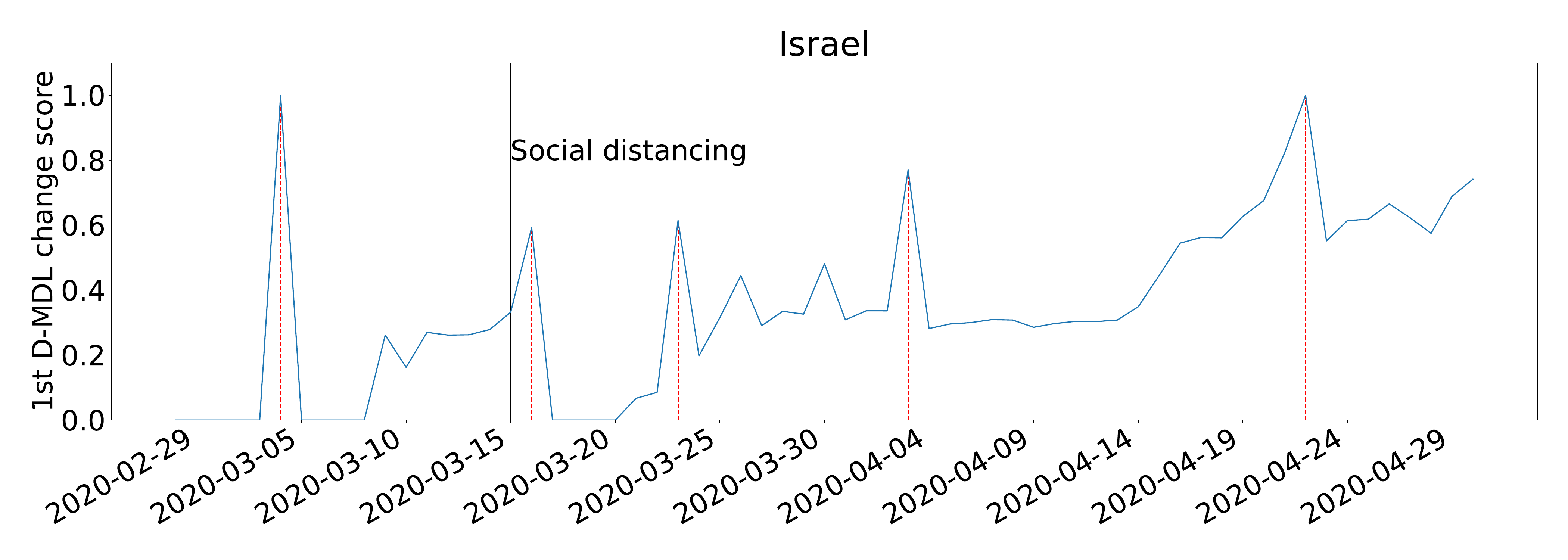} \\
		    \vspace{-0.35cm}
			\textbf{e} & \includegraphics[keepaspectratio, height=3.3cm, valign=T]
			{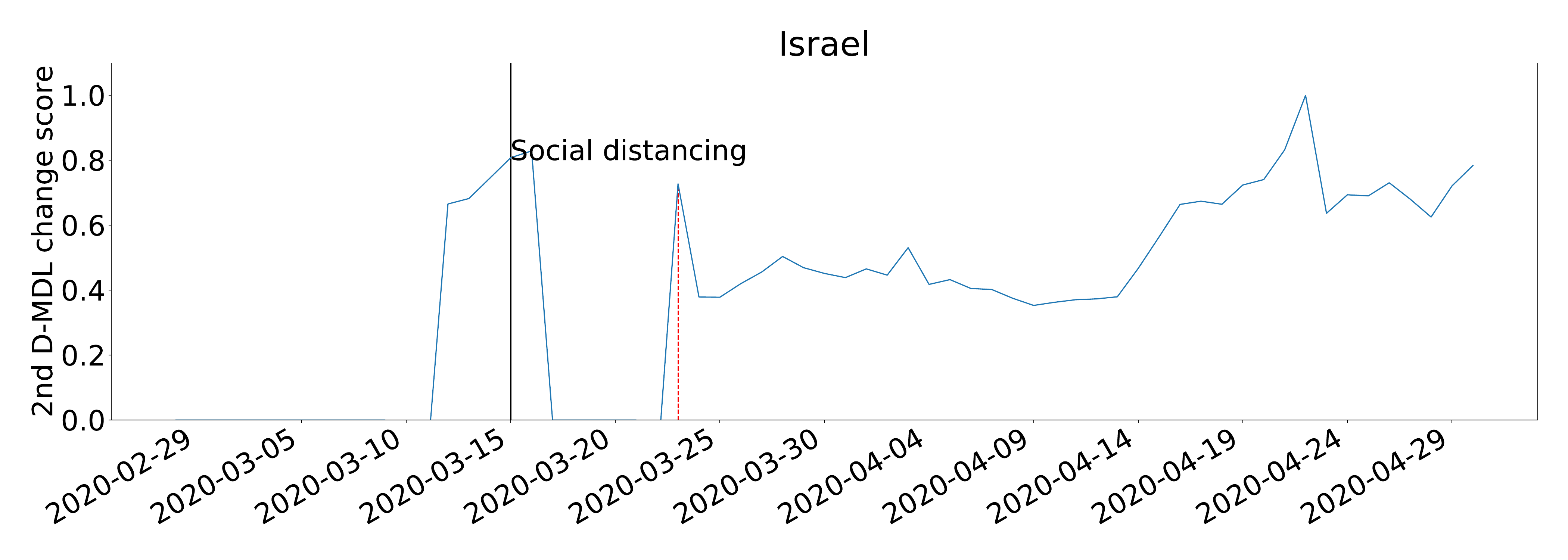} \\
		\end{tabular}
			\caption{\textbf{The results for Israel with Gaussian modeling.} The date on which the social distancing was implemented is marked by a solid line in black. \textbf{a,} the number of daily new cases. \textbf{b,} the change scores produced by the 0th M-DML where the line in blue denotes values of scores and dashed lines in red mark alarms. \textbf{c,} the window sized for the sequential D-DML algorithm with adaptive window where lines in red mark the shrinkage of windows. \textbf{d,} the change scores produced by the 1st D-MDL. \textbf{e,} the change scores produced by the 2nd D-MDL.}
\end{figure}

\begin{figure}[H]  
\centering
\begin{tabular}{cc}
			\textbf{a} & \includegraphics[keepaspectratio, height=3.3cm, valign=T]
			{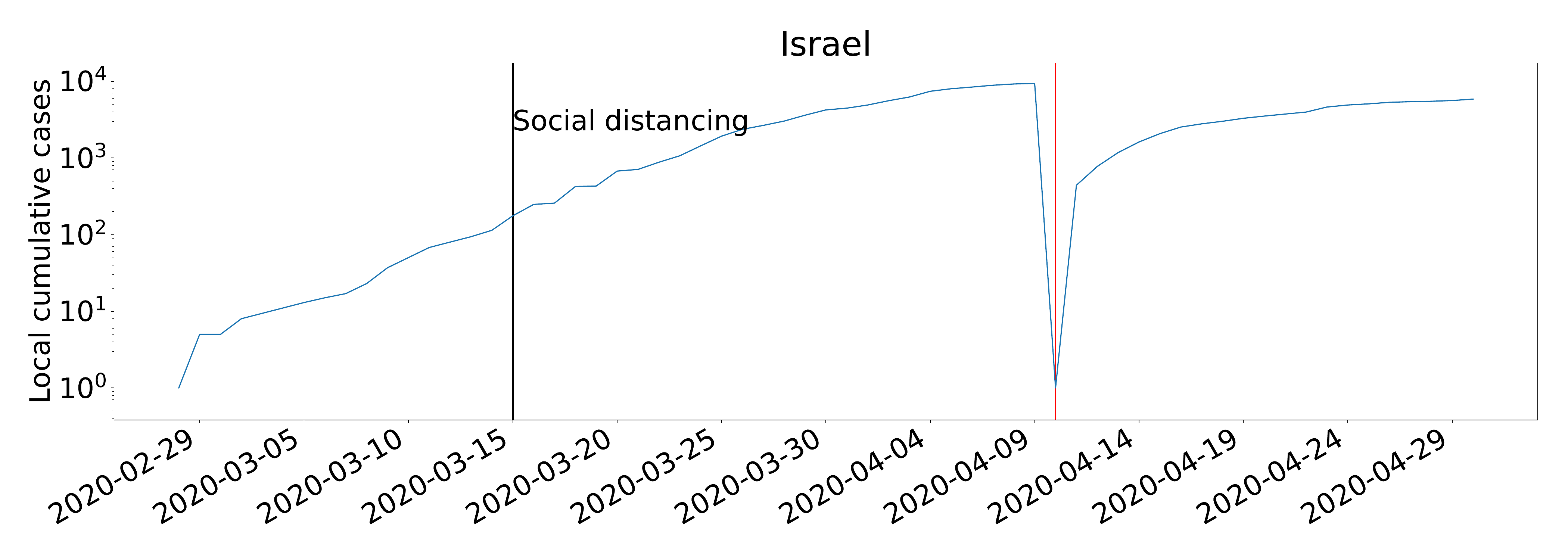} \\
	        \vspace{-0.35cm}
            \textbf{b} & \includegraphics[keepaspectratio, height=3.3cm, valign=T]
			{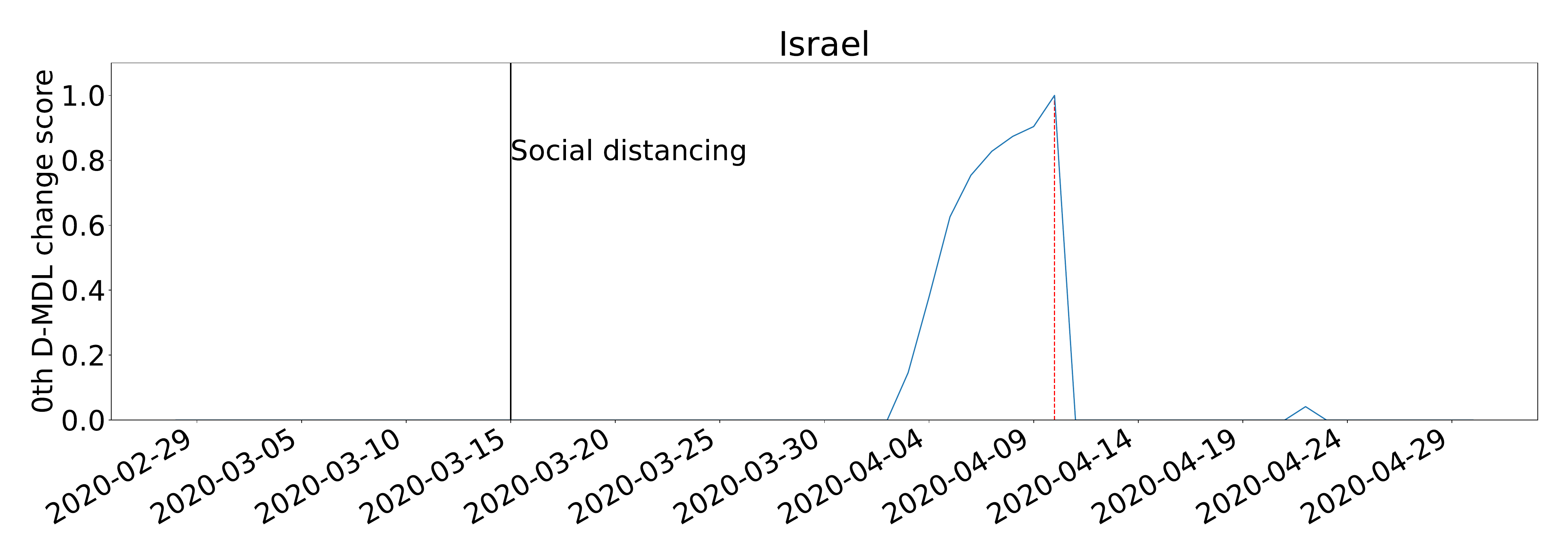}   \\
            \vspace{-0.35cm}
            \textbf{c} & \includegraphics[keepaspectratio, height=3.3cm, valign=T]
			{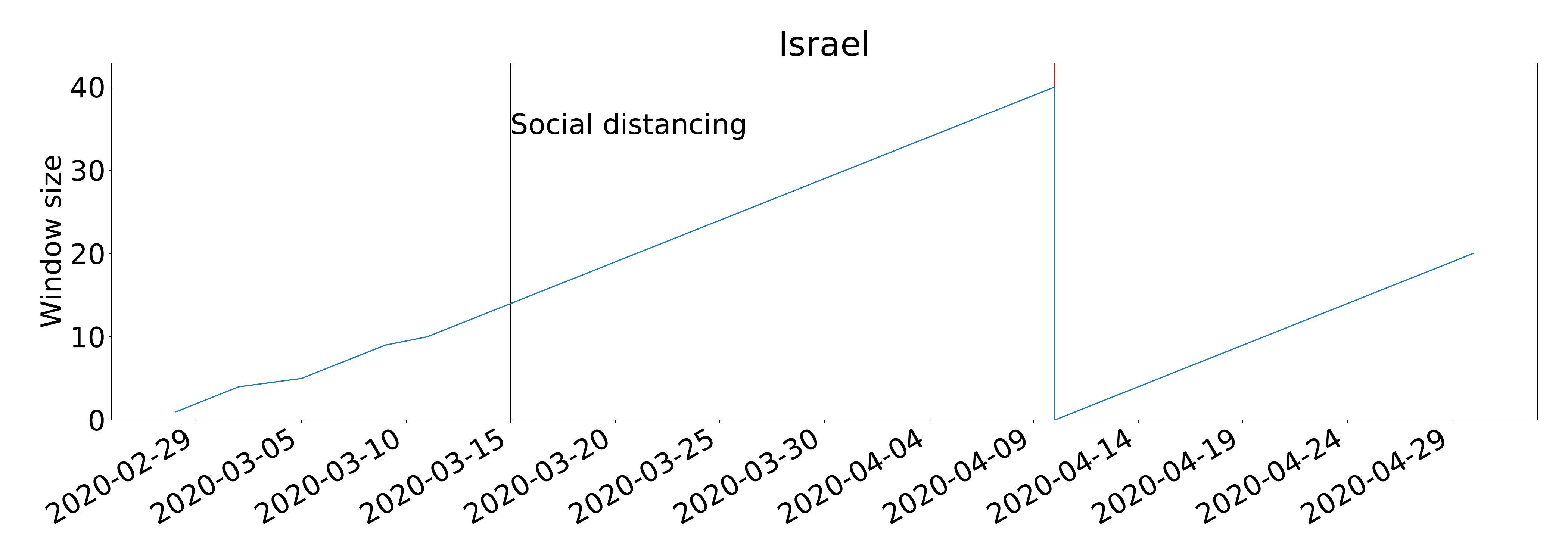} \\
			\vspace{-0.35cm}
			\textbf{d} & \includegraphics[keepaspectratio, height=3.3cm, valign=T]
			{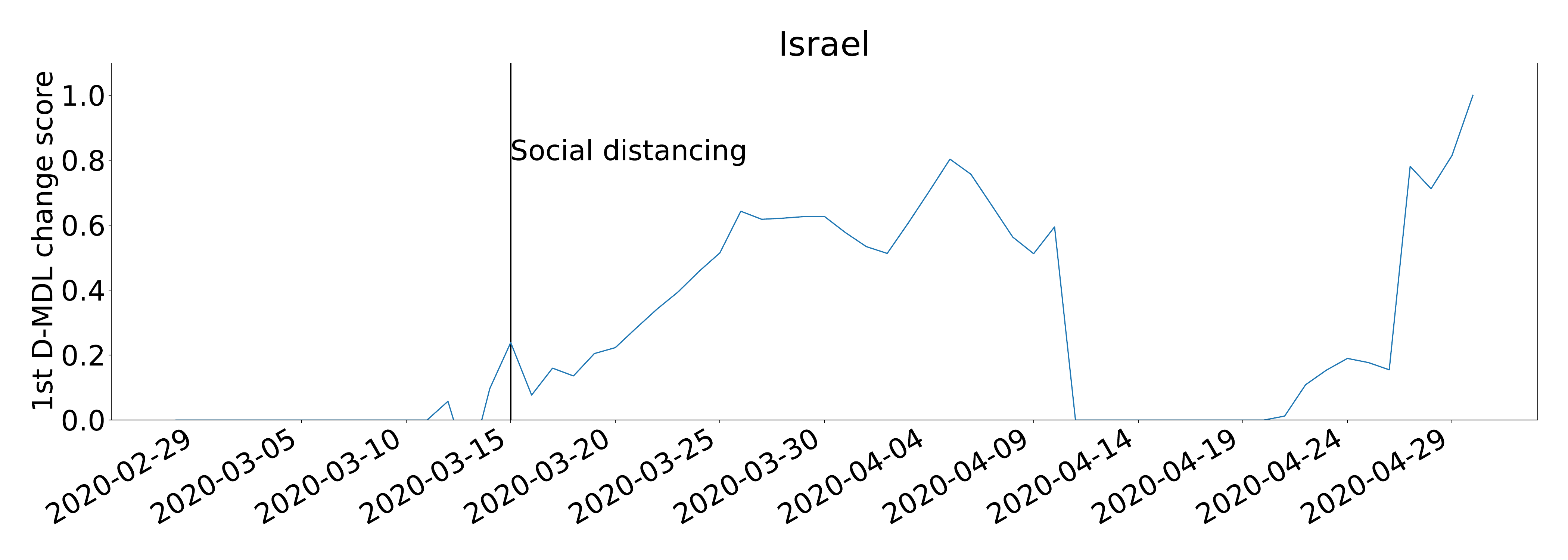} \\
			\vspace{-0.35cm}
			\textbf{e} & \includegraphics[keepaspectratio, height=3.3cm, valign=T]
			{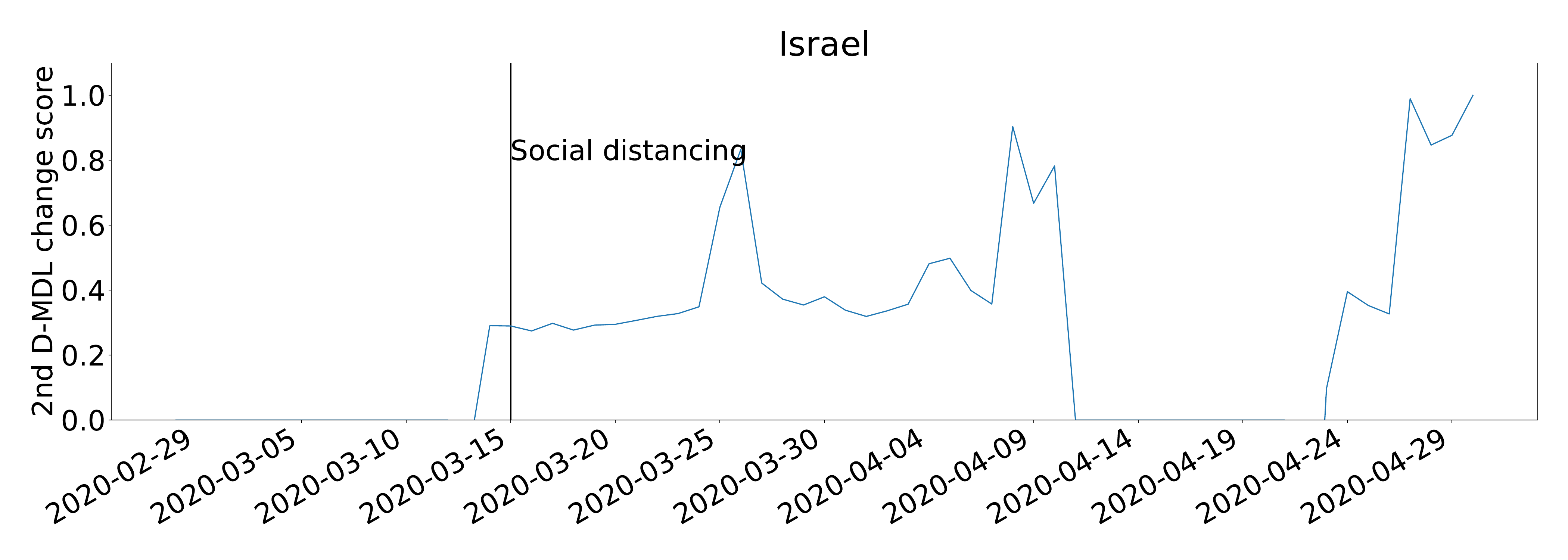} \\
		\end{tabular}
			\caption{\textbf{The results for Israel with exponential modeling.} The date on which the social distancing was implemented is marked by a solid line in black. \textbf{a,} the number of cumulative cases. \textbf{b,} the change scores produced by the 0th M-DML where the line in blue denotes values of scores and dashed lines in red mark alarms. \textbf{c,} the window sized for the sequential D-DML algorithm with adaptive window where lines in red mark the shrinkage of windows. \textbf{d,} the change scores produced by the 1st D-MDL. \textbf{e,} the change scores produced by the 2nd D-MDL.}
\end{figure}

\begin{figure}[H] 
\centering
\begin{tabular}{cc}
		 	\textbf{a} & \includegraphics[keepaspectratio, height=3.3cm, valign=T]
			{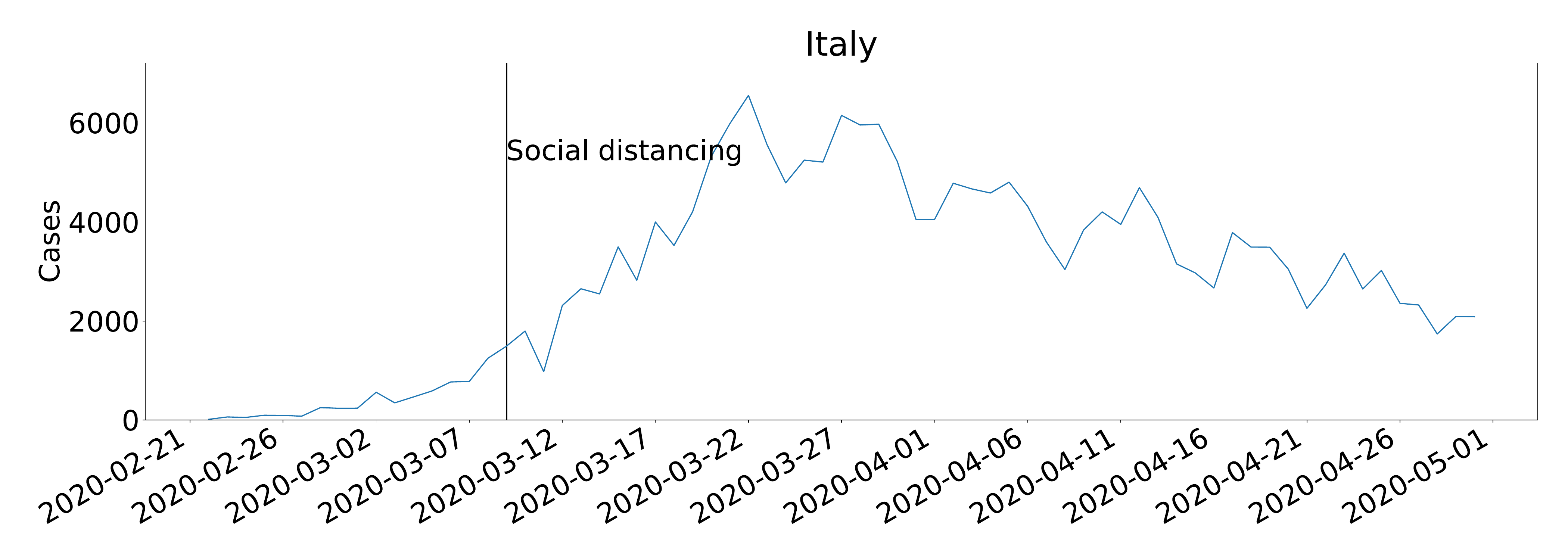} \\
			\vspace{-0.35cm}
	 	    \textbf{b} & \includegraphics[keepaspectratio, height=3.3cm, valign=T]
			{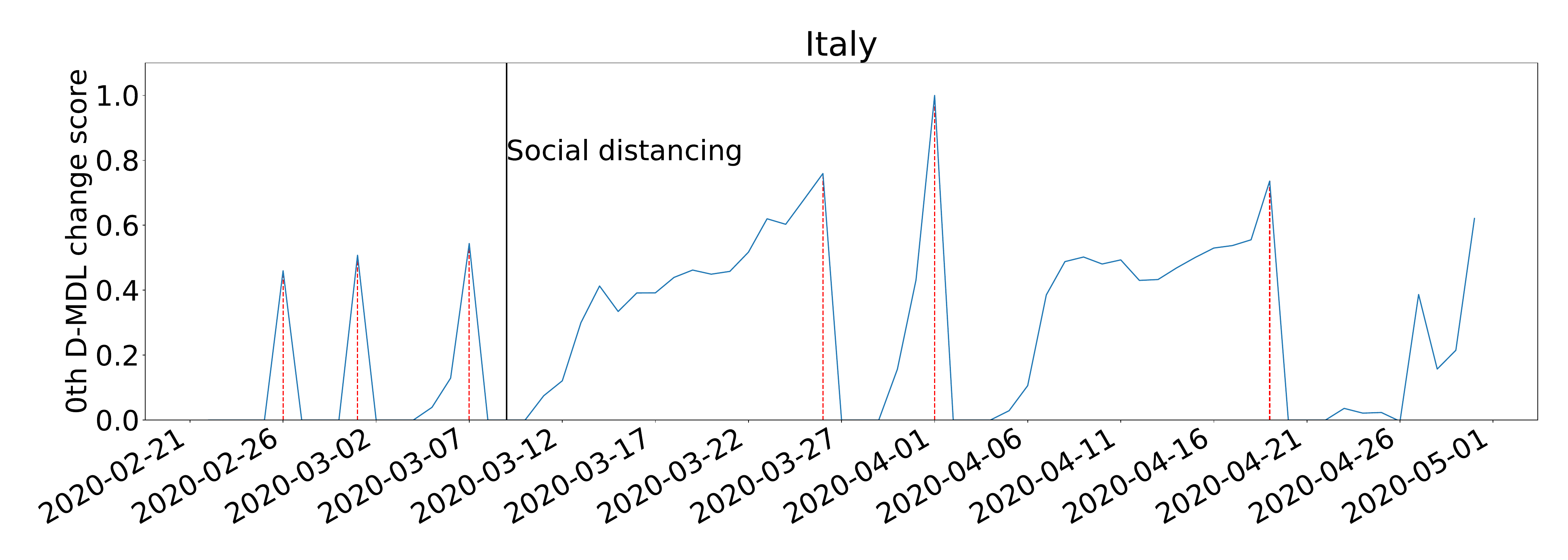}   \\
	        \vspace{-0.35cm}
			\textbf{c} & \includegraphics[keepaspectratio, height=3.3cm, valign=T]
			{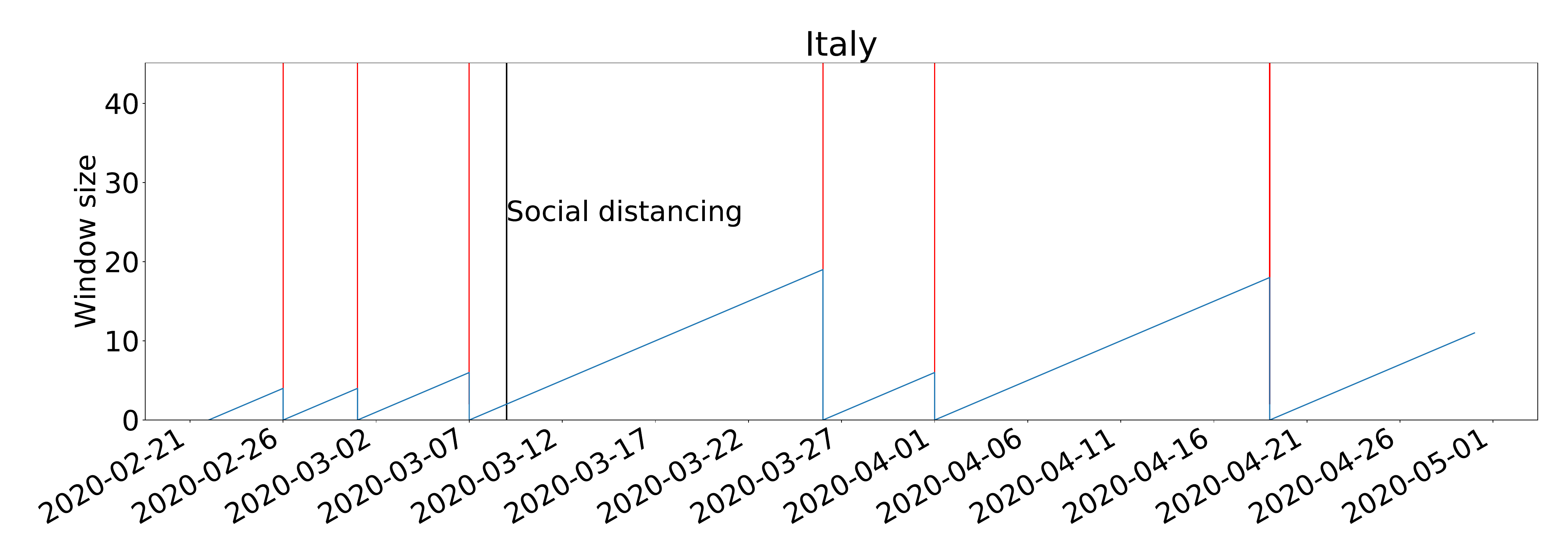} \\
		    \vspace{-0.35cm}
			\textbf{d} & \includegraphics[keepaspectratio, height=3.3cm, valign=T]
			{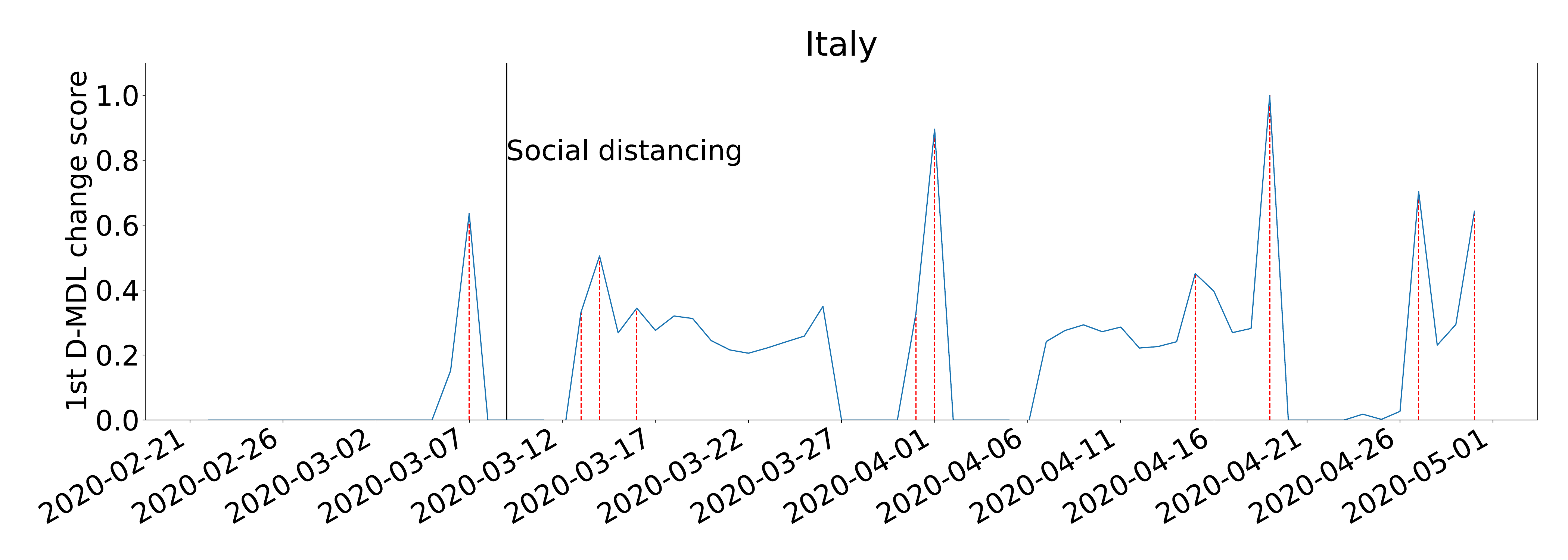} \\
		    \vspace{-0.35cm}
			\textbf{e} & \includegraphics[keepaspectratio, height=3.3cm, valign=T]
			{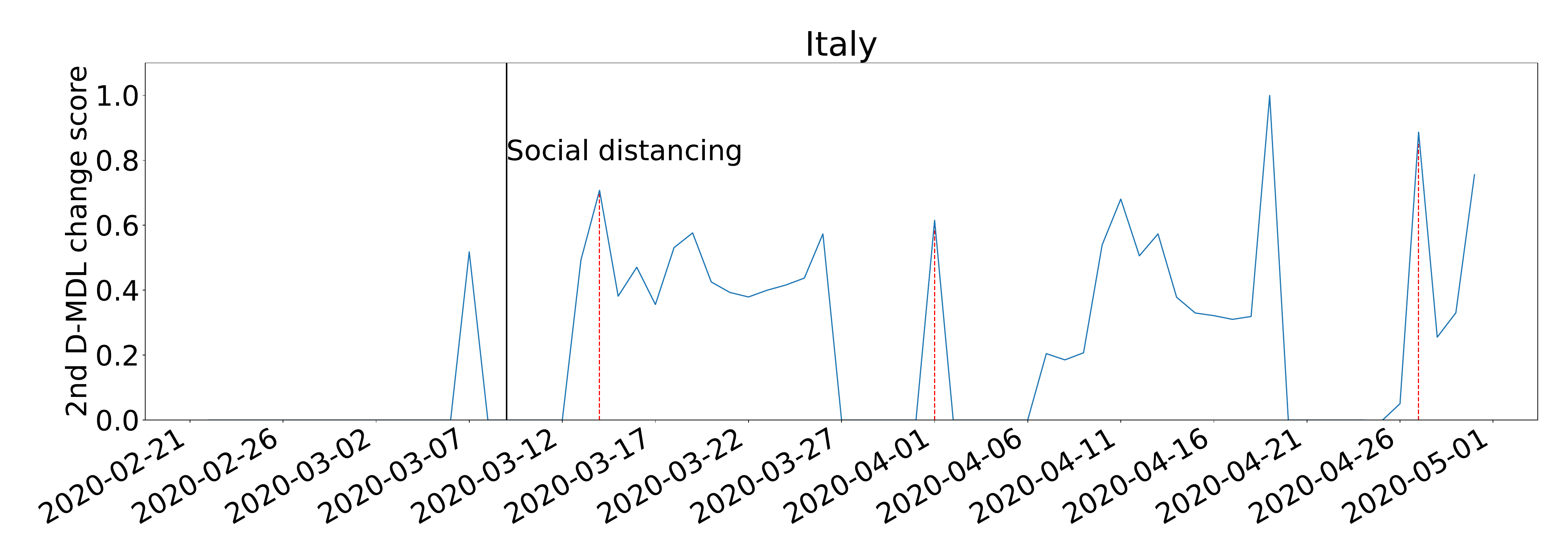} \\
		\end{tabular}
			\caption{\textbf{The results for Italy with Gaussian modeling.} The date on which the social distancing was implemented is marked by a solid line in black. \textbf{a,} the number of daily new cases. \textbf{b,} the change scores produced by the 0th M-DML where the line in blue denotes values of scores and dashed lines in red mark alarms. \textbf{c,} the window sized for the sequential D-DML algorithm with adaptive window where lines in red mark the shrinkage of windows. \textbf{d,} the change scores produced by the 1st D-MDL. \textbf{e,} the change scores produced by the 2nd D-MDL.}
\end{figure}

\begin{figure}[H]  
\centering
\begin{tabular}{cc}
			\textbf{a} & \includegraphics[keepaspectratio, height=3.3cm, valign=T]
			{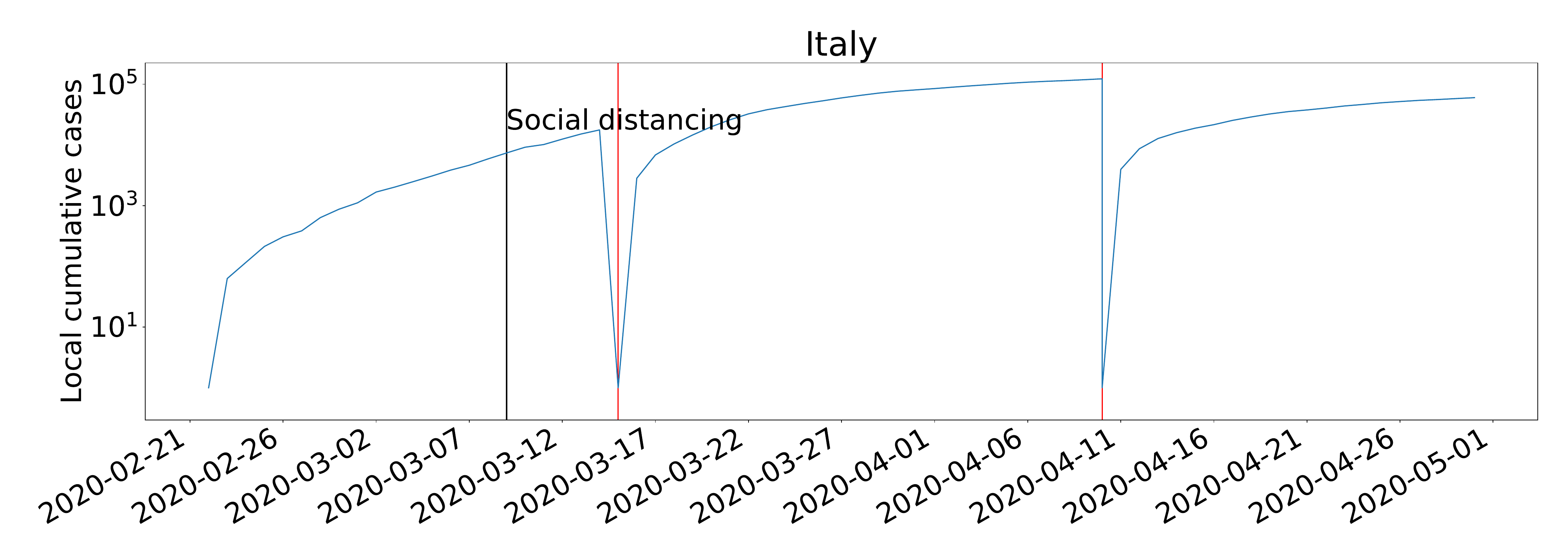} \\
	        \vspace{-0.35cm}
            \textbf{b} & \includegraphics[keepaspectratio, height=3.3cm, valign=T]
			{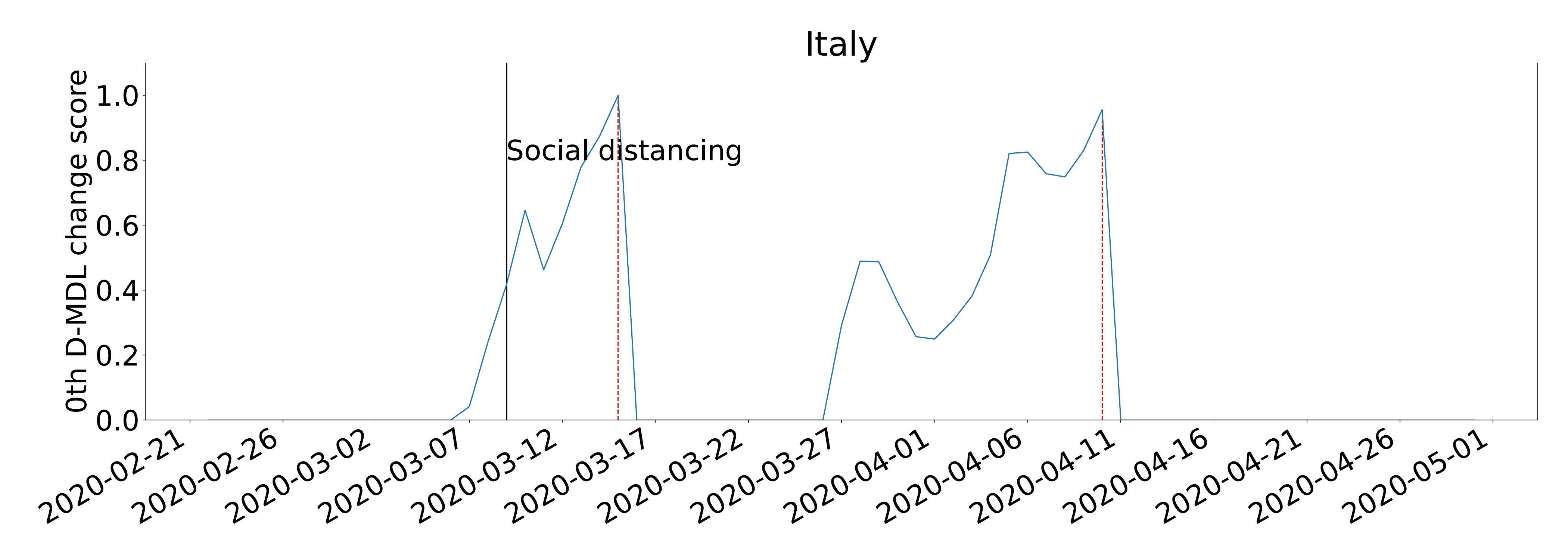}   \\
            \vspace{-0.35cm}
            \textbf{c} & \includegraphics[keepaspectratio, height=3.3cm, valign=T]
			{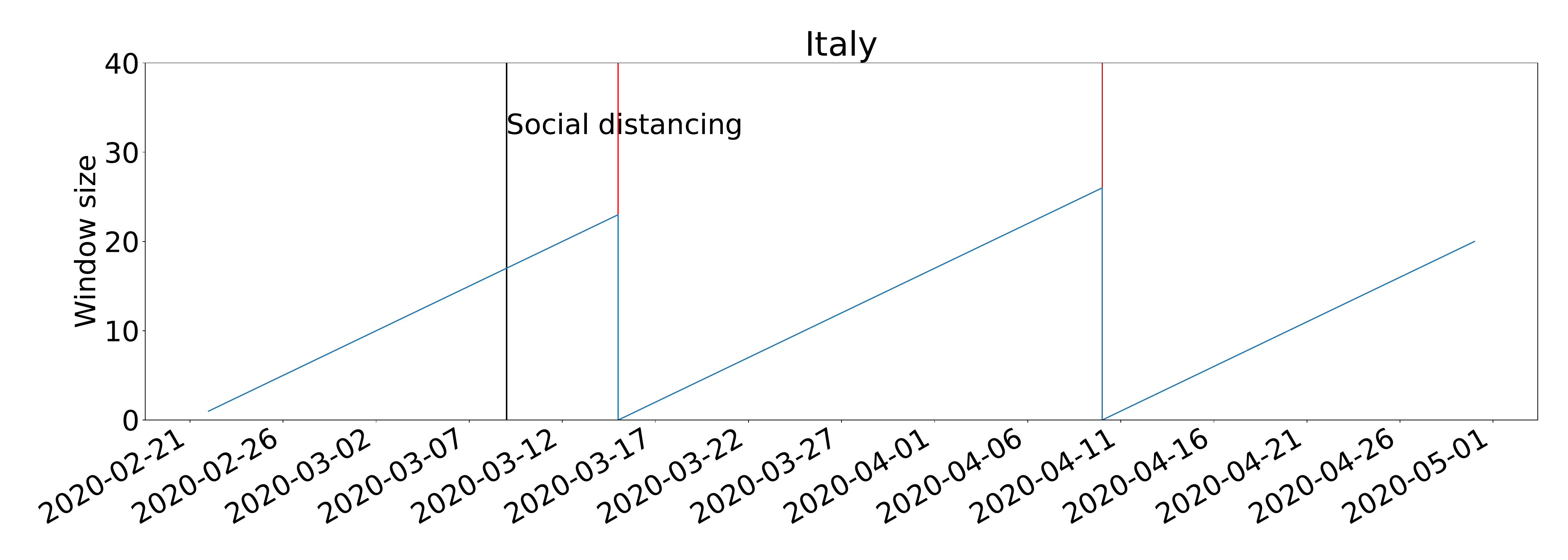} \\
			\vspace{-0.35cm}
			\textbf{d} & \includegraphics[keepaspectratio, height=3.3cm, valign=T]
			{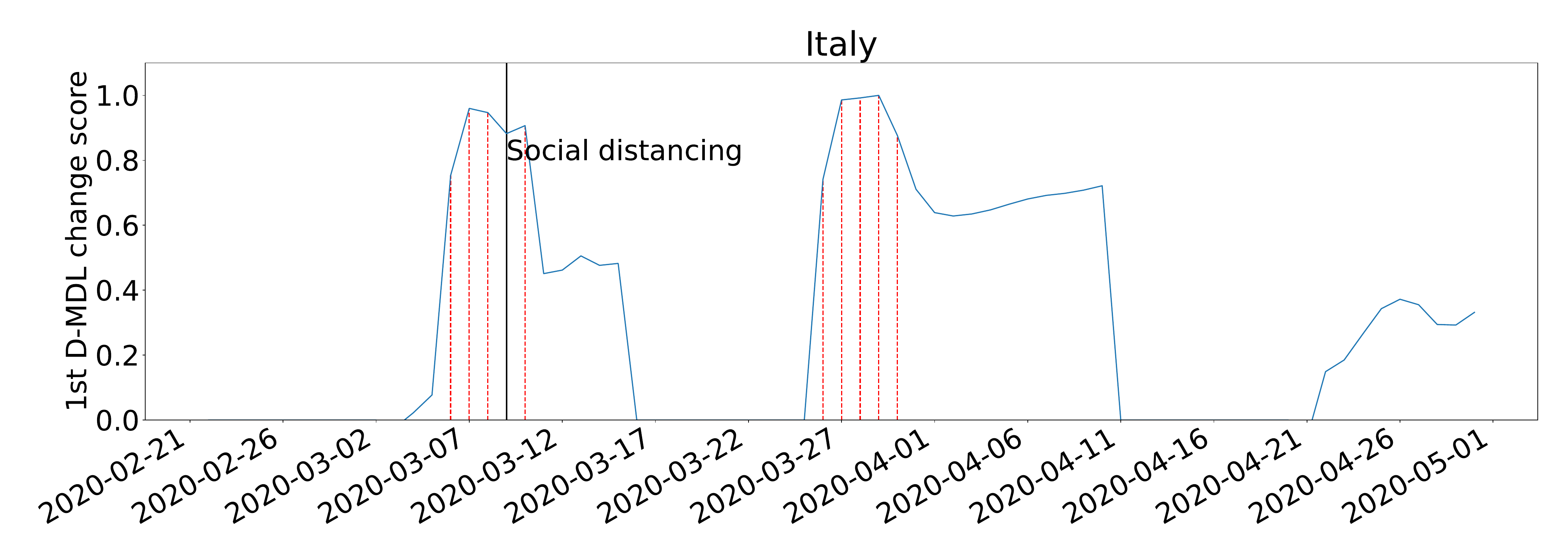} \\
			\vspace{-0.35cm}
			\textbf{e} & \includegraphics[keepaspectratio, height=3.3cm, valign=T]
			{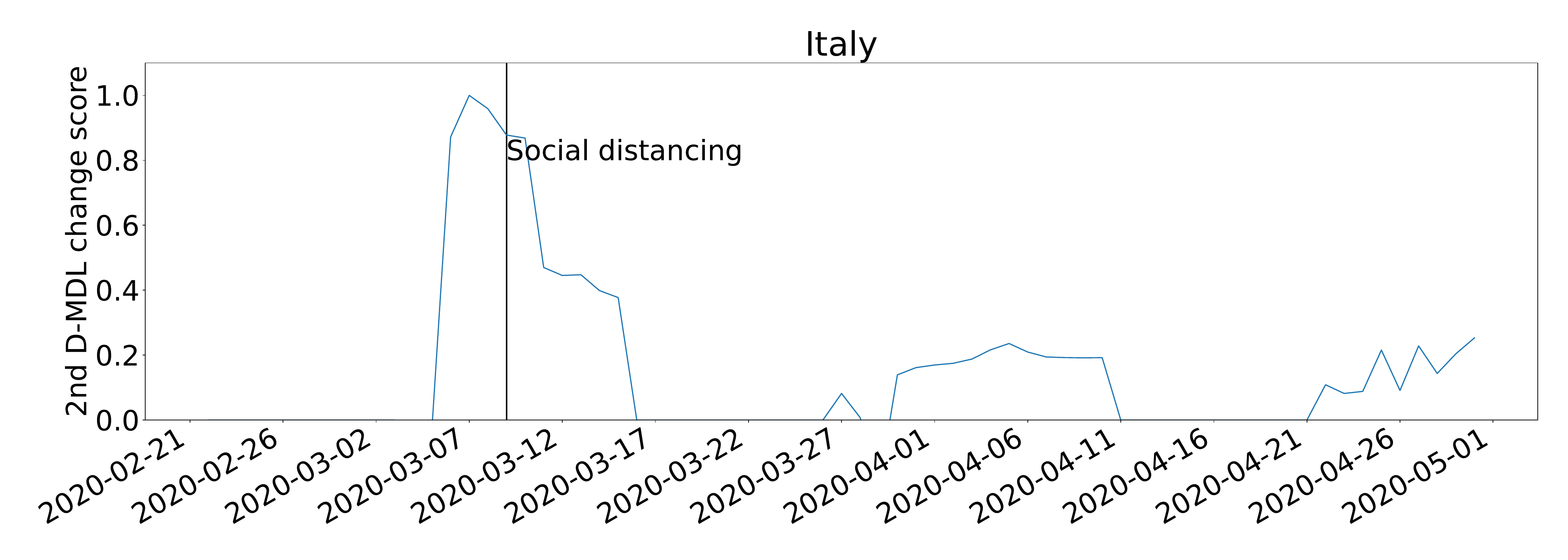} \\
		\end{tabular}
			\caption{\textbf{The results for Italy with exponential modeling.} The date on which the social distancing was implemented is marked by a solid line in black. \textbf{a,} the number of cumulative cases. \textbf{b,} the change scores produced by the 0th M-DML where the line in blue denotes values of scores and dashed lines in red mark alarms. \textbf{c,} the window sized for the sequential D-DML algorithm with adaptive window where lines in red mark the shrinkage of windows. \textbf{d,} the change scores produced by the 1st D-MDL. \textbf{e,} the change scores produced by the 2nd D-MDL.}
\end{figure}

\begin{figure}[H] 
\centering
\begin{tabular}{cc}
		 	\textbf{a} & \includegraphics[keepaspectratio, height=3.3cm, valign=T]
			{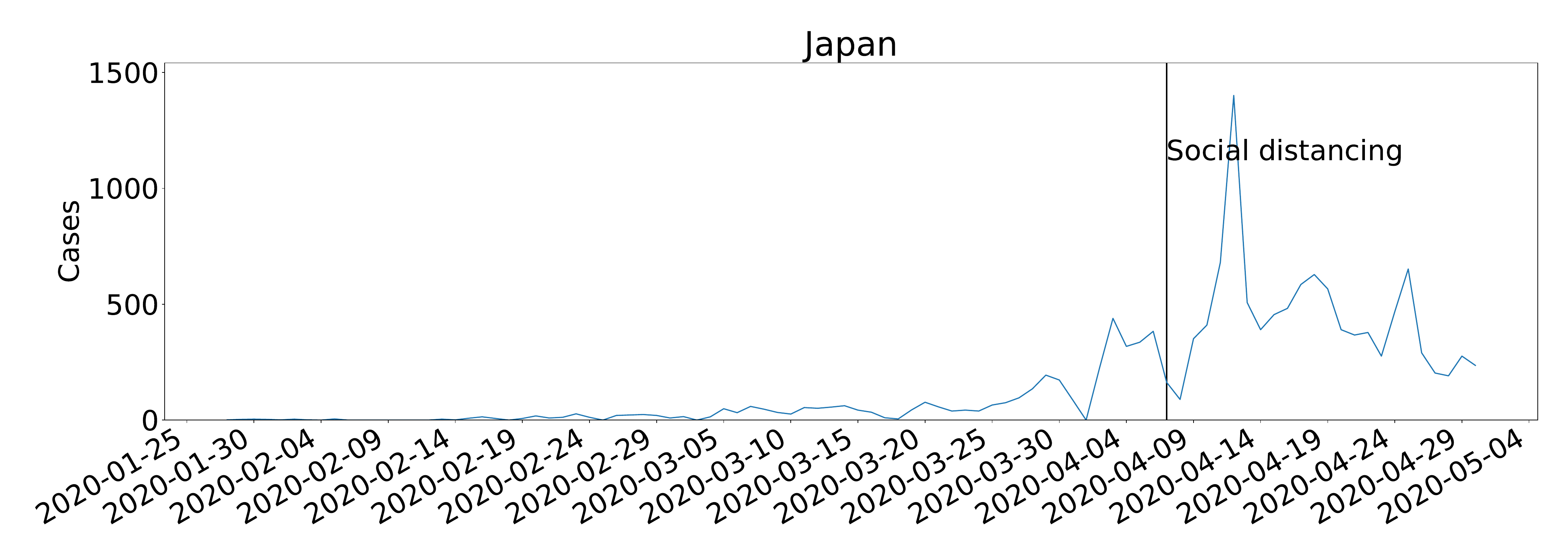} \\
			\vspace{-0.35cm}
	 	    \textbf{b} & \includegraphics[keepaspectratio, height=3.3cm, valign=T]
			{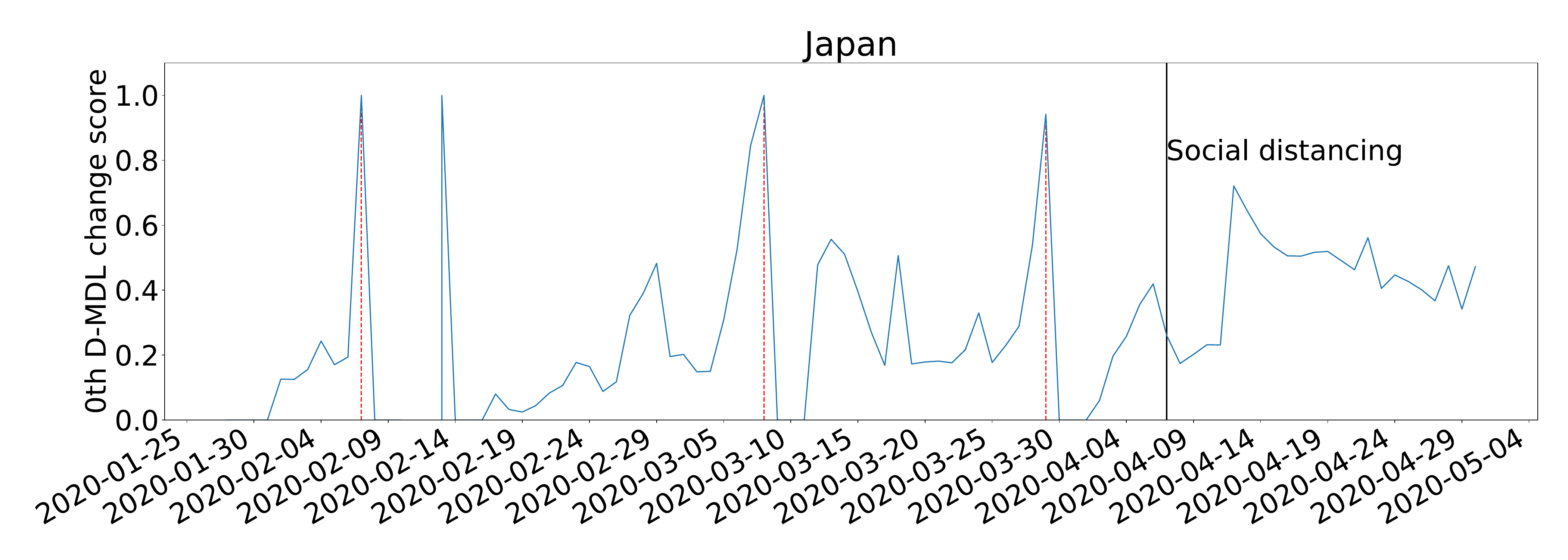}   \\
	        \vspace{-0.35cm}
			\textbf{c} & \includegraphics[keepaspectratio, height=3.3cm, valign=T]
			{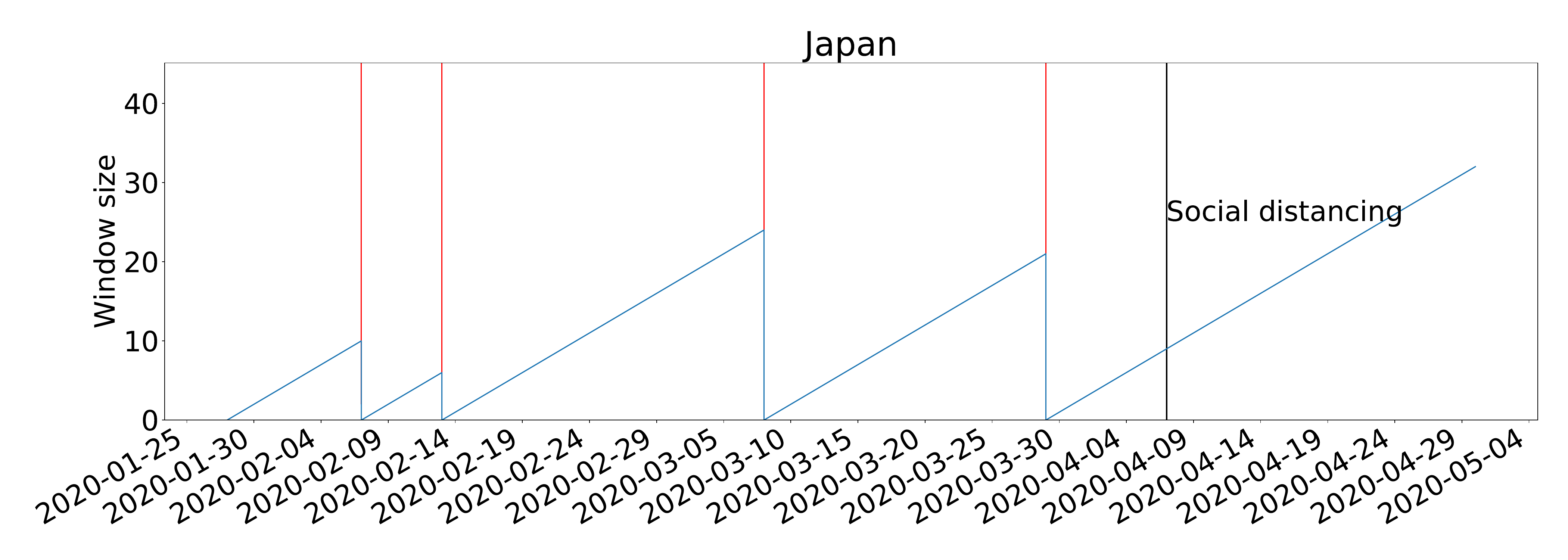} \\
		    \vspace{-0.35cm}
			\textbf{d} & \includegraphics[keepaspectratio, height=3.3cm, valign=T]
			{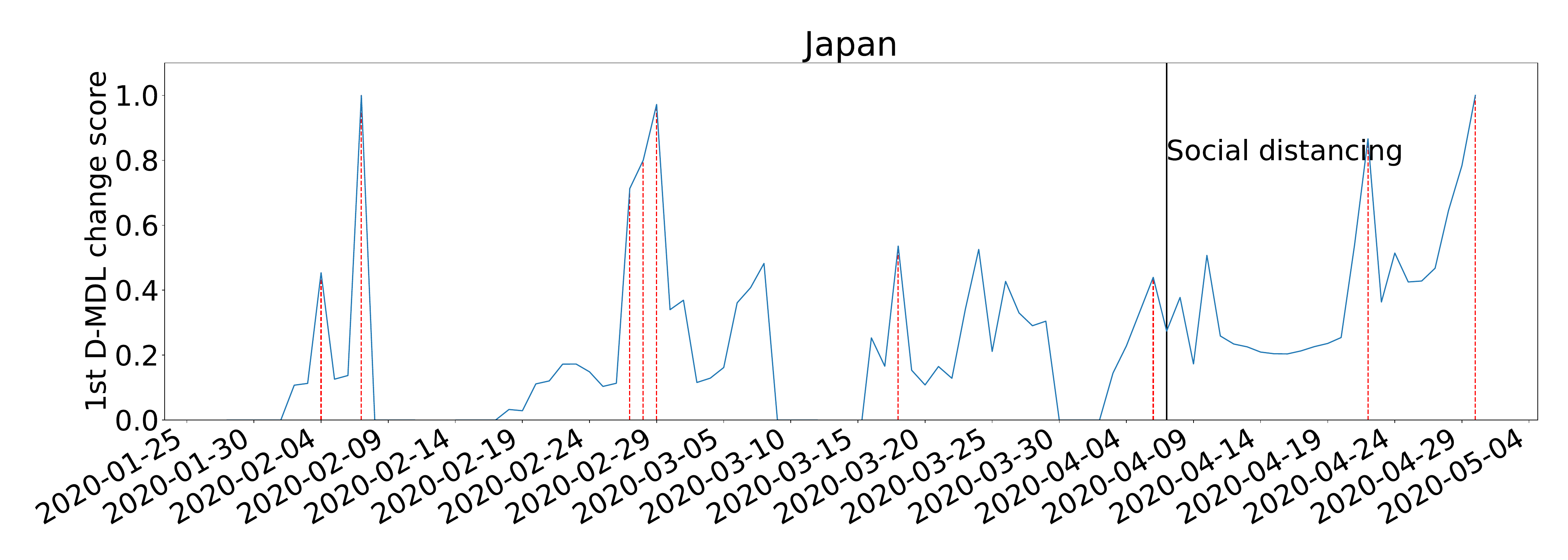} \\
		    \vspace{-0.35cm}
			\textbf{e} & \includegraphics[keepaspectratio, height=3.3cm, valign=T]
			{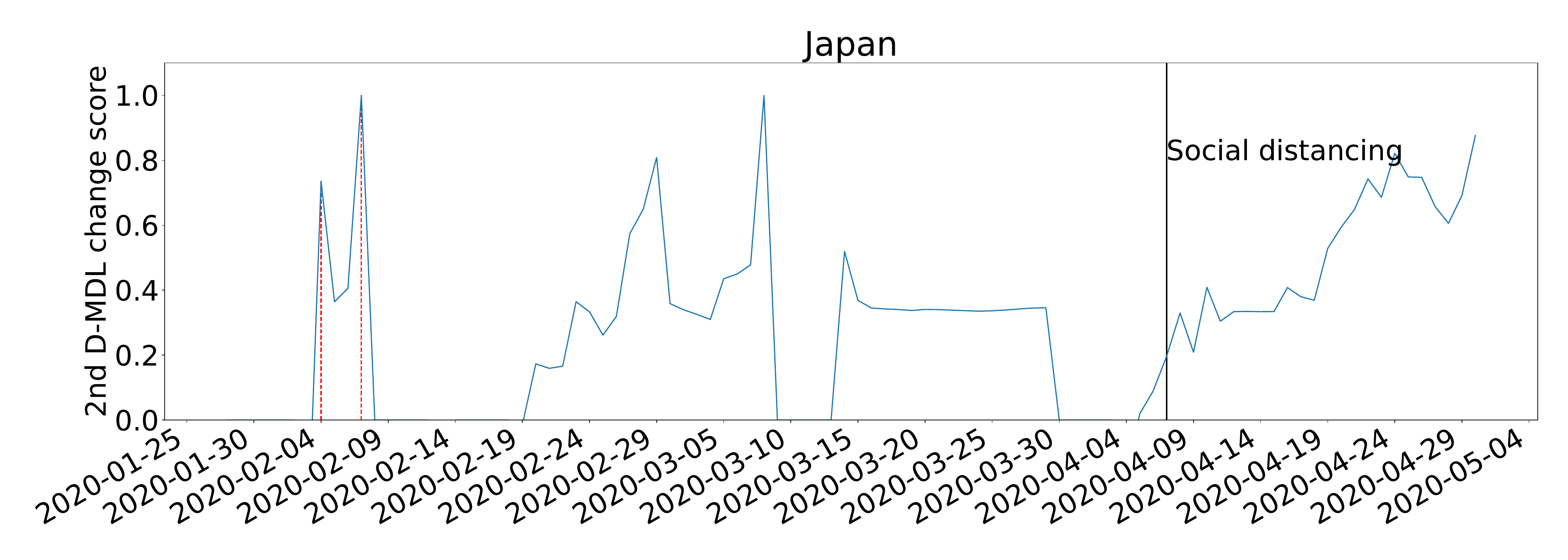} \\
		\end{tabular}
			\caption{\textbf{The results for Japan with Gaussian modeling.} The date on which the social distancing was implemented is marked by a solid line in black. \textbf{a,} the number of daily new cases. \textbf{b,} the change scores produced by the 0th M-DML where the line in blue denotes values of scores and dashed lines in red mark alarms. \textbf{c,} the window sized for the sequential D-DML algorithm with adaptive window where lines in red mark the shrinkage of windows. \textbf{d,} the change scores produced by the 1st D-MDL. \textbf{e,} the change scores produced by the 2nd D-MDL.}
\end{figure}

\begin{figure}[H]  
\centering
\begin{tabular}{cc}
			\textbf{a} & \includegraphics[keepaspectratio, height=3.3cm, valign=T]
			{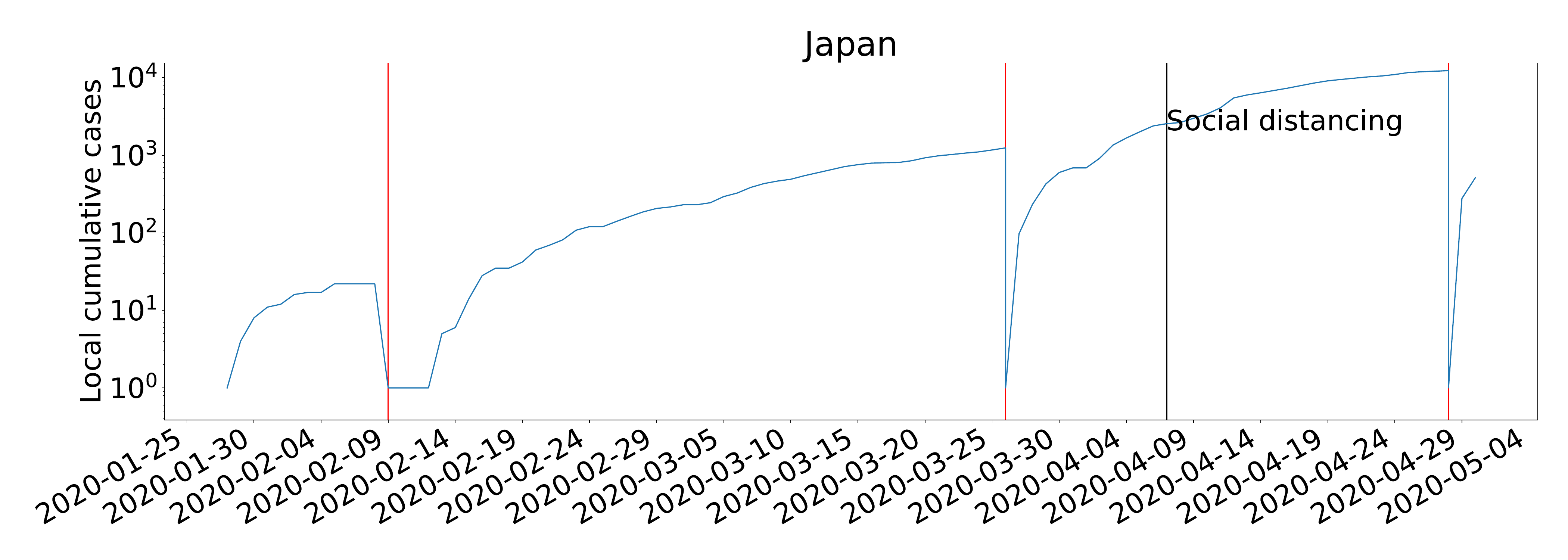} \\
	        \vspace{-0.35cm}
            \textbf{b} & \includegraphics[keepaspectratio, height=3.3cm, valign=T]
			{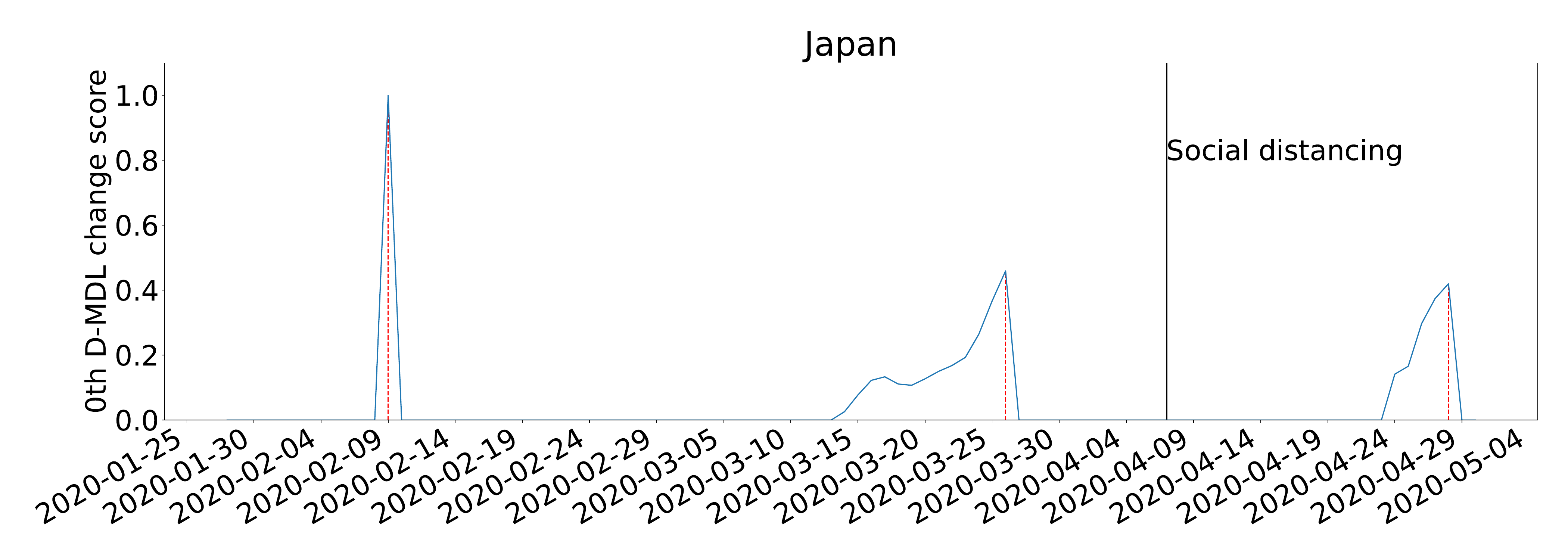}   \\
            \vspace{-0.35cm}
            \textbf{c} & \includegraphics[keepaspectratio, height=3.3cm, valign=T]
			{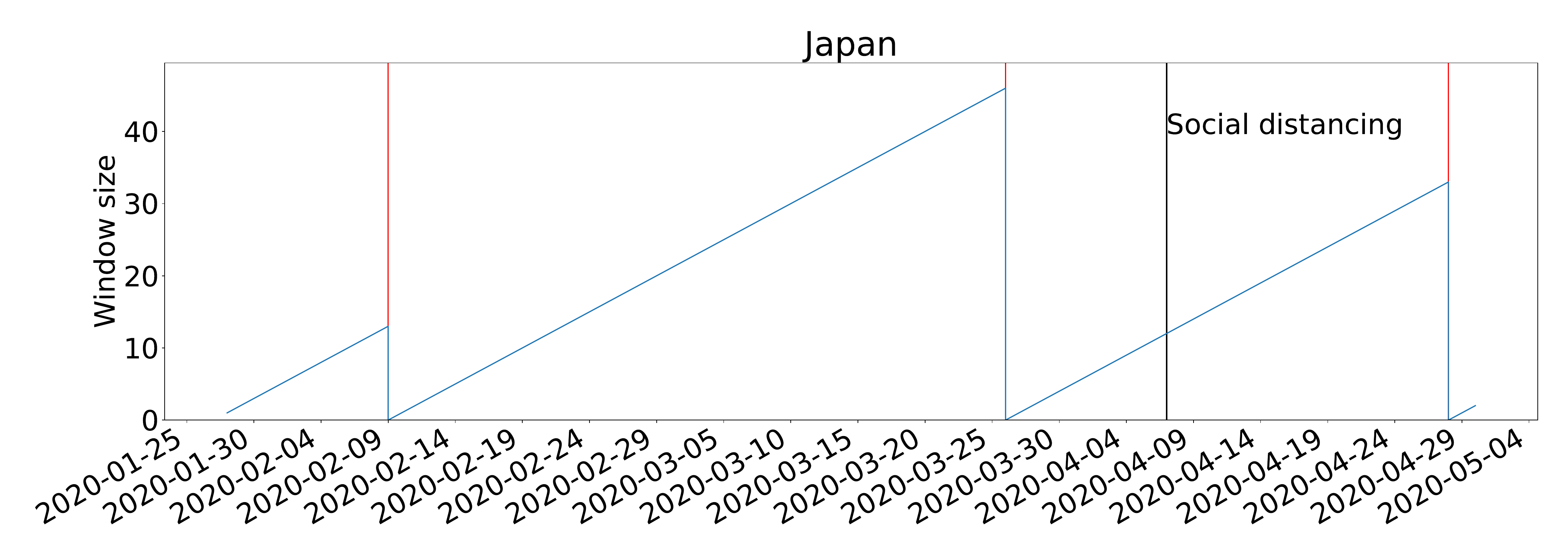} \\
			\vspace{-0.35cm}
			\textbf{d} & \includegraphics[keepaspectratio, height=3.3cm, valign=T]
			{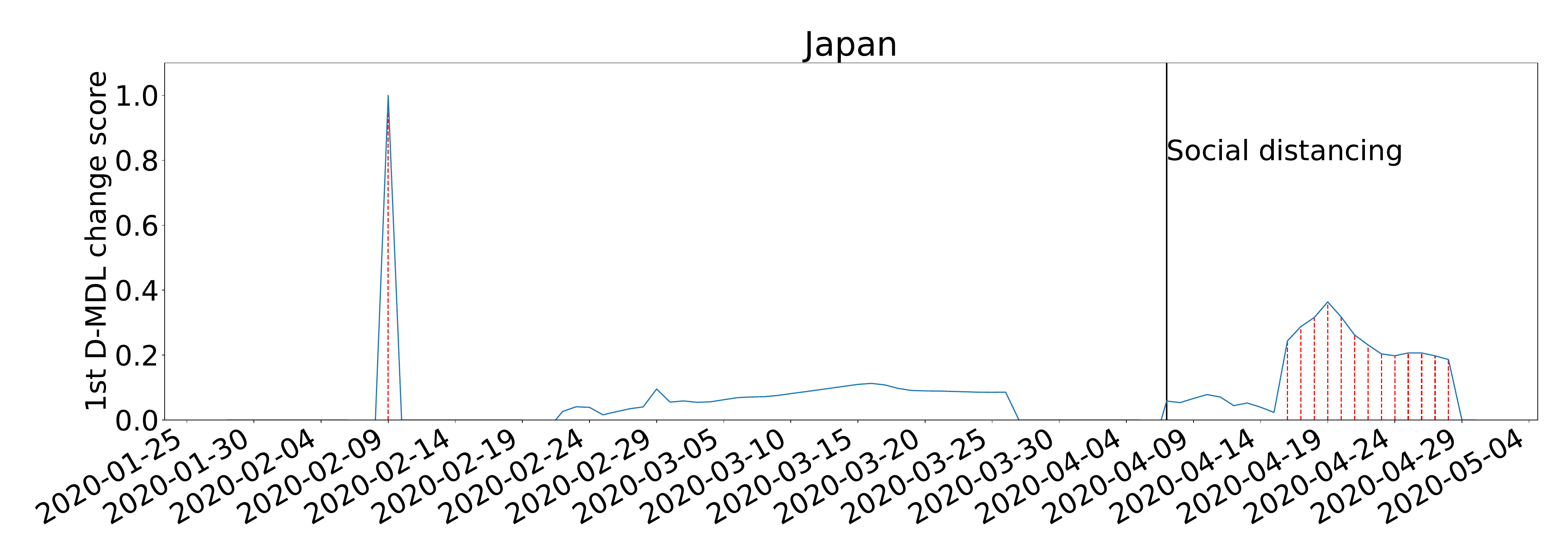} \\
			\vspace{-0.35cm}
			\textbf{e} & \includegraphics[keepaspectratio, height=3.3cm, valign=T]
			{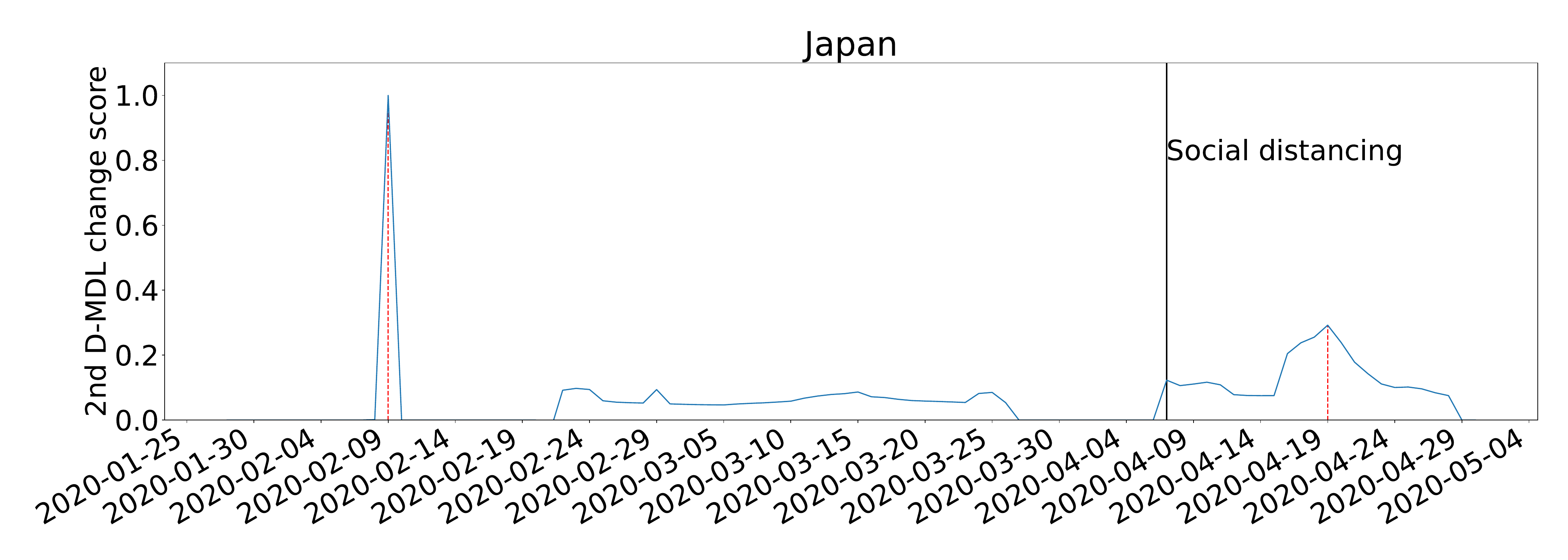} \\
		\end{tabular}
			\caption{\textbf{The results for Japan with exponential modeling.} The date on which the social distancing was implemented is marked by a solid line in black. \textbf{a,} the number of cumulative cases. \textbf{b,} the change scores produced by the 0th M-DML where the line in blue denotes values of scores and dashed lines in red mark alarms. \textbf{c,} the window sized for the sequential D-DML algorithm with adaptive window where lines in red mark the shrinkage of windows. \textbf{d,} the change scores produced by the 1st D-MDL. \textbf{e,} the change scores produced by the 2nd D-MDL.}
\end{figure}

\begin{figure}[H] 
\centering
\begin{tabular}{cc}
		 	\textbf{a} & \includegraphics[keepaspectratio, height=3.3cm, valign=T]
			{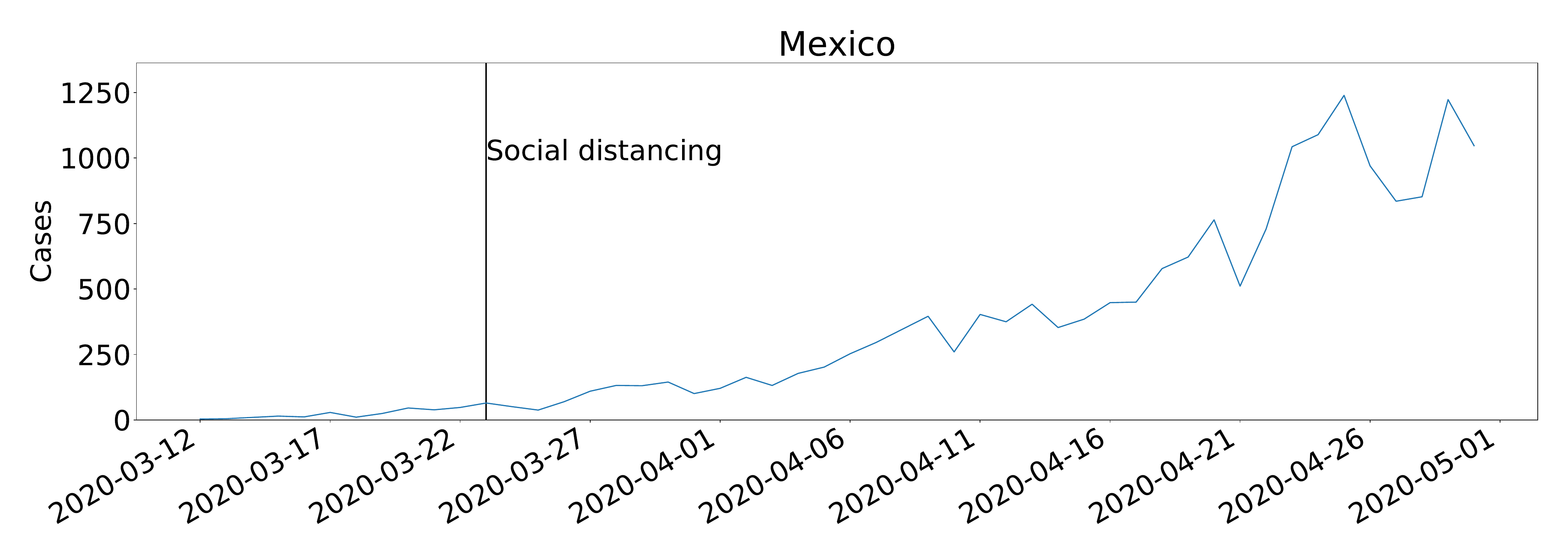} \\
			\vspace{-0.35cm}
	 	    \textbf{b} & \includegraphics[keepaspectratio, height=3.3cm, valign=T]
			{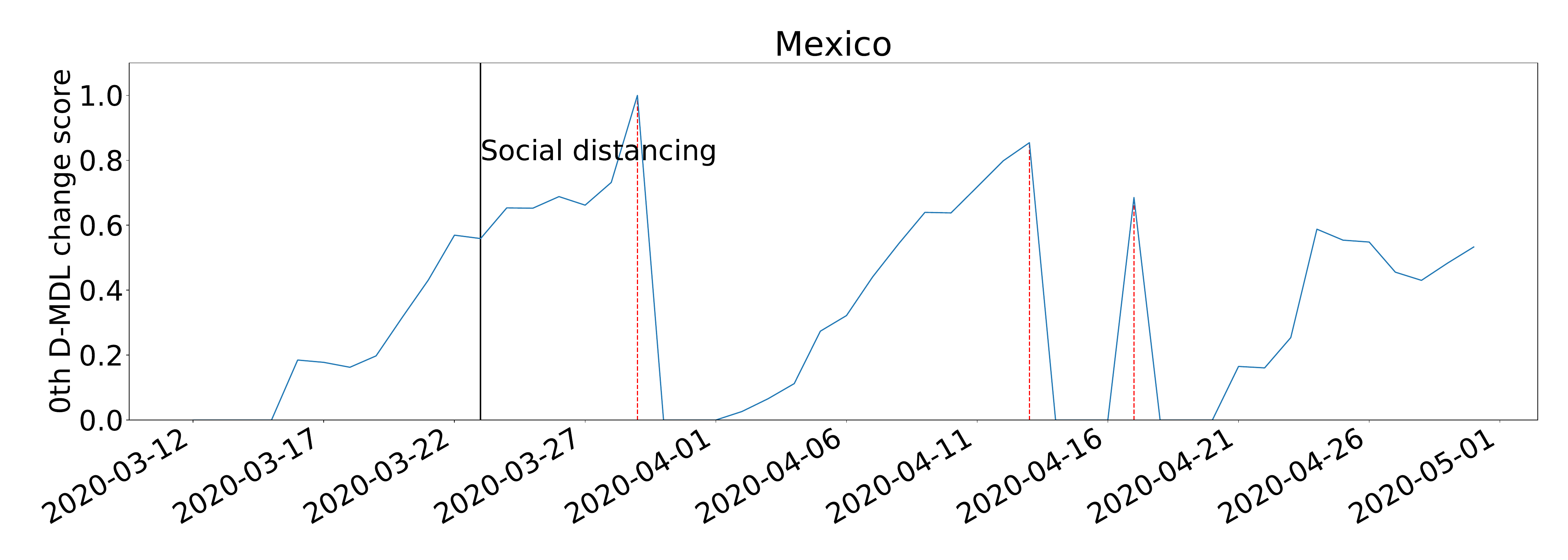}   \\
	        \vspace{-0.35cm}
			\textbf{c} & \includegraphics[keepaspectratio, height=3.3cm, valign=T]
			{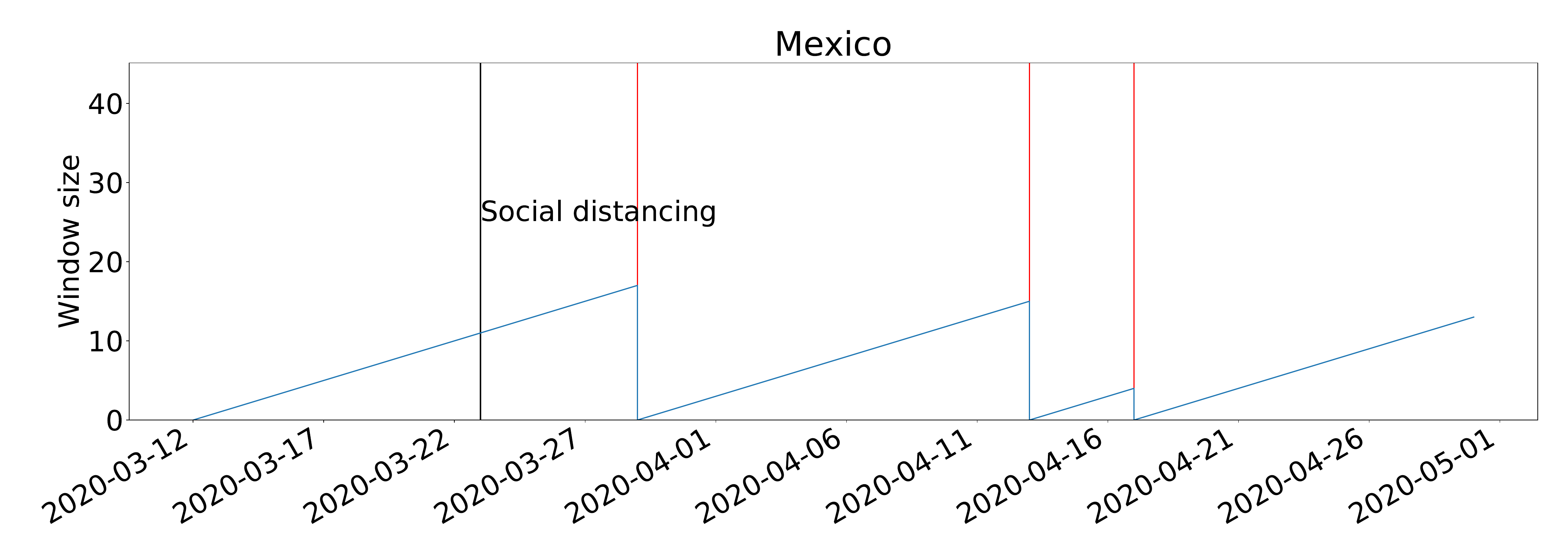} \\
		    \vspace{-0.35cm}
			\textbf{d} & \includegraphics[keepaspectratio, height=3.3cm, valign=T]
			{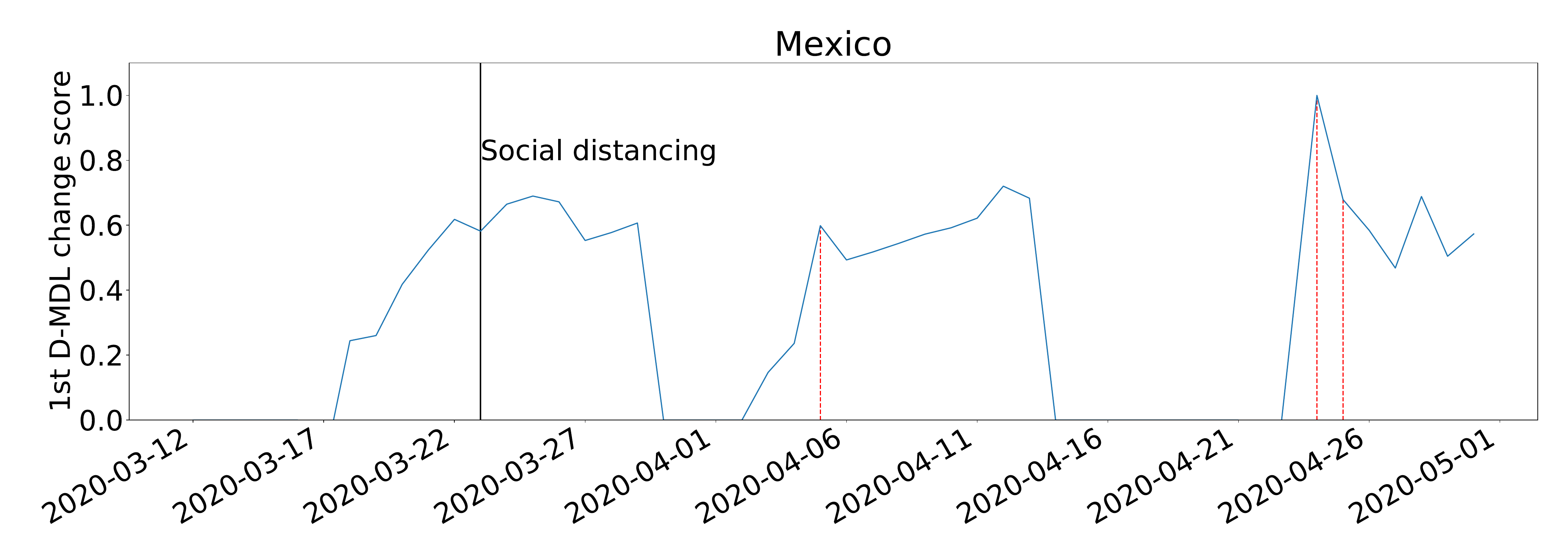} \\
		    \vspace{-0.35cm}
			\textbf{e} & \includegraphics[keepaspectratio, height=3.3cm, valign=T]
			{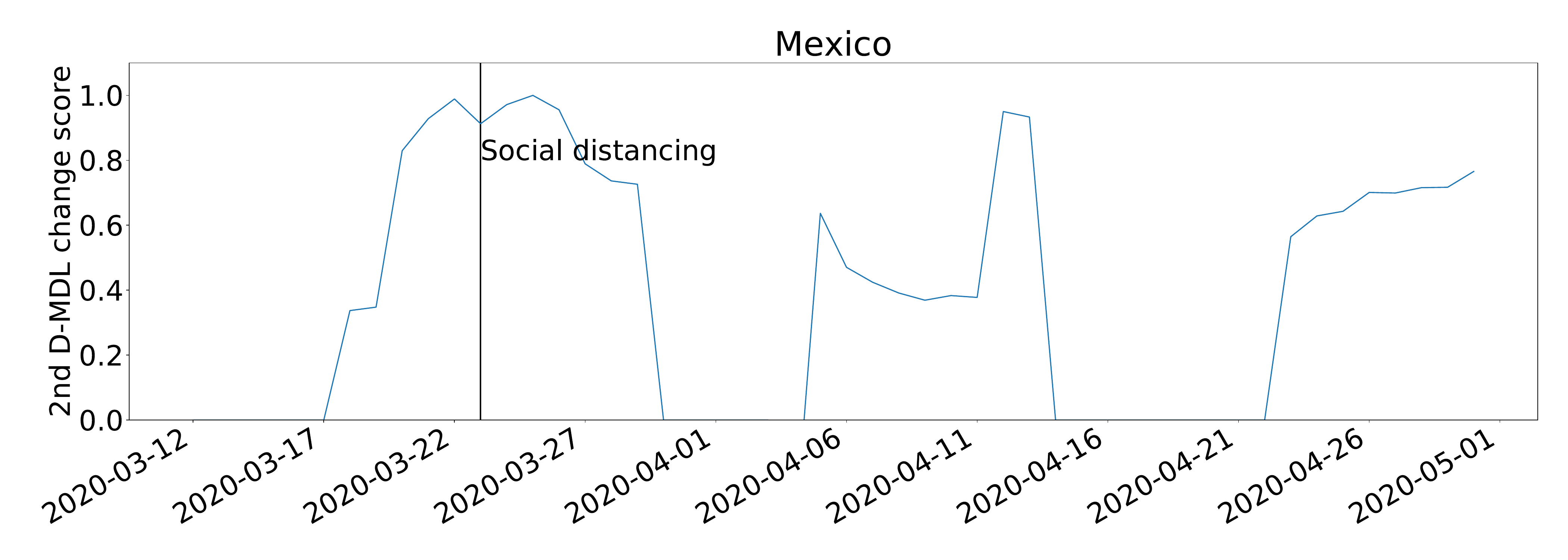} \\
		\end{tabular}
			\caption{\textbf{The results for Mexico with Gaussian modeling.} The date on which the social distancing was implemented is marked by a solid line in black. \textbf{a,} the number of daily new cases. \textbf{b,} the change scores produced by the 0th M-DML where the line in blue denotes values of scores and dashed lines in red mark alarms. \textbf{c,} the window sized for the sequential D-DML algorithm with adaptive window where lines in red mark the shrinkage of windows. \textbf{d,} the change scores produced by the 1st D-MDL. \textbf{e,} the change scores produced by the 2nd D-MDL.}
\end{figure}

\begin{figure}[H]  
\centering
\begin{tabular}{cc}
			\textbf{a} & \includegraphics[keepaspectratio, height=3.3cm, valign=T]
			{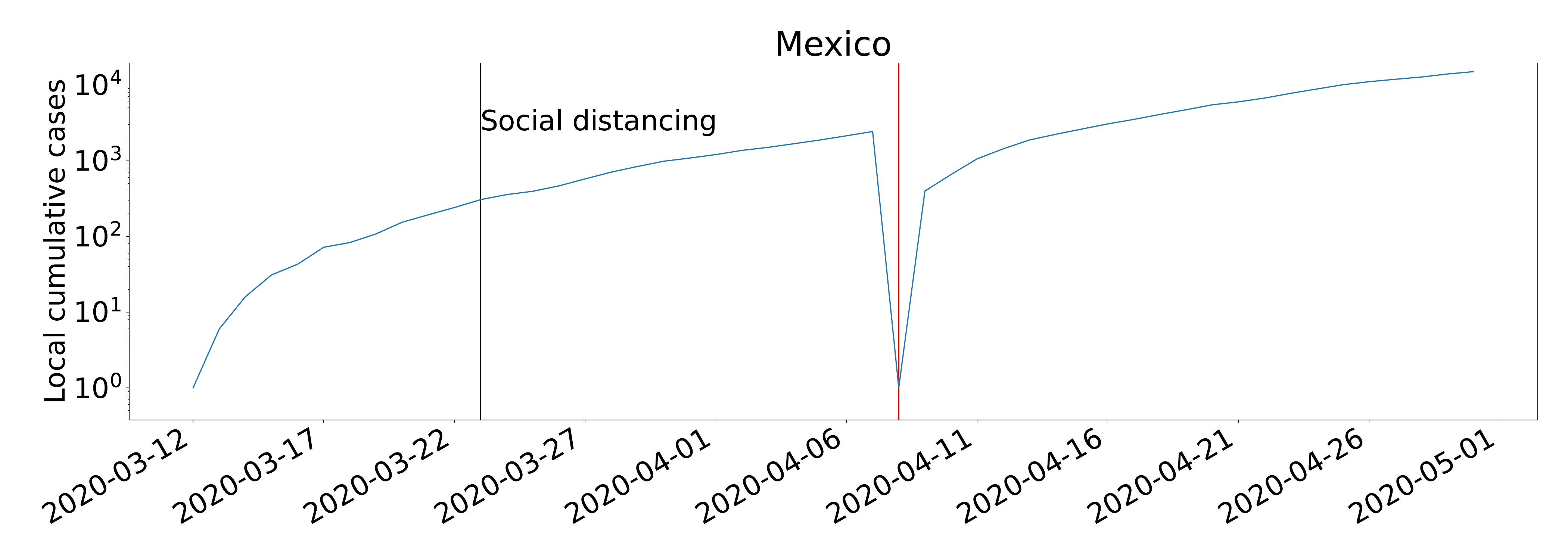} \\
	        \vspace{-0.35cm}
            \textbf{b} & \includegraphics[keepaspectratio, height=3.3cm, valign=T]
			{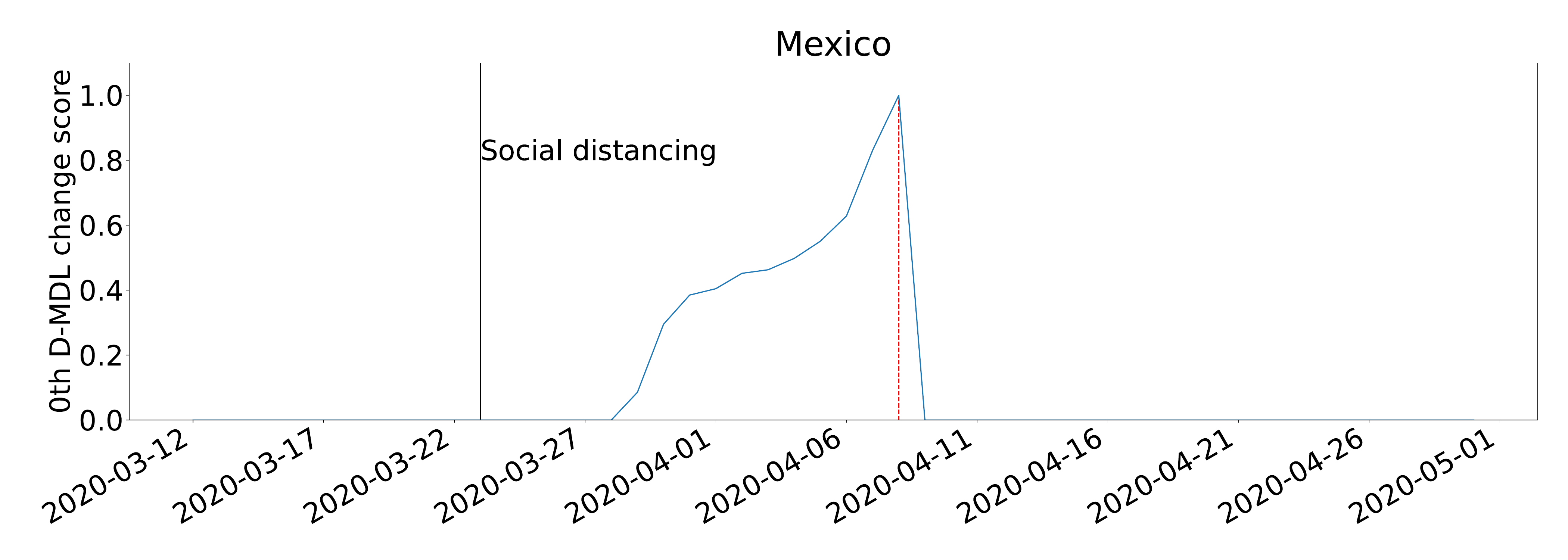}   \\
            \vspace{-0.35cm}
            \textbf{c} & \includegraphics[keepaspectratio, height=3.3cm, valign=T]
			{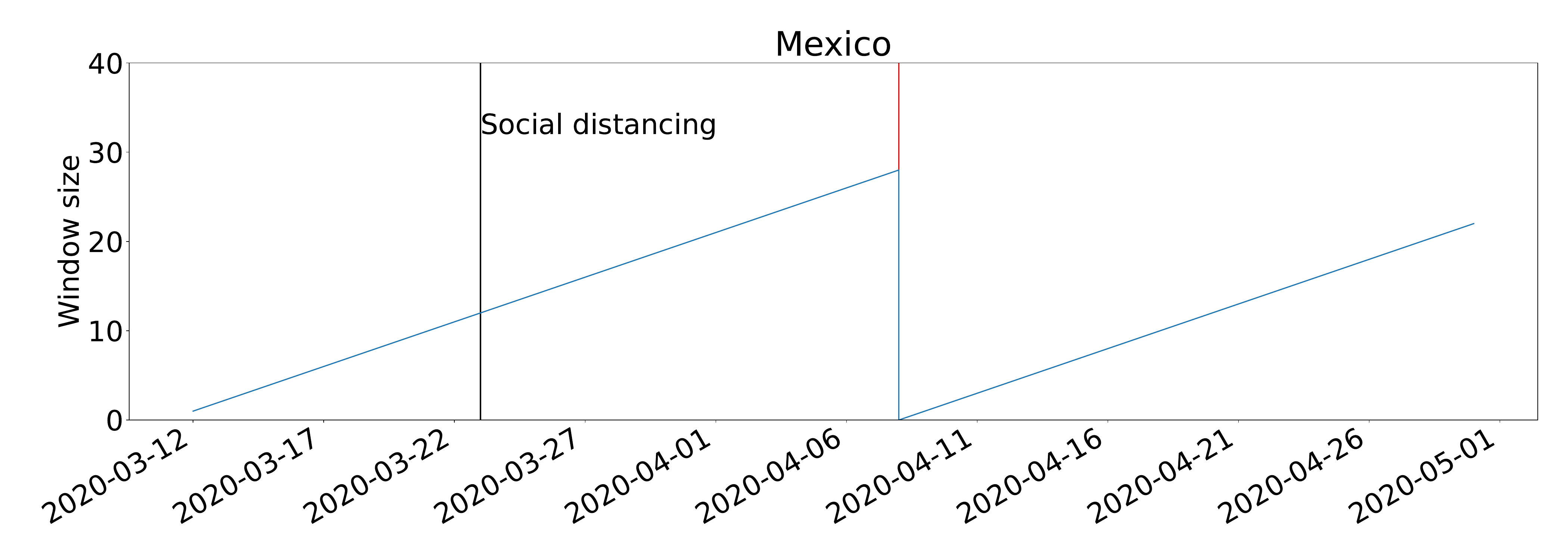} \\
			\vspace{-0.35cm}
			\textbf{d} & \includegraphics[keepaspectratio, height=3.3cm, valign=T]
			{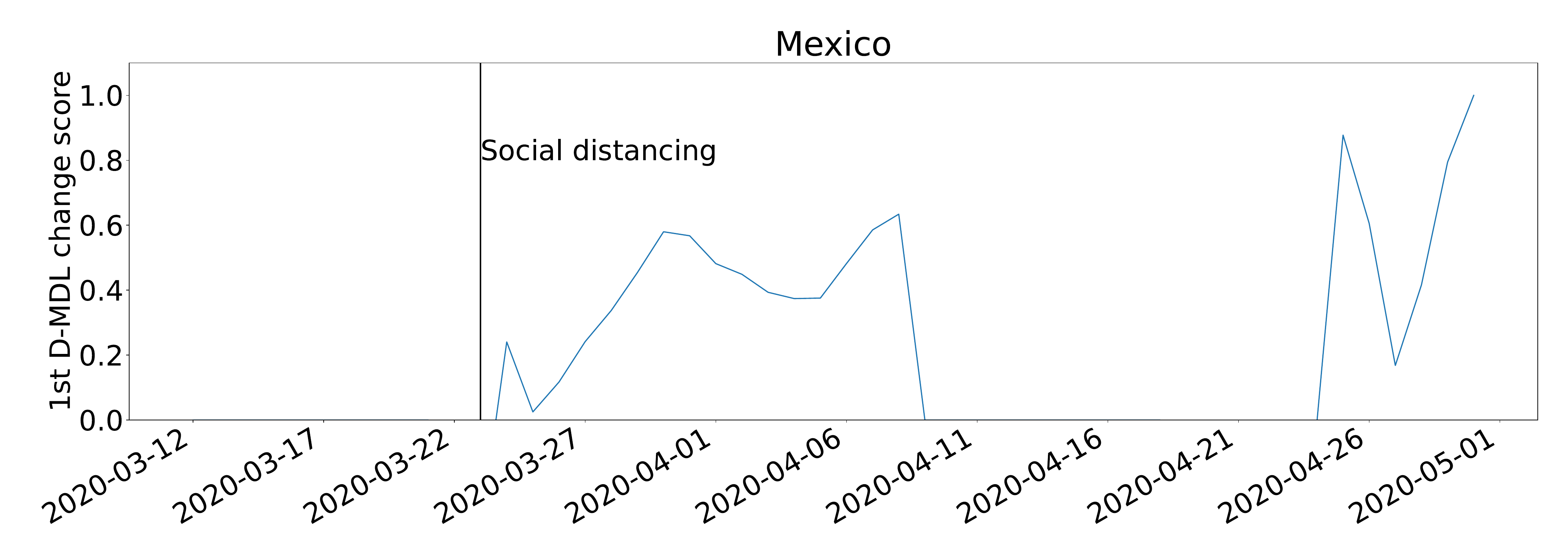} \\
			\vspace{-0.35cm}
			\textbf{e} & \includegraphics[keepaspectratio, height=3.3cm, valign=T]
			{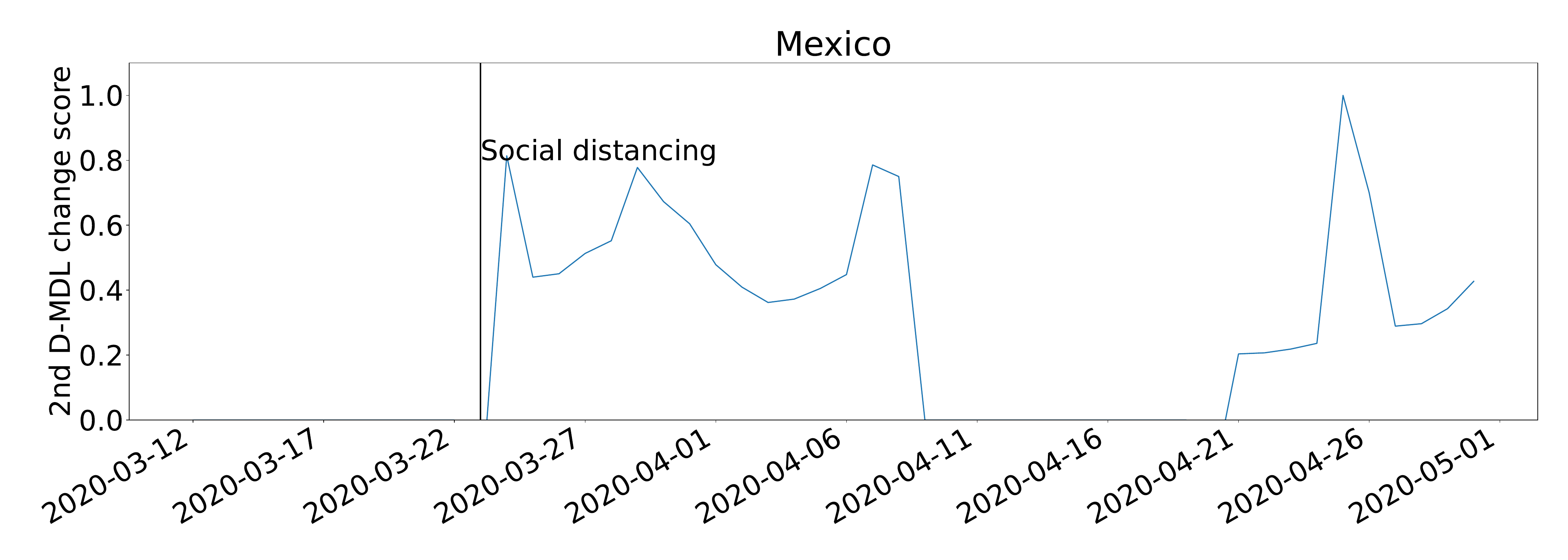} \\
		\end{tabular}
			\caption{\textbf{The results for Mexico with exponential modeling.} The date on which the social distancing was implemented is marked by a solid line in black. \textbf{a,} the number of cumulative cases. \textbf{b,} the change scores produced by the 0th M-DML where the line in blue denotes values of scores and dashed lines in red mark alarms. \textbf{c,} the window sized for the sequential D-DML algorithm with adaptive window where lines in red mark the shrinkage of windows. \textbf{d,} the change scores produced by the 1st D-MDL. \textbf{e,} the change scores produced by the 2nd D-MDL.}
\end{figure}

\begin{figure}[H] 
\centering
\begin{tabular}{cc}
		 	\textbf{a} & \includegraphics[keepaspectratio, height=3.3cm, valign=T]
			{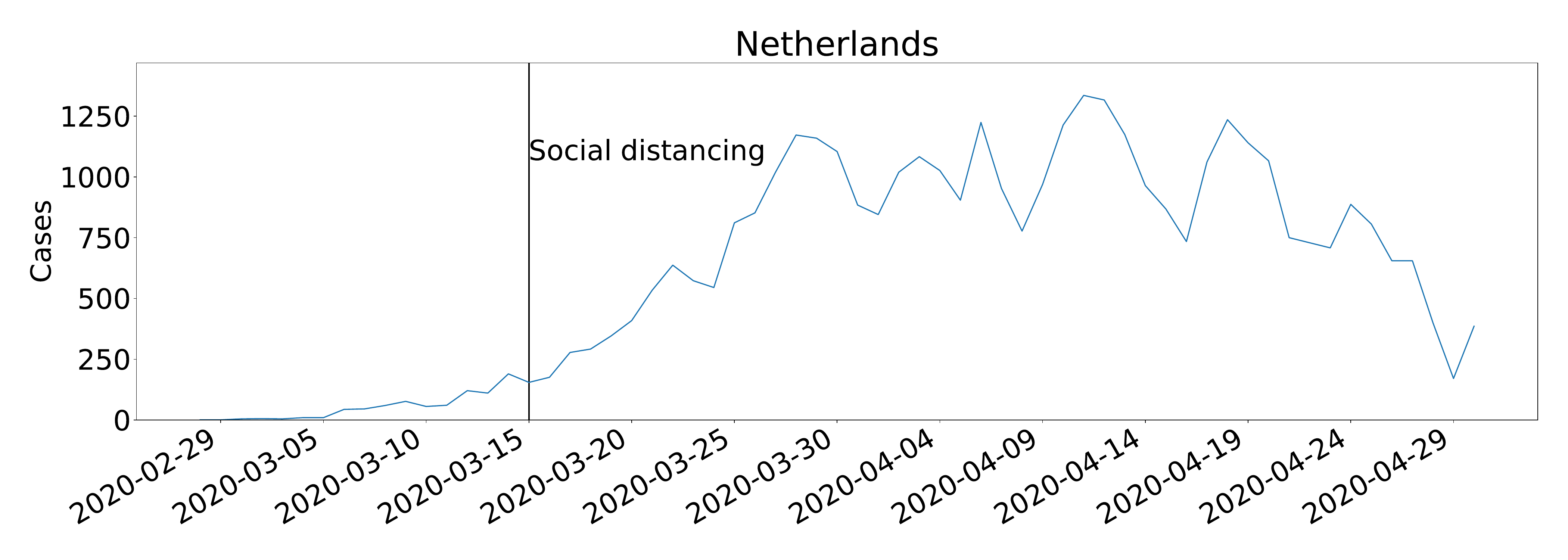} \\
			\vspace{-0.35cm}
	 	    \textbf{b} & \includegraphics[keepaspectratio, height=3.3cm, valign=T]
			{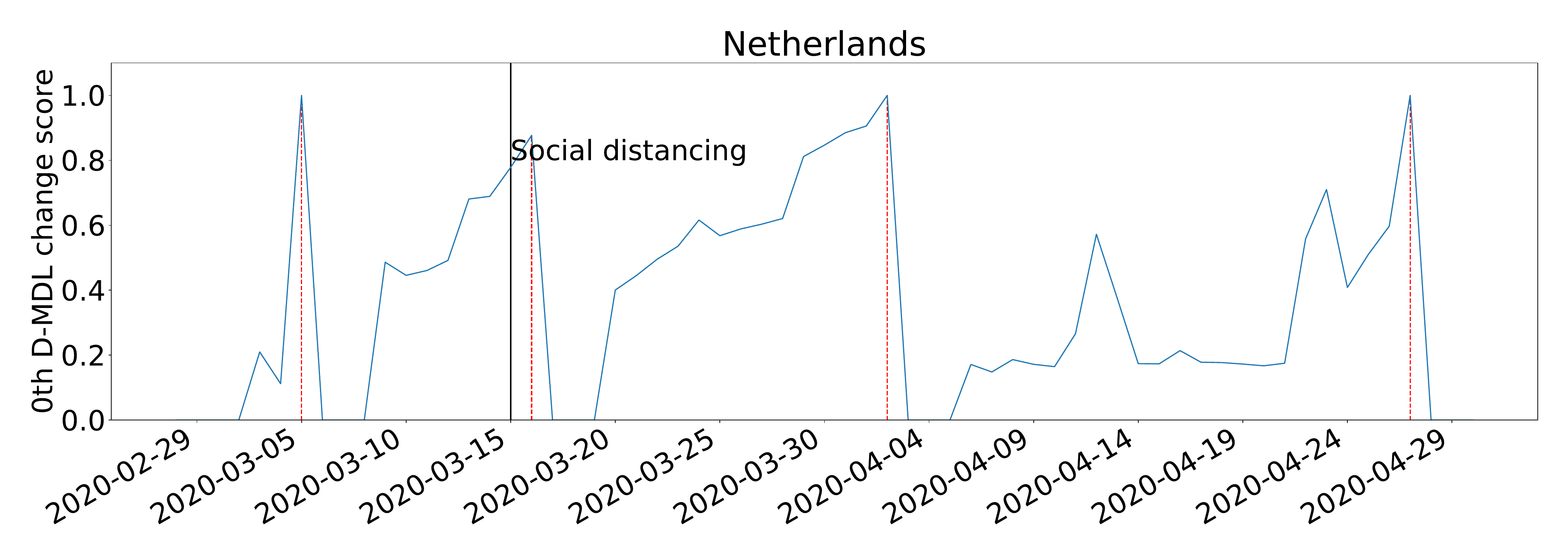}   \\
	        \vspace{-0.35cm}
			\textbf{c} & \includegraphics[keepaspectratio, height=3.3cm, valign=T]
			{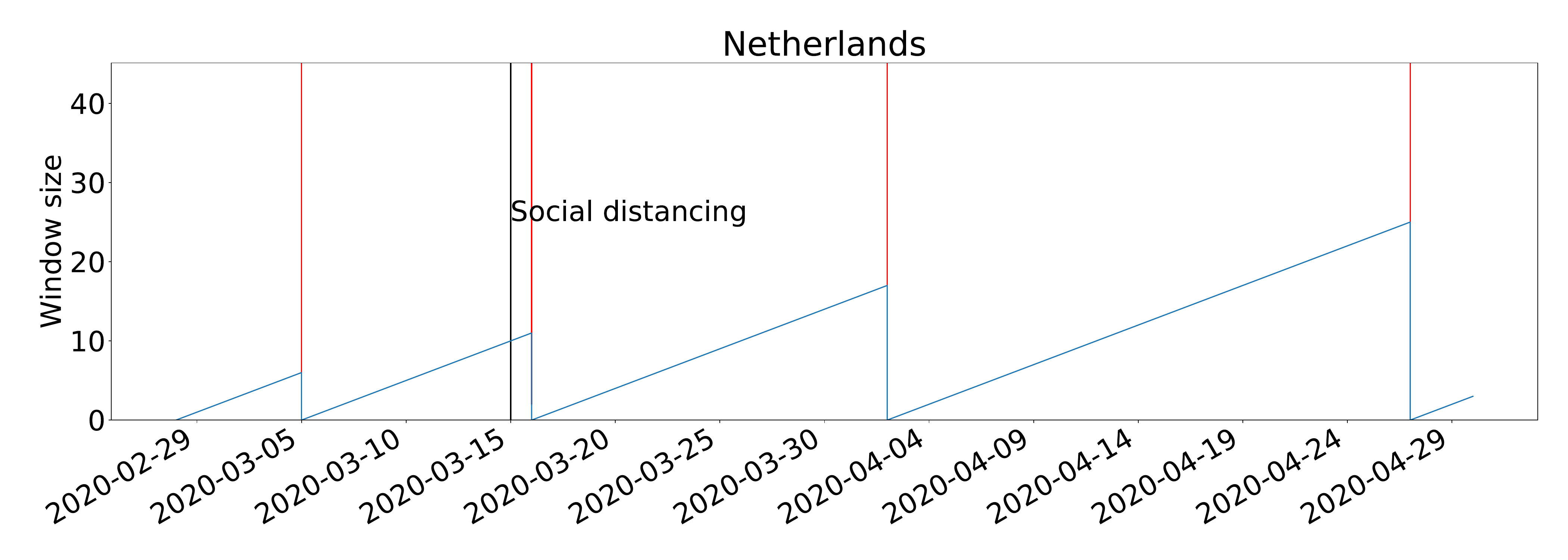} \\
		    \vspace{-0.35cm}
			\textbf{d} & \includegraphics[keepaspectratio, height=3.3cm, valign=T]
			{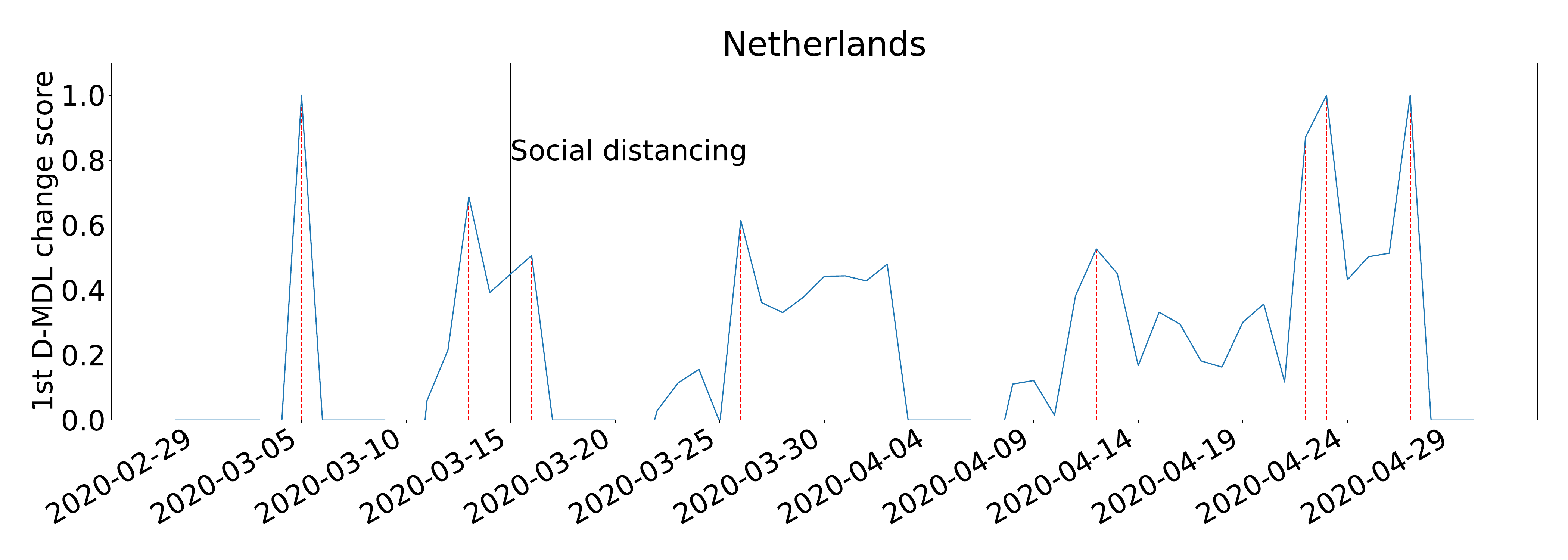} \\
		    \vspace{-0.35cm}
			\textbf{e} & \includegraphics[keepaspectratio, height=3.3cm, valign=T]
			{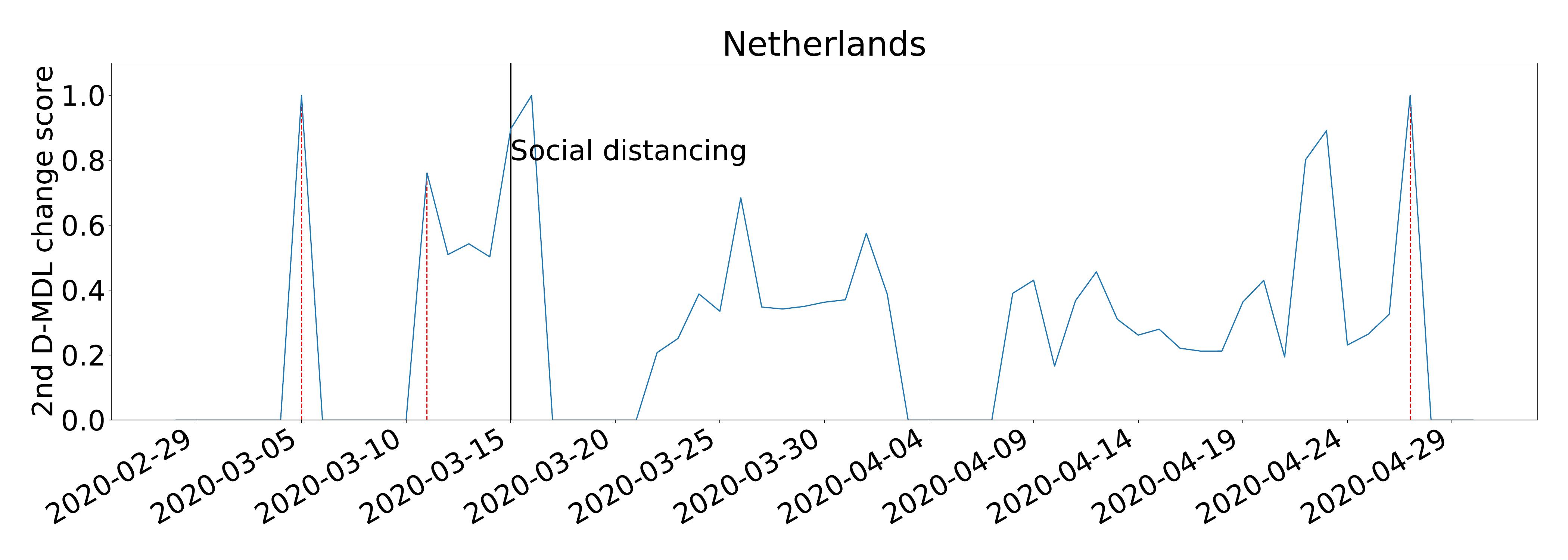} \\
		\end{tabular}
			\caption{\textbf{The results for Netherlands with Gaussian modeling.} The date on which the social distancing was implemented is marked by a solid line in black. \textbf{a,} the number of daily new cases. \textbf{b,} the change scores produced by the 0th M-DML where the line in blue denotes values of scores and dashed lines in red mark alarms. \textbf{c,} the window sized for the sequential D-DML algorithm with adaptive window where lines in red mark the shrinkage of windows. \textbf{d,} the change scores produced by the 1st D-MDL. \textbf{e,} the change scores produced by the 2nd D-MDL.}
\end{figure}

\begin{figure}[H]  
\centering
\begin{tabular}{cc}
			\textbf{a} & \includegraphics[keepaspectratio, height=3.3cm, valign=T]
			{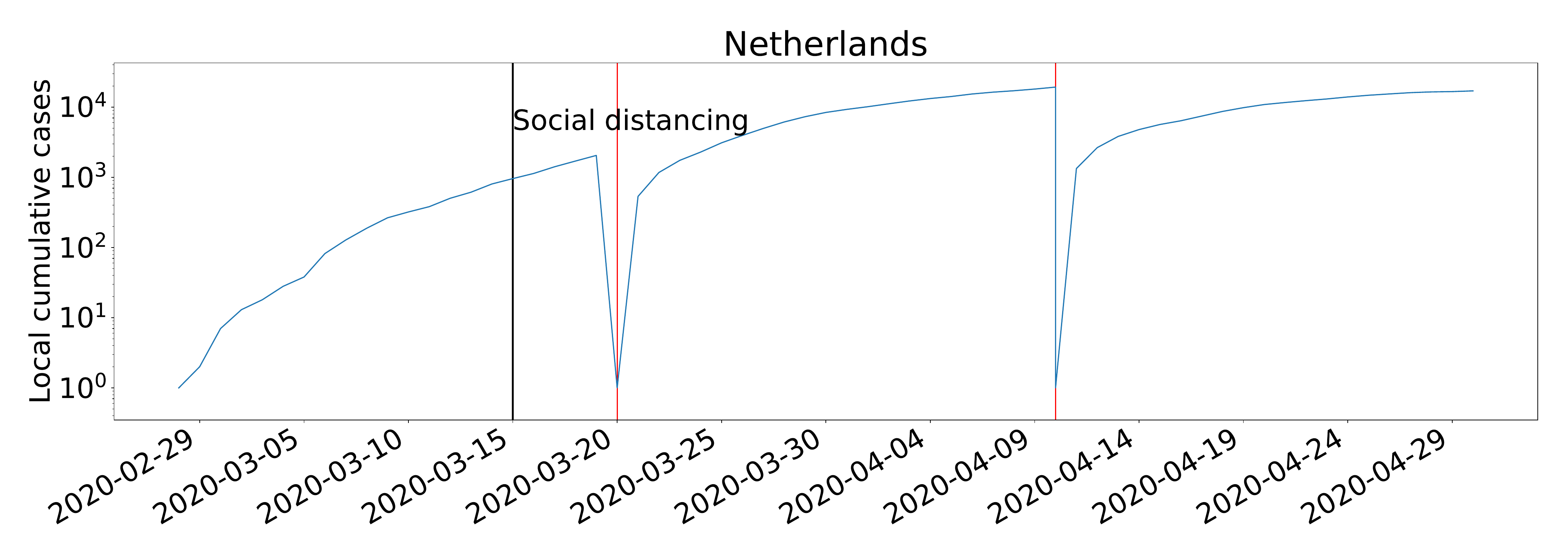} \\
	        \vspace{-0.35cm}
            \textbf{b} & \includegraphics[keepaspectratio, height=3.3cm, valign=T]
			{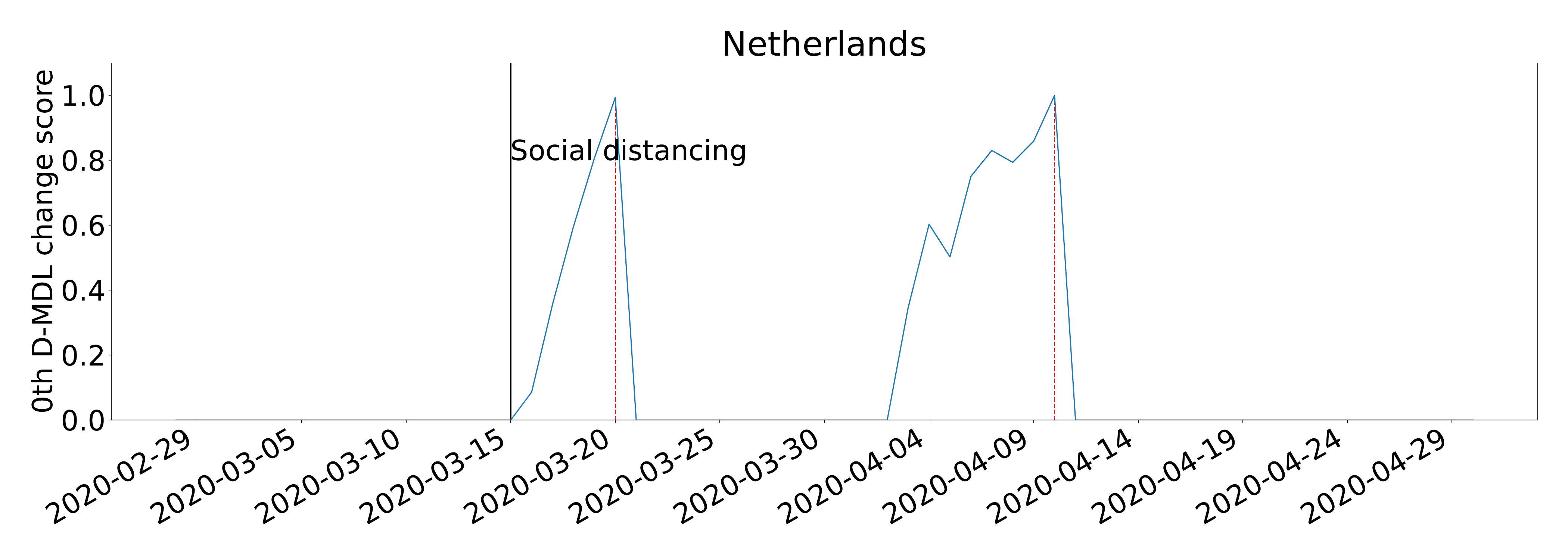}   \\
            \vspace{-0.35cm}
            \textbf{c} & \includegraphics[keepaspectratio, height=3.3cm, valign=T]
			{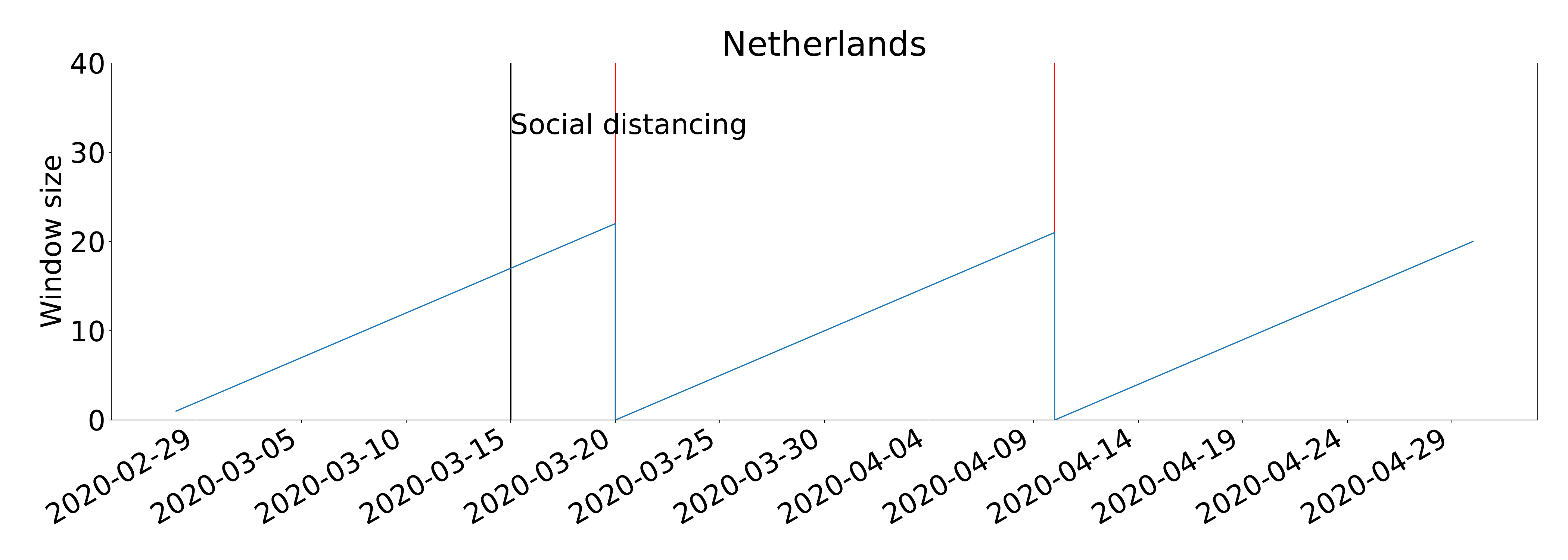} \\
			\vspace{-0.35cm}
			\textbf{d} & \includegraphics[keepaspectratio, height=3.3cm, valign=T]
			{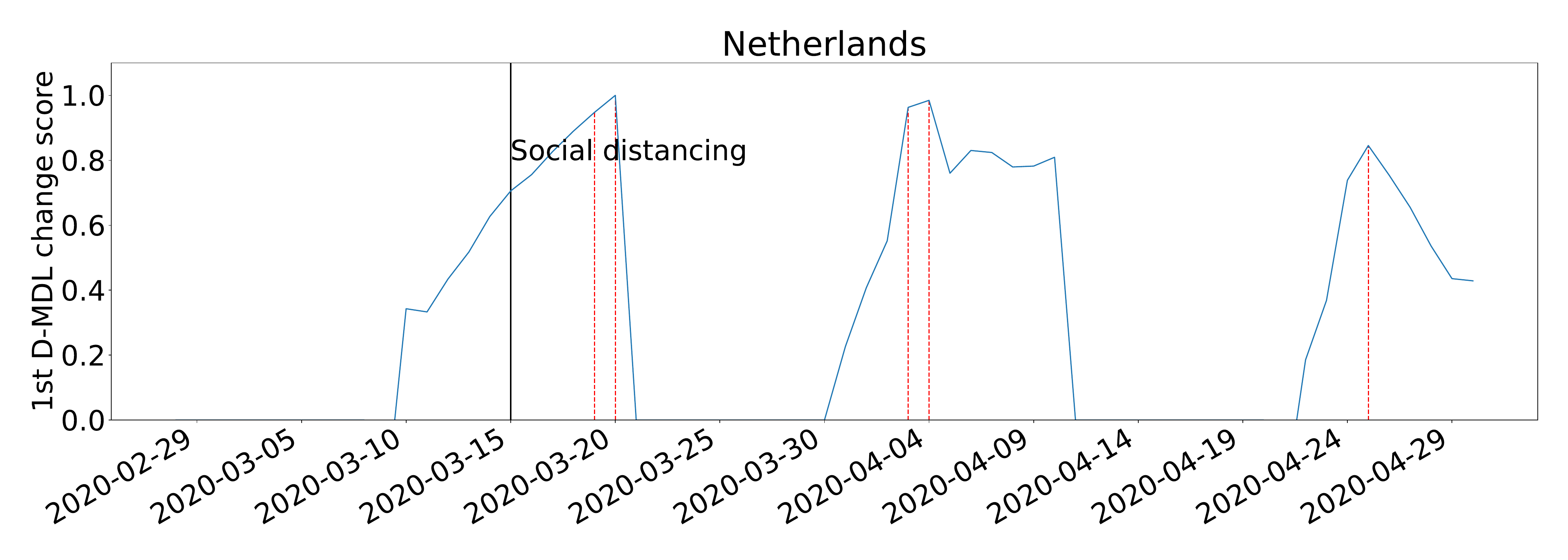} \\
			\vspace{-0.35cm}
			\textbf{e} & \includegraphics[keepaspectratio, height=3.3cm, valign=T]
			{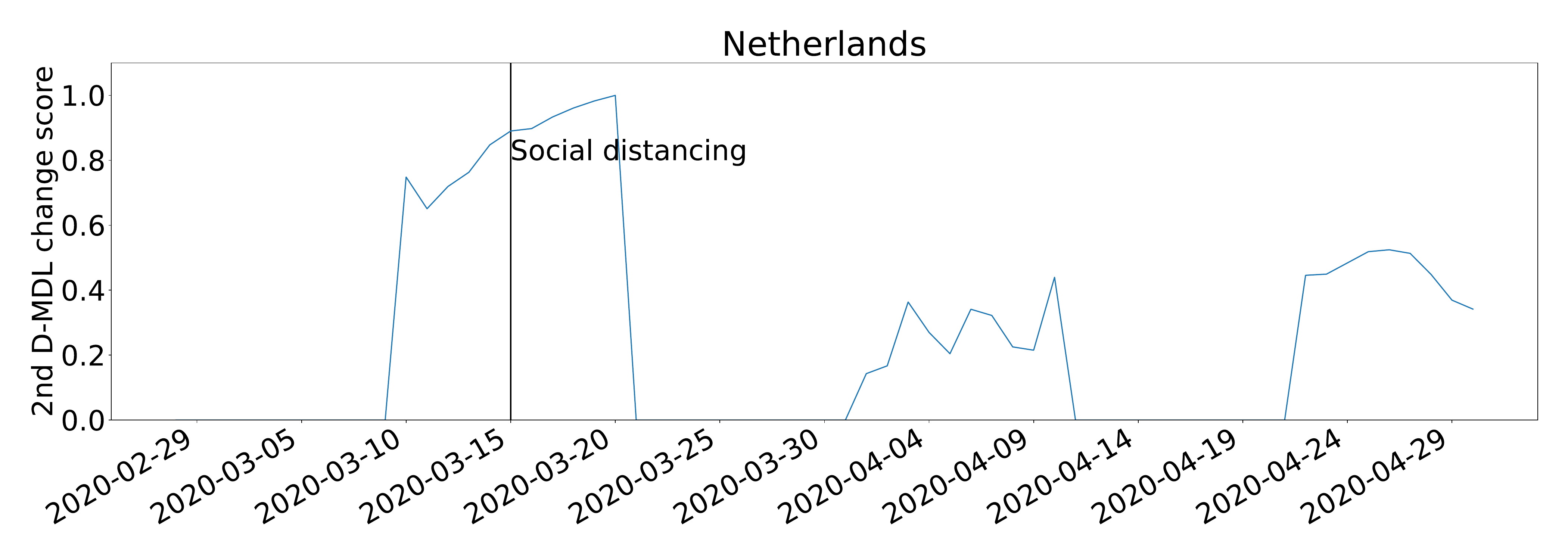} \\
		\end{tabular}
			\caption{\textbf{The results for Netherlands with exponential modeling.} The date on which the social distancing was implemented is marked by a solid line in black. \textbf{a,} the number of cumulative cases. \textbf{b,} the change scores produced by the 0th M-DML where the line in blue denotes values of scores and dashed lines in red mark alarms. \textbf{c,} the window sized for the sequential D-DML algorithm with adaptive window where lines in red mark the shrinkage of windows. \textbf{d,} the change scores produced by the 1st D-MDL. \textbf{e,} the change scores produced by the 2nd D-MDL.}
\end{figure}

\begin{figure}[H] 
\centering
\begin{tabular}{cc}
		 	\textbf{a} & \includegraphics[keepaspectratio, height=3.3cm, valign=T]
			{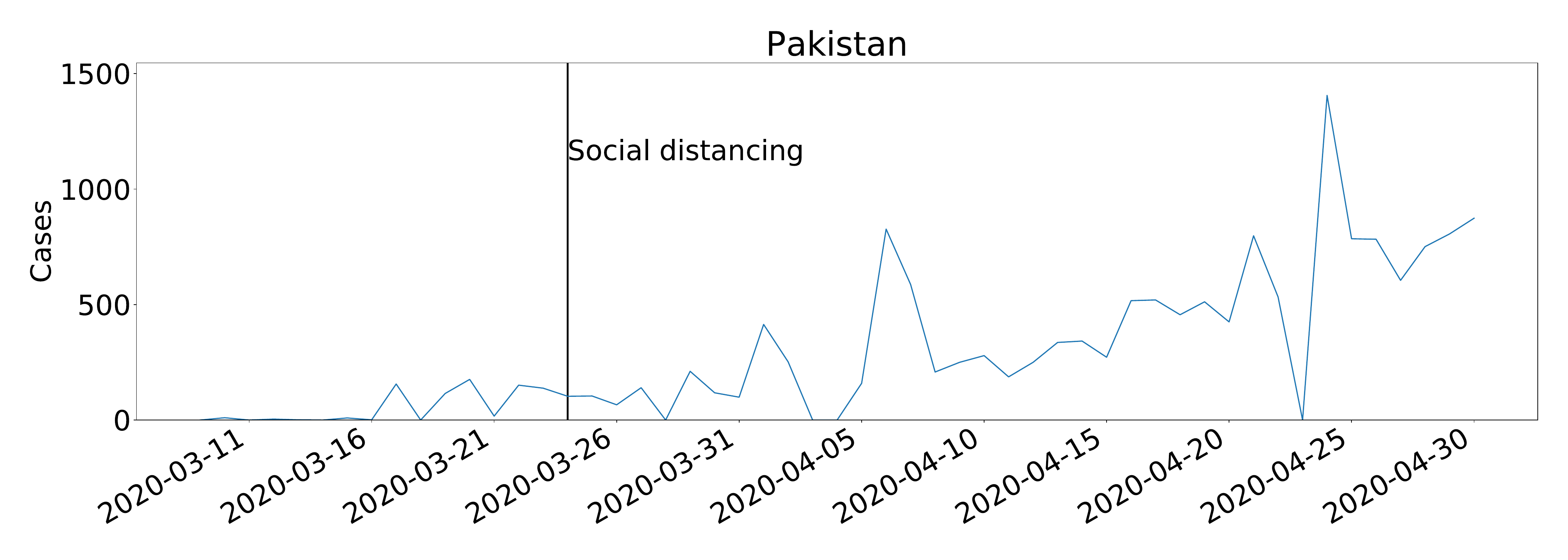} \\
			\vspace{-0.35cm}
	 	    \textbf{b} & \includegraphics[keepaspectratio, height=3.3cm, valign=T]
			{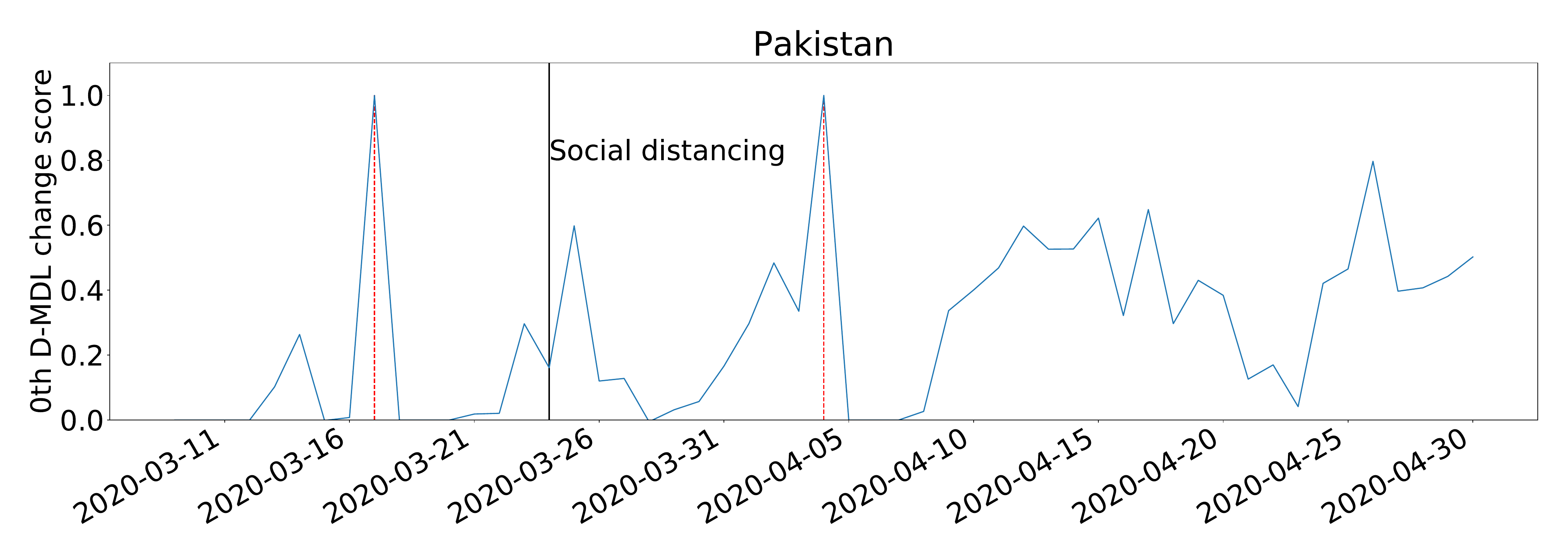}   \\
	        \vspace{-0.35cm}
			\textbf{c} & \includegraphics[keepaspectratio, height=3.3cm, valign=T]
			{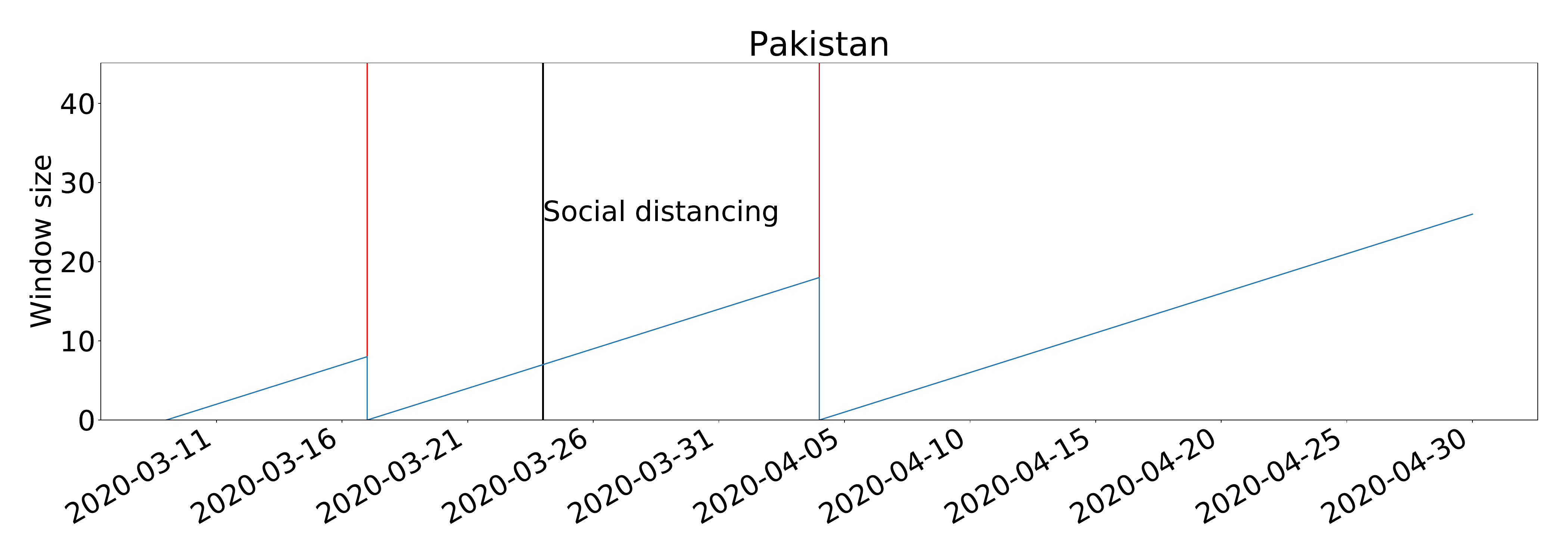} \\
		    \vspace{-0.35cm}
			\textbf{d} & \includegraphics[keepaspectratio, height=3.3cm, valign=T]
			{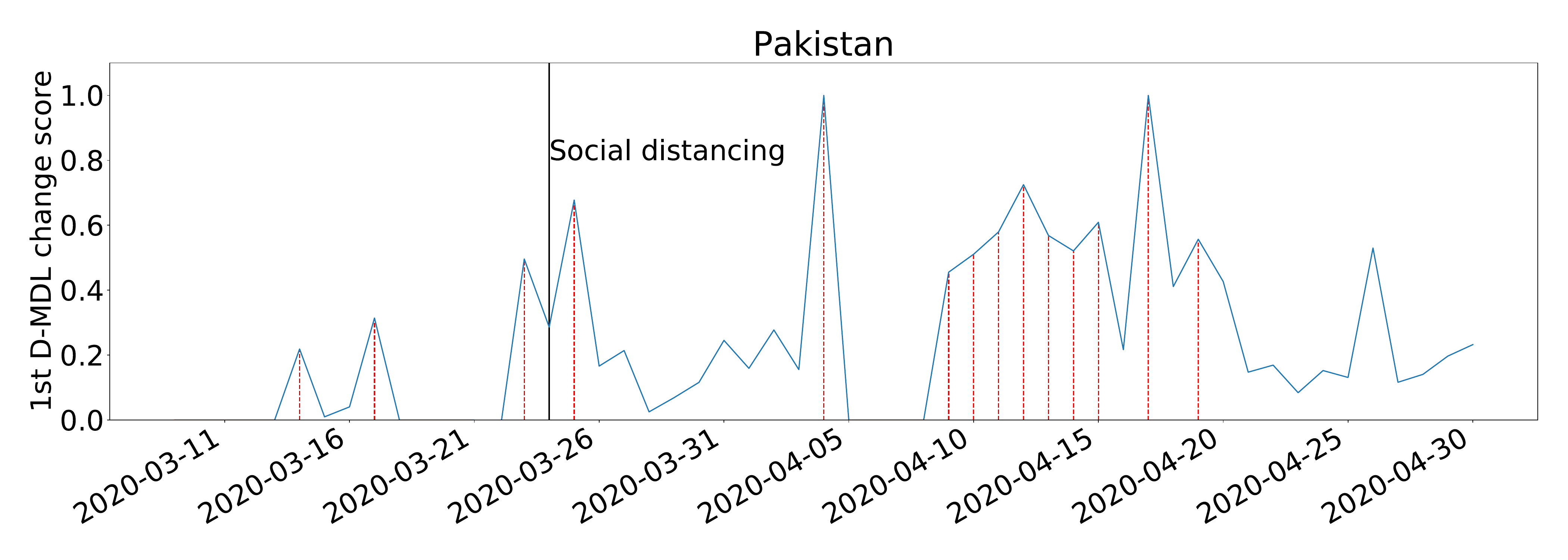} \\
		    \vspace{-0.35cm}
			\textbf{e} & \includegraphics[keepaspectratio, height=3.3cm, valign=T]
			{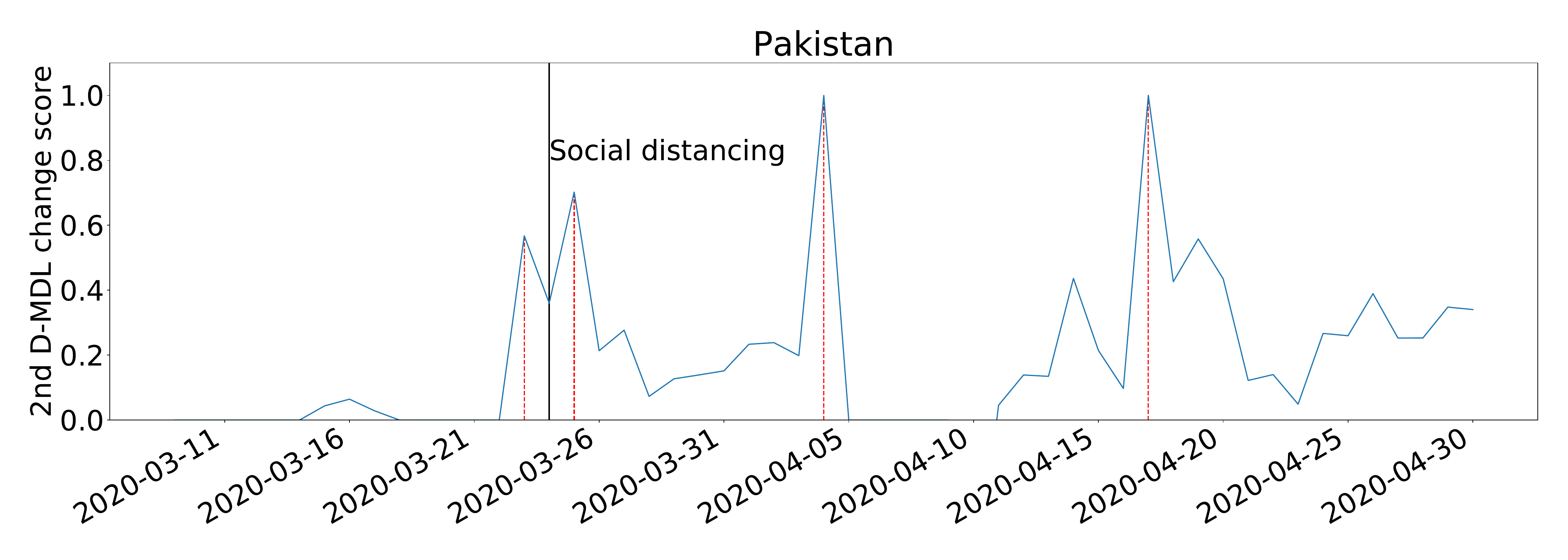} \\
		\end{tabular}
			\caption{\textbf{The results for Pakistan with Gaussian modeling.} The date on which the social distancing was implemented is marked by a solid line in black. \textbf{a,} the number of daily new cases. \textbf{b,} the change scores produced by the 0th M-DML where the line in blue denotes values of scores and dashed lines in red mark alarms. \textbf{c,} the window sized for the sequential D-DML algorithm with adaptive window where lines in red mark the shrinkage of windows. \textbf{d,} the change scores produced by the 1st D-MDL. \textbf{e,} the change scores produced by the 2nd D-MDL.}
\end{figure}

\begin{figure}[H]  
\centering
\begin{tabular}{cc}
			\textbf{a} & \includegraphics[keepaspectratio, height=3.3cm, valign=T]
			{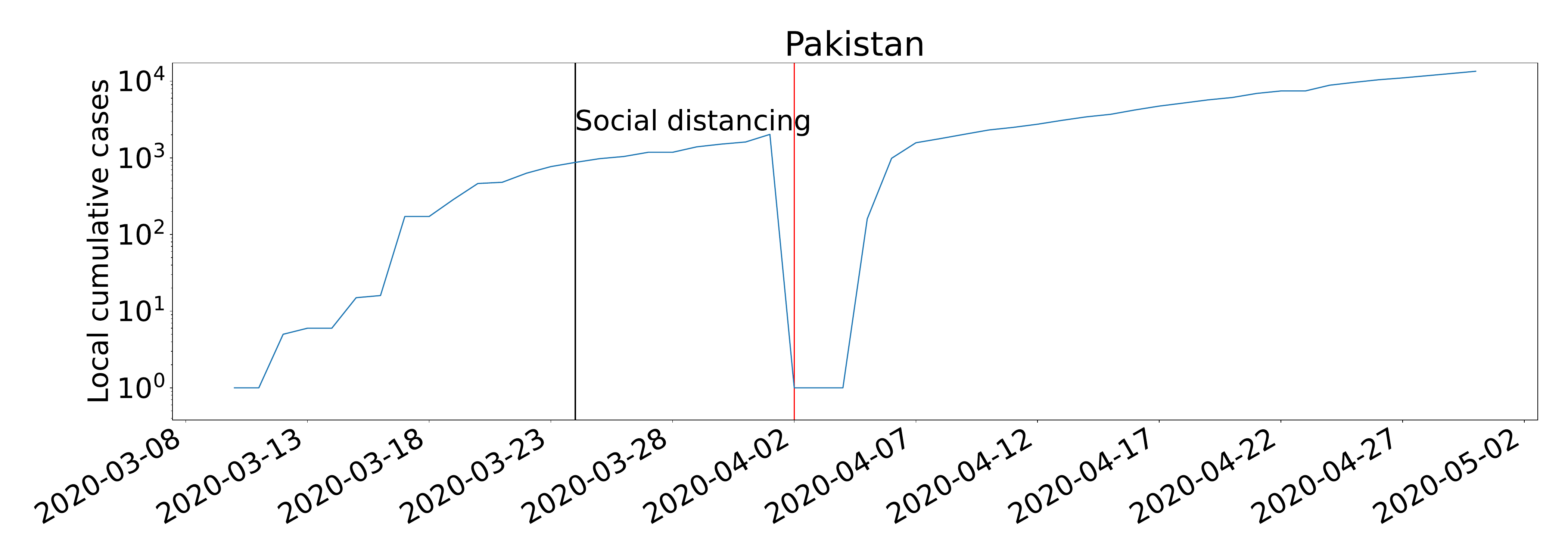} \\
	        \vspace{-0.35cm}
            \textbf{b} & \includegraphics[keepaspectratio, height=3.3cm, valign=T]
			{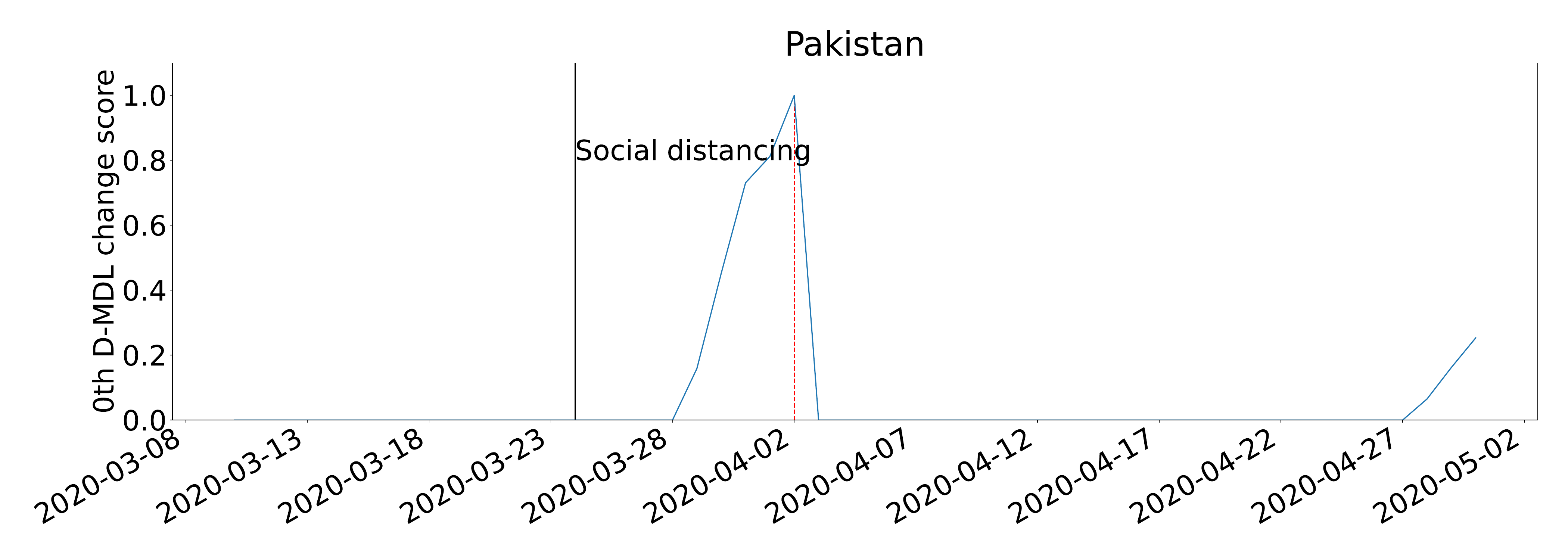}   \\
            \vspace{-0.35cm}
            \textbf{c} & \includegraphics[keepaspectratio, height=3.3cm, valign=T]
			{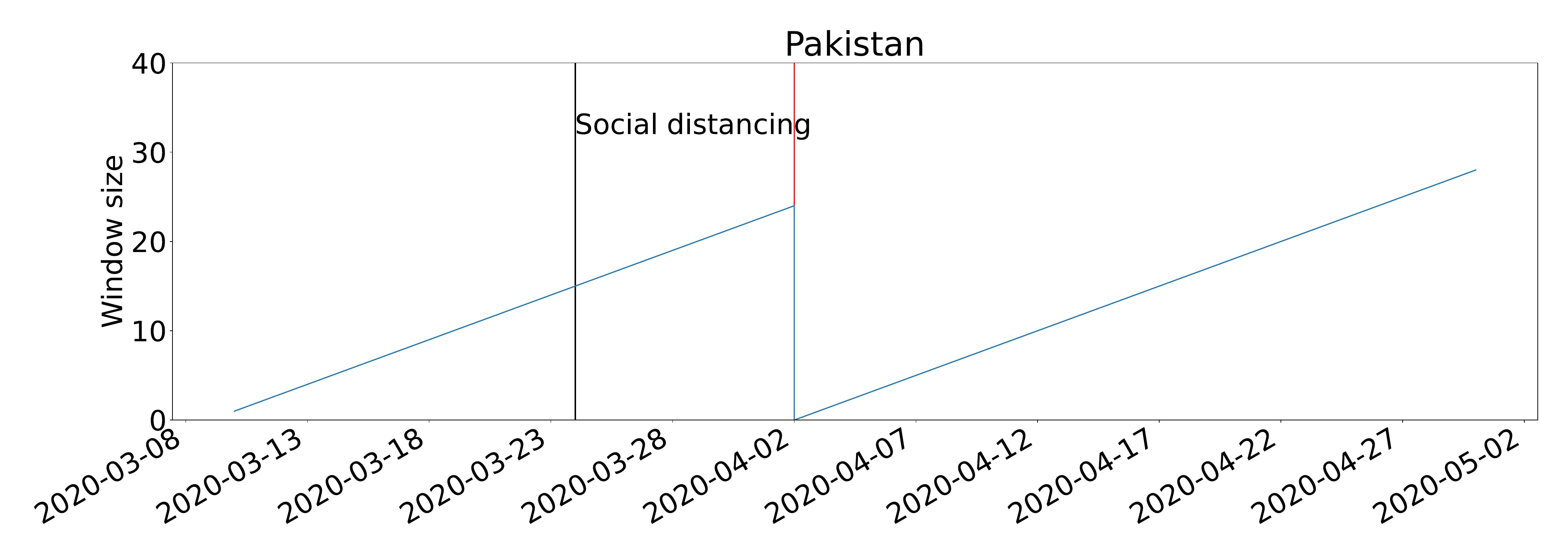} \\
			\vspace{-0.35cm}
			\textbf{d} & \includegraphics[keepaspectratio, height=3.3cm, valign=T]
			{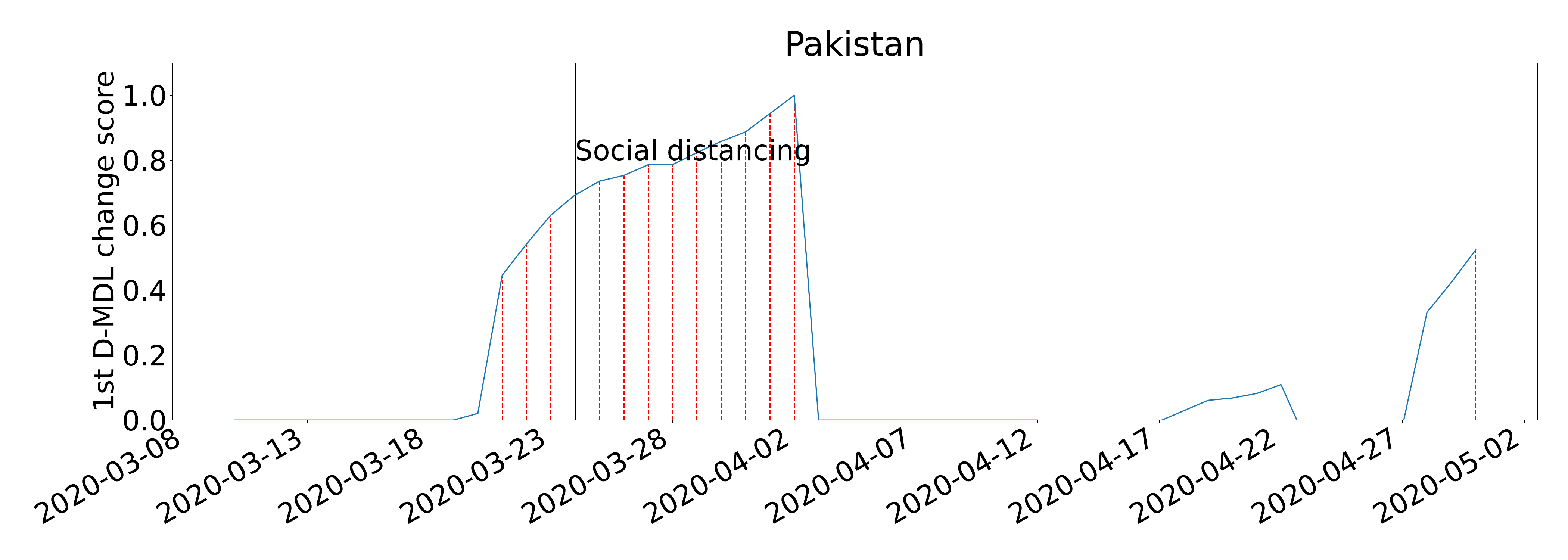} \\
			\vspace{-0.35cm}
			\textbf{e} & \includegraphics[keepaspectratio, height=3.3cm, valign=T]
			{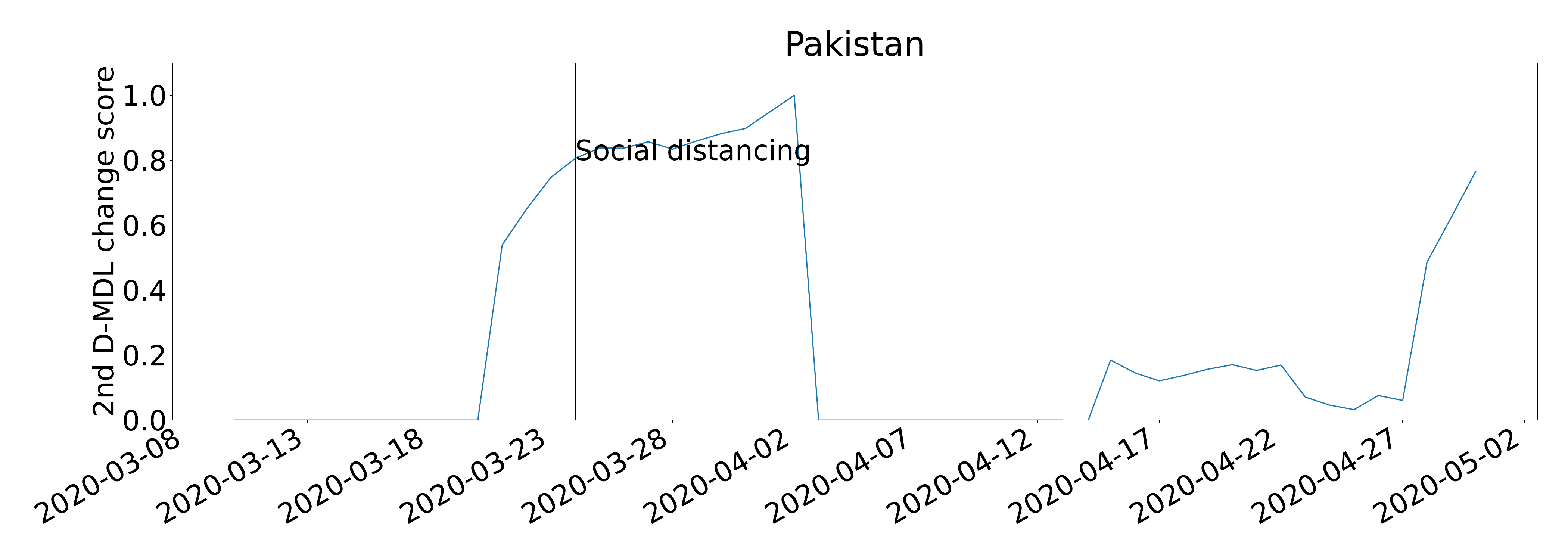} \\
		\end{tabular}
			\caption{\textbf{The results for Pakistan with exponential modeling.} The date on which the social distancing was implemented is marked by a solid line in black. \textbf{a,} the number of cumulative cases. \textbf{b,} the change scores produced by the 0th M-DML where the line in blue denotes values of scores and dashed lines in red mark alarms. \textbf{c,} the window sized for the sequential D-DML algorithm with adaptive window where lines in red mark the shrinkage of windows. \textbf{d,} the change scores produced by the 1st D-MDL. \textbf{e,} the change scores produced by the 2nd D-MDL.}
			\label{exp:pakistan}
\end{figure}

\begin{figure}[H] 
\centering
\begin{tabular}{cc}
		 	\textbf{a} & \includegraphics[keepaspectratio, height=3.3cm, valign=T]
			{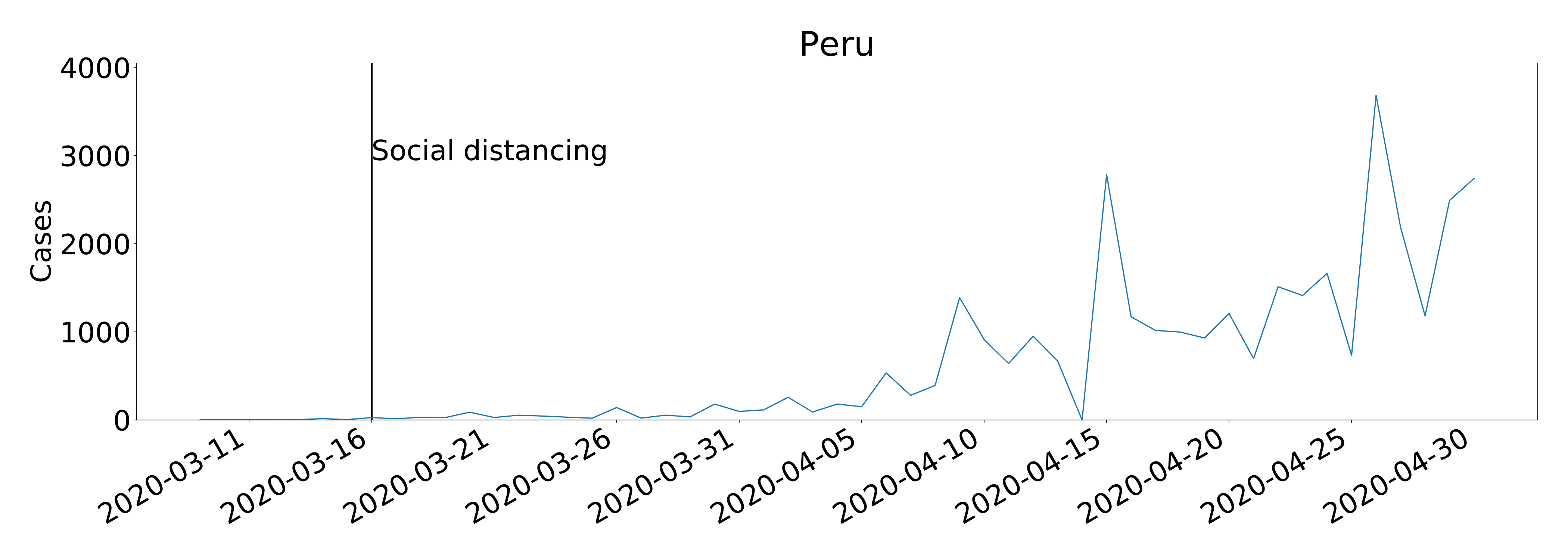} \\
			\vspace{-0.35cm}
	 	    \textbf{b} & \includegraphics[keepaspectratio, height=3.3cm, valign=T]
			{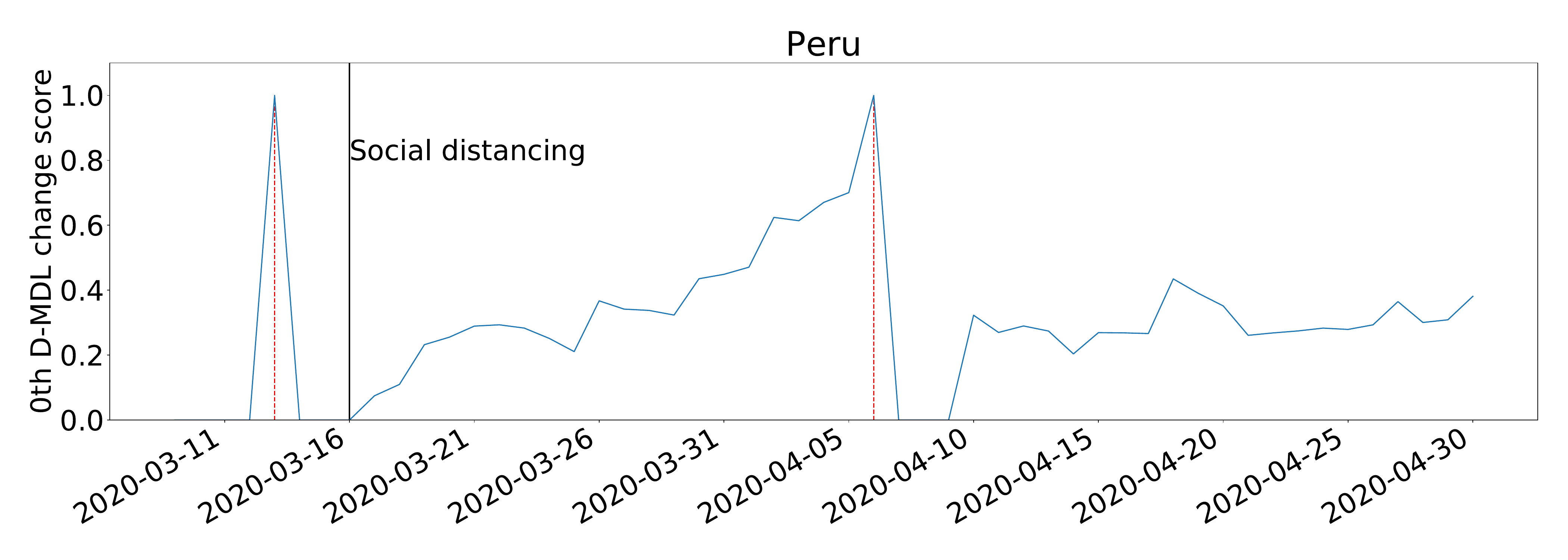}   \\
	        \vspace{-0.35cm}
			\textbf{c} & \includegraphics[keepaspectratio, height=3.3cm, valign=T]
			{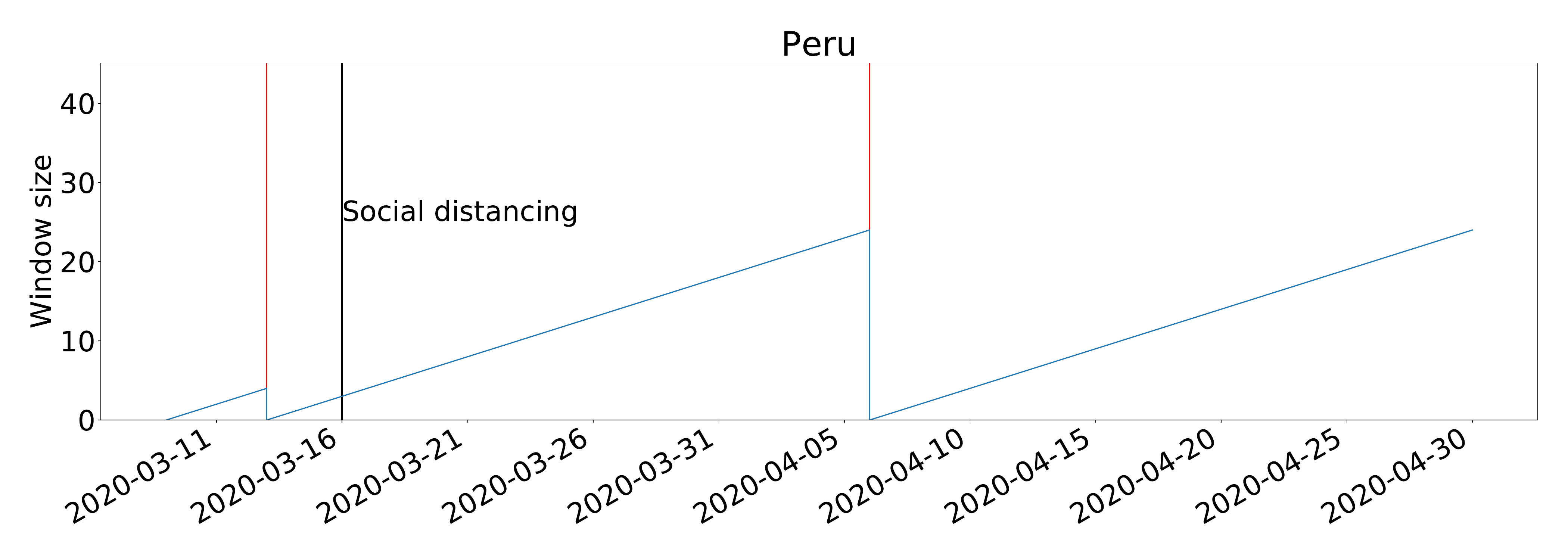} \\
		    \vspace{-0.35cm}
			\textbf{d} & \includegraphics[keepaspectratio, height=3.3cm, valign=T]
			{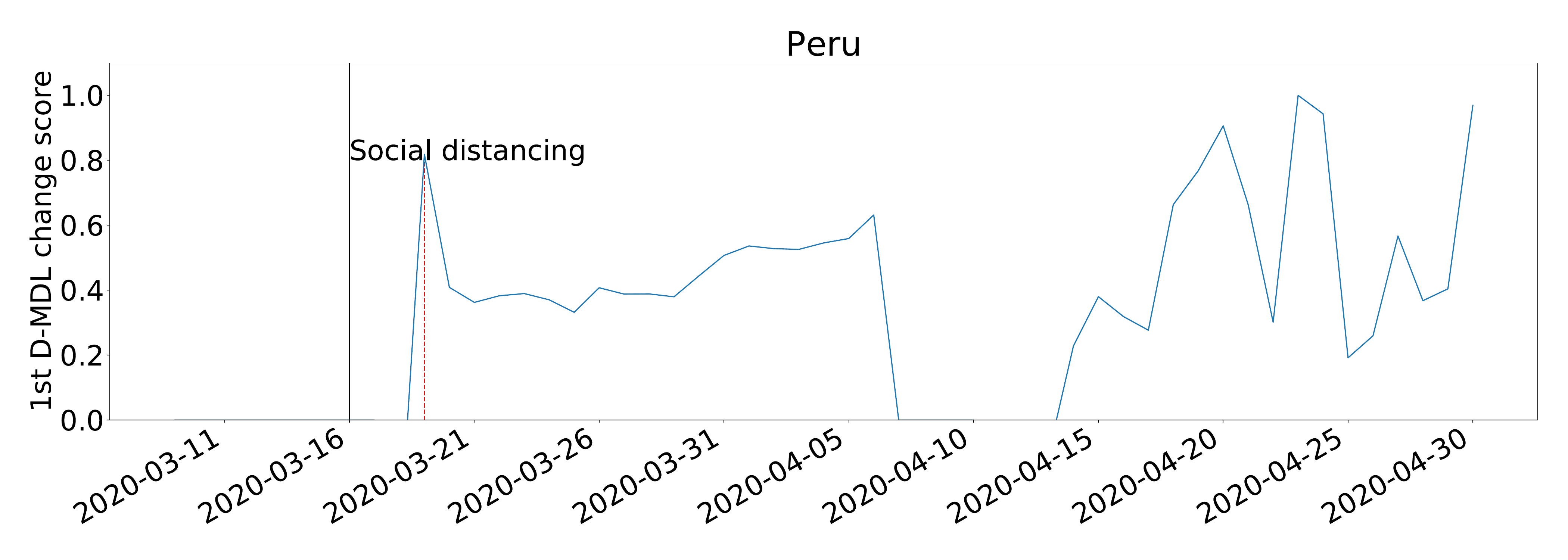} \\
		    \vspace{-0.35cm}
			\textbf{e} & \includegraphics[keepaspectratio, height=3.3cm, valign=T]
			{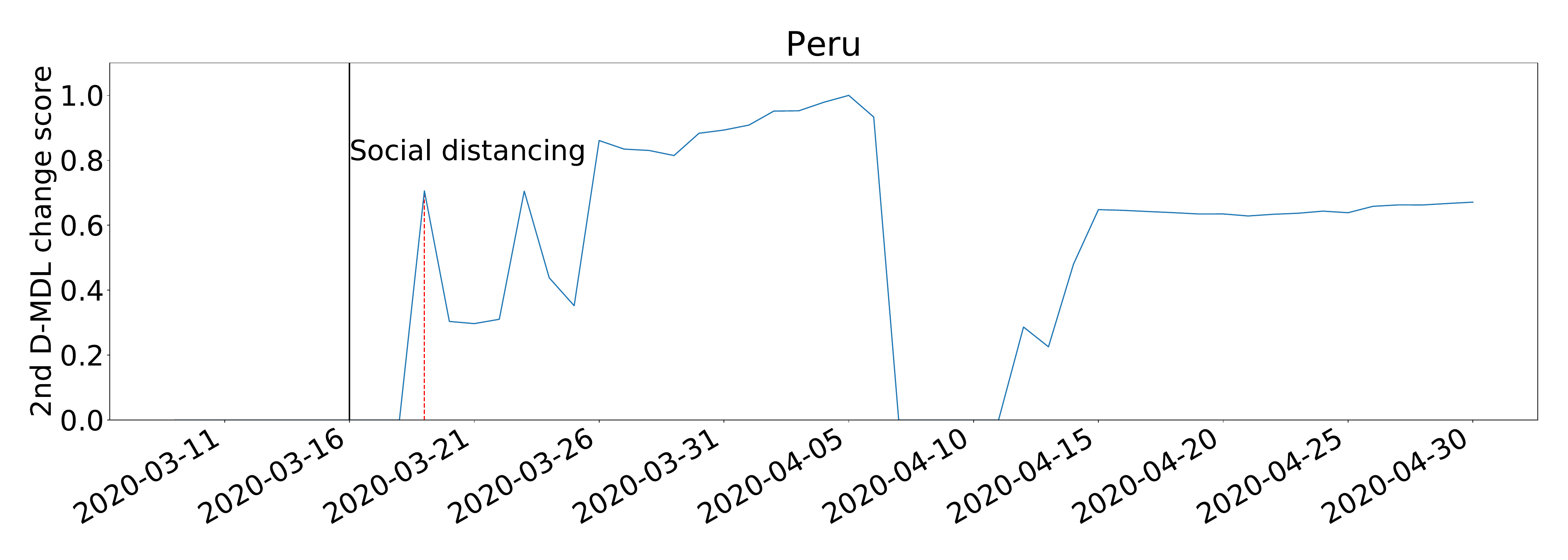} \\
		\end{tabular}
			\caption{\textbf{The results for Peru with Gaussian modeling.} The date on which the social distancing was implemented is marked by a solid line in black. \textbf{a,} the number of daily new cases. \textbf{b,} the change scores produced by the 0th M-DML where the line in blue denotes values of scores and dashed lines in red mark alarms. \textbf{c,} the window sized for the sequential D-DML algorithm with adaptive window where lines in red mark the shrinkage of windows. \textbf{d,} the change scores produced by the 1st D-MDL. \textbf{e,} the change scores produced by the 2nd D-MDL.}
\end{figure}

\begin{figure}[H]  
\centering
\begin{tabular}{cc}
			\textbf{a} & \includegraphics[keepaspectratio, height=3.3cm, valign=T]
			{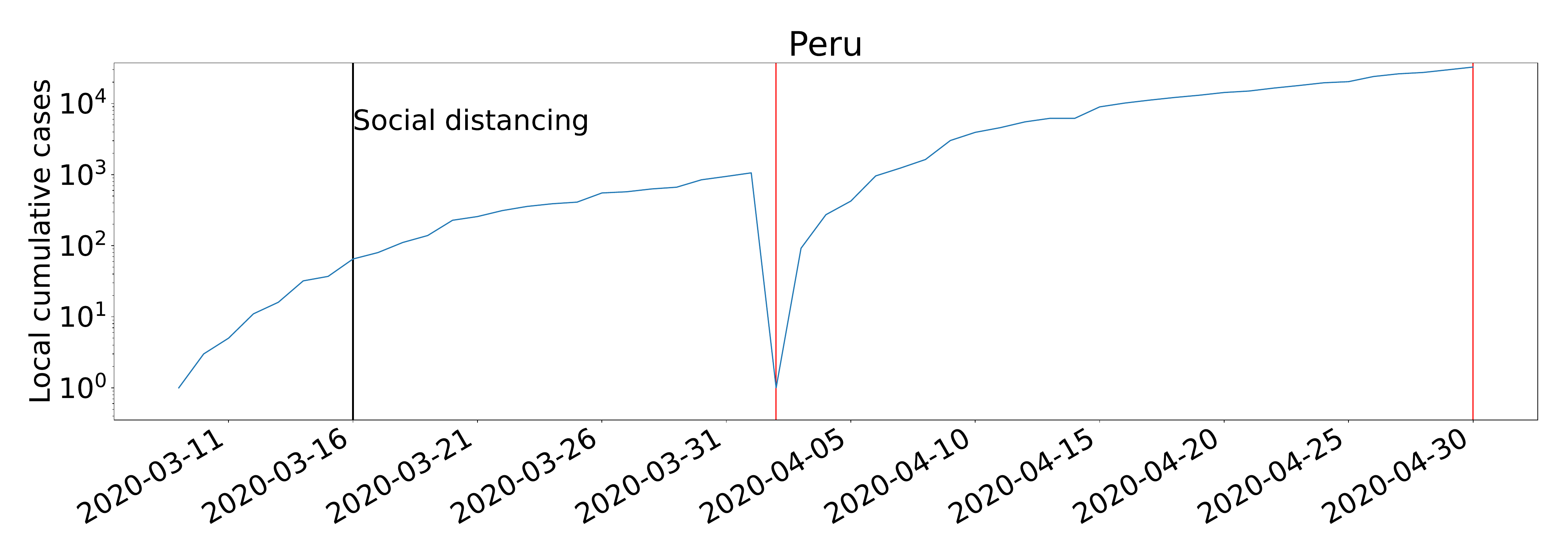} \\
	        \vspace{-0.35cm}
            \textbf{b} & \includegraphics[keepaspectratio, height=3.3cm, valign=T]
			{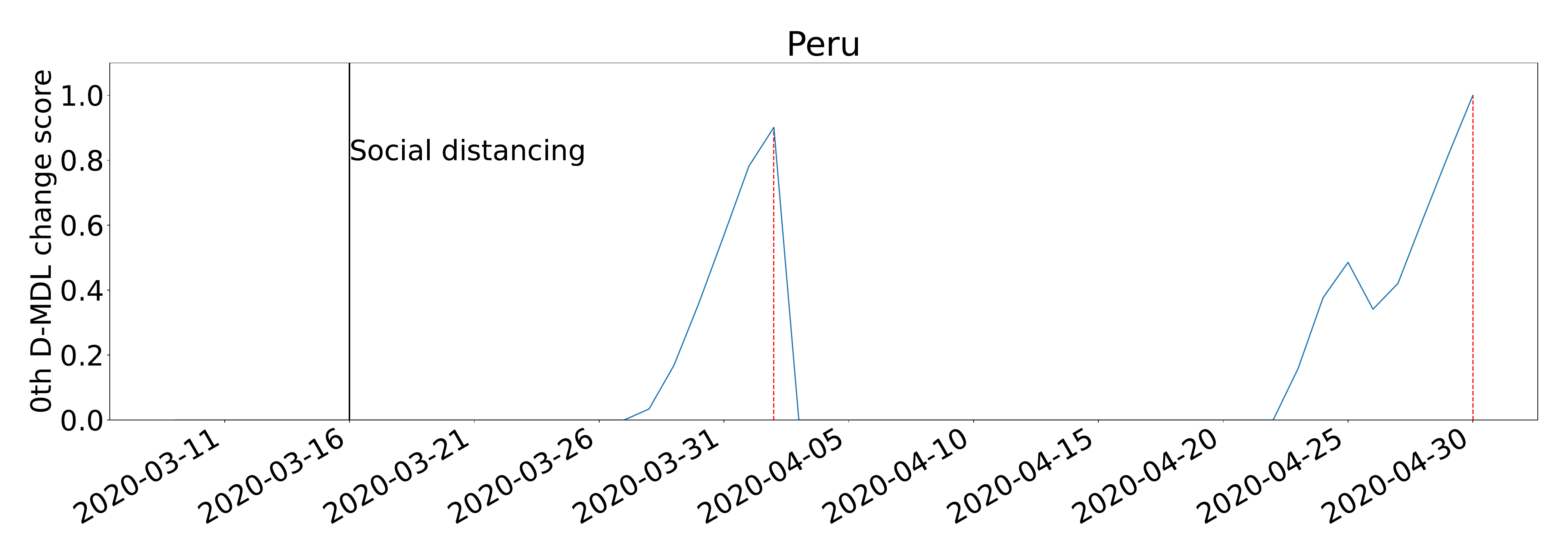}   \\
            \vspace{-0.35cm}
            \textbf{c} & \includegraphics[keepaspectratio, height=3.3cm, valign=T]
			{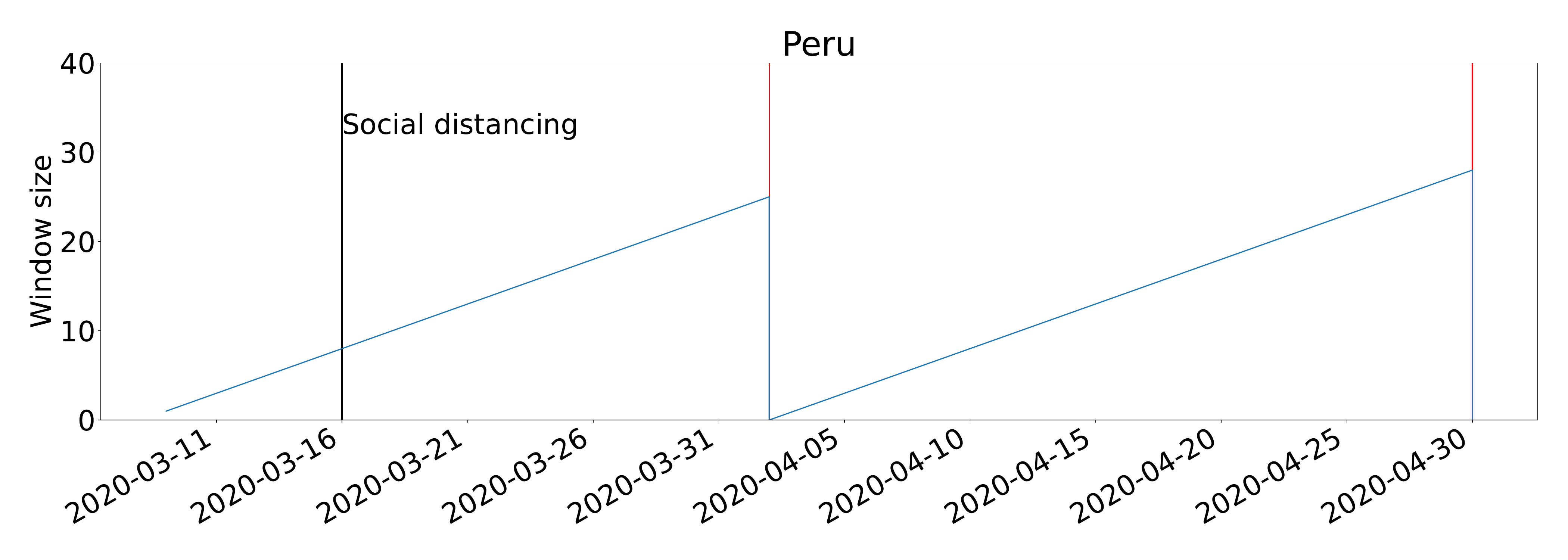} \\
			\vspace{-0.35cm}
			\textbf{d} & \includegraphics[keepaspectratio, height=3.3cm, valign=T]
			{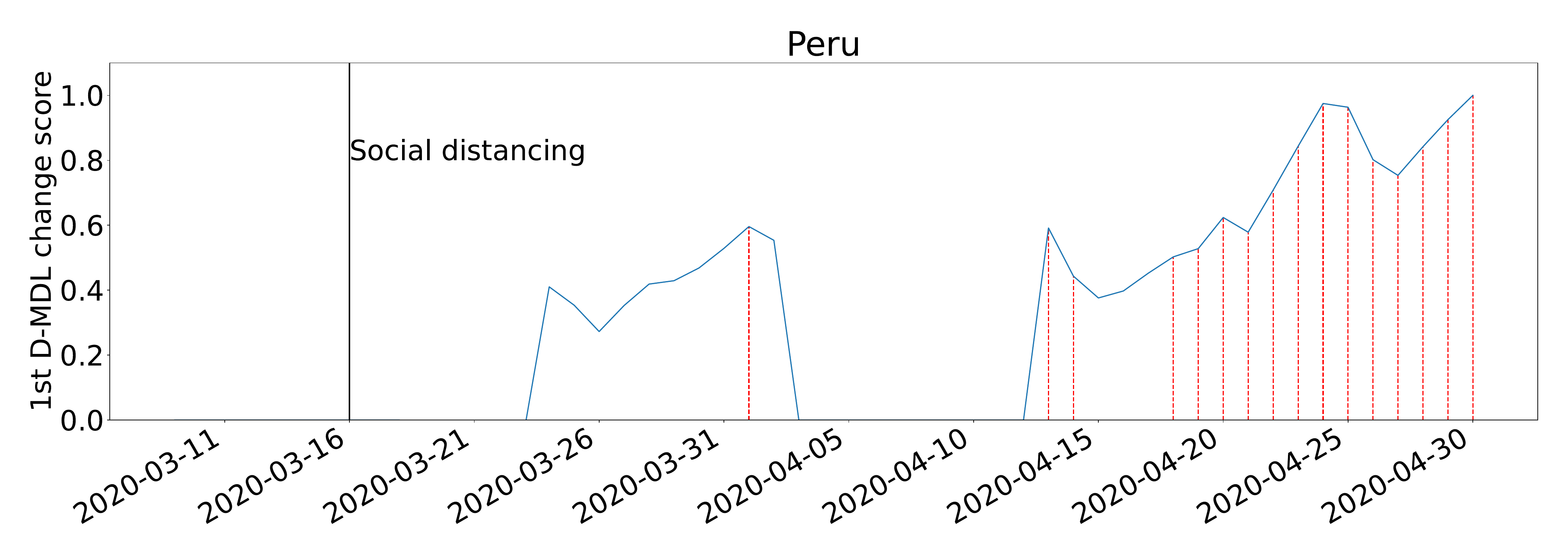} \\
			\vspace{-0.35cm}
			\textbf{e} & \includegraphics[keepaspectratio, height=3.3cm, valign=T]
			{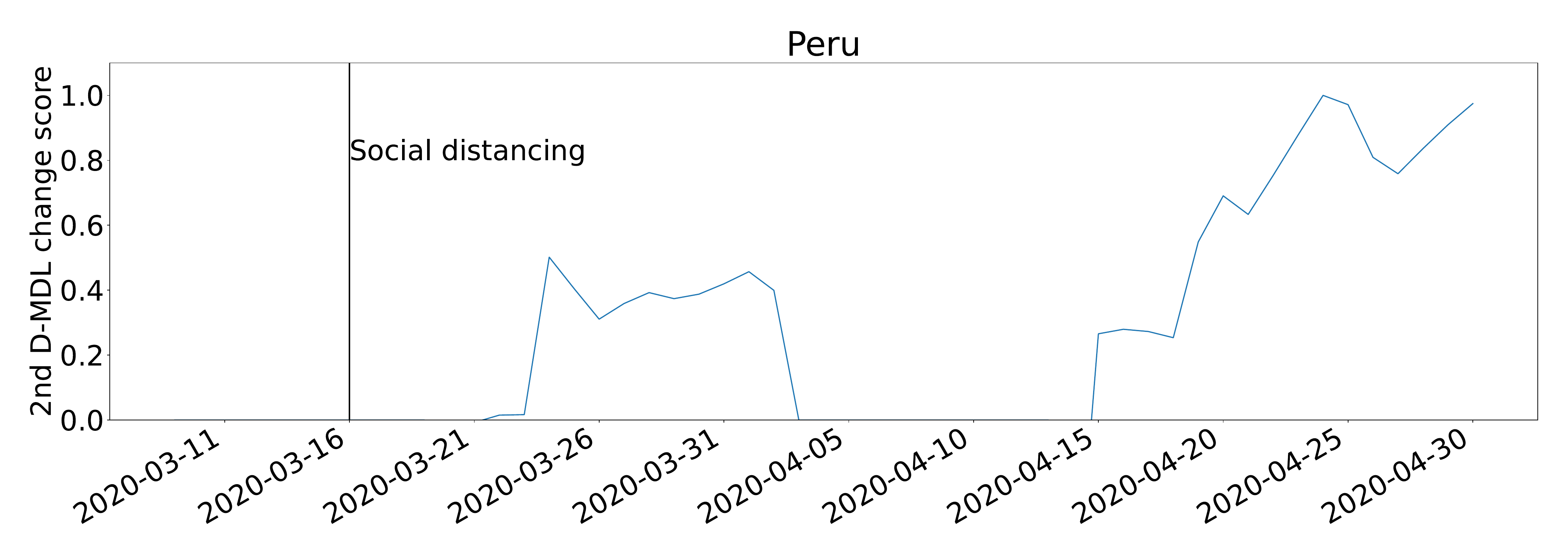} \\
		\end{tabular}
			\caption{\textbf{The results for Peru with exponential modeling.} The date on which the social distancing was implemented is marked by a solid line in black. \textbf{a,} the number of cumulative cases. \textbf{b,} the change scores produced by the 0th M-DML where the line in blue denotes values of scores and dashed lines in red mark alarms. \textbf{c,} the window sized for the sequential D-DML algorithm with adaptive window where lines in red mark the shrinkage of windows. \textbf{d,} the change scores produced by the 1st D-MDL. \textbf{e,} the change scores produced by the 2nd D-MDL.}
\end{figure}

\begin{figure}[H] 
\centering
\begin{tabular}{cc}
		 	\textbf{a} & \includegraphics[keepaspectratio, height=3.3cm, valign=T]
			{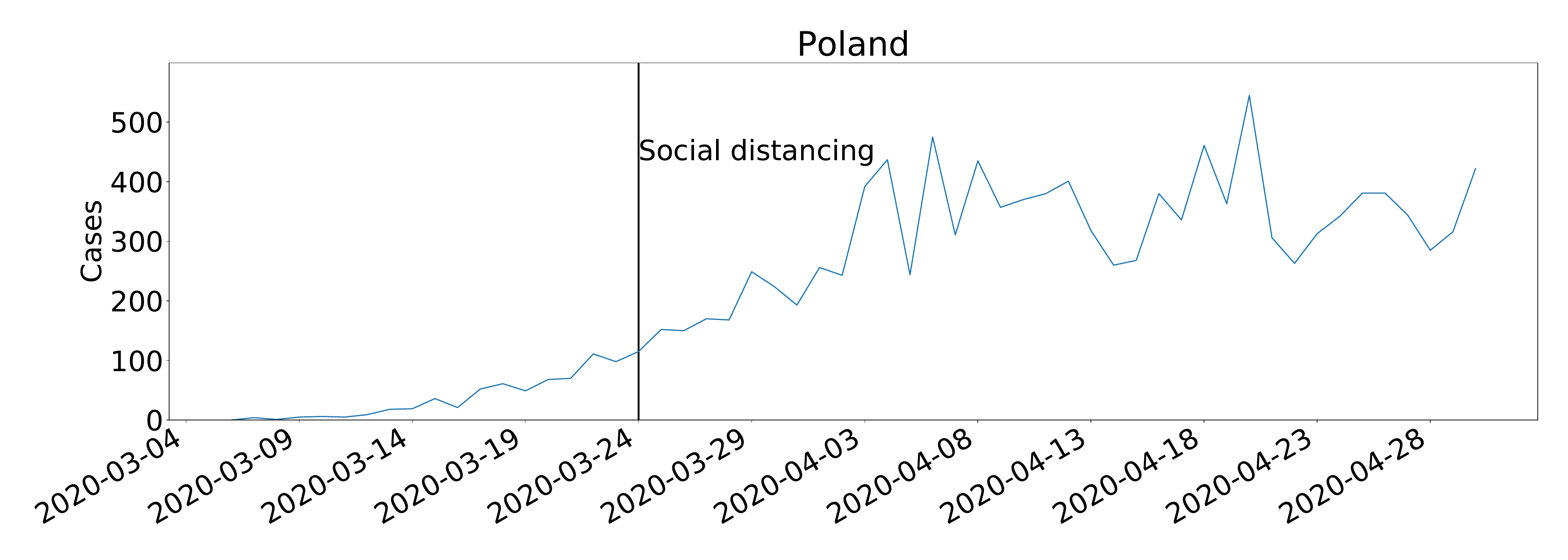} \\
			\vspace{-0.35cm}
	 	    \textbf{b} & \includegraphics[keepaspectratio, height=3.3cm, valign=T]
			{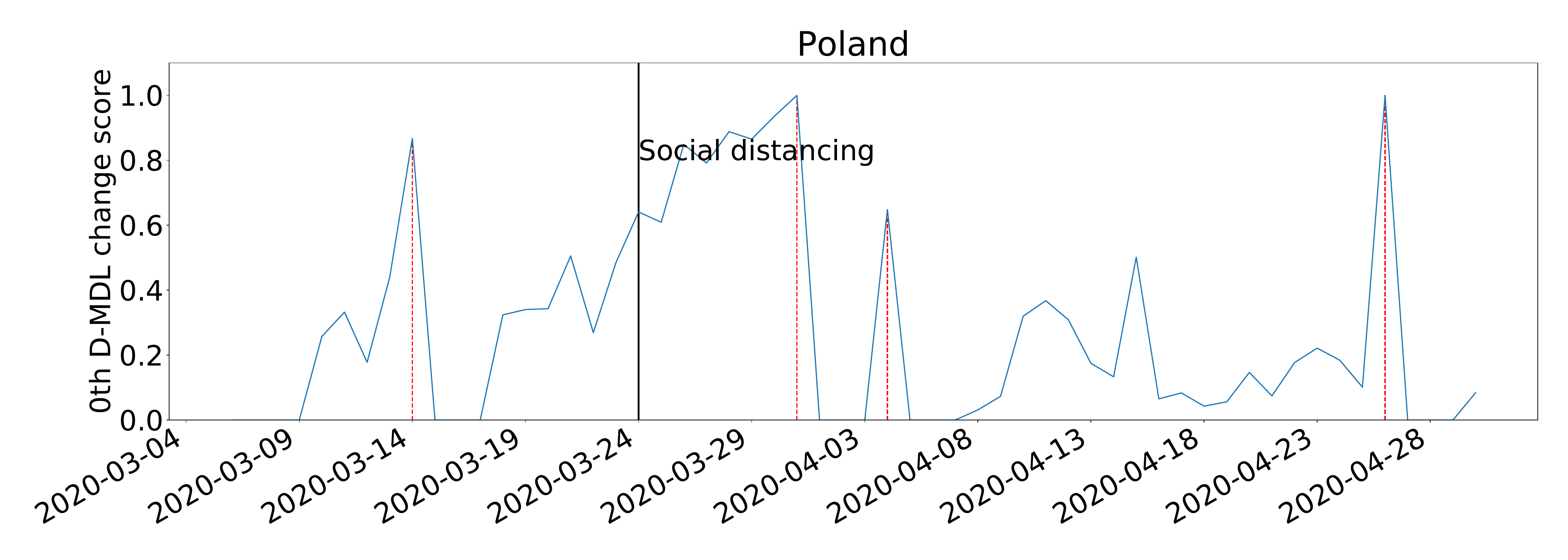}   \\
	        \vspace{-0.35cm}
			\textbf{c} & \includegraphics[keepaspectratio, height=3.3cm, valign=T]
			{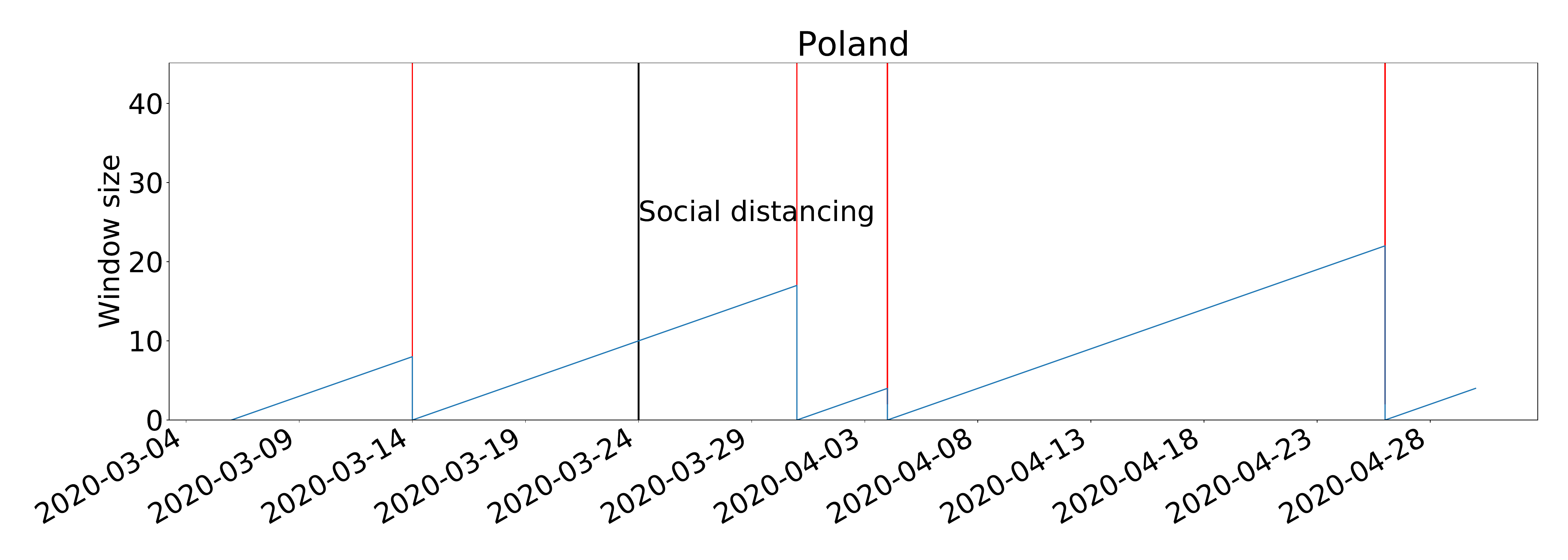} \\
		    \vspace{-0.35cm}
			\textbf{d} & \includegraphics[keepaspectratio, height=3.3cm, valign=T]
			{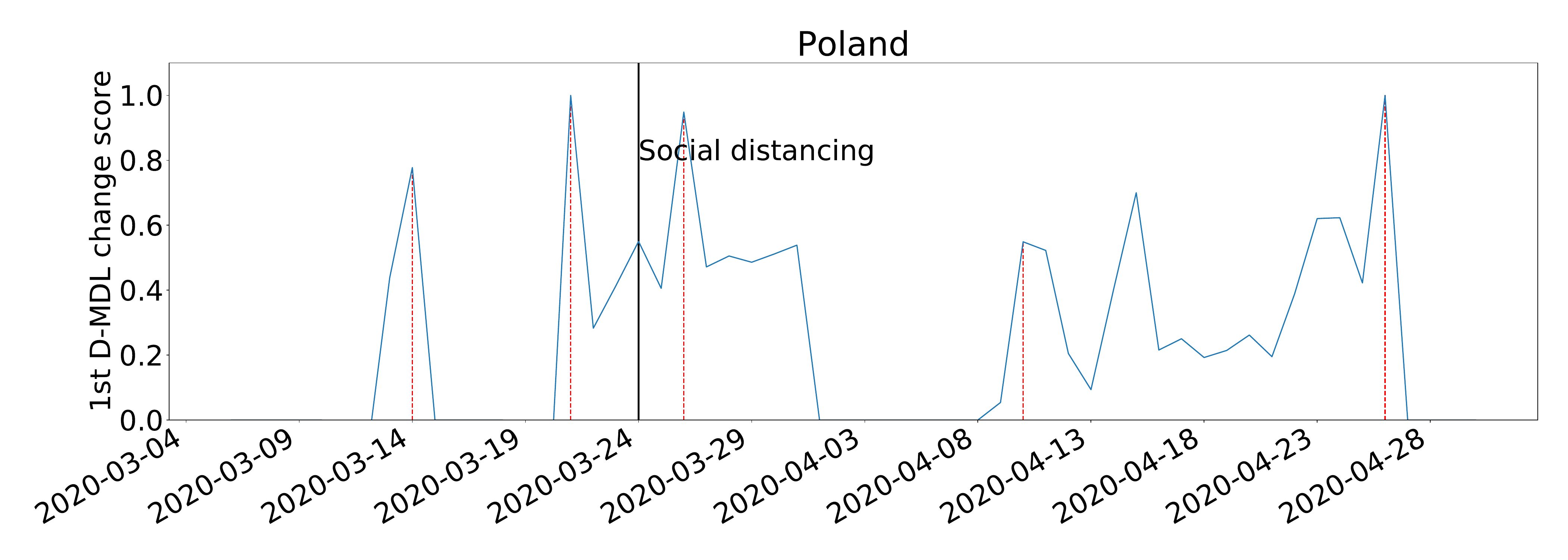} \\
		    \vspace{-0.35cm}
			\textbf{e} & \includegraphics[keepaspectratio, height=3.3cm, valign=T]
			{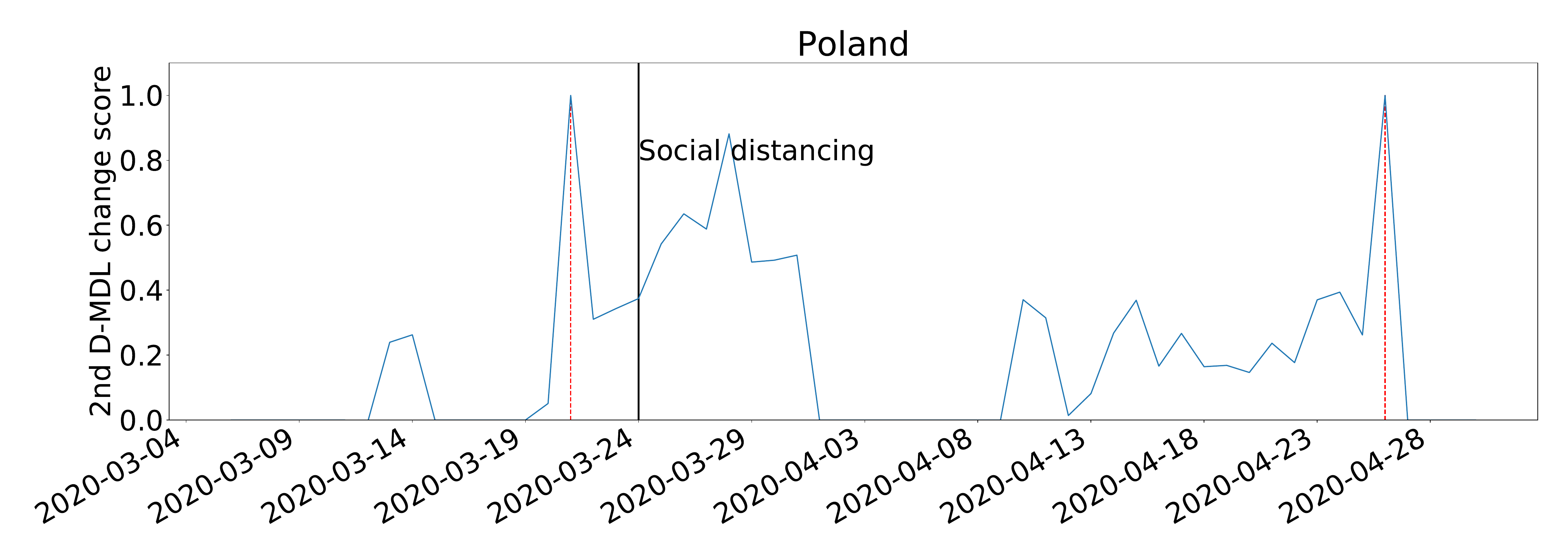} \\
		\end{tabular}
			\caption{\textbf{The results for Poland with Gaussian modeling.} The date on which the social distancing was implemented is marked by a solid line in black. \textbf{a,} the number of daily new cases. \textbf{b,} the change scores produced by the 0th M-DML where the line in blue denotes values of scores and dashed lines in red mark alarms. \textbf{c,} the window sized for the sequential D-DML algorithm with adaptive window where lines in red mark the shrinkage of windows. \textbf{d,} the change scores produced by the 1st D-MDL. \textbf{e,} the change scores produced by the 2nd D-MDL.}
\end{figure}

\begin{figure}[H]  
\centering
\begin{tabular}{cc}
			\textbf{a} & \includegraphics[keepaspectratio, height=3.3cm, valign=T]
			{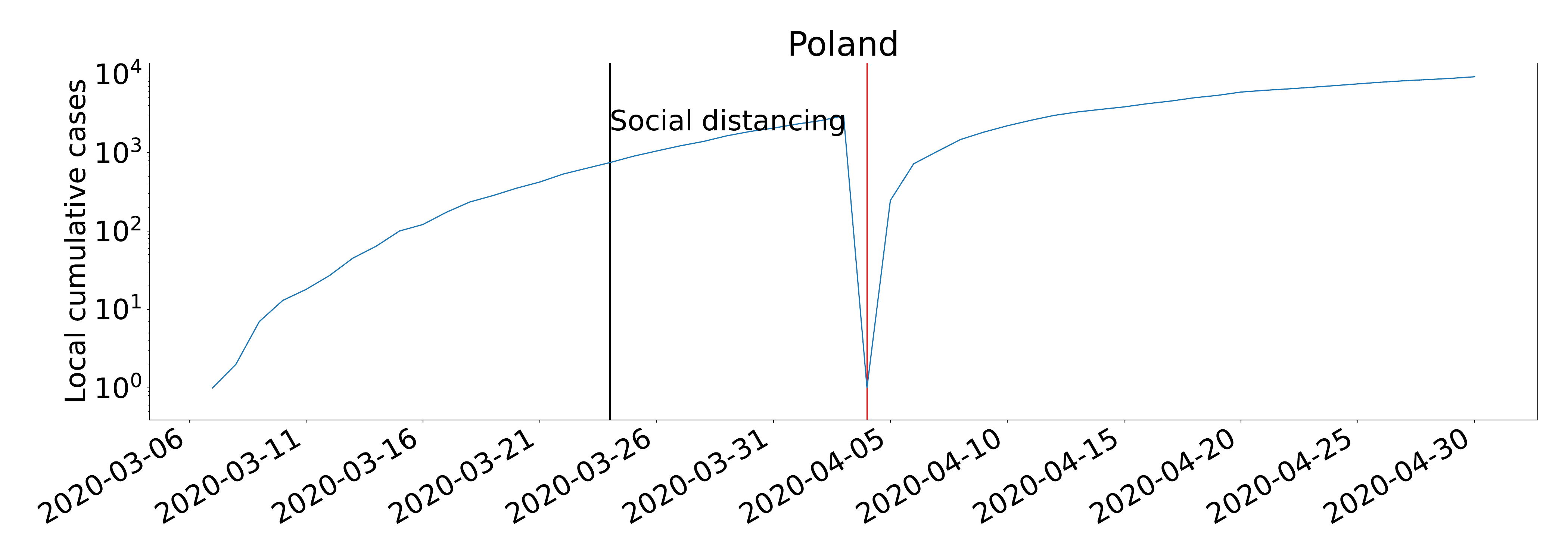} \\
	        \vspace{-0.35cm}
            \textbf{b} & \includegraphics[keepaspectratio, height=3.3cm, valign=T]
			{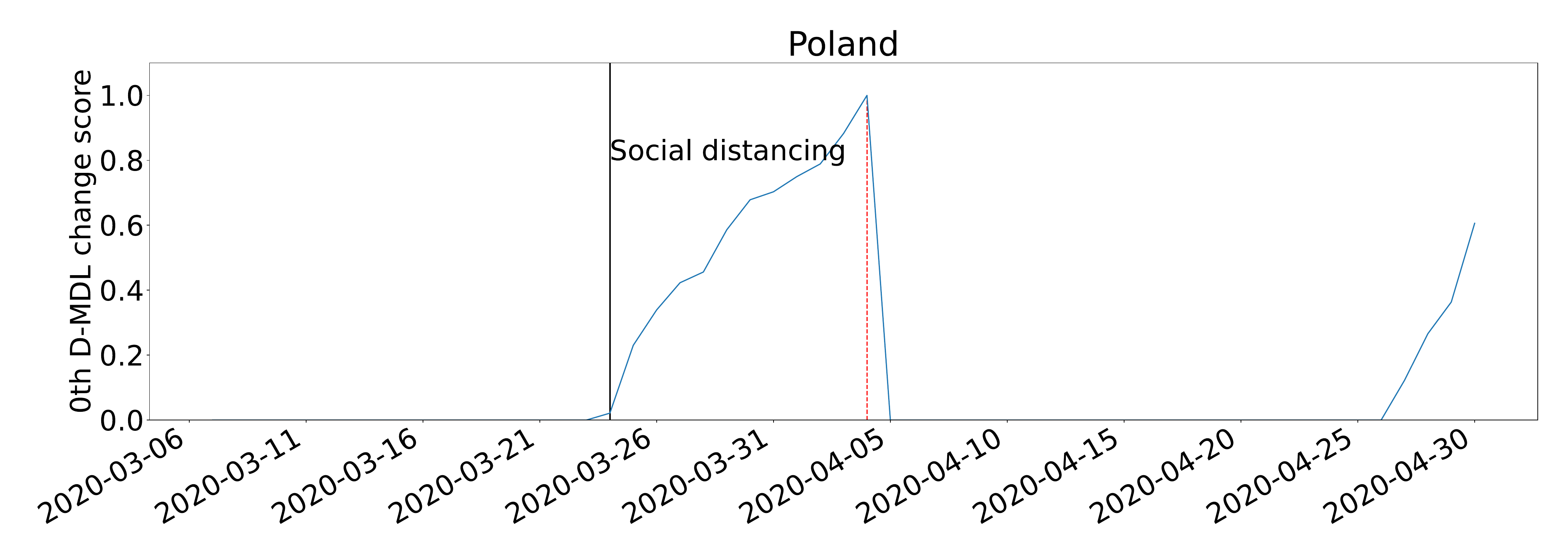}   \\
            \vspace{-0.35cm}
            \textbf{c} & \includegraphics[keepaspectratio, height=3.3cm, valign=T]
			{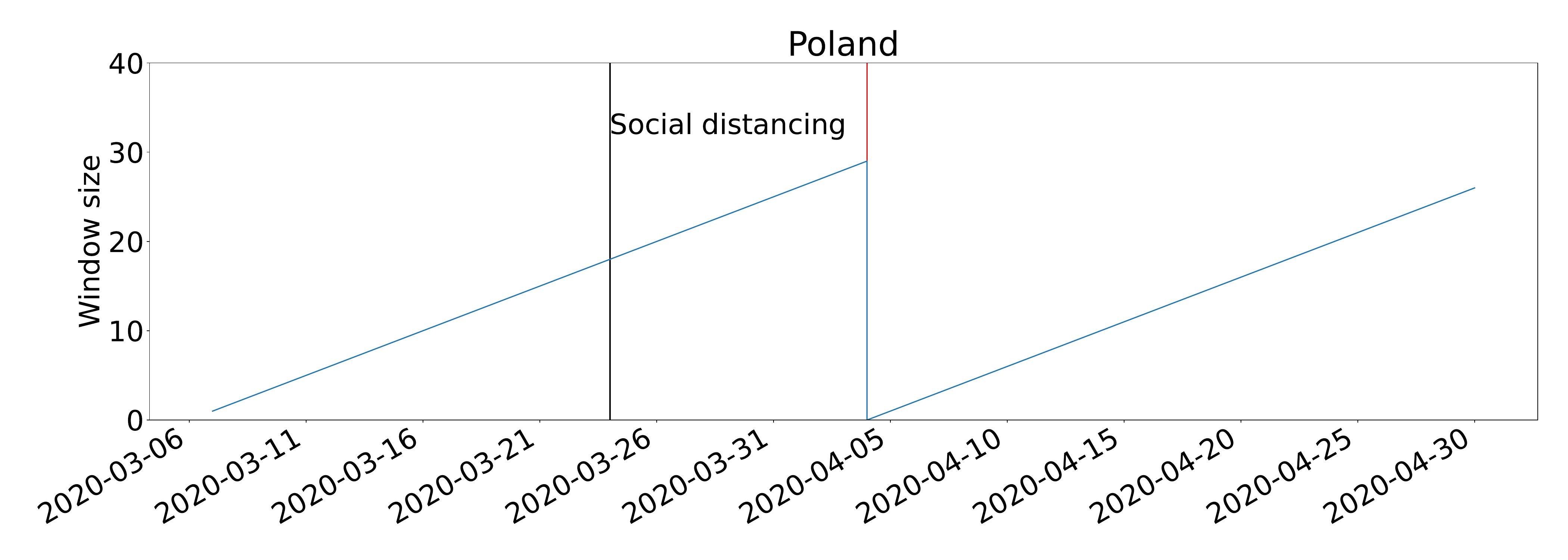} \\
			\vspace{-0.35cm}
			\textbf{d} & \includegraphics[keepaspectratio, height=3.3cm, valign=T]
			{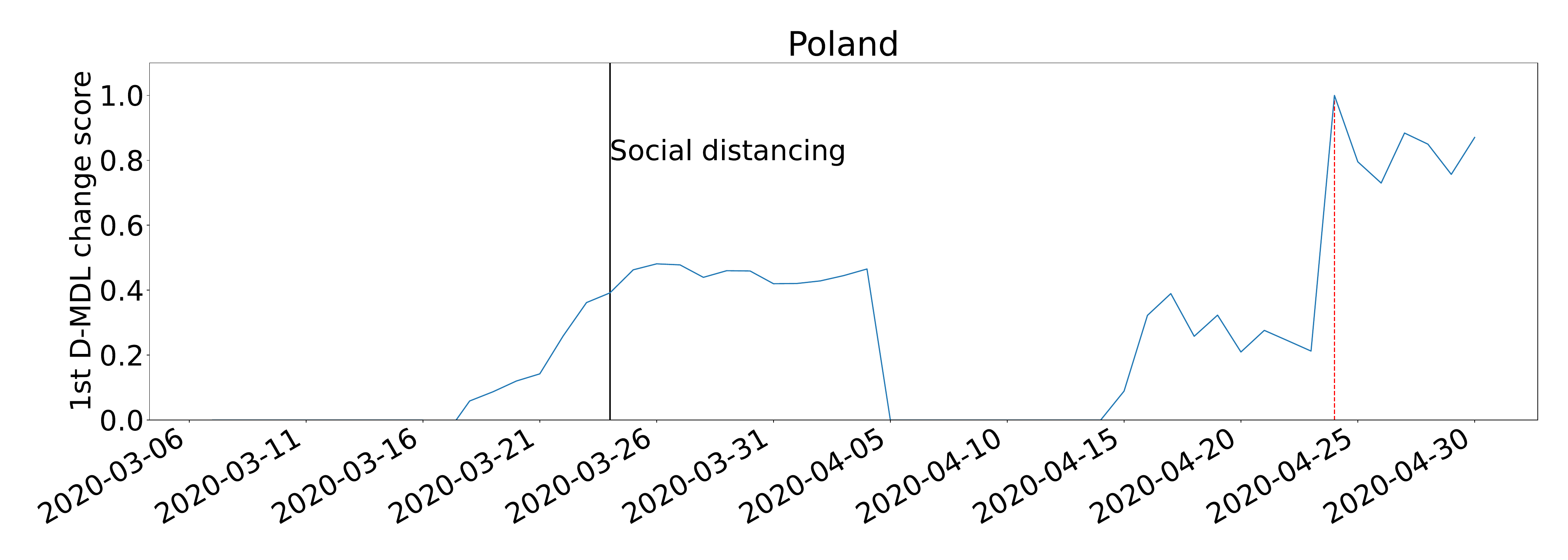} \\
			\vspace{-0.35cm}
			\textbf{e} & \includegraphics[keepaspectratio, height=3.3cm, valign=T]
			{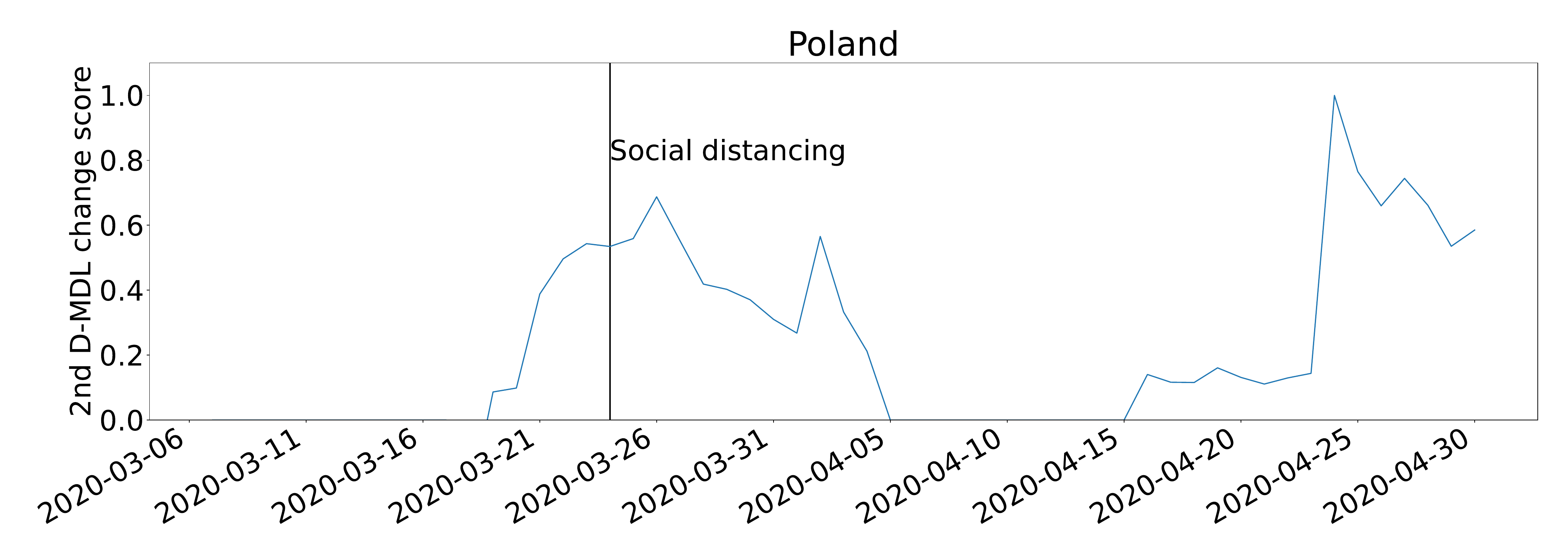} \\
		\end{tabular}
			\caption{\textbf{The results for Poland with exponential modeling.} The date on which the social distancing was implemented is marked by a solid line in black. \textbf{a,} the number of cumulative cases. \textbf{b,} the change scores produced by the 0th M-DML where the line in blue denotes values of scores and dashed lines in red mark alarms. \textbf{c,} the window sized for the sequential D-DML algorithm with adaptive window where lines in red mark the shrinkage of windows. \textbf{d,} the change scores produced by the 1st D-MDL. \textbf{e,} the change scores produced by the 2nd D-MDL.}
\end{figure}

\begin{figure}[H] 
\centering
\begin{tabular}{cc}
		 	\textbf{a} & \includegraphics[keepaspectratio, height=3.3cm, valign=T]
			{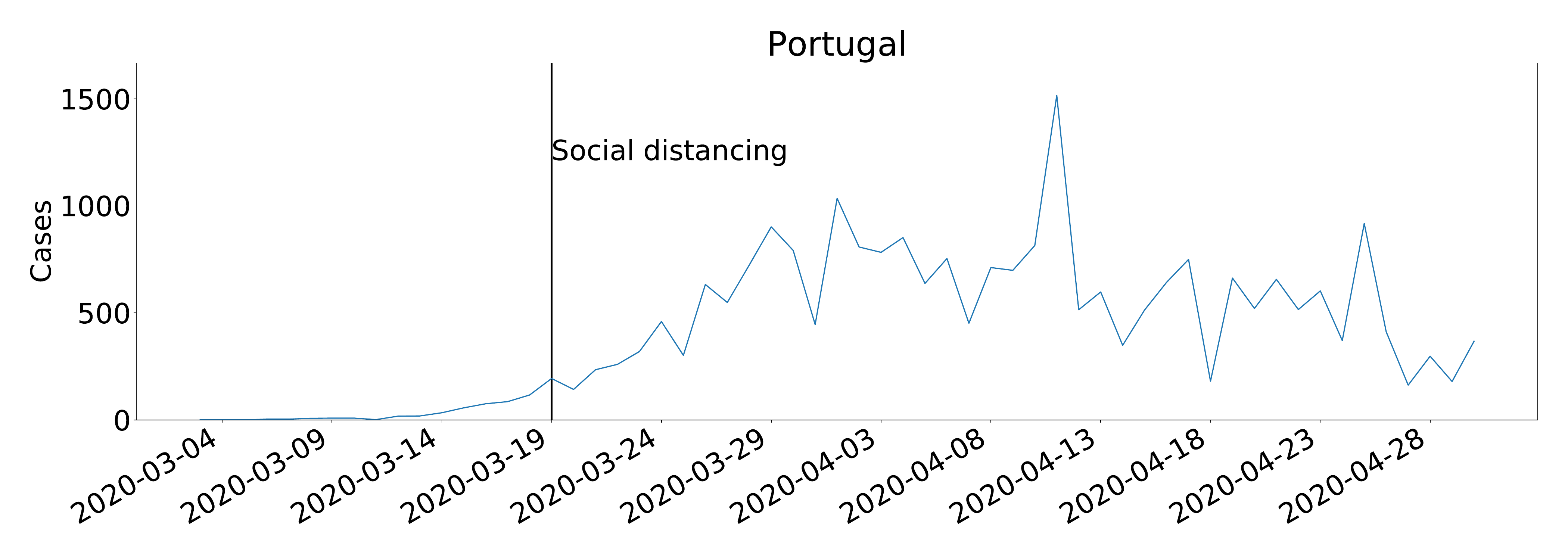} \\
			\vspace{-0.35cm}
	 	    \textbf{b} & \includegraphics[keepaspectratio, height=3.3cm, valign=T]
			{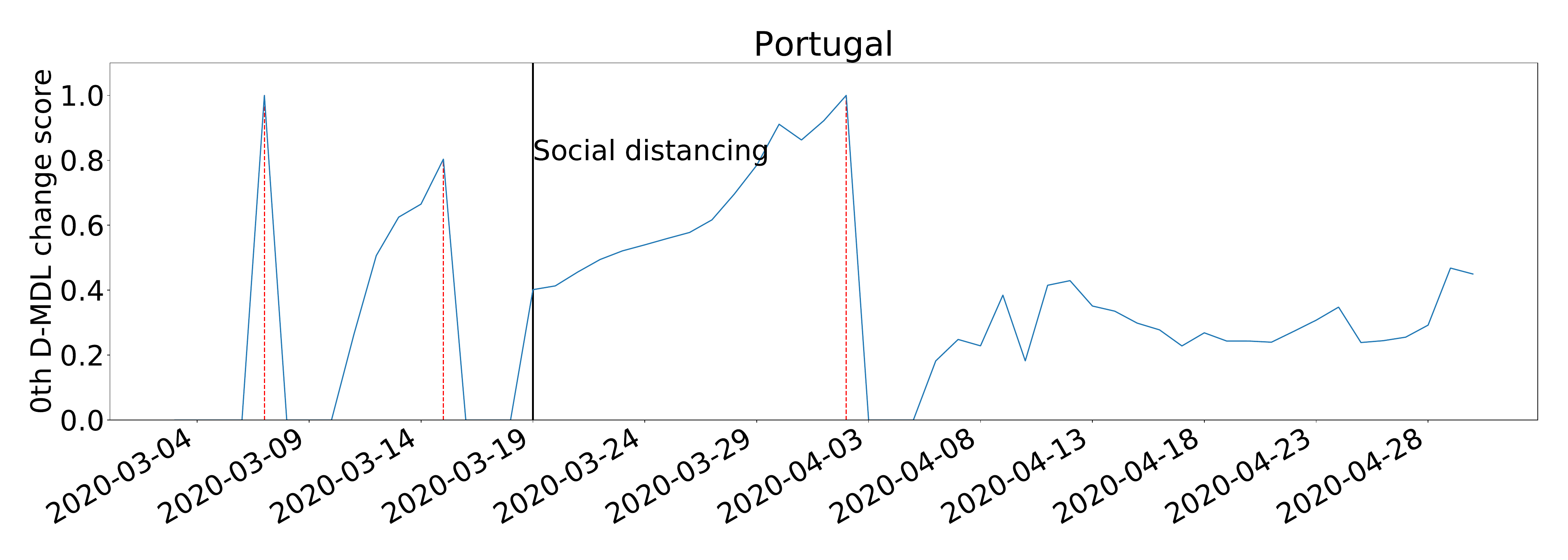}   \\
	        \vspace{-0.35cm}
			\textbf{c} & \includegraphics[keepaspectratio, height=3.3cm, valign=T]
			{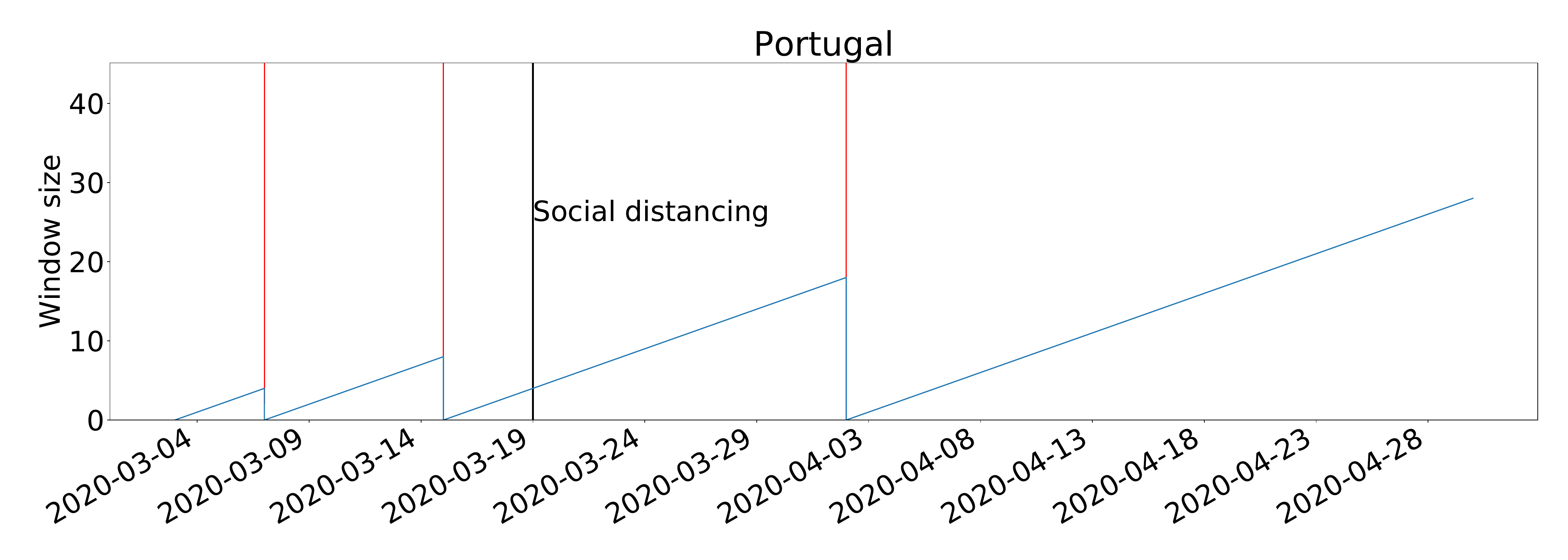} \\
		    \vspace{-0.35cm}
			\textbf{d} & \includegraphics[keepaspectratio, height=3.3cm, valign=T]
			{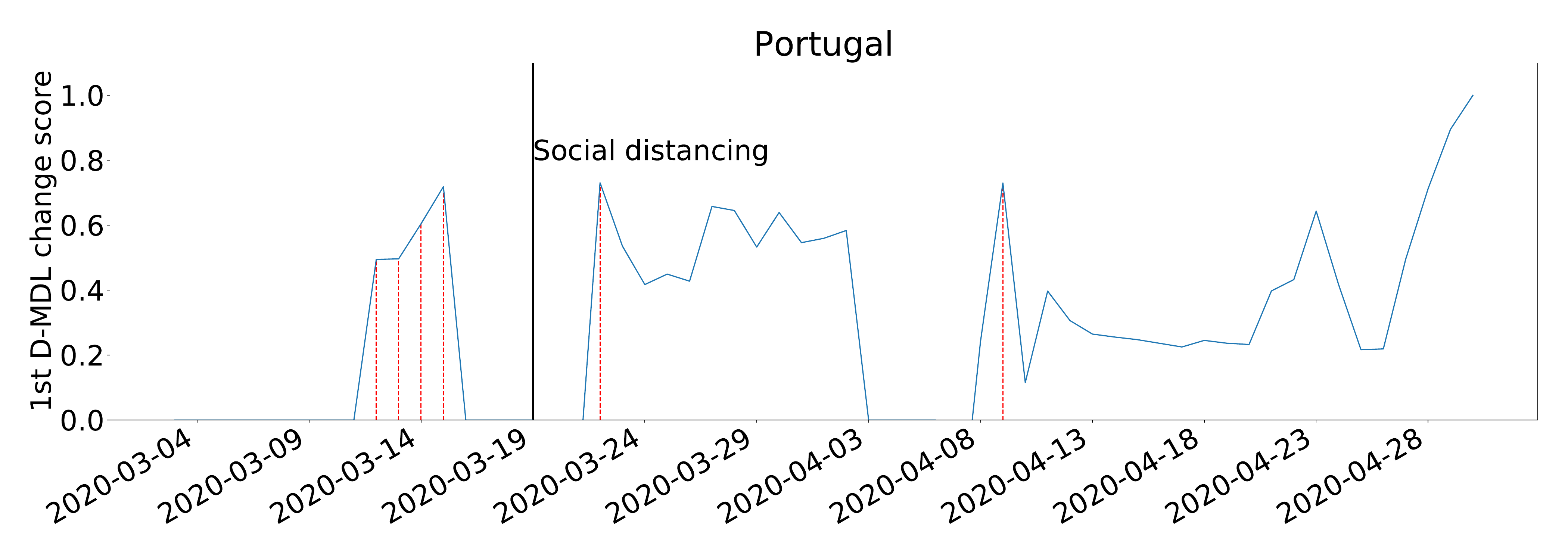} \\
		    \vspace{-0.35cm}
			\textbf{e} & \includegraphics[keepaspectratio, height=3.3cm, valign=T]
			{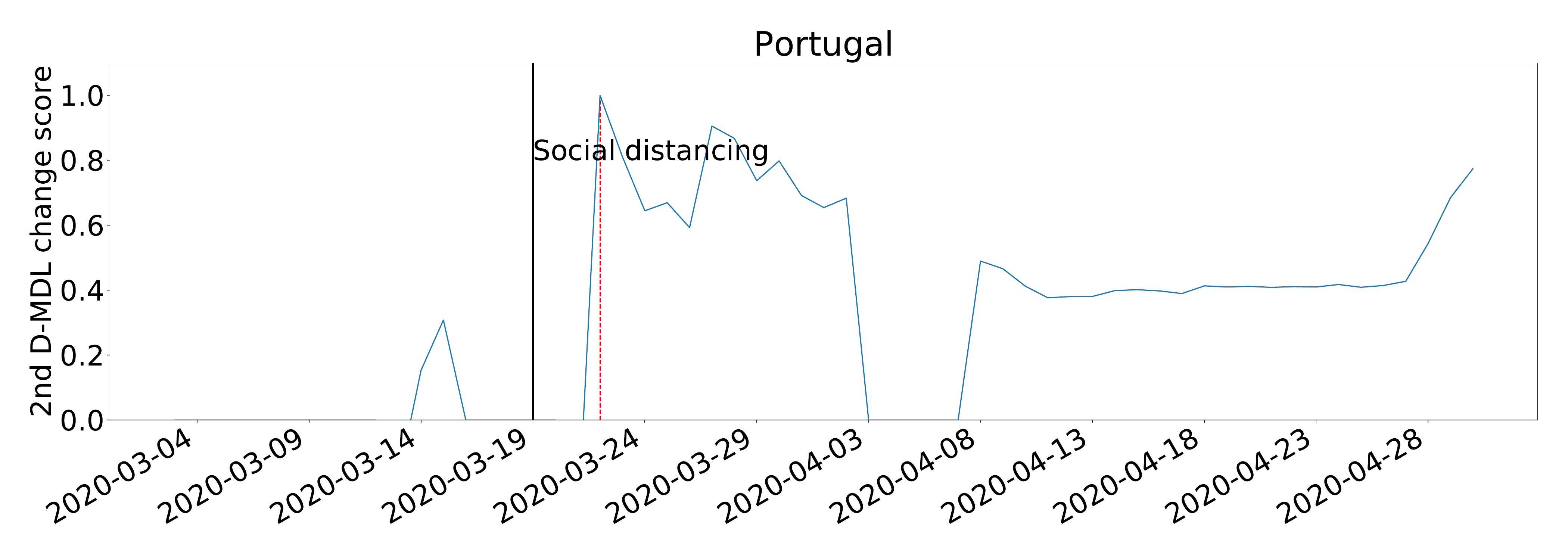} \\
		\end{tabular}
			\caption{\textbf{The results for Portugal with Gaussian modeling.} The date on which the social distancing was implemented is marked by a solid line in black. \textbf{a,} the number of daily new cases. \textbf{b,} the change scores produced by the 0th M-DML where the line in blue denotes values of scores and dashed lines in red mark alarms. \textbf{c,} the window sized for the sequential D-DML algorithm with adaptive window where lines in red mark the shrinkage of windows. \textbf{d,} the change scores produced by the 1st D-MDL. \textbf{e,} the change scores produced by the 2nd D-MDL.}
\end{figure}

\begin{figure}[H]  
\centering
\begin{tabular}{cc}
			\textbf{a} & \includegraphics[keepaspectratio, height=3.3cm, valign=T]
			{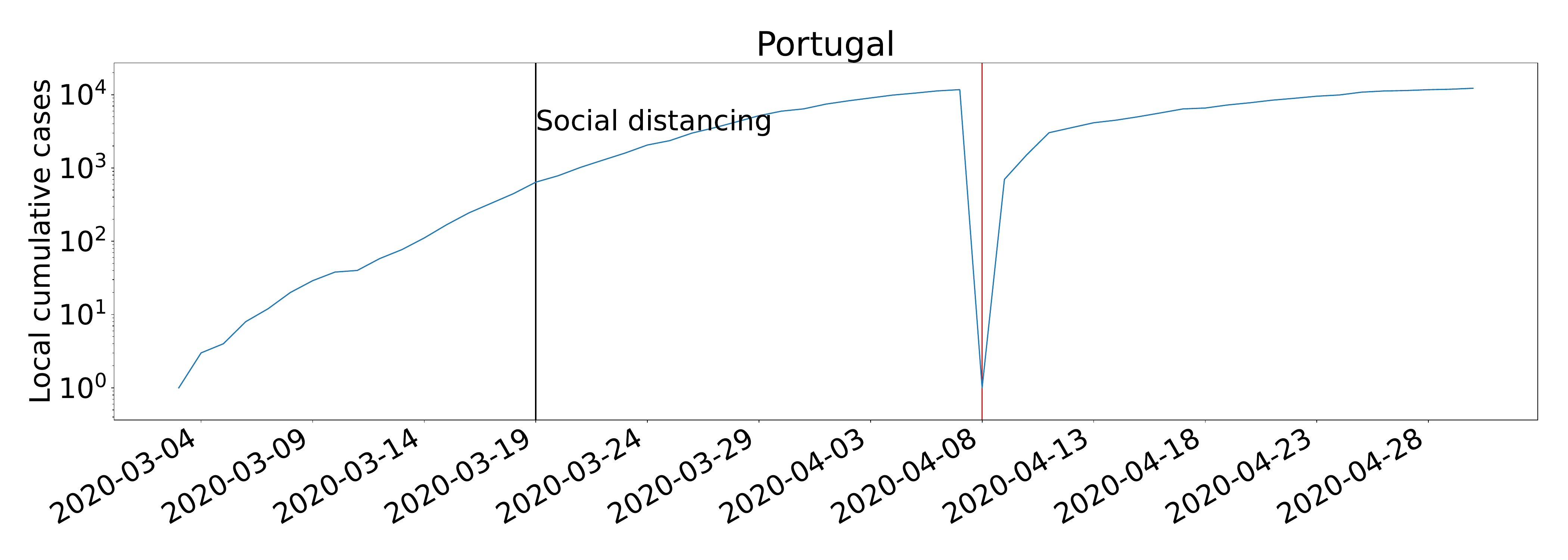} \\
	        \vspace{-0.35cm}
            \textbf{b} & \includegraphics[keepaspectratio, height=3.3cm, valign=T]
			{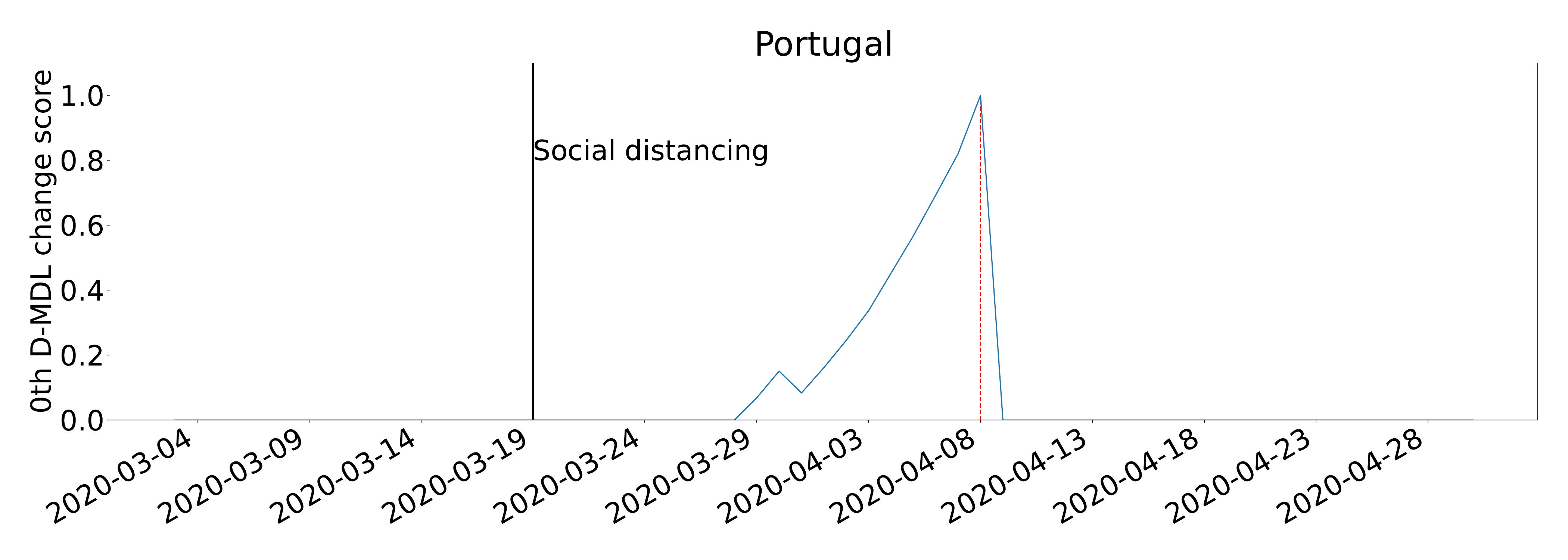}   \\
            \vspace{-0.35cm}
            \textbf{c} & \includegraphics[keepaspectratio, height=3.3cm, valign=T]
			{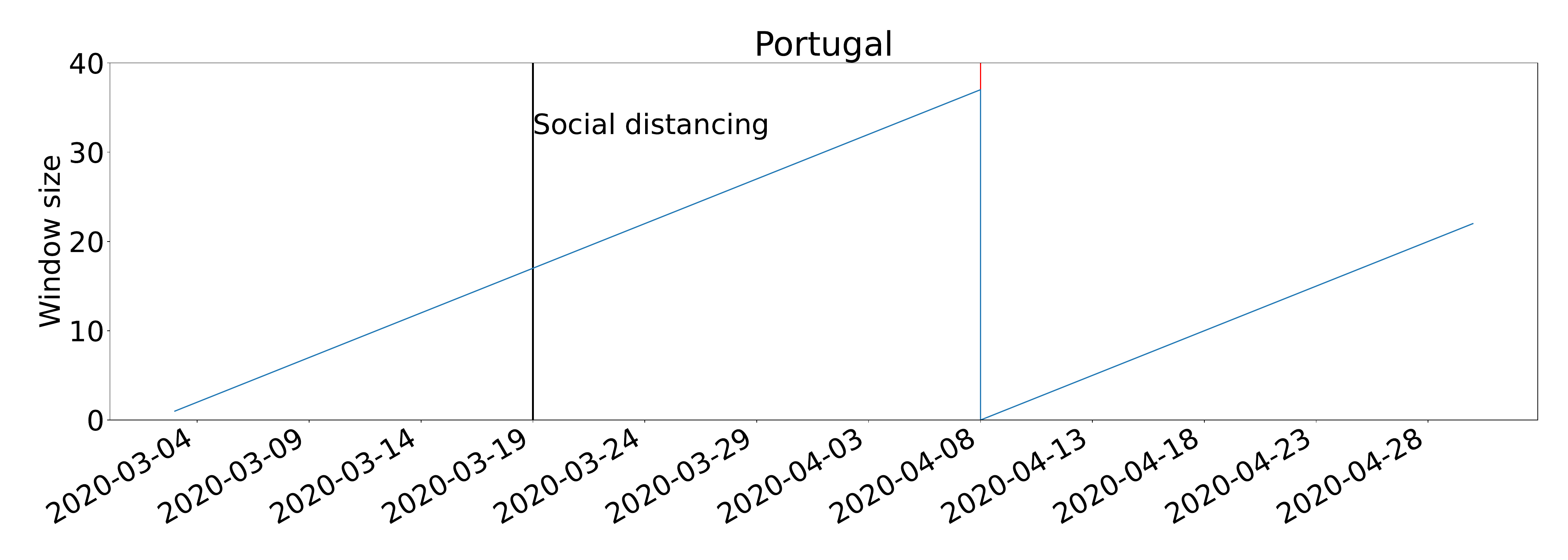} \\
			\vspace{-0.35cm}
			\textbf{d} & \includegraphics[keepaspectratio, height=3.3cm, valign=T]
			{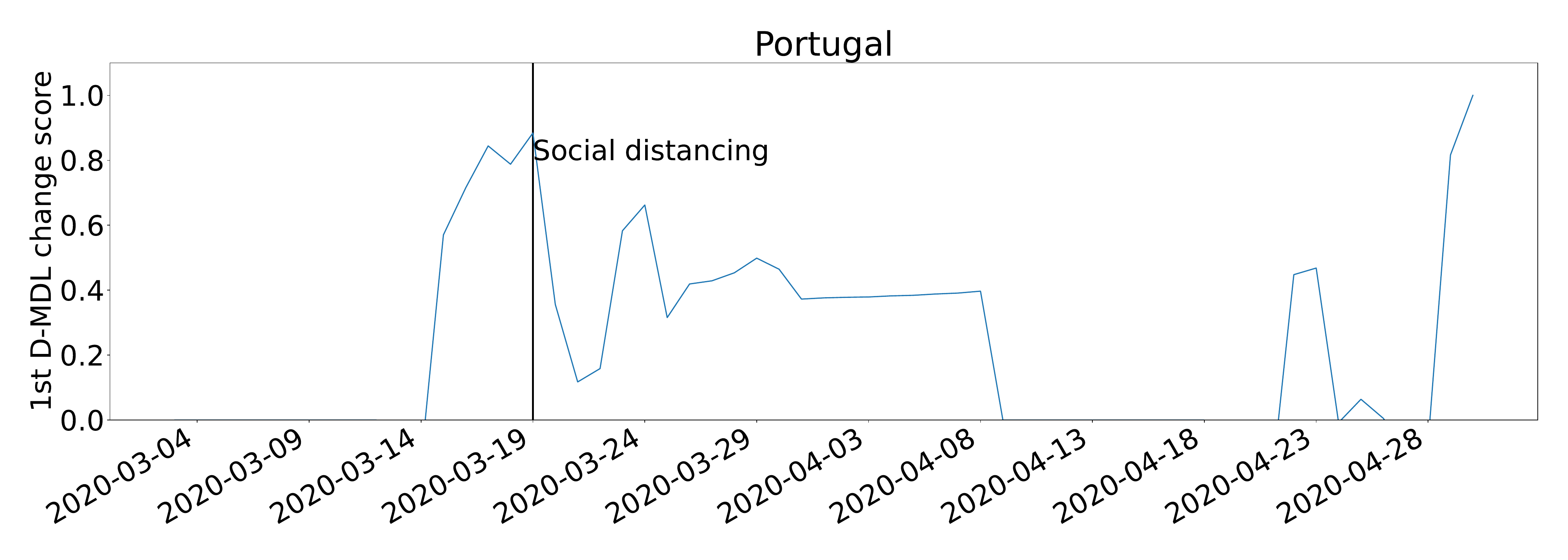} \\
			\vspace{-0.35cm}
			\textbf{e} & \includegraphics[keepaspectratio, height=3.3cm, valign=T]
			{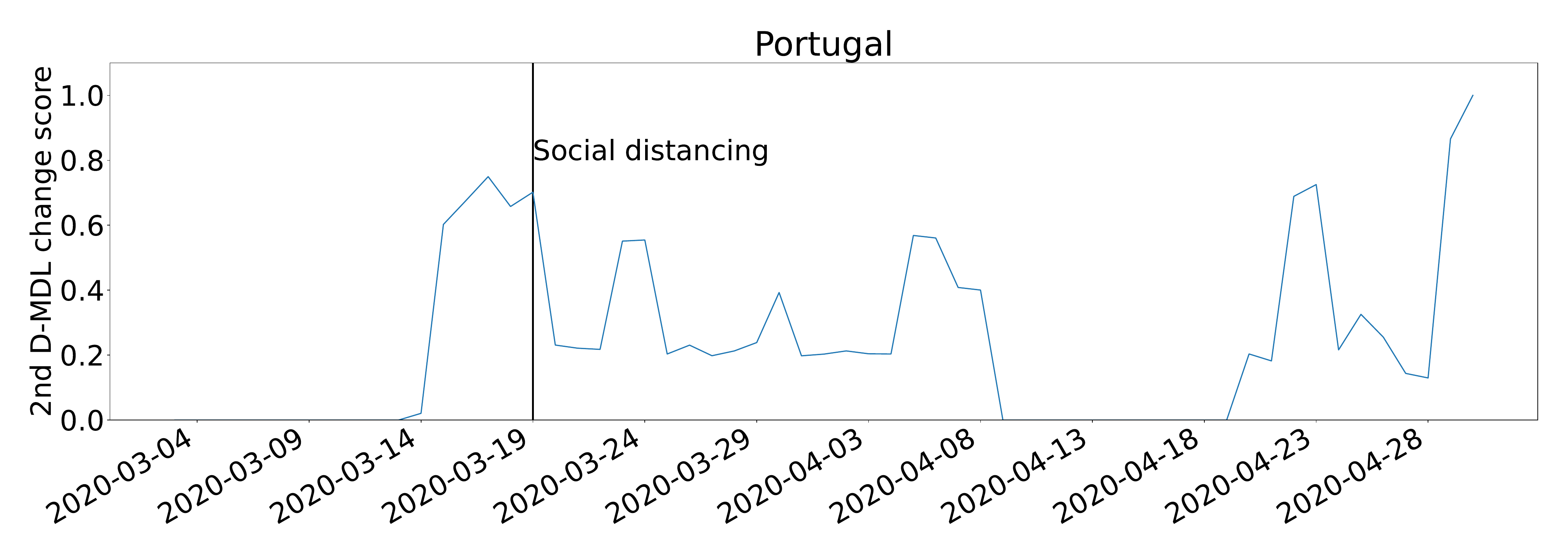} \\
		\end{tabular}
			\caption{\textbf{The results for Portugal with exponential modeling.} The date on which the social distancing was implemented is marked by a solid line in black. \textbf{a,} the number of cumulative cases. \textbf{b,} the change scores produced by the 0th M-DML where the line in blue denotes values of scores and dashed lines in red mark alarms. \textbf{c,} the window sized for the sequential D-DML algorithm with adaptive window where lines in red mark the shrinkage of windows. \textbf{d,} the change scores produced by the 1st D-MDL. \textbf{e,} the change scores produced by the 2nd D-MDL.}
\end{figure}

\begin{figure}[H] 
\centering
\begin{tabular}{cc}
		 	\textbf{a} & \includegraphics[keepaspectratio, height=3.3cm, valign=T]
			{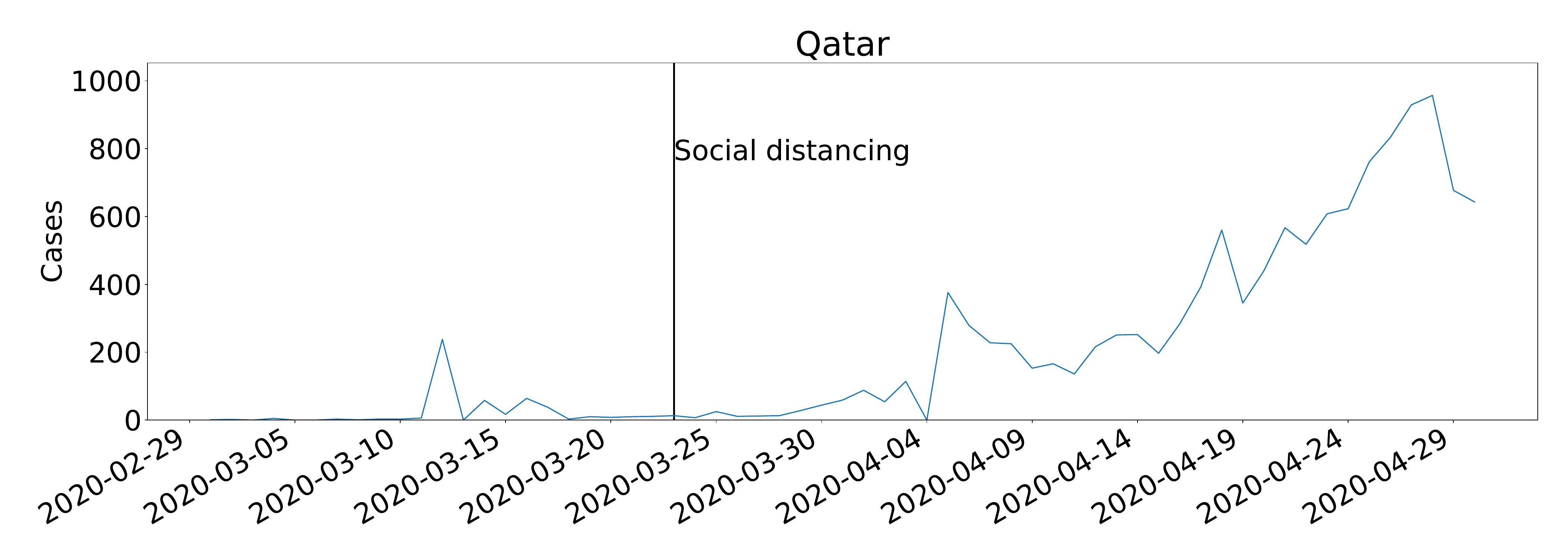} \\
			\vspace{-0.35cm}
	 	    \textbf{b} & \includegraphics[keepaspectratio, height=3.3cm, valign=T]
			{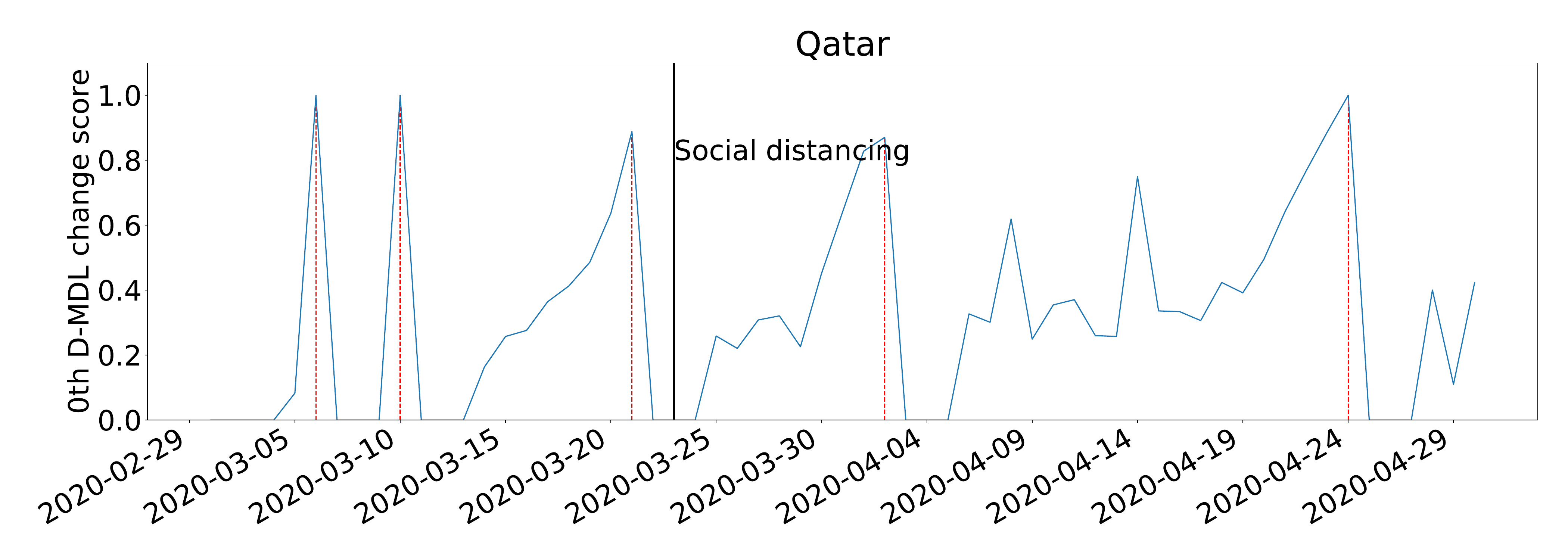}   \\
	        \vspace{-0.35cm}
			\textbf{c} & \includegraphics[keepaspectratio, height=3.3cm, valign=T]
			{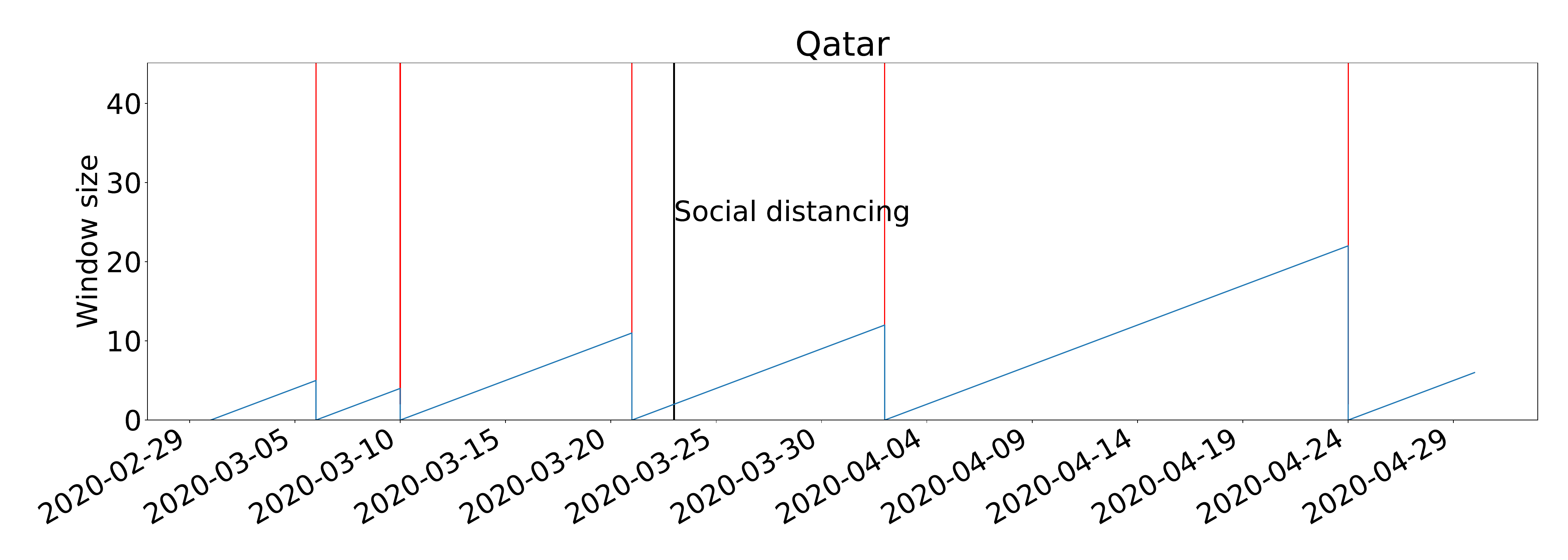} \\
		    \vspace{-0.35cm}
			\textbf{d} & \includegraphics[keepaspectratio, height=3.3cm, valign=T]
			{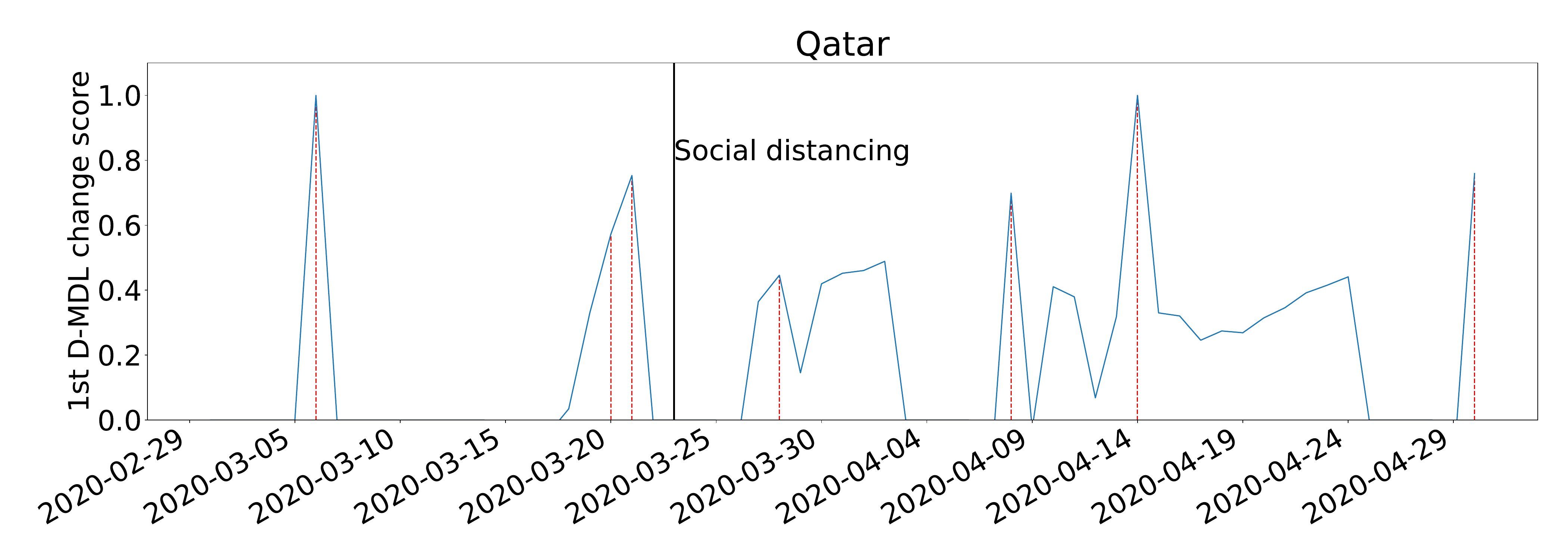} \\
		    \vspace{-0.35cm}
			\textbf{e} & \includegraphics[keepaspectratio, height=3.3cm, valign=T]
			{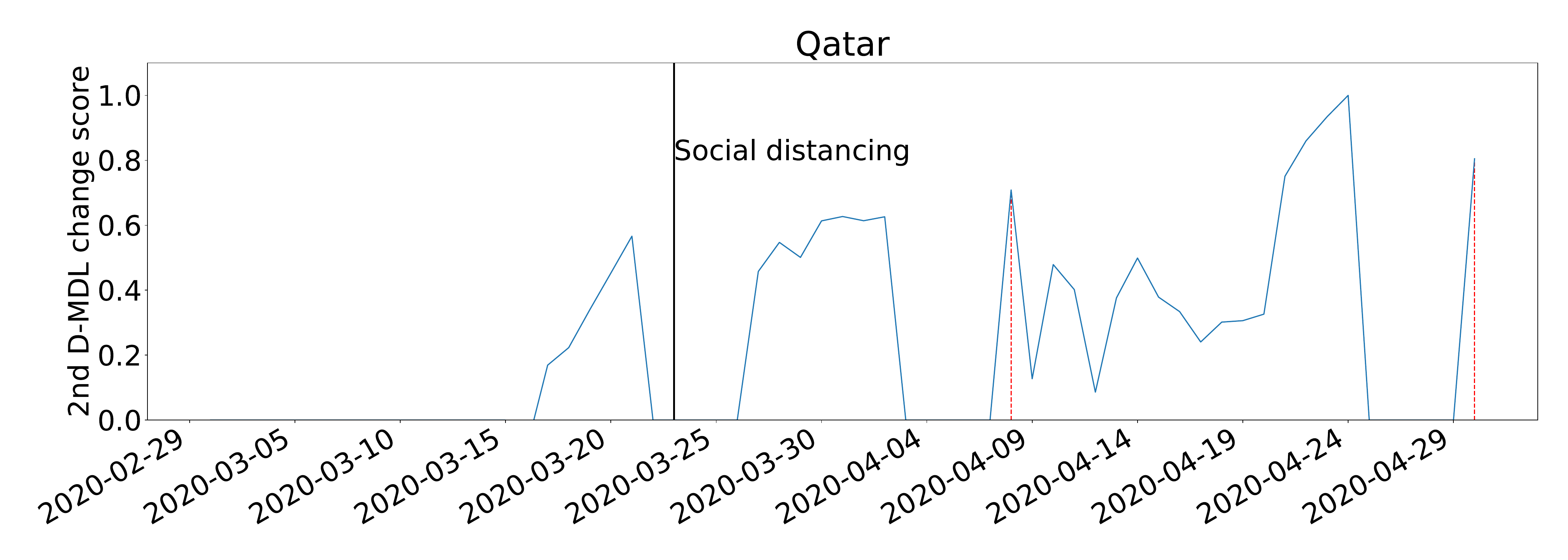} \\
		\end{tabular}
			\caption{\textbf{The results for Qatar with Gaussian modeling.} The date on which the social distancing was implemented is marked by a solid line in black. \textbf{a,} the number of daily new cases. \textbf{b,} the change scores produced by the 0th M-DML where the line in blue denotes values of scores and dashed lines in red mark alarms. \textbf{c,} the window sized for the sequential D-DML algorithm with adaptive window where lines in red mark the shrinkage of windows. \textbf{d,} the change scores produced by the 1st D-MDL. \textbf{e,} the change scores produced by the 2nd D-MDL.}
\end{figure}
\clearpage
\begin{figure}[H]  
\centering
\begin{tabular}{cc}
			\textbf{a} & \includegraphics[keepaspectratio, height=3.3cm, valign=T]
			{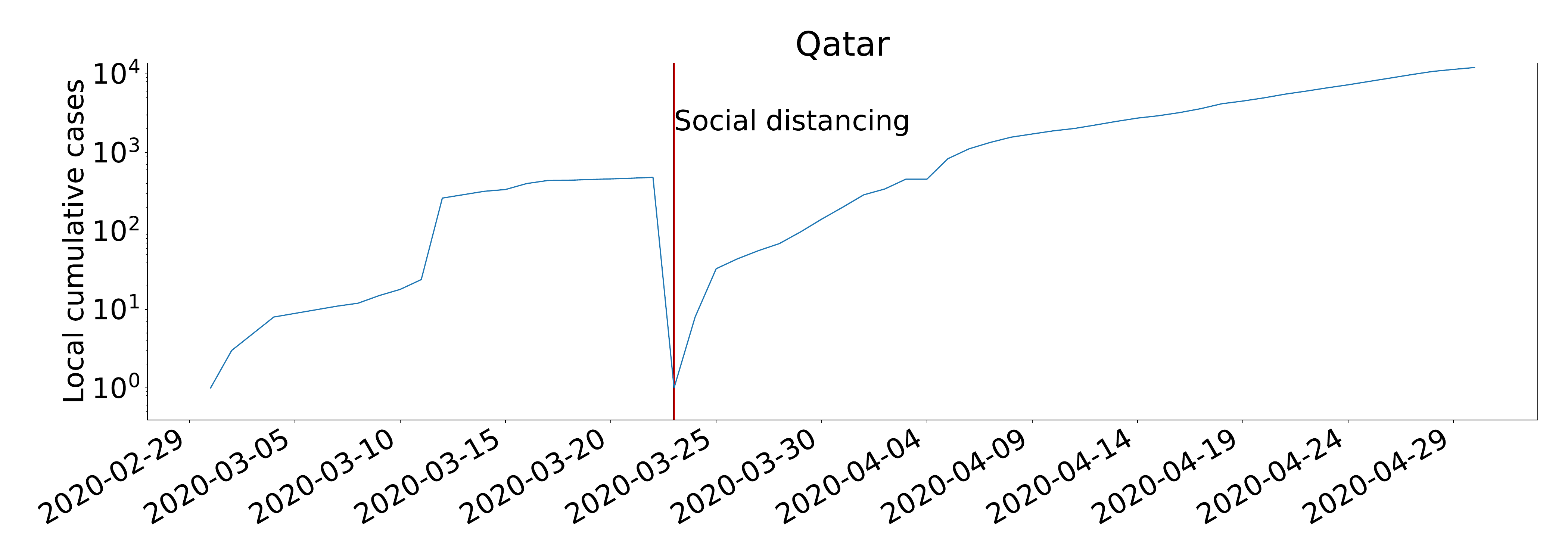} \\
	        \vspace{-0.35cm}
            \textbf{b} & \includegraphics[keepaspectratio, height=3.3cm, valign=T]
			{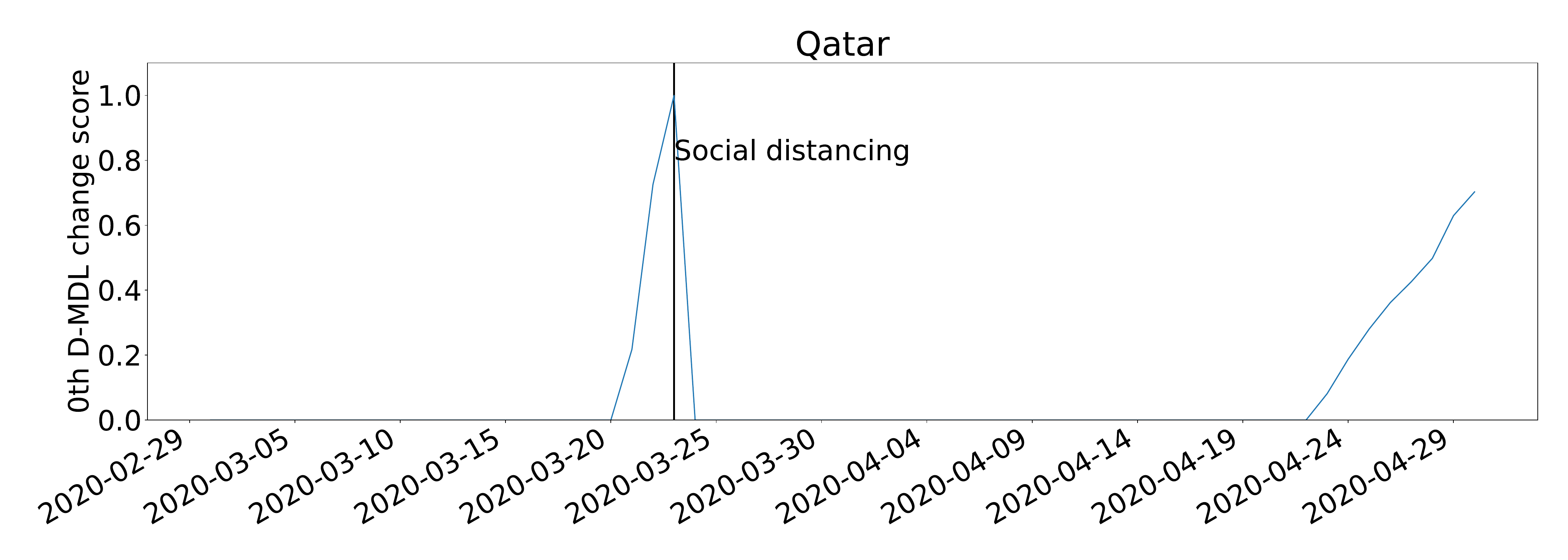}   \\
            \vspace{-0.35cm}
            \textbf{c} & \includegraphics[keepaspectratio, height=3.3cm, valign=T]
			{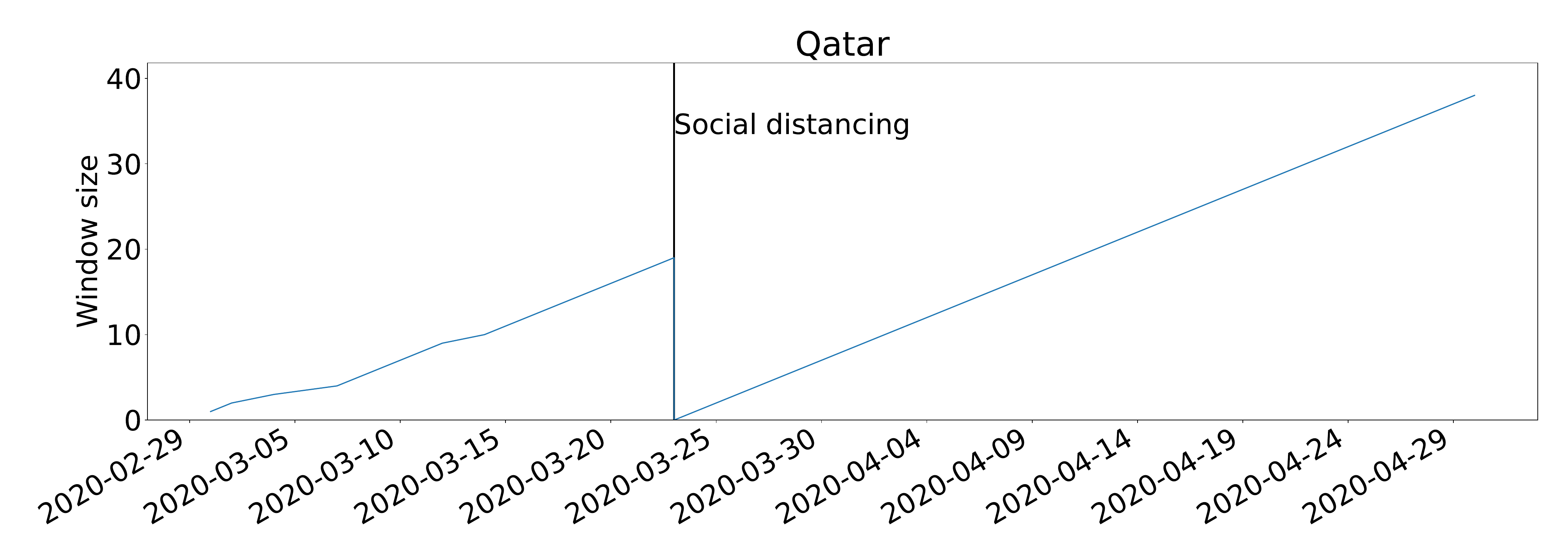} \\
			\vspace{-0.35cm}
			\textbf{d} & \includegraphics[keepaspectratio, height=3.3cm, valign=T]
			{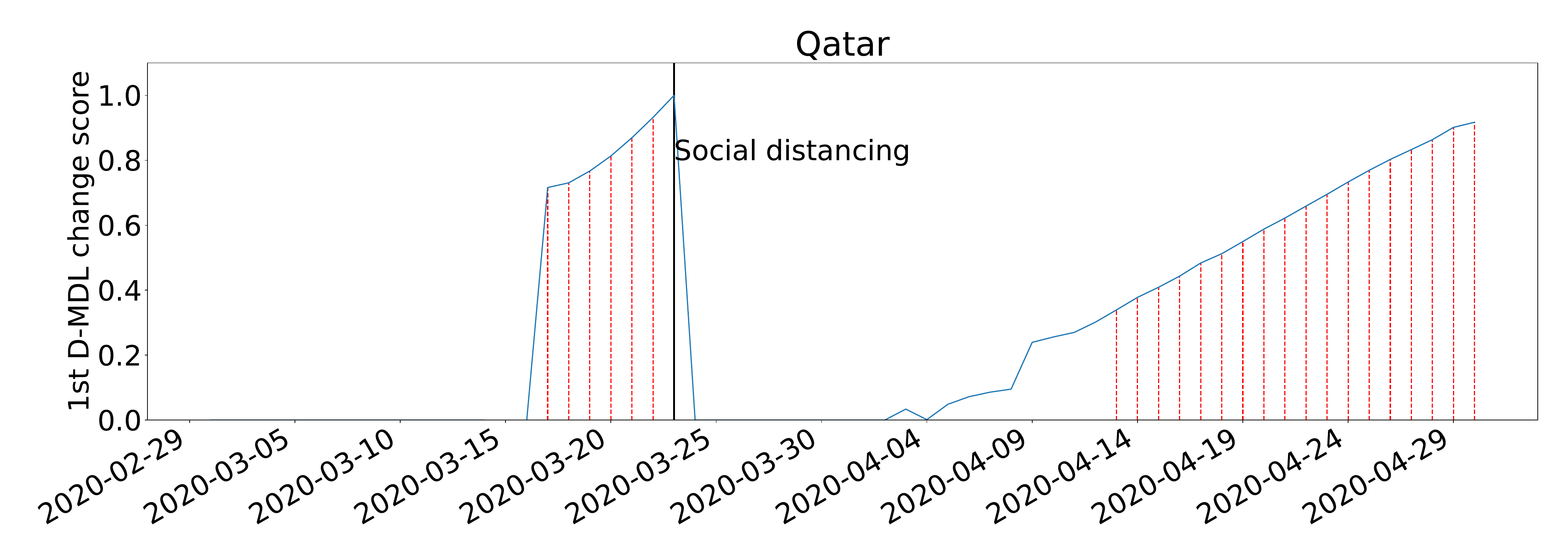} \\
			\vspace{-0.35cm}
			\textbf{e} & \includegraphics[keepaspectratio, height=3.3cm, valign=T]
			{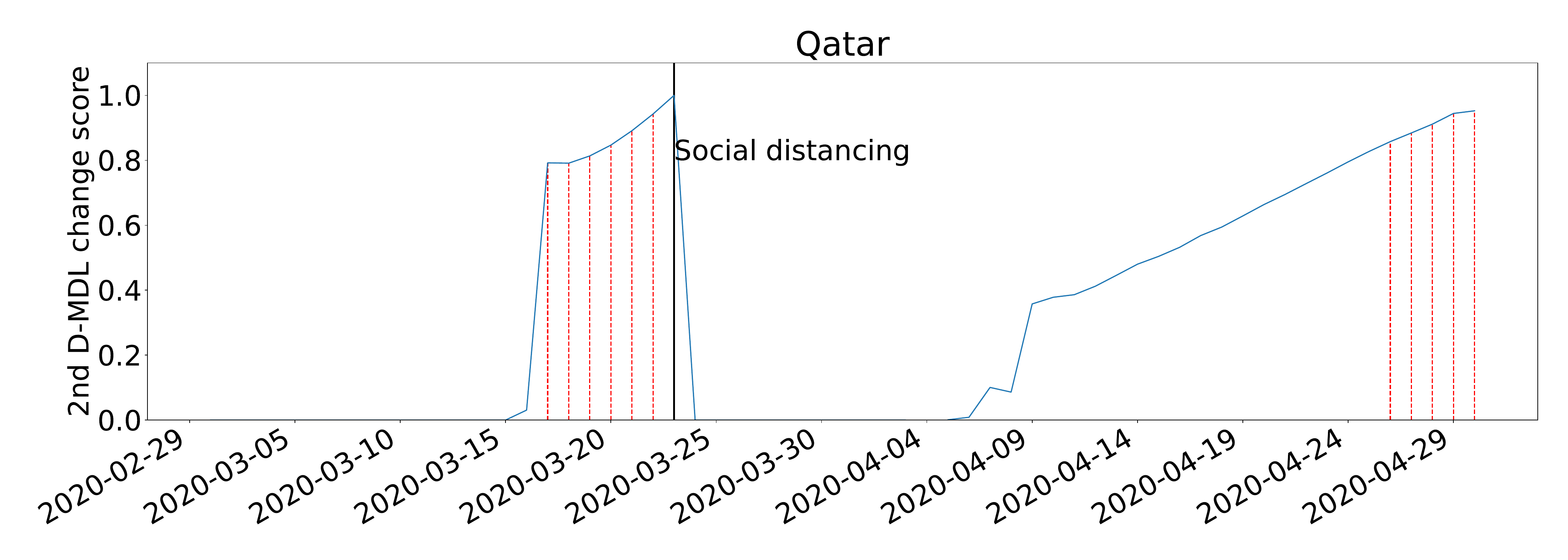} \\
		\end{tabular}
			\caption{\textbf{The results for Qatar with exponential modeling.} The date on which the social distancing was implemented is marked by a solid line in black. \textbf{a,} the number of cumulative cases. \textbf{b,} the change scores produced by the 0th M-DML where the line in blue denotes values of scores and dashed lines in red mark alarms. \textbf{c,} the window sized for the sequential D-DML algorithm with adaptive window where lines in red mark the shrinkage of windows. \textbf{d,} the change scores produced by the 1st D-MDL. \textbf{e,} the change scores produced by the 2nd D-MDL.}
			\label{exp:qatar}
\end{figure}

\begin{figure}[H] 
\centering
\begin{tabular}{cc}
		 	\textbf{a} & \includegraphics[keepaspectratio, height=3.3cm, valign=T]
			{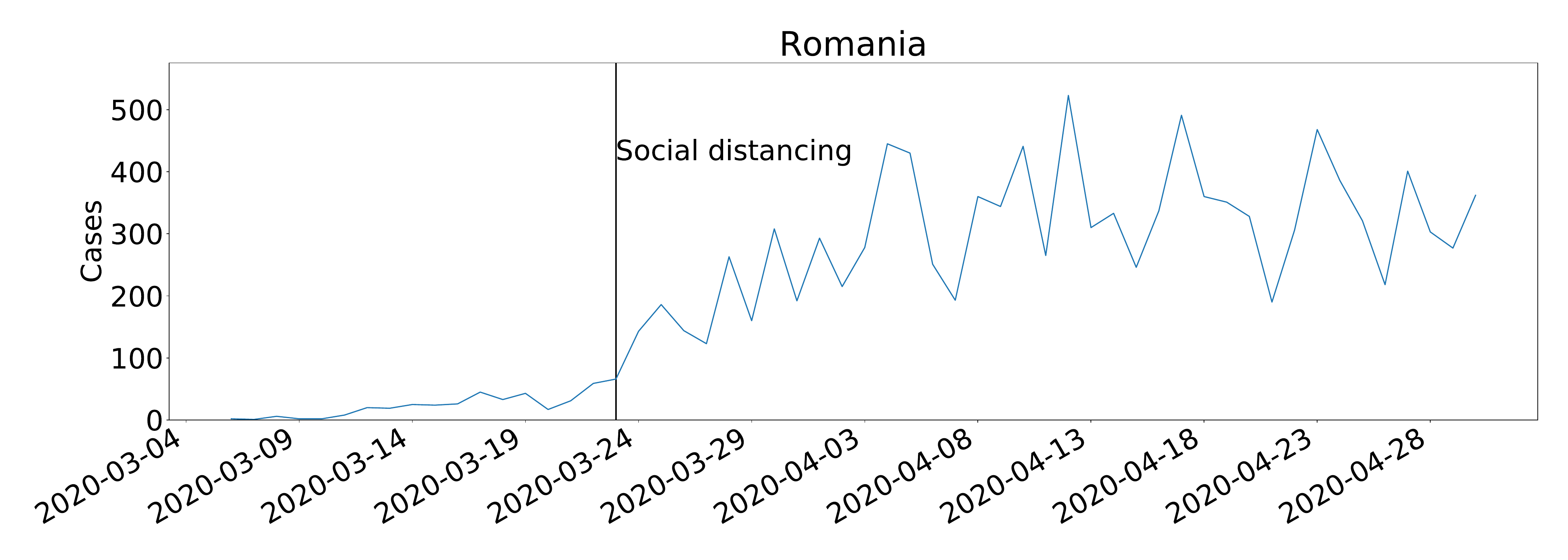} \\
			\vspace{-0.35cm}
	 	    \textbf{b} & \includegraphics[keepaspectratio, height=3.3cm, valign=T]
			{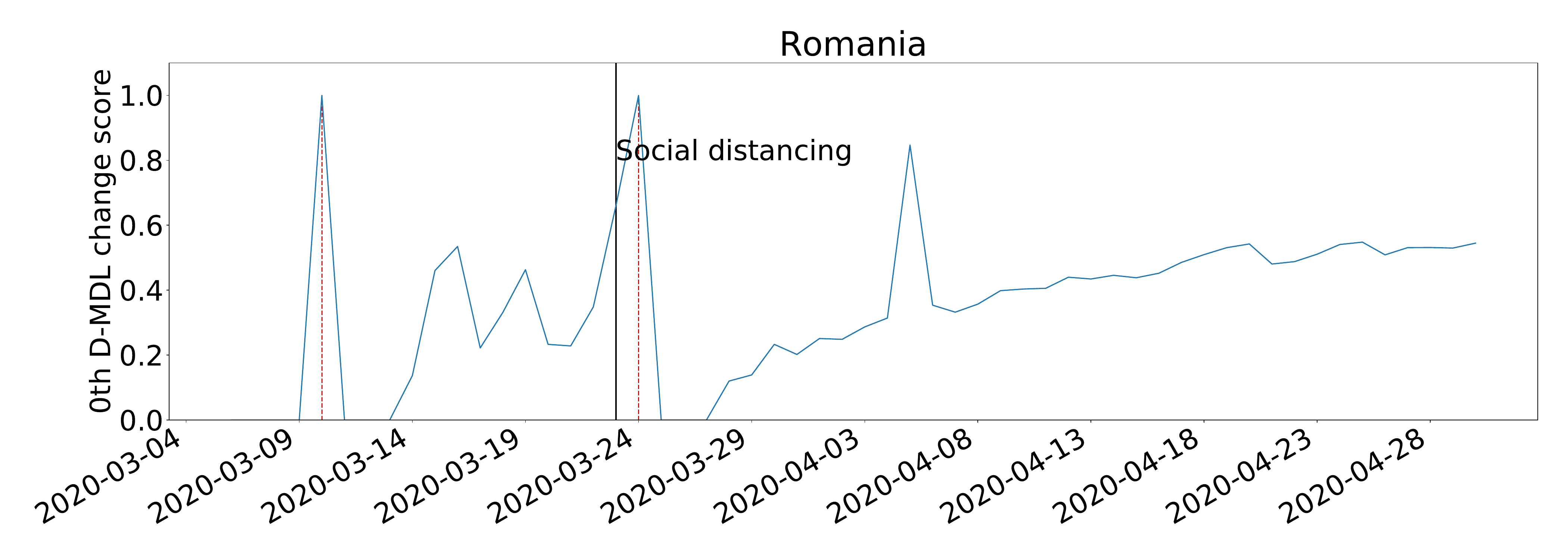}   \\
	        \vspace{-0.35cm}
			\textbf{c} & \includegraphics[keepaspectratio, height=3.3cm, valign=T]
			{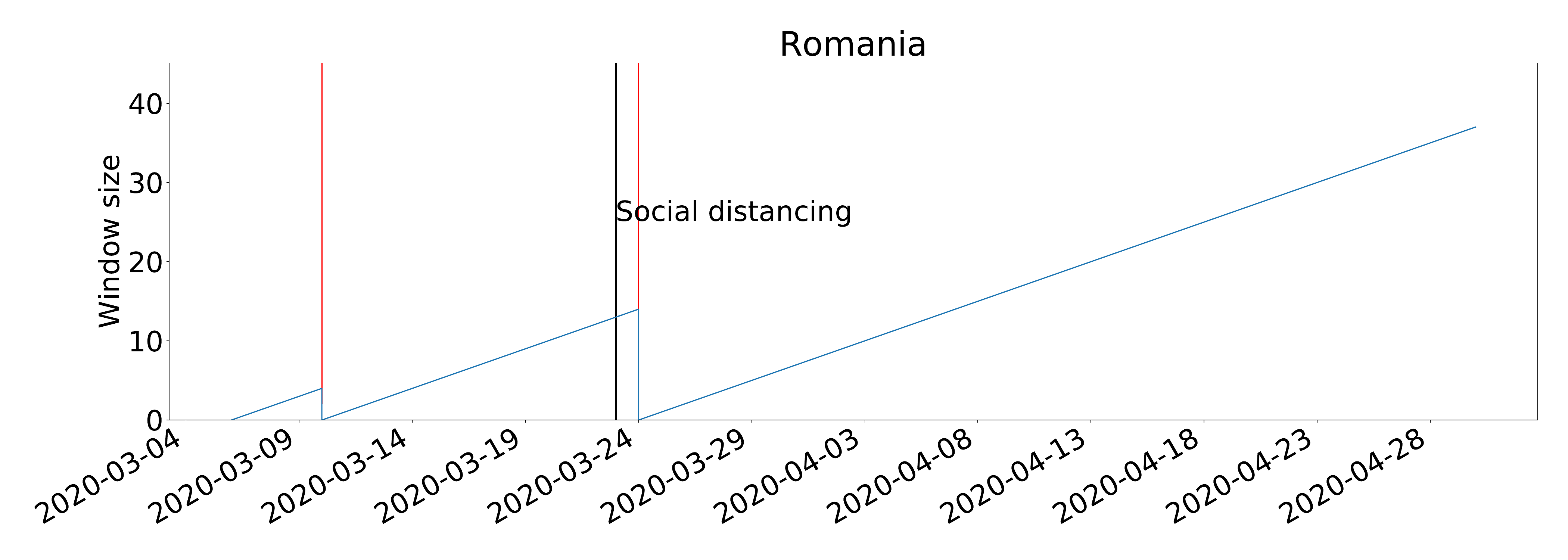} \\
		    \vspace{-0.35cm}
			\textbf{d} & \includegraphics[keepaspectratio, height=3.3cm, valign=T]
			{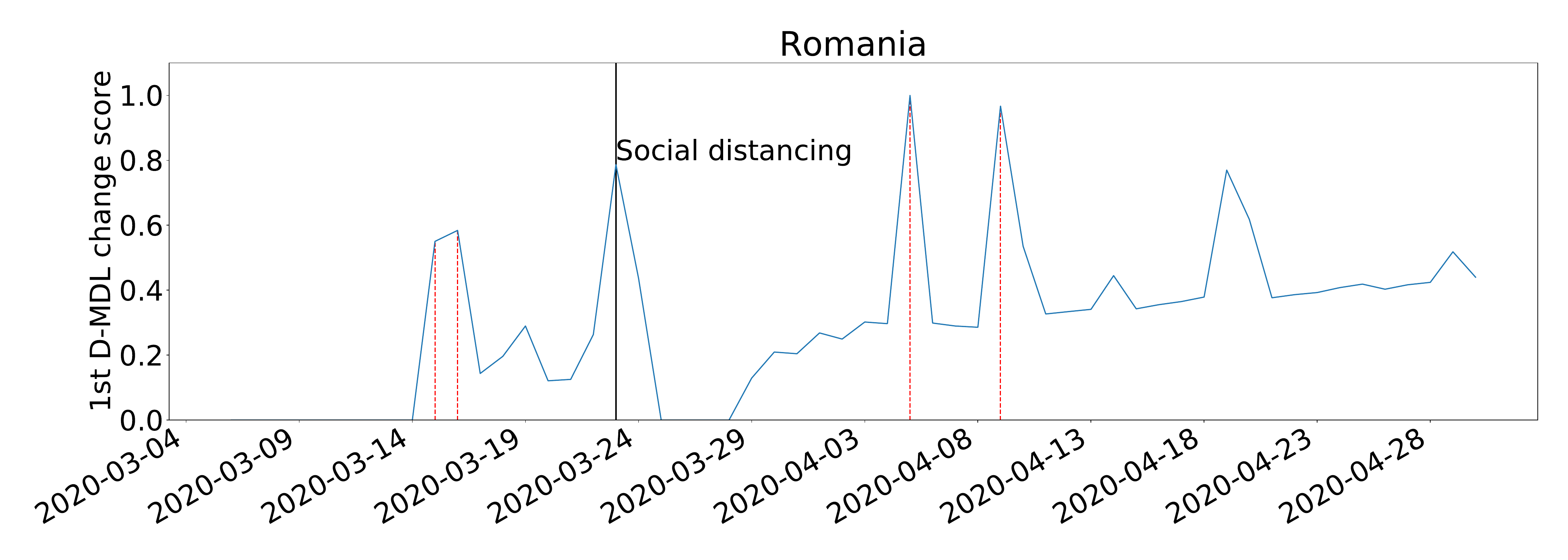} \\
		    \vspace{-0.35cm}
			\textbf{e} & \includegraphics[keepaspectratio, height=3.3cm, valign=T]
			{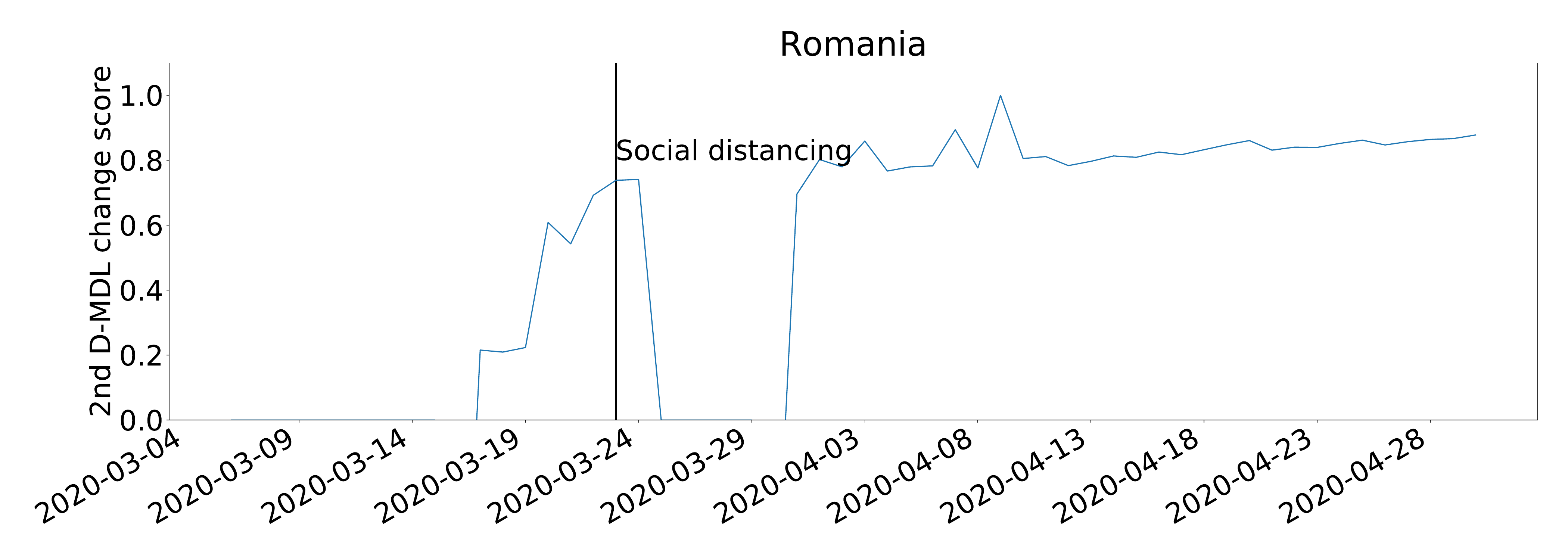} \\
		\end{tabular}
			\caption{\textbf{The results for Romania with Gaussian modeling.} The date on which the social distancing was implemented is marked by a solid line in black. \textbf{a,} the number of daily new cases. \textbf{b,} the change scores produced by the 0th M-DML where the line in blue denotes values of scores and dashed lines in red mark alarms. \textbf{c,} the window sized for the sequential D-DML algorithm with adaptive window where lines in red mark the shrinkage of windows. \textbf{d,} the change scores produced by the 1st D-MDL. \textbf{e,} the change scores produced by the 2nd D-MDL.}
\end{figure}

\begin{figure}[H]  
\centering
\begin{tabular}{cc}
			\textbf{a} & \includegraphics[keepaspectratio, height=3.3cm, valign=T]
			{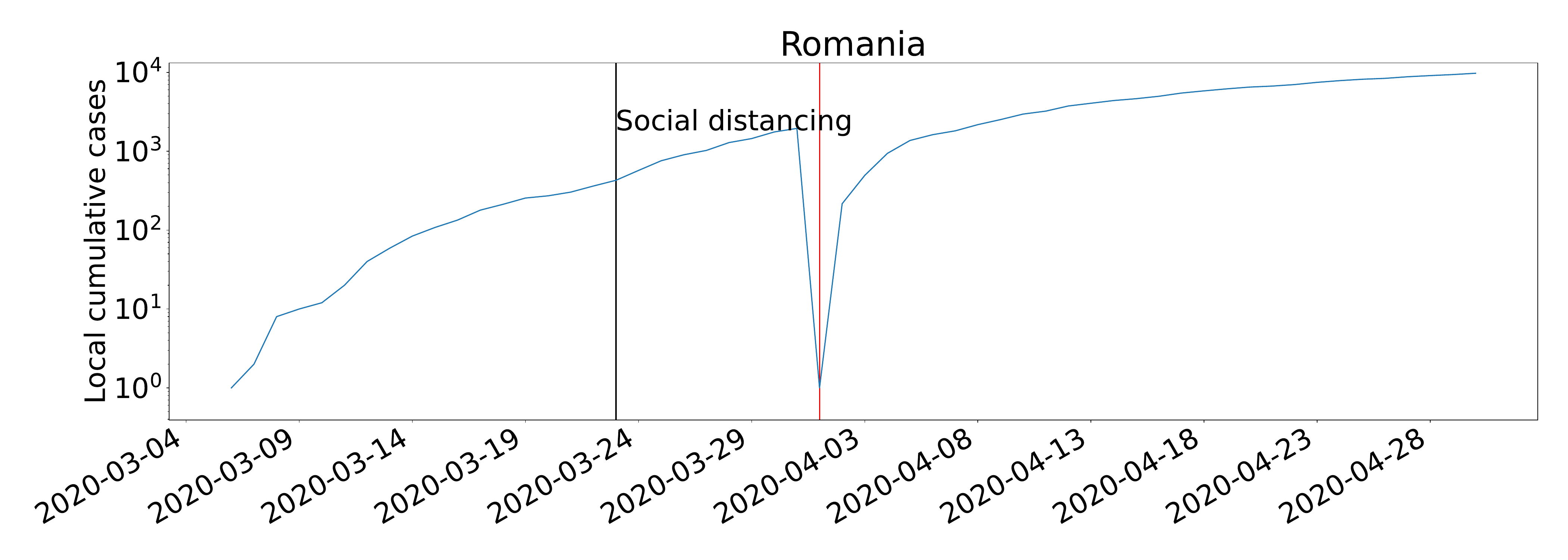} \\
	        \vspace{-0.35cm}
            \textbf{b} & \includegraphics[keepaspectratio, height=3.3cm, valign=T]
			{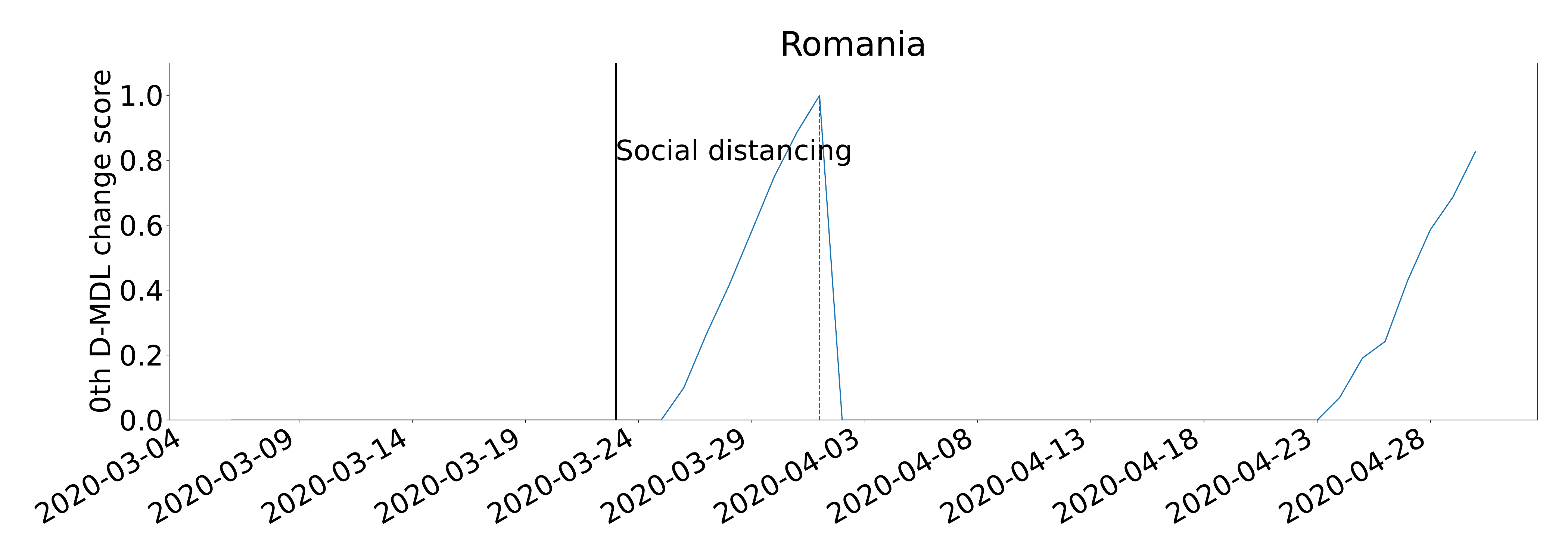}   \\
            \vspace{-0.35cm}
            \textbf{c} & \includegraphics[keepaspectratio, height=3.3cm, valign=T]
			{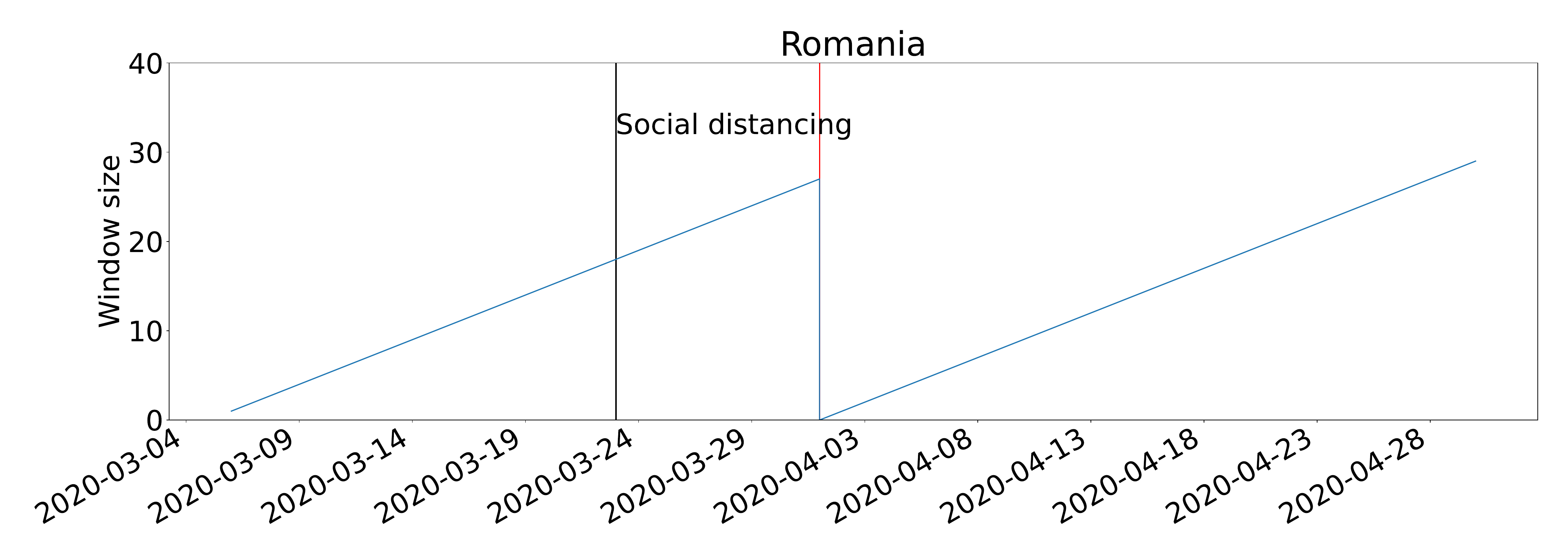} \\
			\vspace{-0.35cm}
			\textbf{d} & \includegraphics[keepaspectratio, height=3.3cm, valign=T]
			{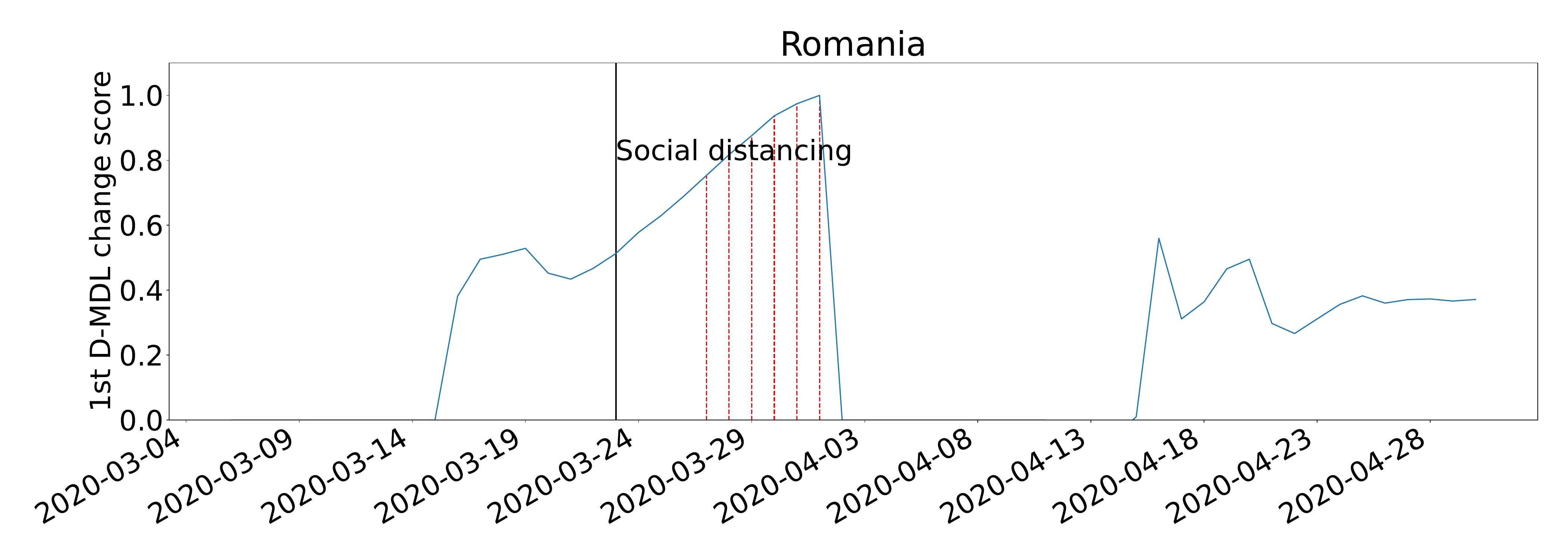} \\
			\vspace{-0.35cm}
			\textbf{e} & \includegraphics[keepaspectratio, height=3.3cm, valign=T]
			{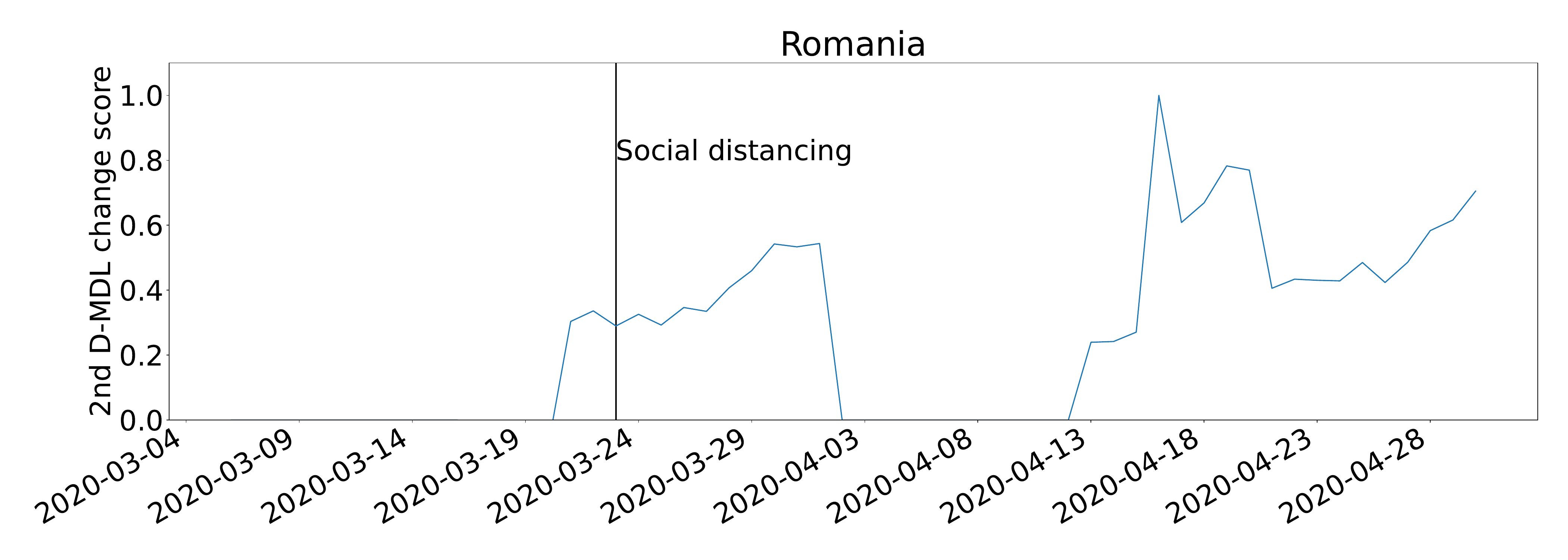} \\
		\end{tabular}
			\caption{\textbf{The results for Romania with exponential modeling.} The date on which the social distancing was implemented is marked by a solid line in black. \textbf{a,} the number of cumulative cases. \textbf{b,} the change scores produced by the 0th M-DML where the line in blue denotes values of scores and dashed lines in red mark alarms. \textbf{c,} the window sized for the sequential D-DML algorithm with adaptive window where lines in red mark the shrinkage of windows. \textbf{d,} the change scores produced by the 1st D-MDL. \textbf{e,} the change scores produced by the 2nd D-MDL.}
\end{figure}

\begin{figure}[H] 
\centering
\begin{tabular}{cc}
		 	\textbf{a} & \includegraphics[keepaspectratio, height=3.3cm, valign=T]
			{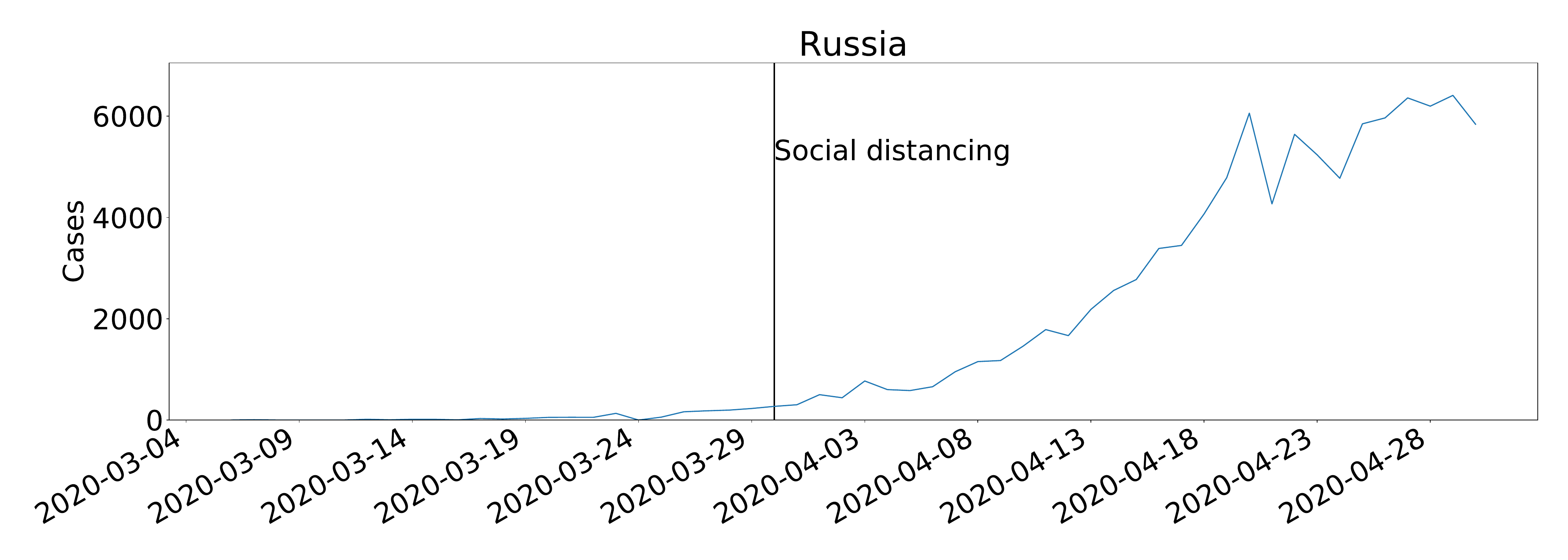} \\
			\vspace{-0.35cm}
	 	    \textbf{b} & \includegraphics[keepaspectratio, height=3.3cm, valign=T]
			{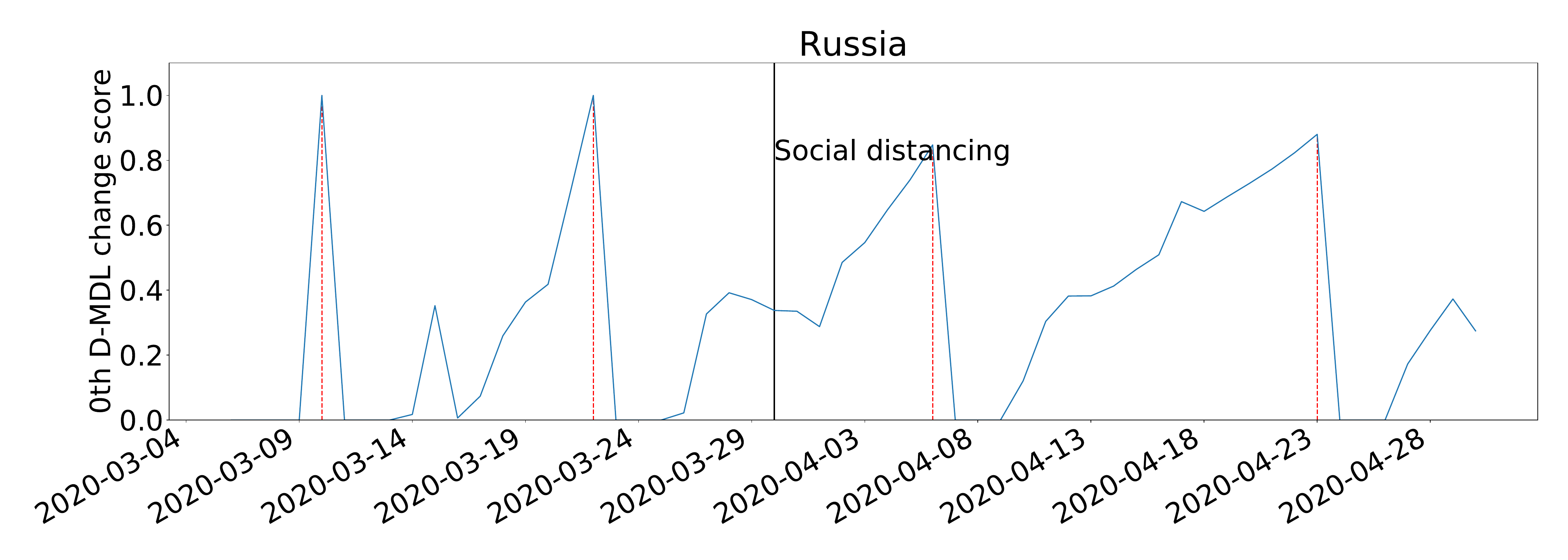}   \\
	        \vspace{-0.35cm}
			\textbf{c} & \includegraphics[keepaspectratio, height=3.3cm, valign=T]
			{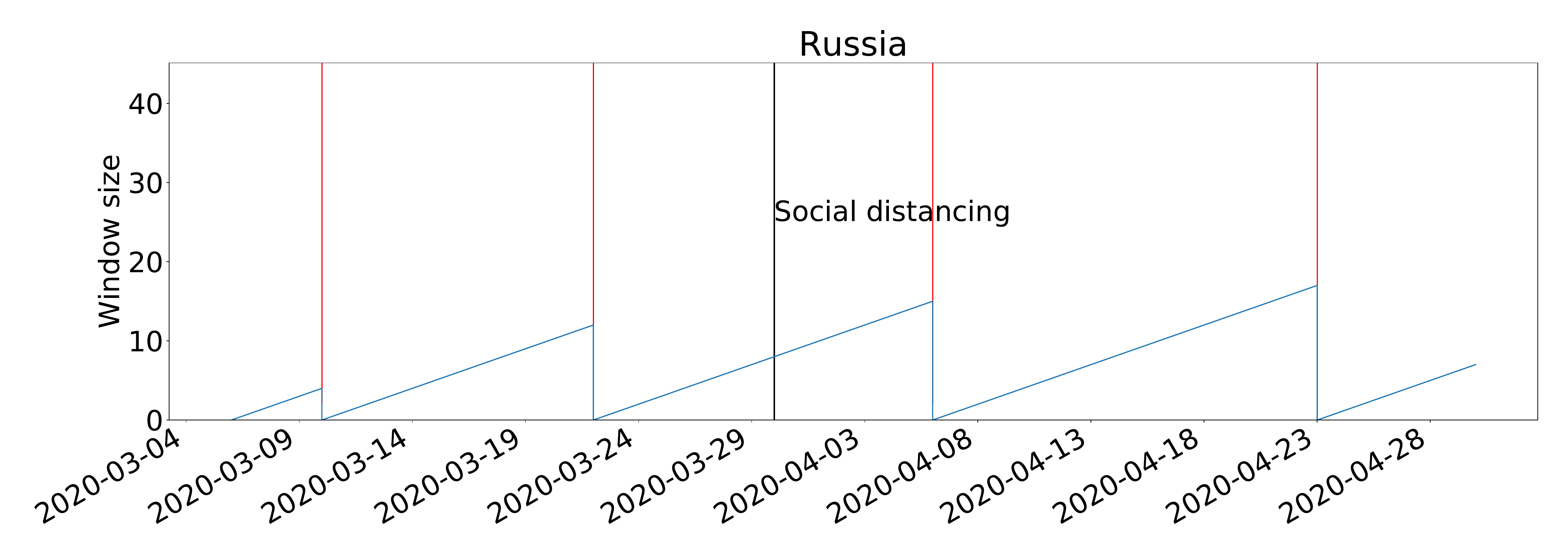} \\
		    \vspace{-0.35cm}
			\textbf{d} & \includegraphics[keepaspectratio, height=3.3cm, valign=T]
			{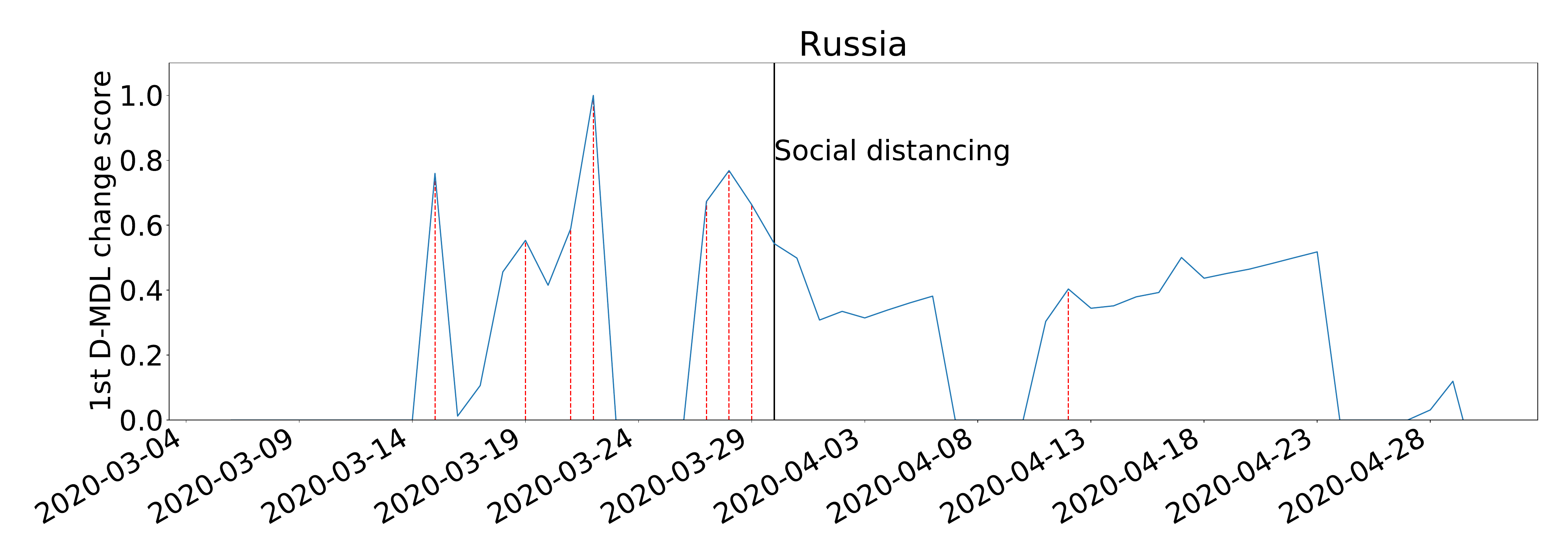} \\
		    \vspace{-0.35cm}
			\textbf{e} & \includegraphics[keepaspectratio, height=3.3cm, valign=T]
			{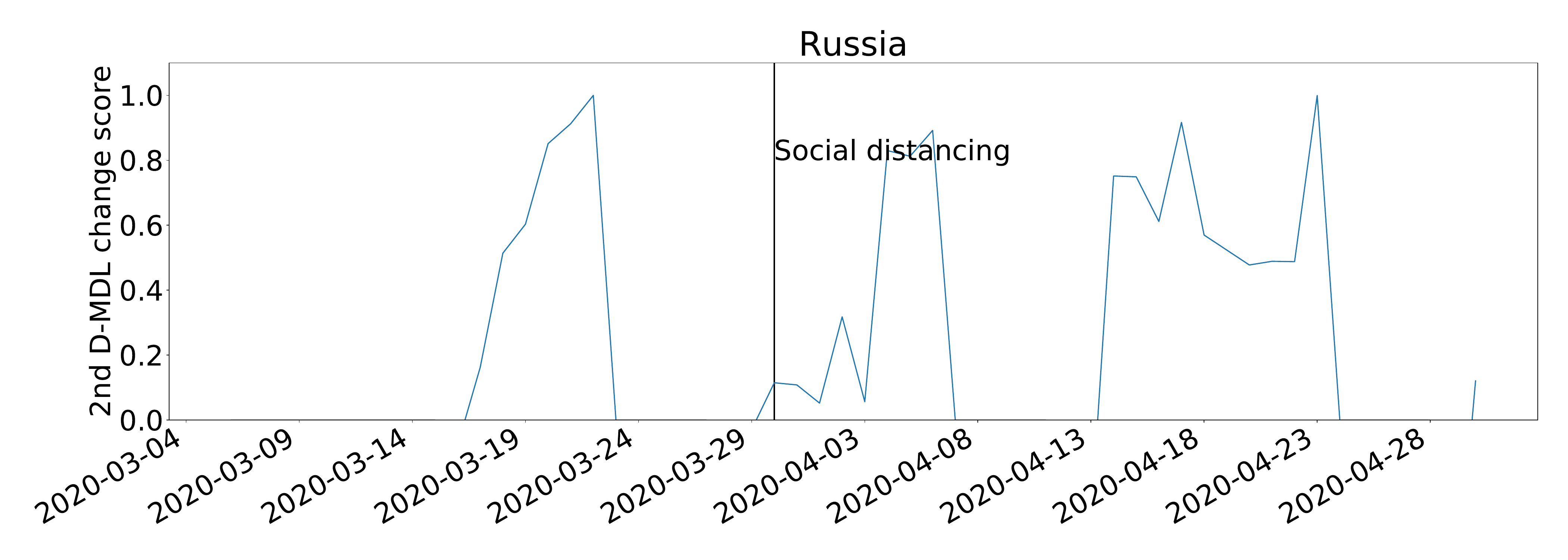} \\
		\end{tabular}
			\caption{\textbf{The results for Russia with Gaussian modeling.} The date on which the social distancing was implemented is marked by a solid line in black. \textbf{a,} the number of daily new cases. \textbf{b,} the change scores produced by the 0th M-DML where the line in blue denotes values of scores and dashed lines in red mark alarms. \textbf{c,} the window sized for the sequential D-DML algorithm with adaptive window where lines in red mark the shrinkage of windows. \textbf{d,} the change scores produced by the 1st D-MDL. \textbf{e,} the change scores produced by the 2nd D-MDL.}
\end{figure}

\begin{figure}[H]  
\centering
\begin{tabular}{cc}
			\textbf{a} & \includegraphics[keepaspectratio, height=3.3cm, valign=T]
			{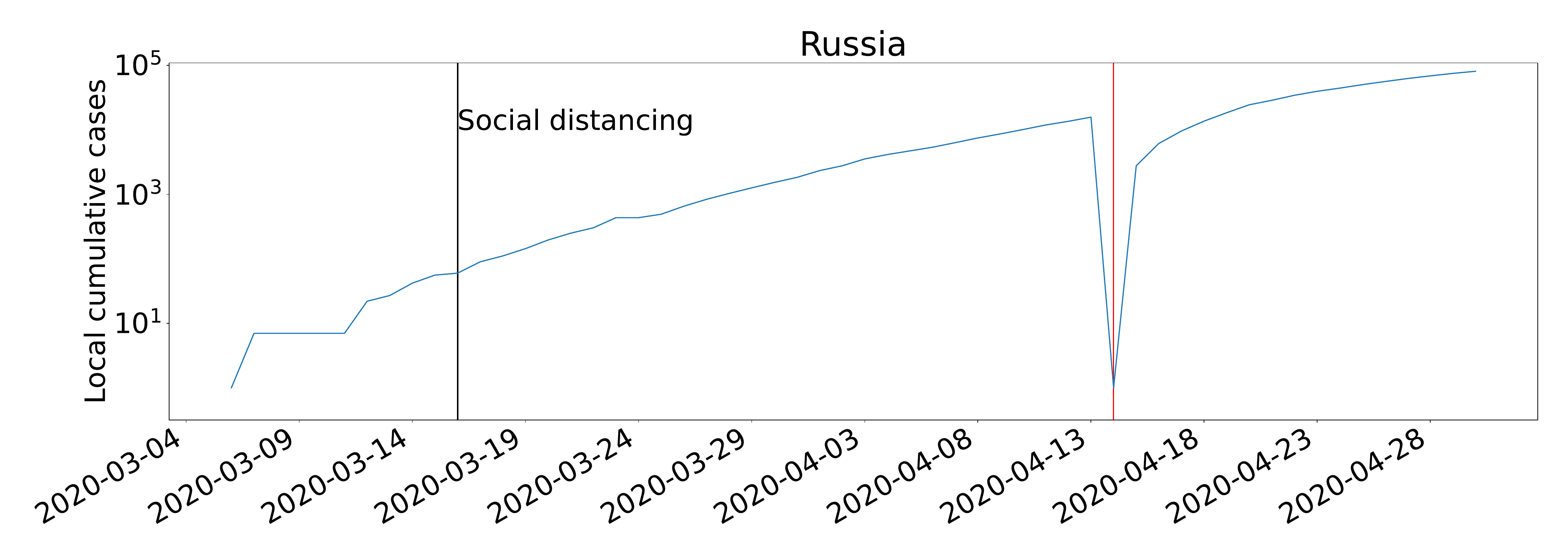} \\
	        \vspace{-0.35cm}
            \textbf{b} & \includegraphics[keepaspectratio, height=3.3cm, valign=T]
			{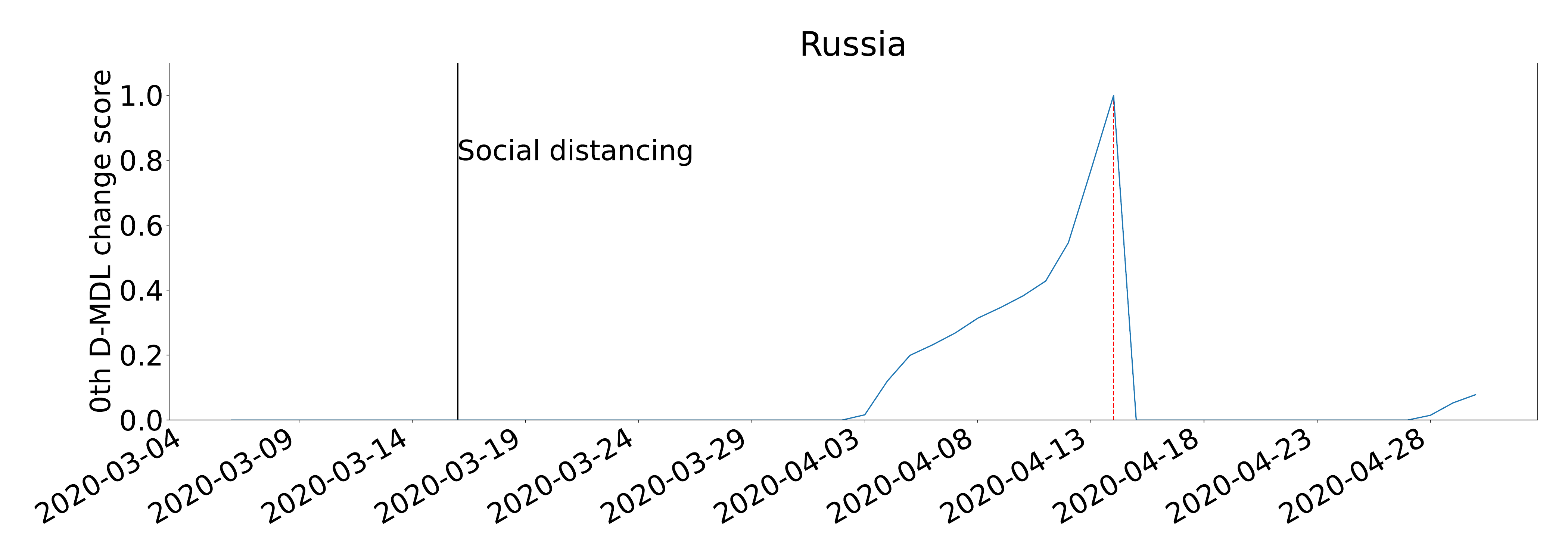}   \\
            \vspace{-0.35cm}
            \textbf{c} & \includegraphics[keepaspectratio, height=3.3cm, valign=T]
			{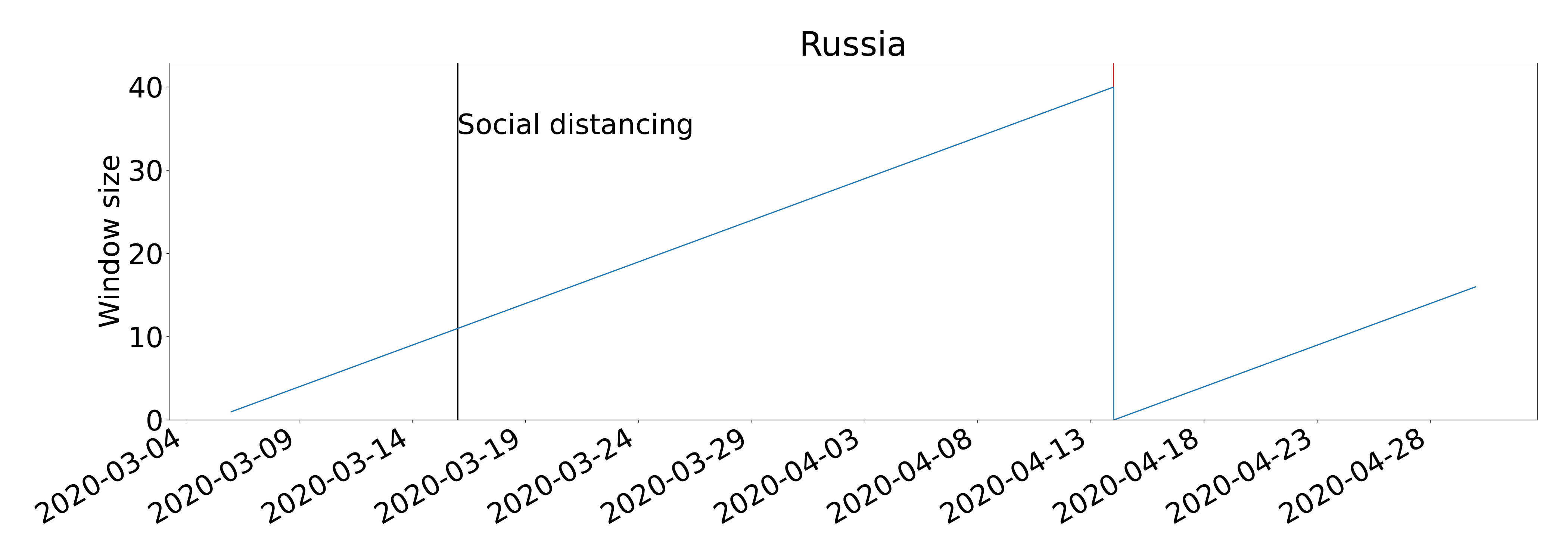} \\
			\vspace{-0.35cm}
			\textbf{d} & \includegraphics[keepaspectratio, height=3.3cm, valign=T]
			{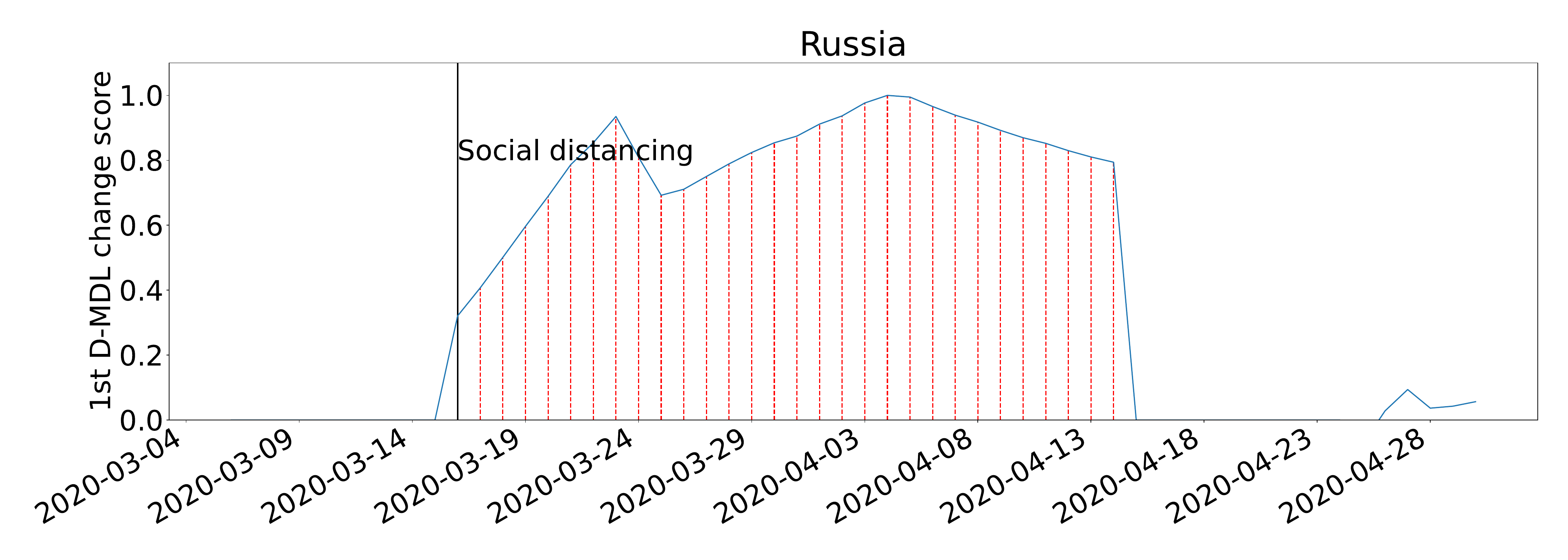} \\
			\vspace{-0.35cm}
			\textbf{e} & \includegraphics[keepaspectratio, height=3.3cm, valign=T]
			{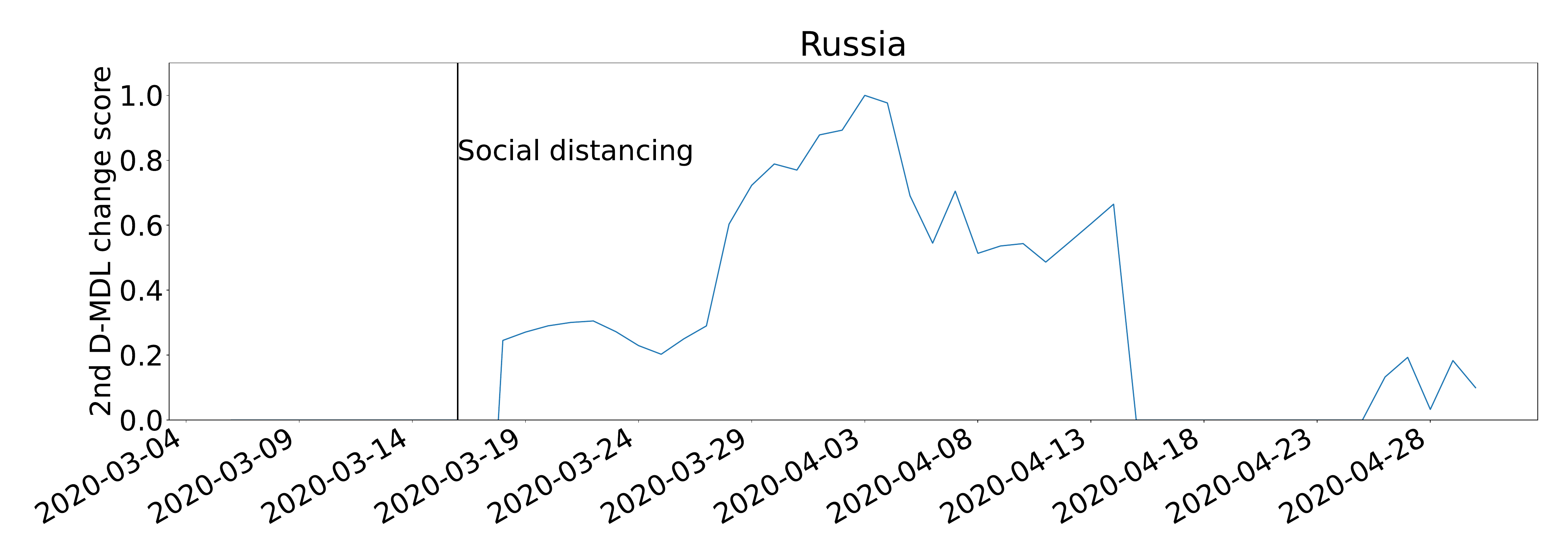} \\
		\end{tabular}
			\caption{\textbf{The results for Russia with exponential modeling.} The date on which the social distancing was implemented is marked by a solid line in black. \textbf{a,} the number of cumulative cases. \textbf{b,} the change scores produced by the 0th M-DML where the line in blue denotes values of scores and dashed lines in red mark alarms. \textbf{c,} the window sized for the sequential D-DML algorithm with adaptive window where lines in red mark the shrinkage of windows. \textbf{d,} the change scores produced by the 1st D-MDL. \textbf{e,} the change scores produced by the 2nd D-MDL.}
\end{figure}

\clearpage
\begin{figure}[H] 
\centering
\begin{tabular}{cc}
		 	\textbf{a} & \includegraphics[keepaspectratio, height=3.3cm, valign=T]
			{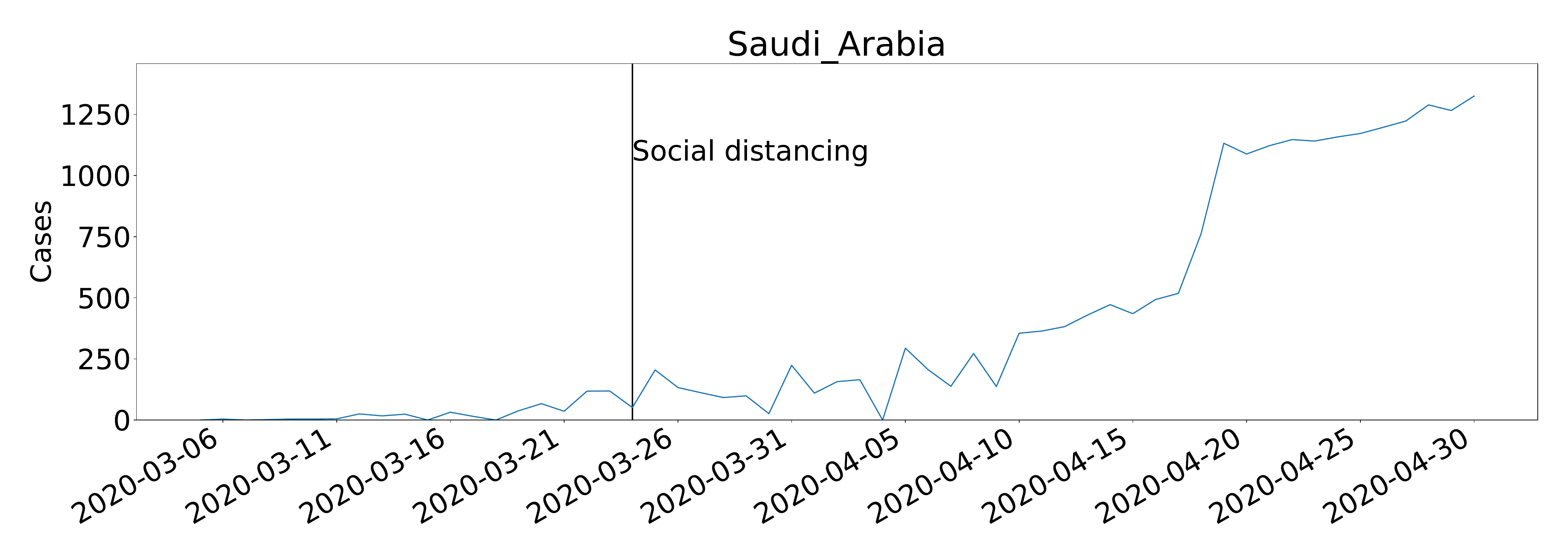} \\
			\vspace{-0.35cm}
	 	    \textbf{b} & \includegraphics[keepaspectratio, height=3.3cm, valign=T]
			{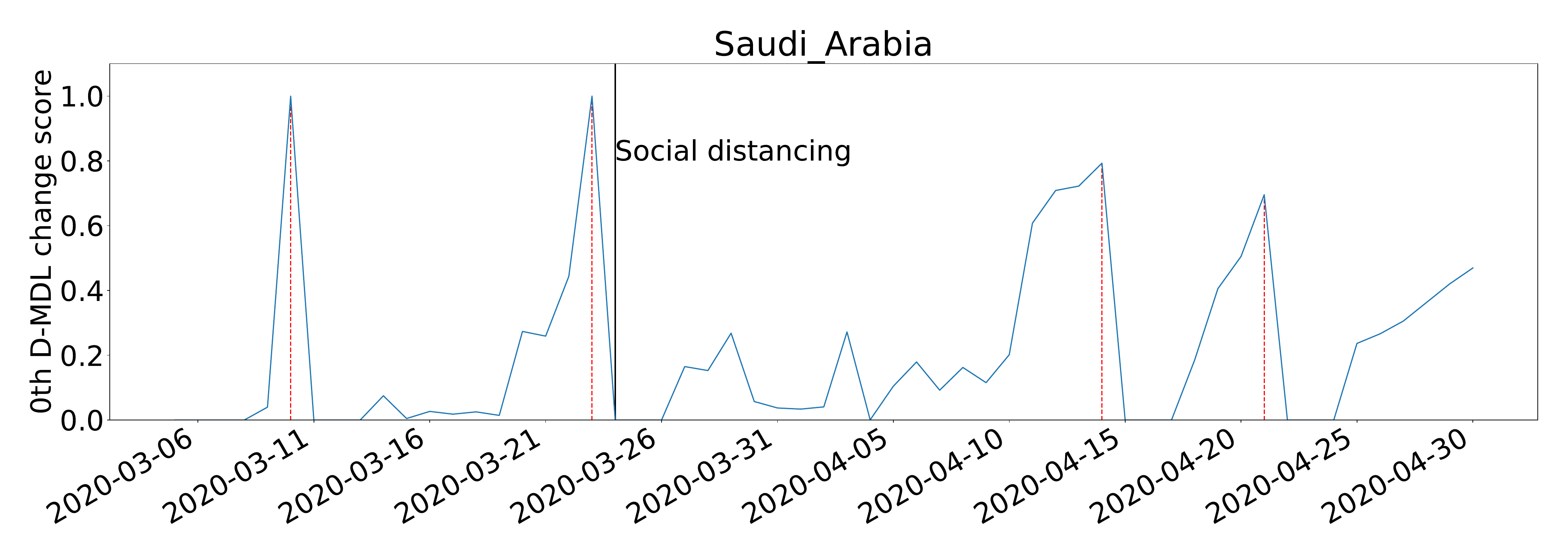}   \\
	        \vspace{-0.35cm}
			\textbf{c} & \includegraphics[keepaspectratio, height=3.3cm, valign=T]
			{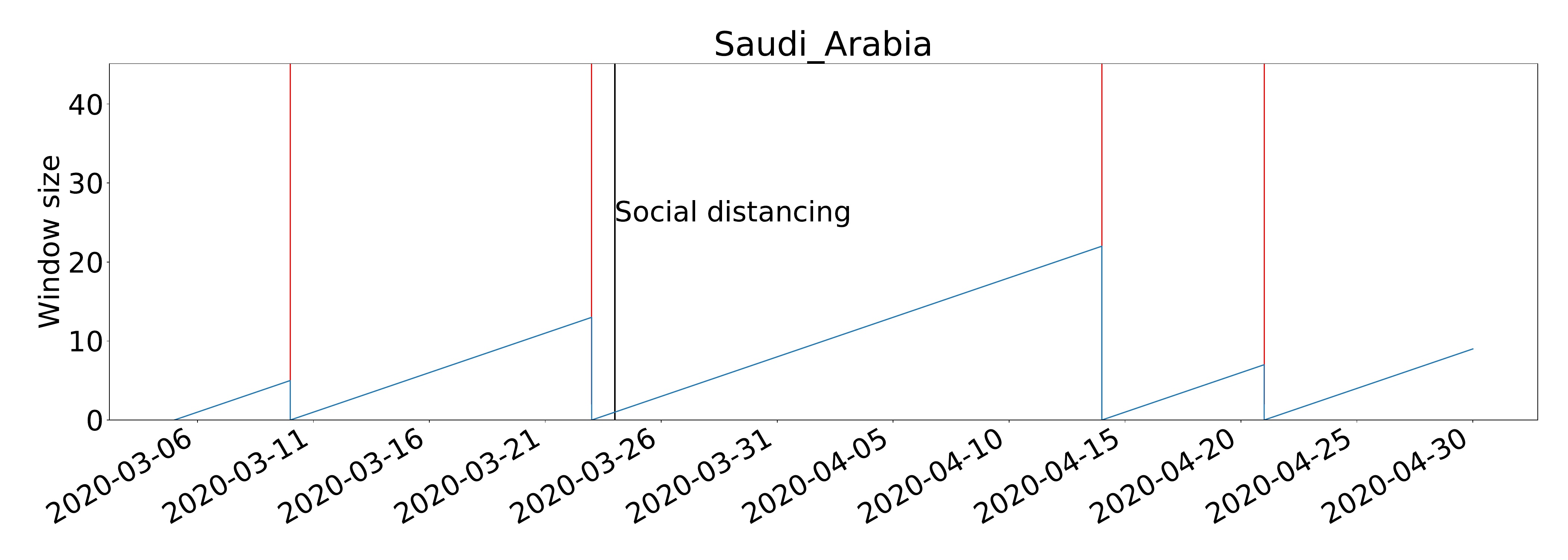} \\
		    \vspace{-0.35cm}
			\textbf{d} & \includegraphics[keepaspectratio, height=3.3cm, valign=T]
			{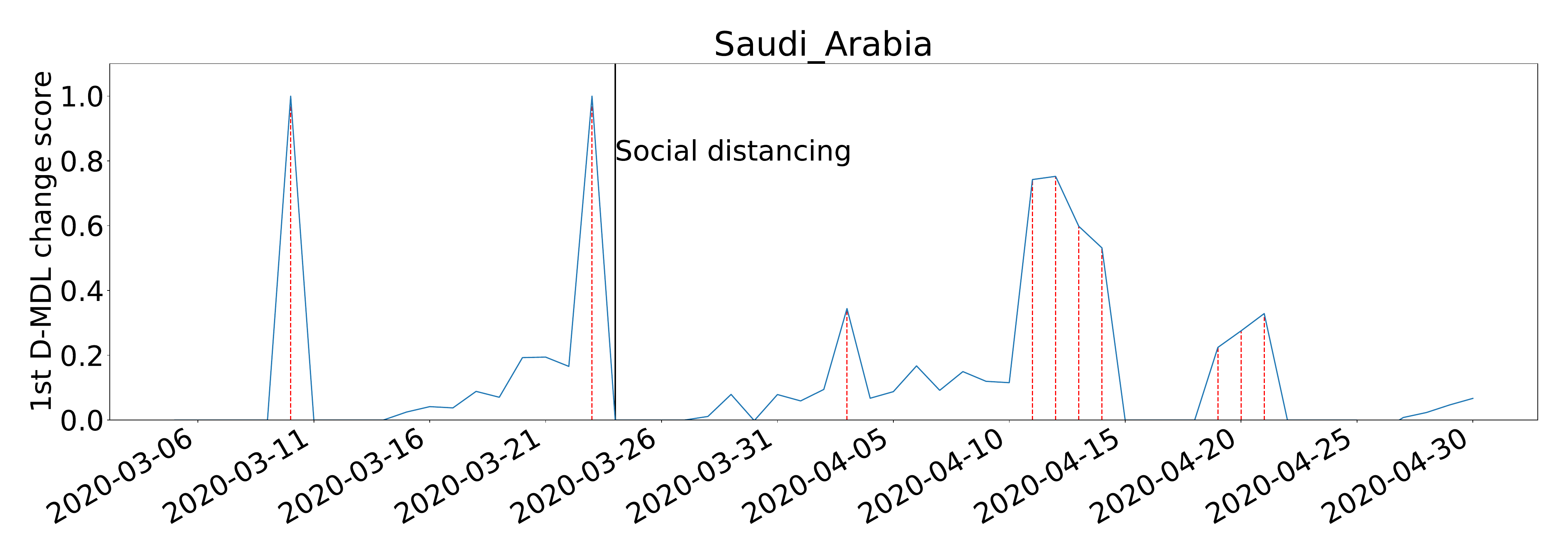} \\
		    \vspace{-0.35cm}
			\textbf{e} & \includegraphics[keepaspectratio, height=3.3cm, valign=T]
			{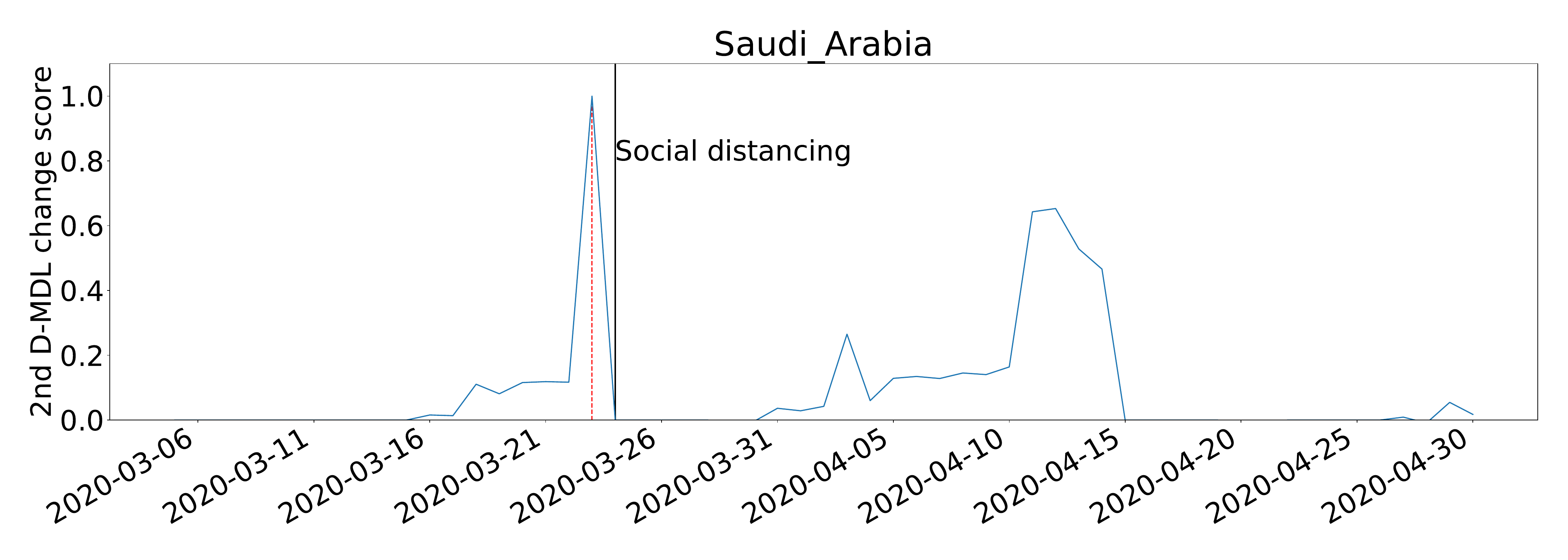} \\
		\end{tabular}
			\caption{\textbf{The results for Saudi Arabia with Gaussian modeling.} The date on which the social distancing was implemented is marked by a solid line in black. \textbf{a,} the number of daily new cases. \textbf{b,} the change scores produced by the 0th M-DML where the line in blue denotes values of scores and dashed lines in red mark alarms. \textbf{c,} the window sized for the sequential D-DML algorithm with adaptive window where lines in red mark the shrinkage of windows. \textbf{d,} the change scores produced by the 1st D-MDL. \textbf{e,} the change scores produced by the 2nd D-MDL.}
\end{figure}
\clearpage
\begin{figure}[H]  
\centering
\begin{tabular}{cc}
			\textbf{a} & \includegraphics[keepaspectratio, height=3.3cm, valign=T]
			{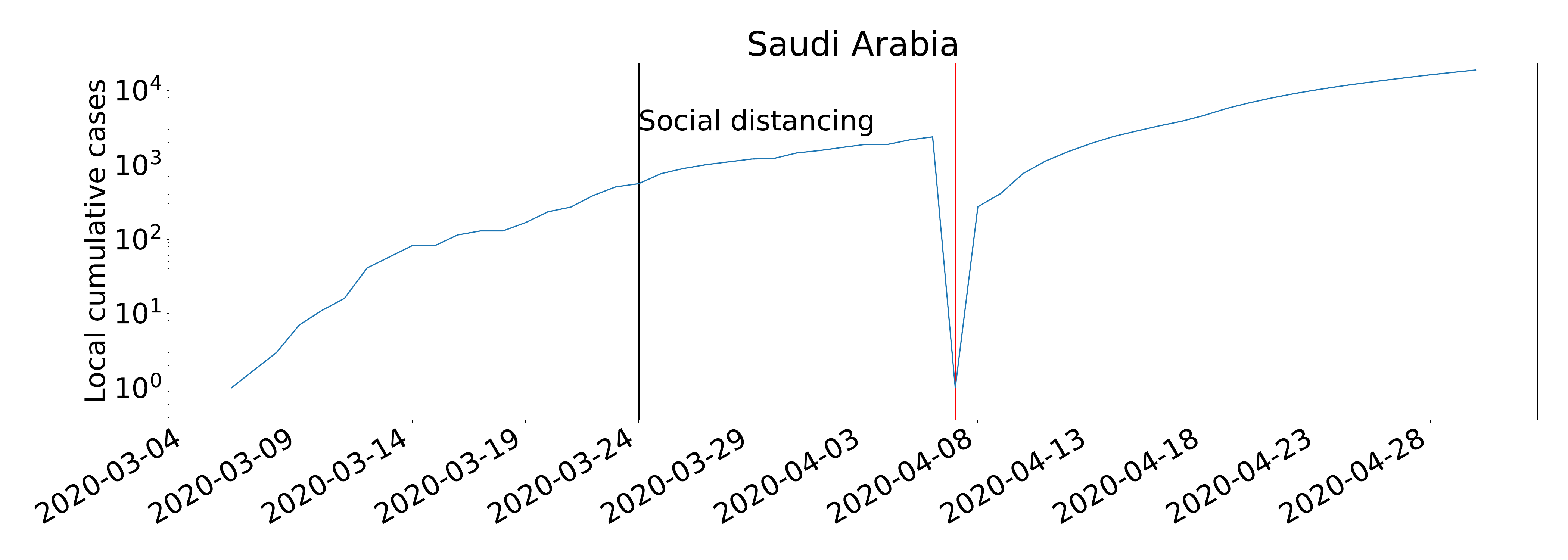} \\
	        \vspace{-0.35cm}
            \textbf{b} & \includegraphics[keepaspectratio, height=3.3cm, valign=T]
			{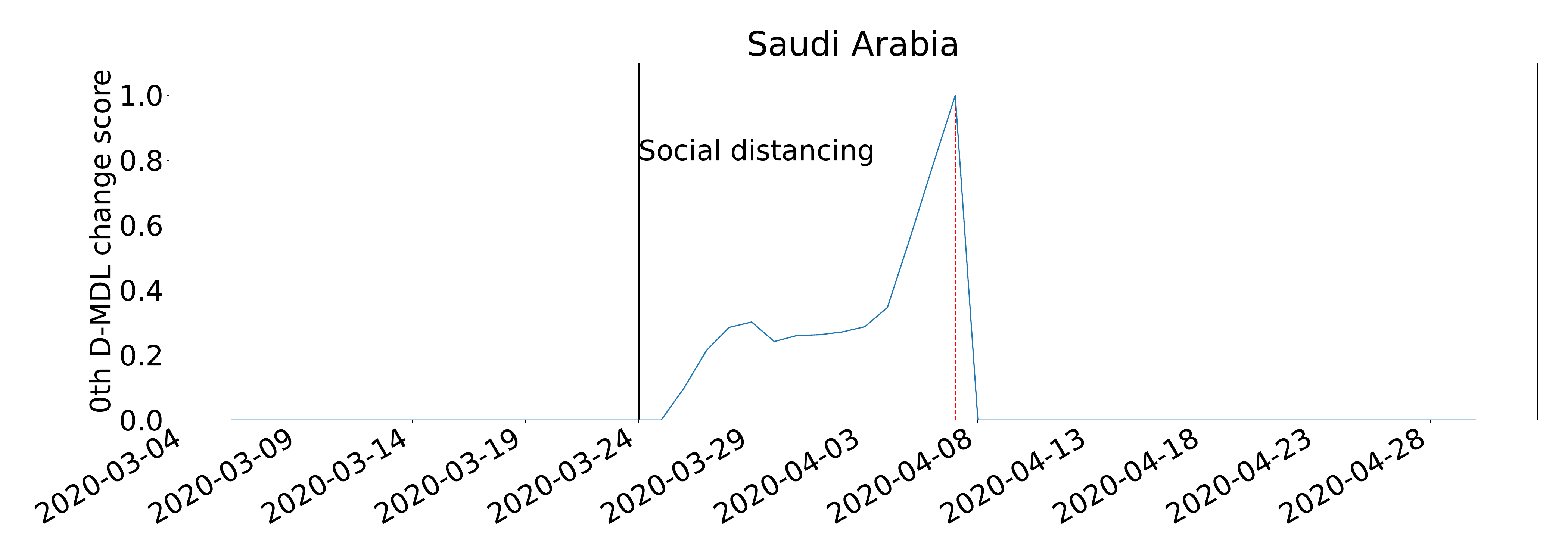}   \\
            \vspace{-0.35cm}
            \textbf{c} & \includegraphics[keepaspectratio, height=3.3cm, valign=T]
			{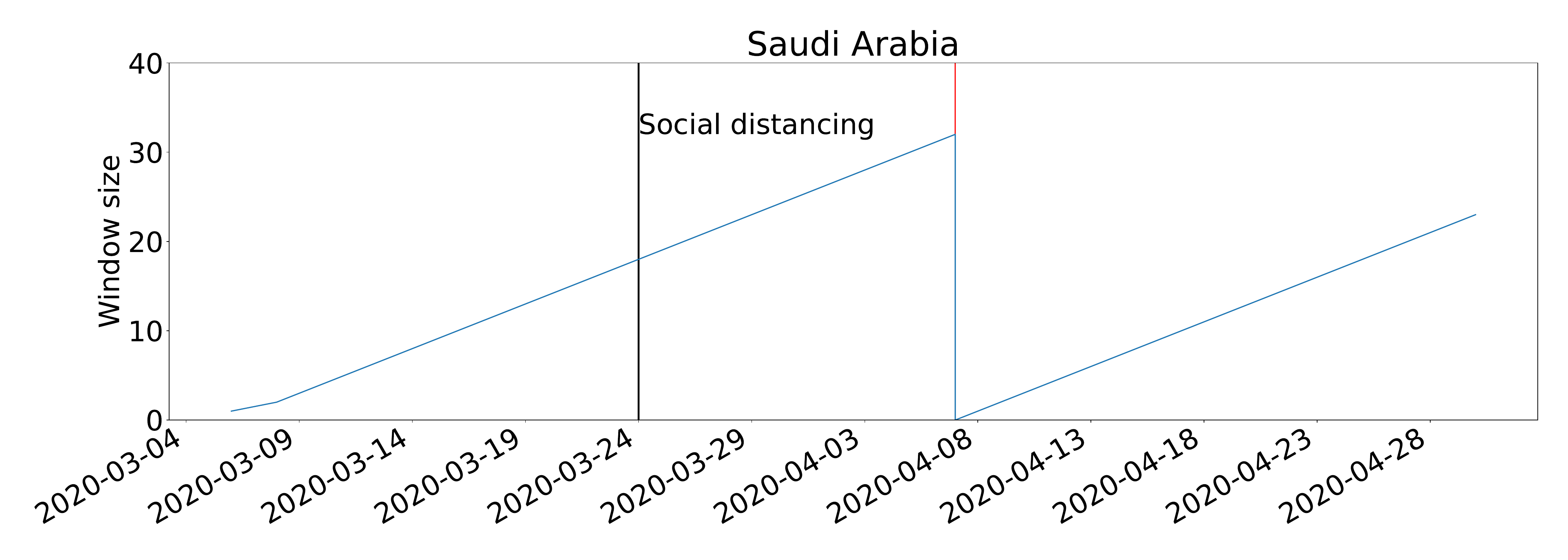} \\
			\vspace{-0.35cm}
			\textbf{d} & \includegraphics[keepaspectratio, height=3.3cm, valign=T]
			{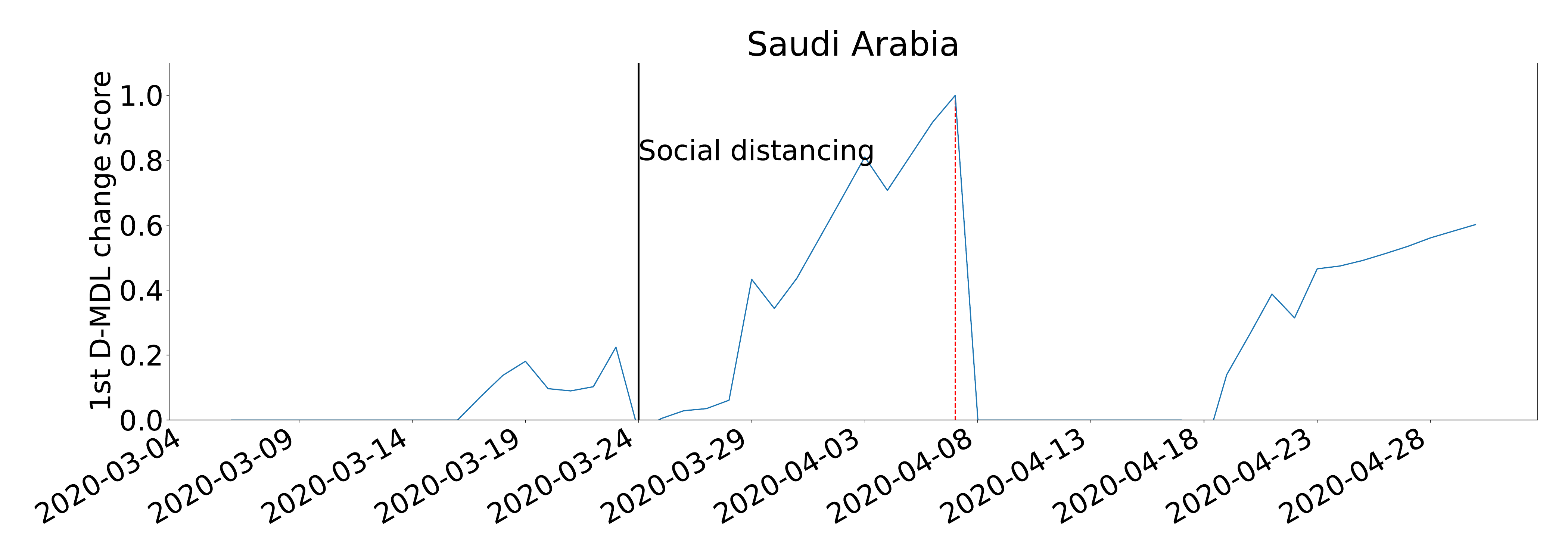} \\
			\vspace{-0.35cm}
			\textbf{e} & \includegraphics[keepaspectratio, height=3.3cm, valign=T]
			{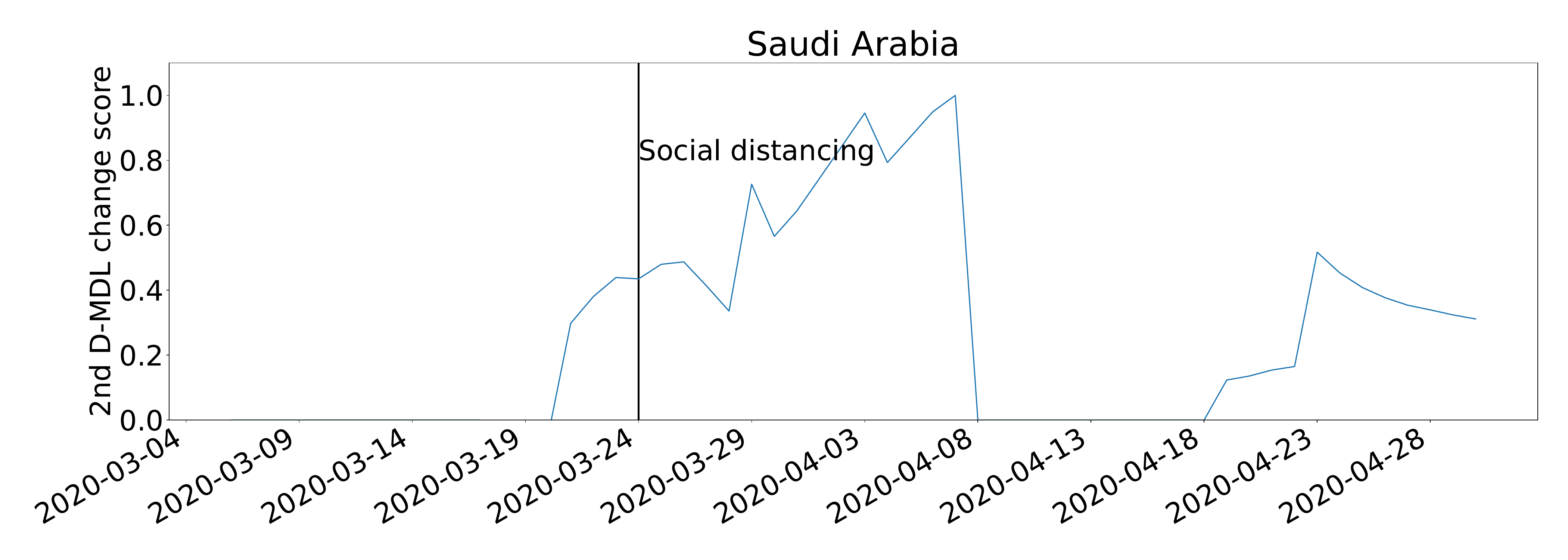} \\
		\end{tabular}
			\caption{\textbf{The results for Saudi Arabia with exponential modeling.} The date on which the social distancing was implemented is marked by a solid line in black. \textbf{a,} the number of cumulative cases. \textbf{b,} the change scores produced by the 0th M-DML where the line in blue denotes values of scores and dashed lines in red mark alarms. \textbf{c,} the window sized for the sequential D-DML algorithm with adaptive window where lines in red mark the shrinkage of windows. \textbf{d,} the change scores produced by the 1st D-MDL. \textbf{e,} the change scores produced by the 2nd D-MDL.}
\end{figure}

\begin{figure}[H] 
\centering
\begin{tabular}{cc}
		 	\textbf{a} & \includegraphics[keepaspectratio, height=3.3cm, valign=T]
			{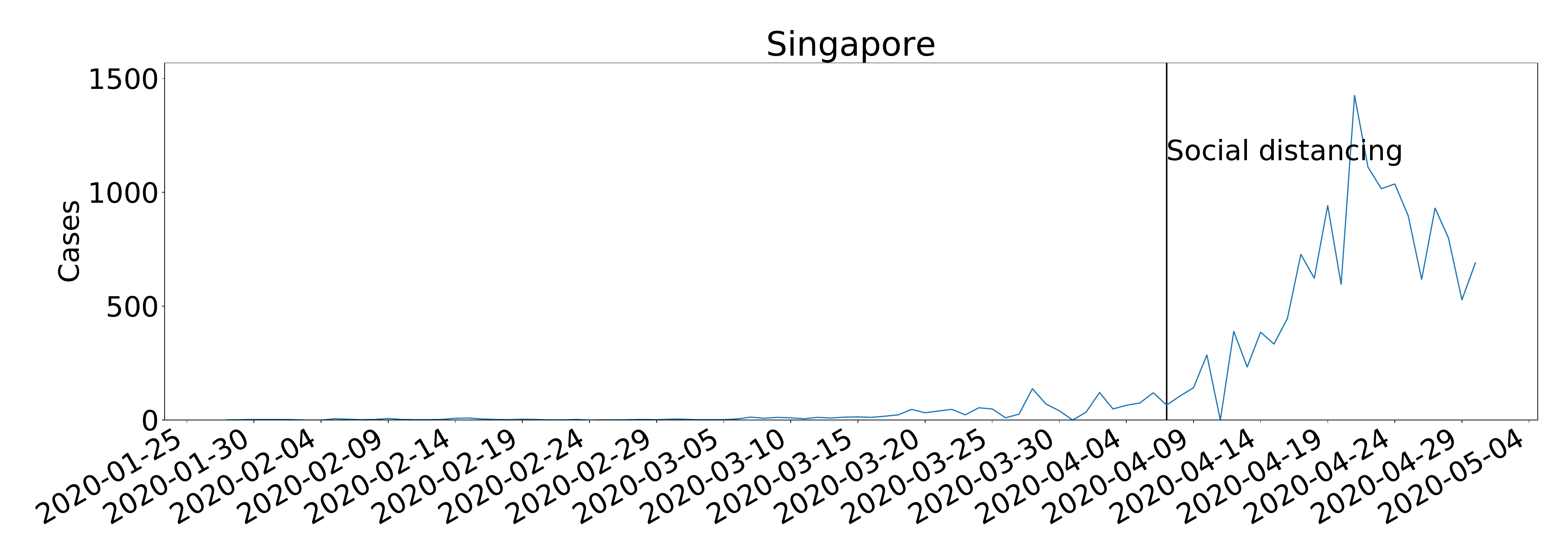} \\
			\vspace{-0.35cm}
	 	    \textbf{b} & \includegraphics[keepaspectratio, height=3.3cm, valign=T]
			{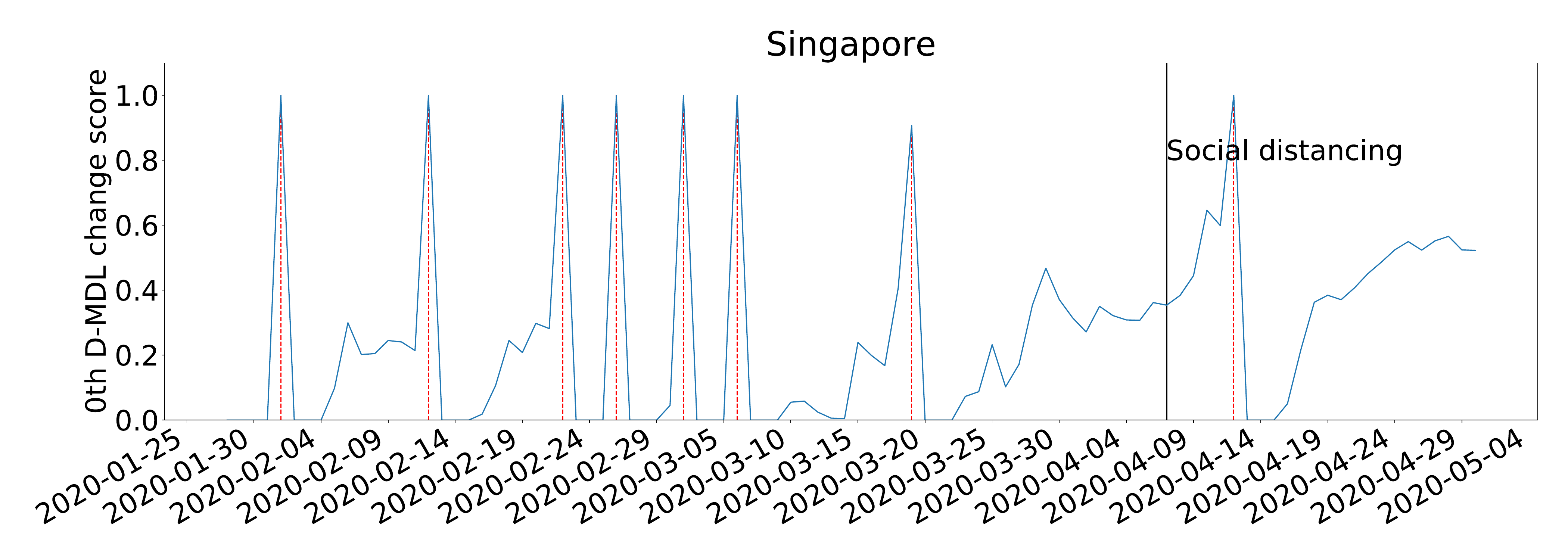}   \\
	        \vspace{-0.35cm}
			\textbf{c} & \includegraphics[keepaspectratio, height=3.3cm, valign=T]
			{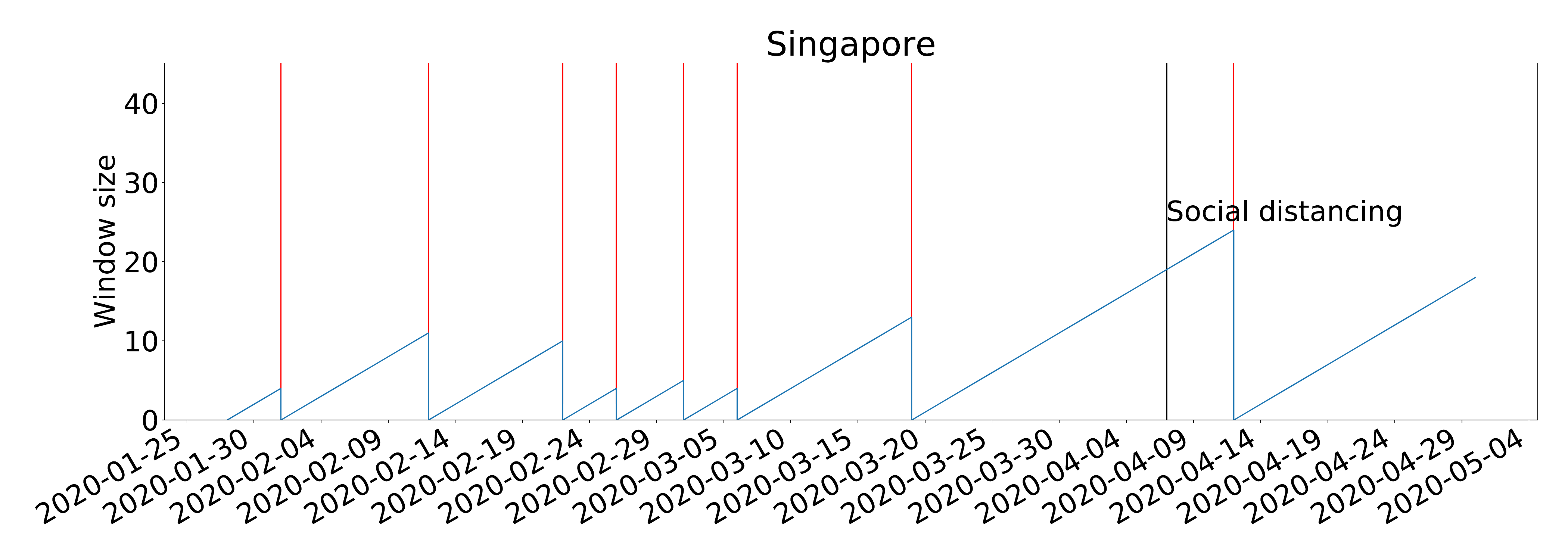} \\
		    \vspace{-0.35cm}
			\textbf{d} & \includegraphics[keepaspectratio, height=3.3cm, valign=T]
			{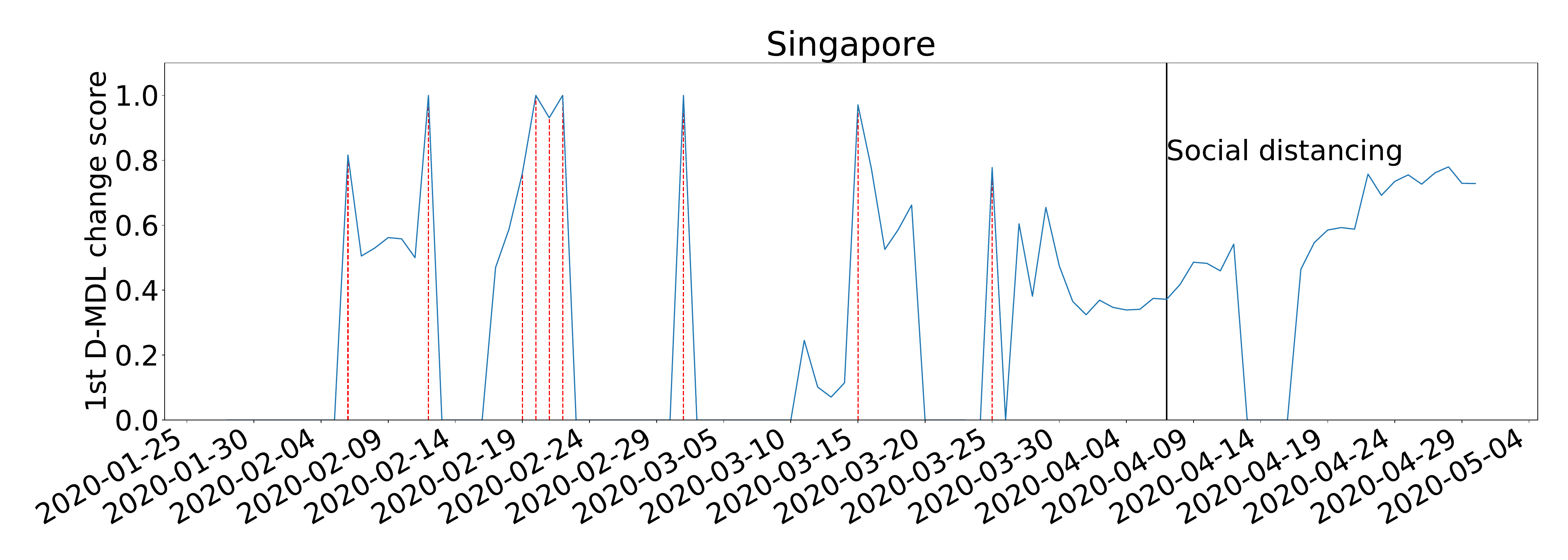} \\
		    \vspace{-0.35cm}
			\textbf{e} & \includegraphics[keepaspectratio, height=3.3cm, valign=T]
			{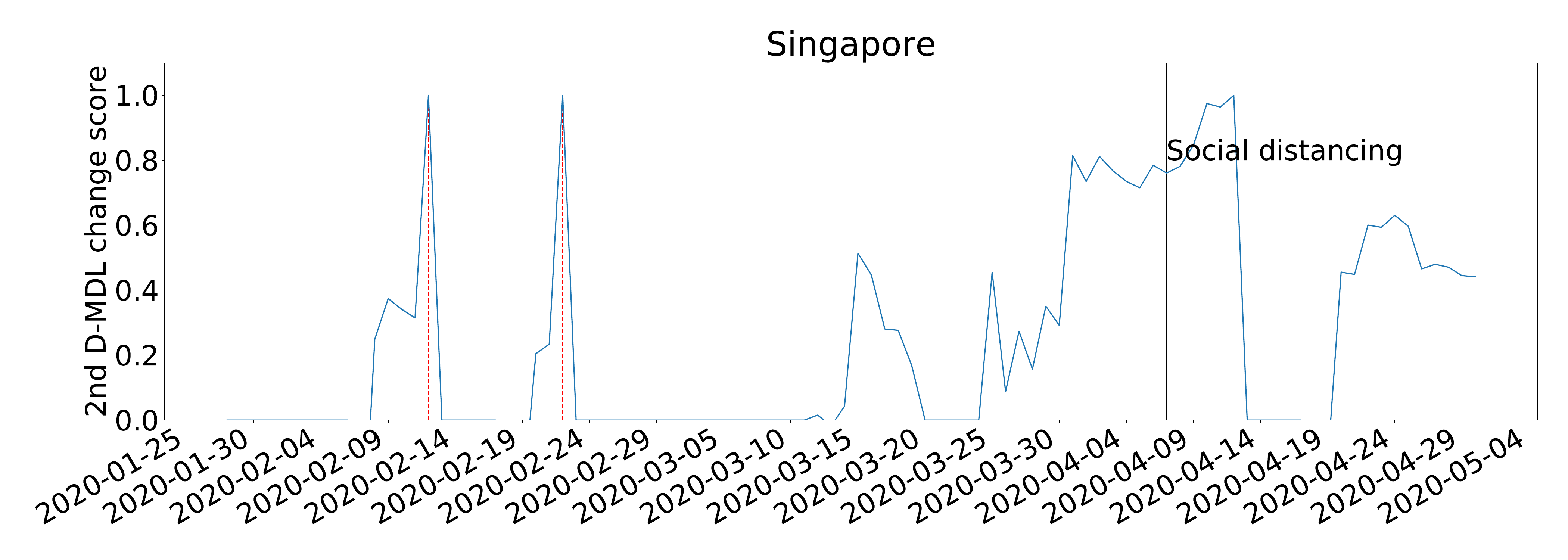} \\
		\end{tabular}
			\caption{\textbf{The results for Singapore with Gaussian modeling.} The date on which the social distancing was implemented is marked by a solid line in black. \textbf{a,} the number of daily new cases. \textbf{b,} the change scores produced by the 0th M-DML where the line in blue denotes values of scores and dashed lines in red mark alarms. \textbf{c,} the window sized for the sequential D-DML algorithm with adaptive window where lines in red mark the shrinkage of windows. \textbf{d,} the change scores produced by the 1st D-MDL. \textbf{e,} the change scores produced by the 2nd D-MDL.}
\end{figure}

\begin{figure}[H]  
\centering
\begin{tabular}{cc}
			\textbf{a} & \includegraphics[keepaspectratio, height=3.3cm, valign=T]
			{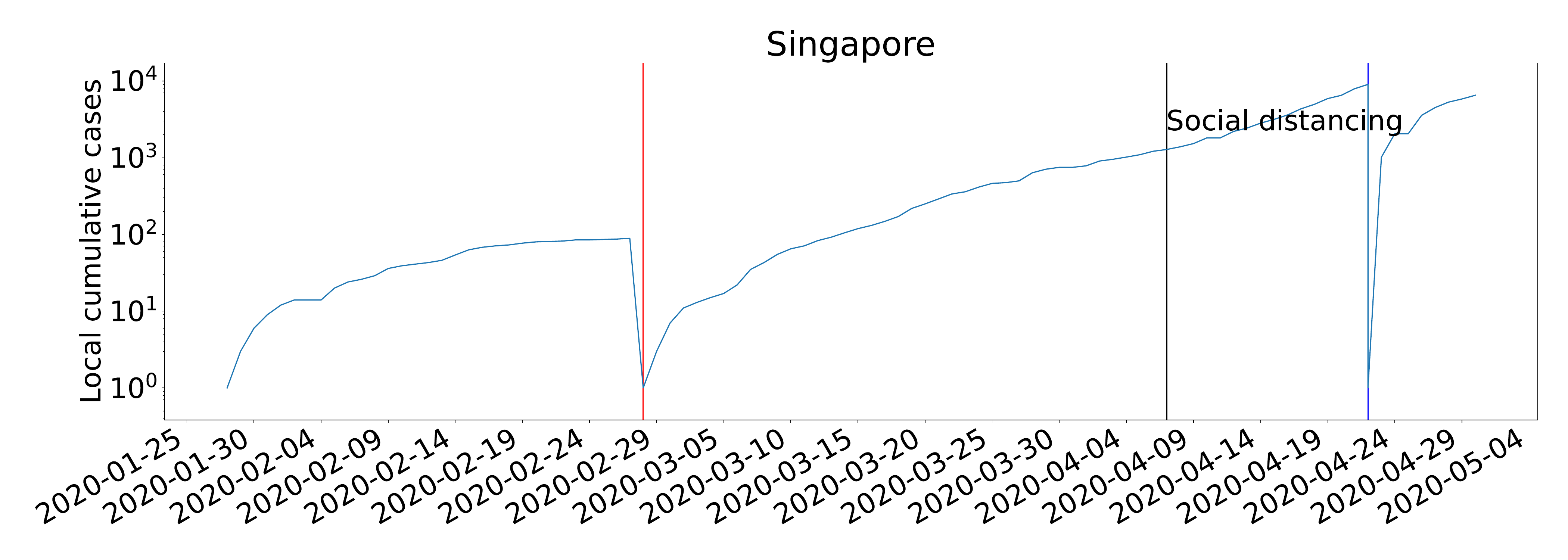} \\
	        \vspace{-0.35cm}
            \textbf{b} & \includegraphics[keepaspectratio, height=3.3cm, valign=T]
			{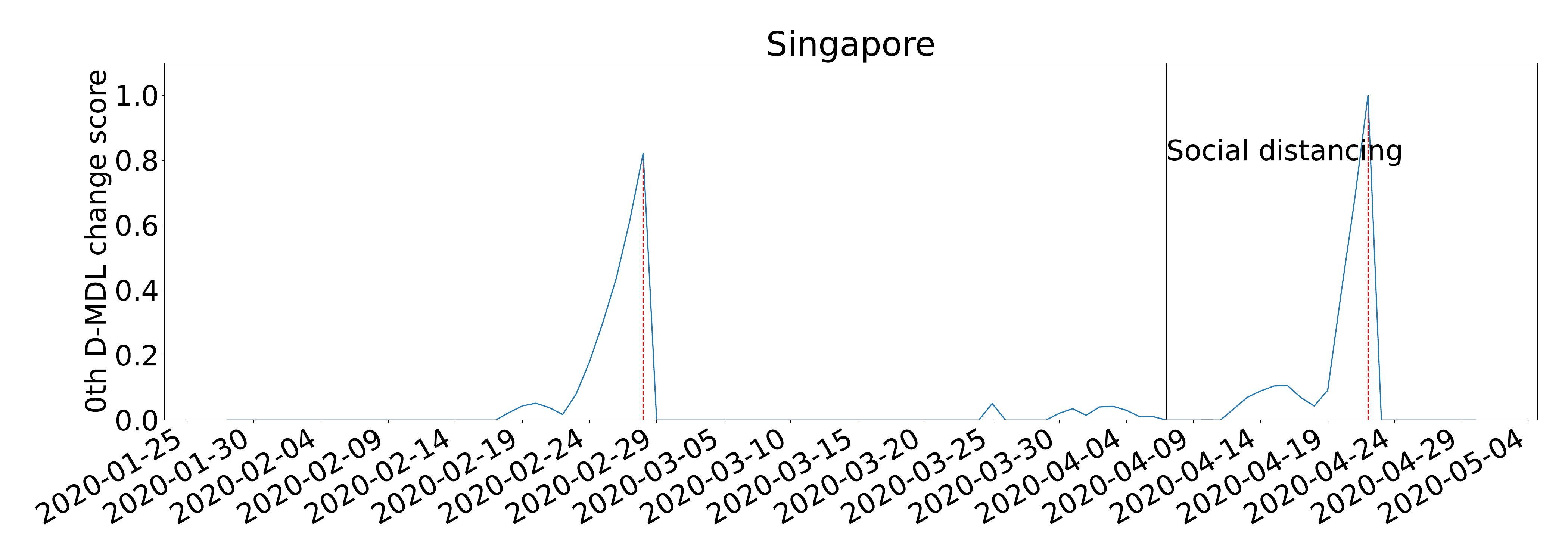}   \\
            \vspace{-0.35cm}
            \textbf{c} & \includegraphics[keepaspectratio, height=3.3cm, valign=T]
			{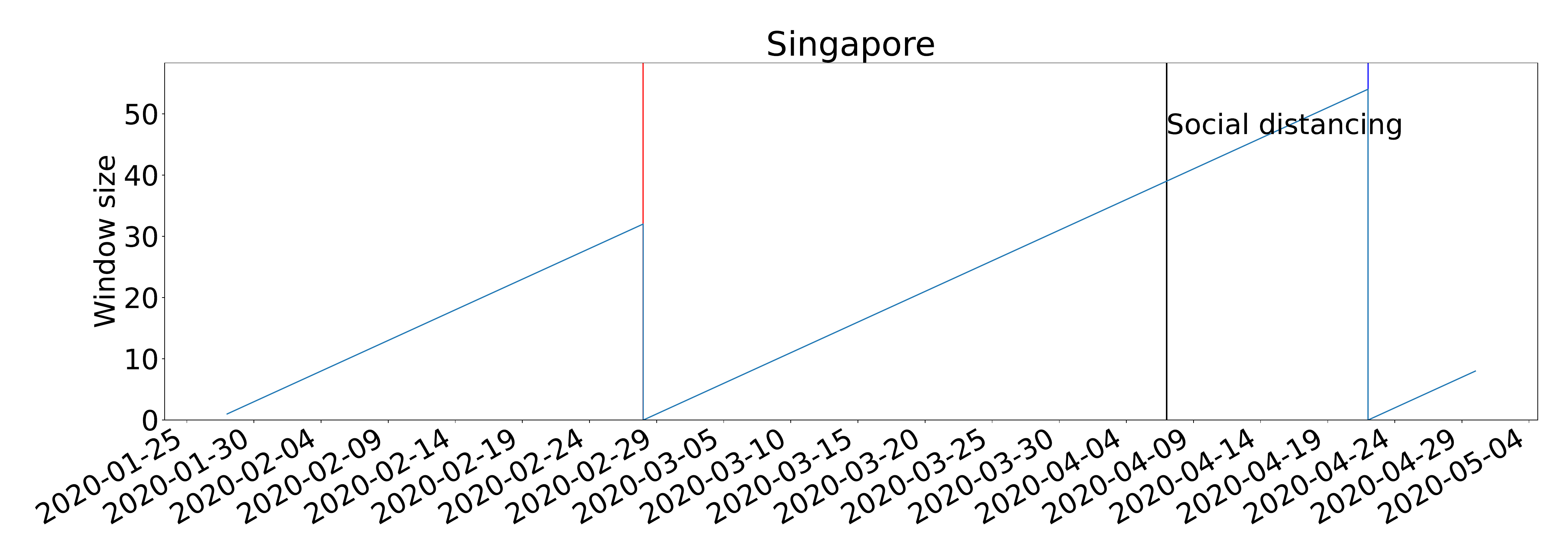} \\
			\vspace{-0.35cm}
			\textbf{d} & \includegraphics[keepaspectratio, height=3.3cm, valign=T]
			{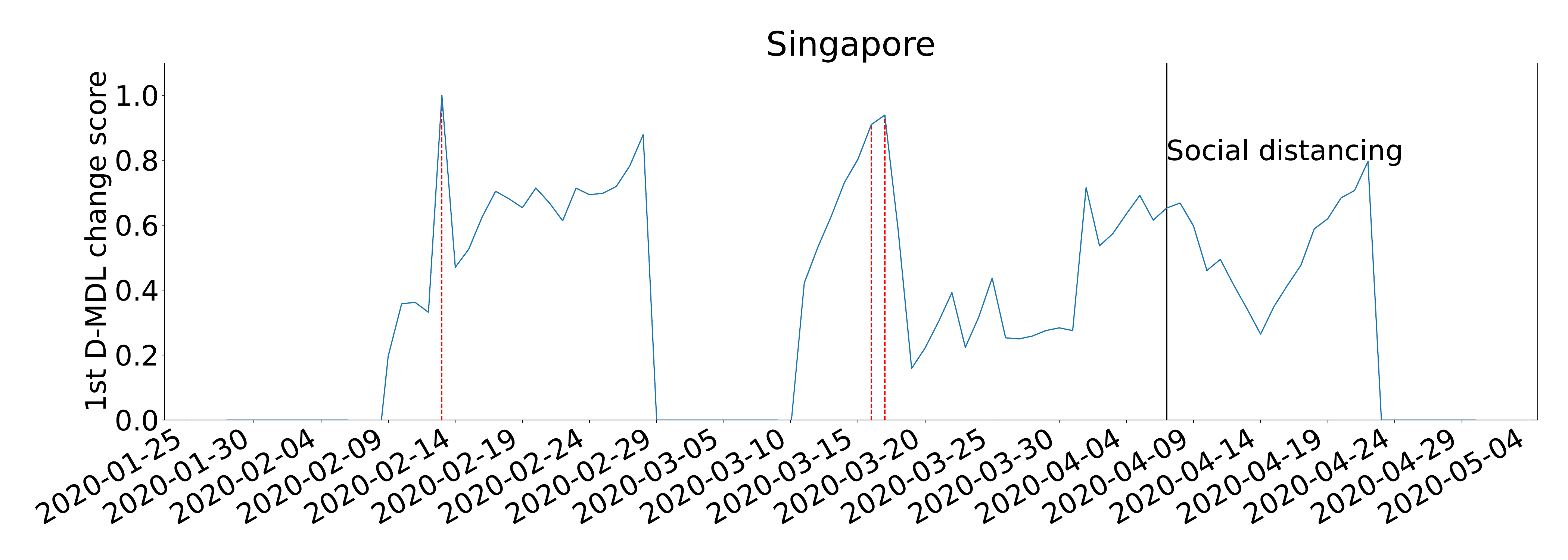} \\
			\vspace{-0.35cm}
			\textbf{e} & \includegraphics[keepaspectratio, height=3.3cm, valign=T]
			{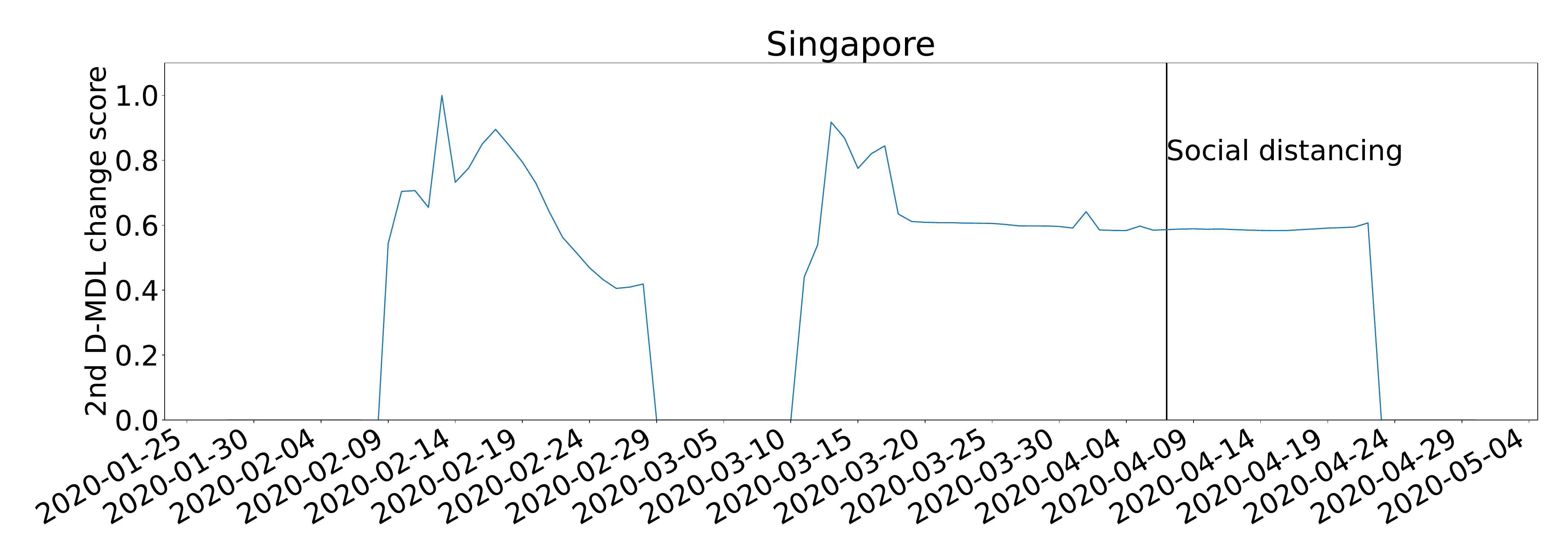} \\
		\end{tabular}
			\caption{\textbf{The results for Singapore with exponential modeling.} The date on which the social distancing was implemented is marked by a solid line in black. \textbf{a,} the number of cumulative cases. \textbf{b,} the change scores produced by the 0th M-DML where the line in blue denotes values of scores and dashed lines in red mark alarms. \textbf{c,} the window sized for the sequential D-DML algorithm with adaptive window where lines in red mark the shrinkage of windows. \textbf{d,} the change scores produced by the 1st D-MDL. \textbf{e,} the change scores produced by the 2nd D-MDL.}
\end{figure}

\begin{figure}[H] 
\centering
\begin{tabular}{cc}
		 	\textbf{a} & \includegraphics[keepaspectratio, height=3.3cm, valign=T]
			{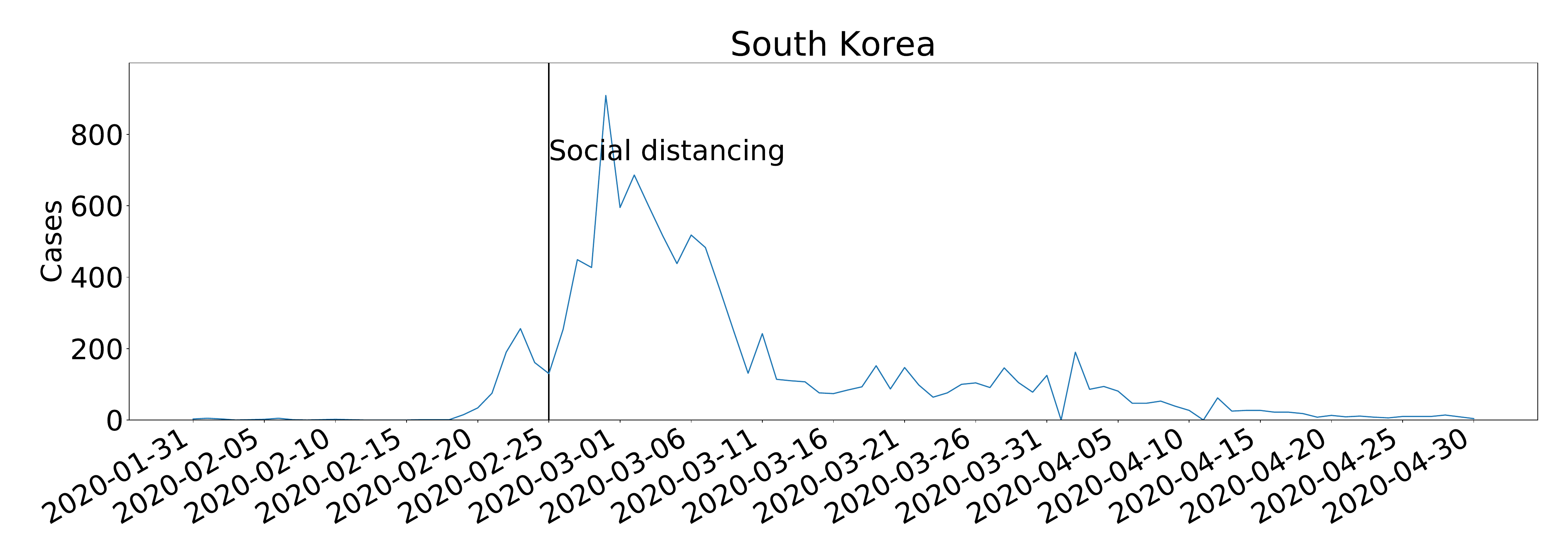} \\
			\vspace{-0.35cm}
	 	    \textbf{b} & \includegraphics[keepaspectratio, height=3.3cm, valign=T]
			{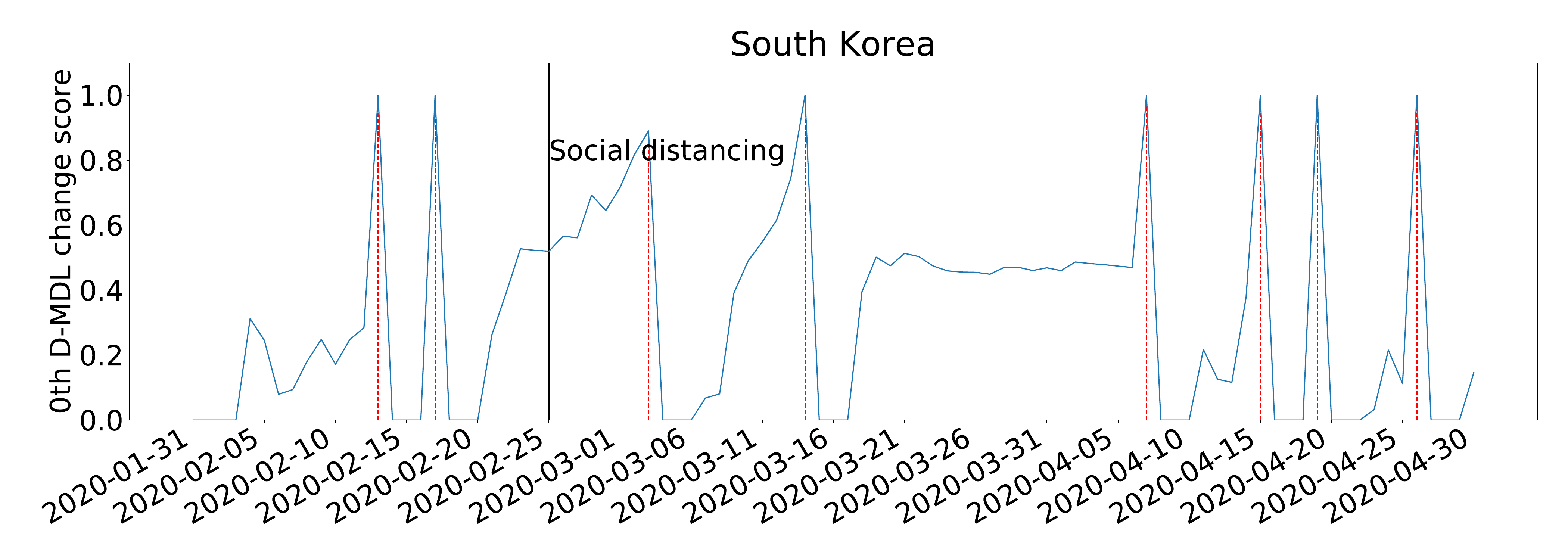}   \\
	        \vspace{-0.35cm}
			\textbf{c} & \includegraphics[keepaspectratio, height=3.3cm, valign=T]
			{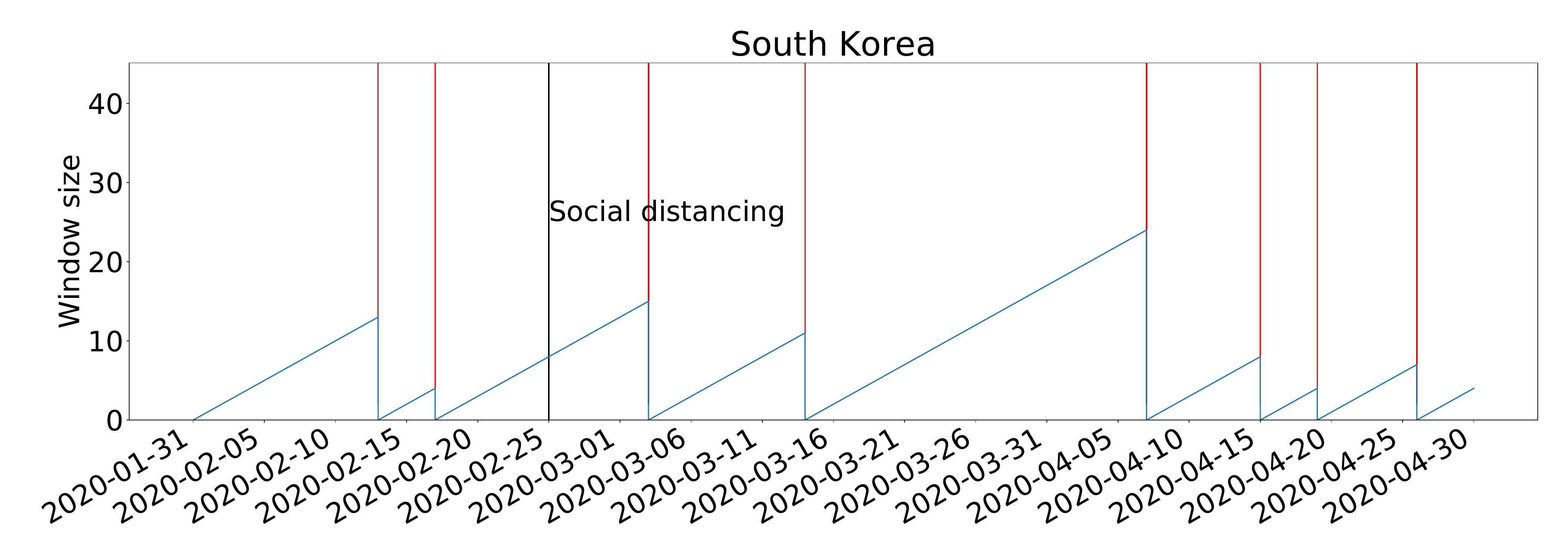} \\
		    \vspace{-0.35cm}
			\textbf{d} & \includegraphics[keepaspectratio, height=3.3cm, valign=T]
			{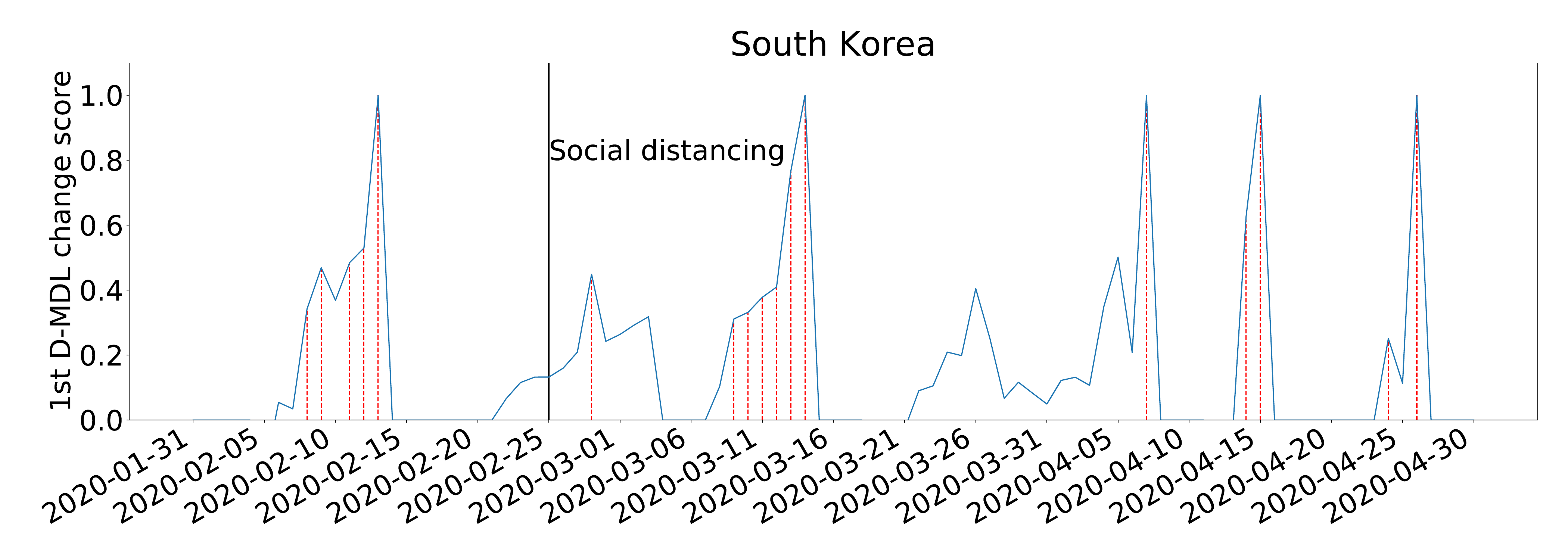} \\
		    \vspace{-0.35cm}
			\textbf{e} & \includegraphics[keepaspectratio, height=3.3cm, valign=T]
			{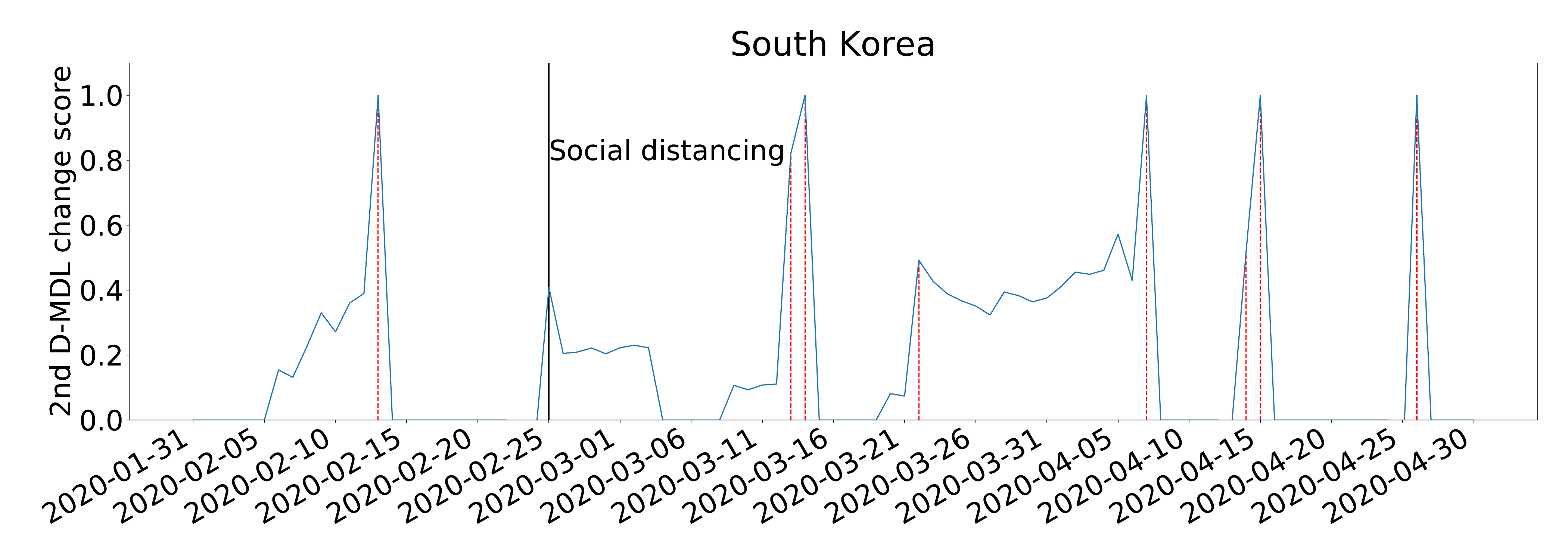} \\
		\end{tabular}
			\caption{\textbf{The results for South Korea with Gaussian modeling.} The date on which the social distancing was implemented is marked by a solid line in black. \textbf{a,} the number of daily new cases. \textbf{b,} the change scores produced by the 0th M-DML where the line in blue denotes values of scores and dashed lines in red mark alarms. \textbf{c,} the window sized for the sequential D-DML algorithm with adaptive window where lines in red mark the shrinkage of windows. \textbf{d,} the change scores produced by the 1st D-MDL. \textbf{e,} the change scores produced by the 2nd D-MDL.}
\end{figure}

\begin{figure}[H]  
\centering
\begin{tabular}{cc}
			\textbf{a} & \includegraphics[keepaspectratio, height=3.3cm, valign=T]
			{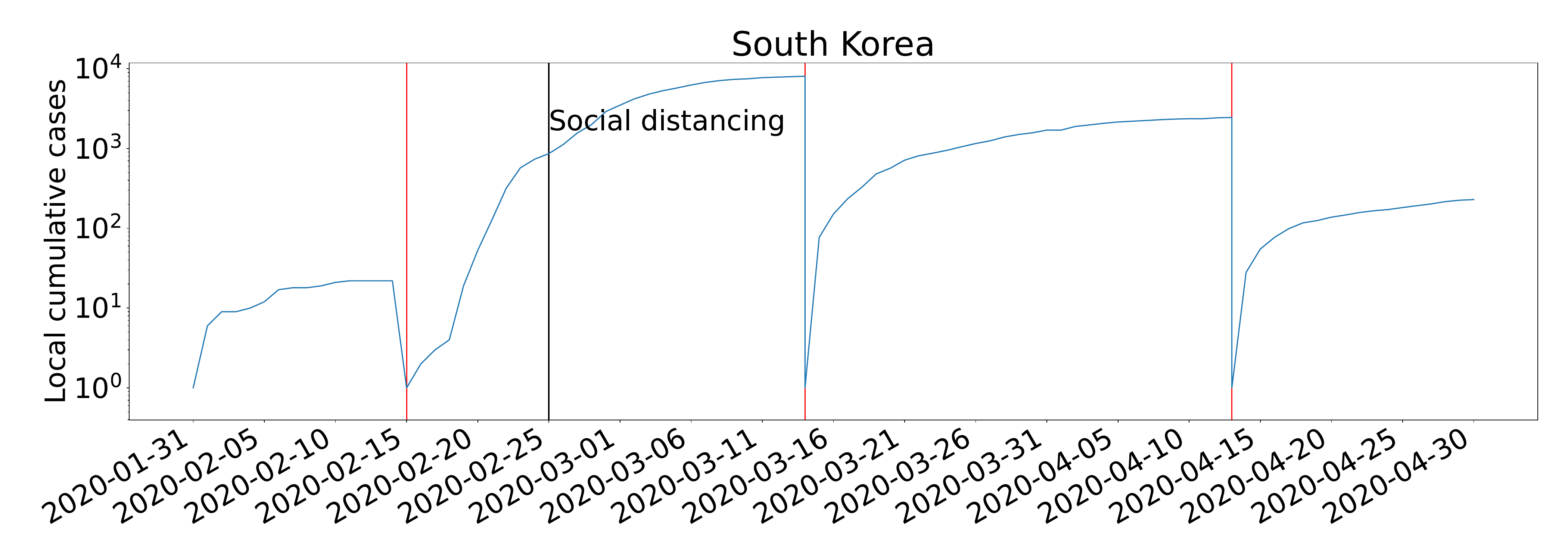} \\
	        \vspace{-0.35cm}
            \textbf{b} & \includegraphics[keepaspectratio, height=3.3cm, valign=T]
			{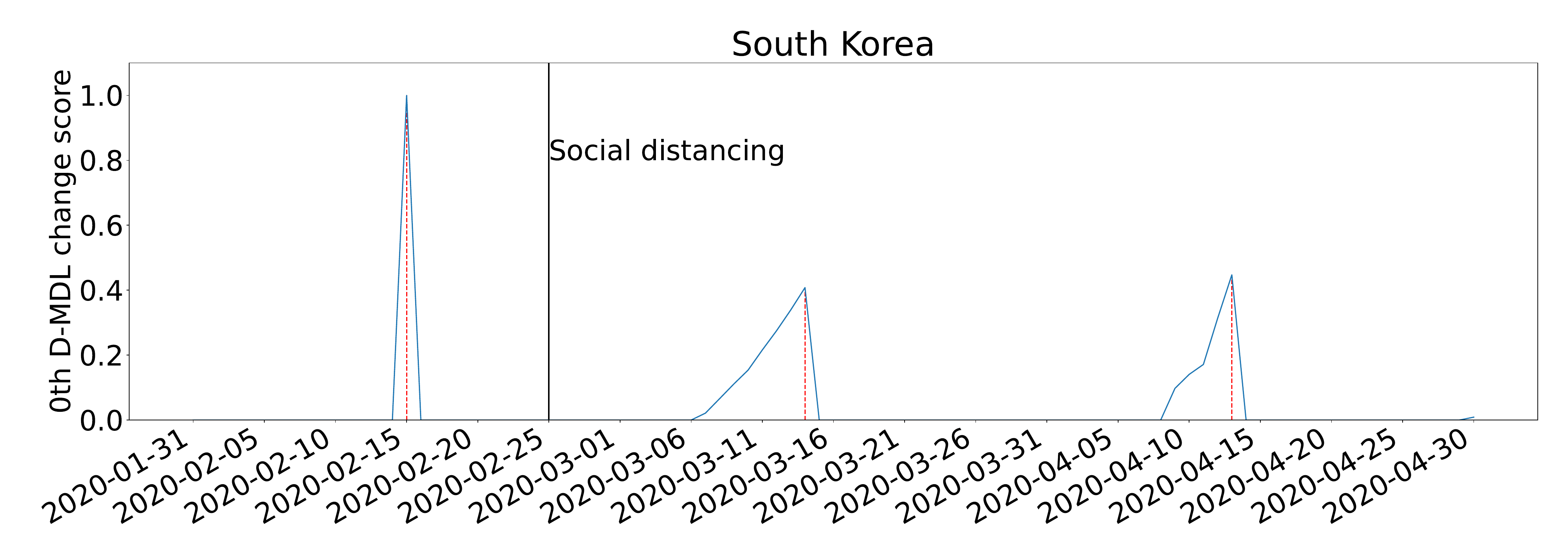}   \\
            \vspace{-0.35cm}
            \textbf{c} & \includegraphics[keepaspectratio, height=3.3cm, valign=T]
			{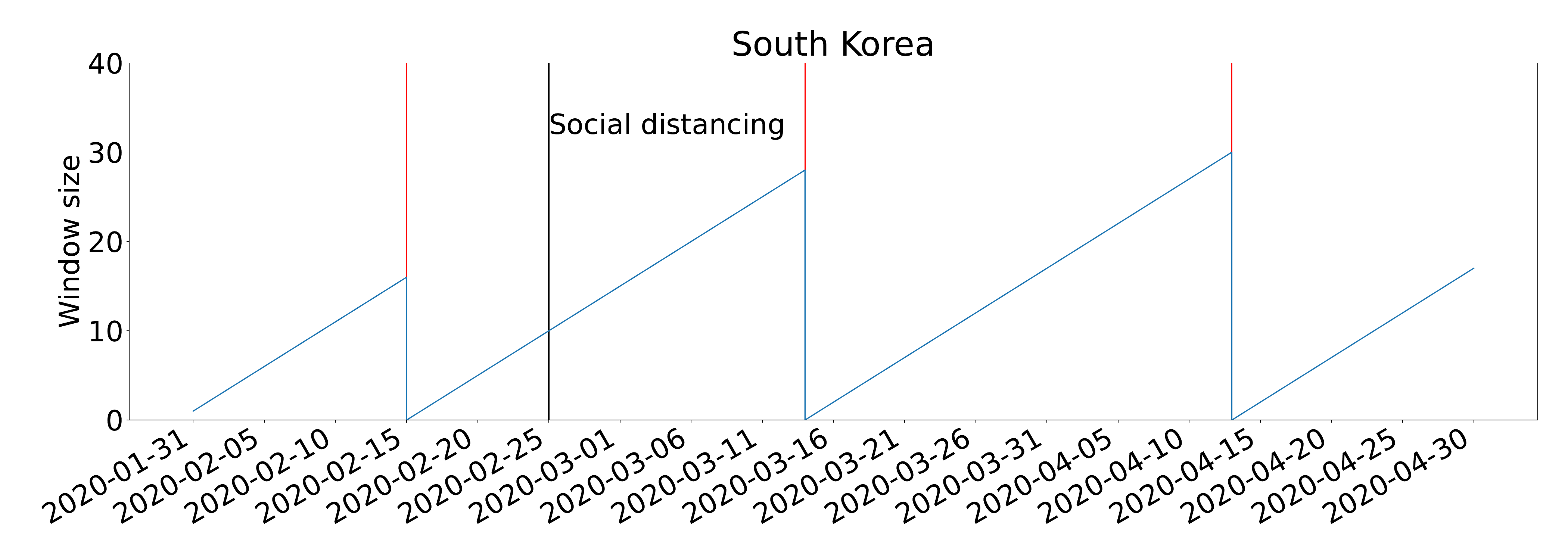} \\
			\vspace{-0.35cm}
			\textbf{d} & \includegraphics[keepaspectratio, height=3.3cm, valign=T]
			{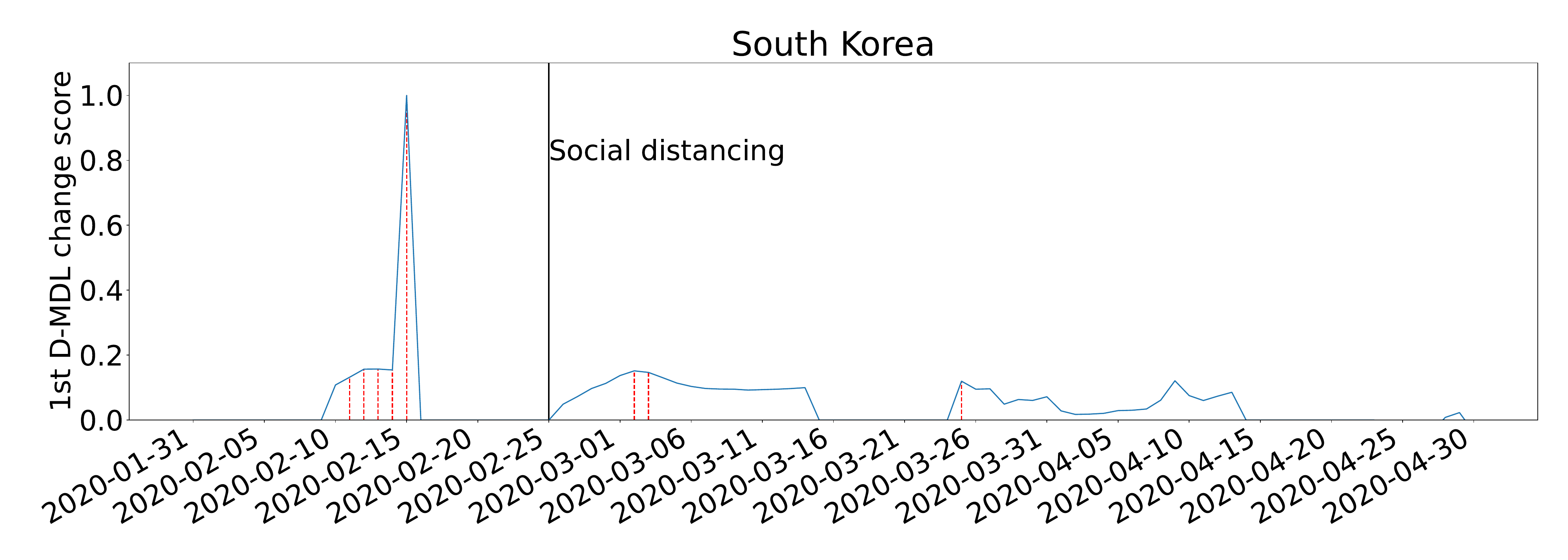} \\
			\vspace{-0.35cm}
			\textbf{e} & \includegraphics[keepaspectratio, height=3.3cm, valign=T]
			{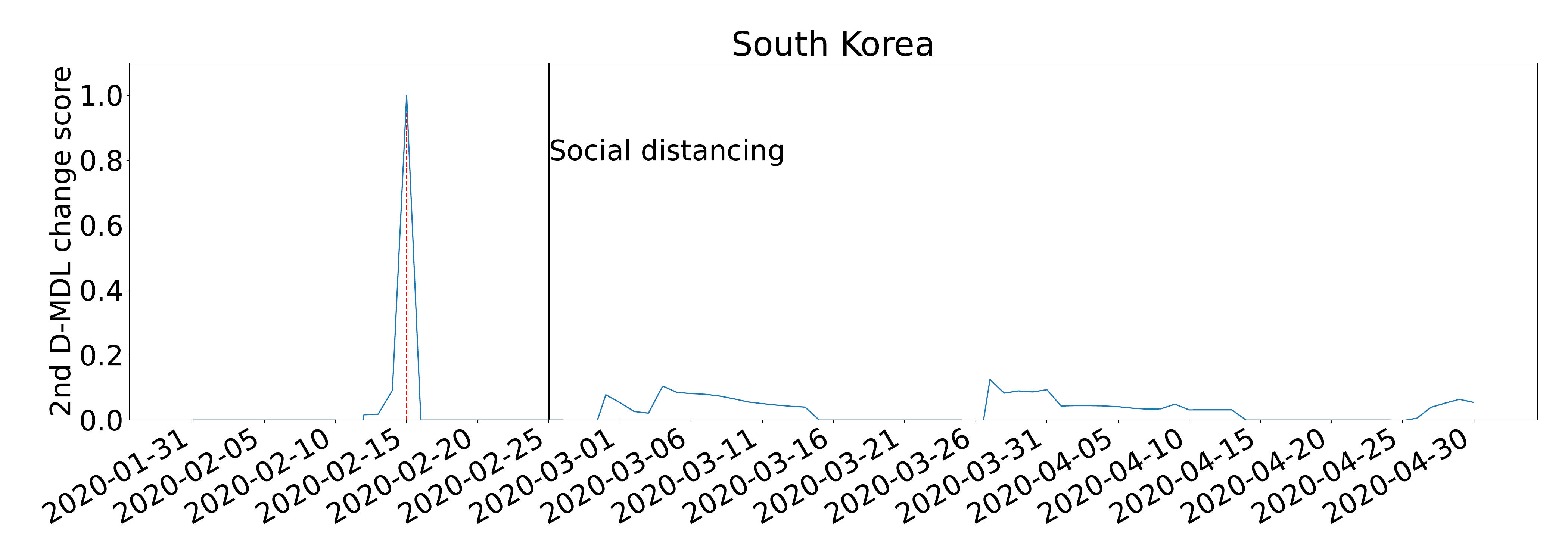} \\
		\end{tabular}
			\caption{\textbf{The results for South Korea with exponential modeling.} The date on which the social distancing was implemented is marked by a solid line in black. \textbf{a,} the number of cumulative cases. \textbf{b,} the change scores produced by the 0th M-DML where the line in blue denotes values of scores and dashed lines in red mark alarms. \textbf{c,} the window sized for the sequential D-DML algorithm with adaptive window where lines in red mark the shrinkage of windows. \textbf{d,} the change scores produced by the 1st D-MDL. \textbf{e,} the change scores produced by the 2nd D-MDL.}
\end{figure}

\begin{figure}[H] 
\centering
\begin{tabular}{cc}
		 	\textbf{a} & \includegraphics[keepaspectratio, height=3.3cm, valign=T]
			{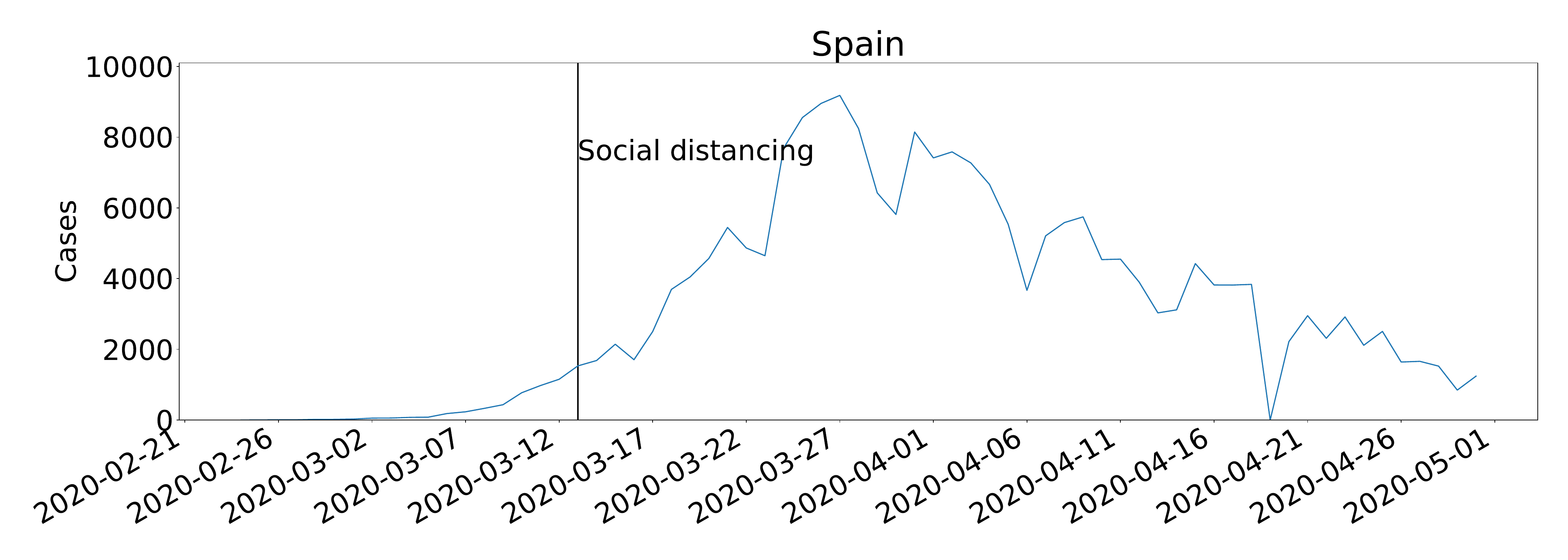} \\
			\vspace{-0.35cm}
	 	    \textbf{b} & \includegraphics[keepaspectratio, height=3.3cm, valign=T]
			{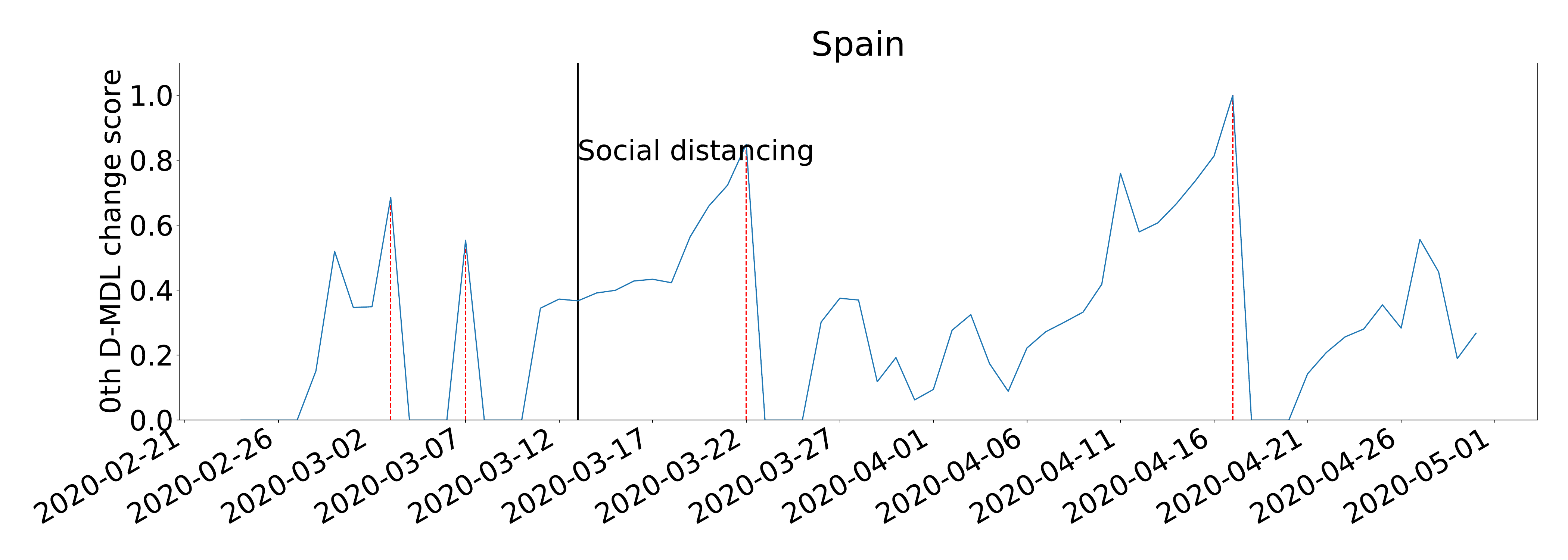}   \\
	        \vspace{-0.35cm}
			\textbf{c} & \includegraphics[keepaspectratio, height=3.3cm, valign=T]
			{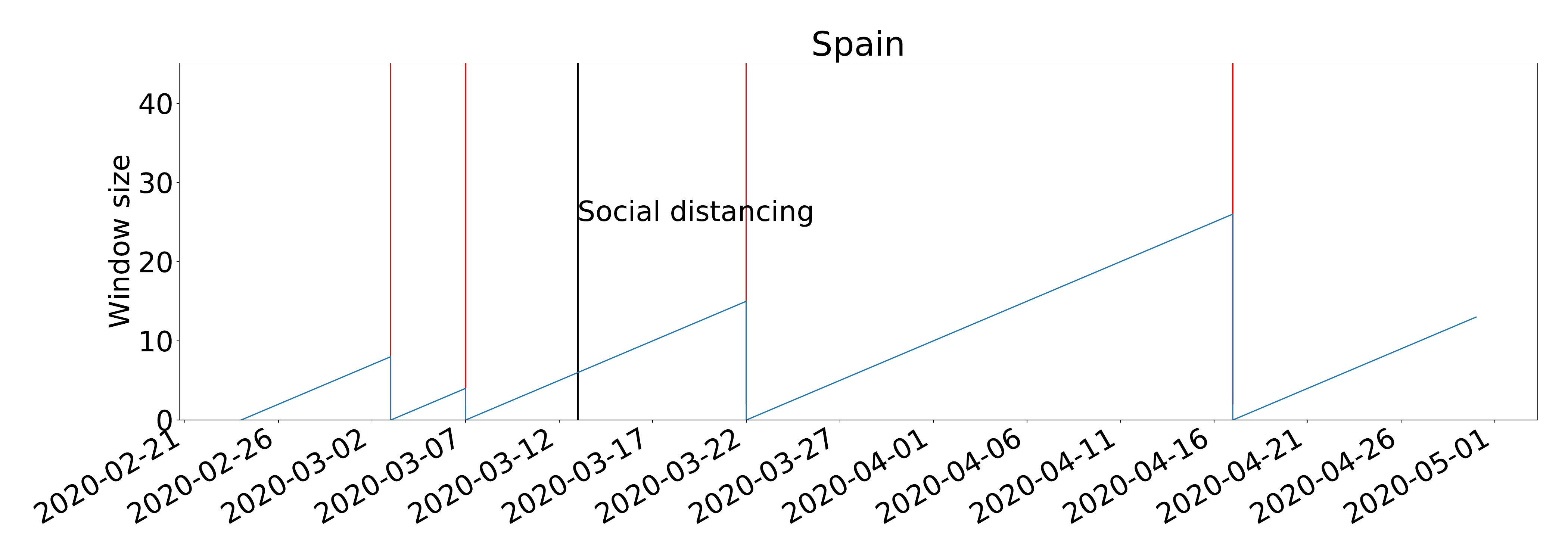} \\
		    \vspace{-0.35cm}
			\textbf{d} & \includegraphics[keepaspectratio, height=3.3cm, valign=T]
			{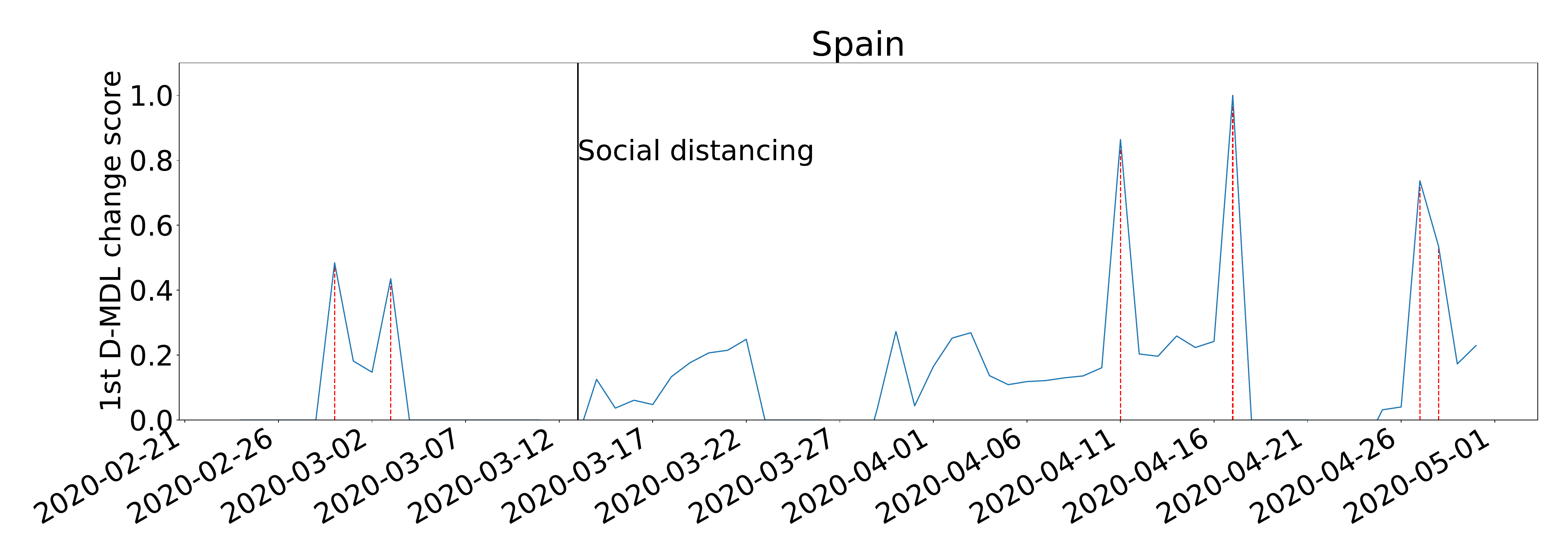} \\
		    \vspace{-0.35cm}
			\textbf{e} & \includegraphics[keepaspectratio, height=3.3cm, valign=T]
			{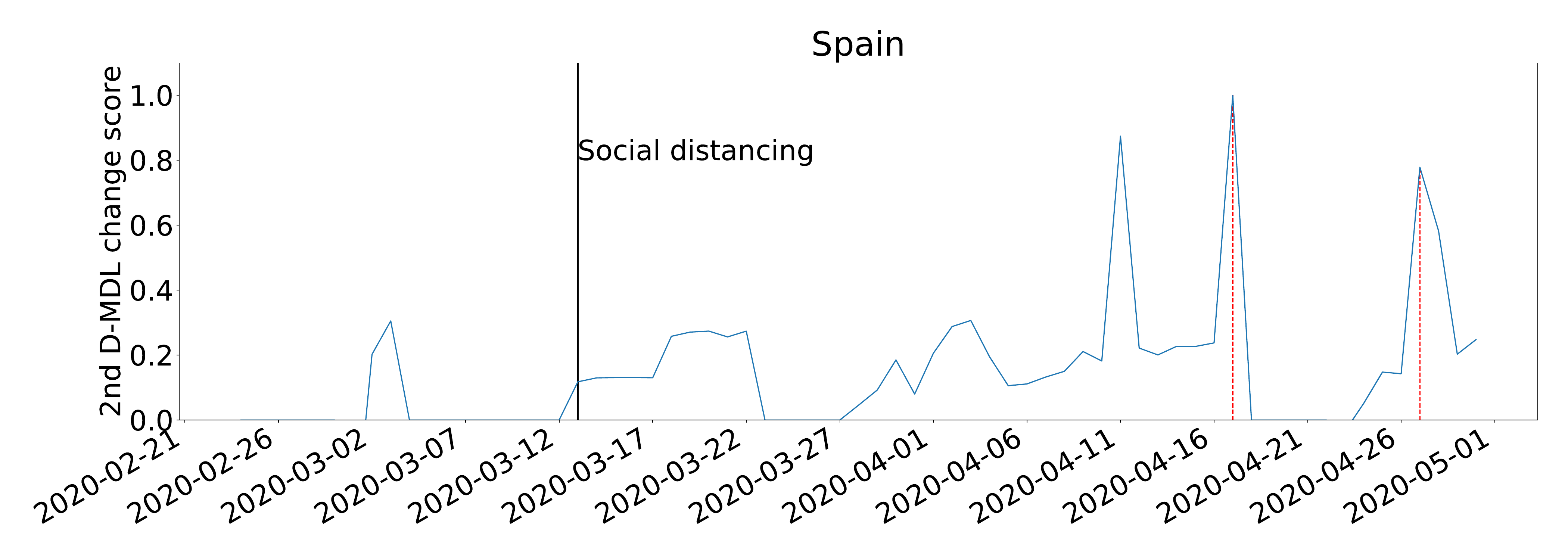} \\
		\end{tabular}
			\caption{\textbf{The results for Spain with Gaussian modeling.} The date on which the social distancing was implemented is marked by a solid line in black. \textbf{a,} the number of daily new cases. \textbf{b,} the change scores produced by the 0th M-DML where the line in blue denotes values of scores and dashed lines in red mark alarms. \textbf{c,} the window sized for the sequential D-DML algorithm with adaptive window where lines in red mark the shrinkage of windows. \textbf{d,} the change scores produced by the 1st D-MDL. \textbf{e,} the change scores produced by the 2nd D-MDL.}
\end{figure}

\begin{figure}[H]  
\centering
\begin{tabular}{cc}
			\textbf{a} & \includegraphics[keepaspectratio, height=3.3cm, valign=T]
			{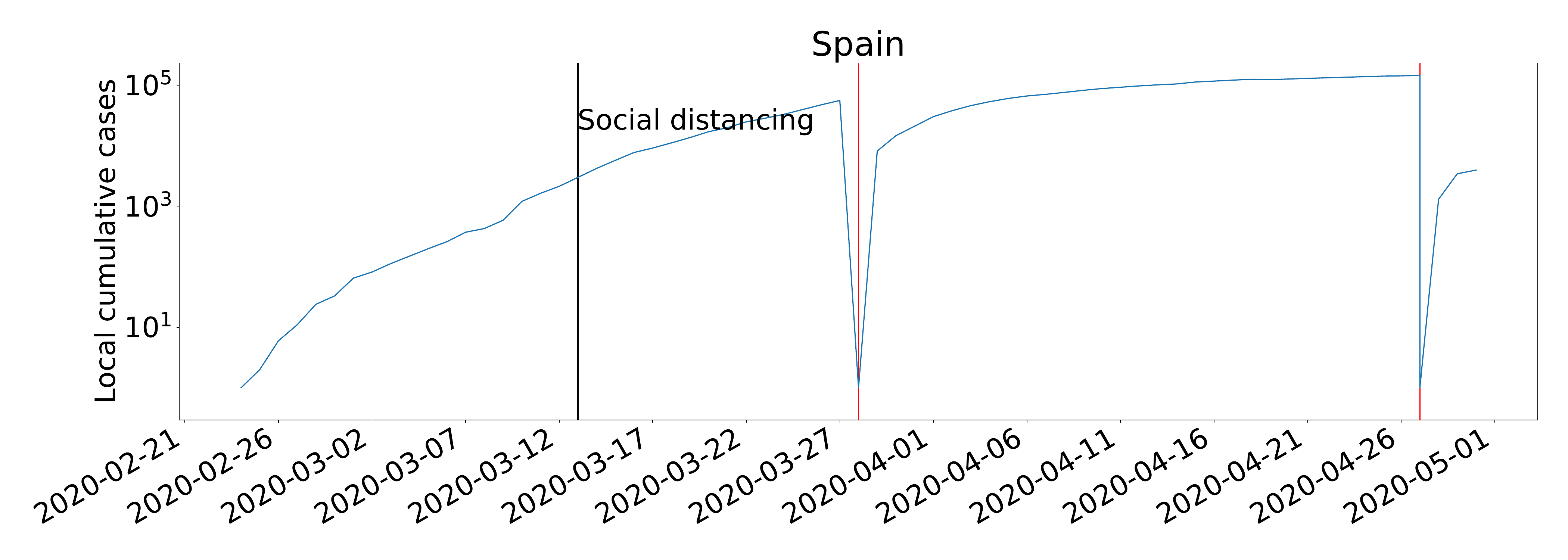} \\
	        \vspace{-0.35cm}
            \textbf{b} & \includegraphics[keepaspectratio, height=3.3cm, valign=T]
			{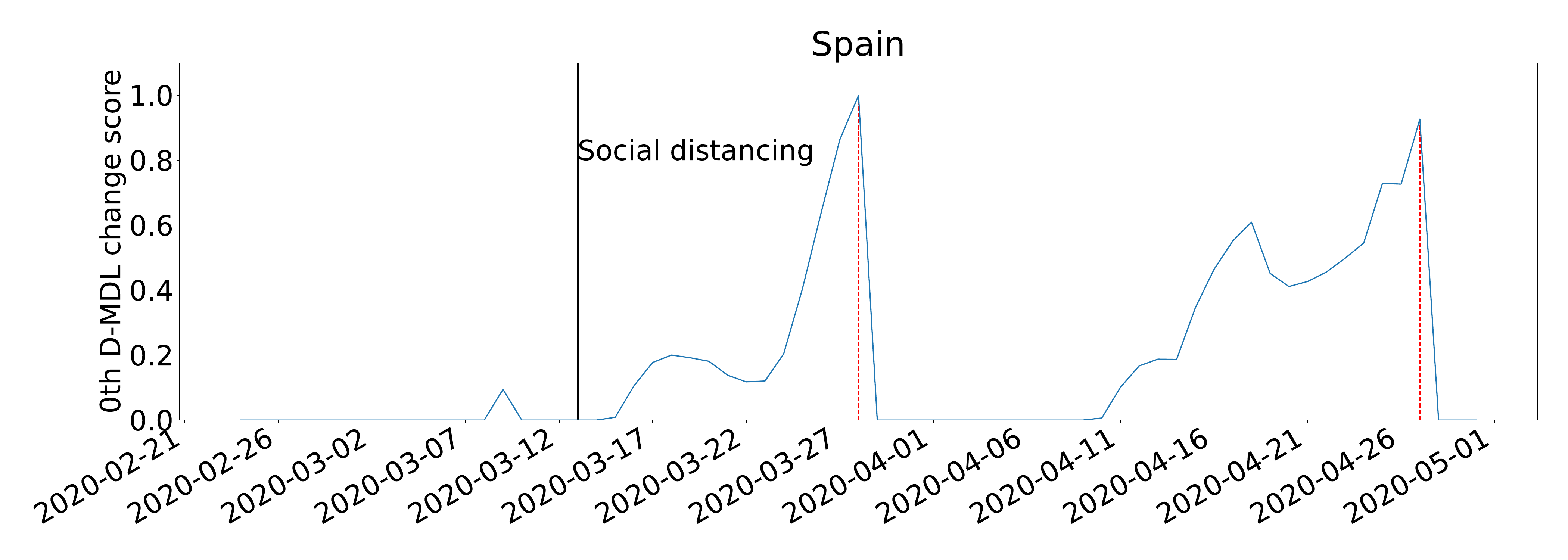}   \\
            \vspace{-0.35cm}
            \textbf{c} & \includegraphics[keepaspectratio, height=3.3cm, valign=T]
			{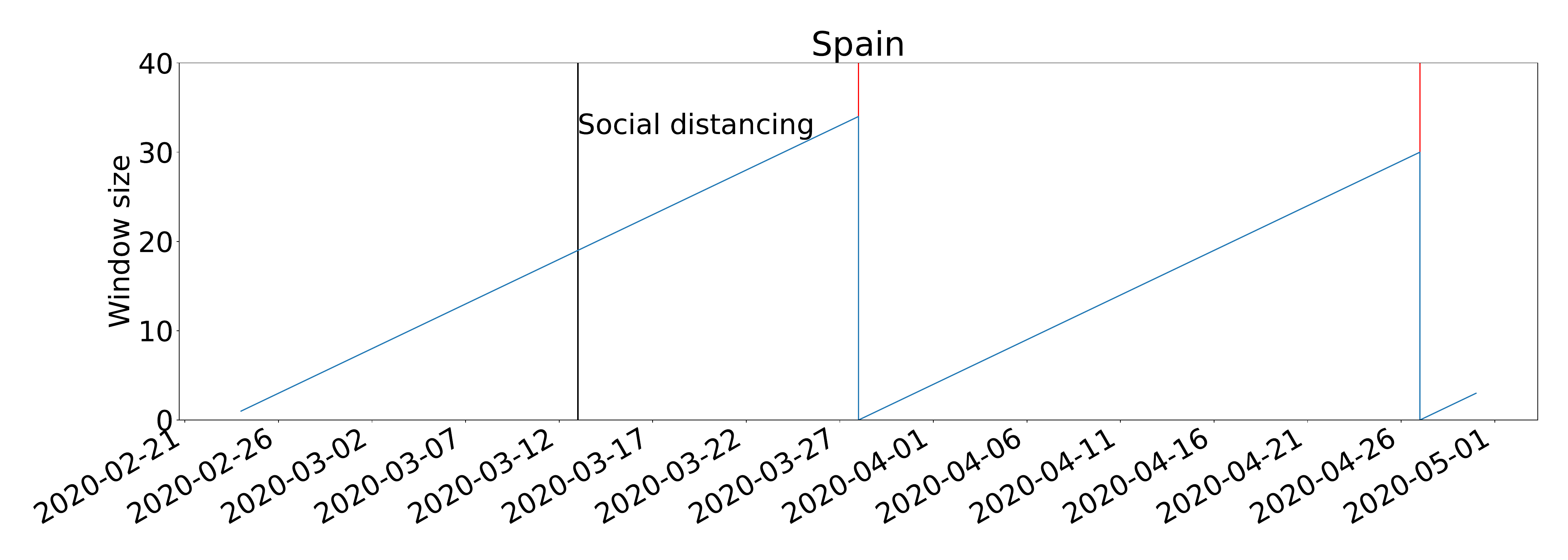} \\
			\vspace{-0.35cm}
			\textbf{d} & \includegraphics[keepaspectratio, height=3.3cm, valign=T]
			{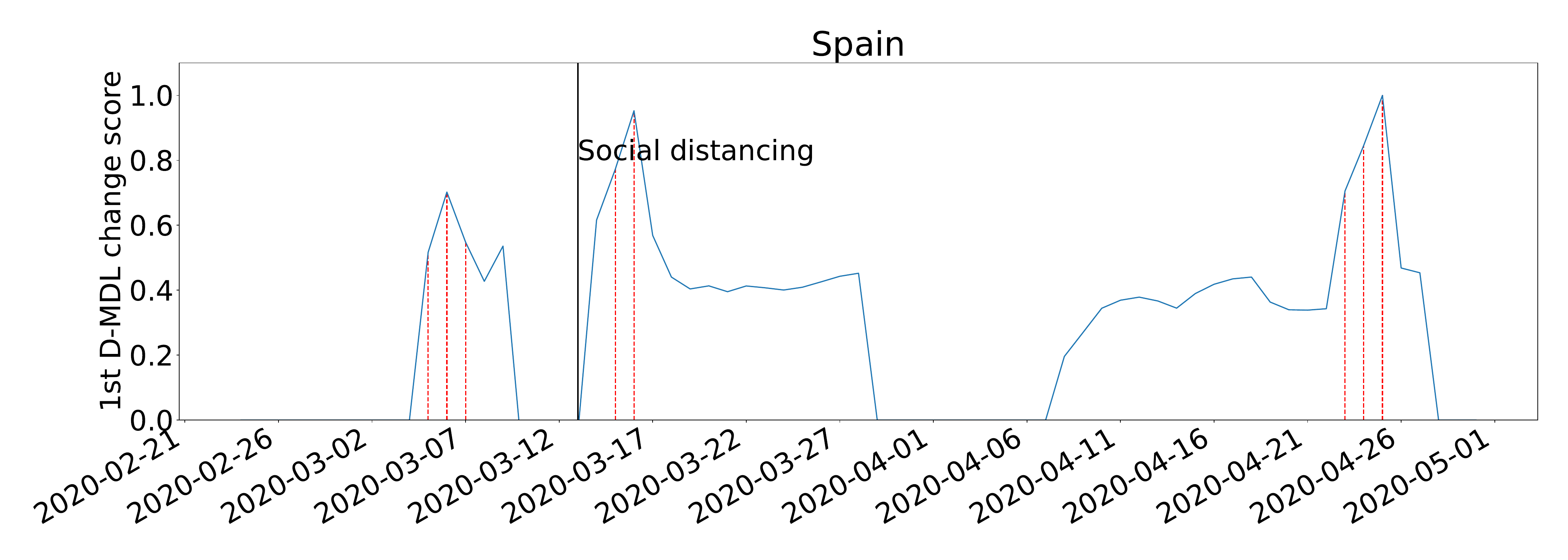} \\
			\vspace{-0.35cm}
			\textbf{e} & \includegraphics[keepaspectratio, height=3.3cm, valign=T]
			{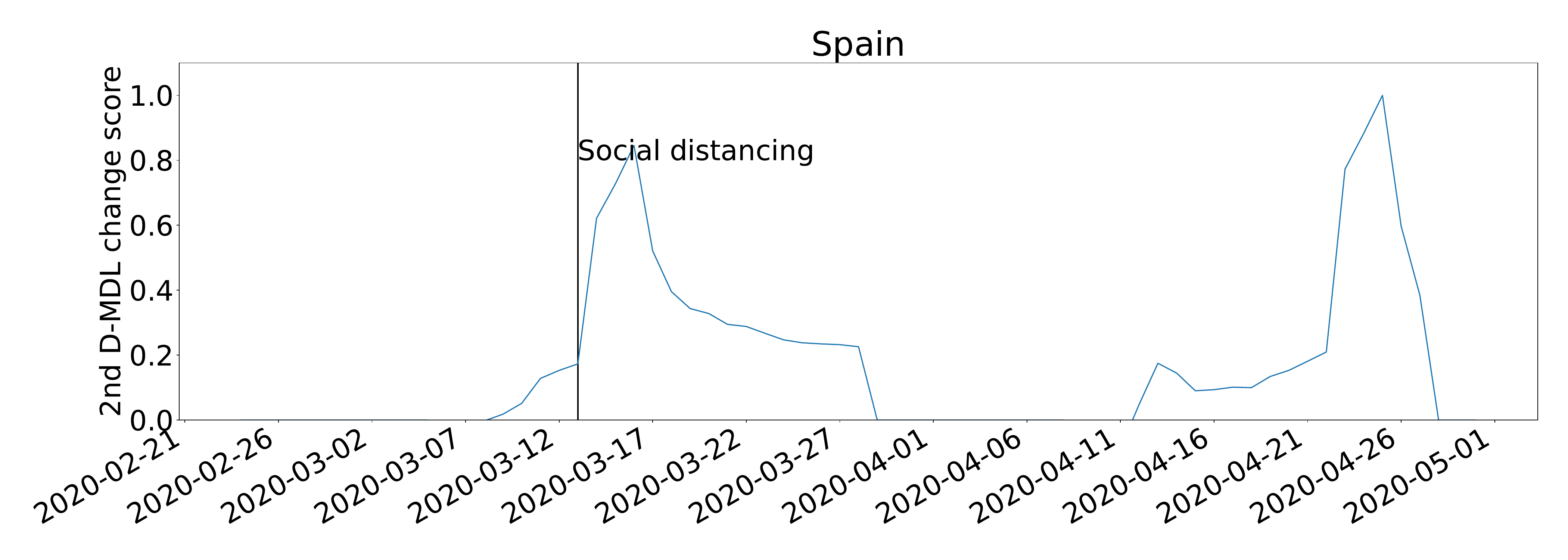} \\
		\end{tabular}
			\caption{\textbf{The results for Spain with exponential modeling.} The date on which the social distancing was implemented is marked by a solid line in black. \textbf{a,} the number of cumulative cases. \textbf{b,} the change scores produced by the 0th M-DML where the line in blue denotes values of scores and dashed lines in red mark alarms. \textbf{c,} the window sized for the sequential D-DML algorithm with adaptive window where lines in red mark the shrinkage of windows. \textbf{d,} the change scores produced by the 1st D-MDL. \textbf{e,} the change scores produced by the 2nd D-MDL.}
\end{figure}

\begin{figure}[H] 
\centering
\begin{tabular}{cc}
		 	\textbf{a} & \includegraphics[keepaspectratio, height=3.3cm, valign=T]
			{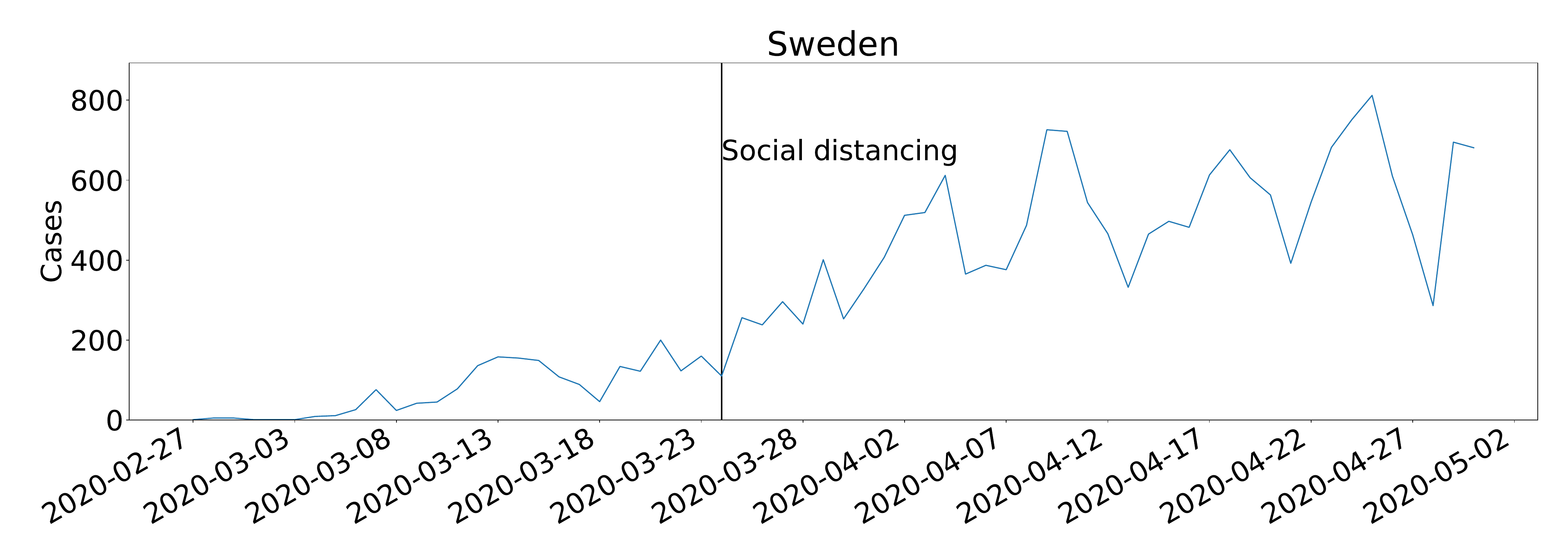} \\
			\vspace{-0.35cm}
	 	    \textbf{b} & \includegraphics[keepaspectratio, height=3.3cm, valign=T]
			{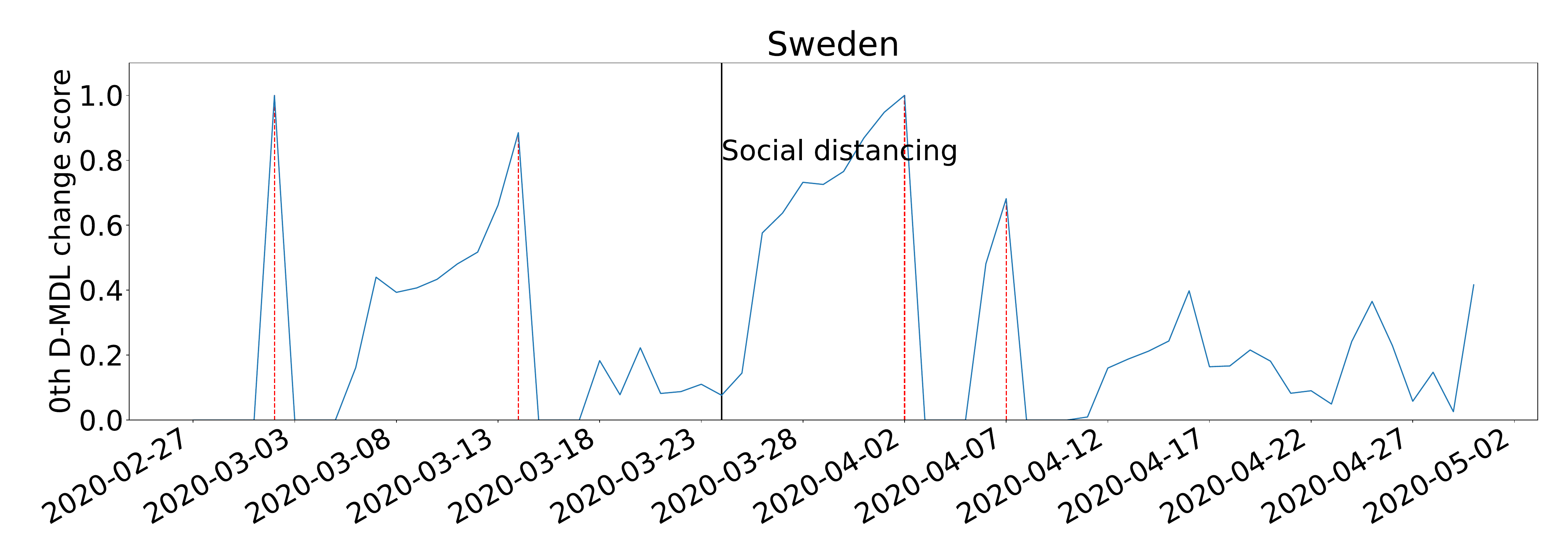}   \\
	        \vspace{-0.35cm}
			\textbf{c} & \includegraphics[keepaspectratio, height=3.3cm, valign=T]
			{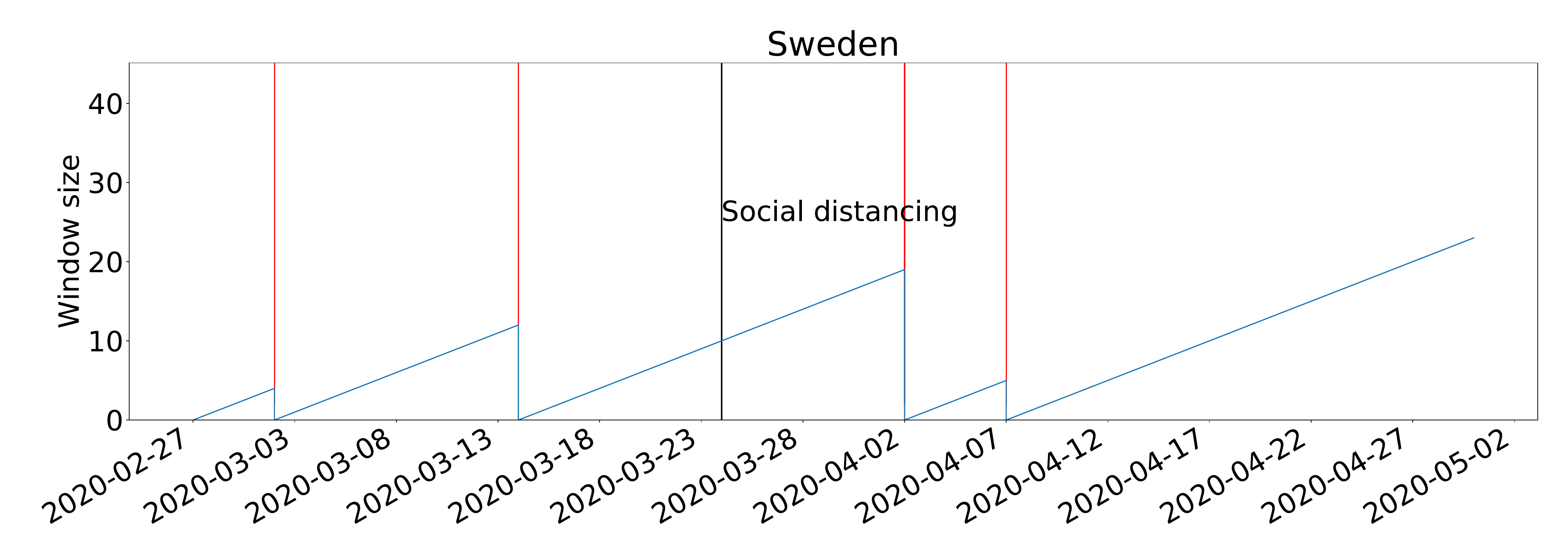} \\
		    \vspace{-0.35cm}
			\textbf{d} & \includegraphics[keepaspectratio, height=3.3cm, valign=T]
			{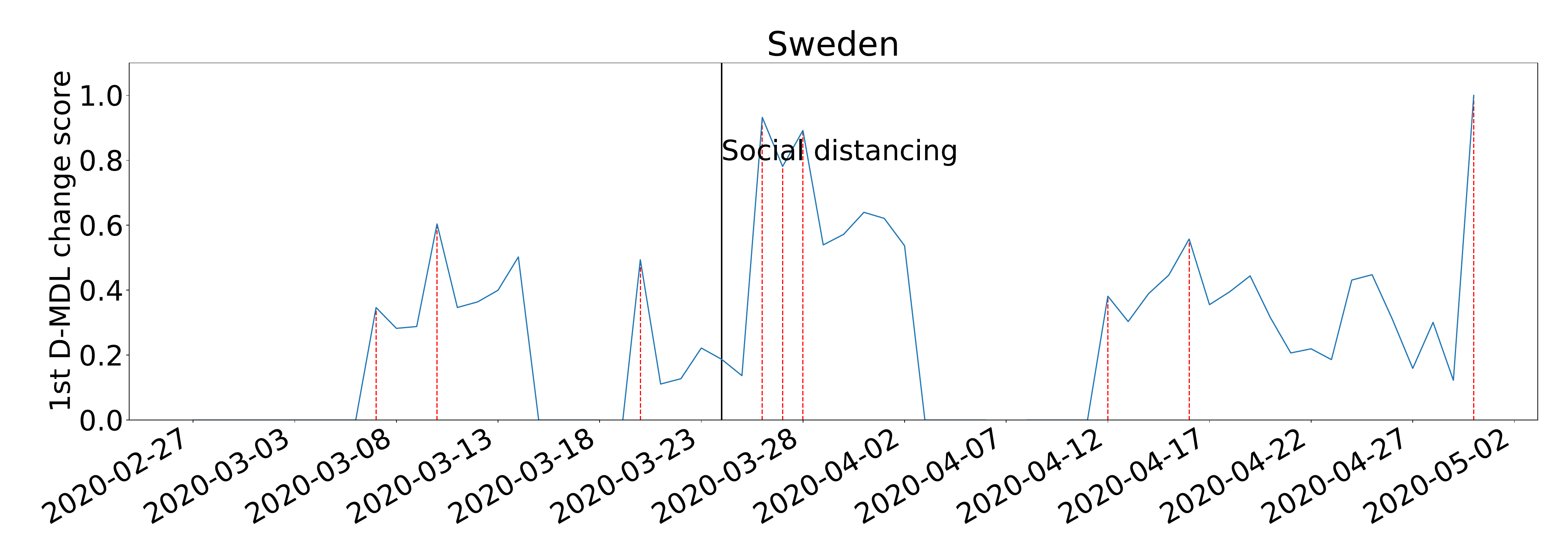} \\
		    \vspace{-0.35cm}
			\textbf{e} & \includegraphics[keepaspectratio, height=3.3cm, valign=T]
			{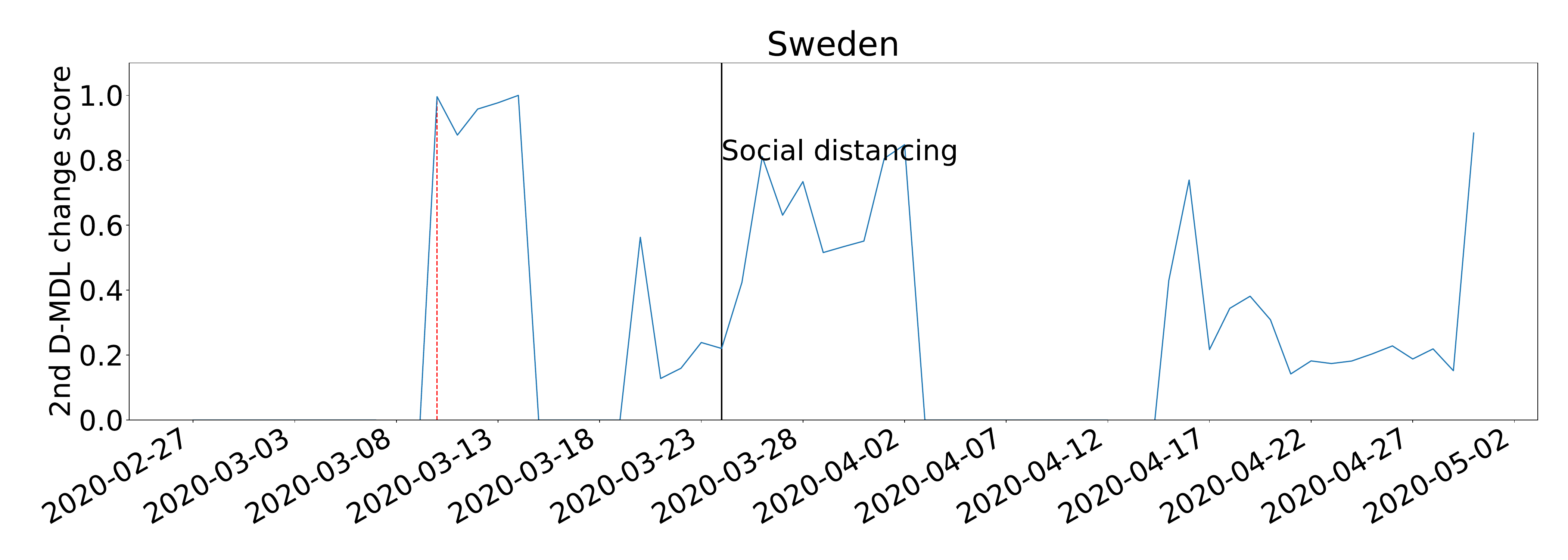} \\
		\end{tabular}
			\caption{\textbf{The results for Sweden with Gaussian modeling.} The date on which the social distancing was implemented is marked by a solid line in black. \textbf{a,} the number of daily new cases. \textbf{b,} the change scores produced by the 0th M-DML where the line in blue denotes values of scores and dashed lines in red mark alarms. \textbf{c,} the window sized for the sequential D-DML algorithm with adaptive window where lines in red mark the shrinkage of windows. \textbf{d,} the change scores produced by the 1st D-MDL. \textbf{e,} the change scores produced by the 2nd D-MDL.}
\end{figure}

\begin{figure}[H]  
\centering
\begin{tabular}{cc}
			\textbf{a} & \includegraphics[keepaspectratio, height=3.3cm, valign=T]
			{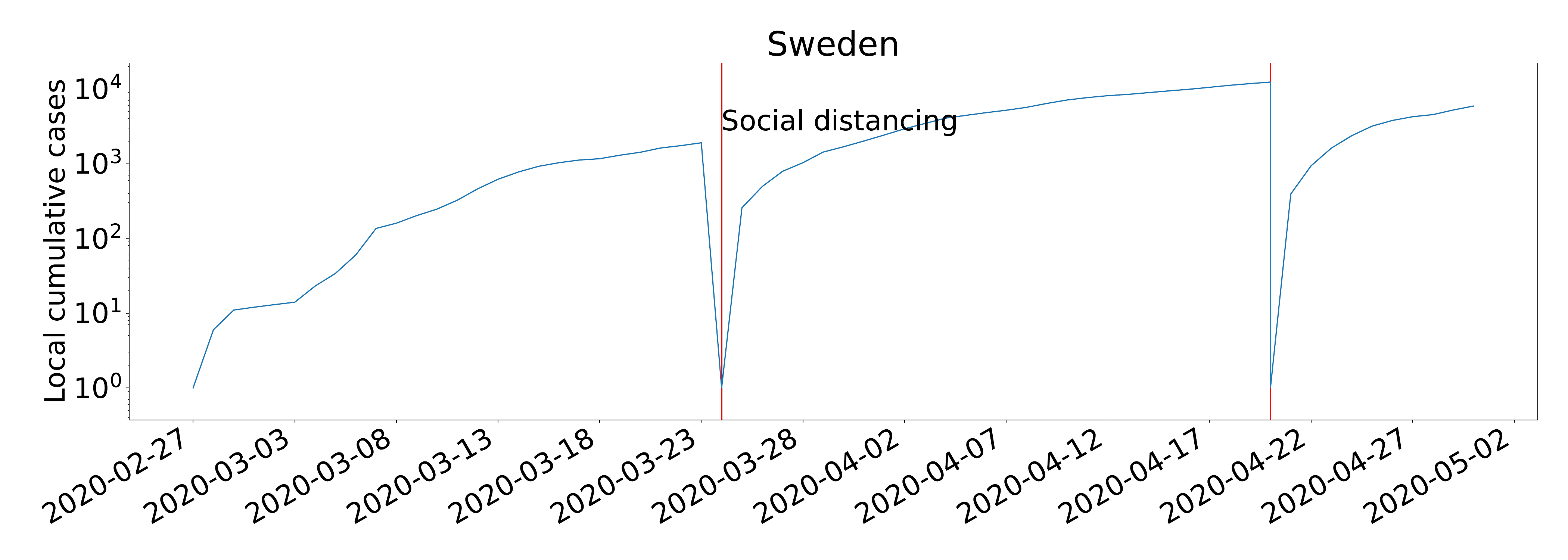} \\
	        \vspace{-0.35cm}
            \textbf{b} & \includegraphics[keepaspectratio, height=3.3cm, valign=T]
			{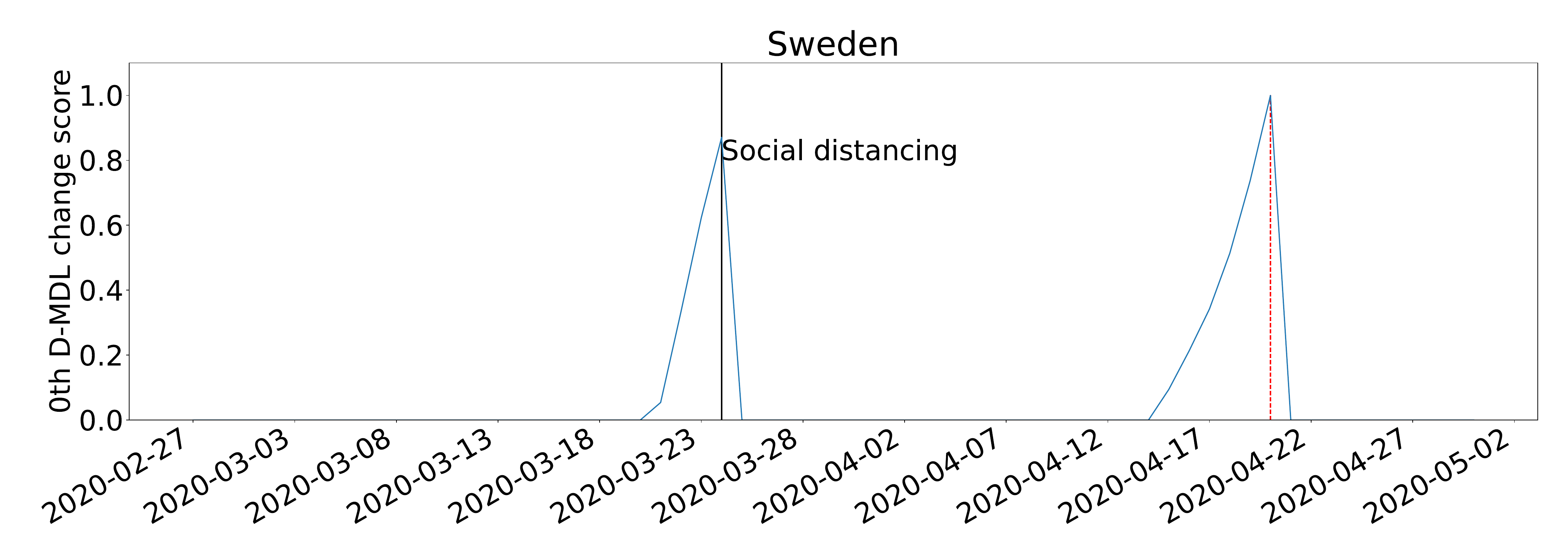}   \\
            \vspace{-0.35cm}
            \textbf{c} & \includegraphics[keepaspectratio, height=3.3cm, valign=T]
			{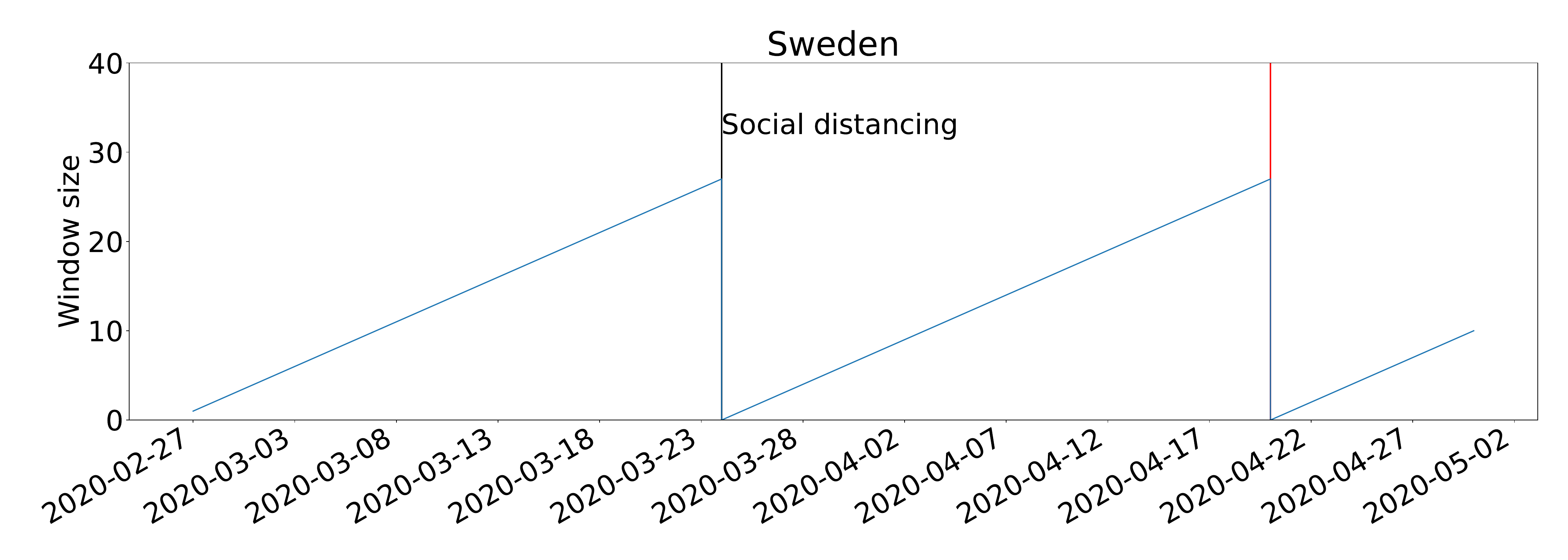} \\
			\vspace{-0.35cm}
			\textbf{d} & \includegraphics[keepaspectratio, height=3.3cm, valign=T]
			{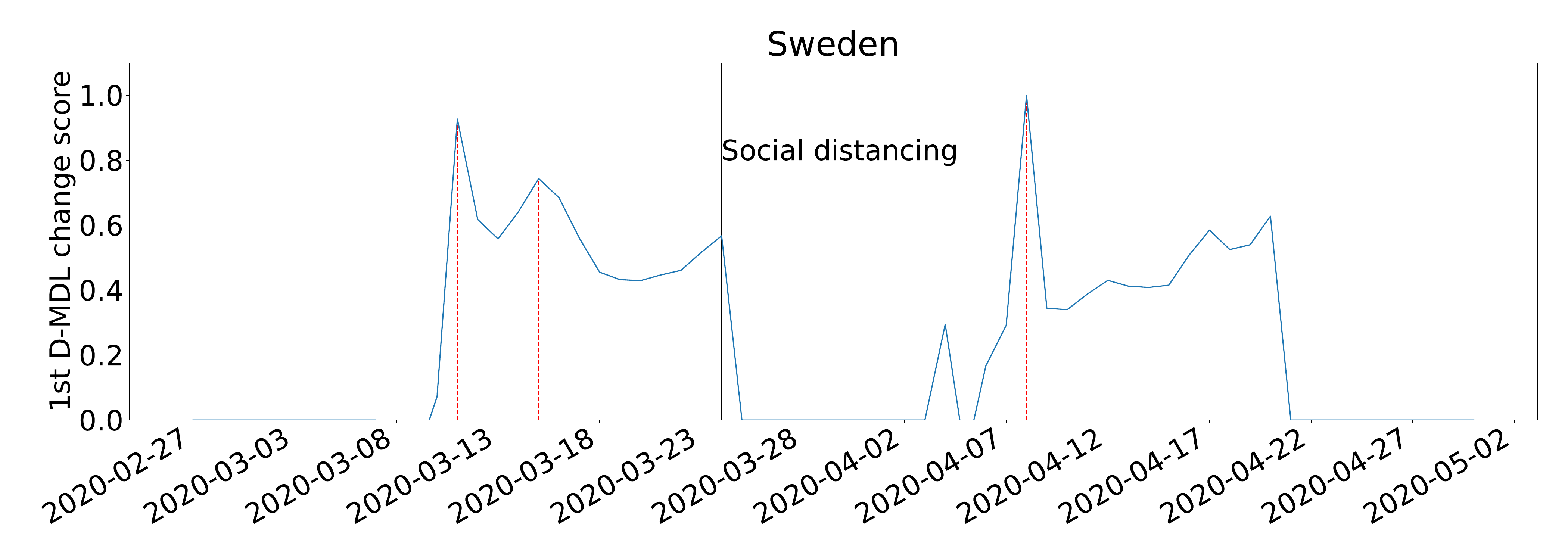} \\
			\vspace{-0.35cm}
			\textbf{e} & \includegraphics[keepaspectratio, height=3.3cm, valign=T]
			{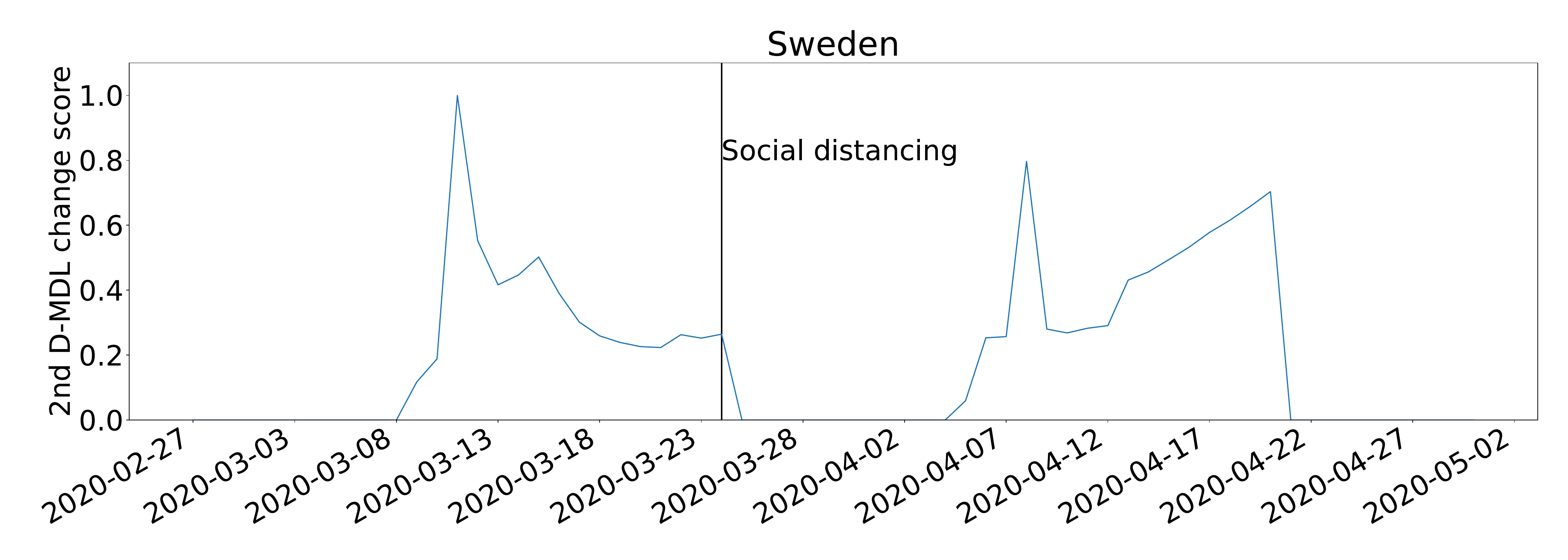} \\
		\end{tabular}
			\caption{\textbf{The results for Sweden with exponential modeling.} The date on which the social distancing was implemented is marked by a solid line in black. \textbf{a,} the number of cumulative cases. \textbf{b,} the change scores produced by the 0th M-DML where the line in blue denotes values of scores and dashed lines in red mark alarms. \textbf{c,} the window sized for the sequential D-DML algorithm with adaptive window where lines in red mark the shrinkage of windows. \textbf{d,} the change scores produced by the 1st D-MDL. \textbf{e,} the change scores produced by the 2nd D-MDL.}
\end{figure}

\begin{figure}[H] 
\centering
\begin{tabular}{cc}
		 	\textbf{a} & \includegraphics[keepaspectratio, height=3.3cm, valign=T]
			{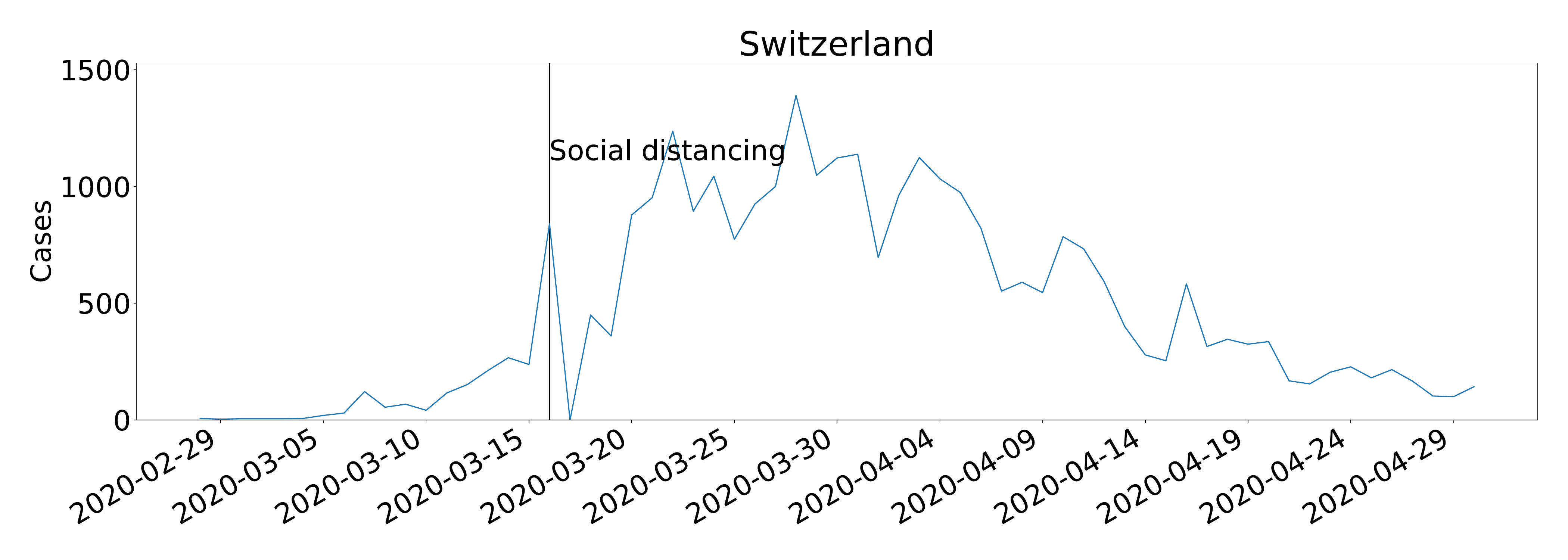} \\
			\vspace{-0.35cm}
	 	    \textbf{b} & \includegraphics[keepaspectratio, height=3.3cm, valign=T]
			{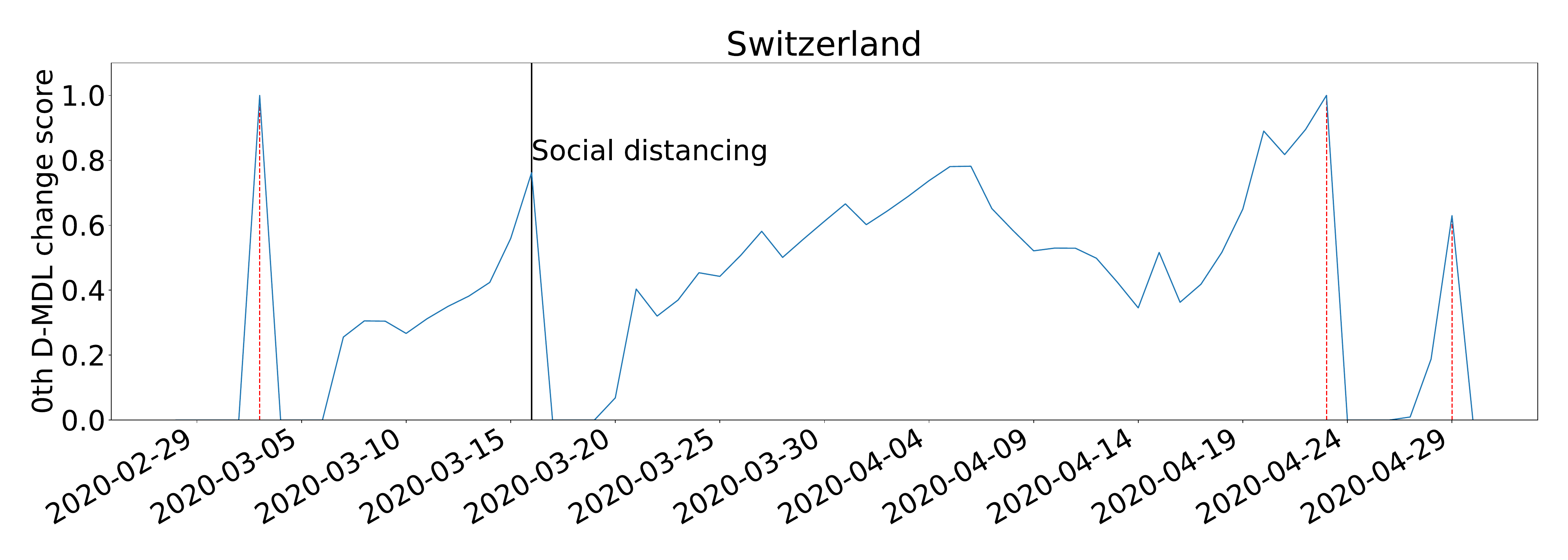}   \\
	        \vspace{-0.35cm}
			\textbf{c} & \includegraphics[keepaspectratio, height=3.3cm, valign=T]
			{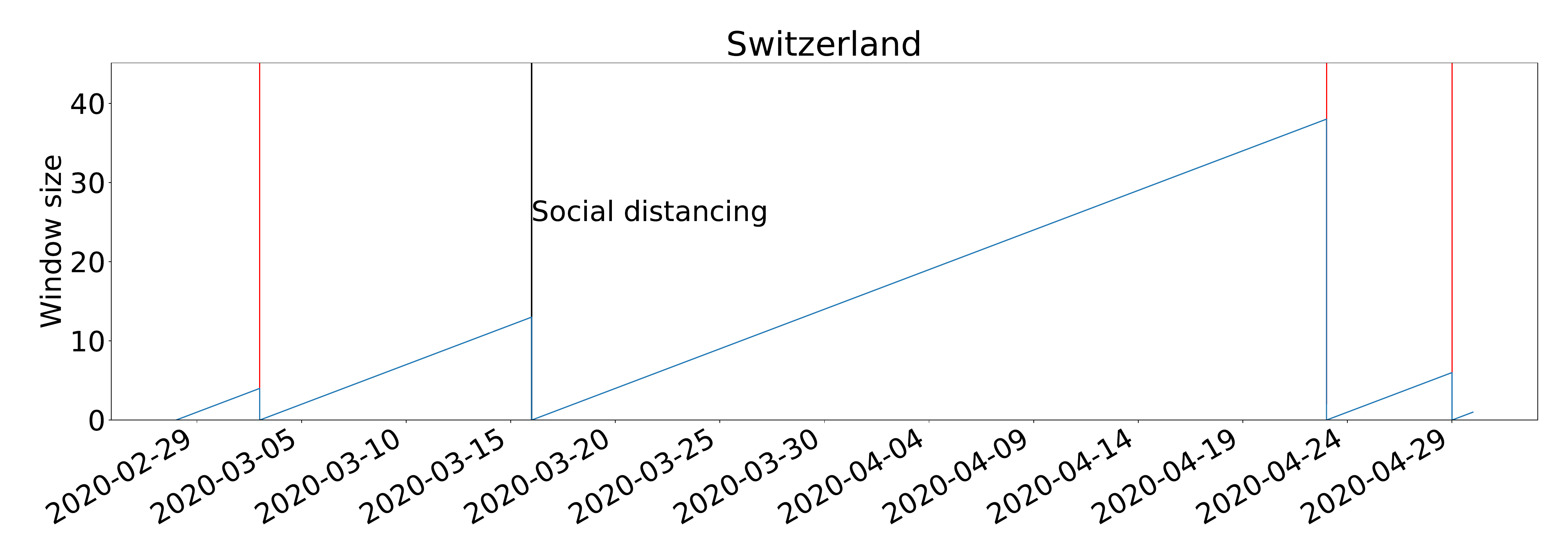} \\
		    \vspace{-0.35cm}
			\textbf{d} & \includegraphics[keepaspectratio, height=3.3cm, valign=T]
			{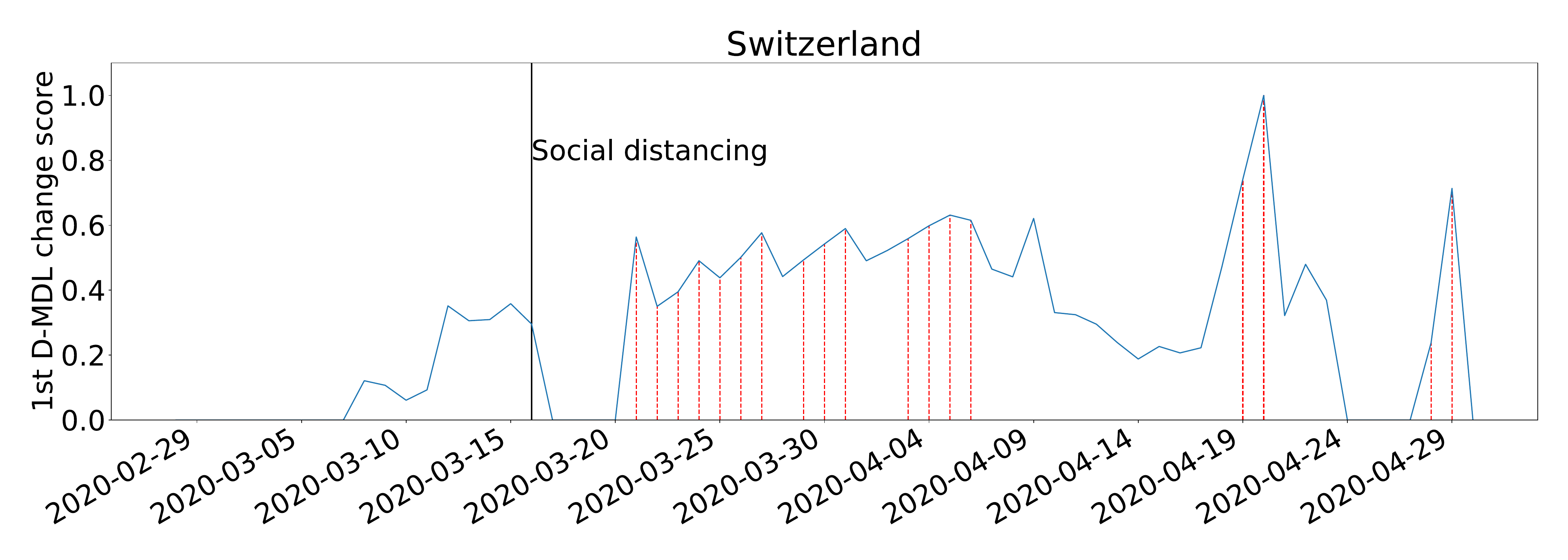} \\
		    \vspace{-0.35cm}
			\textbf{e} & \includegraphics[keepaspectratio, height=3.3cm, valign=T]
			{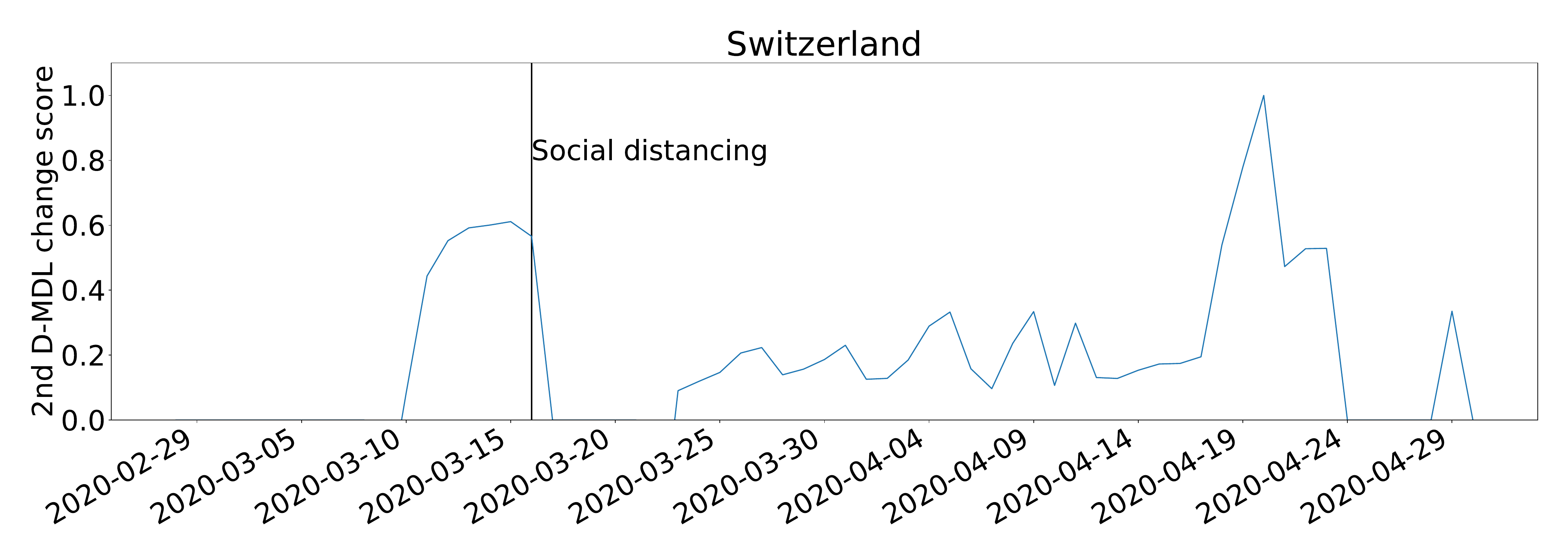} \\
		\end{tabular}
			\caption{\textbf{The results for Switzerland with Gaussian modeling.} The date on which the social distancing was implemented is marked by a solid line in black. \textbf{a,} the number of daily new cases. \textbf{b,} the change scores produced by the 0th M-DML where the line in blue denotes values of scores and dashed lines in red mark alarms. \textbf{c,} the window sized for the sequential D-DML algorithm with adaptive window where lines in red mark the shrinkage of windows. \textbf{d,} the change scores produced by the 1st D-MDL. \textbf{e,} the change scores produced by the 2nd D-MDL.}
\end{figure}

\begin{figure}[H]  
\centering
\begin{tabular}{cc}
			\textbf{a} & \includegraphics[keepaspectratio, height=3.3cm, valign=T]
			{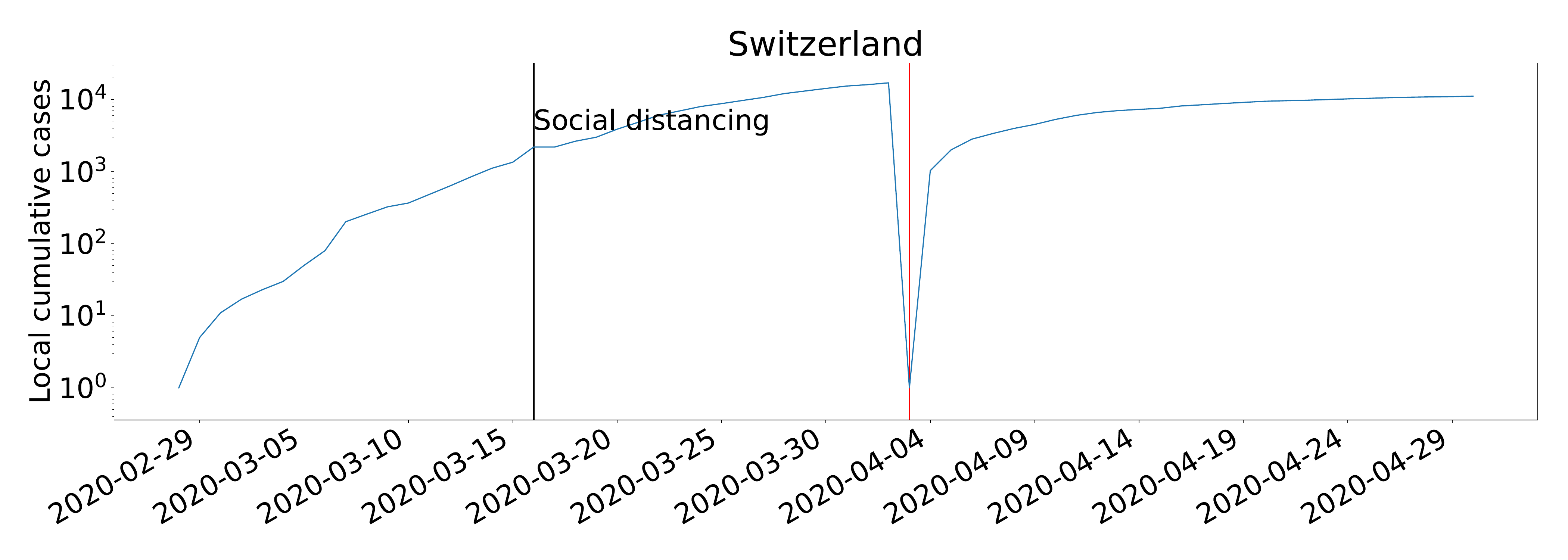} \\
	        \vspace{-0.35cm}
            \textbf{b} & \includegraphics[keepaspectratio, height=3.3cm, valign=T]
			{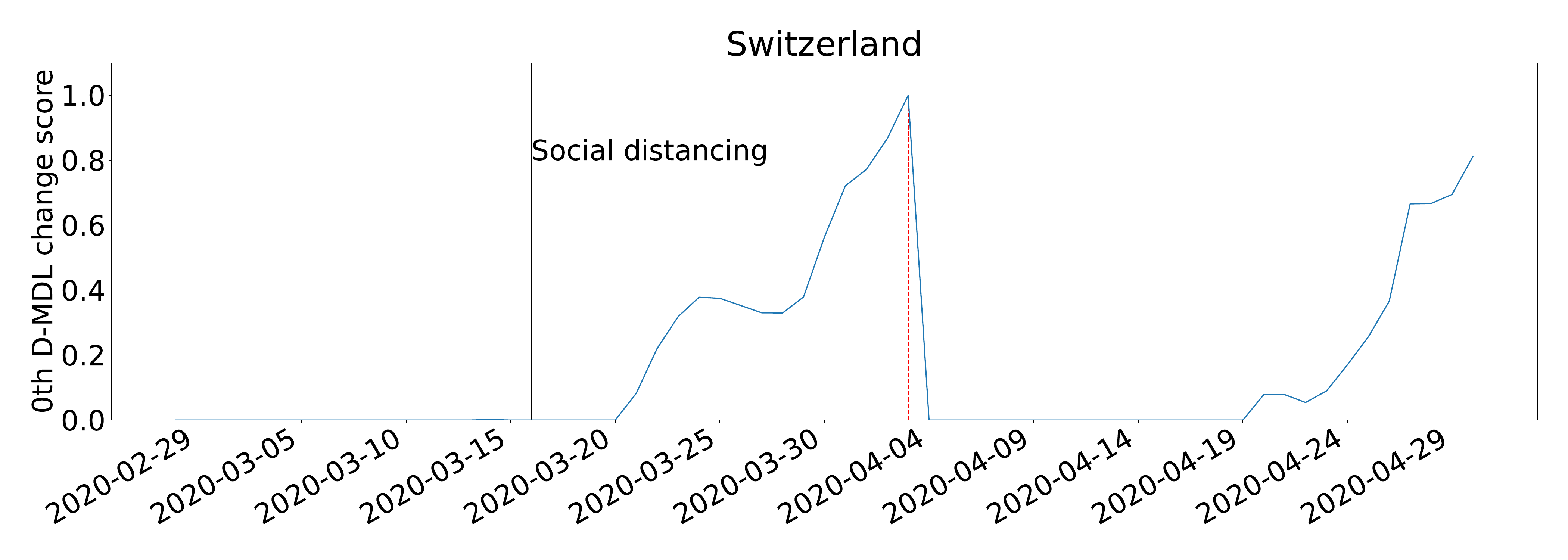}   \\
            \vspace{-0.35cm}
            \textbf{c} & \includegraphics[keepaspectratio, height=3.3cm, valign=T]
			{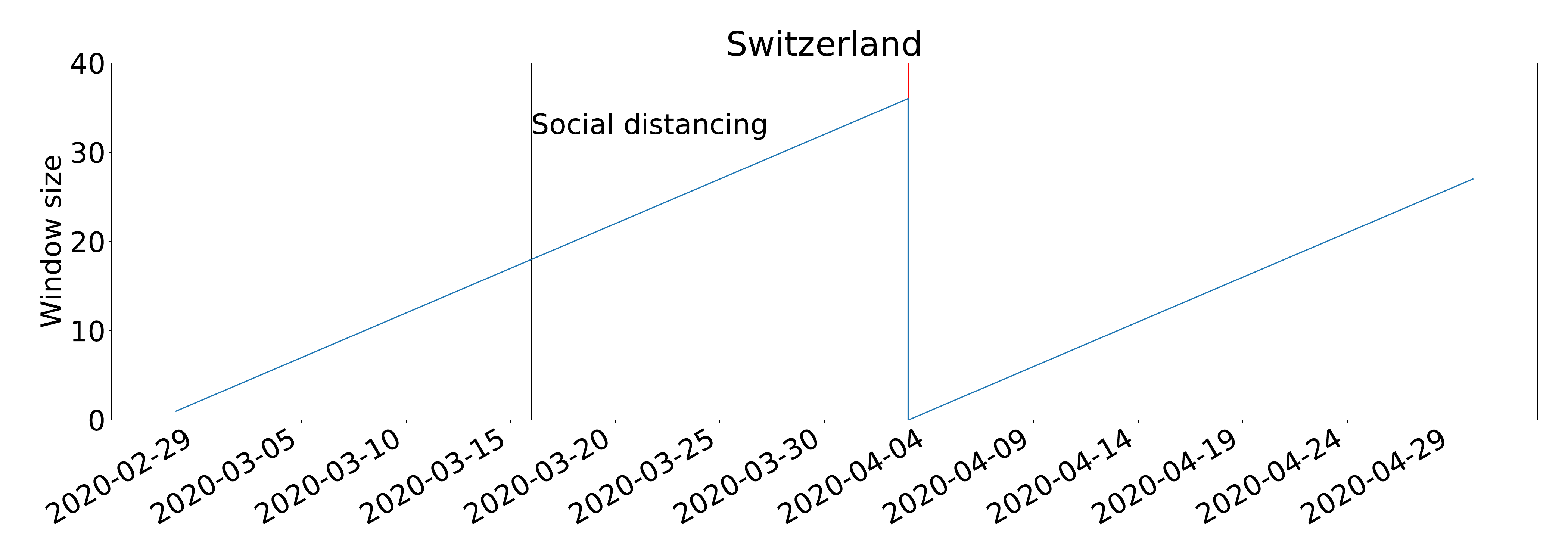} \\
			\vspace{-0.35cm}
			\textbf{d} & \includegraphics[keepaspectratio, height=3.3cm, valign=T]
			{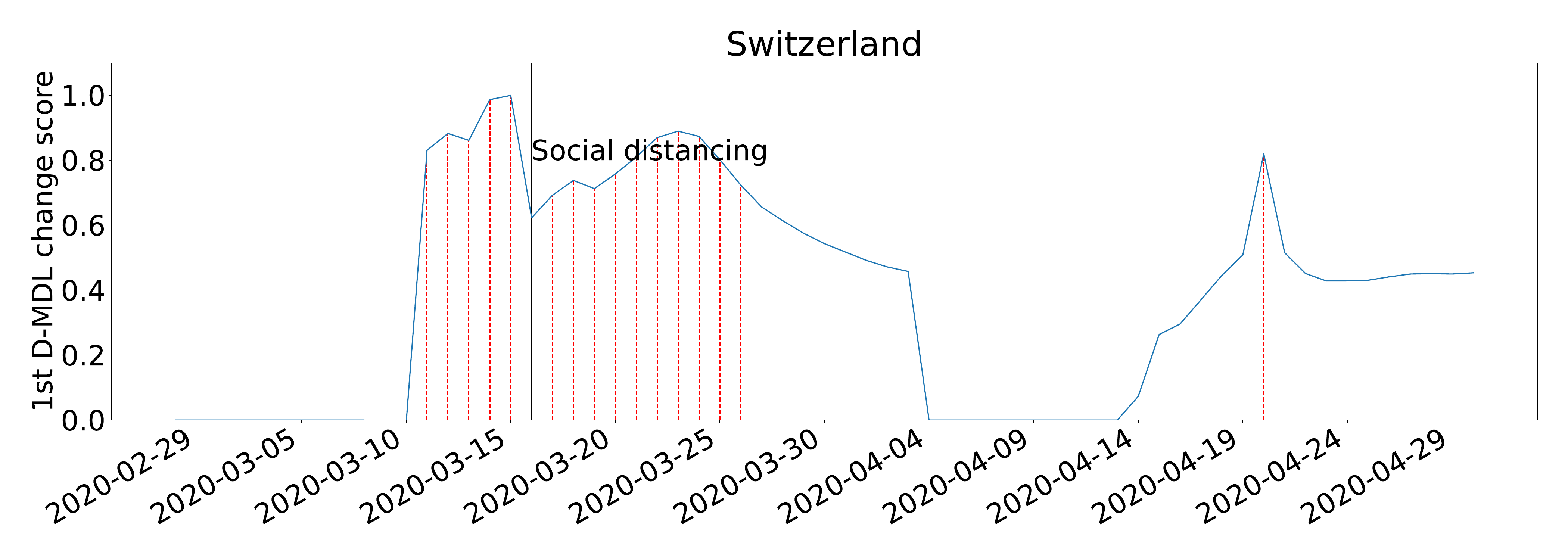} \\
			\vspace{-0.35cm}
			\textbf{e} & \includegraphics[keepaspectratio, height=3.3cm, valign=T]
			{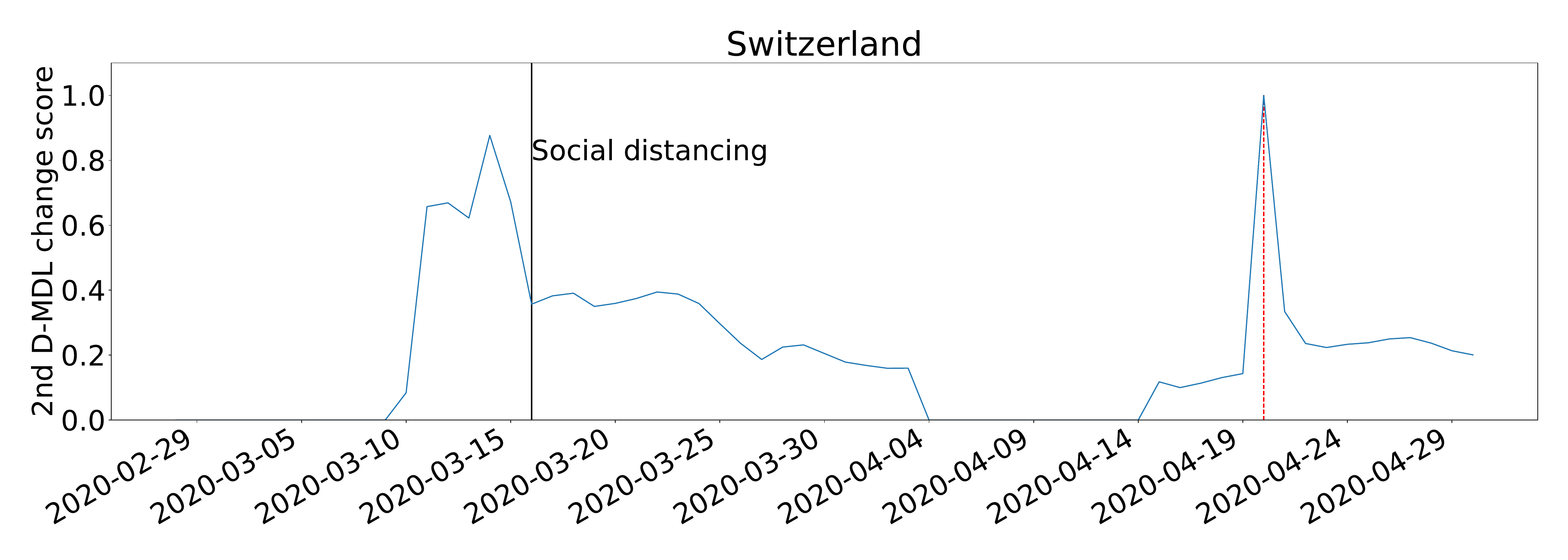} \\
		\end{tabular}
			\caption{\textbf{The results for Switzerland with exponential modeling.} The date on which the social distancing was implemented is marked by a solid line in black. \textbf{a,} the number of cumulative cases. \textbf{b,} the change scores produced by the 0th M-DML where the line in blue denotes values of scores and dashed lines in red mark alarms. \textbf{c,} the window sized for the sequential D-DML algorithm with adaptive window where lines in red mark the shrinkage of windows. \textbf{d,} the change scores produced by the 1st D-MDL. \textbf{e,} the change scores produced by the 2nd D-MDL.}
			\label{exp:switzerland}
\end{figure}

\begin{figure}[H] 
\centering
\begin{tabular}{cc}
		 	\textbf{a} & \includegraphics[keepaspectratio, height=3.3cm, valign=T]
			{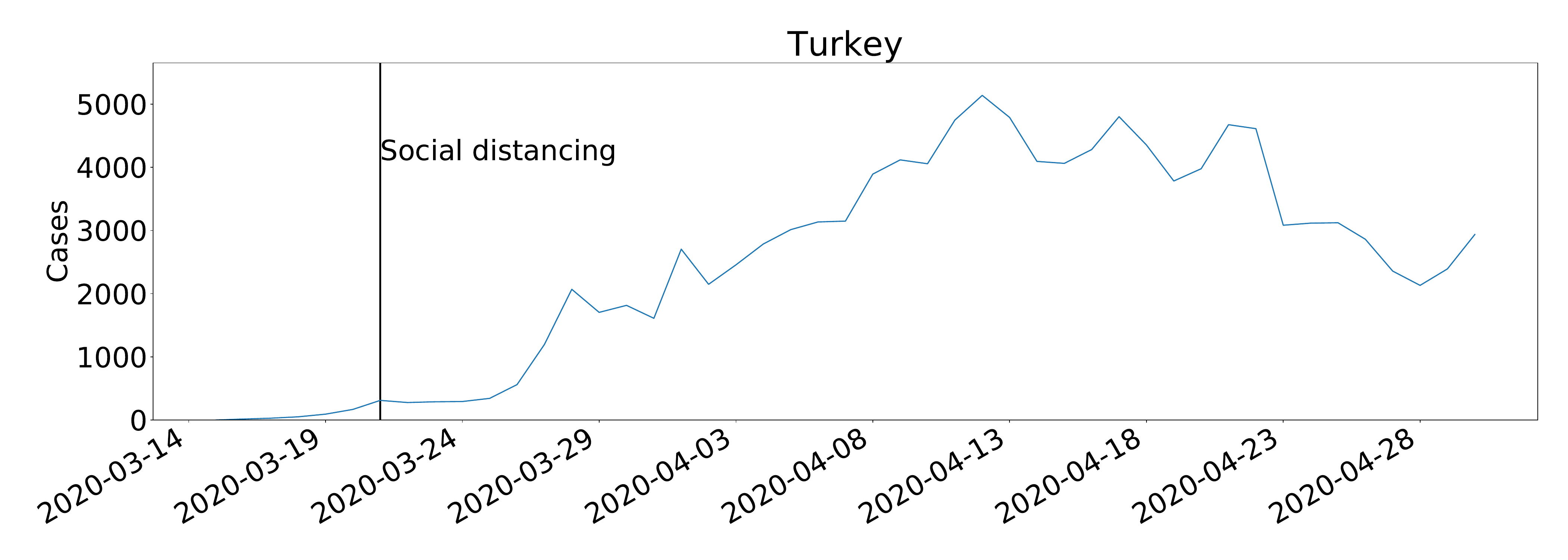} \\
			\vspace{-0.35cm}
	 	    \textbf{b} & \includegraphics[keepaspectratio, height=3.3cm, valign=T]
			{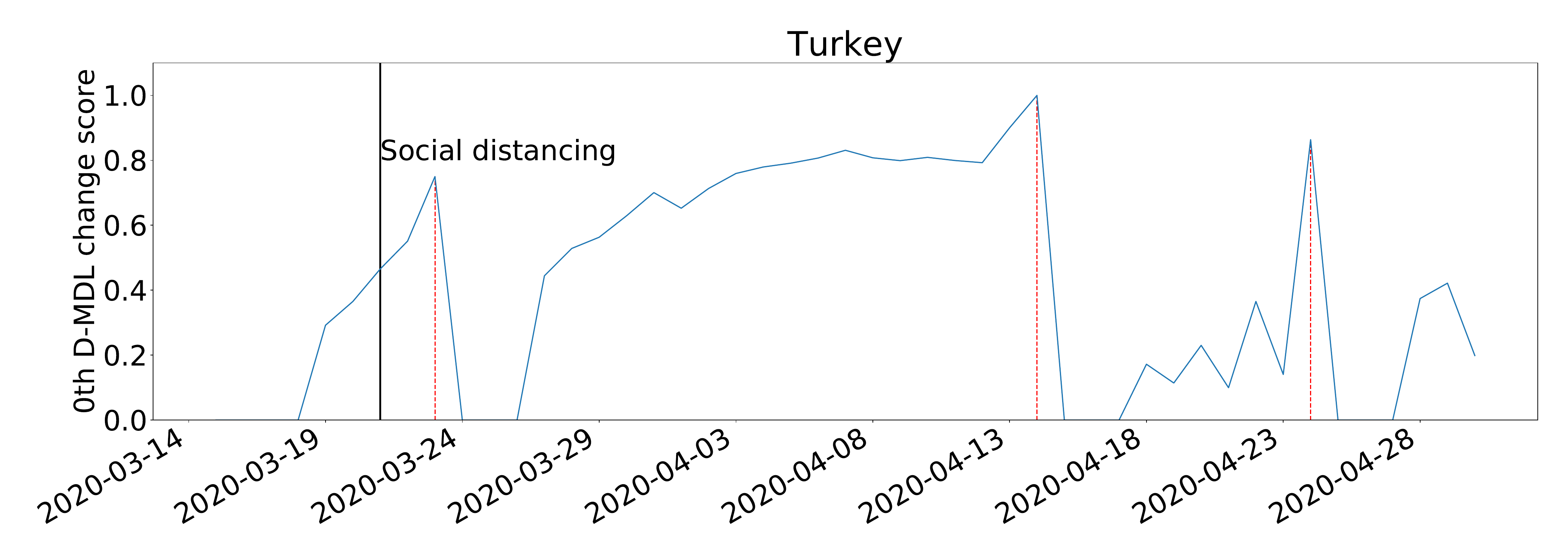}   \\
	        \vspace{-0.35cm}
			\textbf{c} & \includegraphics[keepaspectratio, height=3.3cm, valign=T]
			{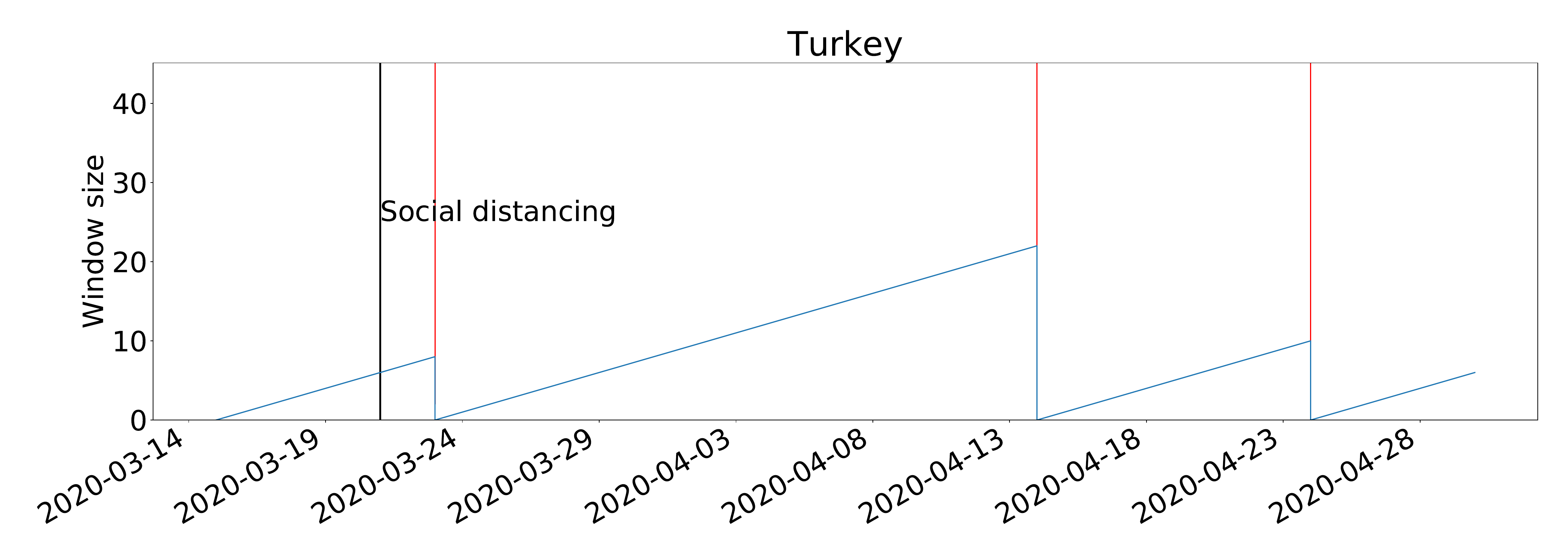} \\
		    \vspace{-0.35cm}
			\textbf{d} & \includegraphics[keepaspectratio, height=3.3cm, valign=T]
			{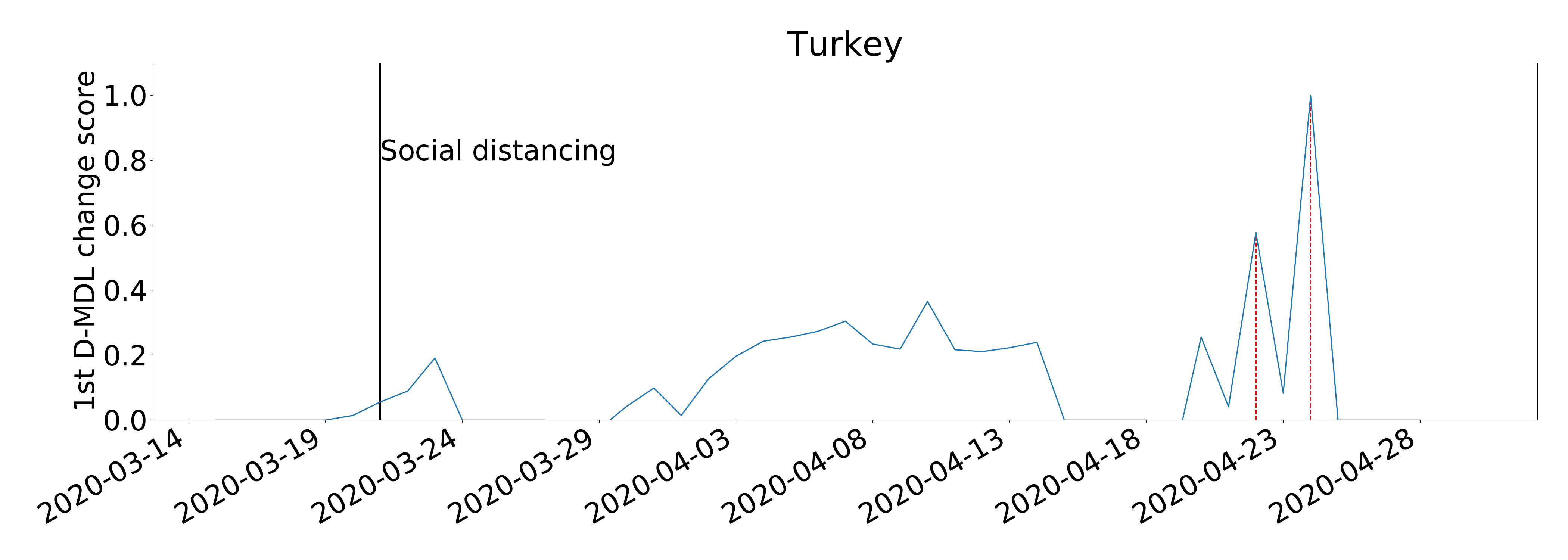} \\
		    \vspace{-0.35cm}
			\textbf{e} & \includegraphics[keepaspectratio, height=3.3cm, valign=T]
			{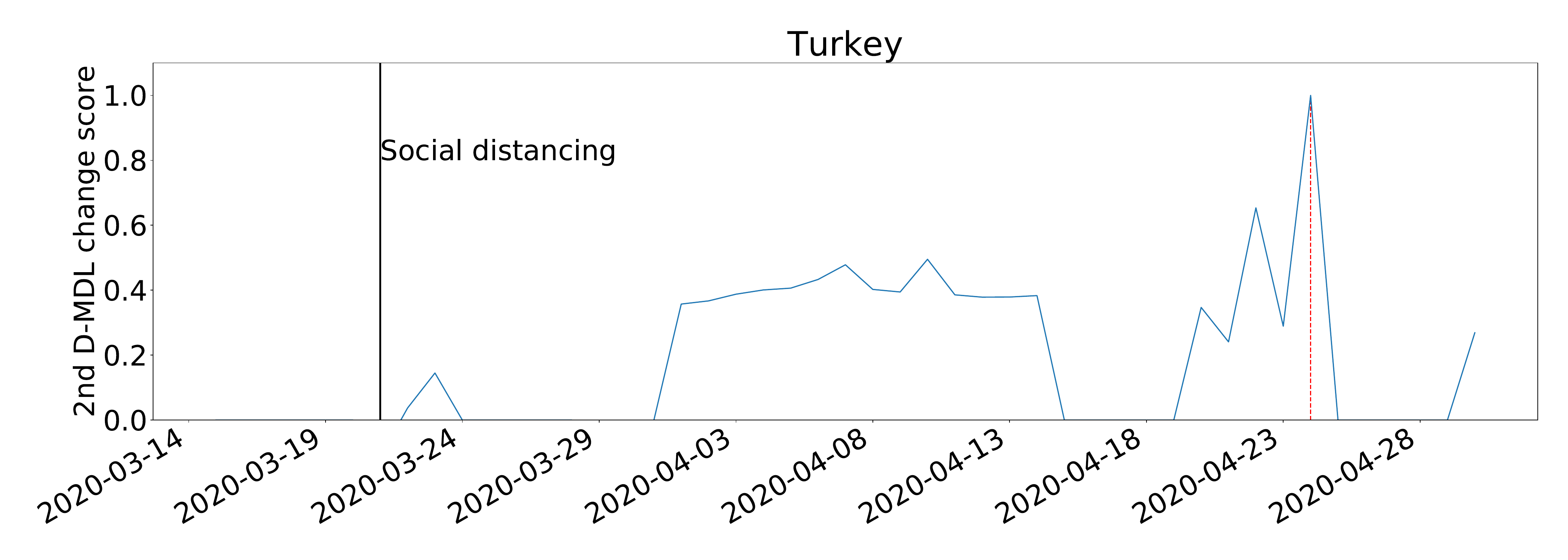} \\
		\end{tabular}
			\caption{\textbf{The results for Turkey with Gaussian modeling.} The date on which the social distancing was implemented is marked by a solid line in black. \textbf{a,} the number of daily new cases. \textbf{b,} the change scores produced by the 0th M-DML where the line in blue denotes values of scores and dashed lines in red mark alarms. \textbf{c,} the window sized for the sequential D-DML algorithm with adaptive window where lines in red mark the shrinkage of windows. \textbf{d,} the change scores produced by the 1st D-MDL. \textbf{e,} the change scores produced by the 2nd D-MDL.}
\end{figure}

\begin{figure}[H]  
\centering
\begin{tabular}{cc}
			\textbf{a} & \includegraphics[keepaspectratio, height=3.3cm, valign=T]
			{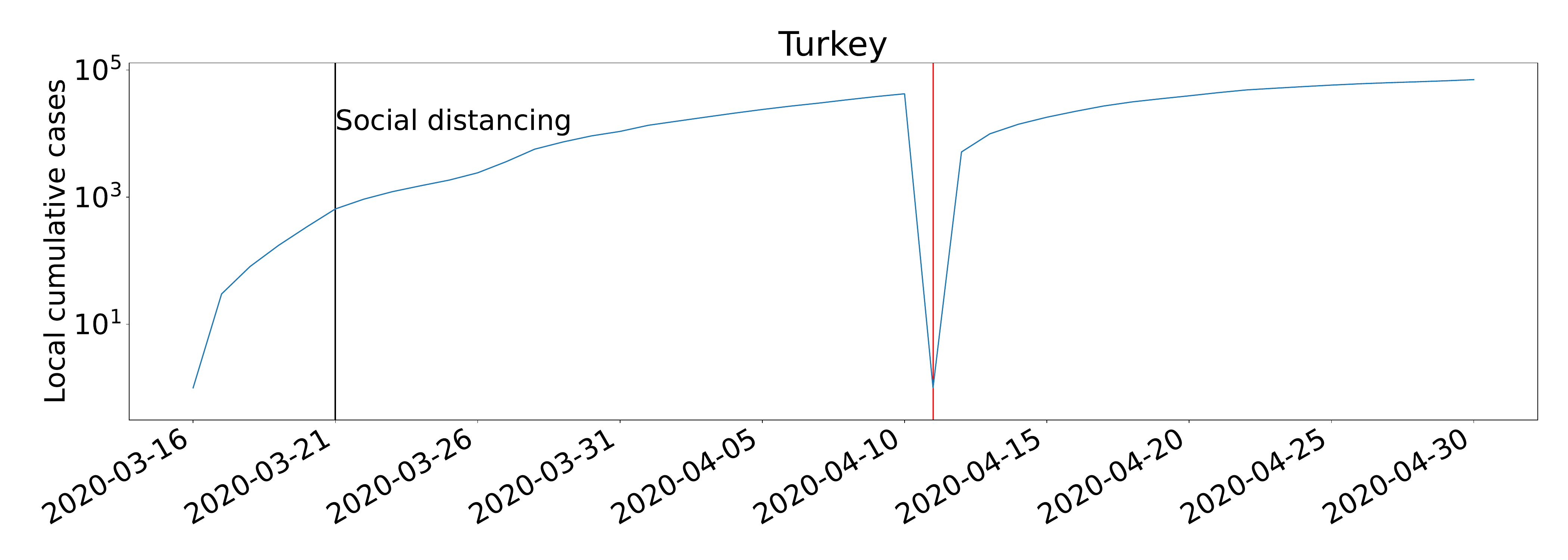} \\
	        \vspace{-0.35cm}
            \textbf{b} & \includegraphics[keepaspectratio, height=3.3cm, valign=T]
			{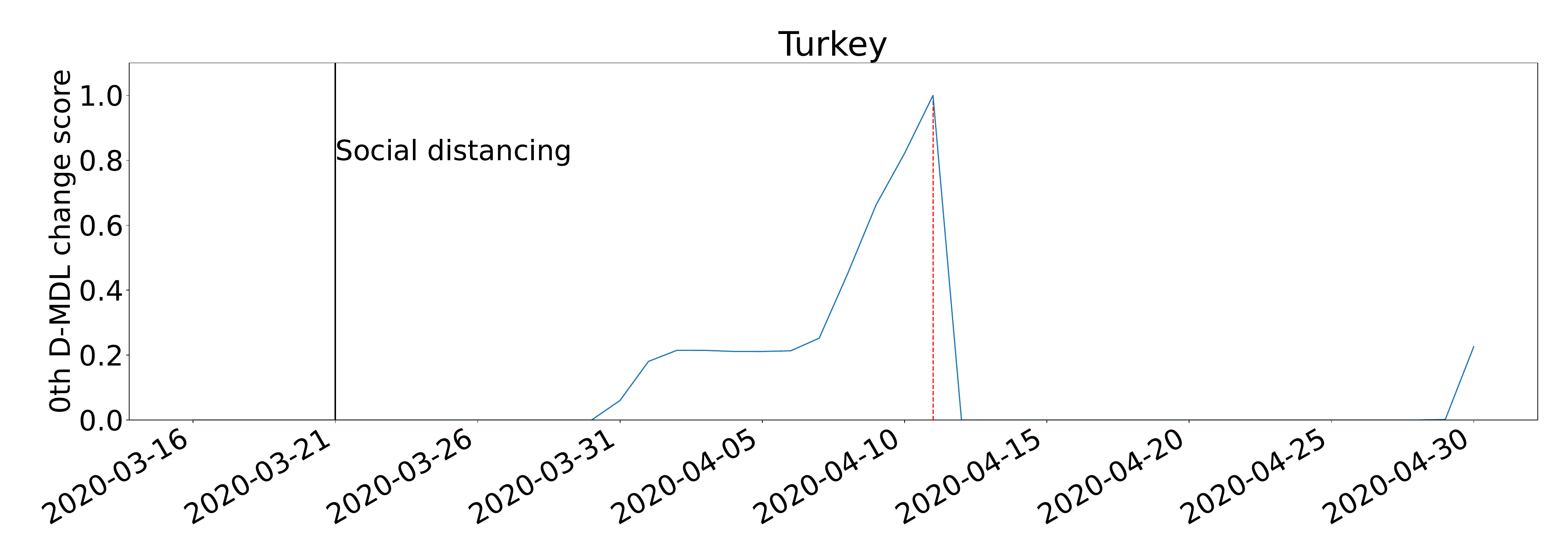}   \\
            \vspace{-0.35cm}
            \textbf{c} & \includegraphics[keepaspectratio, height=3.3cm, valign=T]
			{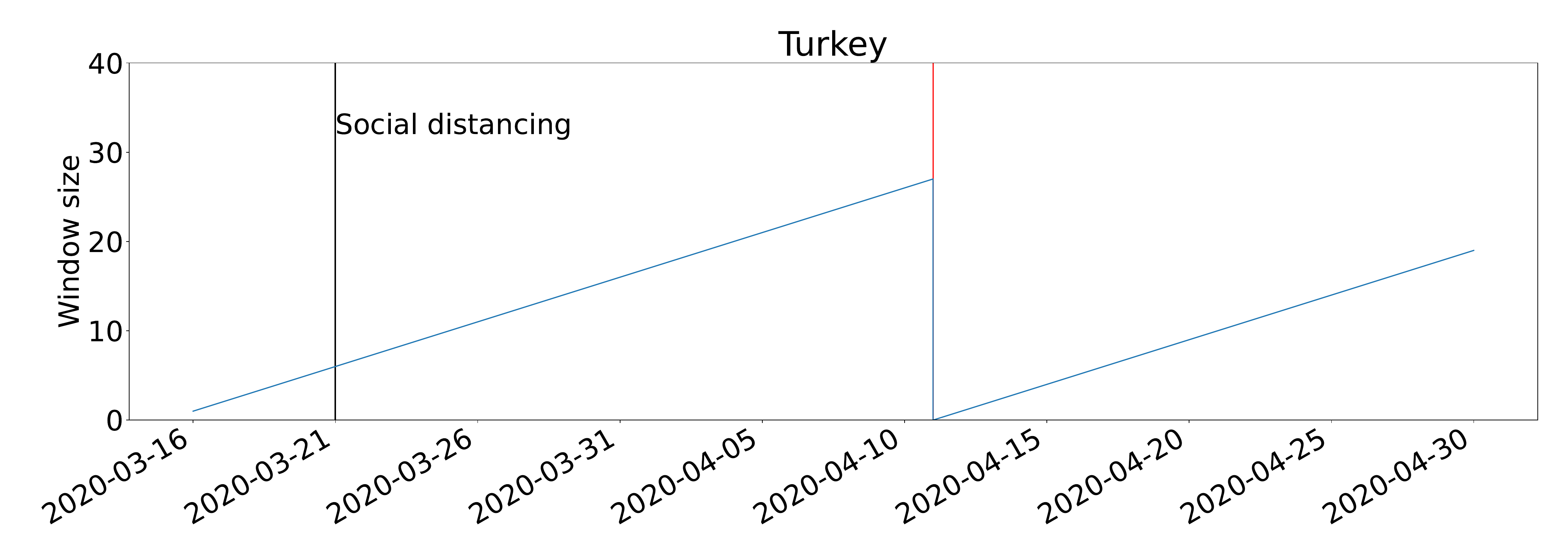} \\
			\vspace{-0.35cm}
			\textbf{d} & \includegraphics[keepaspectratio, height=3.3cm, valign=T]
			{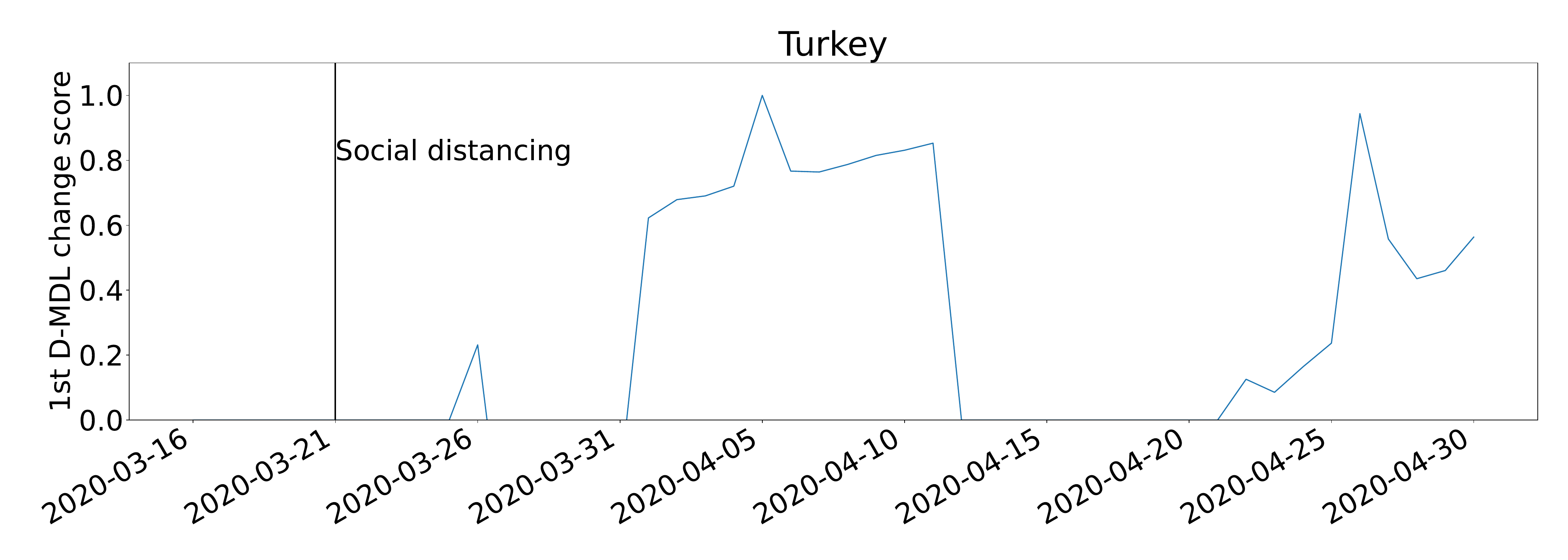} \\
			\vspace{-0.35cm}
			\textbf{e} & \includegraphics[keepaspectratio, height=3.3cm, valign=T]
			{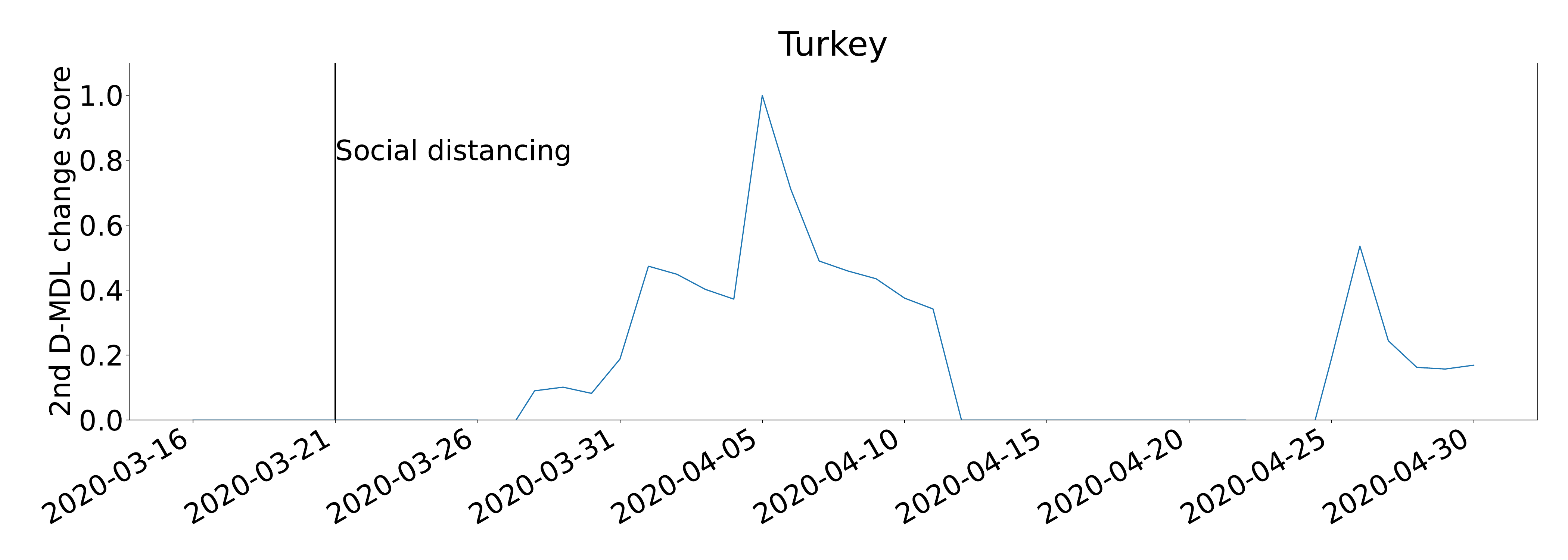} \\
		\end{tabular}
			\caption{\textbf{The results for Turkey with exponential modeling.} The date on which the social distancing was implemented is marked by a solid line in black. \textbf{a,} the number of cumulative cases. \textbf{b,} the change scores produced by the 0th M-DML where the line in blue denotes values of scores and dashed lines in red mark alarms. \textbf{c,} the window sized for the sequential D-DML algorithm with adaptive window where lines in red mark the shrinkage of windows. \textbf{d,} the change scores produced by the 1st D-MDL. \textbf{e,} the change scores produced by the 2nd D-MDL.}
			\label{exp:turkey}
\end{figure}

\begin{figure}[H] 
\centering
\begin{tabular}{cc}
		 	\textbf{a} & \includegraphics[keepaspectratio, height=3.3cm, valign=T]
			{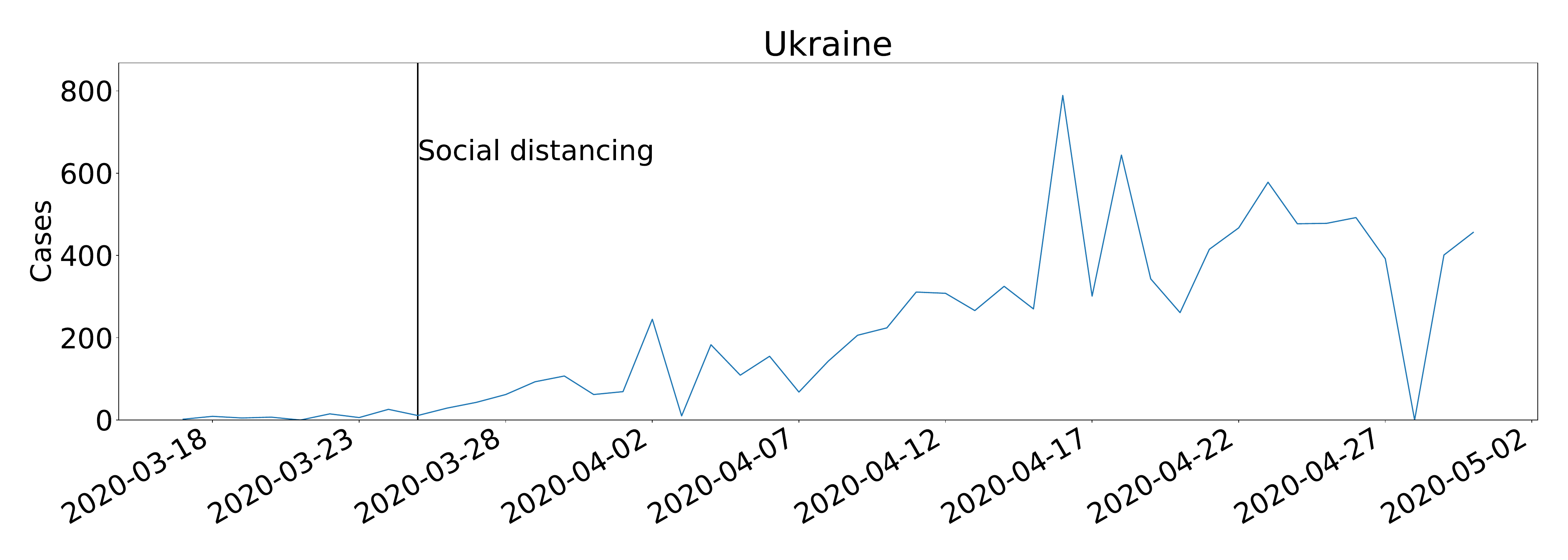} \\
			\vspace{-0.35cm}
	 	    \textbf{b} & \includegraphics[keepaspectratio, height=3.3cm, valign=T]
			{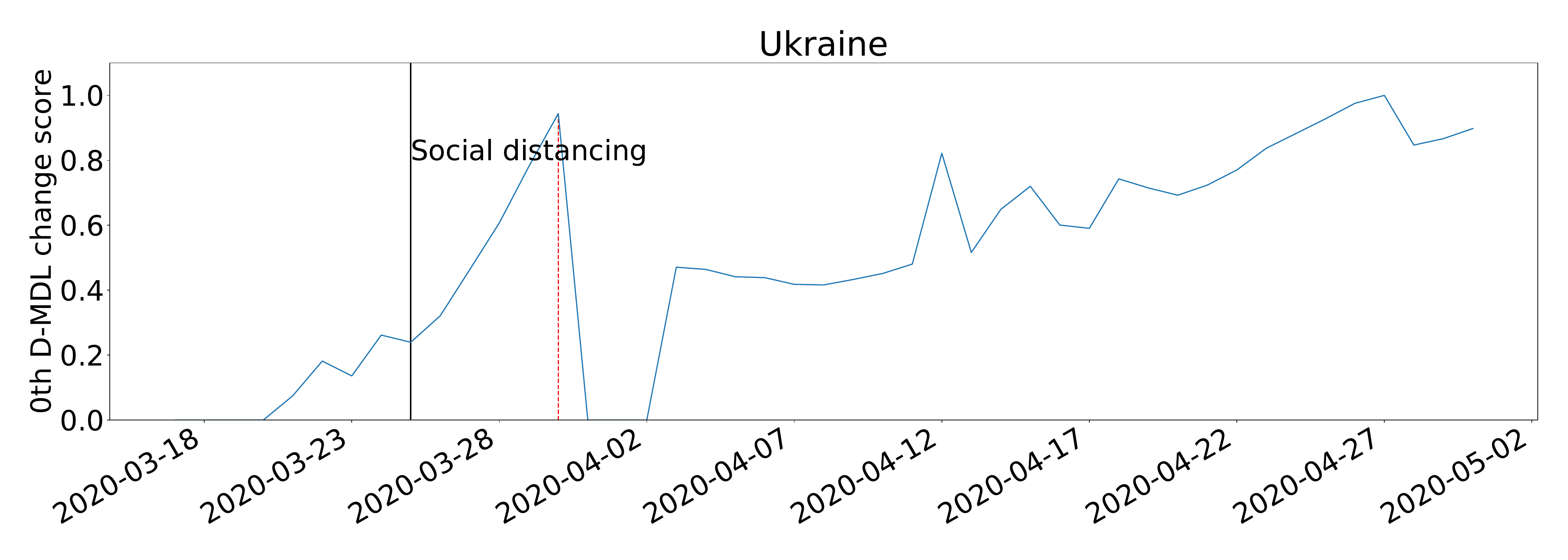}   \\
	        \vspace{-0.35cm}
			\textbf{c} & \includegraphics[keepaspectratio, height=3.3cm, valign=T]
			{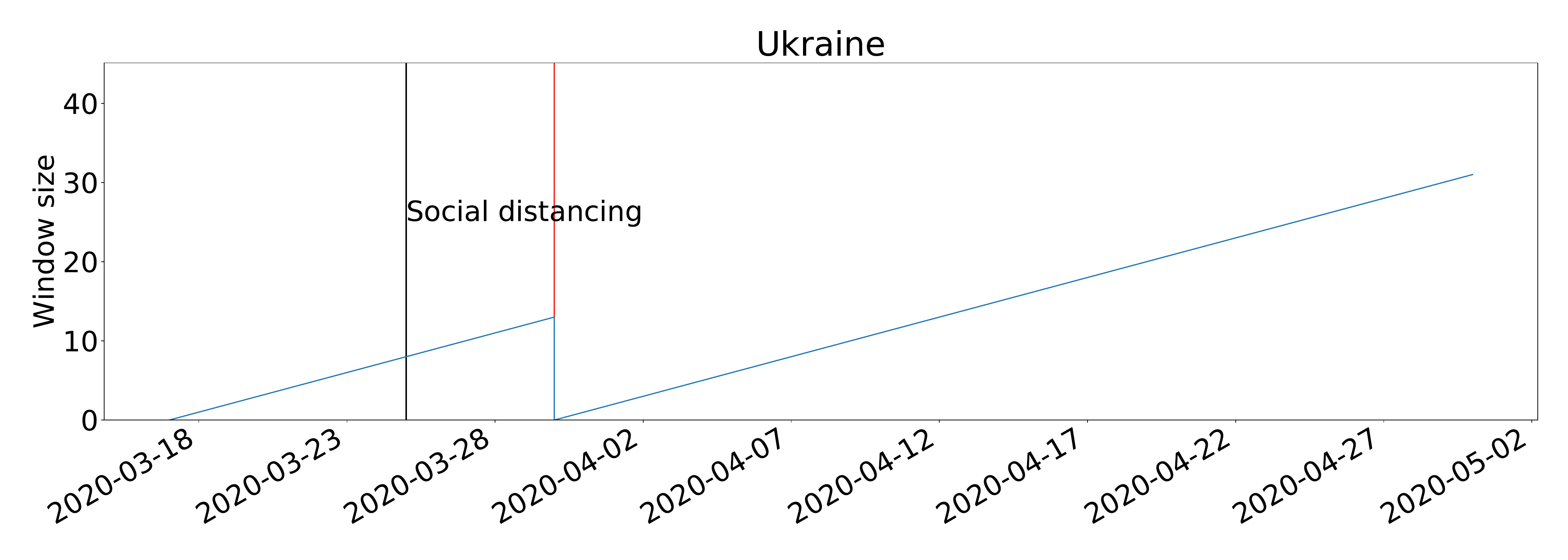} \\
		    \vspace{-0.35cm}
			\textbf{d} & \includegraphics[keepaspectratio, height=3.3cm, valign=T]
			{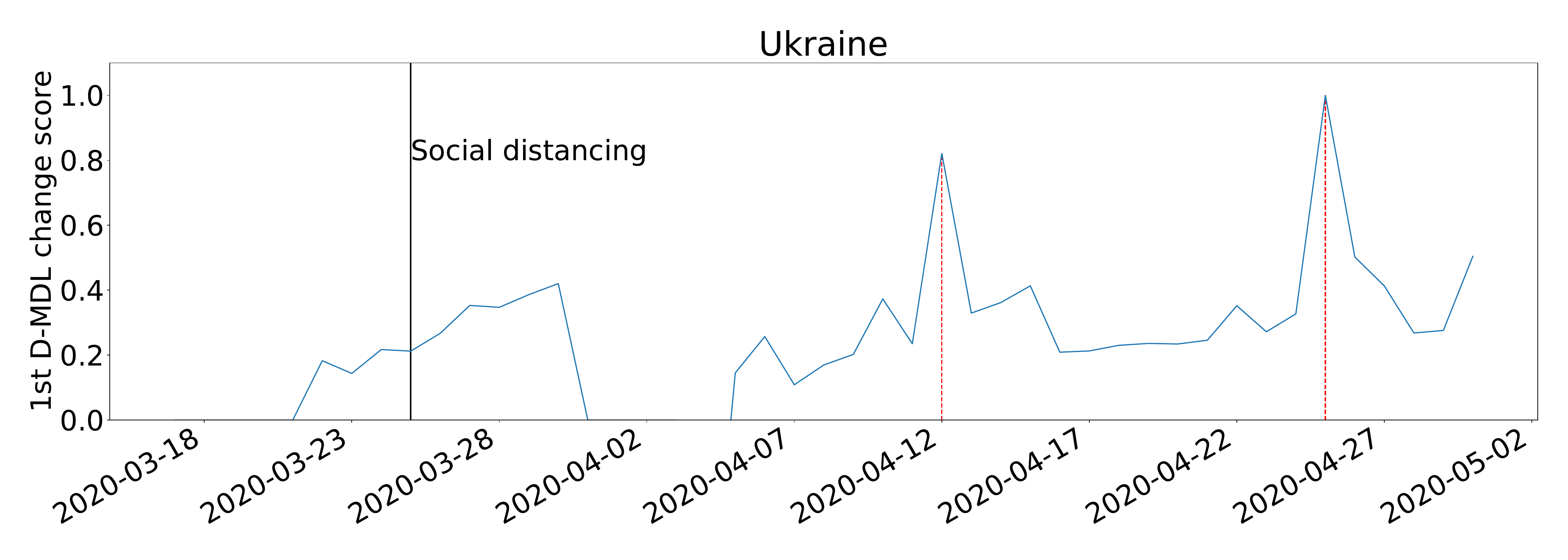} \\
		    \vspace{-0.35cm}
			\textbf{e} & \includegraphics[keepaspectratio, height=3.3cm, valign=T]
			{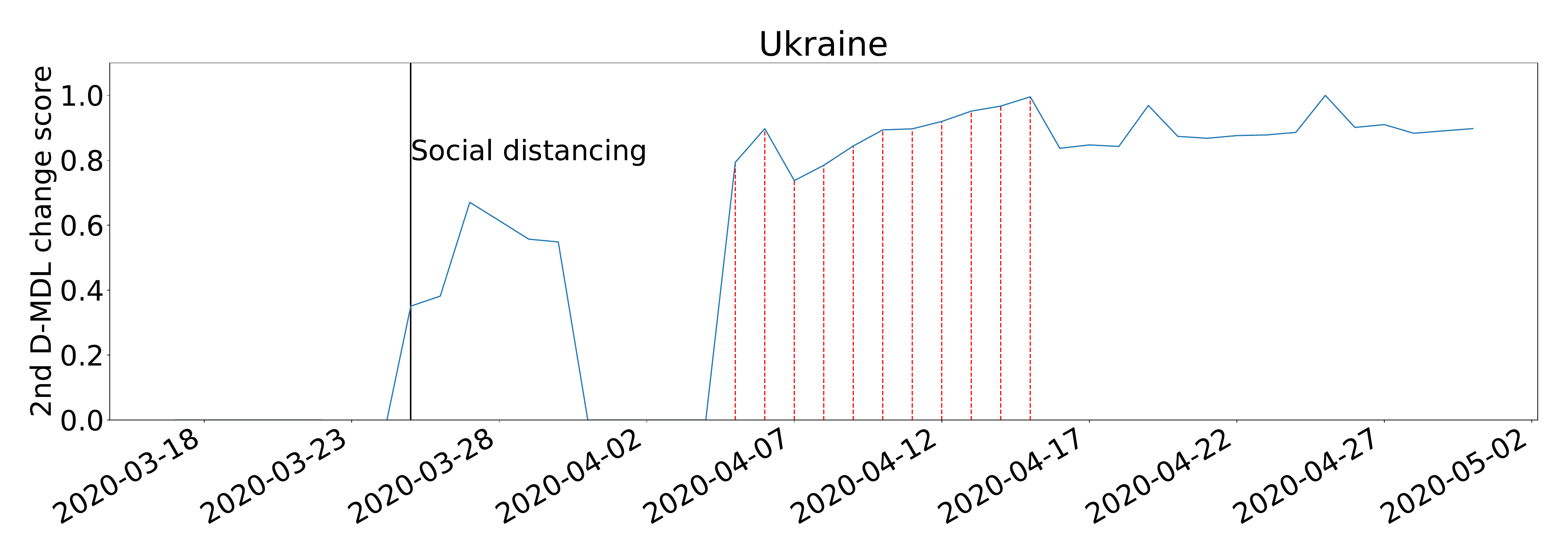} \\
		\end{tabular}
			\caption{\textbf{The results for Ukraine with Gaussian modeling.} The date on which the social distancing was implemented is marked by a solid line in black. \textbf{a,} the number of daily new cases. \textbf{b,} the change scores produced by the 0th M-DML where the line in blue denotes values of scores and dashed lines in red mark alarms. \textbf{c,} the window sized for the sequential D-DML algorithm with adaptive window where lines in red mark the shrinkage of windows. \textbf{d,} the change scores produced by the 1st D-MDL. \textbf{e,} the change scores produced by the 2nd D-MDL.}
\end{figure}

\begin{figure}[H]  
\centering
\begin{tabular}{cc}
			\textbf{a} & \includegraphics[keepaspectratio, height=3.3cm, valign=T]
			{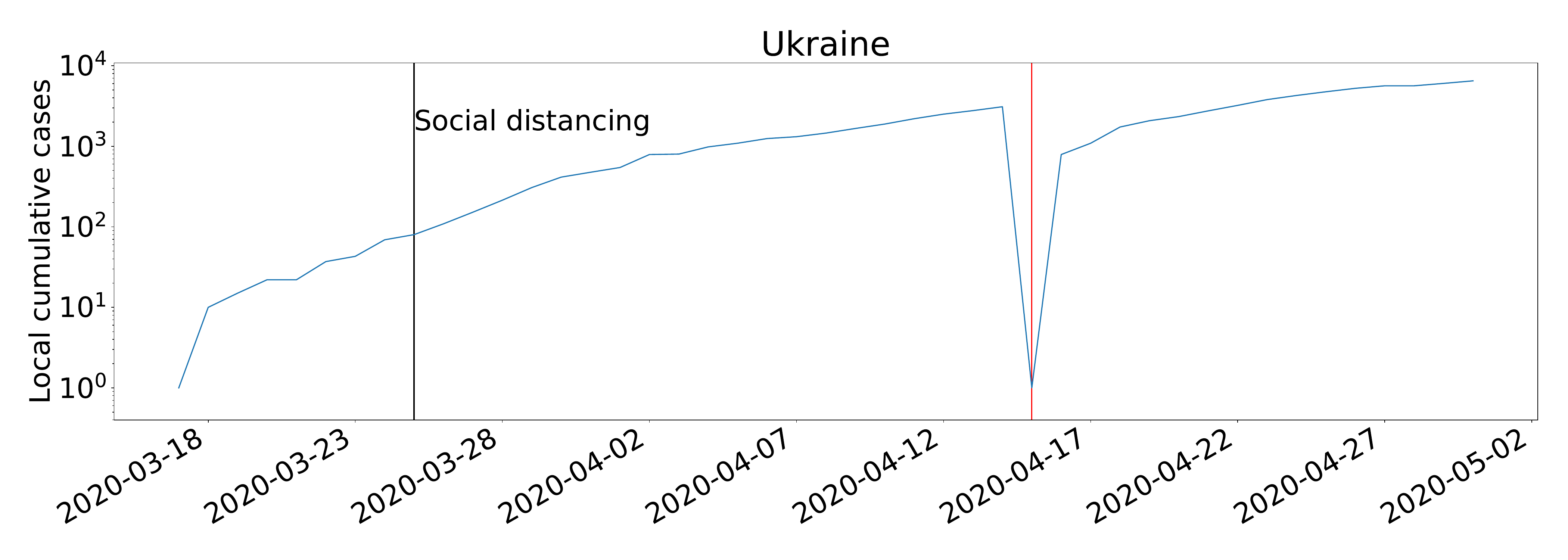} \\
	        \vspace{-0.35cm}
            \textbf{b} & \includegraphics[keepaspectratio, height=3.3cm, valign=T]
			{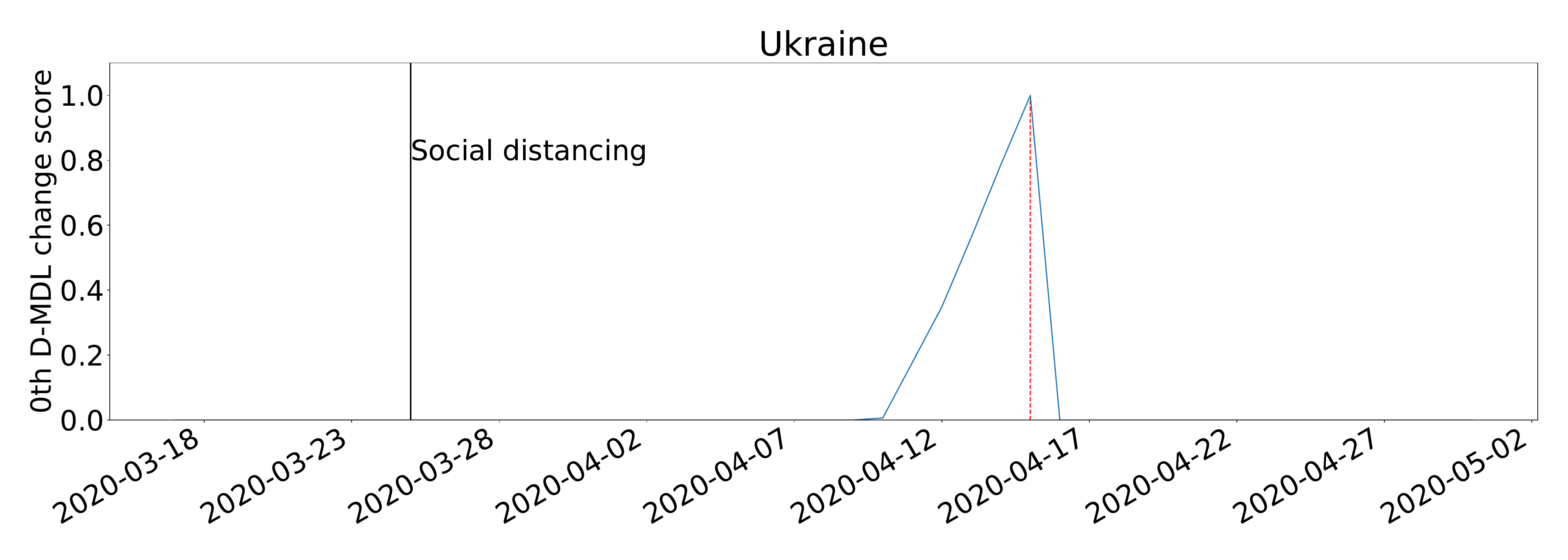}   \\
            \vspace{-0.35cm}
            \textbf{c} & \includegraphics[keepaspectratio, height=3.3cm, valign=T]
			{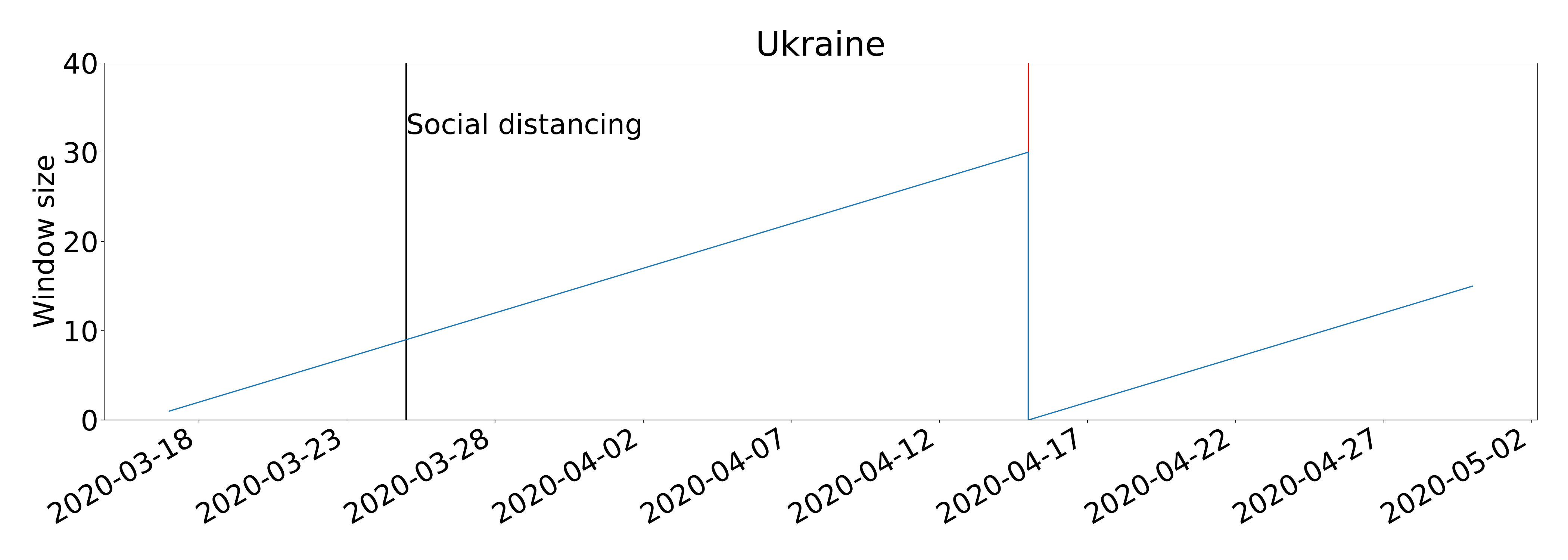} \\
			\vspace{-0.35cm}
			\textbf{d} & \includegraphics[keepaspectratio, height=3.3cm, valign=T]
			{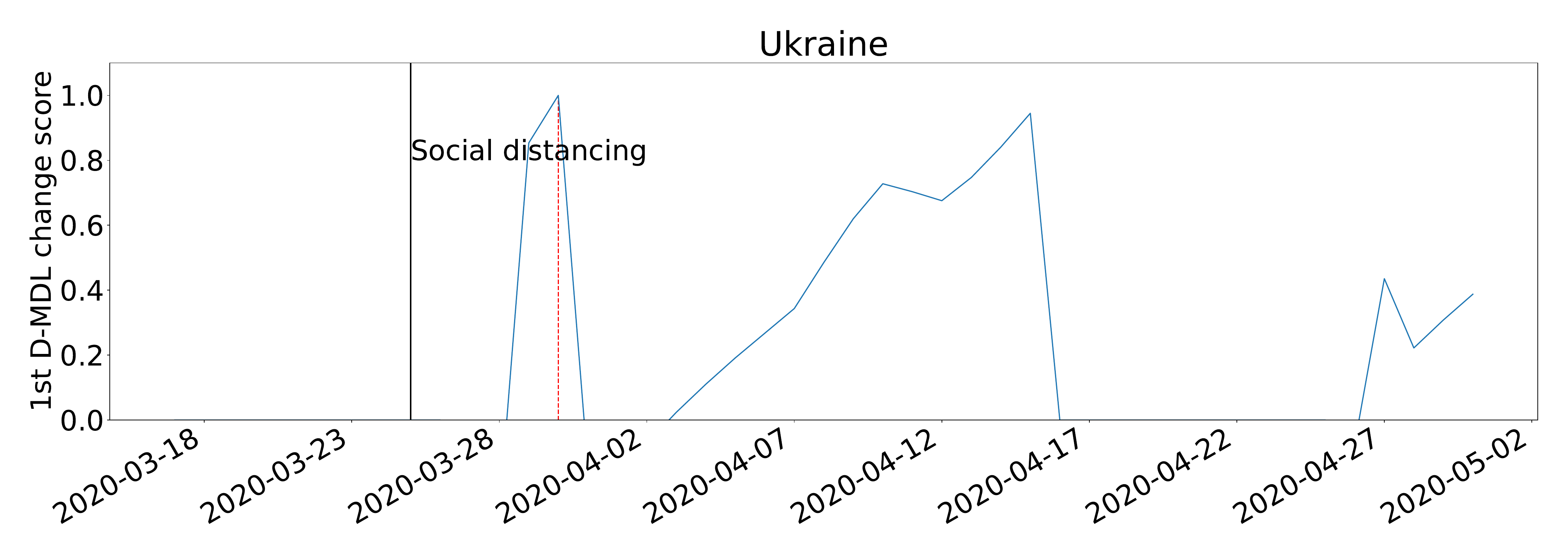} \\
			\vspace{-0.35cm}
			\textbf{e} & \includegraphics[keepaspectratio, height=3.3cm, valign=T]
			{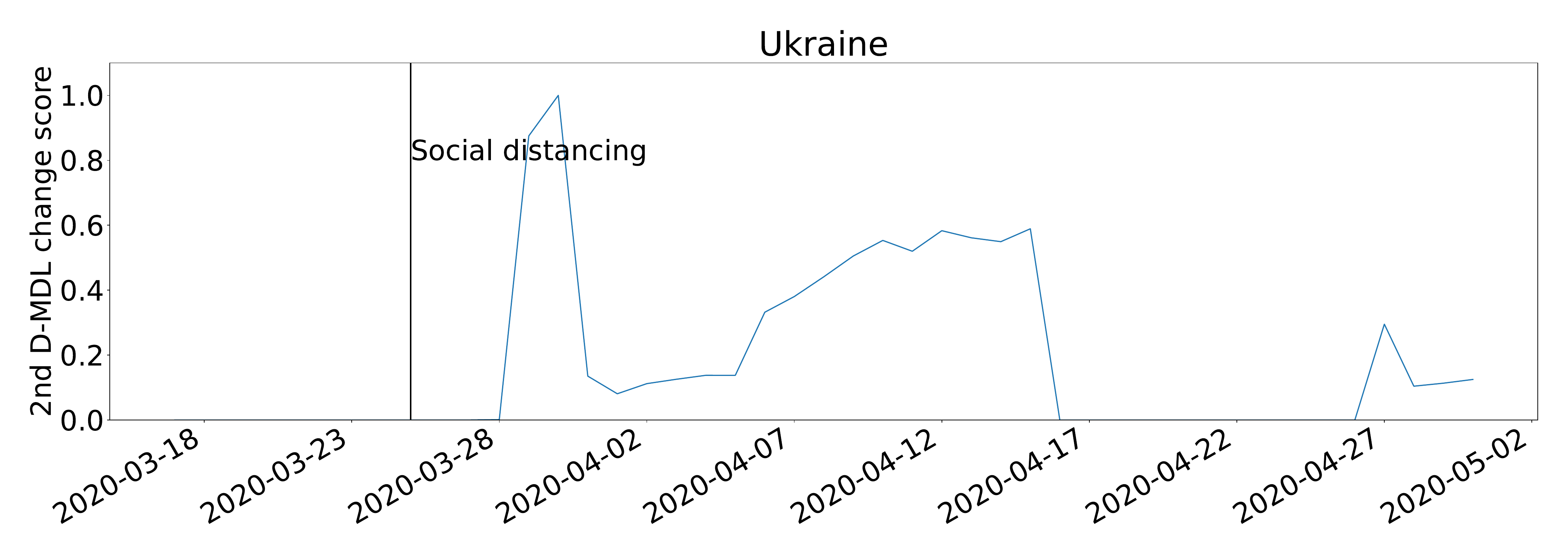} \\
		\end{tabular}
			\caption{\textbf{The results for Ukraine with exponential modeling.} The date on which the social distancing was implemented is marked by a solid line in black. \textbf{a,} the number of cumulative cases. \textbf{b,} the change scores produced by the 0th M-DML where the line in blue denotes values of scores and dashed lines in red mark alarms. \textbf{c,} the window sized for the sequential D-DML algorithm with adaptive window where lines in red mark the shrinkage of windows. \textbf{d,} the change scores produced by the 1st D-MDL. \textbf{e,} the change scores produced by the 2nd D-MDL.}
\end{figure}

\begin{figure}[H] 
\centering
\begin{tabular}{cc}
		 	\textbf{a} & \includegraphics[keepaspectratio, height=3.3cm, valign=T]
			{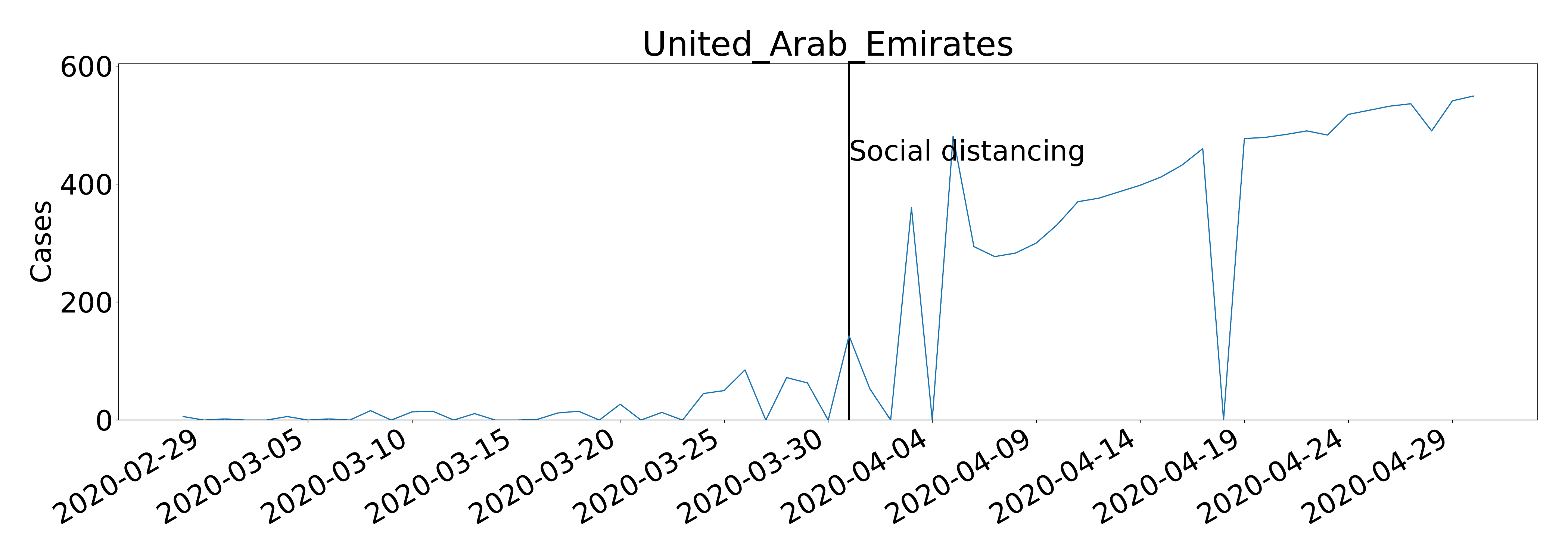} \\
			\vspace{-0.35cm}
	 	    \textbf{b} & \includegraphics[keepaspectratio, height=3.3cm, valign=T]
			{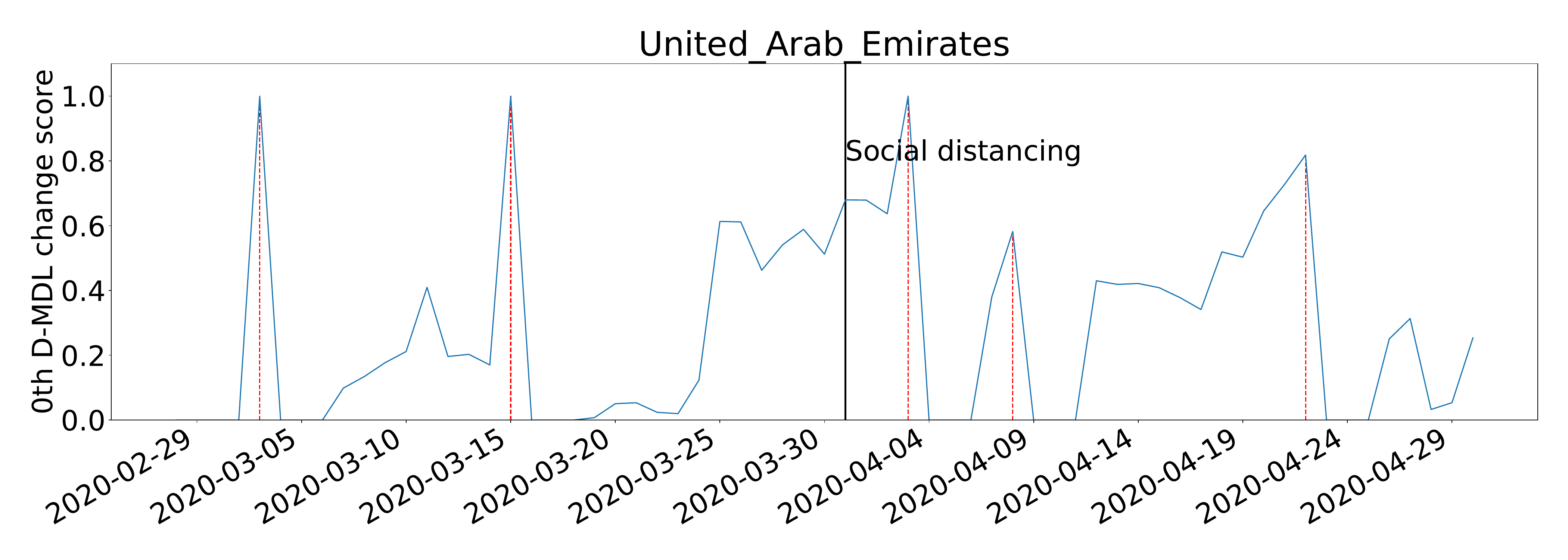}   \\
	        \vspace{-0.35cm}
			\textbf{c} & \includegraphics[keepaspectratio, height=3.3cm, valign=T]
			{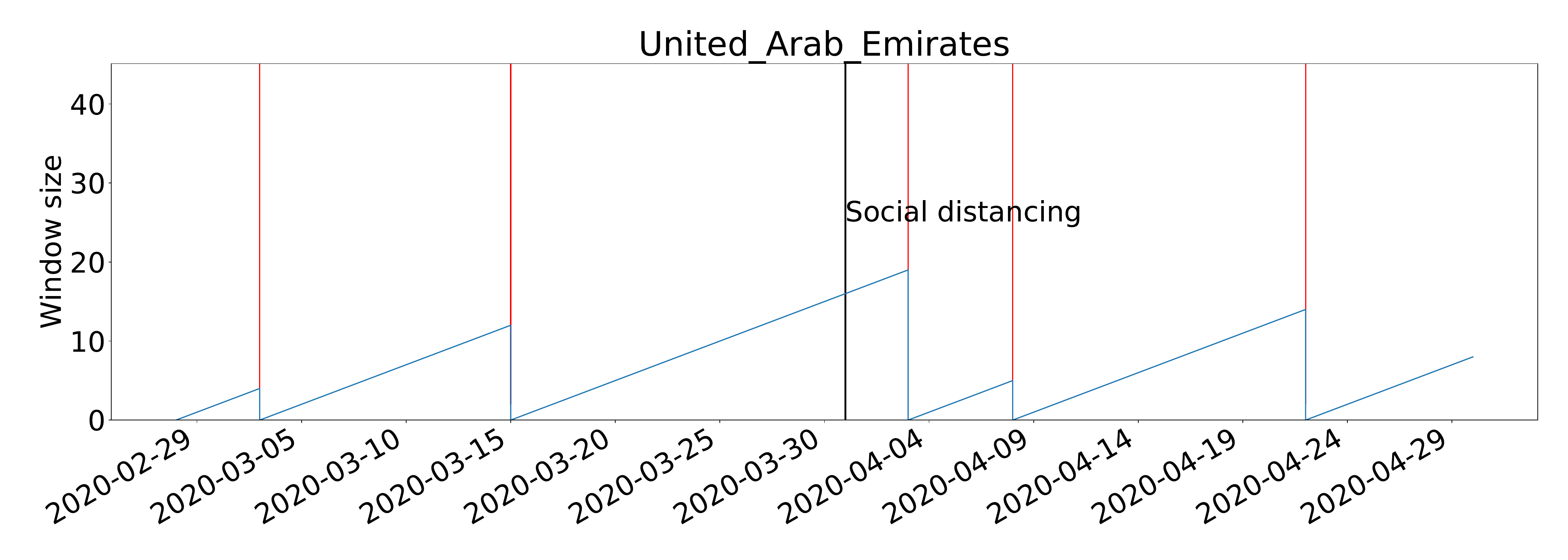} \\
		    \vspace{-0.35cm}
			\textbf{d} & \includegraphics[keepaspectratio, height=3.3cm, valign=T]
			{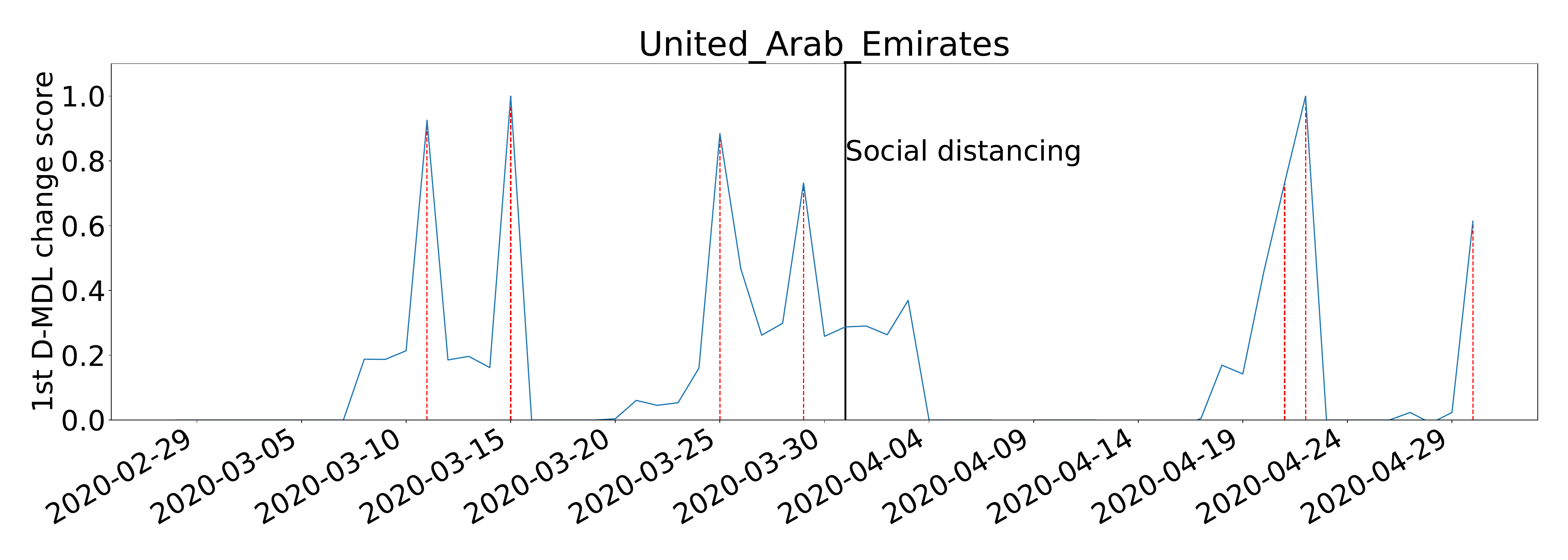} \\
		    \vspace{-0.35cm}
			\textbf{e} & \includegraphics[keepaspectratio, height=3.3cm, valign=T]
			{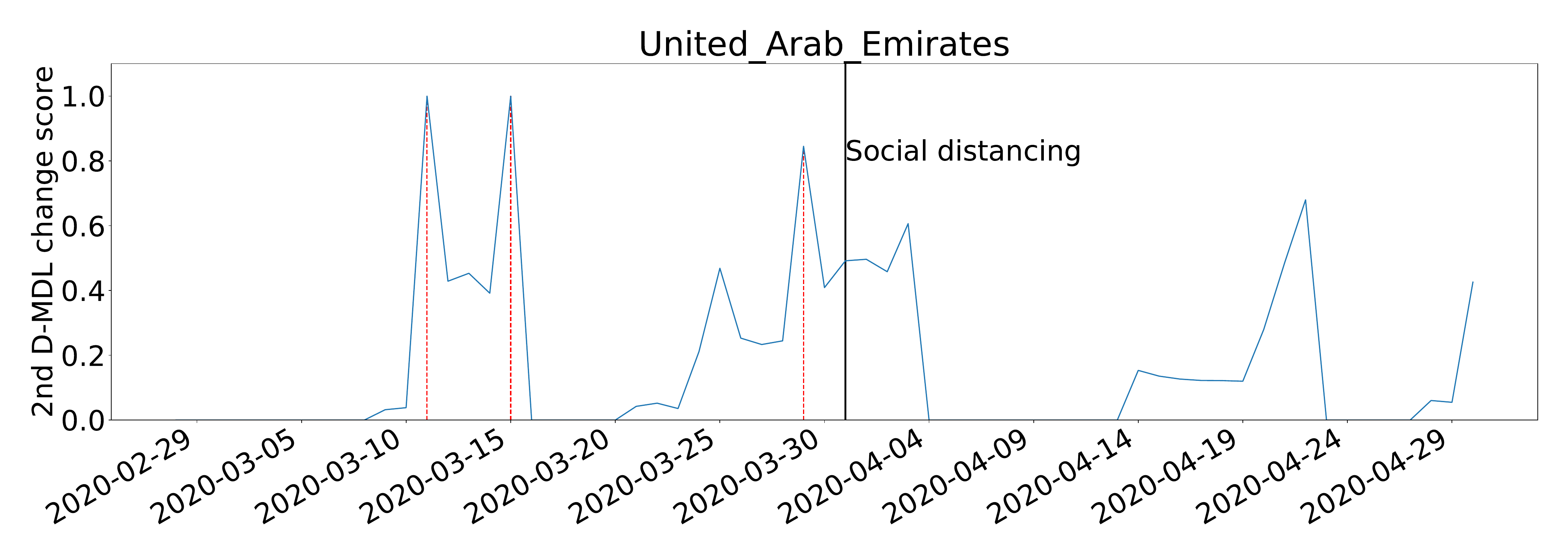} \\
		\end{tabular}
			\caption{\textbf{The results for the United Arab Emirates with Gaussian modeling.} The date on which the social distancing was implemented is marked by a solid line in black. \textbf{a,} the number of daily new cases. \textbf{b,} the change scores produced by the 0th M-DML where the line in blue denotes values of scores and dashed lines in red mark alarms. \textbf{c,} the window sized for the sequential D-DML algorithm with adaptive window where lines in red mark the shrinkage of windows. \textbf{d,} the change scores produced by the 1st D-MDL. \textbf{e,} the change scores produced by the 2nd D-MDL.}
\end{figure}

\begin{figure}[H]  
\centering
\begin{tabular}{cc}
			\textbf{a} & \includegraphics[keepaspectratio, height=3.3cm, valign=T]
			{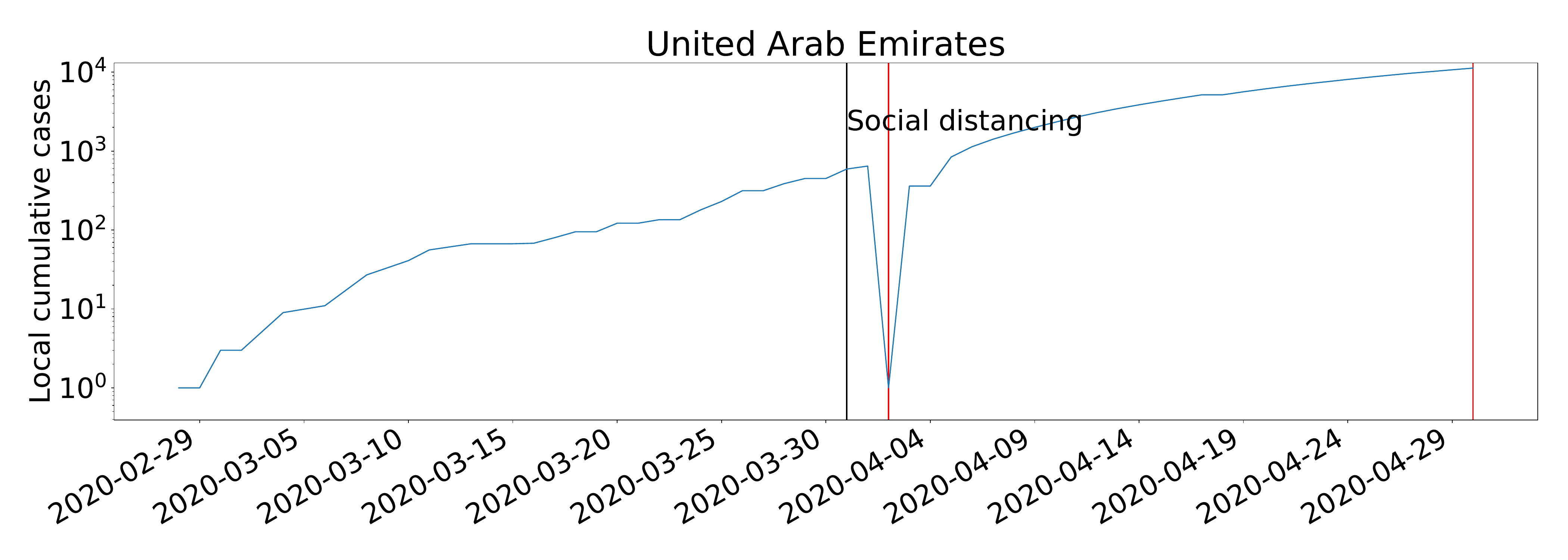} \\
	        \vspace{-0.35cm}
            \textbf{b} & \includegraphics[keepaspectratio, height=3.3cm, valign=T]
			{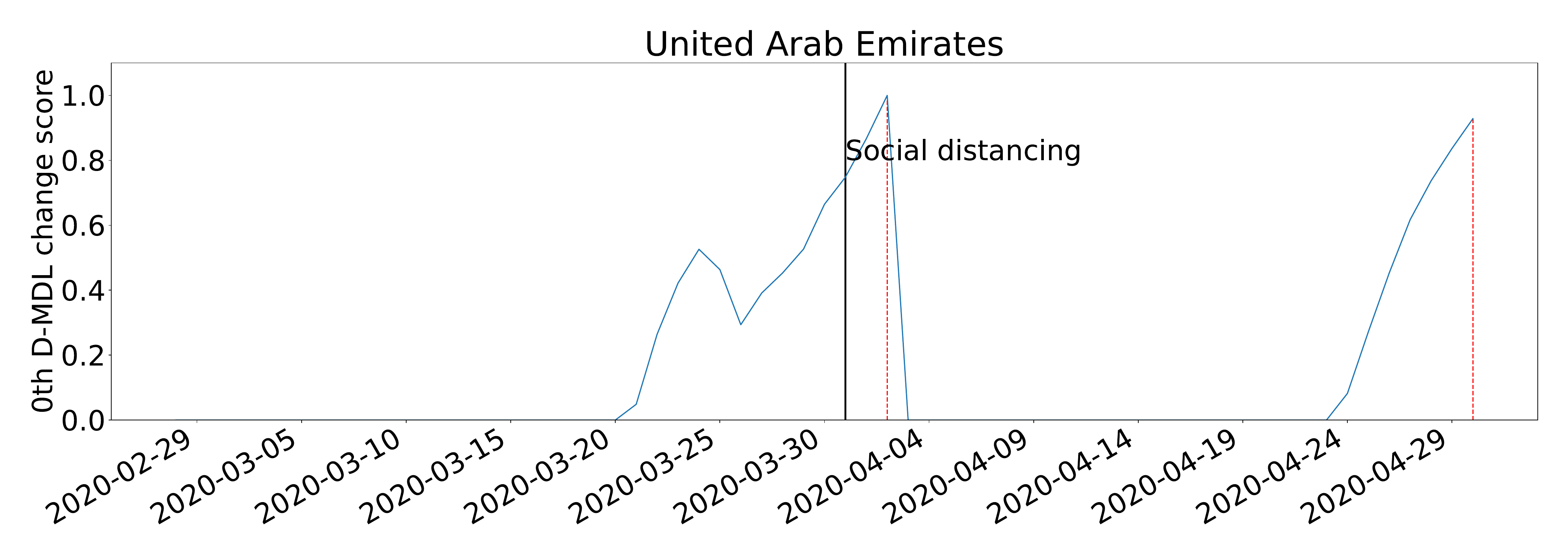}   \\
            \vspace{-0.35cm}
            \textbf{c} & \includegraphics[keepaspectratio, height=3.3cm, valign=T]
			{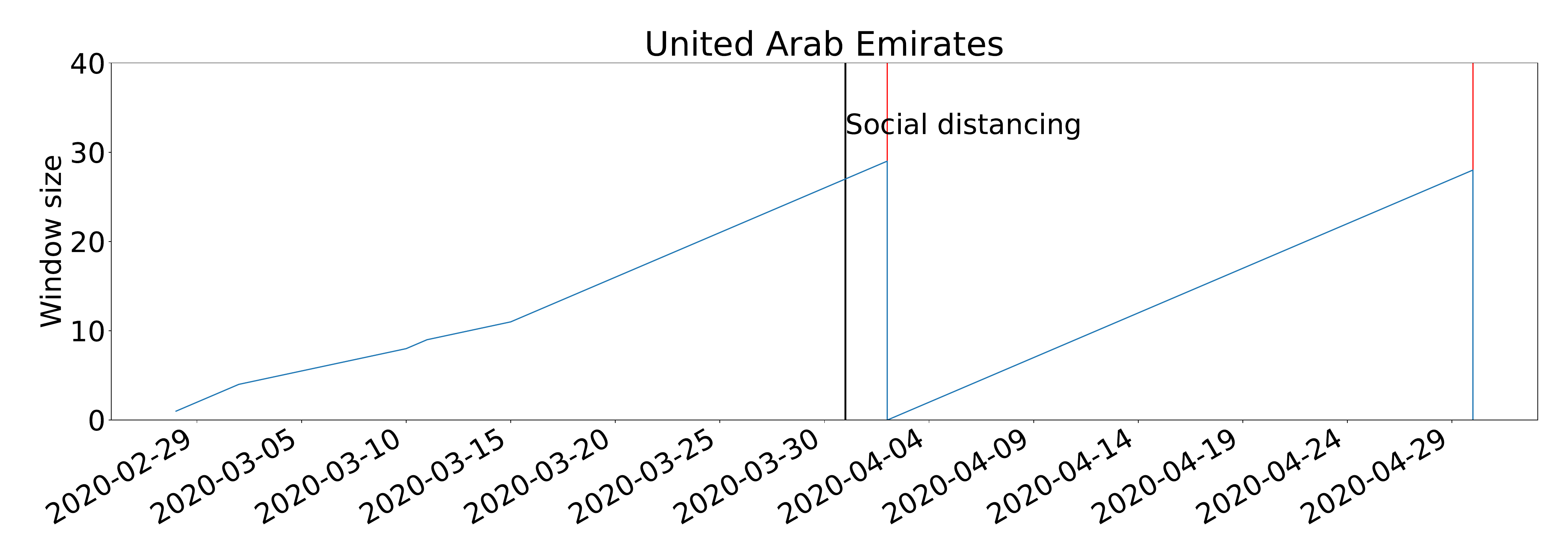} \\
			\vspace{-0.35cm}
			\textbf{d} & \includegraphics[keepaspectratio, height=3.3cm, valign=T]
			{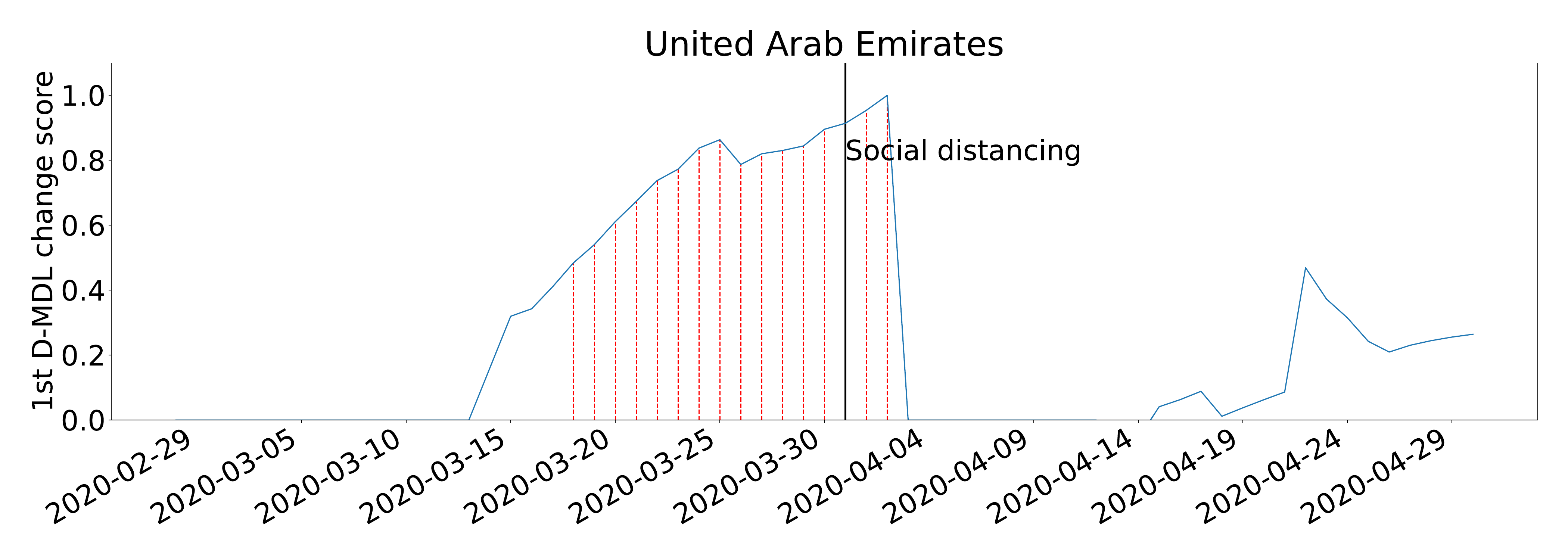} \\
			\vspace{-0.35cm}
			\textbf{e} & \includegraphics[keepaspectratio, height=3.3cm, valign=T]
			{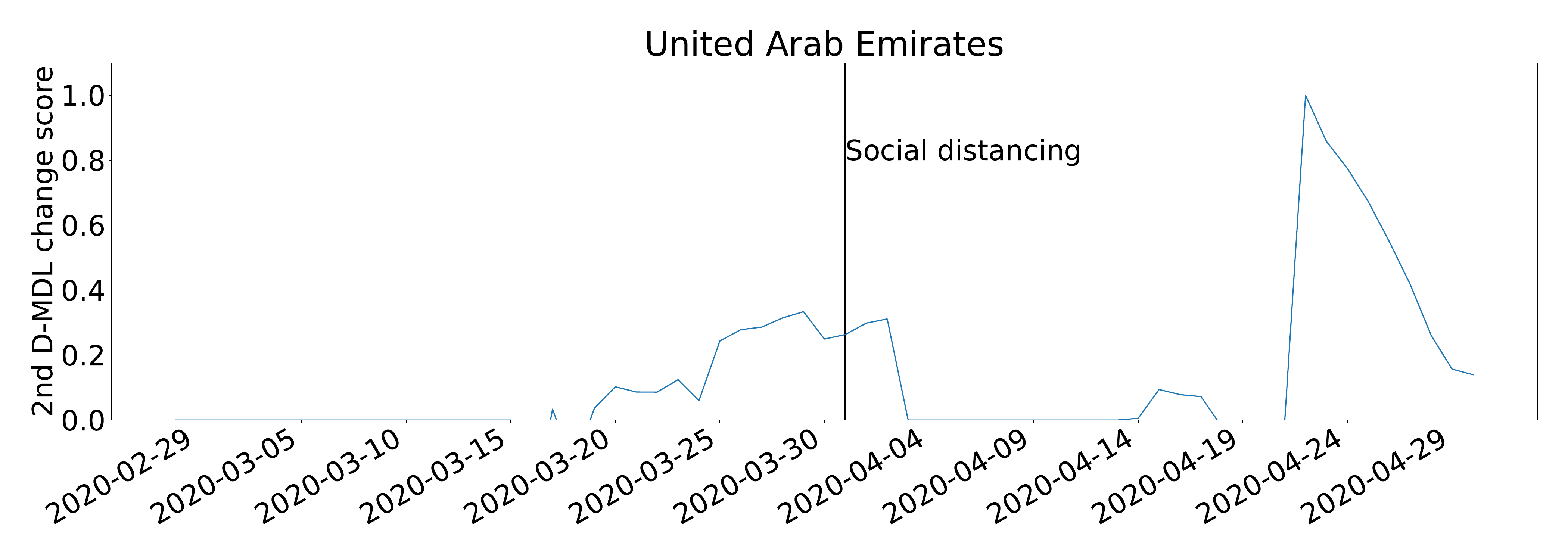} \\
		\end{tabular}
			\caption{\textbf{The results for United Arab Emirates with exponential modeling.} The date on which the social distancing was implemented is marked by a solid line in black. \textbf{a,} the number of cumulative cases. \textbf{b,} the change scores produced by the 0th M-DML where the line in blue denotes values of scores and dashed lines in red mark alarms. \textbf{c,} the window sized for the sequential D-DML algorithm with adaptive window where lines in red mark the shrinkage of windows. \textbf{d,} the change scores produced by the 1st D-MDL. \textbf{e,} the change scores produced by the 2nd D-MDL.}
			\label{exp:united_arab_emirates}
\end{figure}

\begin{figure}[H] 
\centering
\begin{tabular}{cc}
		 	\textbf{a} & \includegraphics[keepaspectratio, height=3.3cm, valign=T]
			{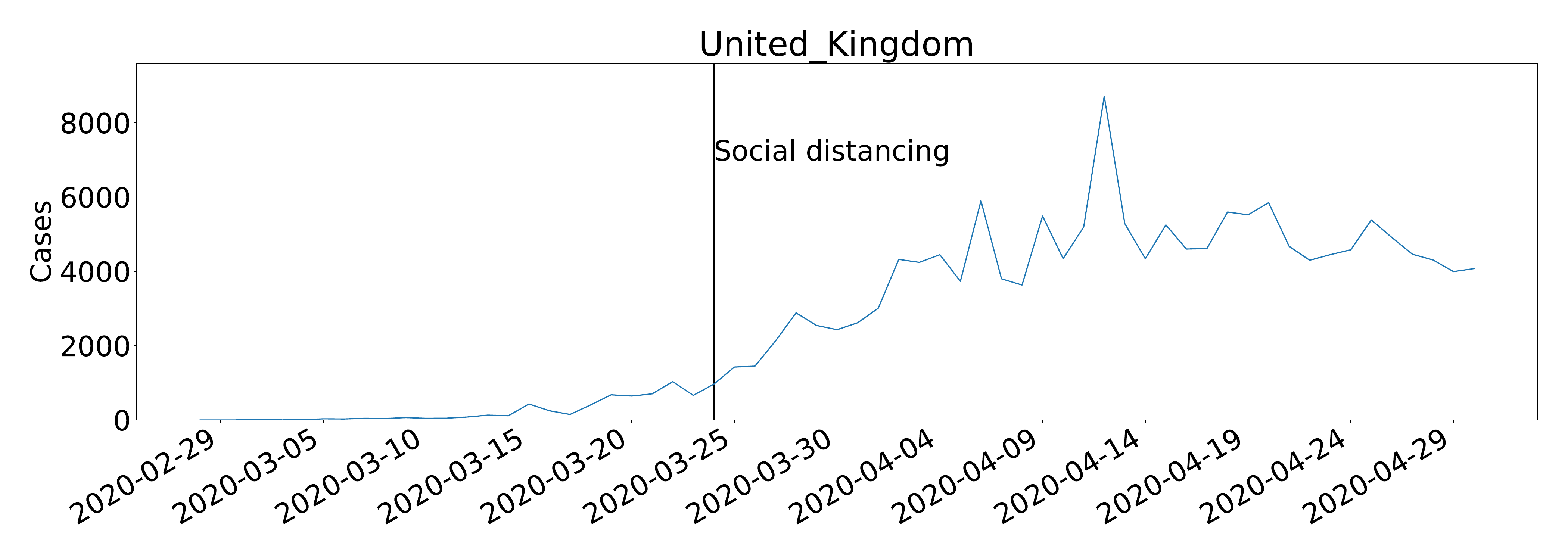} \\
			\vspace{-0.35cm}
	 	    \textbf{b} & \includegraphics[keepaspectratio, height=3.3cm, valign=T]
			{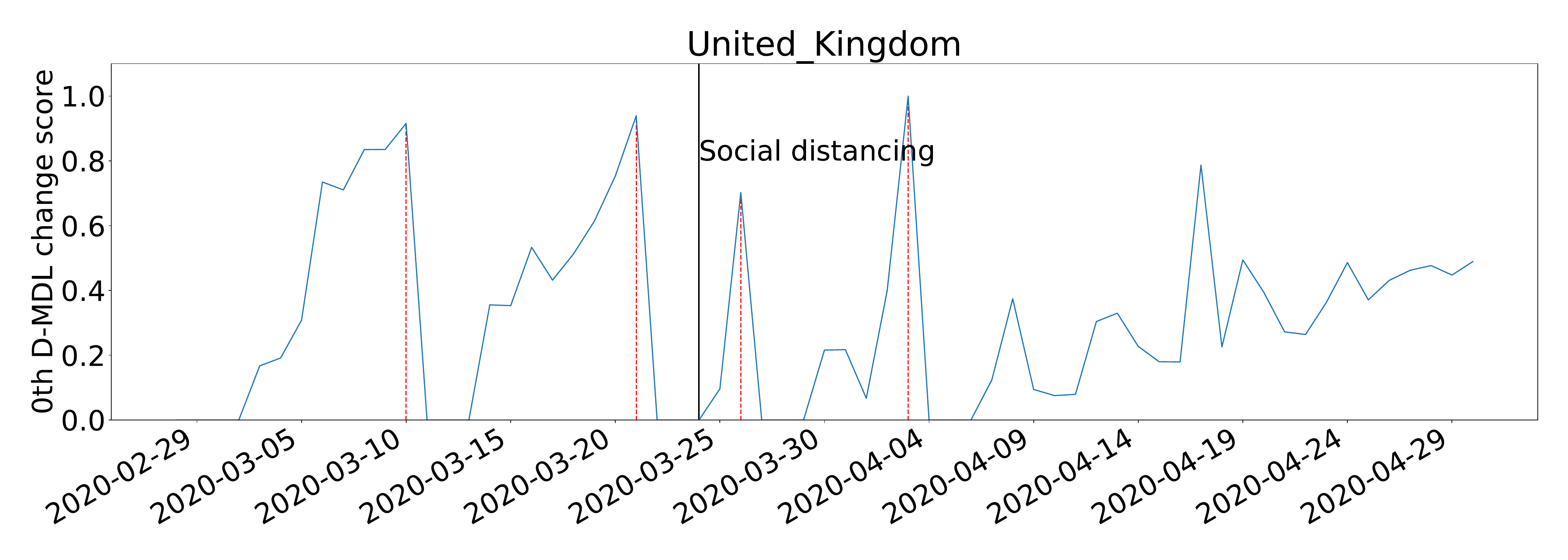}   \\
	        \vspace{-0.35cm}
			\textbf{c} & \includegraphics[keepaspectratio, height=3.3cm, valign=T]
			{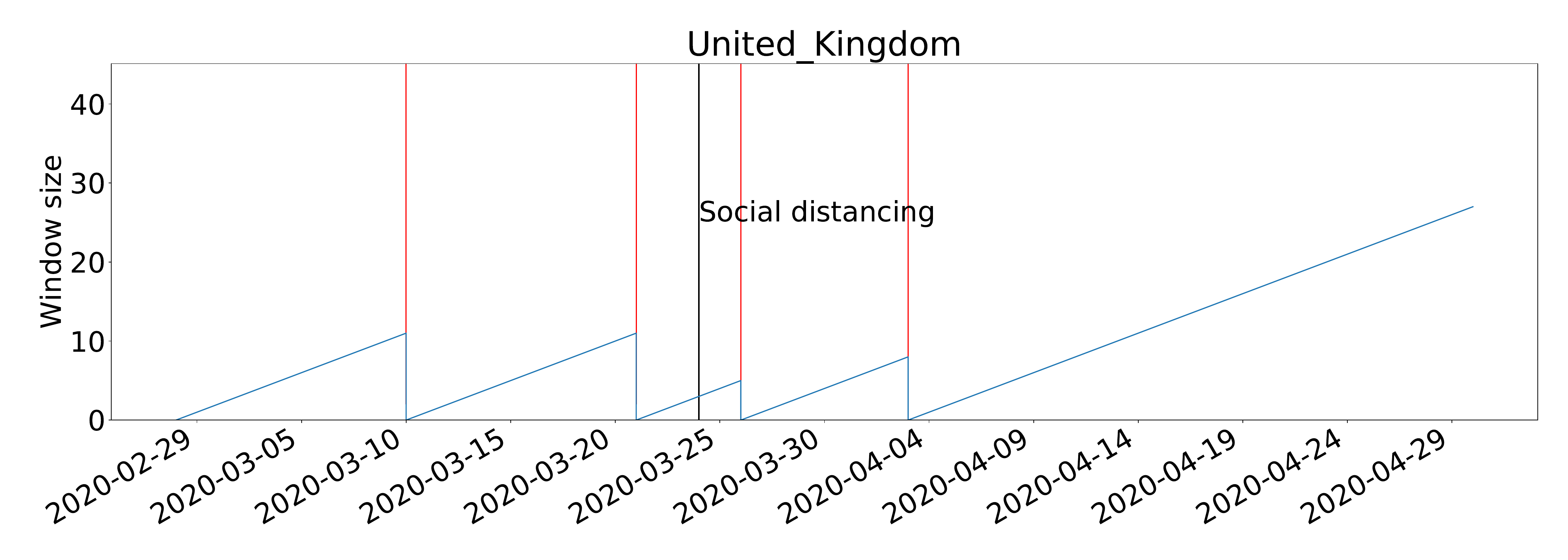} \\
		    \vspace{-0.35cm}
			\textbf{d} & \includegraphics[keepaspectratio, height=3.3cm, valign=T]
			{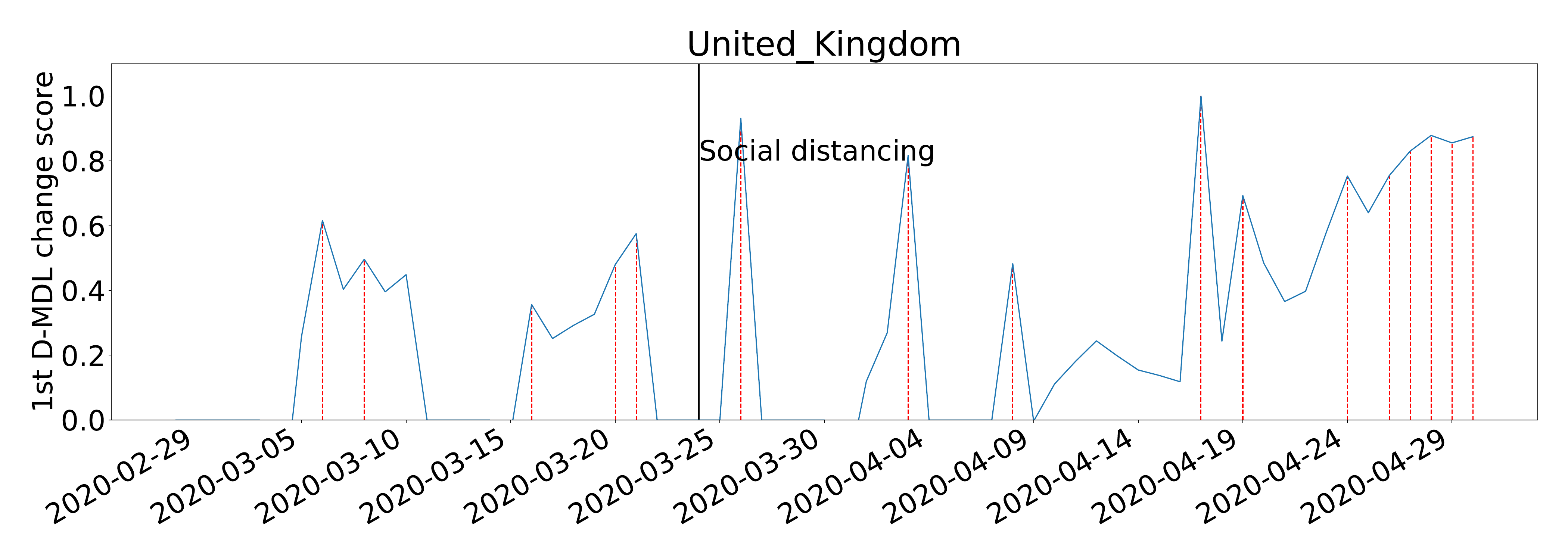} \\
		    \vspace{-0.35cm}
			\textbf{e} & \includegraphics[keepaspectratio, height=3.3cm, valign=T]
			{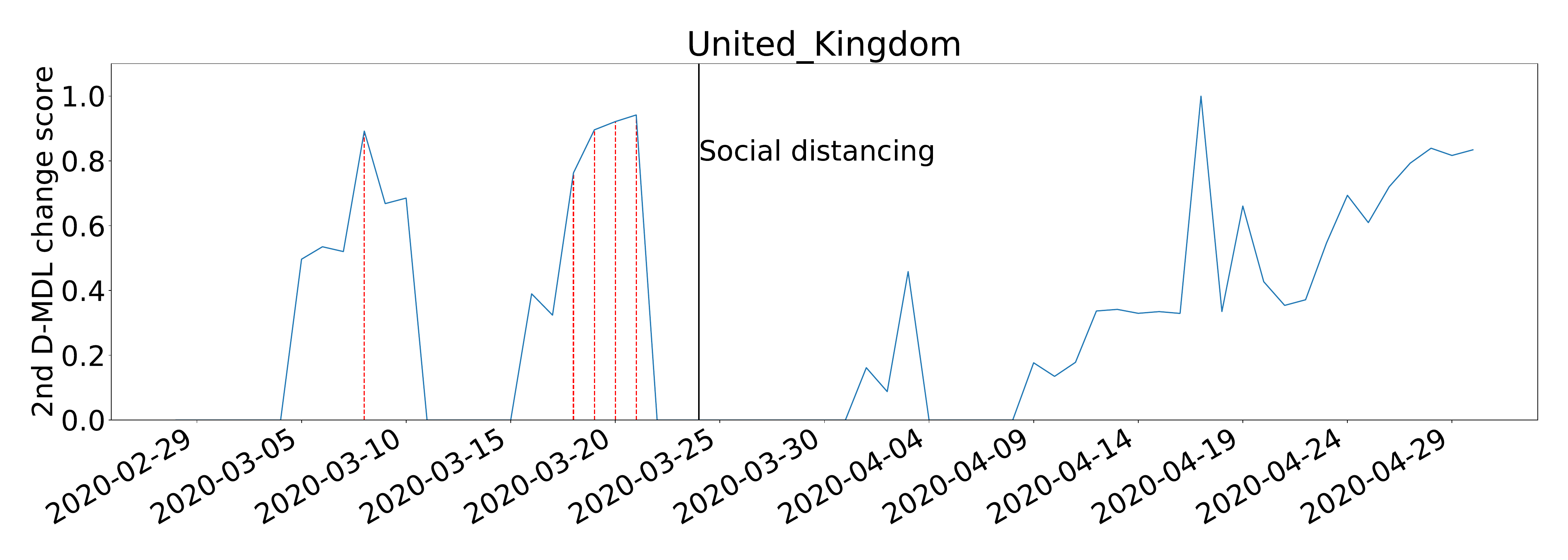} \\
		\end{tabular}
			\caption{\textbf{The results for the United Kingdom with Gaussian modeling.} The date on which the social distancing was implemented is marked by a solid line in black. \textbf{a,} the number of daily new cases. \textbf{b,} the change scores produced by the 0th M-DML where the line in blue denotes values of scores and dashed lines in red mark alarms. \textbf{c,} the window sized for the sequential D-DML algorithm with adaptive window where lines in red mark the shrinkage of windows. \textbf{d,} the change scores produced by the 1st D-MDL. \textbf{e,} the change scores produced by the 2nd D-MDL.}
\end{figure}

\begin{figure}[H]  
\centering
\begin{tabular}{cc}
			\textbf{a} & \includegraphics[keepaspectratio, height=3.3cm, valign=T]
			{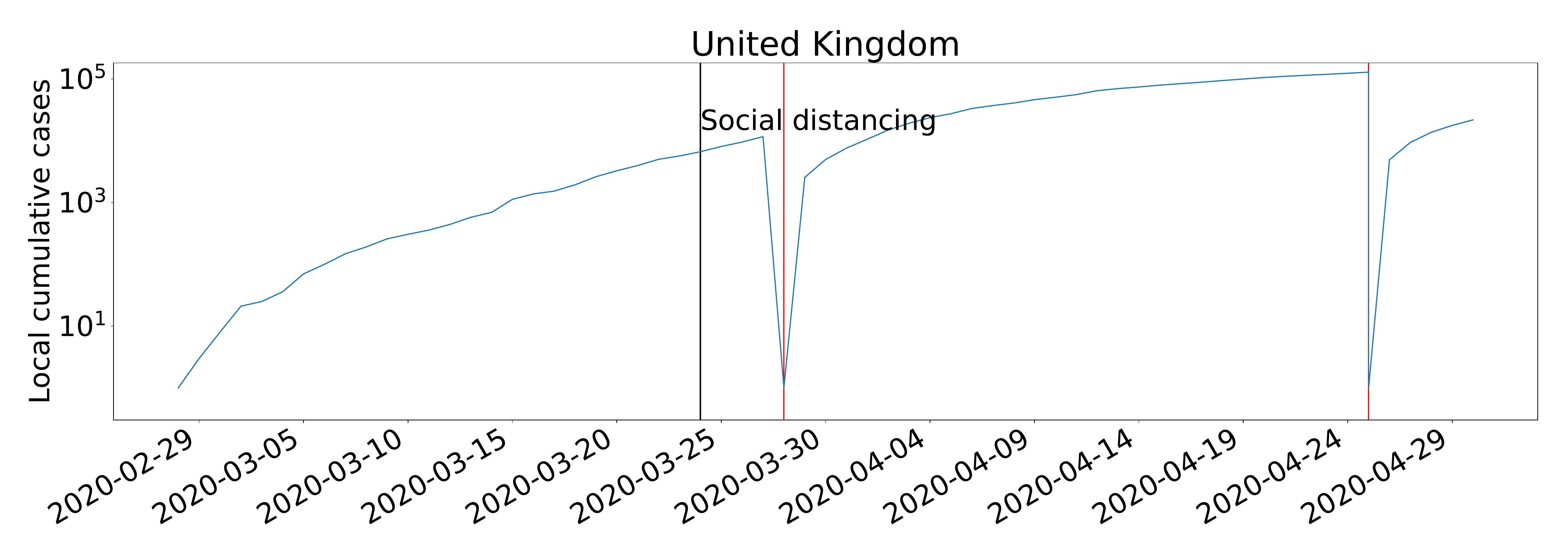} \\
	        \vspace{-0.35cm}
            \textbf{b} & \includegraphics[keepaspectratio, height=3.3cm, valign=T]
			{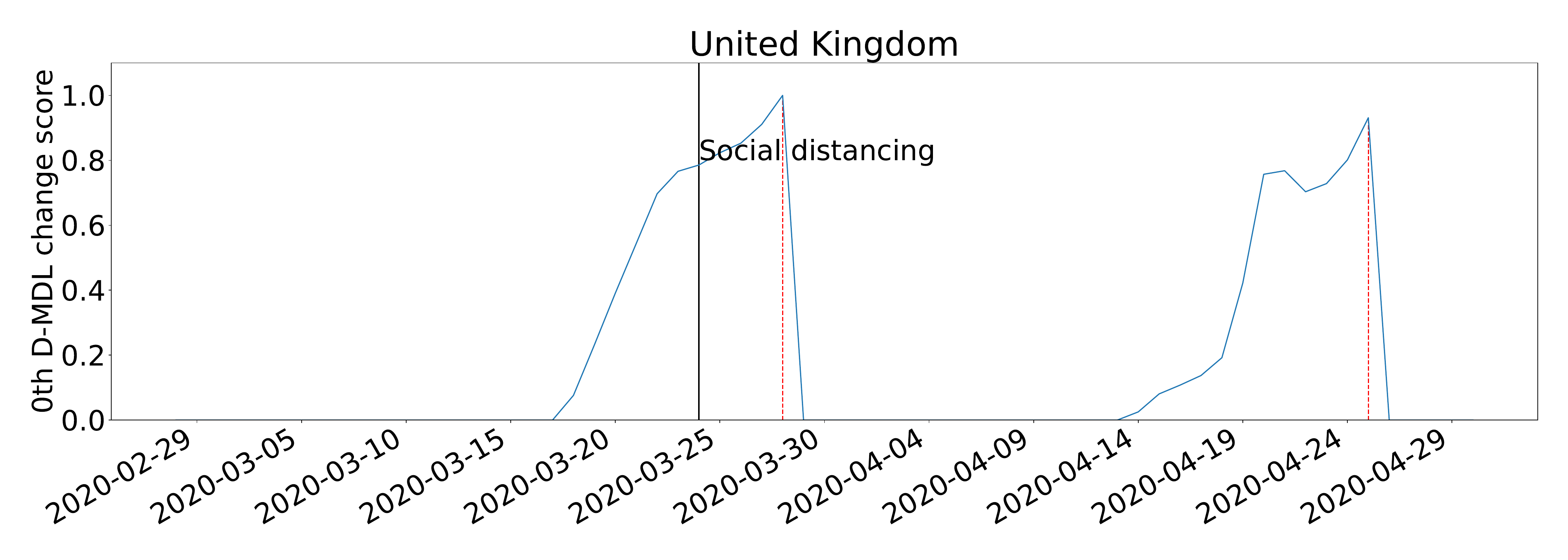}   \\
            \vspace{-0.35cm}
            \textbf{c} & \includegraphics[keepaspectratio, height=3.3cm, valign=T]
			{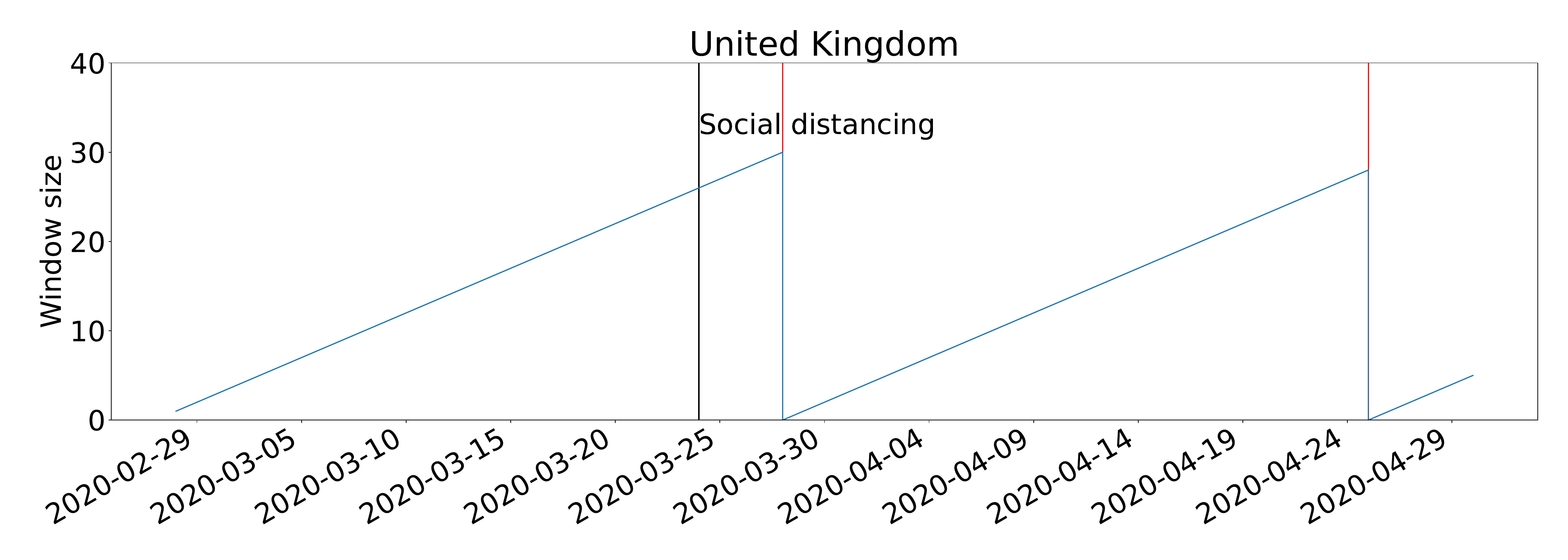} \\
			\vspace{-0.35cm}
			\textbf{d} & \includegraphics[keepaspectratio, height=3.3cm, valign=T]
			{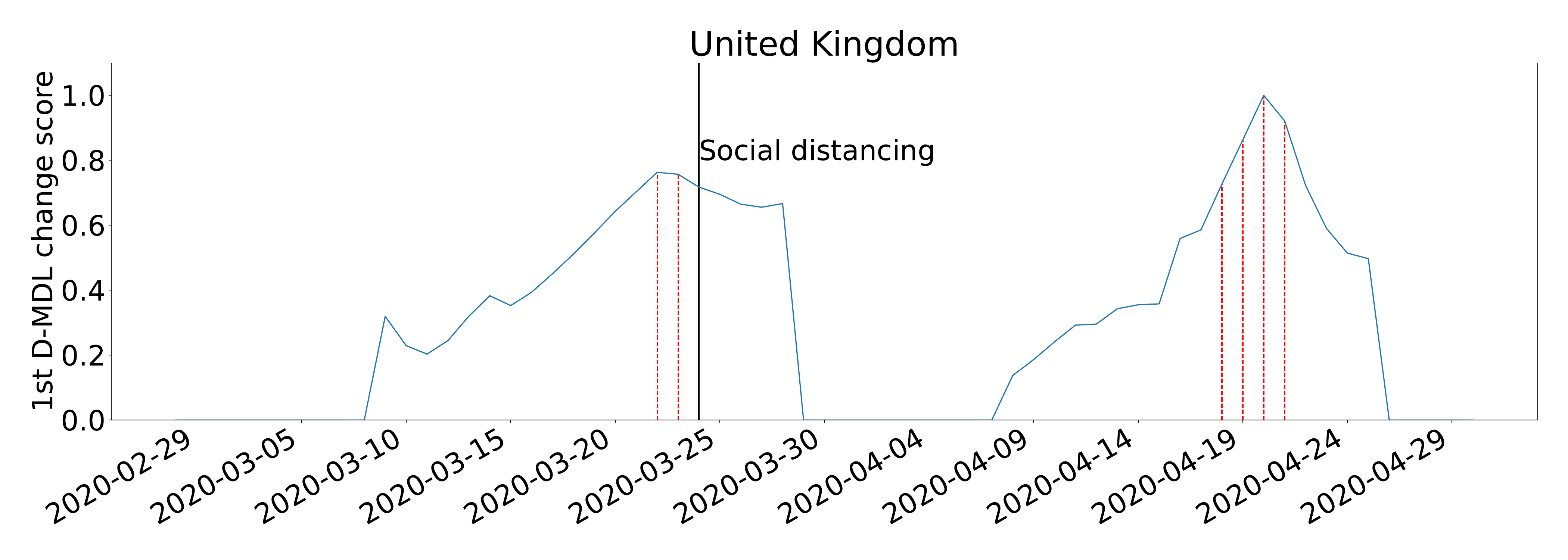} \\
			\vspace{-0.35cm}
			\textbf{e} & \includegraphics[keepaspectratio, height=3.3cm, valign=T]
			{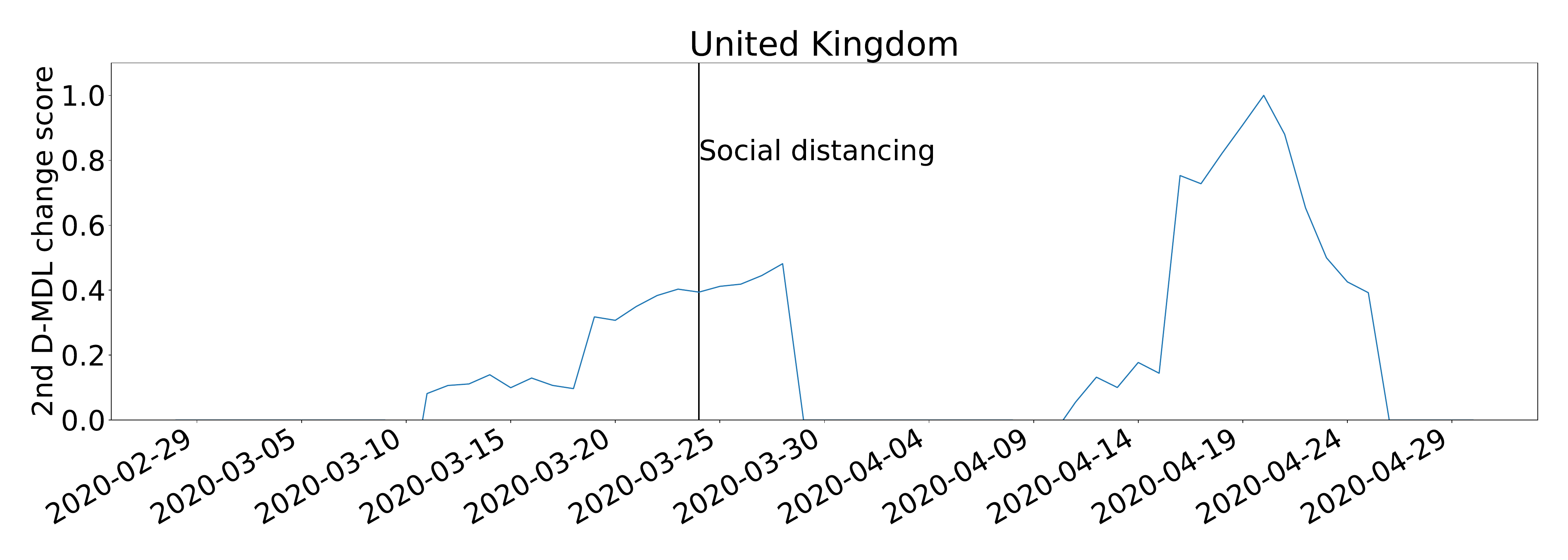} \\
		\end{tabular}
			\caption{\textbf{The results for United Kingdom with exponential modeling.} The date on which the social distancing was implemented is marked by a solid line in black. \textbf{a,} the number of cumulative cases. \textbf{b,} the change scores produced by the 0th M-DML where the line in blue denotes values of scores and dashed lines in red mark alarms. \textbf{c,} the window sized for the sequential D-DML algorithm with adaptive window where lines in red mark the shrinkage of windows. \textbf{d,} the change scores produced by the 1st D-MDL. \textbf{e,} the change scores produced by the 2nd D-MDL.}
\end{figure}

\begin{figure}[H] 
\centering
\begin{tabular}{cc}
		 	\textbf{a} & \includegraphics[keepaspectratio, height=3.3cm, valign=T]
			{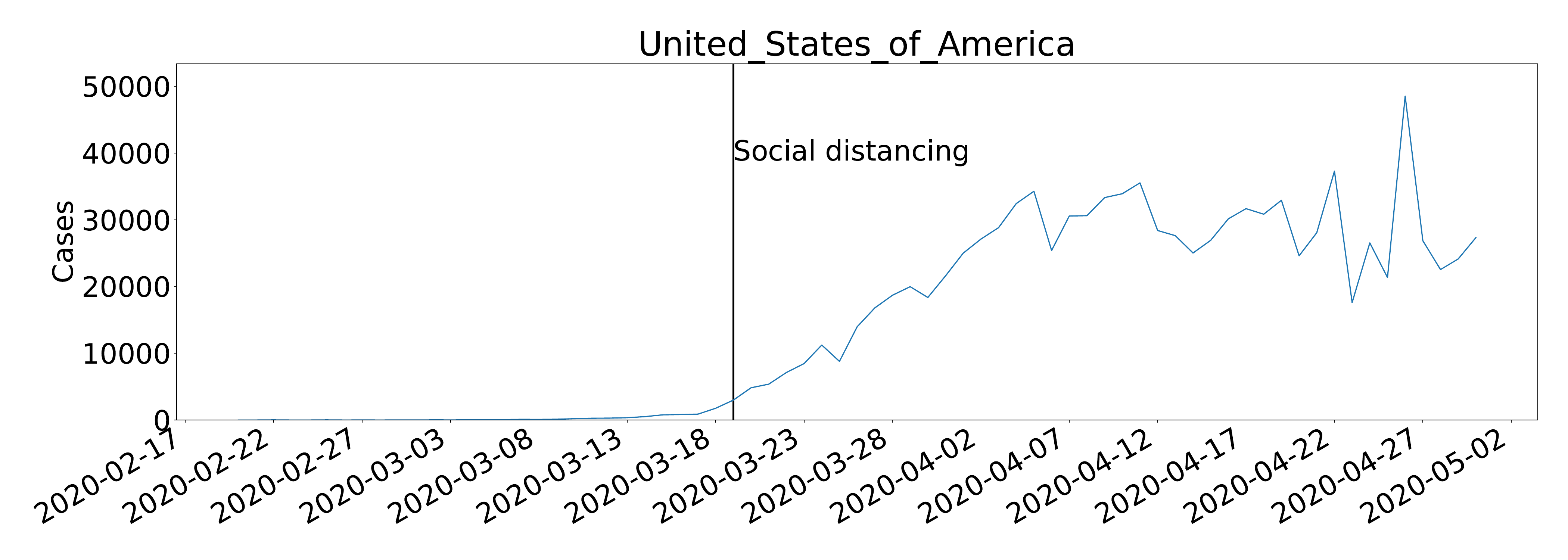} \\
			\vspace{-0.35cm}
	 	    \textbf{b} & \includegraphics[keepaspectratio, height=3.3cm, valign=T]
			{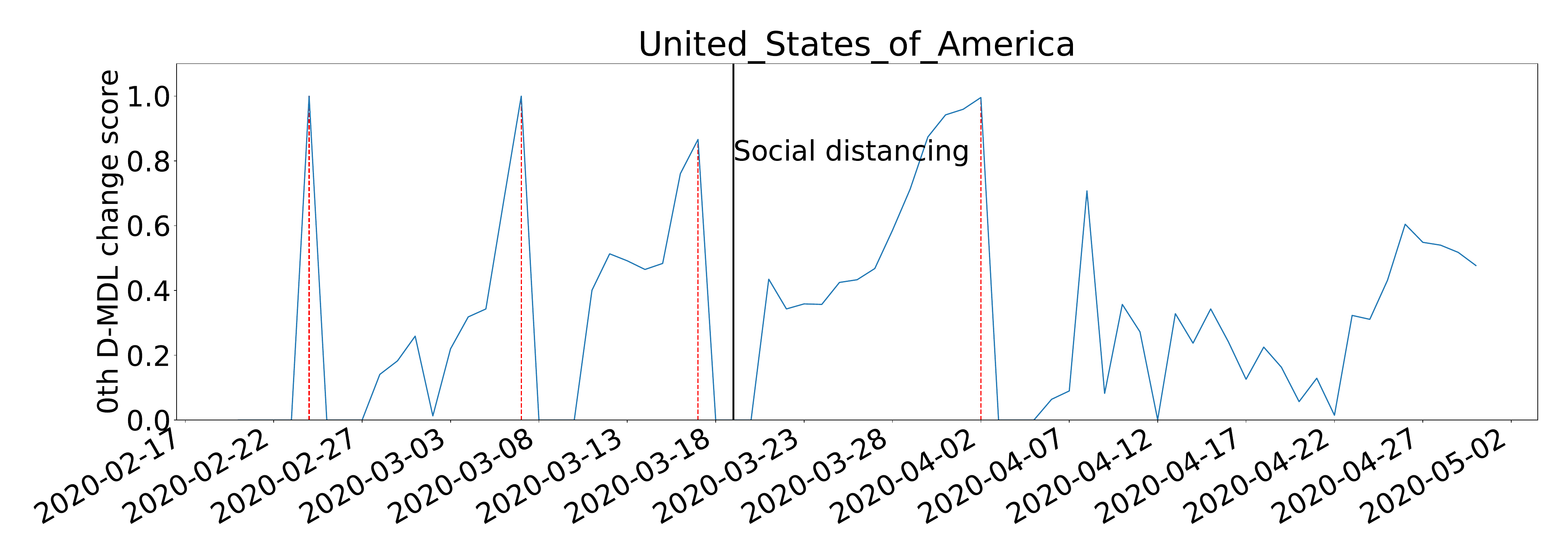}   \\
	        \vspace{-0.35cm}
			\textbf{c} & \includegraphics[keepaspectratio, height=3.3cm, valign=T]
			{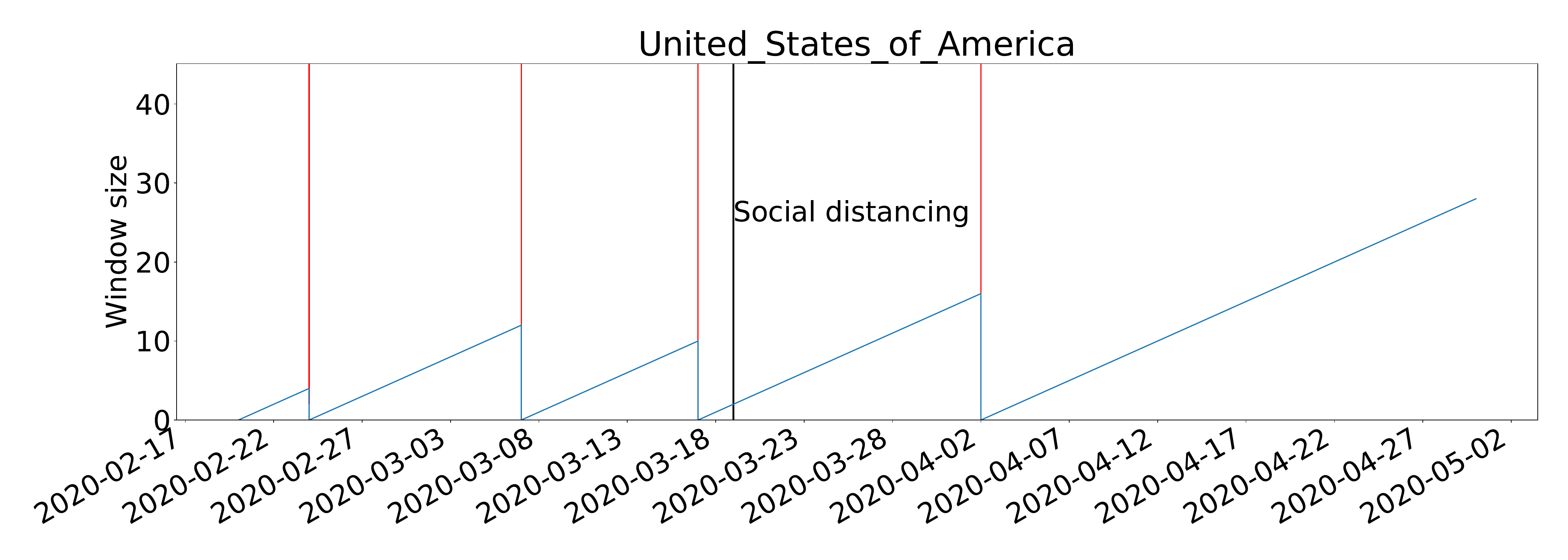} \\
		    \vspace{-0.35cm}
			\textbf{d} & \includegraphics[keepaspectratio, height=3.3cm, valign=T]
			{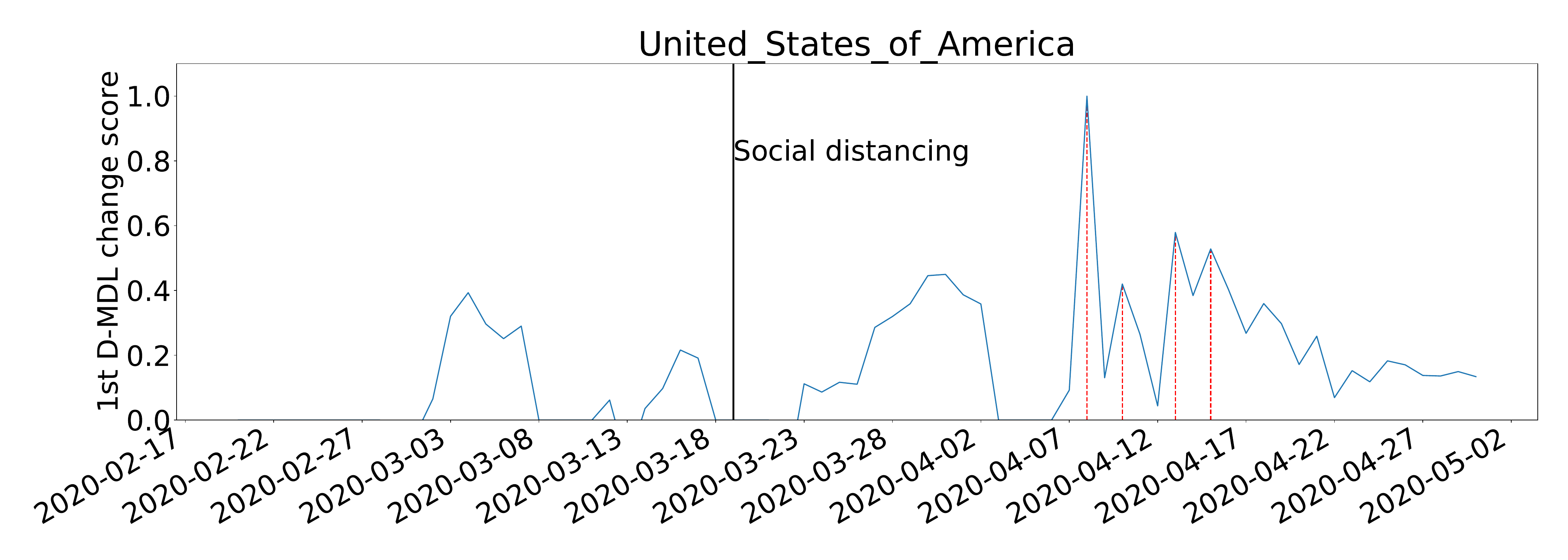} \\
		    \vspace{-0.35cm}
			\textbf{e} & \includegraphics[keepaspectratio, height=3.3cm, valign=T]
			{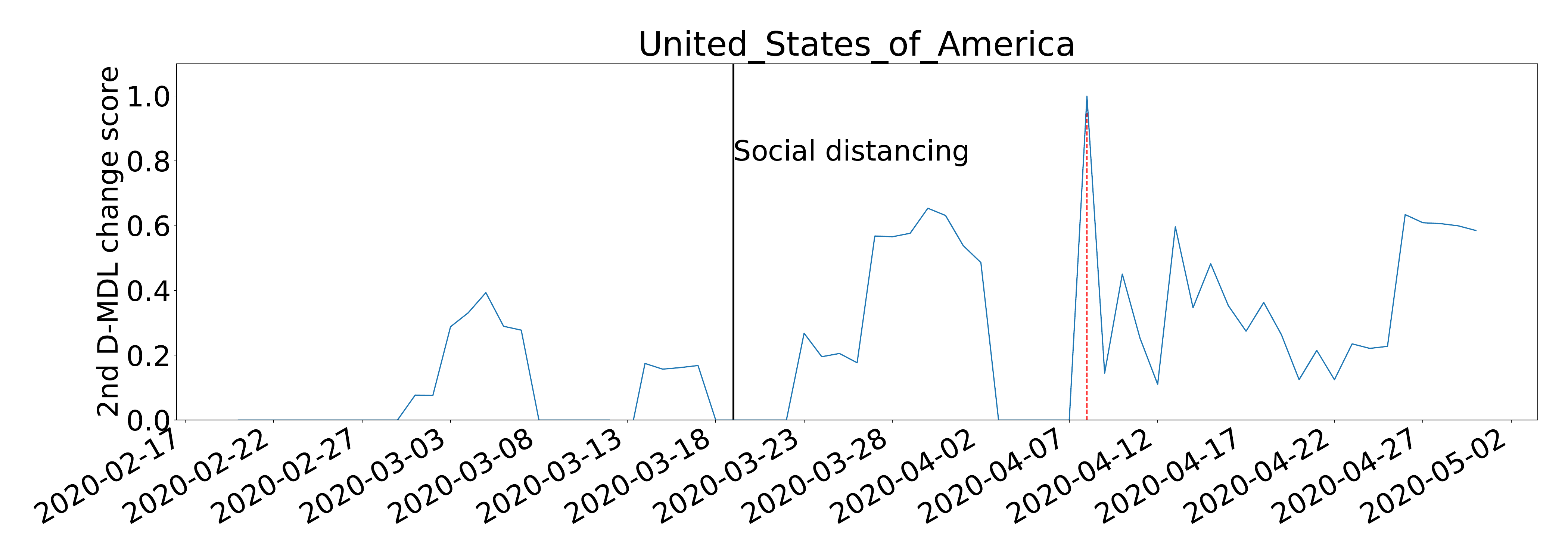} \\
		\end{tabular}
			\caption{\textbf{The results for the United States of America with Gaussian modeling.} The date on which the social distancing was implemented is marked by a solid line in black. \textbf{a,} the number of daily new cases. \textbf{b,} the change scores produced by the 0th M-DML where the line in blue denotes values of scores and dashed lines in red mark alarms. \textbf{c,} the window sized for the sequential D-DML algorithm with adaptive window where lines in red mark the shrinkage of windows. \textbf{d,} the change scores produced by the 1st D-MDL. \textbf{e,} the change scores produced by the 2nd D-MDL.}
\end{figure}

\begin{figure}[H]  
\centering
\begin{tabular}{cc}
			\textbf{a} & \includegraphics[keepaspectratio, height=3.3cm, valign=T]
			{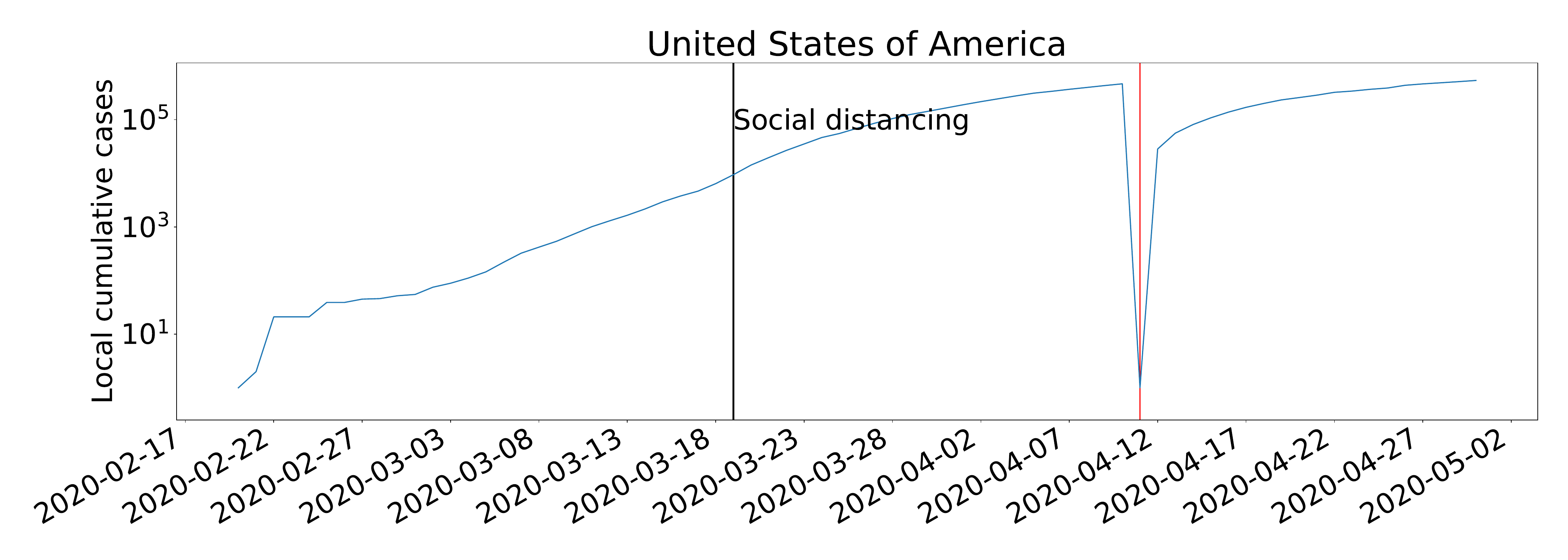} \\
	        \vspace{-0.35cm}
            \textbf{b} & \includegraphics[keepaspectratio, height=3.3cm, valign=T]
			{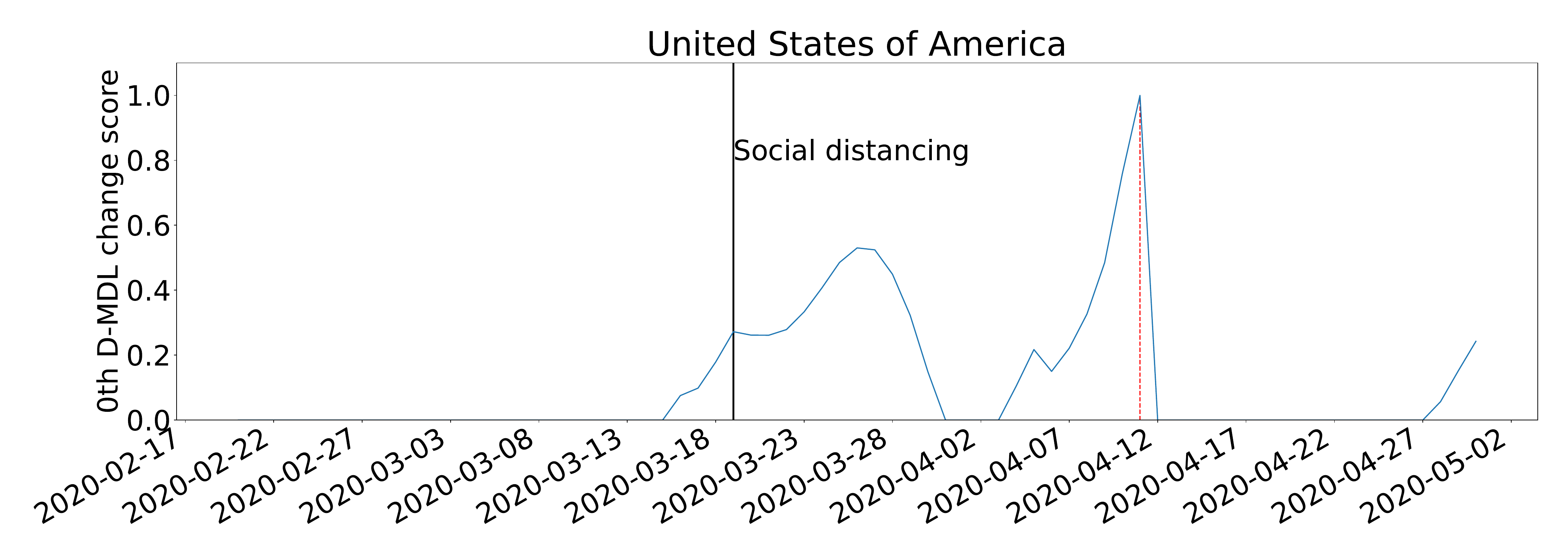}   \\
            \vspace{-0.35cm}
            \textbf{c} & \includegraphics[keepaspectratio, height=3.3cm, valign=T]
			{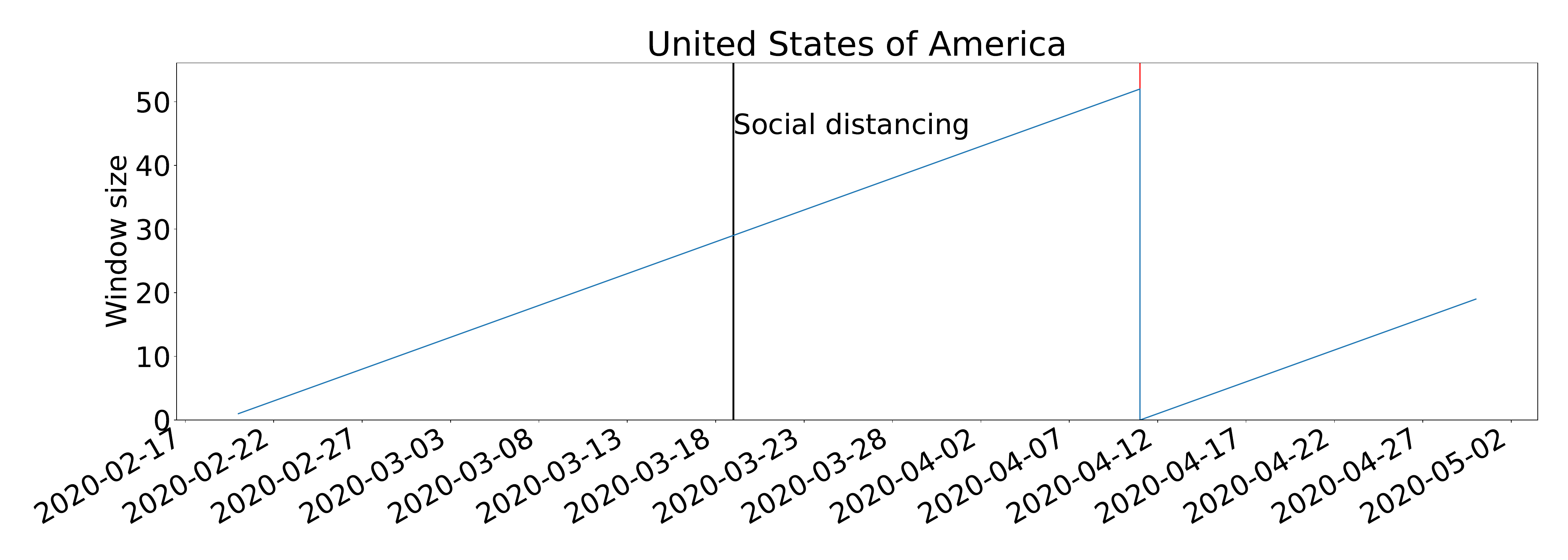} \\
			\vspace{-0.35cm}
			\textbf{d} & \includegraphics[keepaspectratio, height=3.3cm, valign=T]
			{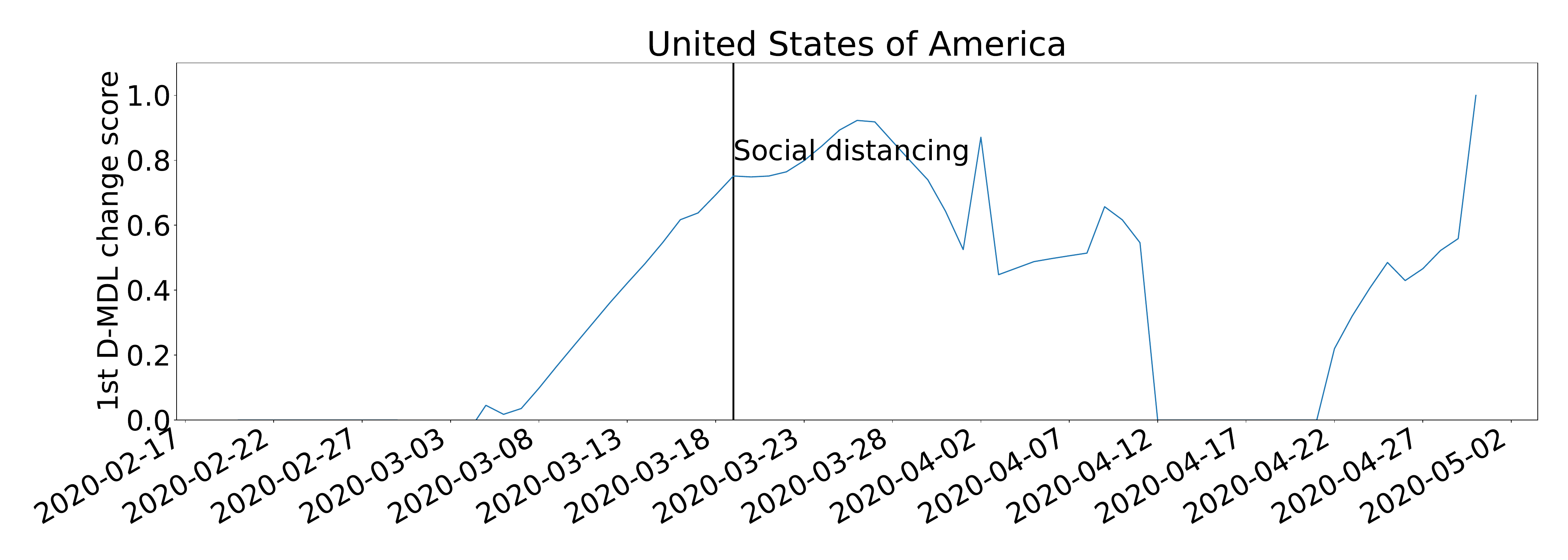} \\
			\vspace{-0.35cm}
			\textbf{e} & \includegraphics[keepaspectratio, height=3.3cm, valign=T]
			{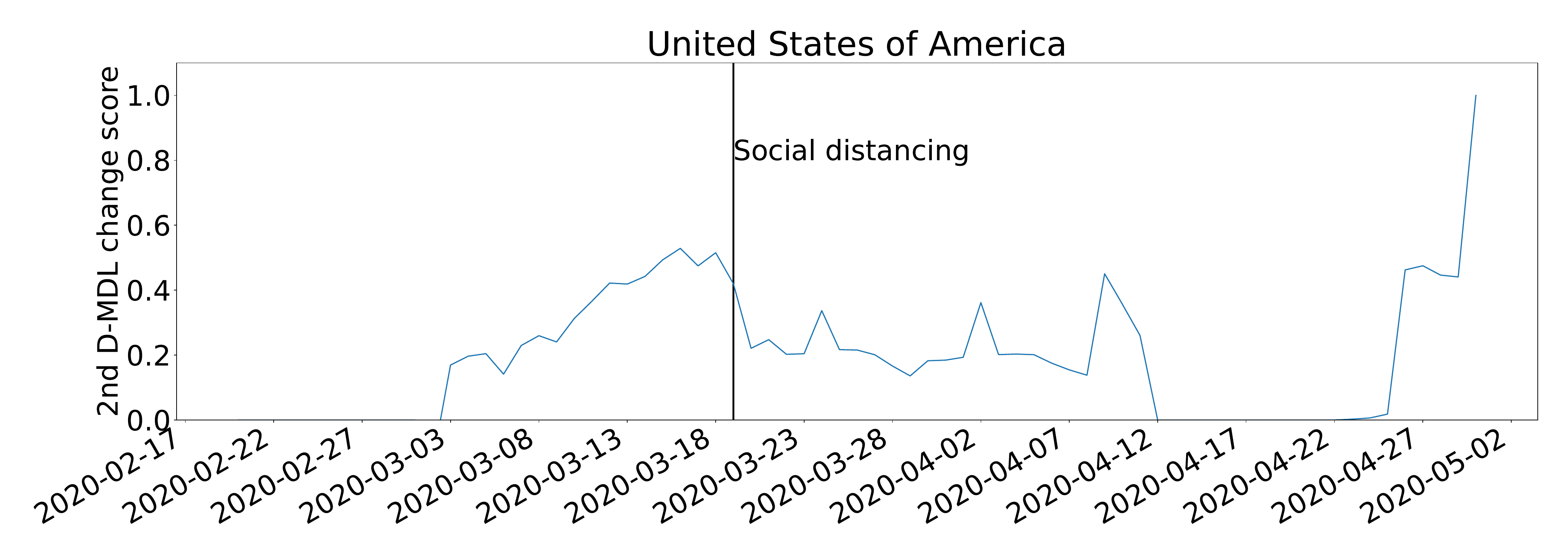} \\
		\end{tabular}
			\caption{\textbf{The results for the United States of America with exponential modeling.} The date on which the social distancing was implemented is marked by a solid line in black. \textbf{a,} the number of cumulative cases. \textbf{b,} the change scores produced by the 0th M-DML where the line in blue denotes values of scores and dashed lines in red mark alarms. \textbf{c,} the window sized for the sequential D-DML algorithm with adaptive window where lines in red mark the shrinkage of windows. \textbf{d,} the change scores produced by the 1st D-MDL. \textbf{e,} the change scores produced by the 2nd D-MDL.}
\end{figure}